\documentclass[a4paper,11pt,leqno]{amsbook}
\pdfoutput=1
\usepackage{a4wide,color,array,tikz}

\usepackage{mathtools}
\usepackage{hyperref}

\usepackage[usenames,dvipsnames]{pstricks}
\usepackage{epsfig}

\usepackage[dvipsnames]{xcolor}

\colorlet{monbleu}{black!40!blue}

\colorlet{bleufonce}{black!80!blue}

  \colorlet{monrouge}{black!40!magenta}

\usepackage{cite}

\usepackage{hyperref}
\usepackage[capitalise]{cleveref}

\hypersetup{linktocpage,colorlinks, linkcolor = {monbleu}, citecolor={monrouge}}

\usepackage[bottom]{footmisc}
\crefformat{equation}{(#2#1#3)}

\usepackage{esint,amsmath,amssymb,amsthm}

\usepackage[english]{babel}
\usepackage[utf8]{inputenc}
\usepackage[cyr]{aeguill}

\usepackage{xargs}

\usepackage{lmodern}
\usepackage{bm}

\usepackage{accents}

\newcommand{\Q}{\mathbb{Q}}
\newcommand{\R}{\mathbb{R}}

\newtheorem{theo}{Theorem}[chapter]
\newtheorem*{theo*}{Theorem}
\newtheorem{prop}{Proposition}[chapter]
\newtheorem{lem}[prop]{Lemma}
\newtheorem{coro}[prop]{Corollary}
\newtheorem{remark}[prop]{Remark}
\newtheorem{defi}[prop]{Definition}

\theoremstyle{plain}

\numberwithin{equation}{section}

\numberwithin{equation}{chapter}
\numberwithin{section}{chapter}
\numberwithin{figure}{chapter}
\renewcommand{\theequation}{\arabic{chapter}.\arabic{section}.\arabic{equation}}

\def\t0{\rightarrow 0} 
\def\ti{\rightarrow \infini} 

\newcommand{\f}{\frac}
\newcommand{\infini}{\infty}
\newcommand{\ep}{\varepsilon}

\newcommand{\hal}{\frac{1}{2}}

\newcommand{\supp}{\mathrm{supp }} 
\def\div{\mathrm{div} \, } 
\def\1{\mathbf{1}} 

\def \mc{\mathcal }
\def \ep{\varepsilon}
\def\ux{X}
\def\Cov{\mathrm{Cov}}

\def\meseq{\mu_{V}} 
\def \ZNbeta{Z_{N,\beta}} 

\def\T{\mathbb{T}}
\def\({\left(}
\def\){\right)}
\def\Error{\mathsf{Error}}
\def\Term{\mathrm{Term}}

\def\yg{|y|^\gamma}
\def \W{\mathbb{W}} 

\def\config{\mathcal{X}} 
\def\Elec{\mathrm{Elec}}
\def\Loc{\mathrm{Loc}}

\def \probas{\mathcal{P}}
\def \Pelec{P^{\mathrm{elec}}} 

\def\P{\mathbb{P}} 
\def \Pst{P} 
\def \bPst{\bar{P}} 

\def \PNbeta{\P_{N, \beta}} 
\def \PgN2{\mathbf{P}_{N,2}} 

\def \HN{\mathcal{H}_N}

\def\Esp{\mathbb{E}} 

\def \ERS{\mathsf{ent}} 
\def \bERS{\overline{\mathsf{ent}}} 

\def \Poisson{{\Pi}}
\def\L{\ell}

\def \F{\mathcal{F}} 

\def \conf{\mathrm{Conf}} 

\def \K{\mathcal{K}}

\def\Ani{\mathsf{A}}

\def \dist{d}
\def\l{\ell}

\def \dist{\mathrm{dist}}

\def\muv{\meseq}

\def\I{\mathcal I}

\def\nab{\nabla}

\def\indic{\mathbf{1}}
\def\vp{\varphi}

\def \Var{\mathrm{Var}}

\def \XN{X_N}
\def \YN{Y_N}
\def \L{\mathsf{L}}
\def\K{\mathsf{K}}
\def\F{\mathsf{F}}
\def\FN{\F_N}

\def\G{\mathsf{G}}
\def \Error{\mathsf{Error}}
\def\meseq{\mu_V}
\def\muv{\mu_V}
\def\mut{\mu_{\theta}}

\def\rr{\mathsf{r}}
\def\rrc{\tilde{\mathsf{r}}}
\def\rrh{\hat{\mathsf{r}}}
\def\mn{{\bar{\mathrm{n}}}}

\def\Xint#1{\mathchoice
   {\XXint\displaystyle\textstyle{#1}}%
   {\XXint\textstyle\scriptstyle{#1}}%
   {\XXint\scriptstyle\scriptscriptstyle{#1}}%
   {\XXint\scriptscriptstyle\scriptscriptstyle{#1}}%
   \!\int}
\def\XXint#1#2#3{{\setbox0=\hbox{$#1{#2#3}{\int}$}
     \vcenter{\hbox{$#2#3$}}\kern-.5\wd0}}
\def\dashint{\Xint-}

\def \carr{\square} 

\def \Pelec{\Pst^{\mathrm{e}}}
\def \bPelec{\bPst^{\mathrm{e}}}

\def \Old{\mathcal{O}}
\def \New{\mathcal{N}}

\def \Escr{E^{\rm{scr}}}

\def \Pelec{P^{\mathrm{e}}} 

\def\P{\mathbb{P}} 
\def \Pst{P} 
\def \bPst{\bar{P}} 

\def \PNbeta{\P_{N, \beta}} 
\def \PgN2{\mathbf{P}_{N,2}} 

\def\g{\mathsf{g}}
\def \V{V}
\def \I{\mathcal{E}}

\def \emp{\widehat{\mu_N}}
\def\YN{Y_N}

\def \bEmp{\bar{P}}
\def \C{\mathcal{C}}
\def \bP{\bar{P}}

\def\nab{\nabla}

\def\pa{{\partial}}

\def\eps{\varepsilon}
\def\ep{\varepsilon}

\def\vp{\varphi}
\def\ro{\rho}
\def\loc{\mathrm{loc}}

\def\Fluct{\mathrm{Fluct}}

\def\hal{\frac{1}{2}}

\makeatletter
\def\namedlabel#1#2{\begingroup
    #2%
    \def\@currentlabel{#2}%
    \phantomsection\label{#1}\endgroup
}

\makeatother
\def \k{\mathsf{k}}
\def \d{\mathsf{d}}
\def \s{\mathsf{s}}

\def \f{\mathsf{f}}
\def\fae{\f_{\alpha,\eta}}
\def \p{\partial}

\def \cds{\mathsf{c}_{\d,\s}} 

\def\Rd{\R^\d} 

\def \drd{\delta_{\Rd}}
\def \ent{\mathrm{ent}}
\def \bPstx{\bar{P}^{x}}

\def\cd{\mathsf{c}_{\d}}

\def\Esp{\mathbb{E}} 

\def\x{\mathsf{x}}

\def \be{\begin{equation}}
\def \ee{\end{equation}}
\def \beq*{\begin{equation*}}
\def \eeq*{\end{equation*}}
\def \ba{\begin{eqnarray}}
\def \ea{\end{eqnarray}}
\def \ba*{\begin{eqnarray*}}
\def \ea*{\end{eqnarray*}}
\def\mueqt{\mu_{V_t}}

\def\mub{\mu_\theta}
 \def\fae{\f_{\alpha_i,\eta_i}}
\def\faej{\f_{\alpha_j, \eta_j}}

\def\N{{n_\mathcal{O}}}

\def\mf{f_\d}
\def\rb{\rho_\beta}
\def\id{I}

\def\nut{\nu_\theta^t}
\def\omc{{\overset{\circ}{\Omega}}}
\def\di{d}

\theoremstyle{definition}
\newtheorem{example}[prop]{Example}
\newtheorem{rem}[prop]{Remark}
\newtheorem{defini}[prop]{Definition}

\numberwithin{figure}{chapter}

\newtheorem{conjecture}{Conjecture}[chapter]

\def\veta{\vec{\eta}}

\newcommand{\interieur}{\overset{\circ}}
\renewcommand{\div}{\mathrm{div}\,}

\newcommand{\dd}{\mathrm{d}}

\renewcommand{\V}{{V}}
\newcommand{\ObN}{\mathcal{O}_N}

\newcommand{\Cloc}{\mathcal{C}^{\mathrm{loc}}}

\makeindex

\begin{document}

\title{Lectures on Coulomb and Riesz Gases}

\author{\sc Sylvia Serfaty}

\maketitle

\tableofcontents

\chapter*{Preface}

This book grew out of notes written for graduate courses that I taught at the Courant Institute, at Ecole Normale Sup\'erieure invited by  the Fondation Sciences Math\'ematiques de Paris, and at the 2024  Saint-Flour Probability summer school.  They were meant to reflect the advances made since the previous lecture notes \cite{noteszurich}. 

The goal of this book is not to provide an exhaustive view of the topic but rather a necessarily biased  but  self-contained presentation  of  the approach to Coulomb gases that has emerged from a body of work initiated in collaboration with Etienne Sandier and continued with Nicolas Rougerie, Mircea Petrache, Thomas Lebl\'e, Simona Rota Nodari, Scott Armstrong and Luke Peilen; and which one may characterize as an electric-formulation-based analysis of the statistical mechanics of Coulomb and Riesz gases.

  The book starts by reviewing some standard notions and facts before moving on to the more recent research. It is meant to serve both as a text for researchers interested in learning about the topic, and as a point of reference collecting the various results in one place.  It introduces and analyzes  the main  concepts used in this approach: the modulated energy, the electric formulation, the screening procedure, the renormalized jellium energy,  the transport method, with streamlined and updated definitions and results. For instance all cases of $ \s\in [\d-2, \d)$ are treated, including when $\s \le 0$ and the Coulomb case in dimension one.
  I tried to minimize the assumptions as much as possible and to allow for the broadest temperature regimes possible.
   The text focuses mostly on the analysis of the canonical Gibbs measure of Coulomb and Riesz gases in an external potential, but takes a detour to discuss the application of the tools to the modulated energy method for mean-field limits of the dynamics of such gases. 
I have chosen to present results for both the Coulomb and Riesz cases whenever treating the Riesz case did not add too much complexity, and to restrict to the Coulomb case and refer the reader to the relevant papers when it did, that is, in all instances where the screening procedure needs to be used.

I thank the students who followed my courses for their feedback. I owe much gratitude to Thomas Lebl\'e for his invaluable help all along this project. Many thanks  to  Sungsoo Byun,  Antonin Chodron de Courcel,   Luke Peilen, 
Matt Rosenzweig and Eric Thoma for their careful reading and feedback. Thanks also to Peter Forrester, Yacin Ameur and Paul Bourgade for help with references.

This project was supported by the Simons Foundation through the Simons Investigator program,  by NSF grants DMS-2000205 and DMS-2247846 and by the Fondation Sciences Math\'ematiques de Paris.

\bigskip
\bigskip

\begin{flushright}
Sylvia Serfaty \\
New York, July 2024
\end{flushright}

\chapter{Introduction}
\label{chap:intro}
\label{chap:introduction}

\section{Setting: Coulomb, logarithmic and Riesz cases}

Let $N \geq 1$ and let $\XN$ denote an $N$-tuple of points $(x_1, \dots, x_N)$ in $(\R^\d)^N$, where $\d \geq 1$ is the dimension. We are interested in energy functionals $\HN : (\R^\d)^N \to (-\infty, + \infty]$ of the form:
\begin{equation} \label{HN}
\HN(\XN) := \hal\sum_{1 \leq i \neq j \leq N} \g\left(x_i-x_j\right) +  N\sum_{i=1}^N   \V(x_i),
\end{equation}
where $\g : \R^\d \to (-\infty, + \infty]$ is called the \textit{pair interaction potential}, and $\V : \R^\d \to (-\infty, + \infty)$ is called the \textit{external field} or \textit{confining potential}.

In the cases that we study, $\g$ is given by
\begin{equation}\label{rieszgene}
\g(x)= \begin{cases}  \frac{1}{\s} |x|^{-\s} \quad  & \s \neq 0\\
-\log |x| & \s=0,
\end{cases}\end{equation}
and  we call the first case the Riesz case and the second the logarithmic case. The latter can be obtained as the formal $\s \to 0$ limit of the general Riesz case by noting that $-\log |x|=\lim_{\s\to 0} \(  \frac{1}{\s} |x|^{-\s} -1\)$.  Whenever the parameter $\s$ appears, according to the definition \eqref{rieszgene} it will be with the convention that $\s = 0$ in the logarithmic cases. 

We are particularly interested in the case  $\s\ge 0$ where $\g$ is singular, but can handle some instances of  $\s \le 0$ as well.  The text will focus on the more specific regime 
\be\label{intervalles} \d-2\le \s<\d\ee in all dimensions, which includes the important Coulomb case 
\be\label{coulomb} 
\s=\d-2.\ee

When $\s<\d$, the interaction kernel $\g$ is integrable near $0$, making this the {\it potential case}, for which mean-field theory and potential theory can be applied (see Chapters \ref{chap:eqmeasure} and \ref{chap:leadingorder}). The case $\s\ge\d$, where $\g$ is not integrable near the origin, is called the {\it hypersingular case}. It is not amenable to potential theory, and  behaves much more like a short-range interaction problem. We will not at all discuss that case but refer instead to \cite{borodachovlivre,hlss} for instance.

Let us underline that the interaction is in all cases {\it purely repulsive}, which explains the need for the confining potential $\V$, on which  we will make precise assumptions in the next chapters. In short, we take it fairly smooth and growing sufficiently fast at infinity.
\smallskip

Let $\beta$ be a positive real number called the \textit{inverse temperature}, which may depend on $N$. We let $\PNbeta$ be the probability measure on $(\R^\d)^N$ whose density with respect to the standard Lebesgue measure $d \XN := d x_1 \dots d x_N$ is given by:
\begin{equation}
\label{gibbs}
\dd \PNbeta(\XN) := \frac{1}{\ZNbeta} \exp\left(-\beta  N^{-\frac\s\d}\HN(\XN) \right) d \XN,
\end{equation}
where $\ZNbeta$ is a normalizing constant called the \textit{partition function}:
\begin{equation}
\label{def:ZNbetN}
\ZNbeta := \int_{\left(\R^\d\right)^N} \exp\left(-\beta  N^{-\frac\s\d} \HN(\XN) \right) d \XN.
\end{equation}
The measure $\PNbeta$ is called the \textit{canonical Gibbs measure} \index{Gibbs measure} associated to the energy $\HN$ at inverse temperature $\beta$. The factor $N^{-\frac\s\d}$ is a convenient scaling choice, the reason for which will appear later, but it  does not reduce generality, since $\beta$ may itself  depend on $N$. 

Most of our study is focused on understanding the typical and atypical behavior of particles $\XN$ when randomly distributed
according to $\PNbeta$, or on the deterministic behavior of $\XN$ minimizing $\HN$ which formally corresponds to taking $\beta=+\infty$.

It is also of interest to study  evolutions, in particular the SDE system
\begin{equation}\label{noise1}
d x_i = -\frac{1}{N} \nab_i \HN(x_1, \dots, x_N) dt +\sqrt{\frac{2}{\beta N^{1-\frac\s\d}}}  dW^t_i\end{equation} with $W^t_i$ independent Brownian motions, which is the  overdamped Langevin / Glauber dynamics for \eqref{gibbs}. The measure \eqref{gibbs} can be seen as the invariant measure for this dynamics. Although this is largely open, understanding the features of the evolution  \eqref{noise1} and its convergence to the equilibrium state \eqref{gibbs} is an important statistical mechanics problem of interest in its own right, and can also provide information on the Riesz gas itself. 

Other dynamics are possible and also physically very interesting. A first one is the class of conservative dynamics of the form  
\be\label{noise2}
d x_i = \frac{1}{N}\mathbb{J} \nab_i \HN(x_1, \dots, x_N) dt + \sqrt{\frac{2}{\beta N^{1-\frac\s\d}}} dW_i^t \ee
 where  $\mathbb{J}$ is an antisymmetric matrix, and another is the class of second-order evolution  according to Newton's law 
\be \label{noise3}
d x_i = v_i dt, \qquad dv_i=  -\frac{1}{N} \nab_i \HN(x_1, \dots, x_N) dt +  \sqrt{\frac{2}{\beta N^{1-\frac\s\d}}}  dW_i^t.\ee
Dynamics  will be discussed in Chapter \ref{chap:commutator}.

\paragraph{{\bf Coulomb case.}}
The choice \eqref{coulomb} for the pair interaction potential $\g$ is called the \textit{Coulomb case}, because  $\g$ is then (up to a multiplicative constant) the Coulomb kernel, i.e.~the fundamental solution to the Laplace operator, solving 
\begin{equation}\label{coulombkernel}
-\Delta \g = \cd \delta_0
\end{equation}
where $\delta_0$ is the Dirac mass at the origin and $\cd$ is an explicit constant depending on the dimension, given by: \footnote{Here $\mathbb{S}^{\d-1}$ is the unit sphere in $\R^\d$ and $|\mathbb{S}^{\d-1}|$ denotes its volume for the standard Lebesgue measure.} 
\be
\label{defcd}
\cd= 2\pi \quad\text{if} \ \d =2 \qquad \cd= |\mathbb{S}^{\d-1}| \quad \text{for }\ \d \ge 3.
\ee 
Coulomb interactions are ubiquitous in physics, most notably as the classical electrostatic interaction potential between charged particles. The one-dimensional Coulomb interaction with kernel $\g(x) = -|x|$  is the most ``explicitly solvable", hence the best understood, see \cite{len1,len2,baxter,kunz,brascamplieb,aizm}, however we will still provide new results for this case.

\paragraph{{\bf Logarithmic case.}}
The choice $\s=0$ or $\g (x)=-\log |x|$  for the pair interaction potential is called the \textit{logarithmic case}. It is  important in random matrix theory and several physics models, as we will see below. Note that in the logarithmic case, if two particles $x,y$ are sent to infinity in opposite directions, their pair interaction $\g(x-y)$ tends to $-\infty$. Thus being far away from the origin needs to be penalized by the \textit{confining} potential $\V$, otherwise the integral \eqref{def:ZNbetN} defining $\ZNbeta$ may not converge.
A one-dimensional log gas for arbitrary  $\beta>0$ and general confinement $V$  is also called a {\it  $\beta$-ensemble}.  \index{$\beta$-ensembles} Among all the Coulomb and Riesz gases, the $\beta$-ensembles have probably been the most extensively studied in the mathematical literature. They are also the most ubiquitous, as one encounters them in random matrix theory, quantum mechanics, self-avoiding random walks, random tilings, and even proofs of functional inequalities (see  \cite{dadounzitt} for an example)!

\paragraph{{\bf Riesz cases.}}
The general choices $\d-2\le \s<\d$  in \eqref{rieszgene} are called the \textit{Riesz cases} and the Coulomb case can be considered a  special instance of them. 
As seen above, the logarithmic cases $\s=0$ can be thought of as the  $\s \to 0$ limit of Riesz cases, so by extension we will consider them as included in Riesz cases. 
In the Riesz case with $\d-2<\s<\d$, instead of \eqref{coulombkernel}, $\g$ is known to be the kernel of a {\it fractional Laplacian} operator, in the sense that 
\be\label{fractlapkernel}
(-\Delta)^{\frac{\d-\s}{2}}  \g= \cds \delta_0\ee
for some normalization constant $\cds$ given by 
\be \cds=\begin{cases} 
\frac{2^{\d-\s} \pi^{\d/2} \Gamma (\frac{\d-\s}{2})}{\Gamma(\frac{\s}{2}) }
 & \text{for } \ \s>\max(0, \d-2)\\
\frac{2 \pi^{\d/2} }{\Gamma(\frac{\d}{2})}=|\mathbb{S}^{\d-1}|& \text{if} \ \s=\d-2>0\\
2\pi  & \text{if}\  \s=0, \d=1 \ \text{or } \d=2,\end{cases}\ee see \cite{gelfandsh} for the basis of the computation.
 The fractional Laplacian is  
 a nonlocal operator, it  can be defined via Fourier multipliers or in real space by  
\be \label{deffraclap}(-\Delta)^\alpha f(x) = C_{\d, \alpha} \int_{\R^\d} \frac{f(x)-f(y)}{|x-y|^{\d+2\alpha}} dy,\quad \alpha \in (0,1),\ee
see for instance  \cite{K17}. 
This property of $\g$  will play an important role for us. Note that it is only true in the super-Coulombic range $\s\in [\d-2,\d)$, which is the main reason for our focusing on this regime.

\section{Motivation}
\label{sec1}
\subsection{Statistical and quantum mechanics}\label{sec1.1}
Classical statistical mechanics views  large physical systems of interacting deterministic particles as a random object. If $X \mapsto \mathcal H(X)$ is the interaction energy of the system in a state $X$ and if the ambient temperature is  $\beta^{-1}$ then at equilibrium the probability of observing a given state $X$ is proportional to the Boltzmann factor $\exp\left( - \beta \mathcal H(X) \right)$. This justifies the introduction of the Gibbs measure \eqref{gibbs} from a statistical physics point of view, in order to understand the statistical properties of a  hypothetical system of particles in $\R^\d$ with interaction energy given by \eqref{HN}.  Grand canonical ensembles, i.e.,~where the number of points $N$ is not fixed but also part of the variables, are also considered both in general and in the particular instance of Coulomb and Riesz gases, we refer   to the survey \cite{lewinsurvey}. The canonical ensemble is usually considered more difficult to study than the grand canonical one. 

 Most of the interactions considered here are \textit{singular} at the origin, and  particles  live in $\R^\d$ and not only on a lattice, which would ensure a minimal particle separation. Moreover, in  the  Riesz cases with $\s < \d$ that we consider, the pair interaction is considered \textit{long-range} because it decays slower than $|x|^{-\d}$ at infinity, which gives rise to specific phenomena, see e.g. \cite{bbdr} and references therein. These are the main difficulties of these models.

\paragraph{{\bf Coulomb gas.}} The most physical case is, of course, 
the Coulomb case, since Coulomb is the fundamental electrostatic interaction, with the three-dimensional situation being the most natural. The ensemble given by \eqref{gibbs} in the case \eqref{coulomb} is called a \textit{Coulomb gas} or \textit{one-component plasma} (which refers to the fact that there is only one type of charges, i.e.  positive charges). With a neutralizing background, it is called \textit{jellium}.   Closely related to the jellium is the Uniform Electron Gas (UEG) model which is defined rather via a density constraint, or imposing the one-point marginal. Relations between the jellium and the UEG have been explored in particular recently in  \cite{lls,lauritsen}.

The Coulomb gas   can be seen as a toy model for classical matter, ignoring quantum effects.  
For instance, Gamov's ``liquid drop model'' for the atomic nucleus  (see \cite{choksimuratovreview}) is also a simplified model for electrons and atoms, and in some regime where one phase is in large majority can be reduced to a system of points interacting like \eqref{HN} (see  \cite{ACO,gms,gms2} and references therein).

The Coulomb gas is thus a classical ensemble of statistical mechanics and has been well studied since the 70s, see e.g. \cite{alastueyjancovici,jlm,janco,alastueyAP,sm,PenSm,jlm,LiLe1,LN,Frohlich,FS1,fs2,kiessling,kiesslingspohn}. The quantum Coulomb gas has also been the object of much attention, and some of the techniques developed there also apply to the classical setting \cite{grafschenker,hainzllewinsolovej,lls}. Density functional theory also involves the Coulomb interaction in a crucial way \cite{lieboxford,liebseiringer,lls,lls2,cotarfriesecke,cotarpetrache}, in particular through the ``indirect Coulomb energy" and its bounds via Lieb-Oxford's inequality and $N$-marginal optimal transport with Coulomb costs.

An important instance of  the Coulomb case is  the two-dimensional Coulomb gas, which coincides with the two-dimensional logarithmic case, see e.g. \cite{dyson,mehta,martin,martinyalcin}.  It is also called in the physics literature  \textit{log gas}, \textit{two-dimensional one-component plasma} (which  gets abbreviated as 2DOCP), \textit{two-dimensional jellium}, or \textit{Dyson gas}. 
The dynamics \eqref{noise1} (with proper scaling of $\beta$) is in this two-dimensional logarithmic case called \textit{Dyson Brownian motion}\index{Dyson Brownian motion}.
The 2DOCP is deemed interesting as a natural toy model to do statistical physics in two dimensions in a singular, long-range setting, as a reasonable model for plasmas in astrophysics (in particular \eqref{noise3}) and due to its connection with quantum mechanics, the fractional quantum Hall effect\index{fractional quantum Hall effect} and random matrices, as we will see below.  

We also refer the reader to the very recent physics book  \cite{alastueymartin}, which addresses all statistical mechanics aspects (exact results, correlations, phase transitions) of classical and quantum Coulomb gases.

\paragraph{{\bf Riesz gases.}} The  Riesz case can be seen as a generalization of the Coulomb case. Motivations for studying Riesz gases are numerous in the physics literature (in solid state physics, ferrofluids, elasticity), see for instance \cite{mazars,bbdr,CDR,CDFR,torquato}. For integer $\s < \d$, the pair interaction 
Riesz systems can also be seen as systems with Coulomb interaction constrained to a lower-dimensional subspace, take for example $\d = 2$ and $\s = 1$: one gets the usual three-dimensional Coulomb interaction, only restricted to the two-dimensional plane.    The jellium and Uniform Electron gas have  also been studied for Riesz interactions \cite{lls,cotarpetrache}.
We note that even one-dimensional Riesz systems, possibly with nonsingular repulsion $\s\le 0$, are currently attracting attention \cite{schehr1d,schehr1d2,schehr1d3,schehrcoulomb1d}.
We also refer to the very nice recent survey \cite{lewinsurvey} with many open questions.

\paragraph{{\bf Link with wave functions.}}
From the physics point of view, another motivation for studying the probability measure $\PNbeta$ is that  in several  cases with logarithmic interaction, it happens to be the square of the wave function of certain  quantum systems. Examples corresponding to the one-dimensional logarithmic case are the Tonks-Girardeau model of impenetrable  bosons  \cite{gwt,ffgw}, the Calogero-Sutherland  quantum many-body Hamiltonian  \cite{forresterjm,forrester} and finally the density of the many-body wave function of  non-interacting fermions in a harmonic trap  \cite{ddms,kulkarni}.

Examples in the two-dimensional logarithmic case are  free fermions in a magnetic field in  the lowest Landau level, 
and the Laughlin wave function for the fractional quantum Hall effect\index{fractional quantum Hall effect}~\cite{girvin,stormer}, which is the Gibbs measure \eqref{gibbs} for a 2DOCP. This is called the ``plasma analogy'' in the physics literature, see \cite{laughlin1,laughlin2,laughlin3} and \cite{Rougerie-Elliott,Rougerie-Ency} for an introduction. For recent mathematical progress using this correspondence, see \cite{RSY2,rougerieyngvason,lry}. Moreover, for a certain choice of the inverse temperature ($\beta = 2$), the same measure also arises as the (square of the) wave function of the ground state for $N$ non-interacting fermions confined to a plane with a perpendicular magnetic field, see \cite[Chap. 15]{forrester}.


\subsection{Random matrix theory and related questions}
\label{sec:RMT}
\index{random matrices}
The origins of Random Matrix Theory (RMT) trace back to Hurwitz, with later foundational works by statisticians (Wishart) to understand sample covariance matrices  and physicists (Wigner, Dyson) to understand the spectrum of large atoms and for mathematical curiosity (Ginibre). We refer to \cite{diaconishurwitz} for an historical perspective and extensive references, and to \cite{mehta,agz} for a mathematical introduction. In short, the aim of RMT is to understand the eigenvalues and eigenvectors of (large) matrices drawn at random for certain distributions called \textit{ensembles} or \textit{models}, and it has been the object of a vast mathematical literature.

As noticed early on, for certain natural random matrix ensembles the joint law of the eigenvalues can be computed explicitely and happens to coincide with a Gibbs measure of the form \eqref{gibbs} where $\g$ is chosen to be the logarithmic interaction $\s=0$. In particular,  choosing $\V : x \mapsto \frac{1}{2} |x|^2$ as an external field, the measure \eqref{gibbs} on $\R^N$ or $(\R^2)^N$ coincides in that case  with the joint law of eigenvalues of several important random matrix ensembles:
\begin{description}
\item [GUE] With $\beta = 2$, one gets the law of eigenvalues of an $N\times N$ \textit{Hermitian} matrix with complex Gaussian entries. This distribution of random matrices is called the Gaussian Unitary Ensemble\index{Gaussian unitary ensemble} or GUE.

\item [GOE] With $\beta = 1$, one gets the law of eigenvalues of an $N\times N$ real \textit{symmetric} matrix with real Gaussian entries. This distribution of random matrices is called  the Gaussian Orthogonal Ensemble\index{Gaussian orthogonal ensemble} or GOE.

\item [GSE] With $\beta = 4$, one gets the law of eigenvalues of an $N\times N$ self-dual matrix with quaternionic Gaussian entries. This distribution of random matrices is called  the Gaussian Symplectic Ensemble or GSE.

\item [Tridiagonal models for arbitrary $\beta$] 
In fact, as was realized much later, for every choice  $\beta>0$ (and still $V(x)=x^2$)  there exists a random matrix ensemble whose eigenvalues are distributed on the real line according to the law $\PNbeta$ for $\s=0$. Dumitriu-Edelman and Killip-Nenciu \cite{de,killipnenciu} construct such models, that are sometimes referred to as the \textit{tridiagonal} (or \textit{pentadiagonal} in the case of \cite{killipnenciu}) models.   

\item[Ginibre ensemble] An important addition coming from non-Hermitian random matrix theory is the law of complex eigenvalues of an $N\times N$ matrix where the entries are chosen to be i.i.d. complex Gaussian variables with no symmetry imposed. This law can again be explicitly computed and coincides with \eqref{gibbs} in the two-dimensional logarithmic case $\s=0$, choosing $\V (x)=  |x|^2$ as a confining potential and taking $\beta = 2$. This is called the Ginibre ensemble, after Jean Ginibre, see \cite{ginibre}.  Its eigenvalue statistics arise in quantum chaos \cite{byunforrester}. Quaternionic and real variants are also discussed in \cite{byunforrester}.

\end{description}
We refer to \cite{forrester} for an extensive discussion of this connection
 between random eigenvalues and random particles with logarithmic interaction, which  is  particularly strong in the one-dimensional  case.  
  
This  connection is  among the reasons why the logarithmic case $\s=0$, in particular the one-dimensional one, are by far the most intensively studied among Riesz gases. 
The random matrix correspondence 
provides a physical intuition as well as tools and methods from mathematical physics in order to study the eigenvalues drawn from certain classical ensembles of random matrices. For example, one may readily give a physical interpretation for a well-known phenomenon called ``repulsion of eigenvalues'': eigenvalues are less likely to be close to each other than if they were drawn independently at random, in fact after mapping them onto particles interacting through a repulsive logarithmic potential one may say that they repel each other \textit{logarithmically}.  Conversely, the random matrix models provide access to computing explicitly certain quantities for the log gas in one and two dimensions, see below.

\paragraph{{\bf Determinantal case.}} \index{determinantal point process}
In both the one- and two-dimensional logarithmic cases at the specific temperature\footnote{For $\s=0$ at $\beta = 1$ and $4$, which correspond to the GOE and GSE models mentioned above, there is another specific algebraic structure called \textit{pfaffian process}, for which exact computations are still possible in principle, but less tractable.} $\beta=2$, the Gibbs measure $\PNbeta$ acquires a special algebraic feature, which can be seen by rewriting it as 
\be\label{PNbetadet} d\mathbb{P}_{N,2}(\XN) = \frac{1}{Z_{N,2}} \(\prod_{i<j} |x_i-x_j|\)^2  e^{- 2N \sum_{i=1}^N V(x_i)} d\XN\ee
with $\prod_{i<j} |x_i-x_j|$  equal to the Vandermonde determinant\index{Vandermonde determinant} of the points $x_1, \dots, x_N$.
 This makes the  log gas ensemble in that particular instance belong to the class of   \textit{determinantal} point processes.  An important consequence is  that all of its correlation functions can be obtained explicitly by computing  certain determinants, which allows to give very precise, exact answers to many questions through algebraic computations, see \cite{forrester,byunforrester}.  Again, this is the reason why the log gas with $\beta=2$ is the best understood of all, as we will see with many examples of results. 
 The rewriting \eqref{PNbetadet} 
  also makes the $\beta=2$ log gases a particular instance of orthogonal polynomial ensembles, which are Gibbs measures on $\R^N$  of the form 
  $$ \frac{1}{Z_N} \(\prod_{i<j} |x_i-x_j|\)^2  \prod_{i=1}^N d\mu(x_i), $$
  and form another well-developed field of study \cite{koenig}. 
 We refer to \cite{hkpv,borodin} for more on  determinantal point processes. In the physics literature, one sometimes speaks of systems that are \textit{exactly solvable}.

Understanding the behavior of eigenvalues in certain ensembles can, for some questions, be sufficient because one expects many properties to be \textit{universal} accross a broad family of random matrix models (see for instance \cite{tv2,erdosyaupeche}). Universality has been well understood in determinantal cases $\beta=2$,  see \cite{deiftcpam,joha,deiftg2,ahm} and  the survey \cite{kuijlaarssurvey}.

\paragraph{{\bf Self-avoiding walks in probability and random tilings.}}
Motivated by statistical physics, the analysis of self-avoiding walks and random tilings has been a very active field of probability and integrable probability. At the heart one encounters again  points with  logarithmic-like  repulsion and a discretized log gas on the real line at
general temperature $\beta$ 
\cite{borodingoringuionnet,gorinbook}, leading to similar limit point processes, questions and results as those encountered for the log gas. For reference on these topics as well as a general introduction to the field of integrable probability, see \cite{borodingorin}.

For more on the aspects mentioned in Sections \ref{sec1.1} and \ref{sec:RMT}, we also refer to the very nice recent survey \cite{chafaisurvey}.

\subsection{Vortices in condensed matter physics and fluids}

\begin{figure}[ht!]
\begin{center}
\includegraphics[width=0.3\textwidth]{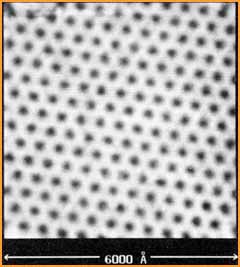}
\caption{Abrikosov lattice, H. F. Hess et al. Bell Labs
{\it Phys. Rev. Lett.} 62, 214 (1989)}
\label{fig32}

\end{center}

\end{figure}

In superconductors with applied magnetic fields, in rotating  superfluids and in Bose-Einstein condensates, one observes  the occurrence of quantized {\it vortices}, which are local point defects of superconductivity or superfluidity, surrounded by a current loop. The vortices repel each other, while being confined together by the effect of the magnetic field or rotation,  and the result of the competition between  these two effects is that, as predicted by Abrikosov \cite{a}, they arrange themselves in a particular {\it triangular lattice} pattern, called {\it Abrikosov lattice}\index{Abrikosov lattice},~cf. Fig. 
\ref{fig32} (for more pictures, see {\tt www.fys.uio.no/super/vortex/}).
Superconductors and superfluids are modelled by the celebrated Ginzburg-Landau energy\index{Ginzburg-Landau energy} \cite{gl}, which in simplified form \footnote{The complete form for superconductivity contains a gauge field, but we omit it here for simplicity.} can be written 
\be \label{gl} 
\int |\nab \psi|^2 + \frac{(1-|\psi|^2)^2}{2\ep^2},\ee
where $\psi$ is a complex-valued unknown function (the ``order parameter'' in physics) and $\ep $ is a small parameter,
and gives rise to the  associated Ginzburg-Landau equation
\be\label{gle}
\Delta \psi+ \frac{1}{\ep^2} \psi(1-|\psi|^2)=0\ee and its dynamical versions, the heat flow 
\be\label{glhf}
\pa_t \psi=\Delta \psi+ \frac{1}{\ep^2} \psi(1-|\psi|^2)\ee
and Schr\"odinger-type flow (called the Gross-Pitaevskii equation in the physics literature)
\be\label{gls}
i\pa_t \psi=\Delta \psi+\frac{1}{\ep^2} \psi(1-|\psi|^2).\ee

When restricting to a two-dimensional situation, it can be shown  rigorously (this was pioneered by \cite{bbh} for \eqref{gl} and extended to the full gauged model \cite{bethuelriviere,livre,ssgl}) that the minimization of \eqref{gl} can be reduced, in terms of the vortices and as $\ep \to 0$, to the minimization of an energy of the form \eqref{HN} in the case $\d=2, \s=0$,  (for a formal derivation, see also \cite{noteszurich}) and  this naturally leads to the question of understanding the connection between minimizers  of \eqref{HN} and the Abrikosov triangular lattice.
Similarly, the dynamics of vortices  under \eqref{glhf} can be formally reduced to the gradient flow of \eqref{HN} which is \eqref{noise1} with $\beta=\infty$, respectively under \eqref{gls} to the Hamiltonian flow associated to \eqref{HN}, \eqref{noise2} with $\beta=\infty$ and $\mathbb{J} $ equal to the matrix of rotation by $\pi/2$. This was established formally for instance in \cite{peresrubinstein,crs,E} and proven for a fixed number of vortices $N$ and  in the limit $\ep \to 0$ in \cite{linvortexdyn,jerrardsonerdyn,collianderjerrard,linxin,linxin2,bethueljerrardsmets}  until the first collision time and in \cite{bos1,bos2,bos3,serfatyjemsdynamics} including after collision.

Vortices also arise in classical fluids, where in contrast with the situation of superconductors and superfluids, their charge is not quantized. In that context the energy \eqref{HN} with $\d=2,\s=0$, is sometimes called the  Kirchhoff energy 
 and the corresponding Hamiltonian system \eqref{noise2} with $\mathbb{J}$ taken to be a rotation by $\pi/2$, known as the point-vortex system, corresponds to the dynamics of idealized vortices in an incompressible fluid whose statistical mechanics analysis was initiated by Onsager, see \cite{eyinksrini}. One of the motivations for studying the gradient flow with additive noise, as in  \eqref{noise2}, is precisely to understand fluid turbulence as he conceived.
  It has thus been quite  studied as such, see \cite{marchioropulvirenti} for further reference. 
  The study of Newton's law \eqref{noise3} with interaction \eqref{HN} is also motivated by plasma physics in which the interaction between ions is Coulombic, see the review  \cite{jabinreview}.

\subsection{Energy minimizers, Fekete points and approximation theory}
\index{energy minimizers}
\paragraph{{\bf Best packings and minimal energy configurations.}}
Finding point configurations that are optimal in some respect is an old, recurrent question in mathematics. One may think of the famous \textit{optimal (sphere) packing problem}: among all possible arrangements of disks, balls, etc.~of fixed radius, which one is the most compact, i.e.,~has the highest density? Since the radius is fixed, only the centers can be chosen, so it is really a problem about point configurations, that can be seen as the $\s\to \infty$ limit of the minimization of the $\s$-Riesz energy $\HN$. In the more general problem of \textit{energy minimization}, one fixes a certain pair potential and asks: among all point configurations of fixed density, which one has the minimal interaction energy? Perhaps surprisingly, these questions are extremely difficult to answer in general, except in dimension 1.  
It is believed that in low enough dimension, many such problems  are minimized by  {\it lattice} point configurations. This is the so-called {\it crystallization conjecture}, we refer to  \cite{blanclewin} for a recent survey.  \index{crystallization}

The solution to the optimal sphere  packing problem is for instance only known  for a handful of dimensions: $\d = 1, 2, 3$ and thanks to very recent progress $\d=8, 24$. This is a special case of the Cohn-Kumar conjecture\index{Cohn-Kumar conjecture} \cite{cohnkumar} (relying on linear programming bounds at the level of Fourier transforms), which asserts that there are some universally minimizing lattices in dimensions $\d=2,8,24$, more precisely the triangular lattice $A_2 $ in dimension 2, the $E_8$ lattice in dimension 8 and the Leech $\Lambda_{24}$ lattice in dimension 24, which minimize not only the sphere packing problem but also all interaction energies which are of the form 
$$\sum_{i,j} f(|x_i-x_j|^2)$$
with $f$ a {\it completely monotonic}\footnote{A function $f$ is said to be completely monotonic when $(-1)^k f^{(k)} \ge 0$ for each $k \geq 0$, an important example $e^{-ct}$ leading to Gaussian interactions. } function.
This conjecture was recently proven for $\d=8, 24$, first for the sphere packing problem in \cite{via,ckrmv}, and then  in \cite{ckrmv2} for all completely monotonic interactions,  and it implies the same result for Coulomb and Riesz interactions in the same dimensions, as shown in \cite{PetSercryst}. The conjecture remains open for $\d=2$. If proven  true, it implies in view of  \cite{PetSercryst}  that points that minimize $\HN$ in the two-dimensional logarithmic or Riesz cases $\d-2\le \s <\d$  arrange themselves along a triangular lattice, the same as the Abrikosov lattice in superconductors.

      In high dimension, where the problem is important for error-correcting codes,  it  is expected that the solution is {\it not} a lattice (in dimension 10 already, the  so-called ``best lattice'', a non-lattice competitor, is known to beat the lattices), see \cite{conwaysloane} for these aspects. We refer to \cite{cohnnotices} for an introduction to this topic, which we will discuss further in Chapter~\ref{chap:renormalized}.


\paragraph{{\bf Fekete points.}}
Fekete points\index{Fekete points} arise in approximation theory as the points minimizing interpolation errors for numerical integration \cite{safftotik}. More precisely, if one is looking for  $N$ interpolation points $\{x_1, \dots, x_N\}$ in some nice compact subset $K \subset \R^\d$ such that the relation
\begin{equation*}
\int_K f(x) dx = \sum_{j=1}^N w_j f(x_j)
\end{equation*}
holds when $f$ is an arbitrary polynomial of degree $\leq N-1$, one needs to compute the coefficients $w_j$ such that $\int_{K} x^k = \sum_{j=1}^N w_j x_j^k$ for $0 \leq k \leq N-1$. This computation turns out to be easy if one knows how to invert the Vandermonde matrix of the $\{x_j\}_{j=1 \dots N}$. The numerical stability of this operation is as large as the \textit{condition number} of the matrix, i.e., as the Vandermonde determinant\index{Vandermonde determinant} of the $x_j$'s. In fact, the $N$-tuple of points that minimize the maximal interpolation error for general functions can be shown to be the Fekete points, defined as those that maximize the Vandermonde determinant\index{Vandermonde determinant} $\prod_{1 \leq i < j \leq N} |x_i - x_j|$, or equivalently minimize the energy $-\frac{1}{2} \sum_{1 \leq i \neq j \leq N} \log |x_i - x_j|$ among all configurations in $K$. Such problems are often studied on compact manifolds, such as the $\mathbb{D}$-dimensional sphere \cite{brauchart,beltranhardy,betermin}.

In Euclidean space, one also considers \textit{weighted Fekete points}, by introducing a weight $\V$ and asking which configuration maximizes the weighted Vandermonde determinant\index{Vandermonde determinant}
$$\prod_{1 \leq i< j \leq N} |x_i-x_j| e^{-N\sum_{i=1}^N \V(x_i)}$$
or equivalently minimizes the logarithmic energy functional
$$-\hal \sum_{1 \leq i\neq j \leq N} \log |x_i-x_j| + N\sum_{i=1}^N V(x_i),$$
which corresponds exactly to the minimization of $\HN$ with $\s=0$.  Fekete points can also be characterized as the zeroes of a family of orthogonal polynomials, see \cite{simon}.

Finally there is also interest in the approximation theory literature in studying Riesz ``$\s$-energies,'' i.e., the minimization of $\sum_{1 \leq i \neq j \leq N} \frac{1}{|x_i - x_j|^\s}$ for all possible $\s$, which provides a motivation for studying the Riesz case \eqref{rieszgene}. For these aspects, we refer to the review papers \cite{saffkuijlaars,bhs} and the recent book \cite{borodachovlivre}.

Let us note that varying $\s$ from $0$ to $+ \infty$ connects Fekete points to the sphere packing problem (which as mentioned above formally corresponds to  $\s=\infty$).

\subsection{Other systems and further motivations}
\paragraph{{\bf Two-component plasma.}}\index{two-component plasma}
The \textit{two-dimensional, two-component plasma} or 2DTCP is a counterpart to the 2DOCP introduced above, which consists in $N$ particles $\XN = (x_1, \dots , x_N)$  of charge $+1$ and $N$ particles $\YN = (y_1, \dots , y_N)$ of charge $-1$ with a logarithmic interaction, which is now attractive for particles with opposite charges. The interaction energy is given by:
$$
\HN(\XN, \YN)  := - \hal \sum_{1 \leq i\neq j \leq N} \log |x_i-x_j|- \hal \sum_{1 \leq i\neq j \leq N} \log |y_i-y_j|+  \sum_{1 \leq i, j \leq N} \log |x_i-y_j|,
$$
and the particles are for instance constrained to a square in $\R^2$. The canonical Gibbs measure associated to the system is again defined as:
$$
\frac{1}{\ZNbeta} \exp\left(-\beta \HN(\XN, \YN) \right) d \XN d \YN,
$$
with obvious notation. 

The energy is no longer bounded below.  In fact $\HN(\XN, \YN) = - \infty$ if two particles with opposite charges happen to have the same position, and more generally, $\HN$ can be very negative if two such particles are very close. However, since such configurations are rare for the Lebesgue measure $d\XN d\YN$, the Gibbs measure is still well defined for high temperatures, i.e. small values of $\beta$. More precisely, the partition function converges for $\beta< 2$, the threshold for convergence of $\int e^{-\beta \log |x-y|} dxdy$. Note that the convergence at high temperature  is only true for attractive interactions that are less singular than the logarithm.

  The 2DTCP is interesting due to its close relation to two important theoretical physics models: the XY model and the sine-Gordon model, which exhibit a Berezinski-Kosterlitz-Thouless phase transition\index{Berezinski-Kosterlitz-Thouless phase transition}~\cite{bietenholzgerber}. We refer to \cite{spencer} for a presentation of this connection, and \cite{Frohlich,DeutschLavaud,FS1,fs2,GunPan} for studies of the 2DTCP in the physics literature. We will not discuss the 2DTCP further in this book, but the techniques developed here can be adapted to a mathematical investigation of its properties, see \cite{LSZ}.

\paragraph{{\bf Multi-component Coulomb gas.}}
Even further, one can consider {\textit multi-component} Coulomb systems, including a mixture of charges of arbitrary integer values with a neutralizing background \cite{martin}.

\paragraph{{\bf Other boundary conditions.}}
\index{Neumann energy}
There has also been interest in considering a Coulomb gas in a domain with boundary conditions, other Dirichlet condition (corresponding to a ``metal wall") \cite{janconew}, or Neumann boundary condition \cite{kiesslingspohn,forresteranalogies,byunkangseo}.  
Predictions are made  for the decay
of the correlations along the boundary (exponentially fast rather
than the algebraic decay with free boundary conditions), and
of the coefficient $D_\beta$ in \eqref{energyexpconj} below.
Authors discuss We also refer to the related discussion in Chapter \ref{chap:screening}.
\paragraph{{\bf Complex geometry and theoretical physics.}}
Coulomb systems and higher-dimensional analogues involving powers of determinantal densities are 
also of interest to geometers as a way to construct K\"ahler-Einstein metrics with negative Ricci curvature on complex manifolds \cite{berman,bbn}.  Another motivation is the construction of Laughlin states for the fractional quantum Hall effect\index{fractional quantum Hall effect} on complex manifolds, which effectively reduces to the study of a two-dimensional Coulomb gas on a manifold, possibly with nontrivial topology. The coefficients in the expansion of the (logarithm of the) partition function have interpretations as geometric invariants and conformal field theories, see for instance \cite{klevtsov,klevtsovlaughlin,klevtsovrandom,klevtsovmamarinescu}, and the end of Section \ref{secfree}. 
     Finally, recent work studying the Coulomb gas on a Jordan curve or a Jordan domain in the determinantal case $\beta=2$ \cite{johansson2023coulomb,courteaut2024partition} highlights the connection with geometry via Grunsky operators and the Loewner energy of Weil-Peterson curves.
          
\section{Questions}\label{sec1.3}
As usual in statistical mechanics \cite{huang,ruelle}, one would like to understand the typical behavior of the system under \eqref{gibbs}  in the \textit{thermodynamic limit} $N \to \infty$, and investigate hypothetical phase transitions and critical phenomena as the inverse temperature $\beta$ varies. 
Compared to many of the topics listed above where exact formulae and determinantal structures play an important role, we will be looking for methods and  results that apply to  {\it all inverse temperatures}  $\beta$ even possibly depending on $N$, and to  the whole class of Coulomb and Riesz interactions constrained by \eqref{intervalles}.  

Let us now list a set of questions that naturally arise, noting that we are still far from able to address them all.

We note that the scaling of the problem is set up so that most particles will be confined to a region of size $O(1)$ of the space, which we call the {\it macroscale}, while the scale at which we see a finite number of particles in a box is the {\it microscale} $N^{-1/\d}$. Intermediate scales are called {\it mesoscales}.



 \paragraph{{\bf Universality, phase portrait, phase transitions.}}
Looking back to the very definition of our systems in \eqref{HN} and \eqref{gibbs}, we see that there are three natural parameters: the pair interaction $\g$, the confining potential $\V$ and the inverse temperature $\beta$. Every time one studies the behavior of, say, a given observable, one may ask:
\begin{itemize}
\item Does it depend on $\V$? A negative answer is usually called a form of \textit{universality}. For example, the global distribution of the particles depends on $\V$ but their local arrangement is expected to be mostly universal, i.e.~be independent of $V$ up to rescalings.  This  has been proven only in the one-dimensional logarithmic case of $\beta$-ensembles in \cite{bey1,bey2}, and in the two-dimensional Coulomb case only for $\beta =2$ \cite{ahm,hwennman}.
\item How much does it depend on $\g$? This kind of universality \textit{with respect to the interaction} is, in fact, not much explored, besides extensions to log-like cases in dimension~1 \cite{borodingoringuionnet,borotguionnetk,venker}. In the present text, we rely crucially on the fact that $\g$ has a very specific form, but it would be very interesting to understand how much of the qualitative properties of Riesz gases are preserved under perturbation and depend only
 on the singularity of $\g$ at the origin and at infinity. 
\item Does it depend on $\beta$? If so, is there any \textit{critical point}, i.e.~value of $\beta$ at which the dependence  as a function of the inverse temperature ceases to be smooth? Usually, this never happens at finite $N$, but may arise in the thermodynamic limit $N \to \infty$. If one may pinpoint a significant \textit{order parameter} (an observable that encodes the order/disorder of the system) which, in the thermodynamic limit, does not depend smoothly on the inverse temperature, then there is a \textit{phase transition}. Drawing the \textit{phase portrait} of a system consists in listing such critical phenomena. It is a very delicate problem and, in a sense, the ultimate goal.
\end{itemize}

\paragraph{{\bf Observables.}}
Many questions about such systems can be formulated as follow. Given a map $\ObN$ from  $(\R^\d)^N$ to some space, we call observable the random variable $\ObN(\XN)$, whose law is the  push-forward of $\PNbeta$ by $\ObN$. 
Natural questions include:
\begin{itemize}
\item Does the observable  have a limit in law as $N \to \infty$? Can we characterize this limit (besides its mere existence)? How does it depend on the scale at which the observable lives?
\item Does the observable concentrate around certain values as $N \to \infty$? If yes, at which speed: can we state concentration inequalities? Can we find a physical or mathematical characterization of those values?
\item If the observable has a typical value in the limit, can we study the fluctuations? Are they Gaussian?
\end{itemize}
By extension, one can ask exactly the same questions about energy minimizers,  for which  observables and their limits are deterministic instead of random. 

A first natural observable  is the {\it empirical measure} $\emp$ with values $\probas(\R^\d)$, defined by
$$
\emp(\XN) := \frac{1}{N} \sum_{i=1}^N \delta_{x_i}.
$$ 
Understanding its limit as $N\to \infty$ corresponds to a Law of Large Numbers and  provides a description of the particles density at the global or macroscale. This is well understood,  see Chapters \ref{chap:eqmeasure} and \ref{chap:leadingorder}.

Another important class of observables consists in counting the number of points in balls (or other more general sets),  or rather its difference with the expected limit,
\index{discrepancy} 
\be \label{defdisc0} D(x,R):= \int_{B(x,R)} d\(\sum_{i=1}^N \delta_{x_i}- N\mu\)\ee
which we call the {\it discrepancy} in the ball $B(x,R)$. We can here let $R$ depend on $N$, allowing to study point discrepancies at the macroscale ($R=1$) down to the microscale  $R=N^{-1/\d}$. 
The discrepancy in the number of points measures how regular a point distribution is, and, together with its variance, is a very important quantity from the point of view of the analysis of point processes, see e.g.  \cite{torquato,torquatosti}.
 Also in  approximation theory, the discrepancy is exactly the measure of the accuracy (or error) in the approximation, see  for instance the surveys \cite{bilyksurvey,simoncoste}.
 We refer to Section \ref{sec:leble} for a discussion of conjectures and results on the discrepancy in the two-dimensional Coulomb case.

A variant of the discrepancy consists in testing $\sum_{i=1}^N \delta_{x_i}- N\mu$, where $\mu=\lim_{N\to \infty} \emp$,  not against indicator functions as in \eqref{defdisc0}, but against more regular (say smooth) functions $\varphi$, and study the size and fluctuations of regular linear statistics $\sum_{i=1}^N \varphi(x_i)$  in the form 
\be \label{deflinearstat}\int_{\R^\d} \varphi\,  d\( \sum_{i=1}^N \delta_{x_i}- N\mu\)= N \int_{\R^\d} \varphi \, d(\emp- \mu).\ee
Again, $\varphi$ can be taken to depend on $N$ and be supported at macro, meso or microscales.  Precise results are known for one and two-dimensional logarithmic cases, see Chapter \ref{chap:clt2}.

\begin{figure}[ht!] 
\begin{minipage}[c]{.46\linewidth}
\begin{center}
\includegraphics[scale=0.18]{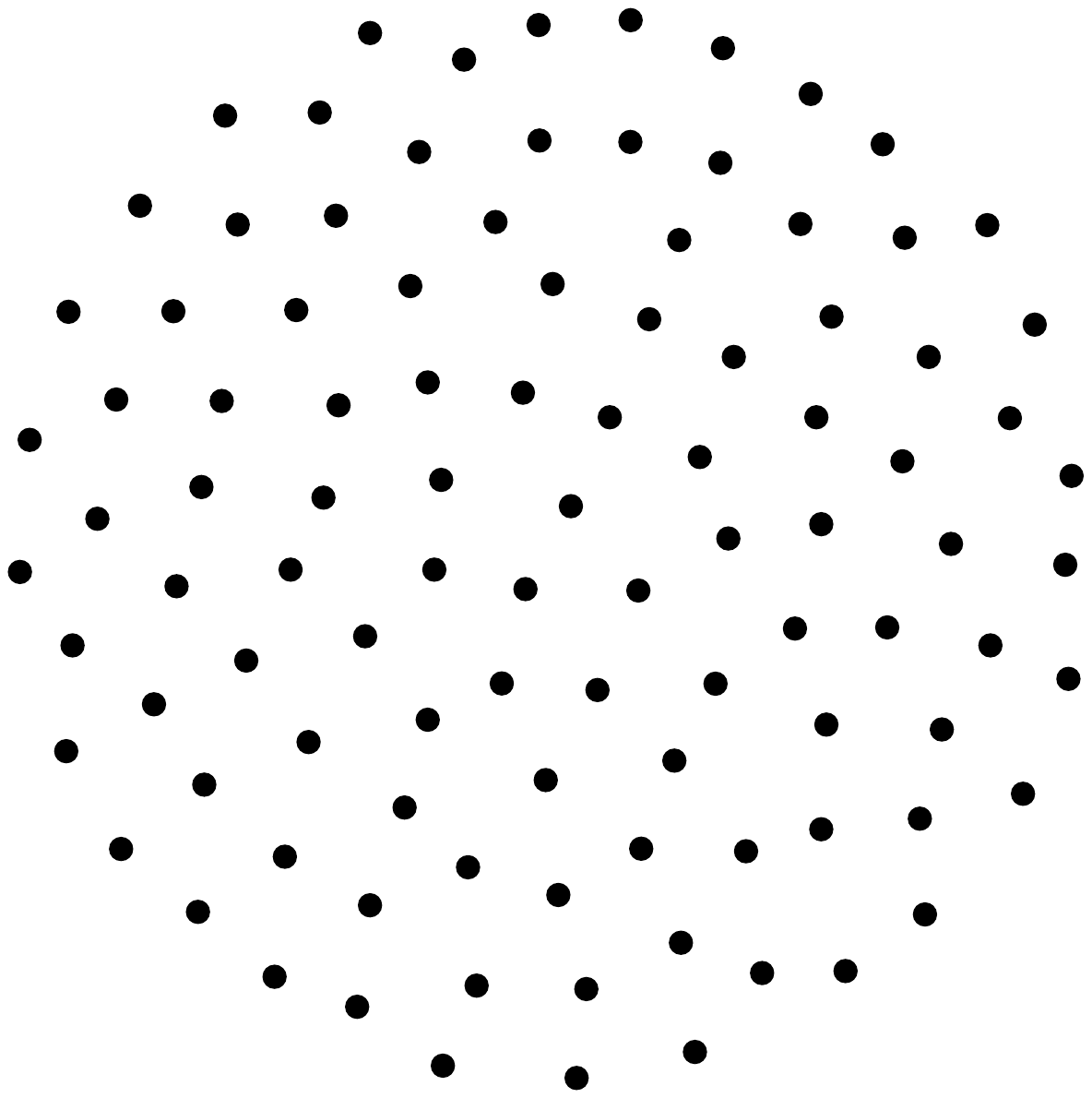} 
\end{center}
\end{minipage}
\begin{minipage}[c]{.46\linewidth}
\begin{center}
\includegraphics[scale=0.19]{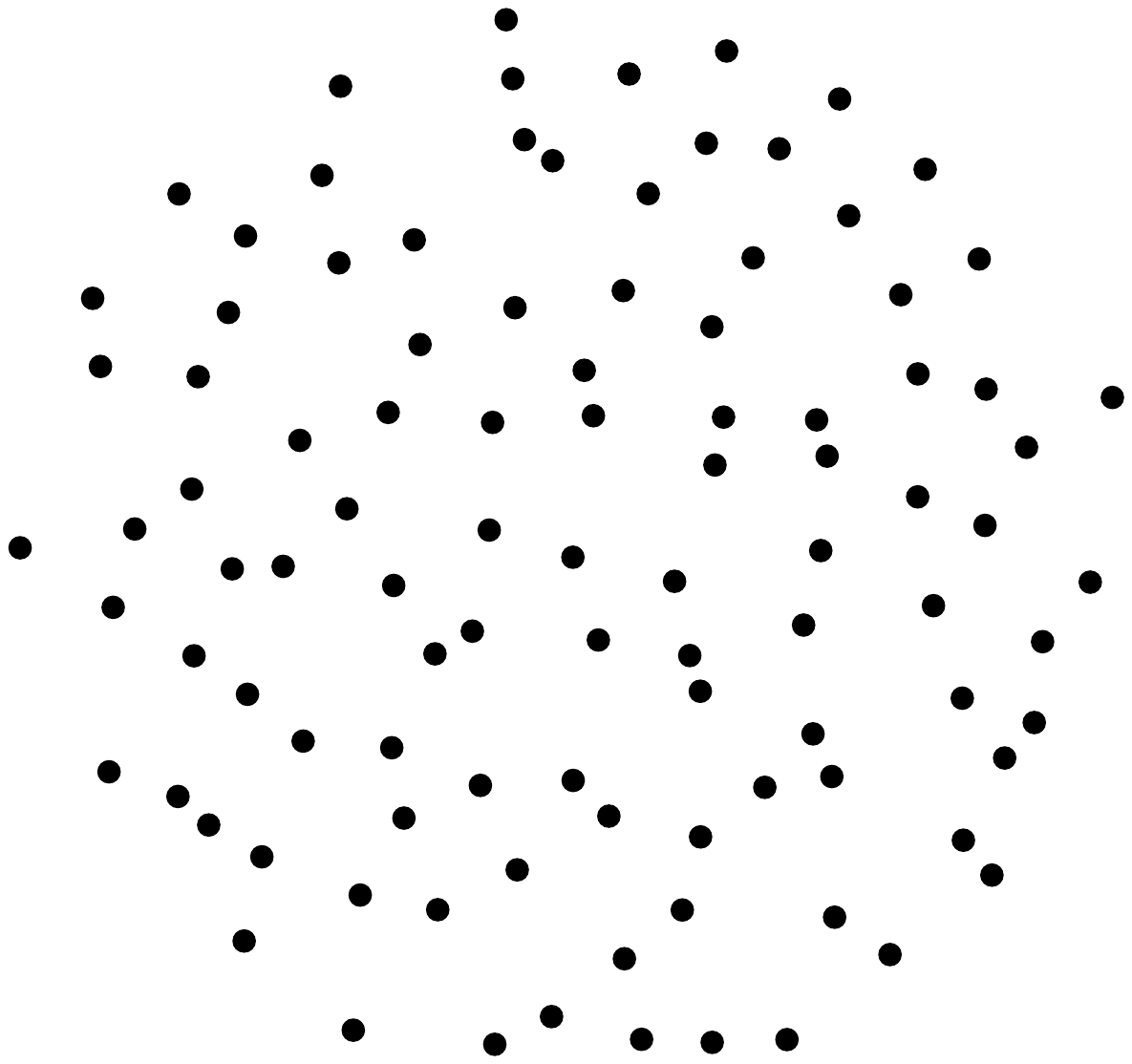} 
\end{center}
\end{minipage}
\vspace{-0.5cm}
\caption{Case $\d=2$, $\s=0$, with $N = 100$ and $\V(x) = |x|^2$, for $\beta = 400$ (left) and $\beta = 5$ (right).}\label{figure1}
\end{figure}

 For other relevant observables, one may think of something that encodes the arrangement of the particles at a certain scale. At the local or microscale, a natural choice would be the local configuration observable $\Cloc_{N, \bar{x}}$ defined by fixing some point $\bar{x}$ in $\R^\d$ and looking at:
\begin{equation}
\label{def:Cloc}
\Cloc_{N, \bar{x}}(\XN) := \sum_{i=1}^N \delta_{N^{1/\d} (x_i - \bar{x})},
\end{equation}
with values in the space  of point configurations.
The law on the limit of $\Cloc_{N, \bar{x}}$ as $N\to \infty$ is called a limiting point process, or local limit,  for the ensemble. Even for energy minimizers, understanding the patterns formed by the points in the limit is very hard, except in dimension~$1$ which is the only one for which the interaction is convex.   In contrast to the macroscopic behavior, several observations, (e.g. by numerical simulations, see Figure \ref{figure1})
 indicate that the behavior of the system at the microscopic scale depends on $\beta $ in a clear, non-trivial way. It is one of our main goals to analyze and characterize that $\beta$-dependence and the microscopic behavior.

\paragraph{{\bf  Limit point processes.}} \index{limit point process}
There are only few instances where the existence and nature of  a limiting point process is known. The central ones are the  case of the logarithmic interaction in dimension 1 with $\beta =2$ for which the process  is the sine-process \cite{mehta,forrester,borodin},  and the case of the logarithmic interaction in dimension 2 with $\beta=2$, or Ginibre ensemble\index{Ginibre ensemble}, for which the limit is the {\it Ginibre point process}\index{Ginibre point process} \cite{mehta,forrester}.
These are all instances of determinantal point processes. \index{determinantal point process}  The special cases $\beta=1 $ and $\beta =4$ also allow a similar treatment as pfaffian point processes.
In the case of the logarithmic interaction in dimension 1 with general $\beta$, the existence of a limiting point process, called the sine-$\beta$ process, was established in \cite{valkovirag,killipstoiciu}. \index{sine-$\beta$ point process}
 Recently \cite{gorinkleptsyn} proposed an approach to defining a $G_\infty E$ ensemble  corresponding to  $\beta=\infty$, and which also appears as a universal limit.
Also  recently, \cite{boursier23a} established the existence of a limit point process in the case  of the one-dimensional circular Riesz gas \cite{boursier23a}.

We will discuss the existence of subsequential limit points, and other ways to characterize the limits of \eqref{def:Cloc}  
in Chapters \ref{chap:derivW} and \ref{chap:ldp}.
Once limiting point processes are obtained, one can ask  for instance whether they satisfy Dobrushin-Lanford-Ruelle equations, whether these point processes are number rigid in the sense of \cite{ghoshperes}, and whether they are hyperuniform
in the sense of \cite{torquato,torquatosti}.  \index{hyperuniformity}
We refer to \cite{osadashirai,dhlm,dereudrevasseur,thoma2,leblehyper} for some results and to \cite{simoncoste} for a survey of questions.

\paragraph{{\bf Correlations and statistics.}}
With a slightly different mindset, one can try to understand the law of $\XN$ itself. The $k$-point correlation function, or $k$-marginal (for $k \geq 1$) is the function $\rho_N^{(k)} : (\R^\d)^k \to [0, + \infty)$ defined by: \index{correlation functions}
\be \label{defrhok}
\rho_N^{(k)}(x_1, \dots, x_k) = \frac{1}{\ZNbeta} 
\int_{(\R^\d)^{N-k} }
\exp\left( -\beta N^{-\frac\s\d} \HN(\XN) \right) d x_{k+1} \dots d x_N.
\ee
It can be thought of as giving the probability density of observing simultaneously a particle at $x_1$,
a particle at $x_2$, $\dots$, and a particle at $x_k$. Correlation functions are a very powerful tool, and getting knowledge of the $\rho_N^{(k)}$'s allows one to answer many of the relevant questions about the system through (possibly challenging) computations. In particular the decay rate as $|x_1-x_2|\to \infty$ of $\rho^{(2)}(x_1, x_2) $, defined as the scaled $N\to \infty$-limit of $\rho_N^{(2)}(x_1, x_2)$ indicates the state of the gas (roughly: solid if no decay, liquid for algebraic decay, and  gas for exponential decay), and transitions in that decay rate depending on $\eta$ correspond to phase-transitions.

 In the  special cases  $\beta=2$ and $\s=0$ mentioned above where the particles form a determinantal point process, the $N\to \infty$ limits of the $\rho_N^{(k)}$ are identified as determinants of the form $\rho^{(k)} = \det [K(x_i,x_j)]_{1\le i,j\le k} $
for  an explicit kernel $K(x,y)$ : the sine-kernel for the case of the sine process, and an exponential kernel for the Ginibre point process. 
Aside from these special cases and other cases where correlation functions can be accessed via kernels expressed via orthogonal polynomials,  no explicit formulae exist in general and  estimating the correlation functions (let alone computing them exactly) or getting a handle on their decay rate is extremely difficult, with the exception of the one-dimensional situations.
  Note that a lot of the statistical physics literature  \cite{martin,martinyalcin}  establishes  ``sum rules" and charge fluctuations estimates  but implicitly assuming properties of $\rho^{(2)}$ at large distances and bootstrapping via BBGKY hierarchies satisfied by the higher order correlation functions, see also comments in Section~\ref{sec:isotropic}. 
  
A weaker form of the question consists in fixing a test function $\varphi : (\R^\d)^k \to \R$ and considering the $k$-point \textit{statistics}
$$
\sum_{1 \leq i_1, \dots, i_k \leq N} \varphi(x_{i_1}, \dots, x_{i_k})
$$
as a random variable whose law we try to understand as $N \to \infty$, which is the generalization of the  linear statistics case $k=1$ of  \eqref{deflinearstat}.

For the one-dimensional Coulomb gas, the behavior of $\rho^{(1)}$ and $\rho^{(2)}$ was elucidated,  proving crystallisation \cite{kunz,brascamplieb,aizm}, see also the discussion in \cite[p. 54]{lewinsurvey}. For the one-dimensional Riesz case with $\s>0$, \cite{boursier23a} proves a rate of decay of correlations of the particles gaps.

\index{crystallization}
For the two and three-dimensional Coulomb gas, a phase transition at finite, non-zero temperature has been conjectured in the physics literature based on numerical  observations~:  in dimension 2 it happens around $\beta = 140$ \cite{caillol1982monte,choquard1983cooperative} and in dimension 3 around $\beta= 175$ \cite{bst,jonesceperley}, see also the review \cite{KK}.
This transition should  correspond to a change of decay of $\rho^{(2)}$  from exponential to algebraic  as $\beta$  crosses the hypothetical transition temperature.
Its existence and precise nature is still disputed in the physics literature,  see \cite{cardosostephan}, and getting any mathematical understanding about it remains a fascinating challenge.

\paragraph{{\bf Free/minimal energy expansions.}}
For energy minimizers, it is natural to understand the value of $\min \HN$ and its asymptotic expansion in $N\to \infty$, if it exists. 

In the statistical mechanics setting, a similar role is played by the partition function defined in \eqref{def:ZNbetN}, and more precisely by the \textit{free energy} $-\frac1{\beta} \log \ZNbeta$. Knowing it is important because it gives access to many physical quantities associated to the system (for e.g.  differentiating $Z_{N,\beta}$ with respect to $\beta$ yields  the average energy, etc), see statistical mechanics textbooks such as \cite{huang}. 
We will also see that evaluating the free energy  is  directly connected with  (hence key to understanding) the fluctuations of linear statistics.

For the one-dimensional Coulomb gas with quadratic confinement, the $N\to \infty$ expansion  of $\log Z_{N, \beta}$ is given in \cite{kunz}, and the coefficients are shown to be analytic in $\beta$.
 For the one-dimensional log gas, the value of $Z_{N, \beta}$ is known explicitly for all $\beta>0$ when $V(x)=x^2$ via 
 the exact computation  of  the integral in \eqref{def:ZNbetN}, which uses so-called {\it Selberg integrals} (see e.g. \cite{mehta}). For more general $V$'s an expansion in $N$ to any order is also known \cite{borotguionnet,shch}. For the two-dimensional Coulomb gas however, no equivalent  of the 
 Selberg integral  exists and the exact value of $Z_{N,\beta}$ is only known for the Ginibre case  $\beta=2$ and $V(x)=|x|^2$ \cite{mehta}, and  a few other determinantal cases mentioned at the end of Chapter~\ref{chapclt}. 
 
 Works of \cite{imbrie,brydgesfeder1,brydgesfeder2} studied the grand canonical case for  small $\beta$  in dimensions 2 or 3,  
 mapping the model to a quantum field theory by sine-Gordon transformation. This leads to believe that the coefficients of $\log Z_{N, \beta}$ should be analytic in $\beta$ when $\beta$ is close enough to $0$.
 
 
There are conjectures on the expansion of $\min \HN$ for Fekete points\index{Fekete points} (energy minimizers for $\s=0$) on the $2$-sphere \cite{RSZ}, see also \cite{BHS2}, as well as conjectures for the expansion of the free energy in the two-dimensional logarithmic case \cite{jancomanificatpisani,tellezforrester,zabrodinwiegmann,cantellez,klevtsov}, in both cases the (free) energy expansion is conjectured to be of the form
\be\label{energyexpconj}
\log \ZNbeta = A_\beta N^2 +\frac14 N\log N + B_\beta N + C_\beta \sqrt{N} + D_\beta \log N + E_\beta+o(1),\ee
with the coefficients having physical and geometric interpretations (we refer to  the end of Section \ref{secfree} for the precise formulas). The existence of the $B_\beta N$ term corresponds to  a  thermodynamic limit   (or limit of free energy per particle) in the statistical mechanics language.  In particular, if these coefficients are found to depend on the inverse temperature in a non-smooth way,  it is again an indication of a phase transition ---  in fact a phase transition can be expected to be manifested  by a joint change of smoothness of the free energy and change of decay rate of the two-point correlation function as described in the previous paragraph.

\paragraph{{\bf Other observables}.}
One can think of other natural observables, such as the minimal distance between two particles \cite{benarousbourgade,fengwei,fengtianwei,ameurrepulsion,ameurromero,thoma}, or the maximal distance to the support of $\mu$ \cite{ameurloc}. 
We will discuss them a bit in Section \ref{sec5.5}. The electrostatic field generated by the particles, and its maximal value, are also observables of interest, see \cite{lewinsurvey}. Closely related is  the characteristic polynomial for $\beta$-ensembles and the two-dimensional  log gas, together with its maximum, which have been  the object of many studies \cite{chhaibimadaule,BMP22,lambertpaquette1,lambertpaquette2,augeributezzeitouni,lambertleblez}.

\section{Plan of the book}

The book starts with the analysis of the macroscopic behavior of the Coulomb or Riesz  gas, which is governed by the Frostman {\it equilibrium measure} $\meseq$, unique minimizer among probability densities of 
\be\label{eqmdef}\I(\mu)=\hal \iint_{\R^\d\times \R^\d} \g(x-y) d\mu(x)d\mu(y)+\int_{\R^\d} V(x) d\mu(x).\ee
We start in Chapter \ref{chap:eqmeasure} with the characterization and description of the equilibrium measure, which are basics of potential theory. We then describe the less well-known connection of this minimization problem with the {\it classical obstacle problem} of calculus of variation in the Coulomb case, and the {\it fractional obstacle problem} in the general Riesz case.  We then examine and describe  the {\it thermal equilibrium measure}, a version of the minimization problem \eqref{eqmdef} with an added entropy cost $\int \mu\log \mu$, for which we present a general existence theorem. Even though the equilibrium measure is what is generally used, 
the thermal measure provides a more accurate description of the particles density, particularly when the inverse temperature gets small.

Chapter \ref{chap:leadingorder} connects the Coulomb/Riesz gas to the (thermal) equilibrium measure,\index{equilibrium measure} via the framework of $\Gamma$-convergence and large deviations, which we start by  reviewing.  
The first main result is  that, for minimizers of $\HN$, the large  $N$ limit of the empirical measure 
$\frac1N \sum_{i=1}^N \delta_{x_i}$ converges to the equilibrium measure $\meseq$. For general configurations the energy $\HN$ can be described in terms of the limit energy \eqref{eqmdef} for the limit  of the empirical measure.
 These facts are standard for logarithmic interactions but here we give a presentation that applies to fairly general interaction $\g$, often left out in the literature. In particular we present a streamlined generalization to all Riesz cases and all temperature regimes of the Large Deviations Principles of \cite{hiaipetz,bz,bg}, asserting roughly that 
$$\PNbeta( \mu_N \simeq \mu) \approx e^{-\beta N^{1-\frac\s\d}(\I(\mu)-\min \I)},$$
i.e.~the probability that the empirical measures limits to something else than the equilibrium measure is exponentially small as $N\to \infty$. This elucidates the macroscopic behavior of the Coulomb/Riesz gas.

The second part of the book is focused on the  next-order electric (called {\it modulated} in the dynamical context) energy.\index{modulated energy} \index{electric formulation} This quantity, defined for any configuration $\XN$ and any reference probability density $\mu$  by 
\begin{align*}\F_N(\XN, \mu) &  = \hal \iint_{\{x\neq y\}} \g(x-y) d\Big( \sum_{i=1}^N \delta_{x_i} -N \mu\Big)(x) d\Big( \sum_{i=1}^N\delta_{x_i}  -N \mu\Big)(x) \\ &= \frac{N^2}{2} \iint_{\{x\neq y\}} \g(x-y) d(\emp-\mu)^{\otimes 2}(x,y),\end{align*}
corresponds to a Coulomb/Riesz interaction of a system discrete point charges and a negative background charge $-\mu$. It can be used as a ``metric" to quantify the convergence of the empirical measure to the reference measure $\mu$. 
In Chapter \ref{chap:nextorder}, the energy is  studied in detail, in particular we present its {\it electric reformulation} as the (weighted) Dirichlet energy of the electric  potential 
\be\label{elecpot}h_N=\g*\( \sum_{i=1}^N \delta_{x_i} -N \mu\),\ee
or square of the electric field $\int |\nab h_N|^2$,  
 which in particular allows to {\it localize} the energy. The electric reformulation involves a {\it renormalization} of the integral,  performed via a {\it truncation} procedure by smearing of the charges on spheres with  {\it point-dependent} radii. We  describe a crucial {\it monotonicity property}, with respect to the truncation radii, of this renormalization, which allows to deduce the  coercivity of the modulated energy and the fact that $\frac1{N^2}\F_N$ is essentially  a  metric $\|\emp-\mu\|^2$ in a weak Sobolev norm, but computed in a renormalized fashion. This way, it controls charge discrepancies as in \eqref{defdisc0}
 or the difference between the empirical measure and the reference measure $\mu$.  

In Chapter \ref{chap:concentrationbounds}, we show how the quantity $\F_N$ appears at the next order in an exact {\it splitting formula} for $\HN$, after subtracting off the energy of either  the equilibrium measure or the  thermal equilibrium measure. More precisely, one checks that  for any configuration 
$$\HN(\XN)= N^2 \I(\meseq)+ \F_N(\XN, \meseq) + \text{effective confinement  energy},$$ 
and a  convenient variant  with respect to the thermal equilibrium measure. This allows to effectively reduce 
the Riesz gas energy to the modulated energy $\F_N$.
Inserting the general lower bound on $\F_N$ found in Chapter \ref{chap:nextorder}, one then obtains an easy  lower bound on the partition function in terms of $N$. It can be complemented with an upper bound, and together these very directly imply {\it concentration bounds} quantifying the closeness of the empirical measure to $\meseq$  thanks to the fact that $\F_N(\XN, \meseq)$ is like a (square) distance between them.
At the end of Chapter \ref{chap:concentrationbounds} we discuss various questions of localization (near the support of $\meseq$), separation, charge excess and discrepancy; on these topics we describe  some results  of Ameur et.~al.~and the isotropic averaging method of Thoma.

Chapter \ref{chap:commutator}  makes a detour through the question of dynamics for systems with Coulomb/Riesz interactions, i.e.~analyzing the large $N$ limit of \eqref{noise1} and \eqref{noise2}. We first present a functional inequality on $\F_N$ that has been termed {\it commutator estimate}.  This functional inequality allows to control {\it derivatives of} $\F_N(\XN, \mu)$ {\it along a generic transport} applied to both the points of the configuration $\XN$ and $\mu$. These derivatives are bounded in terms of $\F_N$ and the Lipschitz norm of the transport map.  This commutator estimate is  the key to  proving quantitative mean-field convergence for  dynamics of systems of points with Coulomb /Riesz interactions, either conservative \eqref{noise2} or gradient flows \eqref{noise1},  via the {\it modulated energy method}, or {\it modulated free energy method} in the case with noise. These convergence results  
are presented in that chapter and of independent interest.
The commutator estimate is also important later for the transport calculus approach to fluctuations of Chapter~\ref{chapclt}.

The third part of the book concerns the mesoscopic behavior of Coulomb  gases (in that part we focus exclusively on the Coulomb case for simplicity) and in particular in proving {\it local laws}, i.e.~exponential moment bounds on the localized version of $\F_N$ introduced in Chapter~\ref{chap:nextorder}.  The idea to do so, presented in Chapter \ref{chap:screening}, is to use a {\it bootstrap on scales} and compare two energy quantities, one defined with Dirichlet boundary conditions and one with Neumann boundary condition, and show that they are close. 
The crucial technical tool to show this is the {\it screening procedure}. \index{screening}
In Chapter \ref{chaploiloc}, we explain how to perform the bootstrap on scales to derive local laws: local laws at the macroscale are easy consequences of the free energy bounds of Chapter \ref{chap:concentrationbounds}, then the screening procedure allows to deduce almost additivity of the free energy   and of the system along slightly smaller than macroscopic boxes, up to surface errors. The local laws then hold down to this slightly smaller scale. The reasoning can then be iterated to get the local law on smaller and smaller scales, until the estimates saturate at a new (temperature-dependent) minimal scale, which is the microscale $N^{-\frac{1}{\d}}$ if $\beta$ is bounded below, $\beta^{-\hal} N^{-\frac1\d}$ otherwise. The local laws being a byproduct of the almost additivity of the free energy, they naturally  come together with free energy expansions with explicit error rates in $N^{1-\frac1\d}$  corresponding to surface errors,  for uniform equilibrium measures.

In Chapter \ref{chapclt}, we introduce the {\it transport method} and transport calculus with a view on studying fluctuations. This method consists in viewing the comparison of the free energy of a  Coulomb gas with an external potential $V$ and that of a Coulomb gas with perturbed potential $V+t\xi$ in terms of a transport map between the corresponding equilibrium measures. This allows in particular to obtain free energy expansions with explicit error rates for varying equilibrium measures, corresponding to the expansion down to the $B_\beta N$ term in \eqref{energyexpconj}, with explicit $o(N)$ error.
In Chapter \ref{chap:clt2}, we combine the transport method of Chapter \ref{chapclt} and the commutator estimates of Chapter \ref{chap:commutator} to obtain bounds on fluctuations of linear statistics, again for Coulomb gases in arbitrary dimension.
When specializing to one and two-dimensional Coulomb gases, we prove a complete central limit theorem, which we now state for dimension~2.
\begin{theo*}\notag If $\d=2$ and $\s=0$, for $\xi$  regular enough, 
$$\int \xi \, d\( \sum_{i=1}^N \delta_{x_i}-N \meseq\)$$ converges as $N \to \infty$ to a Gaussian with explicit mean and variance $ \frac{1}{2\beta} \int |\nab \xi|^2$.
In particular, for $\beta=\infty$, i.e. for minimizers of $\HN$, 
$ \int \xi d\( \sum_{i=1}^N \delta_{x_i}-N \meseq\)$ converges as $N\to \infty$ to an explicit constant.
\end{theo*}
This result can be rephrased as the convergence  of the electric potential \eqref{elecpot} to the 2D Gaussian free field\index{Gaussian free field}. It is also valid for test-functions $\xi$ that are localized on mesoscales, all the way down to large multiples of the microscale.
Chapters \ref{chapclt}-\ref{chap:clt2} are rather independent from what follows.

The last part of the book focuses on  the microscopic description of the configurations via the {\it jellium renormalized energy} $\W$, an infinite volume energy version of the Coulomb / Riesz interaction of an infinite system of  point charges neutralized by a uniform background charge (i.e.~a jellium), defined via the electric formulation of the energy. 
In this part we return to the general setting of Riesz interactions.
This jellium energy is introduced in Chapter \ref{chap:renormalized} and its properties are described. The question of its minimization is connected with crystallization questions and the Cohn-Kumar conjecture\index{Cohn-Kumar conjecture} alluded to above.

Chapter \ref{chap:derivW} derives this energy $\W$ as the $N\to \infty$ limit energy of the next order  energy $\F_N$. This is expressed in terms of {\it empirical fields} or limit point processes, i.e.~probability measures $P$ on point configurations that are obtained as limits of quantities like \eqref{def:Cloc}.  \index{empirical field}
The connection between $\F_N$ and $\W$, again obtained via the electric formulation of the energy,  is first expressed via a general  lower bound on $\F_N$ by the average of $\W$ with respect to the probability $P$. We then obtain the main theorem stating that the empirical fields (i.e. local limits at the microscale) of minimizers of $\HN$ must converge to minimizers of $\W$, thus connecting to the crystallization questions of the prior chapter. This is proven by combining the general lower bound with an upper bound construction relying on the screening procedure.

In Chapter \ref{chap:ldp}, we adapt this to the probabilistic situation of the Coulomb/Riesz gas ensemble. This requires combining the energetic effects (the derivation of $\W$) with the entropic effects, which are accounted for via a Sanov-type theorem at the level of empirical fields. The combination of the two allows to derive a Large Deviations Principle for the empirical field, with a rate function that takes the form of a free energy, as a sum of $\beta$ times the energy $\W$ and the  specific relative entropy of the limit point process with respect to the Poisson point process. This provides a variational interpretation of the limiting point processes, where the energetic (favoring ordered configurations) and entropic (favoring disorder) effects compete.  The LDP, which is for macroscopic averages of the empirical field,  is complemented in the Coulomb case by a local version, for mesoscopic and down to microscopic averages.

\part{Macroscopic behavior}
\chapter{The equilibrium measure(s)}
\label{chap:eqmeasure}
\index{equilibrium measure}

As mentioned in the previous chapter, our setting in this text is that of Riesz interaction potentials
\begin{equation}\label{riesz}
\g(x)= \begin{cases}  \frac{1}{\s} |x|^{-\s} \quad  & \s \neq 0\\
-\log |x| & \s=0
\\ \d-2\le \s<\d .&
\end{cases}\end{equation}
We note that either $\s\ge0$ and $\g$ is singular at the origin, or  $\d=1$ and $
-1\le \s <0.$ In all cases, $\g$ is radial decreasing.

As mentioned previously, the leading order behavior of  the energy $\HN$ is governed by the  functional
\begin{equation} \label{definitionI}
\I(\mu) := \hal\iint_{\R^\d \times \R^\d} \g(x-y) d\mu(x)d\mu(y) + \int_{\R^\d} V(x) d\mu(x)
\end{equation}
defined over the space $\mc{P}(\R^\d)$ of probability measures on $\R^\d$.

Note that $\I(\mu)$ is simply a continuum version of the discrete Hamiltonian $\HN$ defined over all $\mc{P}(\R^\d)$, which may also take the value $+ \infty$. From the point of view of statistical mechanics, $\I$ is the ``mean-field" limit energy of  $\frac{1}{N^2}\HN$, while we will see in the next chapter that from the point of view of probability, $\I$ also plays the role of a {\it rate function}.

Its minimization turns out to be  a classical  problem of electrostatics, that of finding  the  equilibrium distribution of charges in a  capacitor with an external potential also called the ``capacitance problem."  It was historically studied by Gauss and  settled by Frostman \cite{frostman}. It is thus   a fundamental question  in {\it potential theory}, a topic which itself grew out of the study of the electrostatic or gravitational potential, see  e.g. \cite{landkof,adamshedberg,doob,safftotik} and references therein.
The case of $\d=2$ and $\g(x)=-\log |x|$ is precisely treated in \cite[Chap.~1]{safftotik}. Higher dimensional and more general singular interaction potentials are treated for instance in \cite{cgz}. The general case, that we will treat,  is not more difficult.

We will see that, under appropriate assumptions, $\I$ has a unique minimizer $\meseq$, called the {\it equilibrium measure}, or  the Frostman equilibrium measure, or sometimes the extremal measure.   The equilibrium measure arises as the Law of Large Numbers limit of the Gibbs measure \eqref{gibbs}, we will show in the next chapter a stronger version of this result in the form of a Large Deviations Principle.

We will also discuss the {\it thermal equilibrium measure} \index{thermal equilibrium measure} $\mu_\theta$ defined for any given inverse temperature $\theta$ as the
minimizer  in $\mc{P}(\R^\d)$ of the functional $\I$ with an added entropy term, i.e., 
\be \label{defEtheta}
\I_\theta(\mu):= \I(\mu)+ \frac{1}{\theta} \int_{\R^\d} \mu \log \mu.
\ee
This energy functional arises when taking into account temperature effects in the study of \eqref{gibbs}, this was previously used for instance in \cite{CLMP,kiessling,RSY2,bouchaudguionnet}. As we will see throughout the text, when temperature is present, the thermal equilibrium measure, with the choice $\theta = \beta N^{1-\frac\s\d} \gg 1$ in our scaling,  always provides a more precise description of the point distribution (see already Remark \ref{remtem} below), and it will help in our study of regimes of $\beta$ very small when $N \to \infty$. The discussion of the thermal equilibrium measure is new in this level of generality and we have reduced to rather minimal assumptions for its existence.

\section[Existence, uniqueness, characterization]{Existence, uniqueness, and characterization of the equilibrium measure}
\label{sec2.1}

As previously we consider $\g$ given by \eqref{rieszgene} with $\d-2\le\s<\d$.

\begin{lem}[Strict convexity]\label{convexi}Assume $\mu_\pm \in \mc{P}(\R^\d)$ and 
$\iint_{\R^\d\times \R^\d} |\g(x-y)| d\mu_\pm(x) d\mu_\pm(y)<+\infty$. Letting $\mu= \mu_+-\mu_-$, we have 
$$\iint_{\R^\d\times \R^\d} \g(x-y) d\mu(x) d\mu(y) \ge 0,$$   and the 
 functional $\I$ is strictly convex. 
\end{lem}
\begin{proof}
The map $\mu \mapsto \int V\, d\mu$ is linear, so proving  the convexity of $\I$ reduces to  showing that the quadratic function $Q(f)= \iint \g(x-y)\, df(x)\, df(y)$, is convex over probability measures.
Since $$\hal Q(f_1, f_1) + \hal Q(f_2,f_2)-Q(\hal (f_1+f_2))= \frac14 Q(f_1-f_2),$$ it suffices to show that $Q(f_1-f_2) \ge 0$ for $f_1, f_2$ two probability measures, for which it suffices to prove that $Q(f)\ge 0$ if $f$ is a signed Radon measure with $\int f=0$, with equality if and only if $f=0$.

Let us first consider  $f$ to be a Schwarz function with integral $0$, and recall that $\hat \g(\xi)= C_{\d, \s} |\xi|^{\s-\d}\ge 0$ for our choice \eqref{rieszgene} with $\d-2\le \s \le \d$ (see \cite[Theorem 2.4.6]{grafakos})  where $\hat \cdot$ denotes the Fourier transform.\footnote{When $\s\le 0$, $\hat \g$ has an additional Dirac at the origin, but it doesn't affect the argument since we test against $|\hat f|^2$ which vanishes at the origin.}  Also $|\hat f(\xi) |\le C\min( |\xi|, 1) $ since $\int f=0$, while $\hat f$ decays rapidly at infinity,  thus  $\hat \g(\xi)| \hat f|^2(\xi)\in L^1 (\R^\d)$ and 
$$Q(f)= \iint_{\R^\d\times \R^\d} \hat \g(\xi)| \hat f|^2(\xi) d\xi.$$ 

We thus find $Q(f)\ge 0$ in that case. For general $f$, we mollify it by convolution with a Gaussian approximation of the Dirac mass $\delta_0$,  call it $\phi_\ep $, and let $f_\ep = \phi_\ep * f$. Then $\int_{\R^\d}  f_\ep= \int_{\R^\d} f=0$ and  $\hat f_\ep= \hat \phi_\ep \hat f$ decays like a Gaussian, so the previous reasoning applies and we find $Q(f_\ep) \ge 0$.   To conclude we then argue that $Q(f_\ep) \to Q(f)$ by applying the dominated convergence theorem in physical space, and using the assumption. 
 Also  $Q(f) =0$ implies $\hat f=0$ by the same steps, hence $f=0$ and  $Q$ is strictly convex over probability measures. 
 \end{proof}
\begin{rem}
Less restrictive  assumptions on $\g$,  for instance    $\hat{\g}\ge 0$, where $\hat{\g}$ stands for the Fourier transform,  suffice for the above result.\end{rem}

As a consequence of the lemma, there is a unique (if any) minimizer to \eqref{definitionI} among probability measures. 

The existence of a minimizer is a bit more delicate.
To obtain it we make the following assumptions on the potential $V$. 
\begin{description}
\item(\namedlabel{A1}{A1}) $V$ is lower semi-continuous (l.s.c.) and bounded below.
\item (\namedlabel{A2}{A2}) (growth assumption)
$$
   \underset{|x|\to + \infty}{\lim}\( V(x)+ \g(x)\) = + \infty
$$
\end{description}
The first condition is there to ensure the lower semi-continuity of $\I$ and that $\inf \I>-\infty$, the second is made to ensure that $\I$ is coercive. 
Of course, if $\s>0$, condition \eqref{A2} is equivalent to the condition that $V$ tends to $+\infty$ at infinity.

We will repeatedly use the following 
\begin{rem}\label{remarkVg}
If $\g$ is as in \eqref{riesz}, we have
\be \label{gVV}\g(x-y)\ge \g(x)\wedge \g(y)  \indic_{\s= 0}+ (\g(x)+\g(y))\indic_{\s<0} -C , \ee 
for some $C>0$ depending only on $\s$, where $\wedge $ denotes the minimum of two numbers and  $\indic_{A}$ means $1$ if $A$ holds and $0$ otherwise.
Thus, under assumptions \eqref{A1}, \eqref{A2},  the function $\g(x-y)+V(x)+V(y)$ is  bounded below and tends to $+\infty$ as $|x|$ or $ |y| \to +\infty$. 
\end{rem} 
\begin{proof}
For \eqref{gVV} we observe that when $\s=0$,  $\g(x-y) = -\log |x -y| \geq - \log 2 - \log \max(|x|, |y|)$, and 
when $-1\le \s<0$,  we can argue that 
$\g(x-y) \ge \g(x) +\g(y) $. The case $\s>0$ is obvious.
The second assertion then follows from \eqref{gVV} and assumptions \eqref{A1}, \eqref{A2}, after distinguishing the cases $\s>0$, $\s=0$ and $-1\le \s<0$.
\end{proof}

\begin{lem}[Coercivity of $\I$]  \label{coerI}
Assume \eqref{A1} and \eqref{A2} are satisfied, and let $\{\mu_n \}_n$ be a sequence in $\mc{P}(\R^\d)$ such that $\{\I(\mu_n)\}_n$ is bounded. Then, up to extraction of a subsequence,  $\mu_n $ converges to some $\mu\in \mc{P}(\R^\d)$ in the weak sense of probabilities, and 
\be\label{Imun}
\liminf_{n\to \infty} \I(\mu_n) \ge \I(\mu),\ee
and $\inf \I>-\infty$.
In other words, $\I$ is lower semi-continuous, bounded below, and its sub-level sets are compact.
\end{lem}
\begin{proof}
Assume that $\I(\mu_n) \le C_1$ for each $n$. 
By Remark \ref{remarkVg}, given any constant $C_2>0$ there exists a  compact set $K\subset \R^\d$ such that 
\begin{equation}\label{vgrand}
\min_{ (K\times K)^c} \( \g(x-y) + V(x)+V(y)\)>C_2.\end{equation}
Also, by the same remark,  we may write 
\be \label{vgrand2} \g(x-y)+ V(x)+ V(y)\ge - C_3\ee with $C_3>0$.
 Rewriting then  $\I$ as
 \begin{equation}
\I(\mu) = \hal\iint_{\R^\d} \(\g(x-y) + V(x) + V(y)\) d\mu(x) d\mu(y),
\end{equation}
we deduce from \eqref{vgrand} that $\I(\mu) \ge \frac{C_2}{2}$ for all $\mu$, hence $\inf\I >-\infty$, and 
the relations \eqref{vgrand}--\eqref{vgrand2} and our assumption on $\mu_n$ imply that 
$$C_1\ge \I(\mu_n) \ge- C_3+  C_2(\mu_n\otimes \mu_n) ( (K\times K)^c)\ge -C_3 + C_2 \mu_n(K^c).$$
Since $C_2$ can be made arbitrarily large,  $\mu_n(K^c)$ can be made arbitrarily small, which means precisely that $\{\mu_n\}_n$ is a tight sequence of probability measures.
By Prokhorov's theorem, it thus has a convergent subsequence (still denoted $\{\mu_n\}_n$), which  converges to some probability $\mu$.
For any $n$ and any $M > 0$, we may then write
\begin{equation}
\iint \g(x-y) d\mu_n(x) d\mu_n(y) \geq \iint (\g(x-y) \wedge M) d\mu_n(x) d\mu_n(y) 
\end{equation}
where $\wedge$ denotes the minimum of two numbers. 
For each given $M$,
 $\g(x-y) \wedge M$  is  continuous hence $ \g(x-y)\wedge M+ V(x)+V(y)$ is l.s.c and bounded below (by Remark \ref{remarkVg}), and thus the weak convergence of $\mu_n$ to $\mu$, which implies the weak convergence of $\mu_n \otimes \mu_n$ to  $  \mu \otimes \mu$, 
 yields    
\begin{align*}
 \liminf_{n \to + \infty}\I(\mu_n) & 
= \liminf_{n \to + \infty}\hal  \iint (\g(x-y) +   V(x) +V(y) ) d\mu_n (x) d\mu_n(y) 
\\ & \ge   \liminf_{n \to + \infty} \hal \iint (\g(x-y) \wedge M + V(x)+V(y) ) d\mu_n(x) d\mu_n(y)\\
& \ge \hal \iint (\g(x-y) \wedge M + V(x)+V(y) ) d\mu(x) d\mu(y).
\end{align*}
The monotone convergence theorem then allows  to let $M \rightarrow + \infty$, and conclude that \eqref{Imun} holds.

\end{proof}

We have seen above that $\inf \I >-\infty$ thanks to \eqref{A1}. The next question is to see whether  $\inf \I <+\infty$, i.e. that there exist probabilities with finite $\I$'s. For $\s<0$ this is clear. For $\s \ge 0$, this is directly related to the notion of (electrostatic, Bessel, or logarithmic) capacity, whose definition we now give.
 One may find it in \cite{landkof,safftotik,eg,adamshedberg} or \cite[Sec. 11.15]{liebloss}, the formulations differ a bit but are essentially equivalent.  It is usually not formulated this way for $\s=0$ but it can be extended to that case  without trouble.

\index{capacity}
\begin{defini}[Capacity of a set] \label{definitioncapacite1} We define the capacity of a compact set $K\subset \R^\d$ by 
\begin{equation} \label{definitioncapacite2}
\mathrm{cap}(K) : = \Phi\(  \inf_{\mu \in \mc{P}(K)} \iint_{\R^\d} \g(x-y) d\mu(x) d\mu(y)\),
\end{equation}
with $\Phi(t)= e^{-t}$ if $\s=0$ and $\Phi(t)= t^{-1}$ if $\s>0$, and 
where $\mc{P}(K)$ denotes the set of probability measures supported in $K$.  Here  the $\inf$ can be $+\infty$ if there exists no 
 probability measure $\mu\in \mc{P}(K)$ such that
$
\iint_{\R^\d} \g(x-y) d\mu(x) d\mu(y) < + \infty$. 
For a general set $E$, we define $\mathrm{cap}(E)$ as the supremum of $\mathrm{cap}(K)$ over the compact sets $K$ included in $E$. \end{defini}

It is easy to check that capacity is increasing with respect to the inclusion of sets. 

For $\s<0$, in order to have a unified presentation, we use the word ``capacity'' to mean ``cardinality".

A basic fact is that a set of zero capacity also has zero Lebesgue measure (see the references above).
In fact  $\mbox{cap}(E)=0$ is stronger than $|E|=0$, it implies for example that the perimeter of $E$ is also $0$. 
A property is said to hold ``quasi-everywhere" (q.e.) if it holds everywhere except on a set of capacity zero. By the preceding lemma a property that holds q.e.~also holds Lebesgue-almost everywhere (a.e.), whereas the converse is in general not true. In the case $\s<0$, since  by capacity we mean cardinality, saying that a property holds q.e.~just means that it holds everywhere.
\smallskip

For the sake of generality, it is interesting to consider potential $V$'s which can take the value $+\infty$ (this is the same as imposing the constraint that the probability measures only charge  a specific set, the set where $V$ is finite).
We then need to place a third assumption 
$$\text{(\namedlabel{A3}{A3}) } \quad 
 \{ x \in \R^\d , V(x) <+\infty\} \text{ has positive capacity.} $$
 Note that with our notion of capacity, if $\s<0$ this just means that $\{ x \in \R^\d , V(x) <+\infty\} $ is nonempty.
\begin{lem}\label{infiborne}
Under assumptions \eqref{A1}---\eqref{A3}, we have $\inf \I <+\infty$.
\end{lem}
\begin{proof} The case $\s<0$ is obvious, it suffices to take $\mu$ equal to a Dirac mass at a point where $V$ is finite. Let us turn to the case $\s \ge 0$.
Let us define for any $\varepsilon > 0$ the set $\Sigma_{\varepsilon} = \{x\ |\ V(x) \leq \frac{1}{\varepsilon}\}$. Since $V$ is l.s.c.~the sets $\Sigma_{\varepsilon}$ are closed, and it is easy to see that assumption \eqref{A2} implies that they are also bounded, since it implies in all cases that $V(x)$ tends to $+ \infty$ when $|x| \rightarrow + \infty$.

The capacity of $\Sigma_0 = \{x\in \R^\d , V(x) < +\infty \}$ is positive by assumption. It is easily seen that the sets $\{\Sigma_{\eps}\}_{\eps > 0}$ form a decreasing family of compact sets with $\bigcup_{\eps > 0} \Sigma_{\eps} = \Sigma_0$, and by definition (see Definition \ref{definitioncapacite1} or the references given above) the capacity of $\Sigma_0$ is the supremum of capacities of compact sets included in $\Sigma_0$. Hence we have that $\mbox{cap}(\Sigma_{\varepsilon})$ is positive for $\varepsilon$ small enough. Then by definition there exists  a probability measure $\mu_{\eps}$ supported   in $\Sigma_{\eps}$ such that 
\begin{equation}
\iint \g(x-y) d\mu_{\eps}(x) d\mu_{\eps}(y) < + \infty.
\end{equation}
Of course, we also have $\int V d\mu_{\eps} < + \infty$ by definition of $\Sigma_{\eps}$. Hence $\I(\mu_{\eps}) < + \infty$, in particular $\inf \I< + \infty$.

\end{proof}

We may now give the main existence result, together with the characterization of the minimizer.

\begin{rem} In the following  we will not really use  the particulars of the logarithmic and Riesz kernels. While uniqueness essentially requires $\g$ to have positive Fourier transform, the rest of   the theorem still holds for a much more general class of $\g$'s, say $\g$  positive, monotone radial and satisfying $\iint \g(x-y)\, dx\, dy<\infty$.  \end{rem}

\begin{theo}[Frostman \cite{frostman}, existence  and characterization of the equilibrium measure] \mbox{}
\label{theoFrostman} Under the assumptions \eqref{A1}-\eqref{A2}-\eqref{A3}, the minimum of $\I$ over $\mc{P}(\R^\d)$ exists, is finite and is achieved by a unique $\meseq$, which  has a compact support of  positive capacity. In addition $\meseq$ is uniquely characterized by the fact that  there exists a constant $c$ such that 
\begin{equation}
\label{EulerLagrange}
\left\lbrace
\begin{array}{cc} h^{\meseq} +V \geq c & \mbox{q.e.~in } \R^\d  \vspace{3mm} \\ 
 h^{\meseq} + V= c & \mbox{q.e.~in the support of }\meseq \end{array} \right.
\end{equation}
where \be \label{defhmu0}
h^{\meseq}(x) := \int_{\R^\d} \g(x - y) d\meseq(y)
\ee is the electrostatic potential generated by $\meseq$. Moreover, 
the constant $c$ must be 
 \begin{equation}
\label{defc1} c = 2\I(\meseq) - \int_{\R^\d} V(x) d\meseq(x).
 \end{equation}
 \end{theo}
 \begin{defini}\label{def23} From now on, we denote by $\zeta$ the function
\begin{equation} \label{defzeta}
\zeta = h^{\meseq} + V - c.
\end{equation}
We also denote $\omega = \{x\in \R^\d, \zeta(x)=0\}$ and $\Sigma= \supp \, \meseq$.
\end{defini}
\noindent We note that in view of \eqref{EulerLagrange}, $\zeta\ge 0$ a.e. and $\zeta=0$, $\meseq$-a.e. Also,  
\begin{equation} \label{defomega}
\Sigma \subset \{\zeta = 0\}.
\end{equation}
The set $\omega=\{\zeta=0\}$  corresponds to the {\it contact set} or {\it coincidence set} of the obstacle problem, while $\Sigma$ is the set where the obstacle is \textit{active}, sometimes called the {\it droplet}.  
We may place assumptions so that  they coincide. 

Here $h^{\meseq}$ is the  self-generated (electrostatic) potential, while $V$ is the external potential, 
so $h^{\meseq}+V$ corresponds to the total potential; 
 $\zeta$ is thus an effective potential. The theorem says that the equilibrium distribution $\meseq$ is one for which the total potential is minimized, and constant, in the support of the distribution.
 
Moreover, we claim that for every $x\in \R^\d$
\be \label{borninfzeta}
\zeta (x)\ge \g(x)+V(x)-C
\ee where $C$ depends only on $\d,\s$ and $V$. Indeed,
 \be\label{hmg}
 h^{\meseq}(x)-\g(x)= \int_{\R^\d} (\g(x-y)-\g(x))d\meseq(y),\ee and since $\meseq$ is compactly supported the integral reduces to a compact set of $y$, thus using \eqref{gVV} in the case $\s\le 0$, or direct estimate if $\s>0$ we easily deduce that \eqref{hmg} is bounded below. The relation \eqref{borninfzeta} then follows from \eqref{defzeta}.
 
 The effective potential $\zeta$ thus grows like $V+\g$ at infinity.

\begin{rem} \label{rem88} 
It is sometimes of interest to  consider ``weakly confining potentials" for which \eqref{A2} barely fails, and the equilibrium still exists but will fail to be compactly supported, see for instance 
\cite{hardy}.
\end{rem}

\begin{example}[Capacity of a compact set] Let $K$ be a compact set of positive capacity,  and let $V = 0$ in $K$ and $V = + \infty$ in $K^{c}$. In that case the minimization of $\I$ is the same as the computation of the capacity of $K$ as in \eqref{definitioncapacite2}. The support of the equilibrium measure $\meseq$ is contained in $K$, and the associated Euler-Lagrange equation \eqref{EulerLagrange} states that  the  potential  $h^{\meseq} $   is constant  q.e.~on the support of $\meseq$, a well-known result in physics. \end{example}

\begin{rem} \label{rem8} Note that by \eqref{coulombkernel}, if $\s=\d-2$,  the function $h^{\meseq}$ solves :
\begin{displaymath}\label{eqpoisson}
- \Delta h^{\meseq} =\cd \meseq,
\end{displaymath} where $\cd$ is the constant defined in \eqref{defcd}.
In particular in the example above, if $K$ is sufficiently regular and $\s=\d-2$, one finds that $\meseq = 0$ q.e.~in $K$, which indicates  that $\meseq$ is supported on $\partial K$. 
\end{rem}

\begin{example}[$C^{1,1}$ potentials and RMT examples] \label{example2} In the Coulomb case $\s=\d-2$,  using again Remark~\ref{rem8} yields that if $V\in C^{1,1}$, \be \label{densmu0}
 \meseq =\frac{ \Delta V}{\cd} \quad \text{in} \ \overset{\circ}{\Sigma}
\ee
where  $\overset{\circ}{\Sigma}$ is the interior of $\Sigma$, i.e.~the density of the measure on the interior of its support is given by $\frac{\Delta V}{\cd}$. This will be proven in Proposition  \ref{proequivpb}. For example if $V$ is quadratic, then the measure $\meseq $ has constant  density in the interior of its support. 

This corresponds to the important examples of the energies that arise in random matrix theory, more precisely~:

\begin{itemize}
\item in dimension $\d=2$, for $V(x) = |x|^2$, one may check that $\meseq = \frac{1}{\pi} \mathbf{1}_{B_1}$  where $\mathbf{1}$ denotes a characteristic function and $B_1$ is the unit ball, i.e.~the equilibrium measure is the normalized Lebesgue measure on the unit disk (by uniqueness, $\meseq$ should be radially symmetric, and the combination of \eqref{densmu0} with the constraint of being a probability measure imposes the support to be $B_1$). This is known as the {\it circle  law} for the Ginibre ensemble in the context of Random Matrix Theory (RMT). Its derivation  (which can be seen  as a consequence of Section \ref{LDP-sec} below) is attributed to   Ginibre, Mehta, an unpublished paper of Silverstein and  Girko \cite{girko1}.
\item in all Coulomb cases $\s=\d-2$, the same  holds, i.e.  for $V(x)=|x|^2$  we have $\meseq = \frac{\d}{\cd} \indic_{B_{(2/\d)^{1/\d}   }}$ by the same reasoning.
\item in dimension $\d=1$, with $\s=0$ and $V(x) = x^2$, the equilibrium measure is $\meseq (x)= \frac{1}{2\pi} \sqrt{4-x^2} \mathbf{1}_{|x|\leq 2}$, which corresponds in the context of RMT (GUE and GOE ensembles) to {\it   Wigner's semi-circle law}, cf. \cite{wigner,mehta}.
\item Examples of exact computations of equilibrium measures, with some surprising phenomena, are given in \cite{chafaisaff1,chafaisaff2,byun2023planar}.
\end{itemize}
\end{example}
\begin{rem}\label{remvanishingrate}
In the Coulomb case $\s=\d-2$, in view of \eqref{densmu0}  the density $\meseq(x)$ typically has a discontinuity on $\partial \Sigma$.
In contrast, in Riesz cases $\s\in (\d-2, \d)$, the density $\meseq(x)$ is typically continuous and vanishing on $\partial \Sigma$, as  the semi-circle law provides an example of. In fact, the generic vanishing rate in Riesz cases is  in $ \dist(x, \partial \Sigma)^{1-\frac{\d-\s}{2}}$, which can be deduced (see \cite[Appendix]{PeilenSer}) from the connection with  the obstacle problem described in Section \ref{sec2.4} below. 
\end{rem}

We now turn to the proof of Theorem \ref{theoFrostman}, adapted from\cite[Chap. 1]{safftotik}.
\begin{proof}[Proof of Theorem \ref{theoFrostman}]

{\bf Step 1: Existence and uniqueness modulo compact support.} 
The existence of a minimizer $\meseq$ follows directly from Lemmas \ref{coerI} and \ref{infiborne},  its uniqueness from Lemma \ref{convexi} once compactness of the  support is proven.  Indeed, if $\mu_+$ and $\mu_-$  are two probabilities with compact support, to satisfy $\iint |\g(x-y) |\d\mu_\pm(x) d\mu_\pm (y)<+\infty $ and be able to apply Lemma \ref{convexi},  it suffices to know that $\left|\iint \g(x-y) d\mu_\pm(x) d\mu_\pm(y)\right|<+\infty$. When $\s>0$ or $\s<0$ it is clear since $\g$ has a sign, and when $\s=0$ it follows from the fact that $\mu_\pm$ are compactly supported that the negative part of the integral is bounded. Let us now justify that  $-\infty<\iint  \g(x-y) d\meseq(x) d\meseq(y)<+\infty$. 
First, since by \eqref{A1}  $V$ is bounded below and  since $\I(\meseq)<+\infty$ by Lemma \ref{infiborne}, we find that $\iint  \g(x-y) d\meseq(x) d\meseq(y)<+\infty$. Secondly, $\iint  \g(x-y) d\meseq(x) d\meseq(y)>-\infty$ since by definition $\g$ is bounded below on compact sets and $\meseq$ has compact support.

 It remains to show that $\meseq$ has compact support  of finite capacity and that \eqref{EulerLagrange} \smallskip holds.

{\bf Step 2: compact support.} 
  Using Remark \ref{remarkVg}, we may find a compact set  $K$ such that 
 $\g(x-y)+V(x)+V(y) \ge  2(\I(\meseq) +1)$ outside of $K\times K$.

Assume that $\meseq$ has mass outside $ K  $,  i.e. assume $\meseq(K) <1$, and define the new probability measure 
\begin{equation}
\tilde{\mu} := \frac{\meseq\vert_{ K}  }{\meseq(K)}.
\end{equation}
We want to show that $\tilde{\mu}$ has less or equal energy $\I(\tilde{\mu})$ than $\meseq$, in order to get a contradiction. One may compute $\I(\meseq)$ in the following way :
\begin{align*}
\I(\meseq) & = \hal \iint_{K \times K} \left(\g(x-y) + V(x) + V(y)\right) d\meseq (x) d\meseq (y) \\  &+  \hal \iint_{(K \times K)^{c}} \left(\g(x-y) + V(x) + V(y)\right) d\meseq(x) d\meseq(y)  \\ &  \geq
\meseq(K)^2 \I(\tilde{\mu}) + \hal \min_{(K \times K)^{c}}\left(\g(x-y) + V(x) + V(y)\right) \( 1- \iint_{K\times K} d\meseq d\meseq\).
\end{align*}
By choice of $K$, and since we assumed $\meseq(K)<1$, this implies  that  
\begin{equation}
 \I(\meseq) \ge  \meseq(K)^2 \I(\tilde{\mu}) + (1- \meseq(K)^2) (\I(\meseq)  + 1) 
\end{equation} and thus 
$$\I(\tilde{\mu}) \le \frac{\I(\meseq)}{\meseq(K)^2} + \frac{\meseq(K)^2 -1}{\meseq(K)^2}(\I(\meseq)  + 1)  <   \I(\meseq),$$
a contradiction with the minimality of $\meseq$.
 We thus conclude  that $\meseq$ has compact support.
 The fact that the support of $\meseq$ has positive capacity is an immediate consequence of the fact that $\I(\meseq)<\infty$ and the definition of capacity.
\smallskip

{\bf Step 3.}
 We  turn to the proof of the Euler-Lagrange equations \eqref{EulerLagrange}. For  this, we use the ``method of  variations"  which consists in  continuously deforming $\meseq$ into  other admissible probability measures. \\
 Let $\nu$ in $\mc{P}(\R^\d)$ such that $\I(\nu) < + \infty$, and consider the probability measure $(1-t) \meseq + t \nu$ for $t$ in $[0,1]$. Since $\meseq$ minimizes $\I_V$, we have 
\be
\I\((1-t)\meseq + t\nu\) \geq \I(\meseq), \mbox{ for all } t \in [0,1].
\ee
By letting $t \rightarrow 0^+$ and keeping only the first order terms in $t$, one obtains the ``functional derivative" of $\I$ at $\meseq$. More precisely, writing  
\be
\iint \hal \g(x-y) d((1-t)\meseq + t\nu)(x) d((1-t)\meseq + t\nu)(y) + \int V(x) d((1-t)\meseq + t\nu)(x) \geq \I(\meseq),
\ee
one easily gets that 
\begin{multline}
\I(\meseq) + t \left[\hal \iint \g(x-y) d(\nu- \meseq)(x) d\meseq(y) +\hal \iint \g(x-y) d(\nu-\meseq)(y) d\meseq(x) \right. \\\left. + \int V(x) d(\nu - \meseq)(x)\right] + O(t^2) \geq \I(\meseq).
\end{multline}
Here, we may cancel the identical order $0$ terms $\I(\meseq)$ on both sides,  and note that 
in view of \eqref{defhmu0} the expression between brackets can be rewritten as $\int h^{\meseq}(x)d(\nu - \meseq)(x) + \int V(x) d(\nu - \meseq)(x)$. Next, dividing the inequality by $t > 0$, and letting $t \rightarrow 0^+$, we obtain that for all $\nu$ in $\mc{P}(\R^\d)$ such that $\I(\nu) < + \infty$, 
\be
\int h^{\meseq}(x) d(\nu - \meseq)(x) + \int V(x) d(\nu- \meseq)(x) \geq 0,
\ee
or equivalently \be \label{EulerLagrange2}
\int \Big(h^{\meseq}  + V\Big)(x) d \nu(x) \geq \int \Big(h^{\meseq} + V\Big)(x) d\meseq(x).
\ee
Defining  the constant $c$ by 
\begin{multline} \label{defc2} c := 2 \I(\meseq) -  \int V d\meseq=\iint \g(x-y) d\meseq(x)d\meseq(y) +  \int V d\meseq \\
= \int \Big(h^{\meseq} + V\Big) d\meseq ,\end{multline}
(\ref{EulerLagrange2}) asserts that
\begin{equation}
\label{EL3}
 \int \Big(h^{\meseq} + V\Big) d\nu \geq c 
\end{equation}
for all probability measures $\nu$ on $\R^\d$ such that $\I(\nu) < + \infty$. Note that at this point, if we relax the condition $\I(\nu) < + \infty$, then choosing $\nu$ to be a Dirac mass when applying \eqref{EL3}  would yield
\begin{equation} 
\label{EL4}
h^{\meseq} + V \geq c
\end{equation} pointwise. However, Dirac masses have infinite energy $\I$, and we can only prove that \eqref{EL4} holds quasi-everywhere, which we do now. 

Assume not, then there exists a set $K$ of positive capacity such that \eqref{EL4} is false on $K$. By definition of the capacity of $K$ as as supremum of capacities over compact sets included in $K$, we may in fact suppose that $K$ is compact. By definition, this means that there is a probability measure $\nu$ supported in $K$ such that 
\be \label{ene1}
\iint \g(x-y) d\nu(x) d\nu(y) < + \infty.
\ee
Let us observe that $-h^{\meseq}$ is bounded above on any compact set (this is clear if  $\s>0$ because then $\g$  is positive and so is $h^{\meseq}$, and can be easily checked for  $\s\le 0$ because $-\g$ is then bounded above on any compact set and $\meseq$ has compact support). By assumption, equation \eqref{EL4} is false on $K$, that is $V < c -  h^{\meseq}$ on $K$. Integrating this inequality against $\nu$ gives 
\be \label{contrad}
\int V d\nu = \int_K V d\nu < \int_K (c -  h^{\meseq}) \, d\nu< + \infty
\ee
which, combined with \eqref{ene1} ensures that $\I(\nu)$ is finite. But then \eqref{EL3} must hold, which contradicts \eqref{contrad}. We have thus shown that 
\be \label{EL5}
h^{\meseq} +V \geq c\  \mbox{ q.e.},
\ee
which is the first of the relations \eqref{EulerLagrange}. 

For the second one, let us denote by $E$ the set where the previous inequality (\ref{EL5}) fails. We know that $E$ has zero capacity, but since $\meseq$ satisfies $\I(\meseq) < + \infty$, it does not charge sets of zero capacity (otherwise one could restrict $\meseq$ to such a set, normalize its mass to $1$ and get a contradiction with the definition of a zero capacity set). Hence we have \be
h^{\meseq} + V \geq c \quad \meseq\mbox{-a.e. } 
\label{EL6}
\ee
Integrating this relation against $\meseq$  yields 
\be
\int \Big(h^{\meseq} + V\Big) d\meseq \ge  c,
\ee
but in view of \eqref{defc2} this implies that  
 equality must  hold in (\ref{EL6}) $\meseq$-almost everywhere. This establishes the second Euler-Lagrange equation in \eqref{EulerLagrange} and \eqref{defc1} holds by \eqref{defc2}.
 \smallskip
 
  {\bf Step 4.} We show that the relations \eqref{EulerLagrange} uniquely characterize the minimizer of $\I_V$. 
 Assume that $\mu$ is another probability solving \eqref{EulerLagrange} with some constant $c'$,   and set, for $t \in [0,1]$, $\mu_t := t \mu + (1-t)\meseq$, hence $ h^{\mu_t} = t h^{\mu} + (1-t) h^{\meseq}$. We have
 \begin{multline*}
 \I(\mu_t) =\hal  \int \left( t h^{\mu}(x) + (1-t) h^{\meseq}(x) + 2V(x) \right) d\mu_t(x) \\ = \frac{t}{2} \int \left( h^{\mu}(x) + V(x) \right) d\mu_t(x) \\
+ \frac{(1-t)}{2} \int \left( h^{\meseq}(x) + V(x) \right) d\mu_t(x) + \frac{1}{2} \int V(x) d\mu_t(x).
\end{multline*}
By assumption, $h^{\mu} + V \geq c'$ and $h^{\meseq} + V \geq c$ almost everywhere. We thus get that 
\begin{multline}\label{abov}
\I(\mu_t) \geq \hal ( t c' + (1-t) c )+ \frac{1}{2} \int V(x) \left( t d\mu(x) + (1-t) d\meseq(x)\right) \\ = \frac{t}{2} \left(c' + \int V d\mu\right) +\frac{ (1-t)}{2} \left(c + \int V d\meseq \right).
\end{multline}
On the other hand, by \eqref{defc1} we have
\begin{displaymath}
\I(\mu) = \hal \(c' + \int V d\mu \)  \text{ and } \I(\meseq) =\hal \( c + \int V d\meseq\).
\end{displaymath}
Hence \eqref{abov} asserts that $\I(\mu_t) \geq t \I(\mu) + (1-t) \I(\meseq)$, which is impossible by strict convexity of $\I$ unless $\mu = \meseq$. This proves that the  two measures $\mu$ and $\meseq$ must coincide. 
\end{proof}

Henceforth, we will assume that $\meseq$ is a measure that is absolutely continuous with respect to the Lebesgue measure and  we will  use the same notation for a measure $d\mu$ and  its density $\mu(x)$.

\section{A first electric rewriting}
We now reformulate the energy $\I$ in terms of the ``electric" potential of \eqref{defhmu0}. \index{electric formulation}
This electric rewriting  is easy in the Coulomb case and well-known in physics (see for instance \cite{jancomanificatpisani}). It consists in using the fact that $\g$ is the Coulomb kernel and thus satisfies 
\eqref{coulombkernel}, to deduce that  the Coulomb potential defined by 
$$h^\mu = \g*\mu= \int \g(\cdot -y) d\mu(y)$$ 
(whenever the integral is convergent) satisfies 
$$-\Delta h^\mu= \cd \mu,$$ which allows
to rewrite 
\begin{align}\label{carreduchamp}
\iint_{\R^\d\times \R^\d} \g(x-y) d\mu(x)d\mu(y)= 
\int_{\R^\d} h^\mu(x) d\mu(x)=-\frac1{\cd} \int_{\R^\d} h^\mu \Delta h^\mu =\frac1{\cd} \int_{\R^\d} |\nab h^\mu|^2.\end{align}
This rewriting has used Green's formula and assumed that the boundary terms at infinity that it generates vanish. This is not always the case: it is true if $\s>0$ (or $\d \ge 3$) but it is not true for $\s \le 0$ i.e. for the logarithmic case in two dimensions or for the one-dimensional Coulomb kernel, because these do not vanish at infinity. 
In these cases,  \eqref{carreduchamp} will instead  only hold for the differences of two probability measures $\mu= \mu_+-\mu_-$ which are such that $\int_{\R^\d} \mu=0$ and $h^\mu$ decays at higher order at infinity. 
This is closely related to what was done in Lemma~\ref{convexi}.

In order to present  results valid for the general Riesz case, it  is useful to first describe   the {\it dimension extension representation} of fractional Laplacians, that will be used throughout the text.
\subsection{Extension representation of the fractional Laplacian}\label{sec-extension}
The extension method is found in work of Molchanov and Ostrovskii  \cite{molchanov}  in their studies of symmetric stable processes, 
and was made systematic by
 Caffarelli-Silvestre \cite{caffsilvestre}. 
 The difficulty with the fractional Laplacian is that it is a {\it nonlocal operator}.
The extension representation allows to revert to a local operator. It consists in  seeing $\R^\d$  as embedded into $\R^{\d+1}=\{(x,y), x\in \R^\d, y\in \R\}$, by identifying it with $\R^{\d}\times \{0\}$. 
 Let us denote by $\delta_{\R^\d}$ the uniform measure on $\R^\d\times \{0\}$, i.e.~the distribution such that for any smooth $\varphi(x,y) $ (with $x\in \R^\d, y \in \R$) 
we have 
$$\int_{\R^{\d+1}} \varphi \delta_{\R^\d}= \int_{\R^\d  } \varphi(x,0) \, dx.$$

Then,  as observed in \cite{caffsilvestre}, $\g$ can be seen as the restriction to the $\R^\d$ hyperplane of the  kernel of the weighted divergence form operator 
\be -\div (|y|^\gamma \nab \cdot ) \ee
where $y\in \R$ stands for the last (or $x_{\d+1}$) coordinate in $\R^{\d+1}$ and $\gamma$ is given by 
\be \label{gamma} \gamma= \s-\d+1.\ee

In the one dimensional log case $\s=0$, $\g(x)=-\log |x|$ is  the kernel of the half-Laplacian, and it is known that the half-Laplacian can be made to correspond to the Laplacian by adding one extra space dimension. 

In general dimension,  the half-Laplacian is associated to $\d-\s=1$,  hence $\gamma =0$, and it can be viewed as the kernel of a full Laplacian: if one wants to solve 
$$(-\Delta)^{\frac12} u = f\quad \text{in}\  \R^\d$$
it suffices to solve 
$$-\Delta u = f \delta_{\R^\d} \quad \text{in} \ \R^{\d+1}$$
where $\delta_{\R^\d}$ is as above,
and consider the $\R^\d\times \{0\}$ trace of the  solution.
Note that $u$ is automatically harmonic away from $\R^\d\times \{0\}$, and that its Laplacian is, in the distributional sense,  equal to (half) the jump of the normal derivative across $\R^\d\times \{0\}$.


We will use this extension representation in all the book whenever we deal with Riesz interaction. It  allows to view $\g$ as the kernel of a {\it local} operator, $-\div (|y|^\gamma \nab )$. The weight $|y|^\gamma$ does not introduce much change in the computations, and up to adding one dimension, this will allows us to treat the Riesz case  similarly to the Coulomb case in many instances.

Indeed, the cases with extension can be treated in a unified way with the Coulomb cases by working in $\R^{\d+\k}$ where $\k=0$ in the Coulomb cases, $\k=1$ otherwise, 
and setting \be\label{defgamma}\gamma= \s + 2-\k- \d,\ee which is $0$ in the Coulomb case.
From now on we will use this unified formulation. The reader only interested in the Coulomb case can just set $\k=0$ (i.e. remove the dimension extension) and $\gamma=0$, i.e.~remove the $\yg$ weight and replace $\div (\yg \nab \cdot)$ by the Laplacian.

Again, the main feature we will use is that  the extended $\g$ is a fundamental solution for the degenerate elliptic operator $-\frac{1}{\cds}\div(\yg\nabla\cdot)$ in $\R^{\d+\k}$, where $\cds$ is as in \eqref{fractlapkernel}, i.e.,
\begin{equation}\label{eq:Gfs}
-\div\Big(\yg \nabla\g \Big) = \cds\delta_{0} \quad \text{in} \ \R^{\d+\k}
\end{equation}
with equality in the sense of distributions.
This way, for any distribution $f$ on $\R^\d$, the potential $$
\g*f= \int_{\R^\d} \g(x-x')f(x') dx' ,$$  can naturally be extended into a potential 
on $\R^{\d+\k}$
$$h^f(x, y) :=\int_{\R^\d} \g ((x,y)- (x',0)) f(x') dx'$$ satisfying
\be \label{divyg}-\div(\yg \nab h^f)= \cds f \drd \quad \text{in}\  \R^{\d+\k}.\ee

\subsection{First electric rewriting in Riesz cases}\label{sec:elecrewri}
The electric rewriting is another point of view on  the fact that by Plancherel's theorem, since $\hat{\g}=C_{\d,\s} |\xi|^{\s-\d}$, we may rewrite formally
\be\label{formalHs}\iint_{\R^\d\times \R^\d} \g(x-y)d\mu(x)d\mu(x)= \int_{\R^\d} \hat \g (\xi) |\hat{\mu}(\xi)|^2  d\xi= C_{\d,\s} \int_{\R^\d}|\xi|^{\s-\d}|\hat \mu(\xi)|^2d\xi= C_{\d,\s}  \|\mu\|^2_{\dot{H}^{\frac{\s-\d}{2}} }(\R^\d)\ee
where $\dot{H}^{m}(\R^\d)$ denotes the homogeneous Sobolev space defined as the completion of Schwartz functions for the semi-norm
\be \label{normdoth}\|f\|^2_{ \dot{H}^{m}(\R^\d)}= \int_{\R^\d} |\xi|^{2m} |\hat f(\xi)|^2 d\xi.\ee

\begin{prop}[Electric reformulation for Riesz interactions]\label{lemelect1}
If $\s>0$,  for any  bounded Radon measure $\mu$ of finite mass in $\R^\d$ such that $\iint \g(x-y) d\mu(x)d\mu(y)<+\infty$,  letting $h^\mu= \g*(\mu\drd)$ in the sense of distributions,  we have  $h^\mu\in L^1_{\mathrm{loc}}(\R^{\d+\k})$ and 
\be \iint_{\R^\d\times \R^\d} \g(x-y) d\mu(x)d\mu(y)= \frac1{\cds}\int_{\R^{\d+\k}} \yg |\nab h^\mu|^2.\ee

If $\s \le 0$,  then for $\mu_+$ and $\mu_-$ probability measures on $\R^\d$, each satisfying 
\be \label{finitepot} \iint_{\R^\d\times \R^\d}| \g(x-y) | d\mu_\pm (x) d\mu_\pm(y)<+\infty,\ee 
letting $\mu=\mu_+-\mu_-$, the potential
$h^\mu= \g*(\mu\drd)$ is in $ L^1_{\mathrm{loc}}(\R^{\d+\k})$ and we have 
\be \iint_{\R^\d\times \R^\d} \g(x-y) d\mu(x)d\mu(y)= \frac{1}{\cds}\int_{\R^{\d+\k}} \yg |\nab h^\mu|^2<+\infty.\ee
\end{prop}
This also coincides with the fractional Sobolev norm $\|h^\mu \|_{\dot{H}^{\frac{\d-\s}{2}  }}^2$ or $\|\mu\|_{\dot{H}^{\frac{\s-\d}{2}  } }^2$.

\begin{proof}
{\bf Step 1: definition of the potential.} We first check that $h^{\mu}\in L^1_{\mathrm{loc}} (\R^{\d+\k})$.
Let $U$ be a bounded set in $\R^{\d+\k}$, we have $\int_U |\g(x-y)|dx<+\infty$ since $\s<\d$, and thus by Fubini-Tonelli, we may write
\begin{equation}\label{hmubb}
\int_U h^{\mu}=\int_U \int_{\R^\d} \g(x-y) d\mu(y) dx=\int_{\R^\d}\int_U  \g(x-y) dx d\mu(y) <+\infty,\end{equation} hence the claim.

{\bf  Step 2: The case $\s\le 0$, $\mu$ smooth and compactly supported.} Let us first assume that $\mu_\pm$ are both  bounded, Lipschitz and compactly supported densities.  This guarantees that $h^\mu$ is well-defined as $\int \g(x-y) \mu(y)dy $, and the same for $\nab h^\mu $ as 
$\int \g(x-y) \nab \mu(y)dy$.
The fact that $\int\mu_+=\int\mu_-$ implies that $h^\mu$ decays like $\frac{1}{|x|^{\s+1}}$ while $\nab h^\mu$ decays like $\frac{1}{|x|^{\s+2}}$. To see this it suffices to write 
$$h^\mu(x)= \int_{\R^\d} (\g(x-y) -\g(x) ) d\mu_+(y) - (\g(x-y)-\g(x))  d\mu_-(y)$$
and argue that $|\g(x-y)-\g(x)|\le \frac{C}{|x|^{\s+1}}$ for all $x$ large enough when $y$ remains in the compact support of $\mu_\pm$, and similarly for $\nabla h^\mu$. 

Using Green's formula, we have for any $R$,  $B_R$ being the ball centered at $0$ and of radius $R$ in $\R^{\d+\k}$ and $\nu$ the outer unit normal, 
\be\label{grreenfor}  \int_{B_R} \yg |\nab h^\mu|^2=  \int_{\partial B_R} \yg \frac{\partial h^\mu}{\partial \nu}  h^\mu- \int_{B_R} h^\mu  \div(\yg \nab h^\mu).
\ee The boundary term is bounded by $R^{-2\s-3}\int_{\partial B_R} \yg= R^{-2\s-3+\d+\k-1+\gamma}$, and using \eqref{defgamma} this is $R^{-\s-2}$ which tends to $0$ as $R \to \infty$ since $\s \ge -1$ in the cases we consider. Taking the limit, we obtain that 
$$\int_{\R^{\d+\k}}  \yg |\nab h^\mu|^2= - \int_{\R^{\d+\k} } h^\mu  \div(\yg \nab h^\mu)= \cds \int_{\R^{\d}} h^\mu \mu= \iint_{\R^\d\times \R^\d} \g(x-y) d\mu(x)d\mu(y).$$

{\bf Step 3: The general case $\s \le 0$.} Let us first consider, as in the proof of Lemma \ref{convexi},  a Gaussian mollification $\phi_\ep *\mu_\pm$ of $\mu_\pm$, where $\phi_\ep$ is a Gaussian approximation of the identity. Letting $\mu^\ep=\phi_\ep *\mu$, we have $\int \mu^\ep=0$, and   $\mu^\ep \to \mu$ weakly and in $\dot{H}^{\frac{\s-\d}{2}  }$. 
Finally, by standard properties of mollifiers, 
\be\label{limgepp}\lim_{\ep \to 0} \iint_{\R^\d\times \R^\d} \g(x-y) d\mu^\ep(x) d\mu^\ep(y)=  \iint_{\R^\d\times \R^\d} \g(x-y) d\mu(x) d\mu(y).\ee Moreover,
\eqref{formalHs} is correct for Gaussian decaying functions such as $\mu^\ep$, thus by finiteness of $\iint_{\R^\d\times \R^\d} 
 \g(x-y)d\mu(x)d\mu(y) $, the sequence $\{\mu^\ep\}_\ep$ is Cauchy in $\dot{H}^{\frac{\s-\d}{2}  }$ and converges in $\dot{H}^{\frac{\s-\d}{2}  }$ to $\mu$. 

Next, we may consider truncations $\mu^{\ep,n}_\pm$ which are bounded, Lipschitz and compactly supported,  such that $\int \mu^{\ep,n}_+ =\int \mu^{\ep, n}_-$, and $\mu^{\ep, n}_\pm\to \mu^\ep_\pm$ as $n\to \infty$, pointwise and with domination.
For instance, it suffices to restrict the $\mu^{\ep}_\pm$ to   the  compact level sets $\{\mu^\ep_\pm>\frac1n\}$, subtract off $1/n$, and then normalize by a factor that makes the restricted measures probabilities.
The result of Step 2  applies to $\mu^{\ep, n}:=\mu^{\ep, n}_+-\mu^{\ep,n}_-$ and yields that 
\be \label{relapl}
\iint_{\R^\d \times \R^\d} \g(x-y) d\mu^{\ep, n} (x)d\mu^{\ep, n} (y)= \int_{\R^{\d+\k}}\yg |\nab h^{\mu^{\ep,n}}|^2 = \|\mu^{\ep, n}\|_{\dot{H}^{\frac{\s-\d}{2}  }}^2.\ee 
We then need to take the limits $n\to \infty$, then $\ep \to 0$ in this relation. For the left-hand side, we find the desired $n\to \infty$ limit by dominated convergence theorem applied to the positive and negative parts of the integrand, and using the assumptions \eqref{finitepot}. 

Next, with the same argument, we have  for any $n, m$, 
$$ \iint_{\R^\d \times \R^\d} \g(x-y) d(\mu^{\ep, n}-\mu^{\ep, m}) (x)d(\mu^{\ep, n}-\mu^{\ep, m}) (y)= \int_{\R^{\d+\k}}\yg |\nab h^{\mu^{\ep,n}-\mu^{\ep, m}}|^2 = \|\mu^{\ep, n}-\mu^{\ep, m}\|_{\dot{H}^{\frac{\s-\d}{2}  }}^2.$$
Thus for fixed $\ep$, the sequence $\{\mu^{\ep, n}\}_n$ is a Cauchy sequence in $\dot{H}^{\frac{\s-\d}{2}  }$, hence $\{\nab h^{\mu^{\ep, n}}\}_n$ is also a Cauchy sequence in the weighted space $L^2_{\yg}(\R^{\d+\k})$  thus  $\nab h^{\mu^{\ep, n}}$  converges to some vector field $E$. Taking the weak limit in $h^{\mu^{\ep, n}}= \g*\mu^{\ep, n}$
we deduce we must have $E=\nab h^{\mu^\ep}$ and taking the limit in \eqref{relapl} we deduce that
\be\label{relapl2}\int_{\R^{\d+\k}}\yg |\nab h^{\mu^\ep}|^2= \|\mu^\ep\|_{\dot{H}^{\frac{\s-\d}{2}  }}^2= \iint_{\R^\d \times \R^\d} \g(x-y) d\mu^\ep(x)d\mu^\ep (y).\ee
As mentioned above, the sequence $\{\mu^\ep\}_\ep$ converges in $\dot{H}^{\frac{\s-\d}{2}  }$ to $\mu$.
Thus from \eqref{relapl2} applied to $\mu^\ep-\mu^{\ep'}$ we find that $\nab h^{\mu^\ep}$ is Cauchy  $L^2_{\yg}$, hence converges in that space to a limit, which also coincides with $\nab h^\mu$.  Finally, taking the limit $\ep \to 0$ in \eqref{relapl2}, in view of \eqref{limgepp} we obtain the result.
\smallskip

{\bf Step 4: The case $\s>0$.}
Assume first that $\mu$ has a bounded density and is compactly supported.  Then $h^\mu= \int \g(\cdot-y) d\mu(y)$ is finite $\mu$-a.e., moreover, $|h^\mu(x)|\le C |x|^{-\s}$ and $|\nab h^\mu(x)|\le C |x|^{-\s-1}$ for $|x|$ large enough, and 
$$\iint_{\R^\d\times \R^\d} \g(x-y) d\mu(x) d\mu(y)=\int_{\R^{\d+\k}} h^\mu d\mu.$$
Using \eqref{divyg} we may then write 
$$\iint_{\R^\d\times \R^\d} \g(x-y) d\mu(x) d\mu(y)=-\frac1{\cds}\int_{\R^{\d+\k}} h^\mu  \div(\yg \nab h^\mu) .$$
For any $R>0$,  as  in  \eqref{grreenfor}, we have 
$$\int_{B_R} h^\mu  \div(\yg \nab h^\mu)= \int_{\partial B_R} \yg \frac{\partial h^\mu}{\partial \nu}  h^\mu
-\int_{B_R} \yg |\nab h^\mu|^2.$$
Since $h^\mu$ decays like $|x|^{-\s}$ and $\nab h^\mu$ like $|x|^{-\s-1}$, the boundary term is bounded by $R^{\gamma+\d+\k-1 -2\s-1}$ which in view of \eqref{defgamma} is $R^{-\s}$ and tends to $0$ as $R\to +\infty$. Thus we conclude that 
$$ \int_{\R^{\d}} h^\mu  d \mu 
= - \frac{1}{\cds}\int_{\R^{\d+\k}} h^\mu  \div(\yg \nab h^\mu)= \frac1{\cds}\int_{\R^{\d+\k}} \yg |\nab h^\mu|^2.$$
 The case of general $\mu$ then follows by approximation as in Step 3. 
\end{proof}
\section{Linking the equilibrium measure with the obstacle problem in the Coulomb case} 
In Section \ref{sec2.1} we described the characterization of the equilibrium measure minimizing $\I$ via tools of potential theory. In this section, we return to this question and connect it instead to a well-studied problem in the calculus of variations called the {\it obstacle problem} in the Coulomb case and the {\it fractional obstacle problem} in the Riesz case. The connection with the classical obstacle problem is mentioned in passing in \cite{safftotik}.  It allows us to use the rich PDE theory developed for this problem, such as methods based on the maximum principle methods and regularity theory to obtain
additional information on $\meseq$, valid for the Coulomb case in any dimension. In two dimensions complex analytic methods provide a replacement via Sakai's regularity theory \cite{sakai}, providing information on regularity and connectivity \cite{hedenmakarov2,leemakarov}.
Finally,  the connection of the equilibrium measure in the Riesz cases  with the fractional obstacle problem is less well-known and we will describe it below.

\subsection{Short presentation of the obstacle problem}
\index{obstacle problem}
The obstacle problem is generally formulated over a bounded domain $\Omega \in \R^\d$:  given  an $H^1(\Omega)$   function $\psi : \Omega \rightarrow \R$ (called the {\it obstacle}), which is  nonpositive on $\partial \Omega$,  find  the function that achieves 
\begin{equation}
\label{obstaclepb} \min \left\lbrace \int_{\Omega} |\nabla h|^2 , \ h \in H^1_0(\Omega), \ h \geq \psi \right\rbrace.
\end{equation}
In PDE formulation, the solution is characterized by
\be \label{obstaclePDE}
\min( - \Delta h, h-\psi)=0.\ee

For general background and motivation for this problem, see e.g. \cite{kinderlehrerstamp,friedman,ckinder}. 

Here the space $H^1_0(\Omega)$ is the  Sobolev space of trace-zero functions which is the completion of $C^1_c(\Omega)$ ($C^1$ functions with compact support in $\Omega$) under the $H^1$-Sobolev norm $\|h\|_{H^1} = \|h\|_{L^2} + \|\nabla h\|_{L^2}$. The zero trace condition $h \in H^1_0(\Omega)$ may be replaced by different boundary conditions, e.g.~a translation $h \in H^1_0(\Omega) + f$, where $f$ is a given function. Note that the minimization problem \eqref{obstaclepb} is a convex minimization problem under a convex constraint, hence it has at most one  minimizer (it is not too hard to show that the minimum is achieved, hence there actually is a unique minimizer).

 An admissible function for \eqref{obstaclepb} has two options at each point : to touch the obstacle or not (and typically uses both possibilities).  If $h$ is the optimizer, the set $\{x\in \Omega, h(x)=\psi(x) \ q.e.\}$ is closed and called the {\it coincidence set} or the {\it contact set}.  It is unknown (part of the problem), and the obstacle problem thus belongs to the class of so-called {\it free-boundary problems}, see \cite{friedman}.

Trying to compute the Euler-Lagrange equation associated to this problem by perturbing $h$ by a small function, one is led to two possibilities depending on whether $h = \psi$ or $h > \psi$. 
In a region  where $h > \psi$, one can perform infinitesimal variations of $h$ of the form $(1-t)h + tv$ with $v$, say,  smooth  (this still gives an admissible function, i.e.~lying above the obstacle, as soon as $t$ is small enough) which shows that $\Delta h =0$  there (since the ``functional derivative" of the Dirichlet energy is the Laplacian). 
In the set where  $h=\psi$,  only variations of the same form $(1-t)h + tv$  but with $v\ge \psi$ (equivalent to $v\ge h$ there) and $t\ge 0$ provide  admissible functions,  and this  only leads to an inequality $-\Delta h\ge 0$ there. These two pieces of information can be grouped in the following more compact form:
\begin{equation}\label{varineq}
\textrm{for all } v \textrm{ in }  H^1_0 \textrm{ such that } v \geq \psi \textrm{ q.e., }\quad  \int_{\Omega} \nabla h \cdot \nabla (v-h) \geq 0.
\end{equation}
This relation is called a {\it variational inequality}, and 
 it uniquely characterizes the solution to \eqref{obstaclepb}, in particular the coincidence set is completely determined as part of the solution.

In Fig. \ref{fig2} below, we describe a few instances of solutions to   one-dimensional obstacle problems, and in Fig. \ref{fig3} to higher dimensional obstacle problems. 

\begin{figure}[h!]
\begin{center}
\includegraphics[width=9cm]{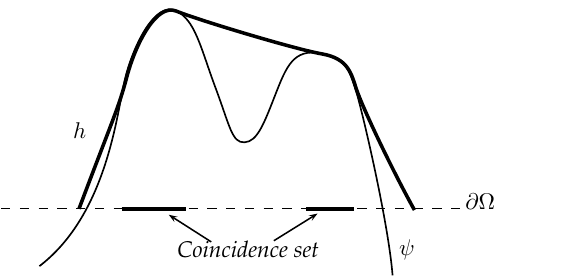}
\includegraphics[scale=1]{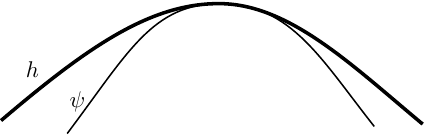}
\caption{The coincidence set for a one-dimensional obstacle problem}
\label{fig2}
\end{center} 
\end{figure}

\begin{figure}[h!]
\begin{center}
\includegraphics[scale=1]{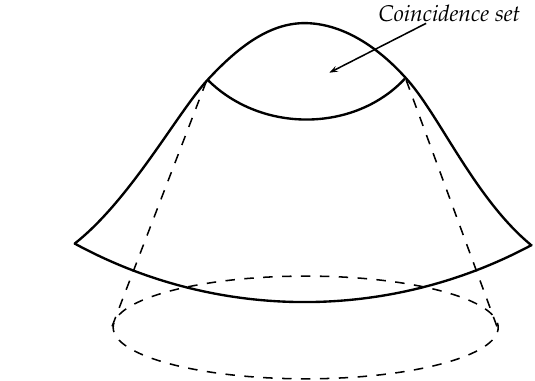}
\end{center}
\caption{A higher-dimensional obstacle problem}\label{fig3}
\end{figure}

\begin{figure}[h!]
\begin{center}
\includegraphics[scale=1]{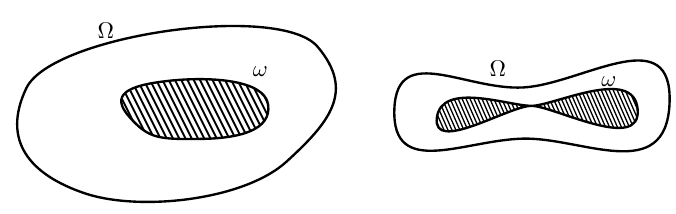}
\caption{Examples of coincidence sets}\label{fig4}
\end{center}
\end{figure}

The regularity theory of the solutions to obstacle problems and of their coincidence sets has been developed for many years,   culminating with the work of Caffarelli   (for a review see \cite{caff}). This sophisticated PDE theory shows, for example, that the solution $h$ is as regular as $\psi$ up to $C^{1,1}$ \cite{frehse}. The boundary $\partial \Sigma$ of the coincidence set is locally a $C^{1,\alpha}$ graph except for cusps \cite{caff}.  These are points of $\partial \Sigma$ at  which, locally, the coincidence set  can fit  in the region between two parallel planes  separated by an arbitrarily small distance (the smallness of the neighborhood depends of course on this desired distance).
Fig. \ref{fig4} gives examples of coincidence sets, a regular one, and one with cusps. Cusps are however nongeneric with respect to the obstacle, as recently shown in \cite{figalliros}.

Moreover, if $\psi$ is $C^{1,1}$, since $\nabla h$ is continuous, the graph of $h$ must lift off from  the coincidence set tangentially. This  formally leads to the following system of equations, where $\omega$ denotes the coincidence set~:
$$\left\{\begin{array}{ll}
-\Delta h =0 & \text{in} \ \Omega\backslash \omega
\\
h=\psi & \text{in} \ \omega\\
\frac{\partial h}{\partial \nu} = \frac{\partial \psi}{\partial \nu} & \text{on} \ \partial \omega\\
h=0 & \text{on} \ \Omega.\end{array}\right.$$
This relation cannot be made rigorous in all cases, because $\omega$ is not an open domain, however it gives the right intuition and is correct when $\omega $ is nice enough. Note that on the boundary of $\Omega \backslash \omega$ we must have a Dirichlet boundary  condition $h=\psi$, together with a Neumann  boundary condition $\frac{\partial h}{\partial \nu} = \frac{\partial \psi}{\partial \nu}$. These two boundary conditions make what is called an {\it overdetermined problem} and this 
overdetermination explains why there is only one possible coincidence set.

\subsection{Connection between the two problems}
The problem we examined, that of the minimization of $\I$, is phrased in the whole space, and not in  a bounded domain. While the minimization problem \eqref{obstaclepb} may not have  a meaning over all $\R^\d$ (because the integral might not converge), the corresponding variational inequality \eqref{varineq} can still be given a meaning over $\R^\d$ as follows~:
given $\psi \in H^1_{\mathrm{loc}}(\R^\d)$ solve for 
\begin{equation}\label{obspbrd}
\forall  v\in \mathcal{K}, \quad  \int_{\R^\d} \nab h \cdot \nab (v-h) \ge 0\end{equation}where 
 $$\mathcal{K}= 
\left\{ v\in H^1_{\mathrm{loc}}(\R^\d) \ \textrm{such that} \   v-h \textrm{ has bounded support and} \  v \ge \psi  \  \textrm{q.e.}\right\}.$$
Solving this is in fact equivalent to the statement that  for every $R>0$, $h$ is the unique solution to 
$$\min\left\{\int_{B_R} |\nab v|^2, v\in H^1(B_R), v -h\in H^1_0(B_R), v\ge \psi \ \textrm{in} \ B_R\right\},$$
which replaces \eqref{obstaclepb}.
The PDE 
\be\label{obspde}
\min(-\Delta h, h-\psi)=0\ee
also still makes sense. We refer to \cite{serser} for more detail on the full space setting. 

The problem \eqref{obspbrd} is easily seen to have a unique solution~: if there are two solutions $h_1$ and $h_2$  it suffices to apply \eqref{obspbrd} for $h_1$, with $h_2$ as a test-function, and then reverse the roles of the two and add the two relations to obtain $h_1=h_2$.

Let us now  compare  the two problems side by side.
\begin{description}
\item[Equilibrium measure]
 $\meseq$ is characterized by the relations 
 \begin{equation} \left\lbrace \begin{array}{cl} h^{\meseq} + V \geq c & \mbox{quasi everywhere} \vspace{1mm} \\ 
h^{\meseq} + V = c & \textrm{q.e. in  the support of } \ \meseq.
\end{array} \right.
\end{equation}

\item[Obstacle problem]
\begin{equation}
\left\lbrace \begin{array}{cl} 
h \geq \psi & \mbox{q.e.}  \vspace{1mm}\\
h = \psi & \mbox{q.e.  in the coincidence set.}
\end{array}\right.
\end{equation}
\end{description}

It is then not surprising to expect a correspondence between the two settings, once one chooses the obstacle to be  $\psi = c - V$. 
\begin{prop}[Equivalence between the minimization of $\I$ and the obstacle problem]\label{proequivpb}
\mbox{}
Assume $\s=\d-2$, $V$ is continuous and satisfies \eqref{A2}. If $\meseq$ is the equilibrium measure associated to the potential $V$ as in Theorem \ref{theoFrostman}, then its potential $h^{\meseq}$, as defined in \eqref{defhmu0},  is the unique solution to the obstacle problem with obstacle $\psi = c - V$ in the sense of \eqref{obspbrd}. 
If in addition  $V\in C^{1,1}$ then $\meseq=( \frac{1}{\cd} \Delta V) \indic_{\omega}$ where $\omega$ is the coincidence set $\{h^{\meseq}= c- V\}= \{\zeta=0\}$.
\end{prop}
Note that the converse might not be true, because a solution of the obstacle problem can fail to provide (via taking $-\frac{1}{\cd}\Delta h$) a {\it probability} measure, however it does in general when shifting $c$ appropriately.

When one works on a bounded domain, this result can be obtained by observing that the problem of minimizing $\I$ and that of minimizing \eqref{obstaclepb} are essentially convex duals of each other (see \cite{brezis,breziss}). When working in an infinite domain, the correspondence is probably folklore and could also be worked out by convex duality,  but we were not able to find it completely written in the literature, except for \cite{hedenmakarov2} which follows a slightly different formulation. Here, for the proof, we follow the approach of  \cite{asz} where the result is established in dimension 2 for the particular case of $V$ quadratic (but with more general constraints). The adaptation to any dimension and to general $V$'s is not difficult. Note that further details on the formulation in the whole space are given in \cite{serser}, where in this setting the stability of the coincidence set  and its  precise dependence with respect to the obstacle function is studied.


\begin{proof}[Proof of Proposition \ref{proequivpb}]
{\bf Step 1.} Let us show that $\nab h^{\meseq}$ is in $L^2_{\mathrm{loc}} (\R^\d)$. This is a consequence of the fact that $\I(\meseq)<\infty $ hence, in view of the assumptions on $V$, $\iint \g(x-y) \, d\meseq(x) \, d\meseq(y)<+\infty$.
In the case $ \s>0$, it is a direct consequence of Proposition \ref{lemelect1}, in fact $\nab h^{\meseq}\in L^2(\R^\d)$. In the case $\s\le 0$, we  consider a reference probability  measure $\bar{\mu}$ for which $h^{\bar\mu}$ is $C^1_{\mathrm{loc}}(\R^2)$. It suffices to consider for example $\bar{\mu}= \frac{1}{\pi } \indic_{B_1}$, the circle law, for which $h^{\bar\mu}$ is radial and can be computed explicitly.
Then, let us  consider $\ro = \meseq - \bar{\mu}$. Since  $\int \, d\rho=0$, and $\iint \g(x-y) \, d\mu(x)\, d\mu(y)<\infty$ holds for both $\mu=\meseq$ and $\mu= \bar\mu$,  we then obtain by Proposition~\ref{lemelect1} that $\nab h^{\meseq}-\nab h^{\bar \mu}\in L^2(\R^\d)$ (we are in the Coulomb case where $\gamma=0$), from which  we deduce $\nab h^{\meseq}$ is also in $L^2_{\loc} (\R^2, \R^2)$, as desired.
\smallskip 

{\bf Step 2.}
Let $v$ be admissible in \eqref{obspbrd}, i.e.~belong to $\mathcal{K}$, and set $\vp= v- h^{\meseq}$.  If $\vp$ is smooth and compactly supported, then, integrating by parts, we have
\be \label{vi1}
\int_{\R^\d}  \nabla h^{\meseq} \cdot \nabla (v - h^{\meseq}) = \cd \int_{\R^\d}   \vp \, d \meseq\ge 0.
\ee 
Indeed, by \eqref{EulerLagrange}, we know that $h^{\meseq} = \psi$ q.e.~in the support of $\meseq$ and by assumption $v \geq \psi$ q.e.~in $\R^\d$. Hence $\vp$ is q.e.~nonnegative on the support of $\meseq$  and the inequality \eqref{vi1} follows, since $\meseq$ does not charge sets of zero  capacity.
To obtain \eqref{vi1} for any   $v\in \mathcal{K}$,  it suffices to show that the subset of $\mathcal{K}$ consisting of $v$'s for which $v-h^{\meseq}$ is smooth and compactly supported is dense in $\mathcal{K}$  for the topology of $H^1$.   Fix some $v$ in the admissible set and $R>1$ such that $v-h^{\meseq}$ is supported in $B_{R/2}$. Let $\eta_\eps$ be a standard mollifier and $\chi_R$ a smooth function supported in $B_{2R}$ with $0\le \chi_R \le 1$ and $\chi_R \equiv 1$ in $B_R$. One may check that  
$$v_{\eps,  \delta}= h^{\meseq}+ (v-h^{\meseq}) * \eta_\eps+  \delta \chi_R$$ satisfies  that $v_{\eps,\delta}- h^{\meseq}$ is smooth and  approximates  $v$ arbitrarily well in $H^1$ when $\delta $ is small enough, and is $\ge \psi$ when $\eps $ is chosen small enough relative to $\delta$.
This concludes the proof of \eqref{vi1}.
\smallskip

{\bf Step 3.} We prove the statements about $\meseq$.
First, since the coincidence set $\omega$ is closed, its complement is open, and the function $h^{\meseq}$ is harmonic on that set. One can note also that in view of   
 \eqref{EulerLagrange} and the definition of the coincidence set $\omega$, the support of $\meseq$ is included in $\omega$ up to a set of capacity $0$.

If we assume that  $V\in C^{1,1}_{\mathrm{loc}}$, then by Frehse's regularity theorem mentioned above \cite{frehse},  $h^{\meseq}$ is also $C^{1,1}_{\mathrm{loc}}$. In particular $h^{\meseq}$ is continuous, and so is $V$, so the relations \eqref{EulerLagrange} hold pointwise and not only q.e. This means that we have 
\be \label{eag} h^{\meseq} + V= c \ \text{ on} \  \omega\ee  and $\supp \,\meseq\subset \omega$.
Also $C^{1,1}_{\mathrm{loc}}= W^{2,\infty}_{\mathrm{loc}}$ hence $\Delta h^{\meseq}$ and $\Delta V$  both make sense as  $L^\infty_{\mathrm{loc}}$ functions, and it suffices to determine $\meseq$ up to sets of measure $0$. We already know that $\meseq=0$ in the complement of $\omega$ since $h^{\meseq}$ is harmonic there, and it suffices to determine it in $\interieur{\omega}$.  But taking the Laplacian on both sides  of \eqref{eag},
since $\meseq= - \frac{1}{\cd} \Delta h^{\meseq}$,  one finds
$$\meseq=  \frac{1}{ \cd} \Delta V \ \text{in} \ \interieur{\omega},$$
and the results follows.

\end{proof}

Since $\meseq$ is a compactly supported probability measure, we have that $h^{\meseq} = \int \g(x-y)\, d\meseq(y)$ asymptotically behaves like $\g(x)$ as $|x|\to \infty$. Since $h^{\meseq}+  V = c $ q.e.~in $\omega$ and since \eqref{A2} holds, it follows that $\omega $ must be a bounded, hence compact, set.

We have seen that $\omega$ contains, but is not always equal to,  the support of $\meseq$, called the droplet.  In \cite{hedenmakarov2}, it is  discussed how these two sets differ in the two-dimensional case: under a logarithmic growth condition for $V$, they are equal except at so-called ``shallow points."

\section{The fractional obstacle problem and link with the equilibrium measure in the Riesz case}\label{sec2.4}
\index{fractional obstacle problem}
We recall that the fractional Laplacian can be  defined via Fourier transform or by 
\eqref{deffraclap}.
\subsection{The fractional obstacle problem}
The study of the fractional obstacle problem is more recent than that of the classical obstacle problem, the first main references are  \cite{caffss,silvestreobstacle}.
The definition of the problem in our setting is
\be\label{eqfractobpb}\begin{cases}
h \ge \psi & \text{in}\ \R^\d\\
(-\Delta)^{\alpha} h \ge 0 & \text{in} \ \R^\d\\
(-\Delta)^{\alpha} h(x) = 0 & \text{for those $x$ such that $ h(x) >\psi(x)$}\\
\lim_{|x|\to \infty} \frac{h}{\g}=1,\end{cases}\ee
which can also be rewritten as 
\begin{equation}\label{eqfractobpb2} \min \( (-\Delta)^\alpha h, h- \psi\) = 0,\end{equation}
playing the analogue role of \eqref{obstaclePDE}.
The problem  can be expressed variationally as 
$$\min_{h \ge \psi}  \int |\nabla^{\alpha}h|^2 := \min_{h \ge \psi}  J(h),  $$
where 
$$J(h)= \iint_{\R^\d\times \R^\d} \frac{|h(x)-h(y)|^2}{|x-y|^{\d+2\alpha} }    dxdy.$$
Again the set $\{h=\psi\}$ is called the coincidence set. 

This problem is directly related to another previously studied problem, which has an easier visual interpretation: the thin obstacle problem or  Signorini problem. 

\subsection{The Signorini or thin obstacle problem}
It  is an obstacle problem in dimension $\d+1$ 
with obstacle in $\R^\d$. Formally, given a function $\psi$  over $\R^\d$, the minimization problem is
$$\min\left\{\int_{\R^{\d+1}} |\nab h|^2,  \   h \in H^1(\R^{\d+1}),\  h(x, 0) \ge \psi (x) \ \text{for all}\ x \in \R^\d    \right\}$$
Solutions are harmonic functions away from $\R^\d\times \{0\}$. The jump of the normal derivative of $h$ across $\R^\d\times \{0\}$ is equal to (twice) the half-Laplacian of $h$. 
This allows to give an interpretation of the fractional obstacle problem when $\alpha=1/2$. For $\alpha \neq 1/2$, one replaces instead $\int |\nab h|^2$ by $\int \yg  |\nab h|^2$ 
where $y$ is the last coordinate in $\R^{\d+1}$ and  $\gamma$ is as in \eqref{gamma}.

Now the characterization of the equilibrium measure \eqref{EulerLagrange}
can be rewritten as 
$$\begin{cases} h^{\meseq} \ge \psi \quad & \text{in} \ \R^\d\\
h^{\meseq} = \psi& \text{in the support of $\meseq$}\\
\end{cases}$$
after setting $\psi= c- V $.
On the other hand, since the $\g$ of \eqref{riesz}
is the kernel of the fractional Laplacian $(-\Delta)^{\frac{\d-\s}{2}}$ up to a constant, we have 
$$(-\Delta)^{\frac{\d-\s}{2}}   h^{\meseq}= \cds \meseq.$$
Thus, setting $\alpha=\frac{\d-\s}{2}$, the potential $h^{\meseq}$ satisfies that either $h^{\meseq}=\psi$ or $h^{\meseq}(x)>\psi(x)$ and $x$ is not in the support of $\meseq$, hence 
$$\min \(  h^{\meseq}- \psi, (-\Delta)^\alpha h^{\meseq}\)=0, \quad \text{in} \ \R^\d$$ and the condition at infinity is also satisfied,  so we have found  exactly 
 \eqref{eqfractobpb2},
or equivalently \eqref{eqfractobpb}.

The thin obstacle problem is easier to visualize than the fractional obstacle problem: one considers an obstacle that lives only on $\R^\d$,  and extends an elastic sheet above it. The sheet is free away from $\R^\d\times \{0\}$, thus leading to a harmonic (or $\div (\yg \nabla )$-free) function away from it. But there is a jump of normal derivative across $\R^\d\times \{0\}$ where the obstacle is touched. The jump of $\yg \nab u$ across $\R^\d\times \{0\}$ is equal to  $-\div (\yg \nab u)$ in the sense of distributions, hence to the fractional equilibrium measure, which is supported on the coincidence set.
One could also write the analogue of Proposition \ref{proequivpb}.

We have seen how the correspondence between the minimization of $\I$ and the obstacle problem thus allows, via the regularity theory of the obstacle problem, to identify the equilibrium measure in terms of $V$ when the former  is regular enough. The known techniques on the obstacle problem \cite{caff} and on the fractional obstacle problem \cite{caffss,caffros} also allow for example  to analyze the rate at which the solution leaves the obstacle:
at regular points of the boundary of the coincidence set $\omega$  \footnote{These regular points are expected to be generic, as seen before this is proved in \cite{figalliros} in  the case $\alpha=1$.} we have 
$$h-\psi = c \, \dist(x, \omega)^{ 1+\alpha} +o(\dist(x, \omega)^{ 1+\alpha}),$$ for some $c>0$, 
 which gives us information on the growth rate of the function $\zeta$ defined in \eqref{defzeta} of the form 
 \be \label{assumpzeta6}
 \zeta(x) \ge c\, \dist ( x, \omega)^{1+\alpha}.\ee
  Also, at regular points, the vanishing of the equilibrium measure is at the rate 
 $$(-\Delta)^\alpha h\sim c'\, \dist(x,  \omega^c )^{1-\alpha},$$
 where $\dist$ denotes the signed distance and $c'$ is equal to $c$ times some constant depending only on $\alpha$.
 For  more details and proofs of these estimates in our particular setting, we refer to \cite{rosotonapp}.

\section{The thermal equilibrium measure}\label{secthermal}\index{thermal equilibrium measure}
We now turn to the study of the minimization of \eqref{defEtheta}, and justify the existence and uniqueness of its minimizer, which 
we call the {\it thermal equilibrium measure}.  For this we will need a thermal version of assumption \eqref{A2}:
$$\text{(\namedlabel{A4}{A4}) } 
\int_{\R^\d} \exp\( - \theta (V(x)+ \g_-(x))\) dx<\infty$$
where $\g_-$ denotes $\min(\g, 0)$. Note that $\g_-=0$ for $\s>0$. We will also denote $\g_+=\max(\g, 0)$, so $\g=\g_++\g_-$.

\begin{prop}[Existence and uniqueness of the thermal equilibrium measure]\label{lem241} For every $\theta>0$, 
if \eqref{A1}--\eqref{A4}  hold, then $\mathcal E_\theta$ is lower semi-continuous, has compact sub-level sets, and  has a unique minimizer.\end{prop}
\begin{proof} Let us start with the case $\s=0$. We use \eqref{gVV} to obtain that  for any probability density $\mu$, we have
\begin{multline*}
 \mathcal E_\theta(\mu) \ge 
 \int_{\R^\d}V  d\mu - \iint_{|x|\ge |y|} \log \max(|x|,1)d\mu(x) d\mu(y)  - C + \frac{1}{\theta} \int_{\R^\d} \mu \log \mu\\
  \ge  \int_{\R^\d} V d\mu - \int_{\R^\d} (\log |x| )_+ d\mu(x)  - C + \frac{1}{\theta} \int_{\R^\d} \mu \log \mu.\end{multline*}
We  obtain in all cases (using the fact that $\g\ge 0$ in the cases $\s>0$, and \eqref{gVV} in the case $\s<0$) 
 \be\label{casdger}
 \mathcal E_\theta(\mu) \ge - C +\int_{\R^\d}( V+\g_- )  d\mu    + \frac{1}{\theta}\int_{\R^\d} \mu \log \mu. 
  \ee  
  We then observe that the function  $\phi(\mu)=   \alpha \mu + \frac{1}{\theta} \mu\log \mu$ achieves its minimum at $u=\exp(-\theta \alpha -1)$ and that the minimum equals $-\frac{1}{\theta} \exp(-\alpha \theta -1)$.
  This leads us to introducing 
  \be\label{defurho}
  u(x)= \exp(-\theta (V+\g_-)(x)) ,\quad \bar u= \int_{\R^\d} u, \quad \rho= \frac{u}{\bar u}\ee which are well-defined  thanks to \eqref{A4}.  
  We may then rewrite 
  \begin{multline}\label{curcE}
 \mathcal E_\theta(\mu) 
= \hal \iint_{\R^\d\times \R^\d}\( \g(x-y)- (\g_-(x)+\g_-(y)) \)  d\mu(x)d\mu(y)  
\\  + \int_{\R^\d} (V+\g_- +\frac1\theta \log \rho) d\mu + \frac1\theta \int_{\R^\d} \mu \log \frac{\mu}{\rho} \\
  =\hal \iint_{\R^\d\times \R^\d}\( \g(x-y)- (\g_-(x)+\g_-(y))  \)  d\mu(x)d\mu(y)  - \frac1\theta \log \bar u
+ \frac1\theta \int_{\R^\d} \mu \log \frac{\mu}{\rho}.\end{multline}
    The first integral in the right-hand side is bounded below  by properties of $\g$ (already seen in the proof of \eqref{gVV}).  The last one is recognized as the relative entropy of $\mu$ with respect to the probability density $\rho$, which is always nonnegative. 
 We deduce that  $\inf \I_\theta >-\infty$. 
 On the other hand we get $\inf \I_\theta<+\infty$,  for instance by taking $\meseq$ as a test probability.
 
 Let now $\{\mu_n\}_n$ be such that $\I_\theta(\mu_n) \le C$ independent of $n$. It is well-known that the relative entropy is lower semi-continuous and its sub-level sets are compact, so we may extract a subsequence (not relabelled) such that $\mu_n\to \mu$  in the weak sense of probabilities.  Hence $\I_\theta$ has compact sublevel sets. 
 Moreover, if $\mu_n\to \mu$ we have 
  \be\label{relentcv}\liminf_{n\to \infty} \int  \mu_n \log \frac{\mu_n}{\rho}\ge \int \mu \log \frac{\mu}{\rho}.\ee
 Using the fact that  $  \g(x-y)- (\g_-(x)+\g_-(y))$ is bounded below,  and the same truncation  procedure as in the proof of Lemma \ref{coerI}, we also have 
 \begin{multline*}\liminf_{n\to \infty}\iint_{\R^\d\times \R^\d}\( \g(x-y)- (\g_-(x)+\g_-(y)) \)  d\mu_n(x)d\mu_n(y)\\ \ge  
 \iint_{\R^\d\times \R^\d}\( \g(x-y)- (\g_-(x)+\g_-(y)) \)  d\mu(x)d\mu(y)  \end{multline*}
 which together with \eqref{relentcv} and \eqref{curcE} implies that  $\liminf_{n\to \infty} \I_\theta(\mu_n) \ge \I_\theta(\mu)$, i.e. we have obtained that $\I_\theta$ is l.s.c.~and admits a minimizer.

  The uniqueness of a minimizer is by strict convexity: $\I$ itself is strictly convex and $\mu \mapsto \int \mu \log \mu$ as well. 
\end{proof}

\begin{rem} The situation is a bit subtle because even though $\I_\theta(\mub)<+\infty$ the entropy $\int_{\R^\d} \mu \log \mu$ may not be finite, but rather it is the relative entropy $\int_{\R^\d} \mu \log \frac{\mu}{\rho}$ that is finite. There is a  possible  cancellation between $V$ and $\frac1\theta \log \mub$ before they get integrated against $\mub$. 
\end{rem}
We are now led to introducing a new assumption
$$\text{(\namedlabel{A5}{A5}) } \quad \exists \epsilon>0 \ \text{such that } \ 
 \int_{\R^\d}\exp\(\epsilon |\g_-| - \theta (V +\g_-)\) <\infty$$
 
\begin{lem} \label{1entraine2}
If \eqref{A1}--\eqref{A5} holds, we have 
\be\label{1elem23}\iint |\g(x-y)| d\mub(x)d\mub(y)<\infty\ee 
and 
\be\label{2elem23}
\int |\g_-|(x) d\mub(x)<+\infty.\ee
\end{lem}
\begin{proof} {\bf Step 1: proof of \eqref{2elem23}.}
We have seen that under \eqref{A4} (which is implied by \eqref{A5}), $\inf \I_\theta<+\infty$ and $\mub$ exists. 
 Returning  to \eqref{curcE} we have  
\begin{multline}\I_\theta(\mub)=\hal \iint_{\R^\d\times \R^\d}\( \g(x-y)- (\g_-(x)+\g_-(y))  \)  d\mub(x)d\mub(y)\\  - \frac1\theta \log \bar u
+ \frac1\theta \int_{\R^\d} \mub \log \frac{\mub}{\rho},\end{multline}
where $\rho$ and $\bar u$ are as in \eqref{defurho}.
As seen in the proof of  Proposition  \ref{lem241} both nonconstant terms are bounded below  and  we deduce that 
\be \label{254}\iint_{\R^\d\times \R^\d}\( \g(x-y)- (\g_-(x)+\g_-(y))  \)  d\mub(x)d\mub(y) <+\infty\ee
and 
\be  \label{256} \int_{\R^\d} \mub \log \frac{\mub}{\rho}<+\infty.\ee

By  the  Donsker-Varadhan lemma (which exploits the fact that the relative entropy and log of exponential moment are dual functions to each other), we can  write that for any $\epsilon>0$
\be \epsilon \int_{\R^\d} |\g_-|d\mub \le \log \int_{\R^\d} e^{\epsilon |\g_-|} d \rho+ \int_{\R^\d} \mub\log \frac{\mub}{\rho} .
\ee
Since by \eqref{256} the relative entropy is bounded, by \eqref{A5} and definition of $\rho$, we obtain  \eqref{2elem23}.
\smallskip

{\bf Step 2: proof of \eqref{1elem23}.}
In the case $\s >0$, $\g_-=0$ so the desired result holds from \eqref{254}.
In the case $\s\le0$, it follows from \eqref{254} that   
\begin{equation}\label{iii}
\iint \g(x-y) d\mub(x)d \mub(y)<+\infty.\end{equation}
Let us show the corresponding lower bound. Using \eqref{gVV}, we may write 
\begin{multline}\label{iiii}  \iint \(\g_-(x-y) d\mub(x)d \mub(y)\ge \iint (\g_-(x)+\g_-(y)) \indic_{\s<0} + \g_-(x) \wedge\g_-(y) \indic_{\s=0} \)d\mub(x) d\mub(y)\\ 
\qquad \ge 2 \int \g_-(x)d \mub(x)>-\infty\end{multline}
in view of \eqref{2elem23}. In the case $\s<0$ since $\g$ is negative we are done. In the case $\s=0$, since $$\iint |\g(x-y)|d\mub(x) d\mub(y)= \iint \g(x-y)d\mub(x)d\mub(y) -2\iint \g_-(x-y) d\mub(x)d\mub(y)$$ the desired result follows from combining \eqref{iii} and \eqref{iiii}.
\end{proof}

As can be guessed formally, when $\theta\to \infty$, $\mu_\theta $ converges to  the (regular) equilibrium measure $\meseq$.  
This convergence is studied in detail and made quantitative in the Coulomb case in  \cite{ascomp} and can serve to approximate the solution to the obstacle problem.

By contrast with $\muv$, $\mub$ is regular and  not compactly supported, but always positive in $\R^\d$ with (typically) exponentially decaying tails. A boundary layer lengthscale equal to $\frac{1}{\sqrt{\theta}}$ appears in the convergence,  as can be seen in Theorem \ref{th1as} below.

A decay rate in $ \exp\(- \theta(V+\g_-)\)$ can be proven by using the maximum-principle-based  proof provided in \cite{ascomp} in the Coulomb case (which can be extended without too much difficulty to the Riesz case).
The starting point of that argument is to exploit the Euler-Lagrange equation solved by $\mub$ minimizing $\I_\theta$, which we now derive. 
\begin{prop}[Characterization of the thermal equilibrium measure]\label{lem242}
Assume $V$ is locally bounded and \eqref{A1} -- \eqref{A5} hold. 
Let $\theta>0$. If 
 $\mu_\theta$ is a probability measure  that minimizes $\I_\theta$, then  it  is a measure with a density  satisfying 
$$\mu_\theta>0\ \text{in} \ \R^\d \quad \text{a.e.}$$
and, for $h^{\mub}= \g* \mub$,  we have that $h^{\mub} \in L^1_{\mathrm{loc}}(\R^\d) $ and 
\be \label{eqhmub} 
h^{\mu_\theta} +V +\frac{1}{\theta}\log \mu_\theta = c_\theta \quad \text{a.e.} \ee for some constant $c_\theta$.
Moreover, the density $\mub$ is bounded above and locally bounded below by positive constants. If in addition $V\in C^2$, then $\mub$ is continuous and  we can get estimates for $\|\mub\|_{L^\infty}$ in terms of $\Delta V$. 
\end{prop}
\begin{rem}
As we will see in the proof, the Euler-Lagrange equation \eqref{eqhmub}  takes an interesting PDE form when differentiated: taking the Laplacian of \eqref{eqhmub} we obtain 
\be\label{eqmub} -\frac1\theta \Delta \log \mub + \cds (-\Delta)^{1+\frac{\s-\d}{2}}\mub-\Delta V =0.\ee
This is also called the Kirkwood-Monroe equation in physics, it is  an elliptic PDE for $\log \mub$, with a right-hand side of lower order of differentiation. It is however not a uniformly elliptic PDE, but once written in divergence form as 
\be\label{eqmubdf}-\frac1\theta\div \frac{\nab \mub}{\mub}= - \cds (-\Delta)^{1+\frac{\s-\d}{2}}\mub+\Delta V,\ee once can apply regularity theory for it in the set where $\mub$ has good bounds from below (which will naturally be $\Sigma$, the support of the equilibrium measure $\meseq$): this is the strategy of \cite{ascomp}.
One should also point out that in the Coulomb cases, rewriting it as an equation for $u_\theta=\log \mu_\theta$, leads to the  elliptic PDE
\be \label{pdeent} -\Delta u = -\cd \theta e^u + \theta \Delta V,\ee 
which  is a  generalization of the Kazhdan-Warner equation for prescribed curvature, and coincides with it if $\Delta V$ is constant.
\end{rem}

\begin{proof}
{\bf Step 1: definition of the potential. } We first check that $h^{\mub}\in L^1_{\mathrm{loc}} (\R^\d)$.
Let $U$ be a bounded set in $\R^\d$, we have
\begin{equation}\label{hmubb2}
\int_U h^{\mub}=\int_{\R^\d}\int_U  \g(x-y) dx d\mub(y).\end{equation}
If $\s>0$ it is then straightforward, by integrability and boundedness of $\g$, that the right-hand side is finite. If $\s\le 0$, this follows from  \eqref{gVV} and  \eqref{2elem23}.
\smallskip

 {\bf Step 2: positivity. } It is standard that the entropy functional is finite only if $\mu$ is a measure which is absolutely continuous with respect to the Lebesgue measure.  To show that $\mub>0$, we follow \cite{RSY2} and  assume by contradiction that there exists a set $S$ of nonzero Lebesgue measure on which $\mub=0$ and set 
$$\nu= \frac{\mub+ \ep \indic_S}{1+\ep |S|}.$$  
 Let us   expand out  
\begin{align*}
&
\I_\theta \left( \frac{\mub+\ep\indic_S}{1+\ep |S|} \right)
 = \mathcal \I_\theta(\mub)
-\ep |S|  \( \iint \g(x-y) d\mub(x) d\mub(y) + \int V d\mub+ \frac{1}{\theta} \mub \log \mub\)  
\\ & \qquad 
+\ep \int_S ( h^{\mub} +V) 
+ \frac{|S|}{\theta} \ep \log \ep + O(\ep^2).
\end{align*}
By Step 1, since  $S$ can be taken as  bounded and $V$ takes finite values,  we have that $\int_S h^{\mub}+ V <\infty$. We deduce that 
\begin{equation*}
\I_\theta \left( \frac{\mub+\ep\indic_S}{1+\ep |S|} \right)
\leq
\I_\theta(\mub) + C\ep + \frac{|S|}{\theta} \ep \log \ep,
\end{equation*}
a contradiction with the minimality of $\mub$ if $|S|>0$ when 
$\ep$ is chosen small enough.
\smallskip

{\bf Step 3: Euler-Lagrange equation.} 
 For every smooth  compactly supported function~$f$ such that~$\int f d\mub=0$ and~$t\in\R$ with $|t|$ sufficiently small, $(1+t f) \mub$ is a probability measure and we may expand 
 $$\I_\theta(\mub) \le \I_\theta ((1+t f)  \mub) $$
to find 
$$t \int_{\R^\d} (h^{\mub}+ V + \frac{1}{\beta}\log \mub) f d\mub  + O(t^2)\ge 0,$$  where $h^{\mub}$  may  take infinite values.
Since this is true for all small enough $|t|$ and any smooth $f$ with $\int fd\mub=0$,  and since $\mub>0$ almost everywhere, it follows that \eqref{eqhmub} holds almost everywhere, for some constant $c_\theta$.
\smallskip

{\bf Step 4: Boundedness from above and below.}
 First, we claim that $h^{\mub}+V$ is bounded below in $\R^\d$. 
Indeed, we may write 
$$h^{\mub}(x)+V(x)=\int_{\R^\d} (\g(x-y)-\g_-(x)) d\mub(y) +\g_-(x)+V(x).$$
On the one hand $\g_- +V$ is bounded below by \eqref{A2} and definition of $\g$. On the other hand, using \eqref{gVV} we have 
$$\int_{\R^\d} (\g(x-y)-\g_-(x)) d\mub(y) \ge \int_{\R^\d} \(\g(x)\wedge\g(y)\indic_{\s=0}+ 
(\g(x)+\g(y) )\indic_{\s<0} -C -\g_-(x)   \) d\mub(y).$$
In the case $\s=0$  we deduce that
$$\int_{\R^\d} (\g(x-y)-\g_-(x)) d\mub(y) \ge \int_{|y|\ge |x|} (\g(y)-\g_-(x)-C )d\mub(y)
\ge \int_{\R^\d} \g_-(y)d\mub(y) -C ,  $$
and the right-hand side is bounded below in view of  \eqref{2elem23}.
In the case $\s<0$, we find  instead
$$\int_{\R^\d} (\g(x-y)-\g_-(x)) d\mub(y) \ge \int_{\R^\d} (\g(y)  -C) d\mub(y) \ge \int_{\R^\d} \g_-(y)  d\mub(y)- C,$$ and we conclude again by \eqref{2elem23}.

From the fact that $h^{\mub}+V$ is bounded below and \eqref{eqhmub} it follows that $\log \mub$, hence $\mub$ is bounded above.

Next, we argue that $h^{\mub}$ is bounded above. 
Using that $\mub$ is bounded above, we may write 
\begin{multline*}
h^{\mub}(x)= \int_{y\in B(x,1)} \g(x-y)d\mub(y)+ \int_{y\notin B(x,1)} \g(x-y)d \mub(y)
\\
\le \|\mub\|_{L^\infty(\R^\d)}\int_{y\in B(0,1)} \g(y)dy + \int_{y\notin B(x, 1)} \g_+(x-y) d\mub(y) 
\le C \|\mub\|_{L^\infty(\R^\d)} + \frac1{\s}\indic_{\s>0} \int d\mub \end{multline*}
where we used that $\g_+(x)  \le1/\s \indic_{\s>0}$ for $|x|\ge 1$ and $\g$ is integrable near the origin.
We thus deduce that $h^{\mub}+V$ is locally bounded above, hence in view of \eqref{eqhmub}, $\mub$ is locally bounded below.
\smallskip

{\bf Step 5: Continuity and upper bound.} Here we give a sketch of how to deduce continuity of $\mub$ and estimates for its maximum. 
 If $V$ has bounded second derivatives, we can take the Laplacian of \eqref{eqhmub} to find  \eqref{eqmub}, and 
once we have shown that~$\mub$ is locally bounded above and below by positive constants, we may rewrite it  as  \eqref{eqmubdf}, a    uniformly elliptic equation in divergence for $\mub$.
By standard elliptic regularity theory (for instance \cite{giltrudinger}) in the Coulomb case $\s=\d-2$, or by fractional elliptic regularity theory otherwise, using that $\s<\d$, we  deduce that $\mub$ is as regular as $V$, in particular $\mub$ is continuous.

Once $\mub$ is continuous,  let $x_0\in B_R$ be a point where it achieves its maximum $m_\theta$, evaluating \eqref{eqmub} at $x_0$, we obtain $$0\ge \frac1\theta \Delta\log \mub(x_0)= \cds(-\Delta)^{1-\frac{\d-\s}{2}} \mub(x_0)-\Delta V(x_0)\ge C_{\d,\s} m_\theta^{\frac{\d+2-(\d-\s)}{\d}}-\max_{B_R}\Delta V$$
by Constantin-Vicol's ``nonlinear maximum principle" \cite{constvicol}, hence we deduce an upper bound for $m_\theta$ in terms of $\Delta V$.
\end{proof}

We now review the results obtained in \cite{ascomp} on the Coulomb $\d\ge 2$ case (which extend without too much trouble to $\d=1$).  They were proven under additional assumptions.

We assume that \eqref{A1}--\eqref{A3} hold so that $\meseq$ exists. We assume in addition  $\pa \Sigma \in C^{1,1}$, where we recall $\Sigma$ is the support of $\meseq$,   and 
\be \label{assumpV1} V \in C^2,\ee
\be
\label{assumpV3} 
\left\{
\begin{aligned}
& \int_{|x|\ge 1} \exp\(-\frac{\theta}{2}  V(x) \)dx<\infty
& \mbox{if} & \ \d\geq 3, \\
& \int_{|x|\ge 1}\( \exp\(-\frac{\theta}{2}  (V(x)-\log |x| ) \) + \exp\(- \theta (V(x)- \log |x|) \) |x| \log^2 |x| \) dx<\infty
& \text{if} & \ \d=2,
\end{aligned}
\right. 
\ee
\be \label{assumpV4} \Delta V \ge\alpha>0 \quad \text{in a neighborhood of }\, \Sigma.\ee
This last assumption \eqref{assumpV4} is placed to use standard results on the obstacle problem \cite{caff} that ensure that 
\be \label{caf}
\zeta(x) \ge \alpha \,\dist(x, \Sigma)^2\quad \text{in a neighborhood of } \Sigma,\ee  with $\zeta$ the function of \eqref{defzeta}, and a corresponding upper bound also holds. In particular it implies that $\omega=\Sigma$, i.e.~the contact set and the droplet coincide. We now assume in addition that 
\be\label{assumpV5} \zeta(x) \ge \alpha \min(\dist(x, \Sigma)^2, 1),\ee  which amounts, up to changing the constant $\alpha>0$ if necessary, to assume 
 that  the solution to the obstacle problem never gets very close to the obstacle, outside of $\Sigma$.
 
 Under assumption \eqref{assumpV5}, the convergence of $\mub$ to $\meseq$ is shown to have {\it a boundary layer of lengthscale $\theta^{-\hal}$ near $\pa \Sigma$} where $\meseq $ is generally discontinuous. This is what we consider as the  lengthscale for macroscopic rigidity.
 

\begin{theo}[Convergence of $\mub$ to $\meseq$, see \cite{ascomp}]
\label{th1as} Let $\d \ge 2$ and $\s=\d-2$. Assume \eqref{A1}--\eqref{A4} and \eqref{assumpV1}--\eqref{assumpV5}. Then \eqref{defEtheta} has a unique minimizer $\mub$. 
Moreover, there exists $C(V,\d) >0$ such that, for every $x\in \R^\d$ and $\theta \in (2,\infty)$, we have
\be 
\label{bornelmub0}
0<\mub(x) \le \begin{cases}\min (C, C\exp\left( -\theta (V(x)   -C)\right) & \text{if }  \ \d \ge 3\\
\min (C, C\exp\left( -\theta (V(x) -\log |x|  -C)\right) & \text{if }  \ \d= 2\end{cases}
\ee 
\be \label{2521} \mub(x) > \frac{1}{C} >0 \quad \text{for} \ x\in \Sigma,\ee
\be   \exp\(-\frac{\theta }{C} \dist(x, \Sigma)^2 - C \)  \le \mub(x)\le \exp\(-\frac{\theta}{C} \dist(x, \Sigma)^2 + C\) \ \text{in a neighborhood of } \Sigma, \ee
 \be \label{ber} \|h^{\mub}- c_\theta - h^{\muv}+c_\infty\|_{L^\infty(\Rd)} \le  \frac{C}{\theta},
\ee
where $c_\infty$ is the constant in \eqref{EulerLagrange}, and 
 \be \label{borneh10}
 \|\nab (h^{\muv}-h^{\mut})\|_{L^\infty(\R^\d)} \le \frac{C }{\sqrt\theta}
 ,\ee 
 \be \label{mhS}
 \mub( \Sigma^c)\le \frac{C}{ \sqrt\theta},\ee
and
\be \label{intromut} \left|\int_{\Sigma^c} \mub\log \mub \right| \le \frac{C}{\sqrt\theta}.\ee

Let  $m $ be an integer $\ge 2$ such that  $V\in C^{2m,\gamma}$ for some $\gamma\in (0,1]$ and letting $f_{k}$ be defined iteratively by 
\be \label{41} f_0= \frac{1}{\cd}\Delta V,\qquad f_{k+1}=\frac{1}{\cd}\Delta V+ \frac{1}{\theta \cd}\Delta \log f_k,\ee
we have  $f_k \in  C^{2(m-k-1),\gamma} (\Sigma)$ and  for every even integer $n\le 2m-4$ and $0\le \gamma'\le \gamma$,  if $\theta$ is large enough
depending on $m$, we have 
\be \label{2318}\|\mub-f_{m-2-n/2}\|_{C^{n,\gamma'}( \Sigma)} \le C \theta^{\frac{n+\gamma'}{2}}\exp\(- C \log^2 (\theta \dist(x, \partial \Sigma)^2)\)  
+ C\theta^{1+n-m+\frac{\gamma'}{2}} .\ee
\end{theo}
The functions $f_k$ provide a sequence of improving approximations to $\mub$ defined iteratively. Spelling out the iteration we easily find the expansion in powers of $1/\theta$
\be\label{corrections}\mub\simeq \frac{1}{\cd}\Delta V+ \frac{1}{\cd \theta} \Delta \log \frac{\Delta V}{\cd} + \frac{1}{\cd \theta^2}\Delta \( \frac{\Delta \log \frac{ \Delta V}{\cd}}{\Delta V}\)+... \quad \text{inside }   \Sigma\ee
up to an order dictated by the regularity of $V$ and the size of $\theta$.
\begin{rem}
\label{remtem}
In the case $\beta=2$ in the two-dimensional Coulomb case $\s=0$, $\d=2$, the large-$N$ expansion  of the one-point density $\rho_N^{(1)}$  (as in \eqref{defrhok}) in $\Sigma$  is known (see  \cite{byunforrester} and references therein) to be 
$$\rho_N^{(1)}(x)= \frac{1}{\cd}\Delta V(x)+ \frac1{\cd 2N}\Delta \log \Delta V(x) + N^{-1}B_2(x)+\dots.$$
Thus comparing with \eqref{corrections} shows that the thermal equilibrium measure with $\theta= \beta N^{1-\frac\s\d}=\beta N$ provides, at least in that case,  the correct next to leading order correction. 
 In particular, the vanishing lengthscale $\frac{1}{\sqrt{\theta}}$, which is the lengthscale of variation of $\mub$ and of decay of its tails away from $\Sigma$ (the support of $\meseq$), can be seen as  corresponding to a lengthscale of macroscopic rigidity and localization near $\Sigma$ for the Coulomb/Riesz gas (compare with Theorem \ref{thameurloc}  for  the two-dimensional log case, in Section \ref{sec53}). 
\end{rem}

In view of \eqref{2318}, letting $\hat \Sigma = \{ x\in \Sigma, \dist (x, \pa \Sigma) \ge \theta^{\ep-1/2}\}$ for some $\ep >0$, we have uniform bounds 
\be\label{boundsmub}\forall \sigma \le 2m+\gamma- 4, \quad \|\mub\|_{C^{0,\sigma} (\hat \Sigma) } \le C .\ee

The proof of the theorem relies on maximum principle arguments, and on observing that $\frac{\mub}{f_k}-1$ solves a divergence form equation  which is uniformly elliptic inside $\Sigma$ then applying regularity theory for elliptic PDE.


\chapter{The leading order behavior  }
\label{chap:leadingorder} 
In this chapter, building on the results of the previous chapter we study  the leading order or ``mean-field" behavior of the Coulomb or Riesz gas energy \eqref{HN} with  $\g$   given by  \eqref{riesz}, and apply it to \eqref{gibbs}.
  The results are quite standard and  in large part adapted from the literature. However, we try to give here a self-contained and general treatment, since results  are a bit scattered between the potential theory literature, the probability and statistical mechanics literature and the analysis literature, and not all situations seem  to be systematically covered.


The beginning of this chapter is devoted to the analysis of $\HN$ only, via $\Gamma$-convergence, leading to the mean-field description of its minimizers and their convergence to the equilibrium measure. Then, 
 in Section \ref{LDPsection} we apply these results to the statistical mechanics model, i.e.~to characterizing the states with nonzero temperature under the law \eqref{gibbs}. This is done again in terms of the equilibrium or thermal equilibrium measure, and via a large deviations principle (LDP).
 We have tried to present the most general LDP result, handling all temperature regimes and all interactions $\d-2\le \s<\d$ at once, under a minimal set of assumptions.

 \section{The case of zero temperature}
 The energy $\HN$ is the sum of  an interaction term $\sum_{i\neq j} \g(x_i-  x_j)$ and a potential  term $N \sum_{i=1}^N V(x_i)$. The first term pushes the points to repel and potentially  escape to infinity, while the second one confines them. The sum of pairwise interactions is expected to scale like the number of pairs of points, i.e. $N^2$, while the sum of the potential terms is expected to scale like $N$ times the number of points, i.e. $N^2$ again. The factor $N$ in front of $V$ in \eqref{HN} is thus precisely so  that the opposing effects of the repulsion and of the confinement balance each other. This is called the ``mean-field scaling". It is the scaling in which  the force acting on each particle is   given in terms of the average field generated by the other particles.
For general reference on mean-field theory, see statistical  mechanics textbooks such as \cite{huang}.

\subsection{$\Gamma$-convergence : general definition} \label{sectiongammaconvergence}
The result we want to show about the leading order behavior of $\HN$ can be formalized in terms of  the notion of $\Gamma$-convergence, in the sense of De Giorgi (see \cite{braides} for an introduction, or \cite{dalmaso} for an advanced reference). It is a notion of convergence for functions (or functionals) which ensures that minimizers tend to minimizers.  This notion is popular in the  calculus of variations  literature  and very used in the analysis of sharp-interface and fracture models, dimension reduction for variational problems, homogenization, evolution problems, etc  \cite{braides,braides3}. Using this formalism here is convenient but not essential.

Let us first give the basic definitions.
\begin{defini}[$\Gamma$-convergence] \label{defgcv}We say that a sequence $\{F_N\}_{N}$ of functions on a metric space $X$ $\Gamma$-converges to a function $F : X \rightarrow (-\infty, +\infty]$ if the following two inequalities hold : 
\begin{enumerate}
\item $(\Gamma$-$\liminf)$ If $x_N \rightarrow x$ in $X$, then $\liminf_{N \to + \infty} F_N(x_N) \geq F(x)$.
\item $(\Gamma$-$\limsup)$ For all $x$ in $X$, there is a sequence $\{x_N\}_N$ in $X$ such that $x_N \rightarrow x$ and $\limsup_{N \to + \infty} F_N(x_N) \leq F(x)$. Such a sequence is called a \textit{recovery sequence}.
\end{enumerate}\end{defini}

The second inequality is  saying that the first one is sharp, since it implies that  there is a particular sequence $x_N \rightarrow x$ for which the equality $\lim_{N \to + \infty} F_N(x_N) = F(x)$ holds.

\begin{rem} \label{gconvcom}
\begin{enumerate}
\item
In practice a compactness assumption is generally needed and sometimes added in the definition, requiring that if $\{F_N(x_N)\}_N$ is bounded, then $\{x_N\}_{N}$ has a convergent subsequence.
A similar compactness requirement also appears in the definition of a good rate function in large deviations theory (see Definition \ref{ratefun} below).
\item
The first inequality is usually proven by functional analysis methods, without making any ``ansatz" on the precise form of $x_N$, whereas the second one is usually obtained by an explicit construction, during which one constructs ``by hand" the recovery sequence such that $F_N(x_N)$ has  asymptotically less energy than $F(x)$. Note also that by a diagonal argument, one may often reduce to  constructing  a recovery sequence for a dense subset of $x$'s.
\item
A $\Gamma$-limit is always lower semi-continuous. (In particular, a function which is not l.s.c.~is a bad candidate for being a $\Gamma$-limit.)
Thus, a functional is not always its own $\Gamma$-limit~: in general $\Gamma$-lim $F = \bar{F}$ where $\bar{F}$ is the l.s.c.~envelope of $F$.
\item 
The notion of $\Gamma$-convergence can be generalized to the situation where $F_N$ and $F$ are not defined on the same space. One may instead  refer to a sense of convergence of $x_N$ to $x$  which is defined via the  convergence of any specific function of $x_N$ to $x$, which may be a nonlinear function of $x_N$,  cf. \cite{ssgcv,jstern} for instances of this.
\end{enumerate}
\end{rem}

We now state the most important property of $\Gamma$-convergence~: $\Gamma$-convergence sends minimizers to minimizers.
\begin{prop}[Minimizers converge to minimizers under $\Gamma$-convergence]\label{gammaconvmini}  Assume $F_N$ $\Gamma$-converges to $F$ in the sense of Definition \ref{defgcv}.
If for every $N$,  $x_N$ minimizes $F_N$,  and if the sequence $\{x_N\}_N$ converges to some $x$ in $X$, then $x$ minimizes $F$, and   moreover, $\lim_{N\to + \infty} \min_X F_N= \min_X  F$.
\end{prop}
\begin{proof}
Let $y\in X$. By the $\Gamma$-$\limsup$ inequality, there is a recovery sequence $\{y_N\}_N$ converging to $y$ such that $F(y) \geq \limsup_{N\to + \infty} F_N(y_N)$. By minimality of $x_N$,  we have $F_N(y_N) \geq F_N(x_N)$ for all $N$ and by the $\Gamma$-$\liminf$ inequality it follows that  $\liminf_{N\to  + \infty} F_N(x_N) \geq F(x)$, hence $F(y) \geq F(x)$.  Since this is true for every  $y$ in $X$, it proves that  $x$ is a minimizer of $F$. The relation $ \lim_{N\to + \infty} \min F_N= \min F$ follows from the previous chain of inequalities applied with $y=x$.
\end{proof}
\begin{rem}\label{gcvcomplete}
An  additional compactness assumption as in Remark \ref{gconvcom} ensures that if $\{\min F_N\}_{N} $ is bounded then a  sequence $\{x_N\}_N$ of minimizers has a limit, up to extraction. That limit  must then be a minimizer of $F$. If moreover it happens that $F$ has a unique minimizer, then the whole sequence $\{x_N\}_N$ must converge to it.
\end{rem}

\subsection{$\Gamma$-convergence of the Coulomb gas energy}\label{seccv}

The example of $\Gamma$-convergence of interest to us here concerns  the sequence of functions $\{\frac{1}{N^2}\HN\}_N$ defined as in \eqref{HN}. The space  $\mc{P}(\R^\d)$  of Borel probability measures on $\R^\d$ endowed with  the topology of weak convergence (i.e. that of the  dual of bounded continuous functions in $\R^\d$), which is metrizable,  will play the role of the metric space $X$ above. We may view $\HN$ as being defined on $\mc{P}(\R^\d)$ through the map 
\begin{equation}\label{defiN} i_N:
\left\lbrace \begin{array}{ccc} (\R^\d)^N & \longrightarrow & \mc{P}(\R^\d) \\
\XN& \mapsto & \emp[\XN]:=  \frac{1}{N} \sum_{i=1}^N \delta_{x_i} \end{array}
\right.
\end{equation}
which associates to any configuration of $N$ points the probability measure
 $\emp$ called the \textit{empirical measure}, or \textit{spectral measure} in the context of random matrices.  More precisely, we can extend the function $\HN$ into a function defined for any $\mu$ in $\mc{P}(\R^\d)$ by \begin{equation}
\HN(\mu) = \left\lbrace \begin{array}{cl} \HN(\XN) & \mbox{ if } \mu \mbox{ is of the form } \frac{1}{N} \sum_{i=1}^N\delta_{x_i} \\
+ \infty & \mbox{otherwise}.
\end{array}
\right.
\end{equation}
The first main result that we prove here is that the sequence $\{\frac{1}{N^2}\HN\}_N$ has the functional $\I$ as its $\Gamma$-limit.

\begin{prop}[$\Gamma$-convergence of $\frac{1}{N^2}\HN$] \label{gammaconvergenceHn} 
If  $V$ satisfies \eqref{A1} and \eqref{A2}, the  sequence $\{\frac{1}{N^2} \HN\}_N$ of functions (defined on $\mc{P}(\R^\d)$ as above)   $\Gamma$-converges as $N \to + \infty$ to  the function $\I$ defined in \eqref{definitionI}. 
\end{prop}


A statement and a proof with $\Gamma$-convergence in dimension~$2$ for $V$ quadratic  appeared  in \cite[Proposition 11.1]{livre}. It is not difficult to adapt them to higher dimensions and more general potentials. Similar arguments are also found in the large deviations proofs of \cite{bg,bz,cgz}.

The  proof of the lower bound uses 
the same ingredients as the proof of  existence of a minimizer of $\I$ in the previous section. 
The proof of the upper bound  can be obtained by constructing a recovery sequence for each measure 
 $\mu$ in $\mc{P}(\R^\d)$.  By density one may reduce to measures which are in
$L^{\infty}(\R^\d)$, supported in a cube $K$ and such that the density $\mu(x)$ is bounded below by $\alpha > 0$ in $K$, one then cuts cubes into subcubes  of size $\ell$, $N^{-1/\d}\ll \ell\ll 1$
 and places the appropriate number of well distributed points in each subcube $K_i$ in such a way that the difference between $N \mu(K_i)$ and the number of points places does not exceed $1$, so that $\emp$ approximates $\mu$. 
  The difficult part is then to show that the contribution of nearby points (or near diagonal terms in $\I$) is not large.  This proof is given in full in \cite[Proposition 2.1]{noteszurich}.
  Here we will use instead a  probabilistic argument:  the result follows from the existence of a whole set of approximating configurations deriving  from the proof of the LDP Theorem~\ref{LDP} below in the case with temperature.

In the whole text, when considering sequences of configurations $(x_1, \dots, x_N)$  we will make  the slight abuse of notation that consists in neglecting the dependency of the points $(x_1, \dots, x_N)$ on $N$, while one should formally write $(x_{1,N}, \dots, x_{N,N})$.

\begin{proof}[Proof of Proposition \ref{gammaconvergenceHn}]
{\bf Step 1. Lower bound.}
From now on, we denote the diagonal of $\R^\d \times \R^\d$ by $\triangle$ and its complement by $\triangle^{c}$.\\
 We need to prove that if $\frac{1}{N} \sum_{i=1}^N \delta_{x_i} \rightarrow \mu\in \mc{P}(\R^\d)$, then $$\liminf_{N\to + \infty} \frac{1}{N^2} \HN(\XN) \geq \I(\mu).$$ Letting $\emp$ denote the empirical measure  $\frac{1}{N} \sum_{i=1}^N \delta_{x_i}$, we may write 
 \begin{equation} \label{Hnpourempiric}
\frac{1}{N^2} \HN(\emp) =\hal \iint_{\triangle^{c}} \g(x-y) d\emp(x) d\emp(y) + \int V(x) d\emp(x).
\end{equation}
In order to handle the singularity of  $\g$, as in the proof of Lemma \ref{coerI}, let us truncate the singularity of $\g$ by writing 
\be 
\iint_{\triangle^{c}} \g(x-y) d\emp(x) d\emp(y) \geq \iint (\g(x-y) \wedge M) d\emp(x) d\emp(y) - \frac{M}{N}
\ee
where $M>0$ and  $\wedge$ still denotes the minimum of two numbers. Indeed one has $\emp \otimes \emp (\triangle) = \frac{1}{N}$ as soon as the points of the  configuration $(x_1, \dots, x_N)$ are simple (i.e. $x_i \neq x_j$ for $i\neq j$). 
We may then write that 
\be \label{troncaturediago1}\frac1{N^2} \HN(\XN) \ge \hal \iint(\g(x-y) \wedge M + V(x)+V(y) ) d\emp(x) d\emp(y)- \frac{M}{N}.\ee

The function $(x,y) \mapsto \g(x-y) \wedge M  $ is continuous hence 
the function $(x, y) \mapsto \g(x-y) \wedge M  +V(x)+V(y)$ is l.s.c. Moreover,  it is bounded below arguing as in    Remark \ref{remarkVg},  and by taking the limit of \eqref{troncaturediago1} as $N \rightarrow + \infty$ one gets,  by weak convergence of $\emp$ to $\mu$ (hence of $\emp \otimes \emp$ to $\mu \otimes \mu$)  that for every $M > 0$ 
\begin{multline*}
\liminf_{N \to + \infty} \hal\iint_{\triangle^{c}} (\g(x-y) + V(x)+V(y) )  d\emp(x) d\emp(y)\\
 \geq\hal \iint (\g(x-y) \wedge M + V(x)+V(y) ) d\mu(x) d\mu(y).
\end{multline*}

By the monotone convergence theorem, the (possibly infinite) limit of the right-hand side as $M \rightarrow + \infty$ exists and equals $\iint (\g(x-y) + V(x)+V(y))  d\mu(x) d\mu(y)$. Combining with \eqref{troncaturediago1}, this concludes the proof of the $\Gamma$-$\liminf$ convergence. 
\smallskip

{\bf Step 2. Upper bound.}
Let $\mu$ be a probability density such that $\I(\mu)<+\infty$. As mentioned in Remark \ref{gconvcom} it suffices to prove the upper bound inequality for a dense subset of probability measures. We may thus assume, without loss of generality, that $\mu$ has compact support. Let us then evaluate
\begin{align*}
& \frac1{N^2}\int_{(\R^\d)^N }\HN(\XN) d\mu(x_1) \dots d\mu(x_N) \\
 &=\frac{1}{N^2}\(  \hal \sum_{i\neq j} \iint_{(\R^\d)^N}
 \g(x_i-x_j) d\mu^{\otimes N} (\XN)
+N \sum_{i=1}^N \int_{(\R^\d)^N } V(x_i) d\mu^{\otimes N} (\XN)\)
\\
&= \frac{1}{N^2} \(\frac{N(N-1) }{2}\iint_{\R^\d\times \R^\d}\g(x-y) d\mu(x)d\mu(y)+ N^2 \int_{\R^\d} Vd\mu\)\\  
& =\I(\mu)- \frac{1}{2N} \iint_{\R^\d\times \R^\d}\g(x-y) d\mu(x)d\mu(y)
\\& \le \I(\mu)- \frac{1}{2N}  \iint_{\R^\d\times \R^\d}\g_-(x-y) d\mu(x)d\mu(y).
\end{align*}
Using the definition of $\g$,  \eqref{gVV} and the fact that $\mu$ is compactly supported, we find that $ \iint_{\R^\d\times \R^\d}\g_-(x-y) d\mu(x)d\mu(y)>-\infty$. 
It follows that, $\emp$ denoting the empirical measure of $\XN$, for any $\ep>0$,
\begin{multline*}
\frac1{N^2}\int_{\emp\in B(\mu, \ep)}\HN(\XN) d\mu(x_1) \dots d\mu(x_N) \\
\le \I(\mu)+ \frac{C}N -\frac{1}{N^2}\int_{\emp\notin B(\mu, \ep)} \HN(\XN) d\mu(x_1) \dots d\mu(x_N) ,\end{multline*}
where $C>0$ depends on $\mu$.
Since $\HN \ge - CN^2$ as a consequence of the $\Gamma$-liminf and the fact that $\inf \I>-\infty$ as seen  in Lemma \ref{coerI}, we may bound 
$$-\frac{1}{N^2}\int_{\emp\notin B(\mu, \ep)} \HN(\XN) d\mu(x_1) \dots d\mu(x_N)\le C \int_{\emp\notin B(\mu, \ep)} d\mu(x_1) \dots d\mu(x_N)$$
while, by the law of large numbers,  $$\int_{\emp\notin B(\mu, \ep)}d\mu(x_1) \dots d\mu(x_N) =o_N(1).$$ 
Assembling all these relations, we have obtained that for any $\ep>0$
$$\frac1{N^2}\int_{\emp\in B(\mu, \ep)}\HN(\XN) d\mu(x_1) \dots d\mu(x_N)\le \I(\mu)+o_N(1)
.$$Thus we may build a sequence of points such that $\emp \to \mu$ as $N \to \infty$ and $$\limsup_{N\to \infty} \frac{1}{N^2} \HN(\XN) \le \I(\mu)$$ as desired.
\end{proof}

  \begin{rem}\label{rem11} Again we do not really need that $\g$ is  a Coulombic (or Riesz) kernel, rather we only used \eqref{A1}--\eqref{A2} and  the facts that $\g$ locally bounded below, locally integrable and l.s.c.~away from the origin. 
\end{rem}


We next derive the consequence of the $\Gamma$-convergence 
 Proposition \ref{gammaconvergenceHn}  given by Proposition    \ref{gammaconvmini}. 
 In order to do so, we must  prove the compactness of sequences with suitably bounded energy, as in Remark \ref{gcvcomplete}.  
 
  \begin{lem}\label{lemcpthn}
 Assume that $V$ satisfies  \eqref{A1}--\eqref{A2}.
 Let $\{\XN\}_N$  be a sequence of configurations in $(\R^\d)^N$, and let $\{\emp\}_N$ be the associated empirical measures.
 Assume $\{\frac{1}{N^2} \HN(\emp) \}_N$ is a bounded sequence. Then  the sequence $\{\emp\}_N$ is tight, and as $N \to \infty$, it converges weakly in $\mc{P}(\R^\d)$ (up to extraction of a subsequence) to some probability measure $\mu$.
 \end{lem}
 \begin{proof}
 The proof is completely analogous to that of Lemma \ref{coerI}.  First, by assumption, there exists a constant $C_1$ independent of $N$ such that $\HN(\XN) \le C_1 N^2$, and in view of \eqref{Hnpourempiric}--\eqref{troncaturediago1} we may write, for every $M>0$, 
 \begin{multline}\label{M1}
 C_1\ge \hal \iint (\g(x-y) \wedge M) \, d\emp(x)\, d\emp(y) - \frac{M}{N} + \int V\, d\emp\\
 =\hal  \iint \(\g(x-y) \wedge M +  V(x) +  V(y) \) \, d\emp(x) \, d\emp(y)- \frac{M}{N}.\end{multline}
Using Remark \ref{remarkVg},  and arguing as in Lemma \ref{coerI}, we get the tightness.
 \end{proof}

 To conclude, we will make the assumptions on $V$ that ensure both the $\Gamma$-convergence result Proposition \ref{gammaconvergenceHn} and the existence result Theorem \ref{theoFrostman}.  

\index{energy minimizers}
 \begin{theo}[Convergence of minimizers and minima of $\HN$] \label{thcvmini} 
  Assume that $V$  satisfies \eqref{A1}--\eqref{A3}.   
 Assume that for each $N$,  $\XN$ is a minimizer of $\HN$. Then, as $N \to \infty$ we have 
\be \label{cvdesmini}\frac1N\sum_{i=1}^N \delta_{x_i} \to \meseq \text{ in the weak sense of probability measures} 
\ee
where $\meseq$ is the unique minimizer of $\I$ as in Theorem \ref{theoFrostman}, and 
\be\label{cvdesmin}
\lim_{N\to + \infty} \frac{\HN(\XN) }{N^2} = \I(\meseq).\ee
\end{theo}

\begin{proof}
Applying the  $\Gamma$-limsup part of the definition of $\Gamma$-convergence,  for example to $\meseq$,  ensures that  $\limsup_{N\to +\infty}\frac{1}{N^2} \min \HN $ is bounded above (by $\I(\meseq)$), hence in particular sequences of minimizers of $\HN$ satisfy  the assumptions of Lemma \ref{lemcpthn}.  It follows that, up to a subsequence, we have 
  $\emp\to \mu$ for some $\mu\in \mc{P}(\R^\d)$.
  By Propositions \ref{gammaconvergenceHn} and \ref{gammaconvmini}, $\mu$ must minimize $\I$, hence, in view of Theorem~\ref{theoFrostman}, it must be equal to $\meseq$. This implies that the convergence must hold along the whole sequence. We also get \eqref{cvdesmin} from Proposition~\ref{gammaconvmini}.
\end{proof}
In the language of statistical mechanics or mean field theory, this  result gives the mean-field behavior or average behavior of ground states, and the functional $\I$ is called the mean-field energy functional. 
It tells us that  points are macroscopically  distributed according to the probability law $\meseq$ as their number tends to $\infty$, and we have the leading order asymptotic expansion of the ground state energy $$\min \HN \sim N^2 \min \I \quad \text{as} \ N \to \infty.$$This is not very precise as it does not give information on the behavior at smaller scales and the patterns followed by the points.
  Understanding this and going beyond the leading order description is the object of the following chapters.

\section{The case with temperature: Large Deviations Principle}\label{LDP-sec}

At this point, we have understood the mean-field behavior of ground states  of the Coulomb/Riesz gas. In this section, we turn for the first time to states with temperature and  derive rather easy consequences of the previous sections for the Gibbs measure \eqref{gibbs}, via the framework of large deviations which allows to  characterize the probability of observing a rare event or a ``non-typical" configuration. 


\index{large deviations principle}
\subsection{Definitions} \label{LDPsection}

Let us first recall the basic definitions in the theory of large deviations, for more reference see the textbooks \cite{denholl,deuschel,dz,seppalainen}.

\begin{defini}[Rate function] \label{ratefun} Let $X$ be a metric space (or  a topological space). A rate function is a l.s.c.~function $I : X \rightarrow [0, + \infty]$, it is called a {\it good rate function} if its sub-level sets $\{x, I(x) \leq \alpha\}$ are compact (see Remark \ref{gconvcom}). 
\end{defini}

\begin{defini}[Large deviations]\label{definiLDP} Let $\{P_N\}_N$ be a sequence of Borel probability measures on $X$ and $\{a_N\}_N$ a sequence of positive real numbers diverging to $+ \infty$. Let also $I$ be a (good) rate function on $X$. The sequence $\{P_N\}_N$ is said to satisfy a large deviation principle (LDP) at speed $a_N$ with (good) rate function $I$ if for every Borel set $E \subset X$ the following inequalities hold 
\be \label{343}
- \inf_{\overset{\circ}{E}}I \leq \underset{N \to + \infty}{\liminf}\, \frac{1}{a_N} \log P_N(E) \leq  \underset{N \to + \infty}{\limsup} \,\frac{1}{a_N} \log P_N(E) \leq - \inf_{\bar{E}} I,
\ee
where $\overset{\circ}{E}$ (resp. $\bar{E}$) denotes the interior (resp. the closure) of $E$ for the topology of $X$.
\end{defini}
Formally, it means that $P_N(E)$   behaves roughly like $e^{-a_N \inf_{E} I}$. The rate function $I$ is the rate of exponential decay of the probability of rare events, and the events with larger probability are the ones on which $I$ is smaller.   For this to make sense, $\inf_X I$ must be zero, and the LDP then asserts that all 
states with $I>\inf_X I$ have exponentially small probability, hence they are rare events and this quantifies their probability.

\begin{defini}[Exponential tightness] We say that the sequence $\{P_N\}_N$ is  exponentially tight (at speed $a_N$)  if  for every $\eps>0$ there exists a compact set $K_\eps$ such that 
$$\limsup_{N\to \infty} \frac{1}{a_N} \log P_N(K_\ep^c)\le -\eps.$$
\end{defini}

\begin{lem} \label{procexptightness}If $\{P_N\}_N$ is exponentially tight at speed $a_N$, then to prove it satisfies a LDP at speed $a_N$ with rate function $I$ it suffices to prove the upper bound in \eqref{343} for every compact set.\end{lem}

The proof is immediate. A corollary  it suffices to prove \eqref{343} for balls if the space $X $ is metrized.
\begin{coro}\label{ldpboules}
If $\{P_N\}_N$ is exponentially tight at speed $a_N$ and if $X$ is metrizable,  then $P_N$ satisfies a LDP at speed $a_N$ with (l.s.c.)~rate function $I$ if and only if for every $x\in X$
  $$ - I(x) \le \liminf_{\ep \to 0}  \liminf_{N\to \infty} \frac1{a_N} \log P_N(B(x,\ep))\le   \limsup_{\ep \to 0} \limsup_{N\to \infty} \frac1{a_N}\log P_N(B(x,\ep))\le - I(x).
  $$
  \end{coro}
  This is a consequence of the fact that a  compact set can be covered by a finite number of balls and also that a sum of exponentials is dominated by the one with the largest exponent (as in the Laplace or stationary phase method).   This is left as an exercise (see also Definitions 2.18, 2.19 in \cite{seppalainen} and comments around them.)

\begin{rem} \label{rem13} At first sight, Definition \ref{definiLDP} looks very close to the $\Gamma$-convergence 
\begin{displaymath}
\frac{\log p_N}{a_N} \overset{\Gamma}{\rightarrow} - I
\end{displaymath} where $p_N$ is the density of the measure $P_N$. However, in general there is no equivalence between the two concepts. For example, in order to estimate the quantity
\be
\log P_N(E) = \log \int_E p_N(x) dx
\ee
it is not sufficient to know the asymptotics of $p_N$, one  also needs to know the size of the volume element $\int_E dx$, which plays a large role in large deviations and usually comes up as an entropy term in the rate function. There are however some rigorous connections between $\Gamma$-convergence and LDP (see\cite{mar}).
\end{rem}

\subsection{Heuristics for the LDP of the Riesz gas}
We now want to specialize to the  Coulomb/Riesz gas  ensemble \eqref{gibbs}.
First, pushing $\mathbb{P}_{N, \beta}$ forward by the map $i_N$ of \eqref{defiN}, we may view it as a probability measure on $\mc{P}(\R^\d)$.

To explain what LDP we can expect for it, let us start by recalling 
the Gibbs variational principle which states that a Gibbs measure minimizes the sum of the average energy and the entropy times the temperature, and whose proof is immediate.
\begin{lem}[Gibbs variational principle]\label{lemgibbsvp}
Let $dP_\beta(x)= \frac{1}{Z} \exp(-\beta F(x))d\lambda(x)$
be a probability measure on a space $X$, where $\lambda$ is a reference measure on $X$.  Then 
$P_\beta$ achieves 
\be
\min_{P \in \mathcal P(X)} \int F(x) dP(x)+ \frac{1}{\beta} \int dP(x)  \log \frac{dP(x)}{d\lambda(x)}  
\ee and the minimum is equal to $- \frac1\beta \log Z$.
\end{lem}
This principle can be applied directly to $\PNbeta$ with $\lambda$ the Lebesgue measure on $(\R^\d)^N$ and  indicates that 
as $N \to \infty$ the Gibbs measure concentrates on states that minimize the sum of a limit energy and an entropy.

Making this more rigorous is done via Large Deviations theory, using 
Sanov's theorem (see for instance \cite{dz,dupuisellis}), which we now recall in the version appropriate to our context. \index{Sanov's theorem}
\begin{theo}[Sanov]  \label{theoremsanov}
Assume $X_1, \dots , X_N$ are i.i.d. random variables with values in $\R^\d$   defined on  a Polish probability space $(X, \mathbb{P})$, with law $\ro$, i.e.~$\mathbb{P}(X_i\in A)= \ro(A)$. Then, setting 
$$P_N(A)= \mathbb{P} \(\frac{1}{N} \sum_{i=1}^N \delta_{X_i} \in A\)$$ for every $A \subset \probas(X)$, 
$\{P_N\}_N$ satisfies a LDP at speed $N$ with good rate function  
$$\ent[\mu| \ro]=\begin{cases} \displaystyle \int_{\R^\d} \mu \log \frac{\mu}{\ro}\quad & \text{if} \ \mu\ll \ro\\
+\infty \ &\text{otherwise}. \end{cases}$$
\end{theo}
Here, $\ent[\mu|\ro]$ is called the {\it relative entropy} of $\mu $ with respect to $\ro$.
Informally, Sanov's theorem states that 
$$\mathbb{P} \( \frac{1}{N} \sum_{i=1}^N \delta_{X_i} \simeq \mu\)\simeq \exp\(-N \int \mu \log \frac{\mu}{\rho}\)$$ if $\mathbb{P}$ has law $\rho$.

Let us now think of $\P$ as corresponding to the normalized Lebesgue measure in the set $\Sigma$, support of the equilibrium measure $\meseq$ and the $X_i$ as the points $x_i$. Then the right-hand side is just a multiple of the regular entropy. This would be the governing function in the case of $\beta\to 0$ where the interaction  between the particles disappear and they just become Poissonian (or Bernoulli). 
In the case with $\beta $ nonzero, then the result of Proposition \ref{gammaconvergenceHn} says that $H_N \simeq N^2 \I$. If $H_N$ were continuous one could conclude directly by ``tilting" or ``Varadhan's lemma" that 
$$\PNbeta (\emp \simeq \mu) \sim \exp\(-\beta N^{-\frac{\s}{\d}} N^2 \I(\mu) -  N \int_{\R^\d} \mu \log \mu\).$$
Thus we see that the energy is in competition with the entropy, the two will be comparable in the regime where $\beta  N^{1-\frac\s\d} $ is proportional to $1$, the energy will dominate as soon as $\beta N^{1-\frac\s\d} \gg 1$ and the entropy will dominate when $\beta N^{1-\frac\s\d} \ll 1$.
This then  leads us to defining an effective temperature that we will use throughout
\be \label{deftheta}\boxed{ \theta  := \beta N^{1-\frac{\s}{\d}} .}
\ee
With this new notation, the quantity to minimize is thus 
$$ -N\( \theta \I(\mu)+ \int_{\R^\d} \mu \log \mu\)= - N\theta\, \I_\theta(\mu),$$ with $\I_\theta$ as in \eqref{defEtheta}.  This  shows that its minimizer $\mub$ (the thermal equilibrium measure) is an even better mean-field approximation of the empirical measure:  even when $\theta \to +\infty$ as $N\to \infty$, it provides a corrected description to $\meseq$.



\subsection{Large Deviations Principle for the Riesz gas}\label{LDPsec}



We may now state the LDP for the Gibbs measure associated to the Coulomb or Riesz gas Hamiltonian. This result was proven in the regime of $\beta $ independent of $N$,   in \cite{hiaipetz} (in dimension 1),  \cite{bg} (in dimension $1$) and \cite{bz} (in dimension 2) for the particular case of a quadratic potential (and $\beta = 2$), see also \cite{berman} for results in a more general (still determinantal) setting of multidimensional complex manifolds. The papers \cite{cgz,liuwu} recently treated more general singular $\g$'s and $V$'s, \cite{liuwu} containing a result for general $k$-point interactions. 
An LDP result for local  empirical observables, in the Coulomb gas case, can be found in \cite{dpg3}.

The LDP implies as a corollary the earlier-known law of large numbers that the empirical measure converges to the equilibrium measure, as was for instance first proven in the determinantal case of random matrix ensembles in \cite{boutet}. 

 We present here the proof for the Riesz gas in any dimension, general potential, and general $\theta \ge 1$, which is not more difficult. 
We will work in the metric space $\mathcal {P}(\R^\d)$ equipped with any distance that metrizes weak convergence. The LDP is stated for the push-forward of the Gibbs measure by the map $i_N$ of \eqref{defiN}, which is a probability on $\mathcal P(\R^\d)$. We recall that the functionals $\I$ and $\I_\theta$ were introduced in \eqref{definitionI} and \eqref{defEtheta}.

\begin{theo}[Large deviations principle at leading order for the Riesz gas] \label{LDP}
\mbox{}\\ \index{large deviations principle} \index{thermal equilibrium measure}
 Let $\PNbeta$ be as in \eqref{gibbs} and $\theta$  as in \eqref{deftheta}. Assume that $V$  is finite-valued and  satisfies \eqref{A1}, \eqref{A2} and \eqref{A5} for $N$ large enough.
\begin{itemize}
\item Assume that $\theta\to + \infty$ as $N \to \infty$.
Then the sequence $\{i_N \# \mathbb{P}_{N, \beta}\}_N$ of probability measures on $\mc{P}(\R^\d)$  satisfies a large deviations principle at speed $N\theta$ with good rate function $\hat{\I}$ where $\hat{\I} = \I - \min_{\mc{P}(\R^\d)}  \I$. Moreover 
\begin{equation}\label{237}
\lim_{N\to + \infty} \frac{1}{N\theta} \log Z_{N, \beta} = - \I(\meseq) = -\min_{\mc{P}(\R^\d)} \I.
\end{equation}
\item Assume that $\theta $ is independent of $N$. Then the sequence $\{i_N \#\mathbb{P}_{N, \beta}\}_N$ of probability measures on $\mc{P}(\R^\d)$  satisfies a large deviations principle at speed $N\theta$ with good rate function $\hat{\I}_\theta$ where $\hat{\I}_\theta = \I_\theta - \min_{\mc{P}(\R^\d)}  \I_\theta$.
 Moreover 
\begin{equation}\label{238}
\lim_{N\to + \infty} \frac{1}{N\theta} \log Z_{N, \beta}= -\I_\theta(\mub) = -\min_{\mc{P}(\R^\d)} \I_\theta.\ee
\end{itemize}\end{theo} 
An analogous result for the case $\theta \to 0$ with the entropy as a rate function also naturally holds,  except that in order to state it one would have to reduce to a bounded subset.

The heuristic reading of the LDP is that 
\be  \label{heurisldp}
\P_{N, \beta}(E) \approx e^{-N\theta (\min_{E} \I - \min \I)},  \ee respectively 
$$ \P_{N, \beta}(E) \approx  e^{-N\theta (\min_E \I_\theta-\min  \I_\theta)} .$$
As a consequence, if $\theta \gg 1$, or equivalently $\beta \gg N^{\frac\s\d-1}$,  the only likely configurations of points (under $\P_{N, \beta}$) are those for  which the empirical measures $\emp = \frac{1}{N} \sum_{i=1}^N \delta_{x_i}$ converge to $\mu=\meseq$, for otherwise $\I(\mu) > \I(\meseq)$ by uniqueness of the minimizer of $\I$, and the probability decreases exponentially fast according to \eqref{heurisldp}. Thus, $\meseq$ is not only the limiting distribution  of  minimizers of $\HN$, but also the limiting distribution for all ``typical" (or likely) configurations.  Moreover, we can estimate the probability under $\P_{N,\beta}$ of the non-typical configurations and see that it has exponential decay at speed $\theta N$. The temperature plays no role in the rate function as long as $\beta \gg N^{\frac\s\d-1}$.

The effect of the temperature is felt at the macroscopic scale 
only if $\theta $ does not tend to $+\infty$, i.e. $\beta \le C N^{\frac\s\d-1}$, which we can consider here a high temperature regime.  
 Recall that the cases of the classic random  matrix ensembles GOE, GUE and Ginibre all correspond  to $\s=0$ and  $\beta =1 ,2$, hence  they all correspond to a low temperature regime from the criterion $\beta \gg N^{\frac\s\d-1}= N^{-1}$.
 
Recalling the corresponding equilibrium measures were given in Example \ref{example2} in Chapter~\ref{chap:eqmeasure}, as a consequence of Theorem \ref{LDP}, we have  a proof of the law of large numbers, i.e.~that the distribution of eigenvalues (more precisely  the spectral or empirical measure) 
 has to follow  Wigner's semi-circle law  $\meseq =\frac{1}{2\pi} \sqrt{4-x^2} \indic_{|x|<2}$ for  the GUE and GOE, and the circle law $\meseq=\frac{1}{\pi}\indic_{B_1} $ for the Ginibre ensemble, as a consequence of the stronger    LDP result.   These are the cases originally treated in \cite{hiaipetz,bz,bg}.
 
In addition, this theorem provides the leading order of the free energy.

The large deviations lower bound  can be proven by showing that given a probability measure $\mu \in \mc{P}(\R^\d)$, there is a sufficiently large volume of empirical measures which approach $\mu$ and whose energy does not asymptotically exceed $\I(\mu)$. This is the proof in the original papers, and can also be found in \cite{noteszurich}. Here, we present a more elegant proof based on Jensen's inequality, following \cite{garciaz} and inspired from the works of Dupuis.  

\begin{proof}
Let us introduce the same notation as in the proof of Proposition \ref{lem241} in \eqref{defurho},  i.e.
$$u(x)= \exp(-\theta (V+\g_-)) ,\quad \bar u= \int_{\R^\d} u, \quad \rho= \frac{u}{\bar u} 
.$$ This is well-defined by \eqref{A4}. Also since \eqref{A4} holds for $N$ large enough, in all cases there exists a $\theta_0>0$ independent of $N$ such that $\int_{\R^\d} \exp(-\theta_0(V+\g_-)) <+\infty$ and thus we may also define
\be \label{defrhzero}
u_0(x) = \exp(-\theta_0 (V+\g_-)) ,\quad \bar u_0= \int_{\R^\d} u_0, \quad \rho_0= \frac{u_0}{\bar u_0} .
\ee

{\bf Step 1: LDP lower bound.} Let us consider $\mu$ a probability  measure with a density which is positive and continuous over $\R^\d$, such that for $N$ large enough, $\mu/\rho$ is bounded below independently of $N$, and such that $\int |\g_-|d\mu<\infty$.

Let us apply the Jensen-based argument.
Starting from \eqref{gibbs} and using \eqref{deftheta}, since $\mu>0$ we may write
\begin{multline}\PNbeta(\emp\in B(\mu, \ep)) \\
= \frac{1}{\ZNbeta} \int_{\emp\in B(\mu ,\ep)} 
\exp\(- \theta N^{-1} \HN(\XN) - \sum_{i=1}^N (\log \frac{\mu}{\rho}) (x_i) -\sum_{i=1}^N \log \rho(x_i) \) d\mu(x_1) \dots d\mu(x_N).\end{multline} Next  we apply Jensen's inequality to the integral to obtain 
\begin{multline}\label{applijensen}
\log \PNbeta(\emp\in B(\mu, \ep))\\ \ge - \log \ZNbeta + \int_{\emp\in B(\mu, \ep)} 
\( - \theta N^{-1} \HN(\XN) - \sum_{i=1}^N (\log \frac{\mu}{\rho}) (x_i)-\sum_{i=1}^N \log \rho(x_i) \) d\mu(x_1) \dots d\mu(x_N).\end{multline}
We then observe that by definition of $\HN$,
\begin{align*} & \int_{\emp\in B(\mu, \ep)} 
\( - \theta N^{-1} \HN(\XN)  - \sum_{i=1}^N (\log\frac{ \mu}{\rho})(x_i) -\sum_{i=1}^N \log \rho(x_i)\)  d\mu(x_1) \dots d\mu(x_N)
\\ &
= -\theta\int_{\emp\in B(\mu, \ep)}\( \frac{1}{2 N}\sum_{i\neq j}  \g(x_i-x_j) +   \sum_{i=1}^N V(x_i)  +\frac1\theta \sum_{i=1}^N (\log \frac{\mu}{\rho})(x_i)  +\frac1\theta\sum_{i=1}^N \log \rho(x_i)\) d\mu(x_1)\dots d\mu(x_N) \\
&=N\log \bar u
 -\theta\int_{\emp\in B(\mu, \ep)}\( \frac{1}{2 N}\sum_{i\neq j}  \g(x_i-x_j) -  \sum_{i=1}^N \g_- (x_i)  +\frac1\theta \sum_{i=1}^N (\log \frac{\mu}{\rho})(x_i)  \) d\mu(x_1)\dots d\mu(x_N) .
\end{align*} We next rewrite the integral in the right-hand side as the difference between the integral over $(\R^\d)^N$ and the integral over $\{\emp \notin B(\mu,\ep)\}$.
For the integral over the whole space, expanding all the terms in the sums, we find that 
\begin{multline}\label{firstline}
 -\theta\int_{(\R^\d)^N}\( \frac{1}{2 N}\sum_{i\neq j}  \g(x_i-x_j) - \sum_{i=1}^N \g_-(x_i) 
+\frac1\theta \sum_{i=1}^N(\log \frac{\mu}{\rho}) (x_i) \) d\mu(x_1)\dots d\mu(x_N)\\=
-\theta\( \frac{N(N-1)}{2N} \iint_{\R^\d\times \R^\d} \g(x-y) d\mu(x)d \mu(y)    - N   \int_{\R^\d} \g_- d\mu +\frac{N}{\theta} \int_{\R^\d} \mu \log \frac{\mu}{\rho} \)\\= - \theta N \I_\theta(\mu) + \frac{\theta}{2} \iint \g(x-y) d\mu(x)d \mu(y) +\theta N \int_{\R^\d}(V+ \g_-) d\mu + N \int \mu \log \rho . 
\end{multline}
If $\s>0$ we bound the term $\iint \g(x-y) d\mu(x)d \mu(y)$  below by $0$, while if $\s \le 0$, we use \eqref{gVV} and the assumption $\int |\g_-|d\mu<\infty$ to bound it  below by $ -N\theta \I_\theta(\mu)-C \theta.$

Using \eqref{gVV} we also find that 
\begin{multline*}
\theta \int_{\emp  \notin B(\mu, \ep)} \( \frac{1}{2 N}\sum_{i\neq j}  \g(x_i-x_j) -  \sum_{i=1}^N \g_- (x_i)  +\frac1\theta \sum_{i=1}^N (\log \frac{\mu}{\rho})(x_i)  \) d\mu(x_1)\dots d\mu(x_N)\\ \ge 
\theta \int_{\emp \notin B(\mu, \ep)}  \sum_{i=1}^N\(   -C +  \frac{1}{\theta} \log\frac{ \mu}{\rho}\)(x_i)   d\mu(x_1) \dots d\mu(x_N)\\
\ge - C \theta N \int_{ \emp \notin B(\mu, \ep)} d\mu(x_1) \dots d\mu(x_N), \end{multline*}
where we used that by  assumption $\mu/\rho$ is bounded below independently of $N$.
By the law of large numbers, we have 
$$
 \lim_{N\to \infty}\int_{ \emp \notin B(\mu, \ep)} d\mu(x_1) \dots d\mu(x_N)=0,  $$
hence the right-hand side  is $ =o(N\theta)$.

Combining all the above elements, we have obtained that 
\begin{align} \label{ldplb}
\qquad  &  \log \PNbeta(\emp\in B(\mu, \ep)) \\
\notag &\qquad \ge - \log \ZNbeta - \theta N \I_\theta(\mu )+ N \log \bar u  +\theta N \int_{\R^\d} (V+\g_-) d\mu + N \int \mu \log \rho 
+ o( \theta N)\\
\notag &\qquad  = - \log \ZNbeta  - \theta N \I_\theta(\mu ) 
+ o( \theta N).
 \end{align}
 
 We then retrace the same steps and use that $\mub>0$ as proven in Proposition \ref{lem242},  as well as the result of Lemma \ref{1entraine2}, to rewrite
 $$\ZNbeta= \int_{(\R^\d)^N} \exp\(-\theta N^{-1} \HN(\XN)- \sum_{i=1}^N (\log \frac{\mub}{\rho})(x_i)- \sum_{i=1}^N \log \rho(x_i) \) d\mub(x_1) \dots d\mub(x_N).$$
 Using Jensen's inequality  we obtain in lieu of \eqref{applijensen}
$$\log \ZNbeta \ge  \int_{(\R^\d)^N} \( - \theta N^{-1} \HN(\XN) - \sum_{i=1}^N( \log \frac{\mub}{\rho} ) (x_i)- \sum_{i=1}^N \log \rho(x_i) \) d\mub(x_1) \dots d\mub(x_N).$$
Proceeding as above and using  \eqref{firstline}, we deduce that 
\be \label{binflogz}
\log \ZNbeta \ge  - \theta N  \I_\theta(\mu_\theta) +o(\theta N) .\ee

By minimality of $\mub$ for $\I_\theta$, we may also write that 
$$\I_\theta(\mub)\le \I_\theta(\meseq) = \I(\meseq)+ O(\theta^{-1})$$
hence we  have obtained that if $\theta \to +\infty$ as $N\to \infty$, 
\be \label{binflogz2}
\log \ZNbeta \ge  - \theta N  \I(\meseq) +o(\theta N) .\ee

{\bf Step 2: large deviations upper bound}.
Let us start with the case $\theta$ fixed.
Let $\mu\in \mc{P}(\R^\d)$ such that $\mu/\rho$  is continuous and bounded below.  By definition of $\HN$ and of $\rho$, we may write  
\begin{multline}\label{dstep1}   \PNbeta(\emp\in B(\mu,\ep))  \\
\le \frac{1}{\ZNbeta} \int_{\emp\in B(\mu, \ep)} \exp\(- \theta \(  \frac1{2N} \sum_{i\neq j}\g(x_i-x_j)   +\sum_{i=1}^N V (x_i)  + \frac{1}{\theta}\sum_{i=1}^N \log \mu(x_i) \)\) d\mu(x_1)\dots d\mu(x_N)\\
= \frac{1}{\ZNbeta} \int_{\emp\in B(\mu, \ep)} \exp\(- \theta \(  \frac1{2N} \sum_{i\neq j} \g(x_i-x_j)  - \sum_{i=1}^N \g_- (x_i)   + \frac{1}{\theta}\sum_{i=1}^N \log \frac{\mu}{\bar u \rho} (x_i) \)\) d\mu(x_1)\dots d\mu(x_N).
\end{multline}
Applying  Proposition \ref{gammaconvergenceHn} to the potential $-\g_- +\frac1\theta \log \frac{\mu}{\rho}$ which is l.s.c.~bounded below,  and the lower semi-continuity of a $\Gamma$-liminf,  we deduce that if $\emp\in B(\mu, \ep)$ we have 
\begin{multline*}\liminf_{N\to \infty} \frac1{N^2}\(\hal   \sum_{i\neq j}\g(x_i-x_j)   +N \sum_{i=1}^N (-\g_-  + \frac1\theta \log \frac{\mu}{\bar  u \rho} )  (x_i) \) \\
\ge 
\hal \iint_{\R^\d\times \R^\d}\g(x-y) d\mu(x) d\mu(y) + \int_{\R^\d} (-\g_- +\frac1\theta \log \frac{\mu}{\bar u \rho}) d\mu+o_\ep(1)= \I_\theta(\mu) +o_\ep(1).\end{multline*}
Therefore,  
\begin{multline*}  \log \PNbeta(\emp\in B(\mu,\ep))  
\\ \le \log \( \frac{1}{\ZNbeta} \int_{\emp\in B(\mu, \ep)} \exp\(- \theta N(  \I_\theta(\mu) + o_{\ep,N}(1) )  \) d\mu(x_1)\dots d\mu(x_N)\) +\theta N o_\ep(1). 
\end{multline*}
Applying  Sanov's theorem (Theorem \ref{theoremsanov}) for i.i.d.~random variables of law $\mu$ we  have on the other hand that 
$$\limsup_{N\to  \infty} \frac1N \log  \(\int_{\emp \in B(\mu, \ep)} \mu(x_1) \dots \mu(x_N) \) \le
- \inf_{\nu \in \bar B(\mu, \ep)} \int \nu \log \frac{\nu}{\mu} \le 0 , $$
thus 
we conclude in that regime ($\theta$ fixed)  that 
\begin{equation}\label{bsupzetc}
\log \PNbeta(\emp\in  B(\mu , \ep) ) \le
   - \log \ZNbeta -   \theta N   \I_\theta(\mu)+o(N) +\theta N o_\ep(1 ).
 \end{equation}
 
 Let us now turn to the regime $\theta \to +\infty$. Using \eqref{defrhzero}, for any event $E$, we may write
 \begin{multline}\label{dstep2}   \PNbeta(\emp\in E)  \\
\le \frac{1}{\ZNbeta} \int_{\emp\in E} \exp\(- \theta \(  \frac1{2N} \sum_{i\neq j}\g(x_i-x_j)   +\sum_{i=1}^N V (x_i)  + \frac{1}{\theta}\sum_{i=1}^N \log \rho_0(x_i) \)\) d\rho_0(x_1)\dots d\rho_0(x_N)\\
= \frac{(\bar u_0)^N}{\ZNbeta} \int_{\emp\in E} \exp\(- \theta \(  \frac1{2N} \sum_{i\neq j} \g(x_i-x_j) +\sum_{i=1}^N  (1-\frac{\theta_0}{\theta}) V(x_i)   - \frac{\theta_0}{\theta}  \g_- (x_i)   \)\) d\rho_0(x_1)\dots d\rho_0(x_N).
\end{multline}
Applying Proposition \ref{gammaconvergenceHn} to the potential $(1-\frac{\theta_0}{\theta}) V$ (it depends on $N$ but this is harmless in the proof) we deduce that if $\emp\in E$ we have
\begin{multline*}\liminf_{N\to \infty} \frac1{N^2}\(\hal   \sum_{i\neq j}\g(x_i-x_j)   +(1-\frac{\theta_0}{\theta} ) \sum_{i=1}^N V  (x_i) \) \\
\ge 
\inf_{\mu \in \bar E} \hal \iint_{\R^\d\times \R^\d}\g(x-y) d\mu(x) d\mu(y) + \int_{\R^\d} V d\mu=\inf_{\mu \in \bar E}  \I (\mu) .\end{multline*}
Therefore,  
$$  \PNbeta(\emp\in E)  
\le \frac{(\bar u_0)^N}{\ZNbeta} \int_{\emp\in E} \exp\(- \theta N( \inf_{\bar E}  \I + o_{N}(1) )  \) d\rho_0(x_1)\dots d\rho_0(x_N).
$$
Applying Sanov's theorem for i.i.d.~random variables of law $\rho_0$ we obtain on the other hand that 
$$\log \( \int_{\emp\in E}d\rho_0(x_1)\dots d\rho_0(x_N)\)\le - N \inf_{\mu \in \bar E} \int_{\R^\d} \mu \log \frac{\mu}{\rho_0} +o(N)\le O(N)=o(\theta N)$$ since $\theta \to +\infty$, 
and we conclude that 
\begin{equation}\label{bsupzetc2}
\log \PNbeta(\emp\in  E ) \le
   - \log \ZNbeta -   \theta N  \inf_{\bar E}  \I+o(\theta N ).
 \end{equation}

{\bf Step 3: exponential tightness.} Let us first consider the case $\theta $ fixed. 
 For any event $E\subset \mc{P}(\R^\d)$, rewriting as above, we have 
\begin{multline*}
\PNbeta(\emp\in E) \\
= \frac{1}{\ZNbeta}\int_{\emp \in E} \exp\(- \theta N^{-1}  \( \hal \sum_{i\neq j} \g(x_i-x_j) + N\sum_{i=1}^N V(x_i) \) - \sum_{i=1}^N \log \rho(x_i)  \) d\rho(x_1) \dots d\rho(x_N)
\\ = \frac{(\bar u)^N}{\ZNbeta}\int_{\emp\in E}\exp\(- \theta   \( \frac1{2N} \sum_{i\neq j} \g(x_i-x_j) - \sum_{i=1}^N \g_-(x_i)  \) \) d\rho(x_1) \dots d\rho(x_N)
 \end{multline*}
 In view of  \eqref{gVV} we may find  that
 \begin{equation*}
  \frac1{2N} \sum_{i\neq j} \g(x_i-x_j) - \sum_{i=1}^N \g_-(x_i)  \\
  \ge \frac1{2N} \sum_{i=1}^N\sum_{j=1}^N \g(x_i-x_j)_-  - \sum_{i=1}^N \g_-(x_i)\ge - C N,\end{equation*} 
so using  \eqref{binflogz} to bound $\ZNbeta$, we have obtained that 
\be \label{Pnjh}
\PNbeta(\emp\in E) \le \exp(C N\theta) \int_{\emp \in E}  d\rho(x_1) \dots d\rho(x_N).\ee
By Sanov's theorem Theorem \ref{theoremsanov} applied  to i.i.d.~random variables of law $\ro$ we  may write that 
$$\limsup_{N\to \infty} \frac{1}{N}\log  \int_{\emp \in E} d\rho(x_1) \dots d\rho(x_N)\le -\inf_{\nu \in \bar E} \int_{\R^\d}
\nu \log \frac{\nu}{\rho}.$$We know that the relative entropy is a good rate function in Sanov's theorem, hence its sub-level sets $K_M= \{\nu\in \mc{P}(\R^\d), \int_{\R^\d}
\nu \log \frac{\nu}{\rho}\le M\}$ are compact. Applying to $E=(K_M)^c$ we thus have that 
$$\limsup_{N\to \infty} \frac{1}{N}\log\PNbeta(\emp\in (K_M)^c )\le - M + C $$
which proves exponential tightness at speed $N$ (or $N\theta$) in this regime.

In the case $\theta \to + \infty$, exponential tightness is in the same way a direct consequence of \eqref{bsupzetc2} since $\I $ has compact sub-level sets, as seen in Lemma \ref{coerI}.

 
  In view of Corollary~\ref{ldpboules}, to get an LDP  it thus suffices to prove the LDP lower and upper bounds for balls.  It also suffices to prove it for balls centered on a dense set, which we  take to be the $\mu$'s such that $\mu>0$ is continuous,  
 $  \frac{\mu}{\rho}$ is 
 bounded below independently of $N$ and $\int |\g_-|d\mu <\infty$.  The LDP upper and lower bound for balls thus follows from Steps 1 and 2. 
\smallskip

{\bf Step 4. Conclusion.}
Let us first consider  the case where $\theta $ is fixed. 
In view of the exponential tightness,  we can extend the results of Steps 1 and 2 to arbitrary sets, and not only balls. In particular, applying \eqref{bsupzetc} to the whole space $\mc{P}(\R^\d)$ we obtain 
\be\label{estimeez2}\log \ZNbeta  \le - \theta N \inf \I_\theta + o(N)+o(\theta N).\ee
Comparing with \eqref{binflogz} we have 
\be \label{estimeez3}
\log \ZNbeta= - \theta N \I_\theta(\mub) + o(N)+o(\theta N), \ee
which proves \eqref{238}. Inserting into \eqref{ldplb} and \eqref{bsupzetc}, we conclude the proof in this regime.

Let us now consider the case where $\theta \to +\infty$ as $N\to \infty$. Then we note that 
by definition of $\I_\theta$, $\I_\theta(\mu)= \I(\mu)+o_N(1)$.
We can then apply \eqref{bsupzetc2} to the whole space $\mc{P}(\R^\d)$ and obtain 
$$\log \ZNbeta  \le - \theta N \min \I  +o(\theta N).$$
Moreover   \eqref{binflogz2} yields  the converse inequality, hence \eqref{237} holds. 
Inserting again into \eqref{ldplb} and \eqref{bsupzetc} we conclude the proof in that regime. 

\end{proof}


\part{Modulated electric energy}

\chapter{The modulated electric energy }

\label{chap:nextorder}
\index{modulated energy}

In this chapter, we  introduce and study  the next order or ``modulated energy'' $\F_N(\XN, \mu) $, which naturally appears  in the study of Coulomb and Riesz gases, but also in dynamics questions, where it acquired its name ``modulated.'' It will play a crucial role in the remainder of the text.  $\F_N$ corresponds to the total Coulomb/Riesz interaction of the system of points at $\XN$ neutralized by the background charge $N\mu$. 
 It appears as a next-order energy after an exact ``splitting" of $\HN$ that we present at the beginning of Chapter \ref{chap:concentrationbounds}, for $\mu$ equal to the equilibrium or thermal equilibrium measure. 
In this chapter, we focus on $\F_N$ and study it for its own sake.
Thanks to an electric rewriting as an integral of the form $\int_{\R^\d} |\nab h|^2$ (for some electric potential $h$)  similar to that of Section \ref{sec:elecrewri}, we show the main results that $\F_N$ is bounded below, behaves as an effective Riesz distance between the empirical measure and the reference measure $\mu$, 
 and controls discrepancies and fluctuations of linear statistics with respect to $\mu$. Moreover, thanks to the electric rewriting, $\F_N$ can be seen as an extensive quantity which {\it can be localized}, as opposed to a sum of pair interactions.
 The main tools, that we present in this chapter, are a {\it renormalization by truncation} of the electric formulation $\int_{\R^\d}|\nab h|^2$, and a {\it monotonicity property} with respect to the truncation parameters.  
The first applications to  statistical mechanics will follow in the next chapter and applications to dynamics in Chapter~\ref{chap:commutator}. 

For $ 0<\alpha \le 1$ we let $|\varphi|_{C^\alpha}$ denote the H\"older semi-norm  
\be \label{defholder} |\varphi|_{C^\alpha} = \sup_{x,y} \frac{|\varphi(x)- \varphi(y)|}{|x-y|^\alpha} ,\ee
 which we will use in all the text.

\section{Definition and electric representation}
\subsection{Definition}

In the whole chapter, we will  consider a reference measure (most often a probability)  $\mu$ and assume that
 \be \label{condmupourFN}\iint_{\R^\d\times \R^\d} |\g(x-y)| d|\mu|(x)d|\mu|(y)<+\infty. \ee
\begin{defi}[The modulated energy]\index{modulated energy} Let $\g$ be as in \eqref{riesz}.
We define for any integrable $\mu$ with $\int_{\R^\d} \mu=1$ satisfying \eqref{condmupourFN} the modulated energy of the configuration $\XN=(x_1,\dots, x_N) \in (\R^\d)^N$ with respect to $\mu$ as
\be\label{def:FN} 
\F_N (\XN,\mu):=\hal \iint_{\R^\d\times \R^\d \backslash \triangle} \g(x-y)\, d\(\sum_{i=1}^N \delta_{x_i} - N \mu\) (x) d\(\sum_{i=1}^N \delta_{x_i} - N \mu\) (y),\ee
where we recall $\triangle$ denotes the diagonal of $\R^\d \times \R^\d$.
\end{defi}

We defined $\F_N(\XN, \mu)$ for a generic   $\mu$  of integral $1$.  We consider this level of generality at this point because for  certain questions pertinent to dynamics and outside the scope of this text,  it is useful to define $\F_N$ for signed measures $\mu$, however in the situations that we will consider here,  $\mu$ will in fact  be a probability density: in the following chapters on Coulomb and Riesz gases, it will either be   the equilibrium measure $\meseq$ or  the thermal equilibrium measure $\mub$,  according to the choice of approach.
We will also soon add  the assumption that the density $\mu$  is  bounded, which is satisfied for the  thermal equilibrium measure by Proposition~\ref{lem242}, and which is true for the equilibrium measure under assumptions of regularity for $V$, see for instance Example \ref{example2}.

The measure $\mu$ should be thought of as a reference measure or neutralizing background charge.
This initial formulation of $\F_N$ shows it as a sum of pair interactions, or total interaction of the system of charges at $x_i$'s and negative charge $-N\mu$. 

We note that another natural normalization of $\F_N$  is to consider $\frac{1}{N^2}\F_N$. This way $\F_N$ can naturally be thought of as a Coulomb or Riesz  (squared) distance  between the empirical measure 
$\emp=\frac{1}{N}\sum_{i=1}^N \delta_{x_i}$ and the measure $\mu$. We will see that even though $\F_N$ is not necessarily positive, this metric interpretation is approximately correct. 
In fact, if the diagonal $\triangle$ was not removed, then  after  Fourier transform $\mathcal F$, $\F_N$ could be seen as 
\begin{multline}\label{4.13} \frac{\F_N(\XN, \mu)}{N^2} = \int_{\R^\d}\mathcal F \g(\xi) |\mathcal F(\emp-\mu)(\xi)|^2= C_{\d, \s} \int_{\R^\d} \frac{1}{|\xi|^{\d-\s} }|\mathcal F(\emp-\mu)(\xi)|(\xi)^2\\ = C_{\d,\s} \|\emp-\mu\|^2_{\dot{H}^{\frac{\s-\d}{2}} (\R^\d)}\end{multline}
for some constant $C_{\d,\s}$, as  in \eqref{formalHs} and \eqref{normdoth}.
Of course this is not really correct, since Dirac masses generally do not belong to the  space $\dot{H}^{\frac{\s-\d}{2}} $. However, this can be given a meaning in a renormalized sense, either by removing the diagonal as done in the definition of $\F_N$, or via truncations as we will see just below.
One could also use the term {\it desingularization}, as in the desingularization of vortices performed in fluid mechanics.

\subsection{Electric formulation in the Coulomb case}\label{sec412} 
\index{electric formulation}
To go further, we use an electrostatic interpretation of the energy $\F_N$, as first used in \cite{ss1}, and  the rewriting of the energy via truncation, as in \cite{rs,PetSer} but using the  nearest-neighbor distance truncation as in \cite{LSZ,ls2}. In this section, we first present it in the easier Coulomb case.

Such a computation allows to replace the sum of pairwise interactions of all the charges and background $\mu$ by an integral (extensive) quantity, which is easier to handle and can be localized. 






Let us first consider the potential generated by the configuration $\XN$ and the background $\mu$, defined by 
\begin{equation}
\label{def:hnmu} 
h^\mu_N[\XN](x) := \int_{\R^\d} \g(x-y) d\(\sum_{i=1}^N \delta_{x_i} -N \mu\)(y).\end{equation}
We will most often omit the dependence in $\XN$ and $\mu$ and simply write $h_N$ when there is no ambiguity.
In the Coulomb case, $\g$ is (up to the constant $\cd$), the fundamental solution to Laplace's equation in dimension $\d$, that is 
$-\Delta \g= \cd \delta_0$,
  so we have 
\be\label{eqhne} -\Delta h_N= \cd\( \sum_{i=1}^N \delta_{x_i} - N \mu\).\ee

We note that as in Proposition \ref{lemelect1} and its proof, even if $\d= 1, 2$ where $\g$ does not tend to $0$ at infinity, $h_N^\mu[\XN]$ always  decays at infinity because $\int \mu=1$ and the system formed by the positive charges at $x_i$ and the  background charge $N\mu$ is neutral. 

We would like to  rewrite $\F_N(\XN,\mu)$  defined in~\eqref{def:FN} via Green's formula  as 
\be\label{formalcomputation}\F_N(\XN, \mu) =\frac{1}{2\cd}  \int_{\R^\d} h_N (-\Delta h_N) = \frac{1}{2\cd} \int_{\R^\d} |\nab h_N|^2,\ee 
exactly as done in Proposition \ref{lemelect1}, 
with  the boundary terms at infinity vanishing thanks to the above noted decay.
This rewriting of the energy of electrostatic charges as the Dirichlet energy of the potential is the well known operation of ``carr\'e du champ" \cite{chafaiblog,hirsch}.

 However this formal computation is not correct due to the singularities of $h_N$ at the points $x_i$, which make the integral diverge if $\s\ge 0$, or equivalently due to the fact that this computation neglects the diagonal.

To remedy this, we use a truncation procedure which allows to give a renormalized meaning to this integral. Giving an electric formulation in a renormalized fashion  was first done in the context of Coulomb gases in \cite{ss1}, inspired by \cite{bbh} in the context of Ginzburg-Landau vortices (see also \cite{noteszurich} for a discussion).
Using a mollification to give sense to this divergent integral is a  natural idea.  In the Coulomb case, what we describe below is another avatar of Onsager's lemma (see \cite{liebseiringer,rs}).  However, here we make use  of two specific points:    first we do not mollify the interaction but only the singular charge distribution (and not the background charge $\mu$). Second we use a regularization scale $\eta_i$ which may depend on the point $x_i$ hence on the whole configuration $\XN$. This will provide more flexibility, in particular when two points get very close. For instance we will use crucially the nearest-neighbor distance as a truncation parameter, as first used in \cite{ls2} and inspired by \cite{GunPan}.
We denote by $\R_+$ the set of nonnegative real numbers.

\begin{defi}[Truncated potentials, Coulomb case] \index{truncation radii}
For any number $\eta>0$, let us denote
\be\label{def:truncation0} 
\f_{\eta} (x) := (\g(x)-\g(\eta))_+, \qquad \g_\eta:= \g - \f_\eta\ee where $(\cdot)_+$ denotes the positive part of a number, and we naturally also view $\g$ as a function of $\R$, i.e. $\g(\eta)$ means $\frac{1}{\s}\eta^{-\s}$ or $-\log \eta$. 

 For any $\vec{\eta}=(\eta_1, \dots, \eta_N)\in \R_+^N$, and any  function $h$ satisfying a relation of the form 
\begin{equation}\label{formu0}
-\Delta h =  \cd\(\sum_{i=1}^N \delta_{x_i}-N \mu\),
\end{equation}
 we define the truncated potential
\begin{equation}\label{formu2}
h_{\vec{\eta}}:= h- \sum_{i=1}^N \f_{\eta_i}(\cdot -x_i).\end{equation}
\end{defi}
Let us point out that $\f_\eta$ is supported in $B(0,\eta)$,  and that $\g_\eta= \min (\g, \g(\eta)) $  is  a truncation of the Coulomb kernel. We note here that we could choose more regular truncated potentials, for instance any $\g_\eta$ which is smooth and radial and coincides with $\g$ outside $B(0, \eta)$. The one we choose has the advantage of being constant in $B(0, \eta)$.

\begin{defi}[Smeared charges, Coulomb case]
We  denote 
\be \label{defdeltaeta0}
\delta_{x_0}^{(\eta)}:=   - \frac{1}{\cd}\Delta \g_\eta(\cdot -x_0). \ee
It is the uniform measure of mass $1$ supported on $\partial B(x_0, \eta)$.\end{defi}
Since $\g_\eta$ is constant in $B(0, \eta)$, and equal to $\g$, which is harmonic, outside $B(0, \eta)$, the distribution $\Delta \g_\eta$ can only be supported on $\partial B(0, \eta)$.  By radial symmetry it must have a uniform density on $\partial B(0, \eta)$. To check that $\int \delta_0^{(\eta)}=1$, one can either compute it explicitly or use Green's formula to say that  for $R >\eta$,
$$\int_{\partial B(0, R)} \frac{\partial \g}{\partial \nu}= \int_{\p B(0, R)} \frac{\partial \g_\eta}{\partial \nu} = \int_{B(0, R)} \Delta \g= \int_{B(0, R)} \Delta \g_\eta,$$
thus in view of \eqref{coulombkernel}  and \eqref{defdeltaeta0}, we conclude that $\int_{\R^\d}  d\delta_0^{(\eta)}=1$ and $\int_{\R^\d} d\delta_{x_0}^{(\eta)} =1$.

 We may next  notice that 
 \be \label{ggeta0} \g_\eta= \g* \delta_0^{(\eta)},\ee  and thus by \eqref{def:truncation},
\be \label{fconv} \f_{\eta}=\g*\( \delta_0-\delta_0^{(\eta)}\).\ee
This is  a rephrasing of what physicists call Newton's theorem: the Coulomb potential generated by a point charge and the Coulomb potential generated by a radial smearing of that point charge coincide outside the smearing region (here,  $\g$ and $\g_\eta$ coincide outside $B(0,\eta)$). Moreover, the Coulomb  potential of the smeared charge is smaller in the smearing region than the original one (here $\g_\eta\le \g$ everywhere). Mathematically, this is a consequence of the mean-value (in)equality for (sub)harmonic  functions.

Taking the Laplacian of  \eqref{fconv},  it follows that 
\be\label{deltaf}-\Delta \f_{\eta}= \cd\(  \delta_0-\delta_0^{(\eta)}\).\ee
  We note that  in view of~\eqref{deltaf}, the function $h_{N,\vec{\eta}}$ defined via \eqref{def:hnmu} and \eqref{formu2}, i.e. 
  $$h_{N,\vec{\eta}}= h_N- \sum_{i=1}^N \f_\eta(\cdot -x_i)$$
   then satisfies 
  $$ h_{N,\vec{\eta}} = \g* \(  \sum_{i=1}^N \delta_{x_i}^{(\eta_i)}-N \mu\)$$ and 
  \be\label{eqhnee0} -\Delta h_{N,\vec{\eta}} = \cd\( \sum_{i=1}^N \delta_{x_i}^{(\eta_i)}-N \mu\).\ee

In the case $\s\le 0$, which we only consider in dimension $\d=1$, the renormalization is not needed at all -- however, we can still do it (for the sake of uniformity of notation).

We can then show the following exact representation for $\F_N$, which may be viewed as a ``renormalized" version of Proposition \ref{lemelect1}.
 \begin{lem}[Electric representation of the modulated energy - Coulomb case]\label{lemelec0}\index{modulated energy}
 Assume $\s=\d-2$.  Let $\mu$ be  integrable such that $\int_{\R^\d} \mu=1$ and   \eqref{condmupourFN} holds.
  For any $\XN$ in $(\R^\d)^N$ pairwise distinct  configuration, we have
\be\label{rewritF0}
 \FN(\XN, \mu) =
 \frac{1}{ 2\cd} \lim_{\eta_i \to 0} \left(\int_{\R^{\d}} |\nab h_{ N,\vec{\eta}}|^2  - \cd\sum_{i=1}^N \g(\eta_i)  \right).
\ee
\end{lem}
\begin{proof} We apply the result of Proposition \ref{lemelect1} to $\mu_+=\frac{1}{N}\sum_{i=1}^N \delta_{x_i}^{(\eta_i)}$ and $\mu_-= \mu$. 
Thanks to assumption \eqref{condmupourFN}, the assumptions of that proposition are satisfied and we deduce that 
\be\label{intdoub0}\iint_{\R^{\d}\times \R^{\d} } \g(x-y)d\left( \sum_{i=1}^N \delta_{x_i}^{(\eta_i)} - N \mu\right)(x) d\left( \sum_{i=1}^N \delta_{x_i}^{(\eta_i)} - N\mu\right)(y)= \frac1{\cd}\int_{\R^{\d}} |\nab h_{N, \vec{\eta}}|^2.\ee
Next, we observe that by translation-invariance
$$\iint \g(x-y) d\delta_{x_i}^{(\eta_i)} (x) d\delta_{x_i}^{(\eta_i)} (y) 
= \iint \g(x-y) d\delta_{0}^{(\eta_i)} (x) d\delta_{0}^{(\eta_i)} (y) 
= \int \g_{\eta_i} d\delta_0^{(\eta_i)}= \g(\eta_i).$$
Here we have used \eqref{ggeta0} and the fact that $\g_\eta=\g$ on $\partial B(0, \eta)$.
We may thus write that
\begin{multline*}
\lim_{\vec{\eta} \to 0 }\Bigg[ \iint_{\R^{\d}\times \R^{\d} } \g(x-y) d\left( \sum_{i=1}^N \delta_{x_i}^{(\eta_i)} - N \mu\right)(x)d \left( \sum_{i=1}^N \delta_{x_i}^{(\eta_i)} - N\mu\right)(y) 
- \sum_{i=1}^N \g(\eta_i)\Bigg] 
\\=
\iint_{\triangle^c} \g(x-y)  d \left( \sum_{i=1}^N \delta_{x_i}-N\mu\right)(x)d \left( \sum_{i=1}^N \delta_{x_i} -N \mu\right)(y) 
 \end{multline*}
and we deduce in view of \eqref{intdoub0} that \eqref{rewritF0} holds.\end{proof}

\subsection{Electric formulation in the Riesz case}\label{secrieszcase} \index{electric formulation}
In the Riesz or one-dimensional logarithmic case, the above representation is not true because $\g$ is not the fundamental solution to the Laplacian. However we can use the extension representation as described in Section \ref{sec-extension} to view $\g$ as the fundamental solution of a divergence-form operator in an extended space. This is the approach that was proposed in  the one-dimensional log case in \cite{ss2}, and in \cite{PetSer} for the more general Riesz cases \eqref{riesz}.
 Let us go through the details, using the notation of Section \ref{sec-extension}. We recall that we use an extension to $\R^{\d+\k}$ where $\k=0$ in the Coulomb case $\s=\d-2$ and $\k=1$ in all other Riesz cases. The most  important fact is that for any distribution $f$ on $\R^\d$, the potential 
 $$\g*f(x)=\int_{\R^\d} \g(x-x')f(x') dx'$$
 can naturally be extended into a potential on $\R^{\d+1}$  (if $\k=1$)
 $$h^f(x,y):=\int_{\R^\d} \g((x,y)-(x',0))f(x') dx'$$
 which satisfies 
 \be \label{etoilegh}
 -\div(\yg \nab h^f)= \cds f\drd\quad \text{in} \ \R^{\d+\k},\ee
 where $\cds$ is defined in \eqref{fractlapkernel} and $\gamma$ is given by \eqref{defgamma}. \index{truncation radii}
 \begin{defi}[Truncated potentials] \label{deftrunc}
 For any number $\eta>0$, we denote
 \be\label{def:truncation} \f_\eta:=(\g(x)-\g(\eta))_+, \qquad \g_\eta:=\g-\f_\eta.\ee
We also naturally extend $\g, \g_\eta$ and $\f_\eta$ into radial functions of $\R^{\d+\k}$.
  For any $\vec{\eta}=(\eta_1, \dots, \eta_N)\in \R_+^N$, and any  function $h$ satisfying a relation of the form 
\begin{equation}\label{formu3}
-\div (\yg \nab h) =  \cds\(\sum_{i=1}^N \delta_{x_i}-N \mu\drd\)\quad \text{in} \ \R^{\d+\k}
\end{equation}
 we define the truncated potential
\begin{equation}\label{formu23}
h_{\vec{\eta}}:= h- \sum_{i=1}^N \f_{\eta_i}(\cdot -x_i)\quad\text{in} \ \R^{\d+\k}.\end{equation}
\end{defi}
Again, $\f_\eta$ is supported in the ball $B(0,\eta)$ of $\R^{\d+\k}$,  and  $\g_\eta= \min (\g, \g(\eta)) $.

\begin{defi}[Smeared charges]\label{defi41}
We  denote 
\be \label{defdeltaeta}
\delta_{x_0}^{(\eta)}:=   - \frac{1}{\cds}\div \( \yg  \nab \g_\eta(x-x_0)\). \ee
It is a   measure of mass $1$ supported on $\partial B(x_0, \eta)$, sphere of  center $x_0$ and radius $\eta$ in $\R^{\d+\k}$.\end{defi}
The fact that $\delta_{x_0}^{(\eta)}$ is a probability measure  supported on $\p B(0, \eta)$ can be argued as in the Coulomb case, or by direct computation: one sees that  its density on the $(\d+\k)$-dimensional sphere is equal to $\frac{1}{\cds} \frac{\yg}{|z|^{\s+1}}= \frac{1}{\cds} \frac{\yg}{\eta^{\s+1}}$, hence it is generally {\it not} uniform.  

Contrary to the Coulomb case, it is not true that $\g_\eta= \g* \delta_0^{(\eta)}$ (the two functions coincide however on $\R^\d \times \{0\}$ as proven just below) because $\delta_0^{(\eta)}$ is not supported in $\R^\d \times \{0\}$, while $\g$ is the fundamental solution to the $-\div (\yg \nab \cdot )$ operator only for functions supported in $\R^\d \times \{0\}$. This corresponds to the fact that Newton's theorem  or the mean-value theorem does not hold for Riesz interactions. 
We will, however, not need this fact.  Instead, we will use that in view of \eqref{eq:Gfs}, \eqref{def:truncation} and \eqref{defdeltaeta} we have 
\be \label{eqpourfeta} 
- \div (\yg \nabla \f_\eta)= \cds \(\delta_0 - \delta_0^{(\eta)} \)\quad \text{in} \ \R^{\d+\k}.
\ee
We now check that as claimed $\g_\eta$ and $ \g* \delta_0^{(\eta)}$ coincide on $\R^\d \times \{0\}$.
\begin{lem}\label{lem613}
If $w,z \in \R^\d$ then 
\begin{equation}
\label{eq:lemme}
\int_{\R^{\d+\k}} \g(x-w) d\delta_z^{(\eta)} (x) = \g_\eta(w-z). \end{equation}
\end{lem}
\begin{proof} By translation invariance, it suffices to consider $z=0$.
By definition \eqref{defdeltaeta} and by \eqref{eq:Gfs}, using integration by parts we have
\begin{align*}
\int_{\R^{\d+\k}}  \g(x-w)d( \delta_0^{(\eta)} -\delta_0) (x) &= -\frac{1}{\cds}\int_{\R^{\d+\k}} \g(x-w) \div (\yg \nabla (\g_\eta-\g))(x)  dx\\ & =  \frac{1}{\cds} \int_{\R^{\d+\k}} \nabla \g(x-w) \cdot \nabla (\g_\eta-\g) (x)\yg dx\\
&= - \frac{1}{\cds} \int_{\R^{\d+\k}} \div (\yg \nabla \g(x-w) ) (\g_\eta-\g) (x) dx \\
& =  \int_{\R^{\d+\k}}( \g_\eta-\g)(x)d\delta_w (x)= (\g_\eta-\g)(w).\end{align*}
In the integration by parts, we can check that the boundary terms vanished at infinity using the same reasoning as in the proof of Proposition \ref{lemelect1}. The result follows.
\end{proof}

As explained above, given $x_1, \dots, x_N \in \R^\d$, if $\k=1$ we identify them with  the  points $(x_1, 0), \dots, (x_N,0)$ in $\R^{\d+\k}$, and
we may then define the potentials $h_N$ and truncated potentials $h_{N,\vec{\eta}}$  in $\R^{\d+\k}$ by 

\be\label{defhNmu}
h_N [\XN]= \g * \left(\sum_{i=1}^N \delta_{x_i} - N\mu \delta_{\R^\d}\right)\qquad h_{N,\vec{\eta}}[\XN]: = h_N -    \sum_{i=1}^N \f_{\eta_i} (\cdot-x_i) \ee
Since $\g$ is naturally extended to a function in $\R^{\d+\k}$, these potentials make sense as functions in $\R^{\d+\k}$.  In view of  \eqref{etoilegh}, \eqref{eq:Gfs} and \eqref{defdeltaeta}, 
 $h_N $ solves 
\be \label{bbe} -\div (\yg \nab h_N)= \cds \left(\sum_{i=1}^N \delta_{x_i} - N\mu \delta_{\R^\d}\right)\quad \text{in} \ \R^{\d+\k},
\ee
while $h_{N,\vec{\eta}}$ solves
\be \label{eqhneta}
-\div (\yg \nab h_{N,\vec{\eta}})= \cds \left(\sum_{i=1}^N \delta_{x_i}^{(\eta_i)} - N\mu \delta_{\R^\d}\right) \quad \text{in} \ \R^{\d+\k}.\ee



\begin{rem}In the case $\d=1$, $\s=0$  we have $ \gamma=0$, and  $h_N$  is nothing else than the harmonic extension to $\R^2$, away from the real axis,  of the potential  defined in dimension $1$ by the analogue of \eqref{def:hnmu}. This is closely related to  the {\it Stieltjes transform}, a commonly used object in Random Matrix Theory, see \cite{agz}. In the cases \eqref{riesz}, the situation is the same, except for the presence of the $|y|^\gamma$ weight. \end{rem}

The electric representation presented in the previous section for the Coulomb case goes through without change, if one replaces 
$$\int_{\R^\d} |\nab h_{N,\vec{\eta}}|^2$$ with $$\int_{\R^{\d+\k}} |y|^\gamma |\nab h_{N,\vec{\eta}}|^2 .$$
As in  Section \ref{sec-extension}, everything can be written in a unified way encompassing Coulomb and Riesz cases \eqref{riesz} by working in $\R^{\d+\k}$ and using \eqref{defgamma}.

 We will frequently use the following bounds.
 \begin{lem}For any $\eta>0$ and any integer $m  \ge 0$ with $m<\d-\s$, we have 
\begin{equation}\label{eq:intf}
\int_{\R^d} |\nabla^{ m}\f_\eta| \le C \eta^{\d-\s-m}
\end{equation}
with $C>0$ depending only on $\d, \s$ and $m$.
\end{lem}
These can be proven using for instance the explicit form 
\begin{equation}\label{4128}
\f_\eta(x) = \begin{cases} \left(-\log\Big(\frac{|x|}{\eta}\Big)\right)_+& {\s=0}, \\ \frac{\eta^{-\s}}{\s}\left(\frac{\eta^{\s}-|x|^\s}{|x|^\s}\right)_+ & {\s>0}\\
-\frac{\eta^{-\s}}{\s} \( \frac{|x|^\s-\eta^\s}{|x|^\s}\right)_+ & {\s<0}.
 \end{cases}
\end{equation}

We may now state the generalization of Lemma \ref{lemelec0}.
\begin{lem}[Electric representation - general Riesz case] 
\label{lemelecriesz}
Assume $\s\in [\d-2, \d)$.  Let $\k=0$ if $\s=\d-2$ and $\k=1$ otherwise. Let  $\gamma$ be given by \eqref{defgamma}.  Assume $\mu$ is a  probability measure satisfying \eqref{condmupourFN}. For any pairwise distinct $\XN$ in $(\R^\d)^N$, we have \be\label{rewritF}
 \FN(\XN, \mu) =
 \frac{1}{ 2\cds} \lim_{\eta_i \to 0} \left(\int_{\R^{\d+\k}}\yg |\nab h_{ N,\vec{\eta}}|^2  - \cds\sum_{i=1}^N \g(\eta_i)  \right).
\ee
\end{lem}
\begin{proof}
Using the assumption \eqref{condmupourFN}, we apply the result of Proposition \ref{lemelect1} to $\mu_+=\frac1{N}\sum_{i=1}^N \delta_{x_i}^{(\eta_i)}  $ and $\mu_-=\mu$ and obtain 
\be\label{intdoub}\iint_{\R^{\d+\k}\times \R^{\d+\k} } \g(x-y)d\left( \sum_{i=1}^N \delta_{x_i}^{(\eta_i)} - N \mu\drd\right)(x) d\left( \sum_{i=1}^N \delta_{x_i}^{(\eta_i)} - N\mu\drd\right)(y) = \frac1{\cds}\int_{\R^{\d+\k}}\yg |\nab h_{N, \vec{\eta}}|^2.\ee
 By definition \eqref{defhNmu} we next rewrite this, via an integration by parts, as 
$$\int_{\R^{\d+\k}}\yg |\nab h_{N, \vec{\eta}} |^2= \cds \int_{\R^{\d+\k}} \( h_{N}-\sum_{i=1}^N \f_{\eta_i} (\cdot -x_i)\) \  d \left(   \sum_{i=1}^N \delta_{x_i}^{(\eta_i)}- N\mu\drd\right)  .$$
Using that $\f_\eta(\cdot -x_i)$ vanishes on $\partial B(x_i, \eta_i)= \supp(\delta_{x_i}^{(\eta_i)})$, we obtain that if $\eta_i$ are small enough that the balls $B(x_i, \eta_i)$ are disjoint, we have
$$\int_{\R^{\d+\k}}\yg |\nab h_{N, \vec{\eta}} |^2= \cds \int_{\R^{\d+\k}}  h_{N} \  d \left(   \sum_{i=1}^N \delta_{x_i}^{(\eta_i)}- N\mu\right)  +N \cds \sum_{i=1}^N   \int_{\R^{\d+\k}} \f_{\eta_i} (\cdot-x_i) \mu \, d \drd.$$ Rewriting $h_N$ again via \eqref{defhNmu}, and using that $\int \g \delta_0^{(\eta)}= \g(\eta)$, we  deduce that  if the $\eta_i$'s are small enough, then
\begin{align}\label{multfinal}& \int_{\R^{\d+\k}}\yg |\nab h_{N, \vec{\eta}} |^2 \\
\notag &\qquad  =
 \cds
\iint_{\R^{\d+\k}\times \R^{\d+\k} } \g(x-y) d\left( \sum_{i=1}^N \delta_{x_i} - N \mu\drd\right)(x)d \left( \sum_{i=1}^N \delta_{x_i}^{(\eta_i)} - N\mu\drd\right)(y) \\  \notag&\qquad 
+ N \cds  \sum_{i=1}^N \int_{\R^{\d+\k}}  \f_{\eta_i} (\cdot-x_i) \mu \, d \drd
\\ \notag &\qquad =
 \cds
\Bigg[ \iint_{\R^{\d+\k}\times \R^{\d+\k} } \g(x-y) d \left( \sum_{i=1}^N \delta_{x_i}^{(\eta_i)} - N\mu\drd\right)(y)  d\left( \sum_{j\neq i} \delta_{x_j} - N \mu\drd\right)(x)
+ \sum_{i=1}^N \g(\eta_i)\Bigg] \\  \notag & \qquad +  N\cds  \sum_{i=1}^N  \int_{\R^{\d+\k}} \f_{\eta_i} (\cdot -x_i) \mu \, d\drd.
 \end{align}
Then, we use that $\f_{\eta_i} (x-x_i) \mu(x)$ converges monotonically to $0$ as $\eta_i\to 0$ to write that $\int_{\R^{\d+\k}} \f_{\eta_i} (\cdot-x_i) \mu \drd \to 0$. 
 Taking the limit of \eqref{multfinal}  as $\eta_i\to 0$, we deduce that  \eqref{rewritF} holds.\end{proof}

The limit in \eqref{rewritF} is not quantitative. We will show next that equality  in  \eqref{rewritF} holds  for $\eta_i$ small enough {\it without having to take a limit}, thanks to a monotonicity property that will bring other consequences. In particular, all fine energy controls will be  obtained below by leveraging this monotonicity.

\section[Monotonicity]{Monotonicity with respect to truncation and consequences}\label{sec:monoto} \index{monotonicity}
The  following proposition, first  proven in \cite[Prop.~2.3]{ls2} and \cite[Prop 3.3]{Serfaty2020},   gives an exact representation for $\F_N$ and shows a 
monotonicity property  that increasing the  truncation parameters $\eta_i$   can only decrease the value of the quantity whose limit is taken in the  right-hand side of~\eqref{rewritF}. The idea is natural when one recalls Newton's theorem in the Coulomb case: when smearing charges radially, the potential they generate can only decrease, hence the interaction energy between smeared charges is smaller than between the discrete charges. Moreover, there is equality in case the smearing balls are disjoint, since the  potential generated by a radially smeared charge in a ball coincides with that of the discrete charge outside of the ball. 
A more precise way to write this is that if $\g$ is Coulomb,  then using the mean-value or Newton's theorem,  we have
\begin{align}\label{rlhs}
\iint \g(x-y) d \delta_{x_1}^{(\eta_1)} (x) d \delta_{x_2}^{(\eta_2)}(y) &\le \iint \g(x-y) d\delta_{x_1} (x)  d \delta_{x_2}^{(\eta_2)}(y)\\ \notag
& \le \iint \g(x-y) d\delta_{x_1} (x)d\delta_{x_2}(y)= \g(x_1-x_2)\end{align}
with equality if and only if $B(x_1, \eta_1)$ and $B(x_2, \eta_2)$ are disjoint.
The choice of truncation procedure made in Definition \ref{deftrunc} allows to have the same property also in the Riesz case, despite the absence of the mean-value theorem.
The following proposition shows how to apply this idea to the context of $\F_N$, moreover it retains a positive term corresponding to the difference between the right and left-hand side in \eqref{rlhs}, which encodes the small-scale interactions that are erased by smearing.

 \begin{prop}[Electric formulation and monotonicity property] \label{prop:monoto}
Assume \eqref{riesz}. Let $\mu$ be a bounded  probability  density on $\R^{\d}$ satisfying \eqref{condmupourFN}, and $\XN$ be in $(\R^{\d})^N$. For any $\veta\in \R_+^N$ we have  
\begin{multline}
\label{fnmeta2}
\hal\sum_{i\neq j}  \(\g(x_i-x_j)- \g(\eta_i)\)_+\\ \le   \F_N(\XN,\mu)   -\(\frac{1}{2\cds} \int_{\R^{\d+\k}} \yg |\nab h_{N,\vec{\eta}}|^2 -\hal\sum_{i=1}^N \g(\eta_i)  -N \sum_{i=1}^N \int_{\R^{\d}} \f_{\eta_i}(x - x_i) d\mu(x)\) 
\end{multline} where $\f_\eta$ is defined in \eqref{def:truncation}. Moreover, there 
\emph{is equality} if the $B(x_i,\eta_i )$'s are all disjoint.
\end{prop} 
In view of \eqref{eq:intf},   if $\eta_i $ is small, we may consider $ \sum_{i=1}^N \int_{\R^{\d}} \f_{\eta_i}(x - x_i) d\mu(x) $ as a small error.
 More precisely, we can use that 
\be\label{fdmu}
\left|\sum_{i=1}^N \int_{\R^\d}\f_{\eta_i} (x-x_i) d\mu(x)\right|\le  C_{\d,\s} \|\mu\|_{L^\infty}  \sum_{i=1}^N\eta_i^{\d-\s}  ,\ee
which follows from \eqref{4128}.

\begin{rem}[The case $\s <0$] \label{remsneg}
In the case $\s <0$, we may take $\eta_i=0$ for all $i$, and we then obtain 
\be\label{casszero} \F_N(\XN, \mu)= \frac{1}{2\cds}\int_{\R^{\d+\k}} \yg |\nab h_N|^2 .\ee
\end{rem}

  We will prove the following more refined statement, from  which Proposition \ref{prop:monoto} follows. 
    \begin{lem}[Monotonicity with respect to the truncation parameter]\label{monoto}
Let $U$ be a domain in $ \R^\d$ and  $u$ solve 
\be\label{eqsu}
-\div (\yg \nab  u)= \cds\(\sum_{i=1}^N \delta_{x_i} - N\mu \drd\) \quad \text{in} \ U \times \R^\k,\ee  and let $u_{\vec{\alpha}}, u_{\vec{\eta}}$ be truncated fields as in~\eqref{formu23}. Assume 
$\alpha_i \le \eta_i$ for each $i$.  Letting $I_N$ denote $\{i, \alpha_i\neq \eta_i\}$, assume that 
for each $i \in I_N$   we have  $B(x_i ,\eta_i) \subset U$.
Then 
\begin{multline}
\label{premono}
\int_{U\times \R^\k} \yg |\nab u_{\vec{\eta}}|^2 - \cds \sum_{i\in I_N} \g(\eta_i) -2N \cds \sum_{i\in I_N} \int_{U}  \f_{\eta_i}(x-x_i)d\mu(x)\\- \( \int_{U\times \R^\k}\yg |\nab u_{\vec{\alpha}}|^2 - \cds \sum_{i\in I_N} \g(\alpha_i) -2N\cds\sum_{i\in I_N} \int_U \f_{\alpha_i}(x-x_i)d\mu(x)\) \le 0.\end{multline}
Moreover, there is equality if the $B(x_i,\eta_i)$'s are disjoint from all the other $B(x_j, \eta_j)$'s for each $i \in I_N$. 
\end{lem}
\begin{proof}
For any $\alpha \le \eta$, let us  denote $\f_{\alpha, \eta}:=\f_{\alpha}-\f_{\eta}$, where $\f$ is as in \eqref{def:truncation}.  Observe that $\f_{\alpha,\eta}$ vanishes outside $B(0, \eta)$ and 
$$\g(\eta)-\g(\alpha) \le \f_{\alpha ,\eta}\le 0,$$ 
while, in view of~\eqref{eqpourfeta},
\begin{equation}\label{eqfae}
- \div (\yg \nab \f_{\alpha, \eta}) = \cds \(\delta_0^{(\eta)}- \delta_0^{(\alpha)}\).\end{equation}
Using the fact that by definition and~\eqref{formu23} we have 
$$\nab u_{\vec{\eta}}(z) -\nab  u_{\vec{\alpha}}(z) = \sum_{i\in I_N} \nab\fae(z-x_i),$$
we compute
\begin{align*}
T:= & \ 
\int_{U\times \R^\k} \yg | \nab u_{\vec{\eta}}|^2 - \int_{U\times \R^{\k} } \yg |\nab u_{\vec{\alpha}}|^2
\\
= &\  2 \int_{U \times \R^\k} \yg (\nab u_{\vec{\eta}}- \nab u_{\vec{\alpha}} ) \cdot \nab u_{\vec{\alpha}}+ \int_{U\times \R^\k}\yg  |\nab u_{\vec{\eta}}- \nab u_{\vec{\alpha}}|^2 \\
= & \  2\sum_{i\in I_N } \int_{U\times \R^\k} \yg\nab \fae(\cdot -x_i)  \cdot \nab u_{\vec{\alpha}}+ \sum_{i,j \in I_N} \int_{U\times \R^\k}\yg   \nab \fae(\cdot -x_i) \cdot \nab \faej (\cdot -x_j).
\end{align*}
If $B(x_i, \eta_i)\subset U$ the function $\fae(\cdot -x_i) $ vanishes on $\partial  ( U\times \R^\k) $, and we can integrate by parts without getting any boundary contribution. 
With the help of~\eqref{eqsu} and~\eqref{eqfae} we thus obtain  
\begin{align}
\label{ii1i2}
T 
=& \ 
2 \cds\sum_{i\in I_N}\int_{U\times \R^\k}  \fae(\cdot-x_i) d\Big( \sum_{j=1}^N \delta_{x_j}^{(\alpha_j)} -  N d\mu\,\drd\Big) 
\\ \notag
&  + \cds \sum_{i,j\in I_N } \int_{U\times \R^\k}   \fae(\cdot-x_i) d\( \delta_{x_j}^{(\eta_j)}-\delta_{x_j}^{(\alpha_j)} \)
\\  \notag
=  & \ 
\cds\sum_{i\in I_N} \int_{U\times \R^\k}   \fae(\cdot-x_i) d\( \sum_{j=1}^N \delta_{x_j}^{(\alpha_j)} + \delta_{x_j}^{(\eta_j)}\)  -2 N\cds\sum_{i\in I_N} \int_U \fae(\cdot-x_i)  d\mu \\  \notag
= & \
 \sum_{j=1}^N\sum_{i\in I_N, i\neq j} \cds \int_{\R^{\d}} \fae (\cdot-x_i) d(  \delta_{x_j}^{(\alpha_j)}+ \delta_{x_j}^{(\eta_j)}) 
+ \cds\sum_{i\in I_N}  \int_{\R^{\d+\k}} \fae(\cdot-x_i) d( \delta_{x_i}^{(\alpha_i)} + \delta_{x_i}^{(\eta_i)} )
\\ & \notag 
- 2N\cds \sum_{i\in I_N} \int_{U} \fae(\cdot-x_i) d \mu.
\end{align}
Since $\f_{\alpha_i, \eta_i} \le 0$, the first term in the right-hand side is nonpositive, and is zero if the $B(x_i, \eta_i)$'s with $i\in I_N$ are disjoint from the other balls. For the diagonal terms, we note that 
  $$  \int_{\R^{\d+\k}}   \fae(\cdot-x_i)   \(  \delta_{x_i}^{(\alpha_i)} + \delta_{x_i}^{(\eta_i)}\)     =  - (\g(\alpha_i)-\g(\eta_i))$$ by definition of $\f_{\alpha,\eta}$ and the fact that $\delta_0^{(\alpha)}$ is a measure of mass $1$ on $\partial B(0,\alpha)$.
Since $\fae=\f_{\alpha_i}-\f_{\eta_i}$, this finishes the proof of~\eqref{premono} after rearranging terms.
\end{proof}  
To prove Proposition \ref{prop:monoto}, we take  advantage of the nonpositive term in the right-hand side of \eqref{ii1i2}.

\begin{proof}[Proof of Proposition \ref{prop:monoto}] 
Let us apply Lemma \ref{monoto} to $u=h_{N}$ on $U=\R^\d$. Let us then return to the nonpositive first term in the right-hand side of~\eqref{ii1i2} and 
 bound it above and below  via
\begin{align}\label{monoplus}& \ \sum_{i\neq j}\( \g_{\eta_i} (|x_i-x_j|+\alpha_j)- \g(|x_i-x_j|-\alpha_j)\)_- \le 
\sum_{i\neq j} \int_{\R^{\d+\k}} \f_{\alpha_i, \eta_i} (x-x_i) d(\delta_{x_j}^{(\alpha_j)}+\delta_{x_j}^{(\eta_j)} )\\ & \notag
\leq 
\sum_{i\neq j} \int_{\R^{\d+\k}}\( \g_{\eta_i}(x-x_i)- \g_{\alpha_i}(x-x_i) \) d\delta_{x_j}^{(\alpha_j)} 
\leq \sum_{i\neq j}
\( \g(\eta_i) - \g_{\alpha_i}(|x_i-x_j|+\alpha_j) \)_-,\end{align}
where we used the definition of $\f_{\alpha_i, \eta_i}$ and the properties of $\g_\alpha$, in particular the  fact that it is radially decreasing.     
Combining the previous relations and \eqref{ii1i2}, we find 
\begin{align*}  
\lefteqn{
 \cds \sum_{ i\neq j}\( \g_{\alpha_i}(|x_i-x_j|+\alpha_j) -\g(\eta_i)\)_+
 } \qquad & 
 \\ & 
\leq 
\(\int_{\R^{\d+\k} }\yg |\nab h_{N,\vec{\alpha}}|^2  -\cds\sum_{i=1}^N  \g(\alpha_i)-2N\cds\sum_{i=1}^N \int_{\R^{\d}} \f_{\alpha_i}(x-x_i)d\mu\) 
\\ & \qquad
-\( \int_{\R^{\d+\k} }\yg |\nab h_{N,\vec{\eta}}|^2 - \cds\sum_{i=1}^N \g(\eta_i)   -2N\cds\sum_{i=1}^N \int_{\R^{\d} } \f_{\eta_i}(x-x_i)d\mu\).
\end{align*}
Letting all $\alpha_i \to 0$, the limit of the first term in the right-hand side is $\F_N(x_N, \mu)$ in  view of~\eqref{rewritF},  while that of the left-hand side is  $\cds \sum (\g(x_i-x_j)-\g(\eta_i))_+ $.  This finishes the proof of \eqref{fnmeta2}.
 \end{proof}

As a corollary, we obtain that even though $\F_N$ has no sign in general, it is bounded below by a constant much smaller than $N^2$.     Choosing $\eta_i$ large  in \eqref{fnmeta2} will make the lower bound in $\g(\eta_i)$ larger, however it will increase the error terms  $\int \f_{\eta_i}(x-x_i) d\mu(x)$ which are bounded by \eqref{fdmu}. Optimizing leads to choosing $\eta_i$ such that $\g(\eta_i)= N \eta_i^{\d-\s}$ which leads to the choice $\eta_i= N^{-\frac1\d}$, which is the natural microscale.  However, if we assume that $\mu \in L^\infty$, a sharper dependence in $\|\mu\|_{L^\infty}$
 is obtained by choosing $\eta_i=\lambda$
  with 
\be \label{deflambda} \lambda:= (N\|\mu\|_{L^\infty} )^{-\frac1\d}, \ee
which a natural distance lengthscale.

 \begin{coro}\label{corminoF} Let $\mu$ be a probability density  satisfying \eqref{condmupourFN}. 
 There exists $C>0$ depending only on $\d$  and $\s$ such that for any $\XN\in (\R^\d)^N$, we have 
 \be \label{minoF2} \F_N(\XN, \mu) + \( \frac{N}{2\d} \log (N\|\mu\|_{L^\infty})  \)\indic_{\s=0}  \ge - C  \|\mu\|_{L^\infty}^{\frac\s\d}  N^{1+\frac{\s}{\d}} \indic_{\s \ge 0}. \ee
 \end{coro}
 \begin{proof} It suffices to apply \eqref{fnmeta2}. Discarding  nonnegative terms and using \eqref{fdmu}, we obtain
\begin{align*} \F_N(\XN,\mu)  & \ge - \hal \sum_{i=1}^N \g(\eta_i) - N \sum_{i=1}^N \int_{\R^\d} \f_{\eta_i} (x-x_i) d\mu(x) \\ & \ge -  \hal \sum_{i=1}^N \g(\eta_i) -CN \|\mu\|_{L^\infty }\sum_{i=1}^N \eta_i^{\d-\s}.\end{align*}
 If $\s \ge 0$, 
 choosing $\eta_i= \lambda$ yields the result.
 If $\s<0$, the result is known from  Remark \ref{remsneg} (or  obtained again by taking instead $\eta_i=0$).
 \end{proof}\begin{rem}
 \label{remc4}
 We will see below  in Corollary \ref{corobenergy} that this estimate is sharp. \end{rem}
\begin{rem}\label{rem2} 
We can work with  $\mu$  less regular than $L^\infty$ as long as we can control the $\int \f_\alpha d\mu$ terms.  For instance suitable $L^p$ integrability  of $\mu$ suffices.
\end{rem}

\begin{rem} 
 We see that the  choice of $\eta_i$ can be refined a bit:  instead of \eqref{fnmeta2}, if we assume $\mu \in C^\alpha$, $\alpha>0$, $\mu \ge 0$ and write $\mu(x)= \mu(x_i) + O(|\mu|_{C^{\alpha}}|x-x_i|^\alpha)$ we have
 $$\left|\int \f_{\eta_i}(x-x_i)  d\mu (x)\right|\le  \mu(x_i) \eta_i^{\d-\s}+ |\mu|_{C^\alpha} \eta_i^{\d-\s+\alpha}.$$
 Optimizing leads to choosing $\eta_i = (N\mu(x_i))^{-\frac1\d} $ which yields
 \be \label{minoF3} \F_N(\XN, \mu) + \( \sum_{i=1}^N\frac{1}{2\d} \log (N\mu(x_i)))  \)\indic_{\s=0}  \ge - C N^{\frac{\s}{\d}} \sum_{i=1}^N \mu(x_i) ^{\frac\s\d}   -N^{\frac{\s-\alpha}{\d}} |\mu|_{C^\alpha} \sum_{i=1}^N  \mu(x_i)^{-\frac{\d-\s+\alpha}\d} . \ee
\end{rem}

We then define an important particular choice of truncation parameters, which we think of as the \textit{nearest-neighbor distance} for $x_i$.
\begin{defi}[Nearest-neighbor distance] If $\XN= (x_1, \dots, x_N)$ is a $N$-tuple of points in $\R^\d$ we denote for all $i=1, \dots, N$,
\begin{equation}\label{defri}
\rr_i := \frac{1}{4} \min\(\min_{j \neq i} |x_i-x_j|,\lambda\),
\end{equation} where $\lambda$ is as in \eqref{deflambda}. 
\end{defi}
\index{truncation radii}
We will denote $h_{N,\rr}$ for $h_{N,\vec{\eta}}$ with the choice $\eta_i=\rr_i$. 

 The next result shows that even though there is a cancellation between the very large terms $\int\yg  |\nab h_{N,\veta}|^2$ and $2\cds \sum\g (\eta_i)$ in the singular case $\s \ge 0$, a very interesting choice of $\eta_i $ is $\rr_i$  because the $\rr_i$'s are small enough that the balls are disjoint and there is still equality in \eqref{fnmeta2}, and large enough that {\it each} of the terms $ \int\yg  |\nab h_{N,\veta}|^2$ and $2\cds \sum\g (\rr_i)$ is {\it separately} controlled by the energy.
A localized version of this result will be later  given  in Proposition~\ref{procontrolelocal}.

\begin{prop}[Minimal distance  and truncated energy controls]\label{34}
Let $\mu $ be a bounded probability density satisfying \eqref{condmupourFN}.  Assume $\s \in [(\d-2)_+, \d).$
Given any pairwise distinct configuration $\ux_N \in (\R^\d)^N$, it holds that  
\be\label{bgr}
\begin{cases}
\displaystyle \sum_{i=1}^N \g(\rr_i) \le C \( \F_N(\XN, \mu)+  \|\mu\|_{L^\infty}^{\frac\s\d} N^{1+\frac\s\d}\)
& \text{if} \ \s> 0\\
\displaystyle\sum_{i=1}^N \g( \frac{\rr_i}{\lambda}) \le 2 \( \F_N(\XN, \mu)+N \frac{\log (N \|\mu\|_{L^\infty})}{2\d} \indic_{\s=0}\)+ C \|\mu\|_{L^\infty}^{\frac\s\d} N^{1+\frac\s\d}
& \text{if} \ \s= 0,\end{cases}
\ee 
and
\be \label{bornehnr}
\int_{\R^{\d+\k}}\yg |\nab h_{N ,{\rr}}|^2\le C\( \F_N(\XN, \mu)+N \frac{\log (N \|\mu\|_{L^\infty})}{2\d} \indic_{\s=0}\) +  C \|\mu\|_{L^\infty}^{\frac\s\d} N^{1+\frac\s\d}\ee
for some $C>0$ depending only on $\d$ and $\s$. \end{prop}
\begin{proof} 
Let us apply \eqref{fnmeta2} with   $\eta_i =\lambda$ from \eqref{deflambda} and observe that, for each $i$, by definition \eqref{defri} there exists $j\neq i$ such that 
$(\g(x_i-x_j)- \g(\lambda))_+ =    (\g(4\rr_i)   - \g(\lambda) )_+ $. Using \eqref{fdmu}, we may thus write  that 
\be\label{lb1}\hal \sum_{i=1}^N ( \g(4 \rr_i)-\g(\lambda) )_+\le  \F_N( \XN, \mu) - \frac{1}{2 \cds}\int_{\R^{\d+\k}}\yg |\nab h_{\vec{\eta}}|^2 +\hal N \g(\lambda ) +  C \|\mu\|_{L^\infty}^{\frac{\s}{\d}} N^{1+\frac\s\d}   . \ee
from which \eqref{bgr} follows after rearranging terms, and noting that in the case $\s=0$ we have $\|\mu\|_{L^\infty}^{\s/\d}N^{1+\s/\d}=N$, which can absorb the $-\sum_{i=1}^N \g(4) $ term.
Let us next choose   $\eta_i=\rr_i$ in \eqref{fnmeta2}. Using that $\rr_i \le \lambda$, this yields
$$ 0 \le \F_N(\XN, \mu) -  \frac{1}{2\cds} \int_{\R^{\d+\k}} \yg |\nab h_{N,{\rr}}|^2 + \hal \sum_{i=1}^N \g(\rr_i) + C \|\mu\|_{L^\infty}^{\frac\s\d}  N^{1+\frac\s\d}  .$$
Combining with \eqref{bgr}, and in the case $\s=0$, writing $\g(\rr_i) = \g(\rr_i/\lambda)+ \g(\lambda)$, \eqref{bornehnr} follows.
\end{proof}

 \section{Coercivity of the electric energy}
 Here, we prove that  the modulated energy does metrize the convergence of $\mu_N$ to $\mu$ and acts as an effective Coulomb or Riesz distance. In the case $ \s <0$, by Remark \ref{remsneg} it is exactly a Coulomb/Riesz distance,  equal to  the (square of the) $\dot{H}^{\frac{\s-\d-\k}{2}}(\R^{\d+\k})$ norm of $\sum_i \delta_{x_i}-N\mu$.
As in \eqref{formalHs} and \eqref{4.13}, in view of  \eqref{intdoub}   and Plancherel's theorem, we have
\begin{align*}
 \int_{\R^{\d+\k}} \yg |\nab h_{N, \rr}|^2& = \cds \iint_{\R^{\d+\k}\times \R^{\d+\k}} \g(x-y) 
d \(\sum_{i=1}^N \delta_{x_i}^{(\rr_i)} - N \mu\drd\)(x)d\( \sum_{i=1}^N \delta_{x_i}^{(\rr_i)} - N \mu\drd\)(y)\\
 & =C_{\d,\s} 
 \left\| \sum_{i=1}^N \delta_{x_i}^{(\rr_i)} - N \mu\drd\right\|_{\dot{H}^{\frac{\s-\d-\k}{2}}(\R^{\d+\k})}^2 ,\end{align*}
where the homogeneous Sobolev semi-norm $\dot H^m(\R^\d)$ is defined by  \eqref{normdoth}.


Since  by \eqref{bornehnr} $\F_N$ controls $\int_{\R^{\d+\k}} \yg |\nab h_{N, \rr}|^2 $, it thus controls this fractional Sobolev semi-norm of $ \sum_{i=1}^N \delta_{x_i}^{(\rr_i)} - N \mu\drd$. 
In particular, in the Coulomb case where $\k=0$ and $\s=\d-2$, we thus  control   $\sum_{i=1}^N \delta_{x_i}^{(\rr_i)} - N \mu$ in  $\dot{H}^{-1}(\R^\d)$.  This is the way in which $\F_N$ can be seen as the square of a Coulomb/Riesz distance.

 To control of 
$ \sum_{i=1}^N \delta_{x_i} - N \mu$, it then suffices to estimate $\sum_{i} \delta_{x_i}^{(\rr_i)}- \delta_{x_i}$ 
which is easily controlled by the fact that the  $\rr_i$'s are small, more precisely smaller than $ \lambda = (N\|\mu\|_{L^\infty})^{-1/\d}$.
Since in general $\sum_i \delta_{x_i} $ does not belong to $\dot{H}^{\frac{\s-\d-\k}{2}}(\R^{\d+\k})$, we can only get a control of $ \sum_{i=1}^N \delta_{x_i} - N \mu$ in a weaker space, which  $\dot{H}^{\frac{\s-\d-\k}{2}}(\R^{\d+\k})$ embeds into.  There are several possible choices.

As a first possibility, we give the following control against test functions, i.e.~bounds on  Lipschitz linear statistics. It is a control in terms of $h_{N,\rr}$ but by \eqref{bornehnr} this amounts to  a control  by  $\F_N$.


\index{fluctuations}
\begin{lem}[The modulated energy controls the fluctuations]\label{prop:fluctenergy}
Let $\varphi$ be a function with bounded support  and assume that $\Omega\subset \R^\d$ contains a $\lambda$-neighborhood of this support. 
 For any configuration $\XN\in (\R^\d)^N$, letting $h_{N}$ be defined as in \eqref{defhNmu}, $\rr_i$ as in \eqref{defri},  and letting $I_\Omega$ denote $\{i,  x_i\in  \Omega\} $ and $\#I_\Omega$ its cardinality,  
for any $0<\alpha\le 1$, we have
\begin{itemize}
\item  in the Coulomb case,
\be\label{coulombfluct}
\left|\int_{\R^\d} \varphi\( \sum_{i=1}^N \delta_{x_i}- N d\mu\) \right|
\le C  \|\nab \varphi\|_{L^2(\Omega)} \|\nab h_{N,\rr}\|_{L^2(\Omega)} +  \# I_\Omega |\varphi|_{C^\alpha} N^{-\frac{\alpha}{\d}}  \|\mu\|_{L^\infty}^{-\frac{\alpha}{\d}} .\ee
\item in the Riesz case,
\be \label{rieszfluct}
\left|\int_{\R^\d} \varphi\( \sum_{i=1}^N \delta_{x_i}- N d\mu\) \right| \le C \|\varphi\|_{\dot{H}^{\frac{\d-\s}{2}}  } \( \int_{\R^{\d+\k}} \yg |\nab h_{N, \rr}|^2 \)^{\hal} + C \# I_\Omega \lambda^{\d-\s}   \| (-\Delta)^{\frac{\d-\s}{2}} \varphi\|_{L^\infty},
\ee
 \item  in a localized way,   for any $\ep \ge \lambda$  as in \eqref{deflambda},
  \begin{align}\label{rieszfluct2}
\left|\int_{\R^\d} \varphi\( \sum_{i=1}^N \delta_{x_i}- N d\mu\) \right|&\le C \( \ep^{\gamma-1} \|\varphi\|_{L^2(\Omega)}^2 + \ep^{\gamma+1}\|\nab \varphi\|_{L^2(\Omega)}^2 \)^\hal  
 \( \int_{\R^{\d+\k}} \yg |\nab h_{N, \rr}|^2 \)^{\hal}\\  \notag & +   \# I_\Omega |\varphi|_{C^\alpha} N^{-\frac{\alpha}{\d}}  \|\mu\|_{L^\infty}^{-\frac{\alpha}{\d}},
  \end{align}
  \end{itemize}where $C>0$ depends only on $\d$ and $\s$.
\end{lem}
\begin{proof}
The Coulomb case is very straightforward.
Integrating~\eqref{eqhnee0} against $\varphi$ and using Green's formula, we  have 
\be\label{relation3}
\left|\int_\Omega   \varphi\,  d  \Big ( \sum_{i=1}^N \delta_{x_i}^{(\rr_i)} -  N \mu\Big)\right|= \frac{1}{\cd}\left| \int_{\Omega} \nabla h_{N,\rr} \cdot \nab  \varphi\right|\le \frac{1}{\cd} \| \nab  \varphi\|_{L^2(\Omega)} 
 \| \nab h_{N,\rr} \|_{L^2(\Omega)}  .
\ee
On the other hand, since by definition $\rr_i\le \frac{\lambda}{4}$ for each $i$, we have
\begin{equation}
\left|\int_\Omega \varphi \, d\Big( \sum_{i=1}^N( \delta_{x_i} - \delta_{x_i}^{(\rr_i)}) \Big)\right|\le \# I_\Omega   | \varphi|_{C^\alpha} \lambda^\alpha,\end{equation}
hence by definition of $\lambda$ we get the result.

Let us now turn to the Riesz case. Let  $\tilde \varphi$ be the $\hal(\d-\s)$-harmonic    extension of $\varphi $ to $\R^{\d+\k}$. Following \cite[Sec. 2.4]{caffsilvestre}, it can be defined by 
$$\tilde \varphi (x,y)= \cds \int_{\R^\d} \frac{| y|^{\d-\s}}{(|x-x'|^2 + |y|^2)^{\frac{2\d-\s}{2}} }\varphi(x') dx'\quad \text{for } (x,y) \in \R^\d\times \R.$$
It is such \cite[(3.7)]{caffsilvestre} that 
\be \label{normeequiv}\int_{\R^{\d+\k}} \yg |\nab \tilde \varphi|^2 \le C \|\varphi\|^2_{\dot{H}^{\frac{\d-\s}{2}} (\R^\d)} \ee
and 
\be\label{1511} -\div (\yg \nab \tilde \varphi)= 2 \((-\Delta)^{\frac{\d-\s}{2}} \varphi \) \delta_{\R^\d}\quad \text{in} \ \R^{\d+1}.\ee
Integrating~\eqref{eqhneta} against $\varphi$ and using Green's formula and Cauchy-Schwarz, we  obtain 
\begin{align} \nonumber
\left|\int_{\R^{\d+\k}} \tilde    \varphi \, d  \Big ( \sum_{i=1}^N \delta_{x_i}^{(\rr_i)} -  N \mu\drd\Big)\right|& = \frac{1}{\cds}\left| \int_{\R^{\d+\k}}\yg \nabla h_{N,\rr} \cdot \nab \tilde \varphi\right|\\
\nonumber  & \le \frac{1}{\cds} \( \int_{ \R^{\d+\k}}\yg |\nab \tilde \varphi|^2\)^{\hal} 
 \( \int_{\R^{\d+\k}}\yg |\nab h_{N,\rr} |^2\)^{\hal}  
 \\  \label{rela3} &  \le  C \|\varphi\|_{\dot{H}^{\frac{\d-\s}{2}}} \( \int_{\R^{\d+\k}}\yg |\nab h_{N,\rr} |^2\)^{\hal}  .
\end{align}
On the other hand, using \eqref{eqpourfeta}, we may write
$$\int_{\R^{\d+\k}} \tilde \varphi \, d\( \sum_{i=1}^N \delta_{x_i}- \delta_{x_i}^{(\rr_i)} \)= -\frac{1}{\cds}\sum_{i=1}^N  \int_{\R^{\d+\k}} 
 \tilde \varphi\, \div (\yg \nab \f_{\rr_i}(x-x_i) ).
 $$
Using integration by parts and \eqref{1511}, it follows that 
\begin{align}\notag\left|\int_{\R^{\d+\k}} \tilde \varphi \, d\( \sum_{i=1}^N \delta_{x_i}- \delta_{x_i}^{(\rr_i)} \)\right| & = \frac{2}{\cds}\left|\sum_{i=1}^N \int_{\R^{\d}}   \f_{\rr_i}(x-x_i)(-\Delta)^{\frac{\d-\s}{2}} \varphi \right|\\ \label{155}
& \le C \sum_{i\in I_\Omega} \rr_i^{\d-\s} \| (-\Delta)^{\frac{\d-\s}{2}} \varphi \|_{L^\infty} ,\end{align}
where we used \eqref{eq:intf}.  Since $\rr_i \le \lambda$, assembling the relations, we deduce that \eqref{rieszfluct} holds.

We next turn to the localized version. For that we define a different extension 
$$\tilde \varphi (x,y) = \varphi (x) \chi(y),$$
where $\chi$ is a cutoff function, equal to $1$ for $|y|<\ep$ and vanishing for  $|y|\ge 2\ep$ with $|\nab \chi| \le \frac{1}{\ep}$, where $\ep$ is any number $\ge \lambda$.
In lieu of \eqref{rela3}, we get
\begin{align}\label{rela4}
& \left|\int_{\R^{\d+\k}} \tilde    \varphi \, d  \Big ( \sum_{i=1}^N \delta_{x_i}^{(\rr_i)} -  N \mu\drd\Big)\right| \le \frac{1}{\cds} \( \int_{ \Omega \times \R^{\k}}\yg |\nab \tilde \varphi|^2\)^{\hal} 
 \( \int_{\Omega\times \R^{\k}}\yg |\nab h_{N,\rr} |^2\)^{\hal}  
\\  \notag& \qquad
\le C \( \int_{\Omega } |\varphi|^2 \int_\ep^{2\ep} \frac{\yg}{\ep^2} dy  + \int_{\Omega }|\nab \varphi|^2 \int_0^{2\ep} \yg dy\)^{\hal}  \( \int_{\Omega\times \R^{\k}}\yg |\nab h_{N,\rr} |^2\)^{\hal}  \\ \notag
&\qquad  \le C \( \ep^{\gamma-1} \|\varphi\|_{L^2(\Omega)}^2 + \ep^{\gamma+1}\|\nab \varphi\|_{L^2(\Omega)}^2 \)^\hal   \( \int_{\Omega\times \R^{\k}}\yg |\nab h_{N,\rr} |^2\)^{\hal} . \end{align}
In lieu of \eqref{155}, since $\ep \ge \lambda$  we simply write 
$$
 \left|\int_{\R^{\d+\k}} \tilde \varphi \, d\( \sum_{i=1}^N \delta_{x_i}- \delta_{x_i}^{(\rr_i)} \)\right|\le\#I_\Omega  |\varphi|_{C^\alpha}    \lambda^{\alpha} .$$
 We conclude that \eqref{rieszfluct2} holds.
\end{proof}

\begin{rem}
Other bounds than \eqref{rieszfluct}, requiring less regularity of $\varphi$ can be obtained by putting more derivatives on $\f_{\rr_i}$ in \eqref{155}. They lead to a worse  power of $\lambda$.
\end{rem}
By duality, the bound \eqref{rieszfluct} allows to deduce a bound on $\sum_{i=1}^N \delta_{x_i}- N \mu$ in a negative Sobolev norm, which shows that $\frac{1}{N^2}\F_N$ does metrize the convergence of $\mu_N$ to $\mu$.
\begin{coro}\label{coro453}
 For any $\sigma >\frac{\d}{2}+\d-\s$,
 \be \label{bornefluctduality}
\left\| \sum_{i=1}^N \delta_{x_i}- N\mu\right\|_{H^{-\sigma}(\R^\d)}
\le C_{\sigma}   \( \int_{\R^{\d+\k}} \yg |\nab h_{N, \rr}|^2 \)^{\hal} + CN^{\frac\s\d} \|\mu\|_{L^\infty}^{-1+\frac{\s}{\d}} 
  ,\ee
  where $H^{-\sigma}$ is the dual of $H^{\sigma}$ (the standard Sobolev space).
In particular, if $\frac{1}{N^2}\F_N(\XN, \mu) \to 0$ as $N \to \infty$, we have that 
\be\frac{1}{N} \sum_{i=1}^N \delta_{x_i} \to  \mu\quad \text{in } H^{-\sigma}(\R^\d).\ee
\end{coro}
\begin{proof}
 Indeed, the Sobolev embedding implies that $H^{\sigma}(\R^\d) \subset C^{\d-\s}(\R^\d)$ and also $H^\sigma (\R^\d)\subset \dot{H}^{\frac{\d-\s}{2}}(\R^\d)$ from which, starting from \eqref{rieszfluct} or \eqref{coulombfluct}  we deduce  \eqref{bornefluctduality}.
 For the second assertion, we  combine \eqref{bornefluctduality} and \eqref{bornehnr} to obtain
 $$\left\| \sum_{i=1}^N \delta_{x_i}- N\mu\right\|_{H^{-\sigma}}^2 \le C \( \F_N(\XN,\mu)+  N\frac{\log (N\|\mu\|_{L^\infty}) }{2\d} \indic_{\s=0}\) + C \|\mu\|^{\frac{\s}{\d}}_{L^\infty} N^{1+\frac{\s}{\d}}
 + C N^{\frac{2\s}{\d}} \|\mu\|_{L^\infty}^{-2+\frac{2\s}{\d}},$$ hence the result after dividing by $N^2$, using that  $\s<\d$.
\end{proof}

We may also obtain $L^p$ bounds on the gradient of the potential $\nab h_N$.
In that respect we have the following estimate, which in turn provides a control of $
\sum_{i=1}^N \delta_{x_i} - N \mu$ in view of \eqref{bbe}. If $\s<0$ this is not needed since we have \eqref{casszero}, hence an $L^2$ control.

\begin{prop}[Control of the electric potential by the modulated energy]\label{procoer} Assume  $\s \ge 0$.  Let $\XN \in (\R^\d)^N$. 
With the same notation as above,
letting  $B_R $ be the ball of radius $R$  centered at $0$ in $\R^{\d+\k}$,  for every  $p$ such that $1\le p<\frac{\d+\k}{\s+1}$ and $p\le 2$, we have, \be\label{resultlp} \|\nab h_N\|_{L^p(B_R)}\\ \le   C_{p,R} 
\( \int_{B_R} \yg |\nab h_{N, \rr}|^2 \)^{\frac{1}{2}}  + C_p (\#I_{B_R})^{\frac{1}{p}}  \lambda^{\frac{\d+\k}{p}-\s-1} .
\ee
\end{prop}
\begin{proof}
Let us start from \eqref{defhNmu}, which gives 
\be\label{depar}\nab h_N= \nab h_{N,\rr}+ \sum_{i=1}^N \nab\f_{\rr_i} (\cdot -x_i).\ee
The function $\f_{\rr_i}$, which is supported in $B(0, \rr_i)$,  has the same singularity at $0$ as  $\g(x)$ (since $\s \ge 0$)  and $\nab \f_{\rr_i}$ a singularity in $\frac{1}{|x|^{\s+1}}$. Thus, $\nab \f_{\rr_i}$ is in $L^p(\R^{\d+\k})$ if and only if $p(\s+1)<\d+\k$, and  if so 
\be  \int_{\R^{\d+\k}} |\nab \f_{\rr_i}|^p \le \int_{B(0, \rr_i)} \frac{dx}{|x|^{p(\s+1)}} \le C_p \rr_i^{\d+\k -p\s-p}  .\ee 
Since the balls $B(x_i, \rr_i)$ are disjoint and $\rr_i \le \lambda$, we deduce from \eqref{depar} that 
\be \|\nab h_N\|_{L^p(B_R)} \le \|\nab h_{N, \rr}\|_{L^p(B_R)} + C_p \( \#I_{B_R} \lambda^{\d+\k-p\s-p}\)^{\frac1p}\ee
In addition, by H\"older's inequality, if $p<\min (2, \frac{\d+\k}{\s+1})$, we have
$$\int_{B_R} |\nab h_{N, \rr}|^p \le \( \int_{B_R} \(\frac{1}{\yg}\)^{\frac{p}{2-p}}  \)^{1-\frac{p}{2}}  \( \int_{B_R} \yg |\nab h_{N, \rr}|^2 \)^{\frac{p}{2}}  .$$

Since $\gamma= \s+2-\k-\d$ in \eqref{defgamma},  we note that the condition $p< \frac{\d+\k}{\s+1}$ always implies (since $\s<\d$) that $\frac{\gamma p}{2-p} <1$ hence the first integral in the right-hand side converges. We deduce that under this condition 
$$\int_{B_R} |\nab h_{N, \rr}|^p \le C_{p,R}  \(\int_{B_R} \yg |\nab h_{N, \rr}|^2 \)^{\frac{p}{2}}  $$
and \eqref{resultlp} follows.
\end{proof}

This result can be localized as we will see in Section \ref{secloc} below.

\section{Discrepancy bounds}\label{sec:discrepancy}
\index{discrepancy}
We now prove that the electric energy $\int \yg |\nab h_{N,\rr}|^2$, hence via \eqref{bornehnr}  the modulated energy $\F_N$, controls the discrepancy (as discussed in Section \ref{sec1.3}), defined as follows.
\begin{defi}
Given a measurable set  $\Omega$ in $\R^\d$, we define the discrepancy in $\Omega$ of the configuration $\XN$ as 
\be \label{defiD} 
D(\Omega):= \int_{\Omega} d\( \sum_{i=1}^N\delta_{x_i} - N \mu\).\ee
\end{defi}

If $\Omega$ is a set of finite perimeter (see e.g. \cite{eg,maggi}), we let  $\di(x)$ denote the signed distance function to $\Omega$, which is positive in the complement of $\Omega$ and negative inside $\Omega$. 
For any $\delta \in\R$ (positive or negative), we let
\be \label{defOmegad}\Omega_\delta = \{ x \in \R^\d, \di(x) <\delta\}.\ee
We denote $|\Omega|$ for the volume of $\Omega$ and $|\partial \Omega|$ for its  perimeter.
The proofs are adapted from \cite[Lemma 4.6]{rs}, \cite[Lemma 2.2]{PetSer}, \cite{ls2}.
\begin{lem}[Control of charge discrepancy, general domain]\label{coronp}
 Let $\XN$ be a configuration in $(\R^\d)^N$, let $h_N$ be associated via~\eqref{def:hnmu}, and let $\Omega$ be a set of finite perimeter. We 
  have, for a constant $C>0$ depending only on $\d$ and $\s$,
  \begin{itemize}
  \item   in the Coulomb case, if $D(\Omega)\ge 0$, for any $\lambda <\delta \le 1$,
  \be\label{disc10}
   \Big( D(\Omega)- N\|\mu\|_{L^\infty} |\Omega_\delta\backslash \Omega|\Big)_+^2 \le C  \frac{|\partial \Omega_\delta|}{\delta}
\int_{\Omega_\delta \backslash \Omega} |\nab h_{N, \rr}|^2  \ee
 \item in the Coulomb case, if $D(\Omega) \le 0$, for any $-1 \le \delta < -\lambda $,
  \be \label{disc1}
  \Big( D(\Omega)+N\|\mu\|_{L^\infty} |\Omega\backslash \Omega_\delta|\Big)_-^2\le  C\frac{|\p \Omega|}{|\delta|} \int_{  \Omega\backslash \Omega_\delta} |\nab h_{N, \rr}|^2   \ee
  \item in the Riesz case, if $D(\Omega) \ge 0$, for any  $\lambda <\delta \le 1$,
  \be \label{4.4.5} \Big( D(\Omega)- N\|\mu\|_{L^\infty} |\Omega_\delta\backslash \Omega| \Big)_+^2\le C
  \( \frac{|\Omega_\delta|}{|\p \Omega_\delta|}\)^\gamma \frac{|\Omega_\delta|}{\delta} 
  \int_{ (\Omega_\delta \backslash \Omega) \times \R^\k}\yg |\nab h_{N,\rr}|^2 
\ee\item in the Riesz case, if $D(\Omega) \le 0$, for any $-1 \le \delta < -\lambda $,
  \be \label{disc1riesz}
  \Big( D(\Omega)+N\|\mu\|_{L^\infty} |\Omega\backslash \Omega_\delta|\Big)_-^2\le  C
  \( \frac{|\Omega|}{|\p \Omega|}\)^\gamma \frac{|\Omega|}{|\delta|} 
   \int_{ ( \Omega\backslash \Omega_\delta)\times \R^\k}\yg |\nab h_{N, \rr}|^2   .\ee
  \end{itemize}
  \end{lem}
  \begin{lem}[Control of charge discrepancy, case of a ball]\label{coronpball}
 Let $B_R$ be a ball of radius $R> 2\lambda$.
  If $D(B_R)\ge 0$ then  either $D(B_R) \le CN^{1-\frac1\d} R^{\d-1} \|\mu\|_{L^\infty}^{1-\frac1\d}$ or
 \be\label{disc3}
\frac{D(B_R)^2}{R^{\s}}\min \(1, \frac{D(B_R)}{ R^\d \|\mu\|_{L^\infty}  }\) \le C \int_{(B_{R+\delta}\backslash B_R) \times \R^\k }\yg |\nab h_{N,\rr}|^2, \ee with 
$$
\delta= \min \(\frac{R}{2}, \Big(R^\d + \frac{D(B_R)}{ 2C_0N \|\mu\|_{L^\infty }}\Big)^{\frac{1}{\d}}-R \),$$
while if  $D(B_R) \le 0$ then either $ |D(B_R)|\le C  N^{1-\frac1\d} R^{\d-1}\|\mu\|_{L^\infty}^{1-\frac1\d} $ or 
 \be\label{disc30}
\frac{D(B_R)^2}{R^{\s}}\left|\min \(1, \frac{|D(B_R)|}{ R^\d  \|\mu\|_{L^\infty}}\) \right|\le C \int_{(B_{R}\backslash B_{R+\delta})\times \R^\k }\yg|\nab h_{N,\rr}|^2, \ee
with 
$$\delta= \max \( -\frac{ R}{2},  \( R^\d + \frac{ D(B_R)}{2C_0N \|\mu\|_{L^\infty} } \)^{\frac1\d}-R \)$$
where $C$, $C_0$ depend only on $\d$ and $\s$.
\end{lem}
Note that $|D(B_R)|$ should be compared to the rescaled volume $NR^\d$ or rescaled perimeter $N^{1-\frac1\d}R^{\d-1}$.

\begin{proof} We will prove both lemmas at once.
 \smallskip

  {\bf Step 1.  The Coulomb positive case}.
Let us first assume $D(\Omega) \ge 0$.  We are going to take advantage of the charge excess by examining the energy outside $\Omega$.

By definition of the $\rr_i$ \eqref{defri}, for $t > \hal \lambda $ we have  
$$\int_{ \di(x) < t} d\(  \sum_{i=1}^N \delta_{x_i}^{(\rr_i)} - N \mu \) \ge D(\Omega)- N\int_{ 0 <\di(x) <t} d\mu .$$
Thus, with Green's formula,  if $\hal \lambda<t<\delta$,
\begin{multline}\label{1518}-\int_{ \di(x)= t} \frac{\partial h_{N, \rr} }{\partial \nu} = - \int_{\di(x)<t} \Delta h_{N, \rr} =\cd \int_{\di(x)<t}  d\(  \sum_{i=1}^N \delta_{x_i}^{(\rr_i)} - N \mu \) \\
\ge\cd  \( D(\Omega)- N\|\mu\|_{L^\infty} |\Omega_{\delta}\backslash \Omega|\) .\end{multline}
On the other hand, the co-area formula (see \cite{eg,maggi}) gives (since $|\nab \di|=1$)  that 
\be \label{1519} \int_{  \hal \lambda <\di(x) < \delta} |\nab h_{N, \rr}|^2 
= \int_{\hal \lambda}^\delta \(\int_{\di(x)= t}  |\nab h_{N, \rr}|^2 \) dt \ge  \int_{\hal \lambda}^\delta \int_{\di(x)= t} \left|\frac{\p h_{N,\rr}}{\p \nu} \right|^2  dt .\ee
 But the 
Cauchy-Schwarz inequality gives 
 \be\label{1520} \int_{\di(x)= t} \left|  \frac{\p h_{N,\rr}   }{\p \nu} \right|^2  \ge   \( \int_{\di(x)=t} \frac{\p h_{N, \rr}}{\p \nu} \)^2 \frac{1}{|\{\di(x)=t\}|}.\ee
Combining \eqref{1518}--\eqref{1520},
we thus obtain 
\be  \label{1521} \int_{  \hal \lambda <\di(x) < \delta} |\nab h_{N, \rr}|^2 
\ge\cd^2 \( D(\Omega)- N\|\mu\|_{L^\infty} |\Omega_\delta\backslash \Omega|\)_+^2 \int_{\hal \lambda}^\delta   \frac{dt}{|\{\di(x)=t\}|} .
\ee
Using that $|\{\di=t\}| \le C |\partial \Omega_\delta|$ for $t \le \delta\le 1$,  we deduce that 
\be  C\int_{\Omega_\delta \backslash \Omega} |\nab h_{N, \rr}|^2 \ge  \( D(\Omega)- N\|\mu\|_{L^\infty} |\Omega_\delta\backslash \Omega|\)_+^2 \frac{\delta}{|\partial \Omega_\delta|}
\ee  if $\delta>\lambda$,
which proves \eqref{disc10}.

Specializing \eqref{1521} to  $\Omega $  a ball of radius $R>2\lambda$, we find 
$$C\int_{B_{R+\delta}\backslash B_R} |\nab h_{N,\rr}|^2 \ge  \( D(B_R) - C_0 N \|\mu\|_{L^\infty} ( (R+\delta)^\d- R^\d) \)_+^2 \int_{\lambda/2}^\delta \frac{dt}{(R+t)^{\d-1}} 
$$ where $C, C_0>0$ depend only on $\d$.
We now choose 
\be \delta:= \min \(\frac{R}{2}, \Big(R^\d + \frac{D(B_R)}{ 2C_0N \|\mu\|_{L^\infty }}\Big)^{\frac{1}{\d}}-R \) \ge 
R \min \( \frac12, C\frac{D(B_R)}{N R^\d \|\mu\|_{L^\infty} } \)
\end{equation}
 so that  the term
$C_0 N  \|\mu\|_{L^\infty} ( (R+\delta)^\d-R^\d) $ appearing above is $<\hal D(B_R)$. Either $\delta \le \lambda$ in which case $D(B_R)\le C \lambda N R^{\d-1} \|\mu\|_{L^\infty}$, or $\delta \ge \lambda$ and  
we then obtain 
$$C\int_{B_{R+\delta}\backslash B_R} |\nab h_{N,\rr}|^2 \ge   D(B_R)^2   \delta  R^{1-\d} \ge   D(B_R)^2     R^{2-\d}\min \(\hal ,  C \frac{D(B_R)}{ N R^\d\|\mu\|_{L^\infty}} \).$$
 This proves \eqref{disc3} in this case.
 
 {\bf Step 2. The Coulomb negative case.}
 Let us next turn to the case $D(\Omega)\le 0$,  still in the Coulomb case. We now find the excess energy inside $\Omega$.
 For $t<-\lambda/2$ we have 
 $$\int_{ \di(x) < t} d\(  \sum_{i=1}^N \delta_{x_i}^{(\rr_i)} - N \mu \) \le D(\Omega)+ N\int_{ t<\di(x) <0} d\mu .$$
Thus 
\begin{multline}\label{1587}-\int_{ \di(x)= t} \frac{\partial h_{N, \rr} }{\partial \nu} = - \int_{\di(x)<t} \Delta h_{N, \rr} =\cd \int_{\di(x)<t}  d\(  \sum_{i=1}^N \delta_{x_i}^{(\rr_i)} - N \mu \) \\
\le D(\Omega)+ N\|\mu\|_{L^\infty} |\{ t <\di(x) <0\}| .\end{multline} Arguing as in the case $D (\Omega ) \ge 0$ we deduce that  for $\delta<- \lambda$
\be  \int_{  \delta <\di(x) < -\hal \lambda} |\nab h_{N, \rr}|^2 
\ge \( D(\Omega)+N\|\mu\|_{L^\infty} |\Omega\backslash \Omega_\delta|\)_-^2 \int_{\delta}^{-\hal \lambda} \frac{dt}{|\{\di(x)=t\}|} .
\ee
We conclude as above that \eqref{disc10} holds.

Specializing now to a ball $B_R$ of radius $R >2 \lambda$  we find 
$$  \int_{B_R \backslash B_{R+\delta }} |\nab h_{N, \rr}|^2 
\ge \( D(\Omega)+C_0 N\|\mu\|_{L^\infty} (R^\d - (R+\delta)^\d)  \)_-^2 \int_{\delta}^{-\hal \lambda} \frac{dt}{(R+t)^{\d-1}} .
$$Choosing 
$$\delta:= \max \(  \( R^\d + \frac{ D(B_R)}{2C_0N \|\mu\|_{L^\infty} } \)^{\frac1\d}-R ,- \hal R\) \le - R \min 
\(\hal , C \frac{|D(B_R)|}{NR^\d\|\mu\|_{L^\infty}}\) $$ we then obtain that either $\delta \ge -  \lambda$ in which case    $|D(B_R)|\le C\lambda  NR^{\d-1} \|\mu\|_{L^\infty}$ or $\delta <- \lambda$ and 
$$C  \int_{B_R \backslash B_{R+\delta }} |\nab h_{N, \rr}|^2 
\ge  D(B_R)^2  
R^{2-\d}\min \(1,  C \frac{|D(B_R)|}{ N\|\mu\|_{L^\infty} R^\d} \),$$
concluding the proof of \eqref{disc30} in the Coulomb case. 
\smallskip 

{\bf Step 3. The Riesz positive case.}
The  relation \eqref{1518} can be replaced by 
\begin{multline}\label{1530}
 - \int_{\p (\{ \di(x)<t\}\times [-m-t,m+ t] ) } \yg \frac{\p h_{N, \rr} }{\p \nu} = -\int_{\{\di(x)<t \}\times [-m-t,m+ t]  } \div (\yg \nab h_{N, \rr}) \drd \\
 \ge  D(\Omega)- N \|\mu\|_{L^\infty} |\Omega_\delta\backslash \Omega| \end{multline} for $t>\hal \lambda$, 
where $m\ge 0$ is to be determined later, and \eqref{1519}--\eqref{1520} are replaced by 
\begin{align*}\label{1532} &  \int_{ \{\hal \lambda <\di(x) <\delta\} \times [-m-\delta,m+ \delta]}\yg  |\nab h_{N, \rr}|^2  \\  & \ge\int_{  \hal \lambda}^\delta\(  \int_{  \p (\{ \di(x)<t\}\times [-m-t,m+ t] )}   \yg \frac{\p h_{N, \rr} }{\p \nu} \)^2 \(\int_{  \partial( \{ \di(x)<t\}\times [-m-t,m+ t] )} \yg \)^{-1} dt\\ &
\ge \( D(\Omega)- N\|\mu\|_{L^\infty} |\Omega_\delta\backslash \Omega| \)_+^2
\int_{  \hal \lambda}^\delta   \(\int_{  \partial( \{ \di(x)<t\}\times [-m-t,m+ t] )} \yg \)^{-1} dt. \end{align*}
We next evaluate that for $t \in [\hal\lambda, \delta]$
\begin{multline*} \int_{  \partial( \{ \di(x)<t\}\times [-m-t,m+ t] )} \yg \\
=  \frac{2}{\gamma+1} (m+t)^{\gamma+1}|\{\di=t\}|+ 2 |\Omega_t| (m+t)^\gamma\le C (m+\delta)^{\gamma+1} |\partial \Omega_\delta| + 2 |\Omega_\delta|(m+t)^\gamma.\end{multline*}
Let us  now take  $m=\frac{|\Omega_\delta|}{|\partial \Omega_\delta|}$ to balance the two terms, and note that $m$  is bounded below independently of $\delta\le 1$. We then obtain
$$ \int_{  \partial( \{ \di(x)<t\}\times [-m-t,m+ t] )} \yg  \le C \(m^{\gamma+1} |\partial \Omega_\delta| +m^\gamma |\Omega_\delta|\) = C\(\frac{|\Omega_\delta|}{|\partial \Omega_\delta|} \)^\gamma   |\Omega_\delta| .$$
Inserting into \eqref{1530},
it follows that 
\be \label{4180} C \int_{ (\Omega_\delta \backslash \Omega) \times \R^\k}\yg |\nab h_{N,\rr}|^2 
\ge \( D(\Omega)- N\|\mu\|_{L^\infty} |\Omega_\delta\backslash \Omega| \)_+^2\frac{  \delta}{  |\Omega_\delta|}  \(\frac{|\Omega_\delta|}{|\partial \Omega_\delta|} \)^{-\gamma }  ,
\ee where we recall that $\gamma= \s-\d+1$ in the non-Coulomb case. This proves \eqref{4.4.5}.

Let us now specialize  to the case of a ball $B_R$, $R>2\lambda$, assuming $D(B_R) \ge 0$. We obtain 
$$
C \int_{ (B_{R+\delta}\backslash B_R)  \times \R^\k}\yg  |\nab h_{N, \rr}|^2  
\ge \( D(B_R)- C_0 N\|\mu\|_{L^\infty}( (R+\delta)^\d- R^\d)\)_+^2 \delta  (R+\delta)^{-\gamma-\d}.$$
Choosing 
$$\delta:= \min \(  \( R^\d + \frac{ D(B_R)}{2C_0N \|\mu\|_{L^\infty} } \)^{\frac1\d}-R , \hal R\) \ge R\min  \(\hal, C \frac{D}{NR^\d \|\mu\|_{L^\infty} } \) $$ 
we then obtain  that either $\delta \le \lambda$ in which case $D(B_R)\le C\lambda  NR^{\d-1} \|\mu\|_{L^\infty}$ or 
$$C \int_{ (B_{R+\delta}\backslash B_R)  \times \R^\k}\yg  |\nab h_{N, \rr}|^2  \ge D(B_R)^2  \delta R^{-\gamma-\d}$$
hence the result \eqref{disc3} follows in this case after noticing that $-\gamma - \d= -\s-1$.
 
The Riesz negative case is similar.

\end{proof}

\section{Localized version of the energy}\label{secloc}
One of the features of $\F_N$ is that once in electric formulation, it can naturally be localized, which will be important for obtaining results at meso-  and microscale later. 
Let us now define this localized version of the energy.

First, for any subset $\Omega$ of $\R^\d$, we define  a variant of the nearest neighbor distance, relative to the set $\Omega$ (we may say relative to $\partial \Omega$):
 \be \label{rrc}  \rrc_i:= \frac14 \left\{
\begin{aligned}
&\min \(\min_{ j\neq i} |x_i-x_j| ,\lambda\) && \text{if} \ \dist(x_i, \partial \Omega ) \ge 2 \lambda,\\
&\lambda  && \text{if} \ \dist(x_i, \partial \Omega ) \le \lambda\\
& t  \min \(\min_{ j\neq i} |x_i-x_j| ,\lambda\) +(1- t)\lambda && \text{if} \ \dist(x_i, \partial \Omega ) = (1+t) \lambda, t\in [0,1].
\end{aligned}\right.\ee
\index{truncation radii}
\begin{defi}[Localized version of the energy]
Given $\Omega\subset \R^\d$, and $\XN\in (\R^\d)^N$, we let $I_\Omega= \{i, x_i \in \Omega\}$
and define the localized modulated energy by 
\begin{equation}
\label{Glocal}
\F_N^{\Omega}(\XN,\mu)  
:=
\frac{1}{2\cds}\( \int_{\Omega\times \R^\k}\yg |\nab h_{N,\rrc}|^2 - \cds \sum_{i\in I_\Omega} \g(\rrc_i) \) -
N\sum_{i\in I_\Omega}\int_{\R^\d} \f_{\rrc_i}(x - x_i) d\mu(x),
\end{equation}where we  denote $h_{N,\rrc}$ for $h_{N,\vec{\eta}}$ with $\eta_i=\rrc_i$ as in \eqref{rrc}. \end{defi}
Note that points that lie outside $\Omega$ but near $\partial \Omega$ affect the truncation in  $h_{N,\rrc}$.

 In view of Lemma \ref{monoto},  changing for all points the radii  $\rr_i$ into $\rrc_i$ relative to $\Omega$ (which is the same as relative to $\Omega^c$) can only decrease the computed value of $\F_N$, hence we find the important subadditivity property 
\be\label{locali} \F_N(\XN, \mu) \ge \F_N^{\Omega}(\XN,\mu) + \F_N^{\Omega^c} (\XN, \mu).\ee

We next give localized versions of some results of the previous sections.

First, we have the following localized version of  Proposition \ref{34}, from \cite{RosenSer2023}.
\begin{prop}[Localized minimal distance  and truncated electric energy controls]\label{procontrolelocal} Assume $\s \in [(\d-2)_+,\d)$.
Let $\Omega \subset \R^\d$ and let  now 
$$\lambda= (N\|\mu\|_{L^\infty(\Omega)} )^{-\frac1\d}.$$   Denote $\hat \Omega =\{x\in \R^\d, \dist(x, \Omega) \le   \frac14 \lambda\}$. For any $\vec{\eta}$ satisfying $ \frac{1}{2} \rrc_i \le \eta_i \le \rrc_i$ for every $1\leq i\leq N$ and $\eta_i=\rrc_i$ if $\dist(x_i, \partial \Omega) \le \eta_i$, it holds that
\begin{multline}\label{eq:11}
2 \left( \F_N^{\Omega}(\ux_N,\mu) -\left(   \frac{\#I_\Omega \log \lambda }{ 2} \right)\indic_{\s=0}\right)  +C \#I_\Omega  N^{\frac{\s}{\d}}\|\mu\|_{L^\infty(\hat{\Omega}) }  \|\mu\|_{L^\infty(\Omega)}^{-1+\frac{\s}{\d}}\\
 \ge
\begin{cases} 
\displaystyle \frac1{ C}\sum_{i\in I_\Omega}\g(\eta_i) & \text{if} \ \s\neq 0\\
\displaystyle \sum_{i \in I_\Omega }\g( \eta_i /\lambda ) & \text{if} \ \s=0\end{cases}
\end{multline}
 and
\begin{multline}\label{eq:14}
\int_{\Omega\times \R^\k}\yg |\nabla h_{N,\vec{\eta}}|^2 \\ 
\le C \(\F_N^{\Omega} (\ux_N, \mu)  - \left(  \frac{ \#I_\Omega\log \lambda }{2} \right)\indic_{\s=0}\) + C\#I_\Omega  N^{\frac{\s}{\d}} \|\mu\|_{L^\infty(\hat{\Omega}) } \|\mu\|_{L^\infty(\Omega)}^{-1+\frac{\s}{\d}},
\end{multline}
where $C$ depends only on $\d$ and $\s$.
\end{prop}

Now that we have \eqref{eq:14} as a localization of \eqref{bornehnr}, we can use it in conjunction with the local controls such as Lemma \ref{prop:fluctenergy} and \ref{coronp}.

In order to prove this proposition let us take another look at the monotonicity property.

\begin{lem}\label{lem:monoto}  For $\Omega \subset \R^\d $, denoting temporarily 
\begin{equation}\label{eq:defFa}
\mathcal F^{\vec{\alpha}}\coloneqq\frac1{2\cds}\Bigg(\int_{\Omega\times \R^\k} \yg |\nabla h_{N,\vec{\alpha}}|^2  -  \cds \sum_{i\in I_\Omega}  \g(\alpha_i)-2N\cds\sum_{i\in I_\Omega}\int_{\R^\d} \f_{\alpha_i}(x-x_i)d\mu(x)\Bigg),
\end{equation} if $\alpha_i = \rrc_i$ for all $i$'s such that $\dist(x_i, \pa\Omega) \le\alpha_i $,  we have
\begin{equation}\label{eq:supplem}  \F_N^\Omega(\ux_N, \mu) -  \mathcal F^{\vec{\alpha}}\ge
\hal \sum_{\substack{ i, j\in I_\Omega, i\neq j \\ \dist(x_i, \pa\Omega) \ge \alpha_i }} 
\(\g(x_i-x_j) - \g(\alpha_i)\)_+ .
\end{equation}
\end{lem}
\begin{proof}Let us consider $\vec{\eta} $ and $\vec{\alpha}$  such that $\alpha_i \le \eta_i$, and $\eta_i=\alpha_i$ except for  $i$ such that $B(x_i,\eta_i)  \subset \Omega$. Then 
 \eqref{ii1i2} can be rewritten as 
\be\label{above}\mathcal F^{\vec{\eta}} - \mathcal F^{ \vec{\alpha}} =  \sum_{i\neq j} \int_{\R^{\d+\k}}\f_{\alpha_i, \eta_i} (z-x_i) d\( 
\delta_{x_j}^{(\alpha_i)}+ \delta_{x_j}^{(\eta_j)}\) (z),\ee  where we recall $\f_{\alpha, \eta}= \f_\alpha-\f_\eta$. The couples $i\neq j$ that contribute to  sum on the right-hand side are those for which $\alpha_i\neq \eta_i$ and $B(x_i, \eta_i)$ intersects $B(x_j, \eta_j)$. Moreover, there is no contribution for points that do not satisfy $B(x_i, \eta_i) \subset \Omega$. 

Applying this to $\eta_i=\rrc_i$ for every $i$, and $\alpha_i=\eta_i$ if $\dist(x_i, \partial \Omega) \le \eta_i$, and $\alpha_i\le \rrc_i$ otherwise, we find that the right-hand side of \eqref{above} vanishes hence 
\be\label{sft}
\mathcal F^{\vec{\alpha}}= \F_N^\Omega(\XN, \mu) \ee
for all such $\vec{\alpha}$. 

Let now $\eta_i$ be arbitrary satisfying  $\eta_i =\rrc_i$ if $\dist(x_i, \partial \Omega) \le \eta_i$, and $\alpha_i\le \eta_i$ with equality if $\dist(x_i, \partial \Omega) \le \eta_i$. 
Using the monotonicity of $\g$ and the definition of $\f_\alpha, \f_\eta$, we may deduce from \eqref{above}, as in \eqref{monoplus}, that
 \begin{equation}
\mathcal F^{\vec{\eta}}-\mathcal F^{\vec{\alpha}} \le 
\frac1{2} \sum_{\substack{i,j\in I_\Omega,  i\neq j \\ { B(x_i, \eta_i) \subset \Omega } } } \left(  \g(\eta_i)-\g_{\alpha_i}(|x_i-x_j|+\alpha_j)\right)_- .
\end{equation}

Letting $\alpha_i\to 0$ for $i$ such that $\dist(x_i, \partial \Omega)> \eta_i$, and $\alpha_i=\eta_i$ otherwise, we have $\mathcal F^{\vec{\alpha}}= \F_N^{\Omega}(\XN, \mu)$  by \eqref{sft} and thus $$\F_N^\Omega(\XN,\mu) -\mathcal F^{\vec{\eta}} \ge \hal \sum_{\substack{i,j\in I_\Omega,  i\neq j \\  B(x_i, \eta_i) \subset \Omega  }} (\g(x_i-x_j)-\g(\eta_i))_+.$$
Replacing the notation $\eta_i$ by $\alpha_i$ we have proved \eqref{eq:supplem}.

\end{proof}

\begin{proof}[Proof of Proposition \ref{procontrolelocal}]
For every $i\in I_\Omega$, let us choose $\alpha_i= \alpha\coloneqq \frac14\lambda$. This choice  satisfies the assumptions of the lemma. Applying inequality \eqref{eq:supplem} and discarding nonnegative terms, we have
\begin{multline}\label{236}
\F_N^\Omega(\XN, \mu) \geq -  \frac{1}{2} \sum_{i\in I_\Omega}  \g(\alpha )- N  \|\mu\|_{L^\infty(\hat\Omega)}\|\f_\alpha\|_{L^1}\# I_\Omega \\
+ \frac{1}{ 2}\sum_{\substack{i,j\in I_\Omega, i\neq j\\ \dist(x_i, \p\Omega) \ge \alpha}} \(\g(x_i-x_j) - \g(\alpha)\)_+.
\end{multline}
From the definition of $\rrc_i$ \eqref{rrc}, we see that either $\min_{j\neq i}|x_i-x_j|>\lambda=4\rrc_i>\alpha$, in which case
\begin{equation}
\forall j\neq i, \qquad \(\g(x_i-x_j)-\g(\alpha)\)_+ = 0>\g(4\rrc_i)-\g(\alpha),
\end{equation}
or there exists $x_j \in \Omega$ such that $4\rrc_i\ge  |x_i-x_j|$. In all cases, there exists $j\neq i$ such that 
\begin{equation}
\(\g(x_i-x_j) - \g(\alpha)\)_+  \geq \g(4\rrc_i)-\g(\alpha).
\end{equation}
Keeping only that $j$ in the sum, 
 it follows from \eqref{236}, in view of \eqref{eq:intf} applied to the factor $\|\f_\alpha\|_{L^1}$ from \eqref{236}, that if $\s \neq 0$,
\begin{equation}
 \hal \sum_{\substack{i\in I_\Omega\\ \dist(x_i, \p\Omega) \ge \alpha}} \g( 4\rrc_i)\le  \F_N^\Omega(\XN, \mu) +  \#I_\Omega\g(\alpha) + CN\|\mu\|_{L^\infty(\hat\Omega)}\#I_\Omega  \alpha^{\d-\s}.
\end{equation}
 Now in view of our choice of $\alpha$ and the definition of $\rrc_i$, if $x_i\in\Omega$ with $\dist(x_i,\p\Omega)<  \alpha$, then $\rrc_i = \alpha$  by definition. Hence, reinserting such points, we have 
\begin{equation}
\hal 
\sum_{i \in I_\Omega} \g( 4\rrc_i) \leq  \F_N^\Omega(\XN, \mu) +  \#I_\Omega\g(\alpha) + C\|\mu\|_{L^\infty(\hat\Omega)}\#I_\Omega N \alpha^{\d-\s} +\hal  \#I_{\Omega}\g( 4\alpha).
\end{equation}
Inserting the definition of $\alpha$ into this inequality and in view of our requirement that $\frac{1}{2}\rrc_i\leq \eta_i\leq \rrc_i$, we conclude that \eqref{eq:11} holds if $\s\neq 0$. If $\s=0$,  using the same reasoning, we  arrive at the inequality
\begin{equation*}
 \sum_{i\in I_\Omega} \g( 4\rrc_i/\alpha)\le  2\Bigg(\F_N^\Omega(\XN, \mu) +  \frac{\#I_\Omega}{2}\g(\alpha) + C\|\mu\|_{L^\infty(\hat\Omega)}\#I_\Omega N \alpha^{\d-\s}\Bigg)+ \#I_\Omega \g(4)
\end{equation*}
and the conclusion follows as well, absorbing the $\g(4)$ term into $\|\mu\|_{L^\infty}^{\s/\d}=1$.

We next turn to showing \eqref{eq:14}. Let us choose $\alpha_i=\eta_i$ with $\eta_i \in [\frac{1}{2} \rrc_i, \rrc_i]$ in \eqref{eq:supplem}, where we replace the right-hand side by $0$. Using that $\rrc_i \le \frac14\lambda$, we deduce, using again \eqref{eq:intf}, that
\begin{multline}
\F_N^\Omega(\XN, \mu)\\
\ge \frac1{2\cds}\(\int_{\Omega\times \R^\k} \yg |\nabla h_{N,\vec{\eta}}|^2 -  \cds \sum_{i\in I_\Omega}  \g(\eta_i) \)  -C  \#I_\Omega  \|\mu\|_{L^\infty(\hat{\Omega})}    \|\mu\|_{L^\infty(\Omega)}^{-1+\frac{\s}{\d}}   N^{\frac{\s}{\d}},
\end{multline}
and in view of \eqref{eq:11}, \eqref{eq:14} follows. In the case $\s=0$, we  write $\g(\eta_i)=\g(  \eta_i /\lambda ) + \g( \lambda )$, and then apply \eqref{eq:11}. 
\end{proof}
%

\chapter[Splittings, concentration, separation]{Splittings,  concentration, and separation estimates}
\label{chap:concentrationbounds}
In Chapter \ref{chap:leadingorder}, we examined the leading order behavior of the Hamiltonian $\HN$ of  \eqref{HN} with $\g$ as in \eqref{riesz}, which can be summarized by~: 
\begin{itemize}
\item The minimal energy $\min \HN$ behaves like $N^2 \min \I$, where $\I$ is the mean-field limit energy, defined on the set of probability measures of $\R^{\d}$.
\item If each $(x_1, \dots, x_N)$ minimizes $\HN$, the empirical measures $\emp=\frac{1}{N} \sum_{i=1}^N \delta_{x_i}$ converge weakly to the unique minimizer $\meseq$ of $\I$, also known as Frostman's equilibrium measure, which can be characterized via an obstacle problem.
\item This behavior also  holds  when $\theta \to \infty$, where $\theta= \beta N^{1-\frac{\s}{\d}}$, except with a very small probability determined by a large deviation principle.
\end{itemize}
The following questions  thus arise naturally~: 
\begin{enumerate}
\item What lies beyond  the term  $N\theta \I(\meseq)$, respectively $N\theta \I_\theta (\mu_\theta)$  in the expansion of $\HN$ or of  the free energy  $-\frac1\beta \log \ZNbeta$ ? 
\item What is the optimal {\it microscopic} distribution of the points ?
\end{enumerate}
To study these questions, we wish to  zoom or blow-up the configurations by the factor $N^{1/\d}$ (the inverse of the typical distance between two points or microscale), so that the points are well-separated, and find a way of expanding the Hamiltonian to next order. 
This very simple ``splitting", or quadratic expansion, was first introduced in the context of Coulomb gases in \cite{ss1}.

 This will directly lead us to the next order (or modulated electric) energy $\F_N$ of Chapter~\ref{chap:nextorder}, whose coercivity properties  we can then exploit.

Henceforth, we  will make the assumption that $V$ satisfies \eqref{A1}--\eqref{A5}, thus the Frostman equilibrium  measure $\meseq$ and the theorem equilibrium measure $\mub$ uniquely exist, and we assume $\meseq$ is absolutely continuous with respect to the Lebesgue measure and has a bounded density. 


We present two splitting formulas: one with respect to the equilibrium measure, which is the appropriate one when studying energy minimizers, and one with respect to the thermal equilibrium measure, more appropriate when studying the Gibbs measure.
They lead, after subtracting off constant leading order terms, to $\F_N$ becoming the main energy of the system. Combining with the results of Chapter \ref{chap:nextorder} on $\F_N$, one easily deduces energy and free energy bounds and first concentration estimates.

Let us point out that the splitting and in particular the electric formulation of the Gibbs measure have been instrumental in a number of studies that go beyond this text: they have served for instance to obtain DLR equations for the one-dimensional log gas in \cite{dhlm}, obtain CLT for the linear statistics of the  sine-beta process \cite{lebleimrn}, study the maximum of the log-gas potential in \cite{lambertleblez,luke2}.
\index{central limit theorem}

At the end of the chapter we will describe separation results for energy minimizers, and mention other approaches to such results in the literature which apply to the Gibbs measure as well, in particular the isotropic averaging method of \cite{thoma}.

\section{Splitting the Hamiltonian} \label{sectionaref} 
\subsection{Splitting with respect to the equilibrium measure}
\index{splitting}

The splitting consists in an exact formula that separates the leading ($N^2$) order term  in $\HN$ from next order terms. 
It suffices to  expand around $\meseq$ by writing $\emp= \meseq+ \( \emp-\meseq\)$.
We note that since $\meseq$ is compactly supported, as seen in the proof of Theorem \ref{theoFrostman}, it satisfies \eqref{condmupourFN} and thus $\F_N(\cdot, \meseq)$ is well defined.

\begin{lem}[Splitting formula]Assume $\meseq $ is absolutely continuous with respect to the Lebesgue measure. 
For any $N$ and any $\XN \in (\R^\d)^N$ we have
\begin{equation}\label{split0}
 \HN(\XN) =  N^2 \I(\meseq) + N \sum_{i=1}^N \zeta(x_i) +\F_N (\XN,\meseq)
\end{equation}
where $\triangle$ denotes the diagonal of $\R^\d\times \R^\d$, $\I$ is as in \eqref{definitionI},  $\zeta$ is as in \eqref{defzeta} and  $\F_N$ is as in \eqref{def:FN}.\end{lem}
\begin{proof} 
We may write \begin{eqnarray}
\nonumber \HN(\XN)  & = &  \hal\sum_{i \neq j} \g(x_i- x_j) + N \sum_{i=1}^N V(x_i)\\
\nonumber & = & \frac{N^2}{2} \iint_{\triangle^c} \g(x-y) d\emp(x)  d\emp(y) + N^2 \int_{\R^\d} V d\emp(x) \\
\nonumber 
& = &  \frac{N^2}{2} \iint_{\triangle^c} \g(x-y) d\meseq(x) d\meseq(y) + N^2 \int_{\R^\d} V d\meseq
\\ 
\nonumber
 & + &  N \iint_{\triangle^c} \g(x-y) d\meseq(x) d \( \sum_{i=1}^N \delta_{x_i}- N\meseq\)(y)
+ N \int_{\R^\d} V d\( \sum_{i=1}^N \delta_{x_i}- N\meseq\) \\ 
\label{finh}
& + & \hal \iint_{\triangle^c} \g(x-y) d\( \sum_{i=1}^N \delta_{x_i}- N\meseq\)(x)d\( \sum_{i=1}^N \delta_{x_i}- N\meseq\)(y).
\end{eqnarray}
We now recall that $\zeta$ was defined in \eqref{defzeta} by
\be
\zeta = h^{\meseq} + V - c = \int_{\R^\d} \g(\cdot-y)\, d\meseq(y)  + V - c.\ee
We may  then rewrite the medium line in the right-hand side of \eqref{finh} as  
\begin{multline*}
N \iint_{\triangle^c} \g(x-y) d\meseq(x) d\( \sum_{i=1}^N \delta_{x_i}- N\meseq\)(y) + N \int_{\R^\d} V d\( \sum_{i=1}^N \delta_{x_i}- N\meseq\) \\
 = N  \int_{\R^\d} (h^{\meseq} + V) d\( \sum_{i=1}^N \delta_{x_i}- N\meseq\) = N  \int_{\R^\d} (\zeta + c) d\( \sum_{i=1}^N \delta_{x_i}- N\meseq\) \\
 = N^2 \int_{\R^\d} \zeta d\emp - N^2 \int_{\R^\d} \zeta d\meseq+  N c \int_{\R^\d}  d\( \sum_{i=1}^N \delta_{x_i}- N\meseq\) = N^2 \int_{\R^\d} \zeta d\emp.
\end{multline*}
The last equality is due to the facts that  $\zeta= 0$ $\meseq$-a.e. as seen in \eqref{EulerLagrange}  and that  $\emp$ and $ \meseq$ are both probability measures.   We also have to notice that since  $\meseq$ is absolutely continuous with respect to the Lebesgue measure, we may include the diagonal  back into the domain of integration.
By that same argument, one may recognize in the first line of the right-hand side of \eqref{finh}, the quantity $N^2 \I(\meseq)$. One then recognizes the last line of \eqref{finh} as $\F_N(\XN, \meseq)$.   This concludes the proof.
\end{proof}

The function $\zeta$ then appears as an effective confinement potential, which only acts outside $\Sigma=\supp\, \meseq$ and localizes the particles to $\Sigma$. We discuss this localization further in Section~\ref{subsec:localization} below. 
With this splitting, we may also rewrite the Gibbs measure \eqref{gibbs} as 
\be\label{rewritegibbs}
d\PNbeta (\XN)=\frac{e^{- \theta N \I(\meseq) }}{\ZNbeta} \exp\(- \beta N^{-\frac\s\d}\(  \F_N(\XN, \meseq) + N\sum_{i=1}^N \zeta(x_i)\) \) d\XN .\ee

This leads us to introducing reduced measures and partition functions as follows.
\be \label{deftildeK}
\tilde \K_{N, \beta}(\mu, \zeta):= \int_{(\R^\d)^N}\exp\(- \beta N^{-\frac\s\d}\(   \F_N(\XN, \mu) + N\sum_{i=1}^N \zeta(x_i)\) \) d\XN\ee
and 
\be \label{deftildeQ} \tilde \Q_{N, \beta}(\mu, \zeta):= \frac{1}{\tilde \K_{N, \beta}(\mu, \zeta)}
\exp\( - \beta N^{-\frac\s\d}\(   \F_N(\XN, \mu) + N\sum_{i=1}^N \zeta(x_i)\) \) d\XN \ee
which make sense for any probability density $\mu$ for which $\F_N(\cdot, \mu)$ is defined,  and $\zeta$ growing fast enough at infinity.
With this notation, we may rewrite 
\be \label{rewritegibbs33}
d\PNbeta(\XN)= d\tilde \Q_{N,\beta} (\meseq, \zeta), \qquad Z_{N,\beta}= e^{-\theta N \I(\meseq)}
\tilde \K_{N,\beta}(\meseq, \zeta) ,\ee
with the $\zeta$ of \eqref{defzeta}.

\subsection{Splitting with respect to the thermal equilibrium measure}
It is advantageous to split with respect to $\mu_\theta$, the minimizer of \eqref{defEtheta}, where $\theta = \beta N^{1-\frac\s\d}$, which may depend on $N$. As explained already in Chapter~\ref{chap:eqmeasure}, when $\theta\gg1$, $\mu_\theta$ can be considered as an $N$-dependent deterministic correction to $\meseq$.

We recall that under \eqref{A1}--\eqref{A5}, in view of Propositions \ref{lem241} and \ref{lem242},   the thermal equilibrium measure $\mub$ minimizing ~\eqref{defEtheta}  exists and
satisfies 
\be \label{eqmb}
h^{\mub}+V + \frac{1}{\theta}\log \mub=c_\theta\quad \text{in} \ \R^\d\ee
where $c_\theta$ is a constant. Moreover, thanks to \eqref{1elem23}, we have that \eqref{condmupourFN} holds and $\F_N(\cdot, \mub)$ is well-defined.
  \begin{lem}[Splitting formula with the thermal equilibrium measure]
For any configuration $\XN \in (\R^\d)^N$,  we have 
\be\label{splitting}
\HN(\XN)= N^2 \mathcal {E}_\theta(\mub)- \frac{ N}{\theta} \sum_{i=1}^N \log \mub(x_i) + \F_N(\XN, \mub),\ee
where  $\I_\theta$ is as in \eqref{defEtheta} and
  $\F_N$ as in~\eqref{def:FN}. 
  \end{lem}
\begin{proof} Similarly as in \eqref{finh},
it suffices to rewrite $\HN(\XN)$ as
$$\HN(\XN)= \hal\iint_{\triangle^c} \g(x-y)d \(  \sum_{i=1}^N \delta_{x_i}\)(x)d \( \sum_{i=1}^N \delta_{x_i}\) (y) + N \int_{\R^\d} V(x) d \(  \sum_{i=1}^N \delta_{x_i}\)(x),$$ and
expand the integral after writing 
$\sum_{i=1}^N \delta_{x_i}= N\mub+ \( \sum_{i=1}^N \delta_{x_i}- N\mub\)$, to find
\begin{eqnarray*}
\nonumber \HN(\XN) &= & N^2 \I(\mub)+ N\iint_{\triangle^c} \g(x-y) d\( \sum_{i=1}^N \delta_{x_i}- N\mub\)(x) d\mub(y)\\& +& N \int_{\R^\d}  V(x) \(  \sum_{i=1}^N \delta_{x_i}- N\mub\)
+\F_N(\XN, \mub)\\
&=& N^2 \I(\mub)+ N\int_{\R^\d}( h^{\mub}+V) d\( \sum_{i=1}^N \delta_{x_i}-N \mub\)+\F_N(\XN, \mub).
\end{eqnarray*} Inserting \eqref{eqmb}, we obtain
$$\HN(\XN)= N^2 \mathcal {E}(\mub)- \frac{ N}{\theta} \(  \sum_{i=1}^N \log \mub(x_i)- N\int_{\R^\d}\mub\log\mub\)  + \F_N(\XN, \mub),$$ hence the result by definition of $\I_\theta $.
\end{proof}
This splitting works out nicely when inserted into the Gibbs measure 
definition: it then yields as an alternate to \eqref{rewritegibbs}
\be \label{rewritegibbs2}
d\PNbeta(\XN) = \frac{e^{-\theta N \I_\theta(\mub)}}{\ZNbeta} 
\exp\( - \beta N^{-\frac\s\d} \F_N(\XN, \mub)\) d\mub(x_1) \dots d\mub(x_N).\ee
The measure is thus made absolutely continuous with respect to the probability measure $\mub^{\otimes N}$ instead of the (nonintegrable) Lebesgue measure $d\XN$ in \eqref{rewritegibbs}, and a confinement potential is no longer needed.
The only disadvantage to this more precise description is that $\mub$ still depends on $N$, but this is remedied by the fact that we can have  precise estimates for $\mub $ in $\Sigma$, as seen in Theorem \ref{th1as}.  
We can thus rewrite 
\be\label{rewritegibbs3}
d\PNbeta(\XN) = \frac{1}{\K_{N,\beta}(\mub)} 
\exp\( - \beta N^{-\frac\s\d} \F_N(\XN, \mub)\) d\mub(x_1) \dots d\mub(x_N),\ee
where we let, for any probability density $\mu$,
\be \label{defKN}
\K_{N,\beta}(\mu):= \int_{(\R^\d)^N} \exp\( - \beta N^{-\frac\s\d} \F_N(\XN, \mu)\) d\mu(x_1) \dots d\mu(x_N).\ee
\begin{defi}
The 
{\it modulated  Gibbs measure}\footnote{by analogy with the modulated energy, but it could also be called a relative Gibbs measure. Its connection with the modulated energy and modulated free energy is explored in \cite{RosenSer4}.} 
 with respect to a  reference probability density $\mu$ is defined by 
\be \label{defQ}
d\Q_{N,\beta}(\mu):= \frac{1}{\K_{N,\beta}(\mu)} \exp\(-\beta N^{-\frac\s\d} \F_N(\XN, \mu)\) d\mu^{\otimes N}(\XN).\ee
\end{defi}

Using the definition, we have thus  obtained
\be \label{splitzk} d\PNbeta = d\Q_{N,\beta}(\mub), \qquad 
\ZNbeta= \exp\(- \theta N  \I_\theta(\mub) \) \K_{N,\beta}(\mub).\ee

In both cases, we have thus reduced to studying $\F_N$, whose main properties were studied in the previous chapter.

When one wishes to go back and forth between the two representations \eqref{splitzk} and \eqref{rewritegibbs3}, one may use that for $\mu>0$,
\begin{align}
&\frac{\tilde \K_{N,\beta}(\mu,\zeta) }{ \K_{N, \beta}(\mu)} \\ \notag
&= 
\frac{1}{\K_{N, \beta}(\mu)} \int_{(\R^\d)^N} \exp \(-\beta N^{-\frac\s\d}\( \F_N(\XN, \mu)  +N\sum_{i=1}^N \zeta(x_i) \) -\sum_{i=1}^N   \log \mu(  x_i)  \)  d\mu^{\otimes N} (\XN)\\ \notag & =
\Esp_{\Q_{N,\beta}(\mu) } \( \exp\( -\beta N^{1-\frac\s\d} \sum_{i=1}^N \zeta(x_i) -\sum_{i=1}^N   \log \mu(  x_i)  \) \)\\  \notag
&= \exp\(-\theta N \int_{\R^\d}\zeta d\mu -N\int_{\R^\d} \mu \log \mu\) 
\Esp_{\Q_{N,\beta}(\mu) } \( e^{\Fluct_\mu( -\theta \zeta-   \log \mu)}  \)
\end{align}
where 
$$\Fluct_\mu(\xi):= \sum_{i=1}^N \xi(x_i)-N\int_{\R^\d} \xi d\mu.$$

A  relative entropy control can be deduced from the rewriting of the Gibbs measure by splitting combined with the lower bound for $\F_N$. It ensures that the Gibbs measure is close to $\mu_\theta^{\otimes N}$ when $\beta \ll 1$. This observation is due to Zhenfu Wang \cite{wangprivate}.

\begin{remark}[Concentration in relative entropy around the thermal equilibrium measure]\label{remwang}
Assume $\s >0$. Let the normalized relative entropy $H_N$ be defined\footnote{It is the quantity $\frac1{N^2} H_N$ which quantifies the convergence of $f_N$ to $\mu^{\otimes N}$, see the discussion at the beginning of Section \ref{sec:dynnoise}.}
 by 
\be  H_N(f_N\vert \mu^{\otimes N})= N \int_{(\R^\d)^N} f_N \log \frac{f_N}{\mu^{\otimes N}} d\XN.\ee
Then we have 
\be
\frac{1}{N^2}  H_N(\Q_{N,\beta}(\mu)| \mu^{\otimes N} )  \le C\beta ,\ee
for $C$ depending only on $\d, \s$ and $\|\mu\|_{L^\infty}$.
In particular, 
\be\label{relentr1} \frac1{N^2} H_N(\PNbeta|\mu_\theta^{\otimes N}) \le C\beta, \ee
where $\mu_\theta $ is the thermal equilibrium measure.
\end{remark}
\begin{proof}
By definition of the relative entropy and  \eqref{defQ}, we have 
$$\frac{\Q_{N,\beta}(\mu)}{ \mu^{\otimes N}}= \frac{1}{\K_{N,\beta}(\mu)} \exp\( - \beta N^{-\frac\s\d} \F_N(\XN, \mu)\)$$
hence by definition \eqref{relentr1} we find that 
\begin{multline*}H_N(\Q_{N,\beta}(\mu)| \mu^{\otimes N} ) + H_N(\mu^{\otimes N}|  \Q_{N,\beta}(\mu))
\\ =-\beta N^{1-\frac\s\d}   \( \int_{(\R^\d)^N} \F_N(\XN, \mu)  d\Q_{N,\beta}(\mu) - \int_{(\R^\d)^N} \F_N(\XN, \mu) d\mu^{\otimes N}\) 
\end{multline*}
 For the second  integral in the right-hand side, we compute as in \eqref{alignfbup} below that 
$$\int_{(\R^\d)^N} \F_N(\XN, \mu) d\mu^{\otimes N} = -\frac{N}{2}\iint \g(x-y) d\mu(x) d\mu(y) \le 0$$
under the assumption $\s>0$, 
while for the first, we insert the lower bound \eqref{minoF2}, written as
$$ \F_N(\XN, \mu)  \ge - C   N^{1+\frac{\s}{\d}} .$$
By nonnegativity of the relative entropy, we thus obtain as claimed
$$H_N(\Q_{N,\beta}(\mu)| \mu^{\otimes N} ) 
 \le  C\beta   N^2 .$$
The second relation follows by \eqref{splitzk}.
\end{proof}

\section{Free energy bounds and concentration}
\subsection{Lower bound for the (free) energy }
The properties of $\F_N$ shown in Chapter \ref{chap:nextorder} 
allow us to immediately deduce a few statements. 
The first is a bound on the free energy obtained from using that $\F_N$ is (almost) bounded below.
Combining \eqref{split0}, Corollary \ref{corminoF}  and the fact that $\zeta \ge 0$, 
we easily obtain 
\begin{coro}[First energy lower bound]\label{corlb}
If $\meseq \in L^\infty(\R^\d)$, then for any configuration $\XN$ in $(\R^\d)^N$, we have
\be\label{flb}
\HN(\XN)\ge
 N^2 \I(\meseq) +N\sum_{i=1}^N\zeta(x_i) -\(  \frac{N}{2\d}\log( N\|\mu_V\|_{L^\infty} ) \) \indic_{\s=0} - C\|\meseq\|_{L^\infty}^{\frac{\s}{\d}} N^{1+\frac\s\d}\indic_{\s\ge 0} ,
 \ee
where $C$ depends only on $\d$ and $\s$.
In particular 
\be \min \HN\ge N^2 \I(\meseq)-\(  \frac{N}{2\d}\log( N\|\mu_V\|_{L^\infty} ) \) \indic_{\s=0} - C\|\meseq\|_{L^\infty}^{\frac{\s}{\d}} N^{1+\frac\s\d}\indic_{\s\ge 0} .
 \ee
 \end{coro}
 
This lower bound easily  translates  into a  lower bound for the free energy $- \frac{1}{\beta} \log \ZNbeta$ or the generic modulated free energy $- \frac{1}{\beta} \log \K_{N,\beta}(\mu)$ of \eqref{defKN}.
\begin{coro}[Upper bound for  the partition function -- thermal equilibrium measure version]\label{boundlogz}
\mbox{}
 Assume that $\mu$ and $\mub$ are $L^\infty$ probability densities. Then for all $\beta>0$, and for $N$ large enough, we have
\be \label{majok} 
\log \K_{N,\beta}(\mu)  \le  \beta  \Big( \frac{N}{2\d}\log( N\|\mu\|_{L^\infty} ) \Big) \indic_{\s=0}+ C\beta N \|\mu\|_{L^\infty}^{\frac{\s}{\d}}   \indic_{\s\ge 0} 
\ee
and
\be\label{logz1}-\frac1\beta \log \ZNbeta  \geq  N^{2-\frac\s\d} \I_\theta(\mub) -  \Big( \frac{N}{2\d}\log( N\|\mub\|_{L^\infty} ) \Big) \indic_{\s=0}- C N \|\mub\|_{L^\infty}^{\frac{\s}{\d}}   \indic_{\s\ge 0} ,\ee
where $C$ depends only on $\s$ and $\d$.
\end{coro}
\begin{proof} It suffices to insert the result of Corollary \ref{corminoF} into \eqref{defKN}  or  \eqref{rewritegibbs2}.
\end{proof}

We next turn to the analogous result expressed with the regular equilibrium measure.
\begin{coro}[Upper bound for  the partition function -- regular equilibrium measure version]\label{boundlogzeqm}
\mbox{}
Assume \eqref{A1}--\eqref{A4} so that $\meseq$ exists and is compactly supported. Assume also that $\meseq$ has a bounded density. Then for all $\beta>0$ and for $N$ large enough, we have
\be\label{logz1eqm}-\frac1\beta \log \ZNbeta  \geq  N^{2-\frac\s\d} \I(\meseq) -  \Big( \frac{N}{2\d}\log( N\|\meseq\|_{L^\infty} ) \Big) \indic_{\s=0}- C N \|\meseq\|_{L^\infty}^{\frac{\s}{\d}}   \indic_{\s\ge 0}- \frac{N}{\beta} C_{\zeta} ,\ee
where $C$ depends only on $\s$ and $\d$, and $C_{\zeta}$ depends on  the integral  in \eqref{A4} and on $\zeta$.
\end{coro}
\begin{proof}
We insert  \eqref{flb} into \eqref{def:ZNbetN} and obtain 
\begin{multline*}
\log \ZNbeta \le - \beta N^{2-\frac\s\d}\I(\meseq) + \beta \frac{N}{2\d} \log (N\|\meseq\|_{L^\infty})  \indic_{\s=0}  + \beta C N \|\meseq\|_{L^\infty}^{\frac\s\d}  \indic_{\s \ge 0}\\
 + \log  \int_{(\R^\d)^N} \exp\(-\theta \sum_{i=1}^N \zeta(x_i)\) d\XN.\end{multline*}
 By separation of variables, 
 $$\log \int_{(\R^\d)^N} \exp\(-\theta \sum_{i=1}^N \zeta(x_i)\) d\XN= N \log \int_{\R^\d} \exp(-\theta \zeta(x))dx.$$ 
In view of  \eqref{borninfzeta}  we have $\zeta \ge \g+V-C $, thus in view of \eqref{A4}  the last integral above converges.
Either $\theta $ is fixed and it is a fixed constant, or $\theta \to +\infty$ as $N \to \infty$ and since $\zeta \ge 0$ the integral in the right-hand side converges monotonically to 
$|\{\zeta=0\}|=|\omega|$.
Thus in all cases, we obtain the result.
\end{proof}
\begin{remark}
As discussed in the proof, $C_\zeta$ is bounded below by a positive constant independent of $N$, thus the estimate is less good than \eqref{logz1}, particularly when $\beta$ is small, illustrating again that the thermal equilibrium measure provides a more precise description than the regular equilibrium measure.
\end{remark}

To obtain a converse bound on the free energy, it will be more convenient to work in the ``blown-up scale" which we now introduce.

\subsection{Rescaling the   energy by blow up}\label{sec:blowup}

We now define potentials and electric fields in blown-up coordinates $x'= N^{1/\d}x$, which turn out to be more natural in several instances. For a measure $\mu$ we let $\mu'(x)= \mu(x N^{-1/\d}) $ and observe that if $\mu$ is a probability density then $\int_{\R^\d} \mu'=N$.
We let 
\be x_i'= N^{1/\d} x_i,\quad  \XN'=(x_1', \dots, x_N')\ee be the blown-up configuration.
Analogously to \eqref{def:FN}, we may define 
\be \label{Fbupfd}
\F(\XN', \mu')= \hal \iint_{\triangle^c} \g(x-y) d \( \sum_{i=1}^N \delta_{x_i'} - \mu'\) (x)  d \( \sum_{i=1}^N \delta_{x_i'} - \mu'\) (y),\ee
where this time the subscript $N$ is absent.
Changing variables in \eqref{Fbupfd} we find 
the rescaling formula for the modulated energy 
\be \label{scalingF} 
N^{-\frac{\s}\d} \( \F_N(\XN, \mu) + \(\frac{N}{2\d}\log N\)\indic_{\s=0}\) = \F(\XN',\mu').\ee
This explains the recurring appearance of the additive term $ \(\frac{N}{2\d}\log N\)\indic_{\s=0}$
 in the logarithmic case $\s=0$, and in that case one should think of the full sum $  \F_N(\XN, \mu) + \(\frac{N}{2\d}\log N\)\indic_{\s=0}$ as being the energy, instead of just $\F_N(\XN, \mu)$. 
 
We have seen in Remark \ref{remc4} that we expect $\F_N(\XN, \mu)+  \(\frac{N}{2\d}\log N\)\indic_{\s=0} 
$ to be of order $
N^{1+\frac\s\d}$, this way in view of \eqref{scalingF}  the blown-up energy $\F(\XN', \mu')$ will  be of order $ N$. 
In other words, $\F(\XN',\mu')$ then becomes proportional to the number of points, or to the volume $N^{1/\d} |\Sigma|$ effectively occupied by the zoomed gas. 

We may now drop the primes and consider the following general definition.
\begin{defi}[Blown-up scale modulated electric energy]
For any configuration $\XN\in (\R^\d)^N$ and any density $\mu$ with $\int_{\R^\d} \mu= N$, the blown-up scale modulated energy is defined by 
\be \label{Fbup}
\F(\XN, \mu):= \hal \iint_{\triangle^c} \g(x-y) d \( \sum_{i=1}^N \delta_{x_i} - \mu\) (x)  d \( \sum_{i=1}^N \delta_{x_i} - \mu\) (y).\ee
\end{defi}
 
We may now define a general blown-up scale partition function and modulated Gibbs measure.
\begin{defi}[Blown-up modulated Gibbs measure and partition function] Let $\mu$ be a bounded density with $\int_{\R^\d}\mu= N$ an integer, we let 
\be \label{defKN2}\mathsf{K}_\beta(\mu):=N^{-N} \int_{(\R^\d)^N} \exp\(-\beta \F(\XN, \mu)\) d\mu^{\otimes N}(\XN)\ee
and 
\be\label{defQbu} d\Q_\beta(\mu):=\frac{1}{N^N\mathsf{K}_\beta(\mu)} \exp\(- \beta\F(\XN, \mu) \) d\mu^{\otimes N}(\XN)
\ee
where we omit $N$ from the notation since we can recover it from $\int_{\R^\d}\mu$.
\end{defi}
In view of \eqref{scalingF} we have 
\be\label{scalingK}
\log  \K_{N,\beta} (\mu)= \log \K_\beta(\mu') +\beta\( \frac{N}{2\d}\log N \) \indic_{\s=0} \ee
 where $\mu'$ is the blown-up of $\mu$ as above.
 
The upper bound of \eqref{majok} becomes in blown-up scale 
\begin{equation}\label{lbmajk}
\log \K_{\beta}(\mu) \le C \beta N\|\mu\|_{L^\infty}^{\frac{\s}{\d}}\indic_{\s\ge 0}.\end{equation}

The choice of temperature/energy scaling with $\beta N^{-\frac\s\d}$ made in \eqref{gibbs} becomes more transparent after blow-up: when \eqref{scalingF} is inserted into \eqref{gibbs}, we may write using the above definition that 
\be\label{rewritePNbeta}
d\PNbeta(\XN)=\frac{1}{N^N \K_\beta(\mub')} \exp\( - \beta \F(\XN, \mub') \) d(\mub')^{\otimes N} (\XN)= d\Q_\beta(\mub').\ee

\medskip

We can give an electric interpretation of $\F$ similarly as what was done in Chapter~\ref{chap:nextorder}.

Analogously to \eqref{def:hnmu}, define
\be\label{HNp} h_N^\mu[\XN] := \int_{\R^\d } \g(\cdot -y) d\( \sum_{i=1}^N \delta_{x_i} - \mu\) (y)\qquad 
h_{N,\vec{\eta}}^\mu[\XN]:= \int_{\R^\d } \g(\cdot -y) d\( \sum_{i=1}^N \delta_{x_i}^{(\eta_i)} - \mu\) (y)
.\ee
Dropping the $[\XN] $ and $\mu$, they  solve (in extended space) 
\be \label{44}-\div (\yg \nab  {h_N})= \cd\( \sum_{i=1}^N \delta_{x_i} - \mu\drd\)\quad \text{in} \ \R^{\d+\k},\ee
\be
-\div (\yg \nab  {h_{N,\vec{\eta}}})= \cd\( \sum_{i=1}^N \delta_{x_i}^{(\eta_i)}- \mu\drd\)\quad \text{in} \ \R^{\d+\k}
.\ee
Similarly as \eqref{defri} we define instead 
\be\label{defrip}
\rr_i:=\frac14\min\(\min_{i\neq j} |x_i-x_j|, \|\mu\|_{L^\infty}^{-1/\d} \).\ee
 \index{truncation radii}
 
 The lower bound for $\F$  deduced from \eqref{minoF2} is in this setting 
 \be\label{lbF}
 \F(\XN, \mu) \ge - C \|\mu\|_{L^\infty}^{\frac\s\d} N \indic_{\s\ge 0},
 \ee with $C$ depending only on $ \d$ and $\s$. It is of order $N$, as announced.
 
 We next record here the equivalent of the discrepancy bounds of Lemma \ref{coronp} at the blown-up scale.
 
 \begin{lem}[Control of charge discrepancy, blown-up scale]\label{coronpbu}\index{discrepancy}
 Let $\XN$ be a configuration in $(\R^\d)^N$, let $h_N$ be associated via~\eqref{44}, and let $\Omega$ be a set of finite perimeter and $\Omega_\delta$ as in \eqref{defOmegad}. Let $D(\Omega)= \int_{\Omega} d(\sum_{i=1}^N \delta_{x_i}- \mu)$. We 
  have, for a constant $C>0$ depending only on $\d$ and $\s$,
  \begin{itemize}
  \item   in the Coulomb case, if $D(\Omega)\ge 0$, for any $\|\mu\|_{L^\infty}^{-1/\d} <\delta \le N^{1/\d}$,
  \be\label{disc10bu}
   \Big( D(\Omega)- \|\mu\|_{L^\infty} |\Omega_\delta\backslash \Omega|\Big)_+^2 \le C  \frac{|\partial \Omega_\delta|}{\delta}
\int_{\Omega_\delta \backslash \Omega} |\nab h_{N, \rr}|^2  \ee
 \item in the Coulomb case, if $D(\Omega) \le 0$, for any $-N^{1/\d} \le \delta < -\|\mu\|_{L^\infty}^{-1/\d} $,
  \be \label{disc1bu}
  \Big( D(\Omega)+\|\mu\|_{L^\infty} |\Omega\backslash \Omega_\delta|\Big)_-^2\le  C\frac{|\p \Omega|}{\delta} \int_{  \Omega\backslash \Omega_\delta} |\nab h_{N, \rr}|^2   \ee
  \item in the Riesz case, if $D(\Omega) \ge 0$, for any  $\|\mu\|_{L^\infty}^{-1/\d} <\delta \le N^{1/\d}$,
  \be  \Big( D(\Omega)- \|\mu\|_{L^\infty} |\Omega_\delta\backslash \Omega| \Big)_+^2\le C
  \( \frac{|\Omega_\delta|}{|\p \Omega_\delta|}\)^\gamma \frac{|\Omega_\delta|}{\delta} 
  \int_{ (\Omega_\delta \backslash \Omega) \times \R^\k}\yg |\nab h_{N,\rr}|^2 
\ee\item in the Riesz case, if $D(\Omega) \le 0$, for any $-N^{1/\d} \le \delta < -\|\mu\|_{L^\infty}^{-1/\d} $,
  \be \label{disc1buriesz}
  \Big( D(\Omega)+\|\mu\|_{L^\infty} |\Omega\backslash \Omega_\delta|\Big)_-^2\le  C
  \( \frac{|\Omega|}{|\p \Omega|}\)^\gamma \frac{|\Omega|}{\delta} 
   \int_{ ( \Omega\backslash \Omega_\delta)\times \R^\k}\yg |\nab h_{N, \rr}|^2   .\ee
  \end{itemize}
  \end{lem}

 \subsection{Upper bound for the free energy, thermal equilibrium version}\label{sec523}
To complement this lower bound, we present an upper bound based on Garcia-Zelada's Jensen argument \cite{garciaz} (as used in the proof of Theorem \ref{LDP}). This is done in the blown-up scale. As an alternate, the proof of Proposition \ref{pro5212} below will be done at the original scale.
\begin{lem}\label{prominok}If $ 0<\s<\d$,  for any bounded nonnegative density $\mu$ such that $\int_{\R^\d} \mu=N$, 
we  have 
\be\label{minok} \log \K_\beta(\mu)  \ge    0.
    \ee
\end{lem}

\begin{proof}
Starting from~\eqref{defKN2}  and using  Jensen's inequality, we may then write
\begin{equation*}
\log  \K_\beta(\mu)\ge - \frac{\beta}{(\int_{\R^\d} \mu)^N}    \int_{(\R^\d)^N}
  \F(\XN, \mu)   
 d\mu^{\otimes N} (X_N).
\end{equation*} 
We next insert \eqref{Fbup}  to obtain, since $\int_{\R^\d} \mu=N$, 
\begin{align}\label{alignfbup}
\lefteqn{\int_{(\R^{\d })^N} \F(\XN, \mu)   d \mu^{\otimes N}(\XN)
} \qquad & 
\\  \nonumber &
= 
\hal \int_{(\R^{\d })^N} \Bigg( \sum_{i\neq j} \g(x_i- x_j) - 2 \sum_{i=1}^N\int_{\R^\d} \g(x_i- y) d\mu(y) 
\\  \nonumber & \qquad\qquad\qquad\qquad\qquad\qquad
+  \iint_{\R^{\d}\times \R^\d} \g(x-y) d\mu(x) d\mu(y)  \Bigg) d\mu^{\otimes N} (\XN)\\ \nonumber
&=\hal \(N(N-1)  N^{N-2} - 2NN^{N-1}   +N^N   \)   \iint_{(\R^\d)^2}\g(x-y) d \mu(x) d\mu(y)\\ 
\nonumber & = -\frac{ N^{N-1}}{2} \iint_{(\R^\d)^2}\g(x-y) d \mu(x) d\mu(y).
\end{align}
It follows that 
\be \log \K_\beta(\mu) \ge \frac{\beta }{2N  } \iint_{(\R^\d)^2}\g(x-y) d \mu(x) d\mu(y).
\ee
Since $\s $ is assumed to be positive, $\g \ge 0$ and  this yields the desired result.
\end{proof}

The case $\s\le 0$ is more delicate. The above proof  can be seen to give (after a change of scales, assuming $\mu$ compactly supported for instance) that $\log \K_\beta(\mu) \ge - C \beta N \g( N^{\frac1\d})   $ which is not the optimal bound. 
A better bound, proportional to $N$, can be obtained by a form of  superadditivity property of $\log \K_\beta$ (see also Chapter \ref{chaploiloc})  after partitioning the support of $\mu$ into regions of size  that should be taken to depend on $\beta$ if one seeks a sharp estimate as $\beta \to 0$.
 We present here a much simpler proof than the one originally  used in \cite{as} and that gives a better result if $\d \ge 3$.

In order to do so we need extra assumptions which ensure that $\mu$'s tails are thin. 
Let us  define 
\be \label{defchibeta}
\chi(\beta)= \begin{cases}  1 & \text{if} \ \s> 0 \ \text{or} \ \beta \ge 1\\
\beta^{\frac{\s}{\d-\s}} |\log \beta|+1 &  \text{if} \ \s\le 0 \ \text{and} \ \beta \le 1,\end{cases}\ee a quantity which will appear repeatedly.

The assumptions are: 
\be\label{assumplbs}
\text{There exists a set $\Lambda$ with piecewise $C^1$ boundary, s.t.  $\mu \ge m >0$ in $\Lambda$}\ee with (if $\mu(\Lambda^c)\neq 0$ and $\s\le 0$)
\be\label{assgmm}\frac{1}{\mu(\Lambda^c)} \iint_{(\Lambda^c)^2} \g(x-y) d\mu(x) d\mu(y) \ge - C\chi(\beta)  N.\ee


In Theorem \ref{th1as} we have seen that, in the Coulomb cases, under suitable assumptions on $V$,  $\Sigma$ being the compact support of $\meseq$,  
from \eqref{2521} and \eqref{mhS} we have $\mub(\Sigma^c) \le \frac{C}{\sqrt{\theta}}= C \beta^{-1/2} N^{\frac{\s-\d}{2\d}}$ and $\mub\ge m>0$ (independent of $N$) in $\Sigma$.
Moreover by scaling, if $\mu$ is the blown-up of $\mub$ and $\Sigma'$ the blown-up of $\Sigma$,  remembering that we are treating the case $\s\le 0$ and using \eqref{gVV}, we have 
\begin{multline}\label{scalingmm}\iint_{((\Sigma') ^c)^2} \g(x-y) d\mu(x) d\mu (y)
= N^{2-\frac\s\d}\iint_{(\Sigma^c)^2} \g(x-y) d\mub(x)d\mub(y)-N^2\log N^{1/\d} \mub(\Sigma^c)^2  \indic_{\s=0} 
\\ 
\ge 2 N^{2-\frac\s\d} \iint_{(\Sigma^c)^2} \(\g(x)_-\wedge \g_-(y) - C\) d\mub(x) d\mub(y) -N^2\log N^{1/\d} \mub(\Sigma^c)^2  \indic_{\s=0} 
\\
\ge - C N^{2-\frac\s\d}\mub(\Sigma^c)- N^2\log N^{1/\d} \mub(\Sigma^c)^2  \indic_{\s=0} 
  , \end{multline}
where we used \eqref{2elem23}, which is true if \eqref{A5} holds. Therefore,
\begin{multline}
\frac{1}{\mu((\Sigma') ^c)} \iint_{((\Sigma') ^c)^2} \g(x-y) d\mu(x) d\mu (y)
\\
=  \frac{1}{N\mub(\Sigma ^c)} \iint_{((\Sigma') ^c)^2} \g(x-y) d\mu(x) d\mu (y)\ge - C N^{1-\frac\s\d} -  C \beta^{-\hal}   N^{\frac{\s-\d}{2\d}+1}   \log N \indic_{\s=0}.\end{multline}

We may check that in view of \eqref{defchibeta}, the right-hand side is $\ge - C \chi(\beta) N$ as long as  $\beta \ge\theta_0 N^{\frac\s\d-1}$ for some $\theta_0>0$,  which is equivalent to $\theta \ge \theta_0>0$, hence we can consider always satisfied.
Thus, these assumptions are verified for the blown-up thermal equilibrium measure with $\Lambda=\Sigma'$, at least in the Coulomb case.

 To partition the support, we may use  the following  tiling lemma, adapted from \cite[Lemma 6.5]{ss2}, that we will use several times in later chapters.
 \begin{lem}[Tiling with quantized cells]\label{tiling} 
 Assume the set $\Lambda $ is such that 
$\partial \Lambda$ is $C^1$, and the density $\mu$ is bounded below in $\Lambda$ by a constant $m>0$. There exists a constant $C>0$ depending on $m$, such that, given any $R\ge 1$, there exists for any $n\in\mathbb N^*$ a collection $\mathcal K$ of closed hyperrectangles in $\Lambda$ with disjoint interiors,  whose sidelengths are between  $ R$ and $R+R^{1-\d} m^{-1}$,  and which are such that
\begin{equation}\label{tile1}
\left\{x\in \Lambda,\dist(x,\p \Lambda)\ge C R \right\} \subset  \bigcup_{K\in\mathcal K} K:= \Lambda_{\rm{int}}  \subset \left\{ x\in \Lambda, \dist(x,\p \Lambda)\ge R \right\},
\end{equation}and
\begin{equation}\label{inttile}
\forall K\in\mathcal K, \quad \int_K \mu \in \mathbb N.
\end{equation}
Moreover, if $\Lambda$ is a hyperrectangle with $\int_\Lambda  \mu$ an integer, then $\mathcal K$ can be constructed so that 
\be \label{tile3} 
 \bigcup_{K\in\mathcal K} K= \Lambda;\ee
 and if $\Lambda$ is not a hyperrectangle, the same tiling into 
 \eqref{tile3} with \eqref{inttile}
  can be performed with $K$'s belonging to the larger class of sets of piecewise $C^1$ boundary  and diameter  bounded by $CR$.
\end{lem}

\begin{proof}
 Let us first look at dimension $\d=1$. Then $\Lambda$ is an interval, say $(a,b)$ over which $\mu\ge m$. We let $t_0=a$, and for each $k \ge 1$, $t_k$ be the minimal value $\ge t_{k-1}+ R$ such that 
$$\int_{t_{k-1}}^{t_k}  \mu(x) dx  \in \mathbb{N}.$$
By definition we have 
$$ m  (t_k- (t_{k-1}+ R) ) \le  \int_{t_{k-1}+ R}^{ t_k} \mu(x) dx \le 1$$
hence  $$R \le t_k -t_{k-1} \le R + \frac1{m}.$$
We take $\mathcal K$ to be the family of intervals $( t_k, t_{k-1})$ and obtain the result. 

Let us now consider $\d\ge 2$. 
For each $\vec{k}$, we let $t_{\vec{k}, 0}=-\infty$ and define by induction $t_{\vec{k}, l}$ to be the smallest $t \ge t_{\vec{k}, l-1}+ R $ such that 
$$\int_{k_1  R}^ {(k_1+1)R} dx_{1} \dots \int_{k_{\d-1} R}^{( k_{\d-1}+1)R}   dx_{{\d-1}}  \int_{ t_{\vec{k}, l-1}}^{t}  du\,   \mu(x_1, \dots, x_{\d-1}, u)\in \mathbb{N}.$$ 
We next consider the hyperrectangles $ [k_1 R, (k_1+1) R]\times \dots \times [k_{\d-1} R, (k_{\d-1}+1) R] \times [t_{\vec{k} , l} , t_{\vec{k}, l+1}]$. 
If such a rectangle $K$ is entirely contained in $\Lambda$, then we  have $\mu(x)  \ge m$ for each $x\in K$, and so by definition of $t_{\vec{k}, l}$ we must have
\begin{multline}\label{arguem}
m R^{\d-1} ( t_{\vec{k}, l} -( t_{\vec{k}, l-1}+ R) ) \\
\le  \int_{k_1   {R}}^ {(k_1+1){R}} dx_{1} \dots \int_{k_{\d-1} R}^{( k_{\d-1}+1)R}   dx_{\d-1} \int_{ t_{\vec{k}, l-1} + R }^{t_{\vec{k}, l} }  du\,   \mu(x_1, \dots, x_{\d-1}, u) \le 1, \end{multline}
hence 
\be\label{sidele}
R \le  t_{\vec{k}, l}- t_{\vec{k}, l-1} \le R +  \frac1{m R^{\d-1}}.\ee

We then define $\mathcal K$ to be the collection of such hyperrectangles which are entirely included in $\{ x\in \Lambda, \dist (x, \partial \Lambda ) >  R\}$.   By construction, \eqref{inttile} is satisfied,  the sidelengths of the rectangle are as desired by \eqref{sidele}, and \eqref{tile1} holds by arguing as above in \eqref{arguem} and using the regularity of $\partial \Lambda$.

The case of $\Lambda$ equal to a hyperrectangle is even easier and follows from the same proof.
If we want to completely tile $\Lambda$, it suffices to split the boundary layer $\Lambda\backslash \Lambda_{\mathrm{int}}$ into cells of diameter $\le CR$, which is possible by the same reasoning (using that $\mu\ge m>0$ and that the layer has ``width" at least $ R$).

\end{proof}

We now proceed to obtaining a lower bound for $\log \K_\beta$ in the case $\s\le 0$.

\begin{prop}\label{pro642}
Assume $\s\le 0$. 
Then, if $\mu$ is a bounded nonnegative density such that  $\int_{\R^\d}\mu=N$ and \eqref{assumplbs}--\eqref{assgmm} hold and if 
\be\label{assumpbeta} \theta \ge \theta_0>0\ee
where $\theta$ is as in \eqref{deftheta},  we have 
\be \label{minologk2d}
\log \K_\beta(\mu) \ge - C \beta \chi(\beta) N\ee
where $\chi$ is as in \eqref{defchibeta}, and $C>0$ depends only on $\d, \s, \|\mu\|_{L^\infty}, \theta_0$ and the constants in the assumptions.
\end{prop}
\begin{proof}
Let us apply Lemma \ref{tiling} to $\Lambda$ as in the assumptions, and let us denote $Q_i$, $i=1,\dots ,p$ the cells of  diameter $\le CR$ constructed via the last statement in the lemma (so they are not necessarily hyperrectangles, and partition all of $\Lambda$) and $n_i=\int_{Q_i} d\mu$ integer. We also let $\mu_i=\mu\indic_{Q_i}$. Finally, we let $\mu_0:=\mu-\sum_{i=1}^p \mu_i= \mu\indic_{\Lambda^c}$, for which $n_0:=\int \mu_0$ must also be integer.

Returning to the definition \eqref{defKN2}, and decomposing  
$\mu= \sum_{i=0}^p \mu_i$ we may write 
$$\mu^{\otimes N} = (\sum_{i=0}^p \mu_i)^{\otimes N} \ge \frac{N!}{n_0 ! \dots n_p!}\prod_{i=0}^p \mu_i^{ \otimes n_i} $$
hence
\begin{eqnarray}\label{commesubad}
\K_\beta(\mu) &  \ge &  N^{-N}  \frac{N!}{n_0 ! \dots n_p!} \int \exp\(-\beta \F(\XN, \mu)\)  d\( \prod_{i=0}^p \mu_i^{\otimes n_i}\) (\XN)\\ \notag
& = &   \frac{N!  \prod_{i=0}^p n_i^{n_i}  }{N^N \prod_{i=0}^p n_i !  } \int \exp\( -\beta \F(\XN, \mu)\)  d\( \prod_{i=0}^p \frac{\mu_i^{\otimes n_i}}{n_i^{n_i}} \) (\XN).
\end{eqnarray}
Thus,  using Jensen's inequality, we find
\begin{equation}\label{lojgk}
\log \K_\beta(\mu)  \ge  \log  \frac{N!  \prod_{i=0}^p n_i^{n_i}  }{N^N \prod_{i=0}^p n_i !  } -\beta   \int  \F(\XN, \mu)   d\( \prod_{i=0}^p \frac{\mu_i^{\otimes n_i}}{n_i^{n_i}} \) (\XN).\ee
First, with Stirling's formula and since $\sum_i n_i=N$,  we have
\begin{equation}\label{stirling1}\log  \frac{N!  \prod_{i=0}^p n_i^{n_i}  }{N^N \prod_{i=0}^p n_i !}   =\sum_{i=0}^p O(\log n_i).\end{equation}
Secondly, we separate the points into batches by  letting  $I_i=\{ n_i, \dots, n_{i+1}-1\}$ so that $\{1, \dots, N\}=\cup_{i=0}^p I_i$ and $\XN^i= \{x_k\}_{k\in I_i}$ and we  may expand $\F$ as
\begin{multline}\label{multfdev}
\F(\XN, \mu) = \sum_{i=0}^p \F(\XN^i, \mu_i)\\
+
\hal \sum_{0\le i\neq j\le p} \Bigg(
  \sum_{k \in  I_i, l\in I_j} \g(x_k-x_l) - \sum_{k\in I_i}  \int \g(x_k-y) d\mu_j(y)\\
     - \sum_{k\in I_j} 
\g(x_k-y) d\mu_i(y)   +  \iint \g(x-y) d\mu_i(x) d\mu_j(y)\Bigg).
\end{multline}
Integrating against $d\( \prod_{i=0}^p \frac{\mu_i^{\otimes n_i}}{n_i^{n_i}} \)$ we find 
\begin{equation}\label{52524} \int   \F(\XN, \mu)   d\( \prod_{i=0}^p \frac{\mu_i^{\otimes n_i} }{n_i^{n_i}} \) (\XN)
= 
\sum_{i=0}^p \frac{1}{n_i^{n_i}} \int  \F(\XN^i, \mu_i)d \mu_i^{\otimes n_i}(\XN^i)\end{equation}
after noticing that the integral of the second line in \eqref{multfdev} vanishes.
Computing as in \eqref{alignfbup}, we thus find 
\begin{equation}\label{52525}\int \F(\XN, \mu) d\( \prod_{i=0}^p \frac{\mu_i^{\otimes n_i}}{n_i^{n_i}} \) (\XN) \ge - \hal  \sum_{i=0}^p \frac1{n_i}
\iint \g(x-y) d\mu_i (x) d\mu_i(y).\end{equation}
Combining with  \eqref{lojgk} and \eqref{stirling1}, we have obtained
$$\log \K_\beta(\mu) \ge  \frac{\beta}{2} \sum_{i=0}^p \frac1{n_i}
\iint \g(x-y) d\mu_i (x) d\mu_i(y) +O(\log n_i) .$$
This can be bounded below by 
\begin{multline*}\log \K_\beta(\mu) \\ \ge \frac{\beta}{2}\frac{1}{n_0} \iint_{(\Lambda^c)^2} \g(x-y) d\mu(x) d\mu(y)
-C \log n_0 +  \sum_{i=1}^p   \(- \frac{\beta}{2} n_i|\g( \mathrm{diam} (\supp \ \mu_i))|-  
C \log n_i \).\end{multline*}
Since the support of $\mu_i$ is of size $\in [R, CR]$ for $i\ge 1$,  and $n_i = O(R^\d)$ since $\mu $ has bounded density,  we have $p\le C\frac{N}{R^\d}$ and using also \eqref{assgmm}, we arrive at 
\begin{multline}\label{mulconc}
\log \K_\beta(\mu) \ge - C \frac{N}{R^\d}\(  \frac{\beta}{2}R^\d|\g( R) |+
C \log R \)-C \beta \chi(\beta) N -  C \log N\\
\ge - C N \(\frac{ \log R}{R^\d}- \beta   \g(R)\)- C \beta \chi(\beta) N - C \log N,\end{multline} where $C$ depends only on $\s\le 0, \d$ and the constants in the assumptions.
Optimizing over $R$ leads to the choice 
\be \label{choixRRR}
R= \begin{cases} 1 & \text{if} \ \beta \ge 1\\
 \beta^{-\frac{1}{\d-\s}} & \text{if} \ \beta \le 1\end{cases}\ee for which we find by definition \eqref{defchibeta}
$$ - C N \(\frac{ \log R}{R^\d}- \beta   \g(R)\)\ge - CN \beta^{ 1 +\frac\s{\d-\s}} |\log \beta| = - C N\beta \chi(\beta).$$ 
On the other hand $\log N $ can be absorbed into $\beta \chi(\beta)N$ because the assumption \eqref{assumpbeta} implies that $C\beta\chi(\beta)\ge \frac{\log N}{N}$.
The conclusion then follows from inserting into \eqref{mulconc}.
\end{proof}
 
This concludes the  proof of the  converse to \eqref{lbmajk} in all cases. 

\begin{coro}[Bound on the reduced free energy]\label{cor:boundlogK} 
Let $\mu$ be a bounded nonnegative density with $\int_{\R^\d} \mu=N$.
If $\s \le 0$, assume \eqref{assumplbs}--\eqref{assgmm} and \eqref{assumpbeta}.
 Then 
\be \label{boundlogK}
\left|\log \K_\beta(\mu)\right|\le C \beta \chi(\beta) N \ee where $C>0$ depends only on  $\s,\d$, $\|\mu\|_{L^\infty}$ and the constants in the assumptions.
\end{coro}

We can translate this into a bound at the original scale, and inserting into \eqref{splitzk}, an approximation for the original free energy.
\index{free energy expansion}
\begin{coro}[First approximation of the free energy] \label{corfirstapp}Assume $\mub$ satisfies the assumptions of the previous corollary, where $\theta$ as in \eqref{defEtheta} \footnote{which we have seen is proven at least in the Coulomb case on the basis of Theorem \ref{th1as}.}
\be\label{logz2b}
\left|\log \ZNbeta + \theta N \I_\theta(\mub) -  \( \frac{\beta}{2\d} N \log N\) \indic_{\s=0}\right|\le C \beta \chi(\beta) N\ee 
where  $C$ depends only on $\d,\s, \|\mub\|_{L^\infty}$ and the constants in the assumptions.
\end{coro}
We emphasize that these bounds are valid for all $\beta$ and $N$ (within the range \eqref{assumpbeta} for $\s\le 0$), so that we may let $\beta$ depend on $N$.

\subsection{Upper bound for the free energy, equilibrium measure version}

We next describe how to obtain the analogous results to the previous subsection for the free energy relative to the standard equilibrium measure as in \eqref{rewritegibbs33}.
\begin{prop}\label{pro5212}
Assume  $\mu $ is a bounded probability density with support $\Sigma$, and that $\zeta=0$ in $\Sigma$. 
Then, letting $\tilde \K_{N,\beta}(\mu, \zeta) $ be as in \eqref{deftildeK}, we have 
\be \label{lbtKmeq}
\log \tilde \K_{N,\beta}(\mu, \zeta) \ge - N \int \mu \log \mu + \( \frac{\beta N}{2\d} \log N  \)\indic_{\s=0}- C \beta \chi(\beta)N, \ee
and in particular, if $\meseq$ exists and has compact support, then for all $\beta$, we have
\begin{equation}\label{lbZcasmeq}
\log Z_{N,\beta} \ge     - \theta N  \I_\theta(\meseq)  + \( \frac{\beta N}{2\d} \log N  \)\indic_{\s=0}- C \beta \chi(\beta)N,
\end{equation}
where $C>0$ depends only on $\d$, $\s$ and $\|\mu\|_{L^\infty}$, resp. $\|\meseq\|_{L^\infty}$.
\end{prop}
\begin{proof}
Starting from \eqref{deftildeK} and using that $\zeta=0$ in $\Sigma$, the support of $\mu$, we have
\begin{align*}
\tilde \K_{N,\beta}(\mu, \zeta) &\ge
 \int_{(\Sigma)^N} \exp \( -\beta N^{-\frac\s\d} \F_N(\XN, \mu)\) d\XN\\
&= \int_{(\Sigma)^N} \exp \( -\beta N^{-\frac\s\d} \F_N(\XN, \mu)-\sum_{i=1}^N \log \mu(x_i) \) d\mu^{\otimes N} (\XN).\end{align*}
Applying Jensen's inequality, we then find
\begin{equation}
\log \tilde \K_{N,\beta}(\mu, \zeta) \ge \int_{\Sigma^N} \( -\beta N^{-\frac\s\d} \F_N(\XN, \mu)-\sum_{i=1}^N \log \mu(x_i) \)d\mu^{\otimes N} (\XN).\end{equation}
Reasoning as in \eqref{alignfbup} (except we are now  working at the usual scale) we obtain
\begin{equation}
\log \tilde \K_{N,\beta}(\mu, \zeta) \ge  -\beta N^{-\frac\s\d}\( -\frac{N}{2} \iint_{\Sigma^2} \g(x-y) d\mu(x) d\mu(y) \) -N\int\mu \log \mu.\end{equation}
If $\s> 0$ then $\g \ge 0$ and the result follows.
If $\s \le 0$, we argue as in the proof of Proposition~\ref{pro642} and split $\mu= \sum_{i=1}^p \mu_i$ where $\mu_i$ is supported in $Q_i$, cell of diameter $\le R N^{-\frac1\d}$, with the $Q_i$'s forming a partition of $\Sigma$, $N \int_{Q_i} \mu_i = n_i$, an integer. We then obtain,  as in \eqref{lojgk}, 
\begin{multline}\label{lojgk2}
\log\tilde \K_{N,\beta}(\mu,\zeta) \\ \ge  \log  \frac{N!  \prod_{i=1}^p( n_i/N)^{n_i}  }{ \prod_{i=1}^p n_i !  }   +\int \( -\beta N^{-\frac\s\d}  \F_N(\XN, \mu)- \sum_{i=1}^N \log \mu(x_i) \)    d\( \prod_{i=1}^p \frac{\mu_i^{\otimes n_i}}{(n_i/N)^{n_i}} \) (\XN).\end{multline}
First, using Stirling's formula and~$\sum_i n_i=N$,  we have
\begin{equation}\label{stirlingbv1}\log \frac{N!  \prod_{i=1}^p( n_i/N)^{n_i}  }{ \prod_{i=1}^p n_i !  } =    \log  \frac{N!  \prod_{i=1}^p n_i^{n_i}  }{N^N \prod_{i=1}^p n_i !}   =\sum_{i=1}^p O(\log n_i).\end{equation}
Separating the points into batches, and integrating, we obtain as in \eqref{52524} and \eqref{52525} that 
\begin{multline}\int  \F_N(\XN, \mu) d\( \prod_{i=1}^p \frac{\mu_i^{\otimes n_i}}{(n_i/N)^{n_i}} \) (\XN) 
=\sum_{i=1}^p \frac{N^{n_i}}{n_i^{n_i}} \int  \F_N(\XN^i, \mu_i)d \mu_i^{\otimes n_i}(\XN^i)
\\
= \hal \sum_{i=1}^p   \frac{N^{n_i}  }{n_i^{n_i}} \(n_i(n_i-1)(\frac{n_i}{N})^{n_i-2} -2 n_i N (\frac{n_i}{N})^{n_i-1}
+ N^2 (\frac{n_i}{N})^{n_i}\)   \iint \g(x-y) d\mu_i(x) d\mu_i(y) \\
=  - \hal \sum_{i=1}^p  \frac{N^2 }{n_i }  \iint \g(x-y) d\mu_i(x) d\mu_i(y).
\end{multline}
Also 
\begin{equation*}\int\sum_{i=1}^N \log \mu(x_i)     d\( \prod_{i=1}^p \frac{\mu_i^{\otimes n_i}}{(n_i/N)^{n_i}} \) (\XN)= \sum_{i=1}^p n_i    \frac{N}{n_i}\int \log \mu \, d\mu_i = N \int \mu \log \mu .  \end{equation*}
It follows that 
\begin{multline}\label{lojgk234}
\log\tilde \K_{N,\beta}(\mu,\zeta) \\ \ge   \sum_{i=1}^p O(\log n_i) +\frac{\beta}{2} N^{-\frac\s\d}  \sum_{i=1}^p  \frac{N^2}{n_i }  \iint \g(x-y) d\mu_i(x) d\mu_i(y) -N\int \mu\log \mu  .
\end{multline}
The sizes of the cells being $RN^{-1/\d}$, and thus $n_i= O(R^\d)$ and $p \le CN R^{-\d}$,  we find
 \begin{multline}\label{lojgk2345}
\log\tilde \K_{N,\beta}(\mu,\zeta) \\ \ge  - C \frac{N}{R^\d}  \log R +\frac{\beta}{2} N^{-\frac\s\d}  \sum_{i=1}^p  \frac{N^2}{n_i }  \frac{n_i^2}{N^2} \( N^{\frac{\s}{\d} } \g(R) +  \frac{1}{\d}  \log N \indic_{\s=0} \)   -N \int \mu \log \mu \\
\ge -N \int \mu \log \mu +\( \frac{\beta N}{2\d} \log N\) \indic_{\s=0}  - C N (  \frac{N}{R^\d}  \log R -\frac{\beta}{2}    \g(R)  )   .
\end{multline}
Optimizing and choosing $R $ as in \eqref{choixRRR}, we obtain the result  and \eqref{lbZcasmeq} follows from \eqref{rewritegibbs33} after grouping terms in the form \eqref{defEtheta}.

\end{proof}

Combining with the converse bound in Corollary \ref{boundlogzeqm}, we obtain the following equilibrium measure  version  of Corollary \ref{corfirstapp}. As explained above, the estimate is less precise that that one.
\begin{coro}[First approximation of the free energy -- equilibrium measure case]\label{cor5214}
Assume \eqref{A1}--\eqref{A4} so that $\meseq$ exists and is compactly supported. Assume also that $\meseq$ has a bounded density. Then for all $\beta>0$, and for $N$ large enough, we have
\be\label{logz1eqmeg}\left|\log \ZNbeta  + \theta   N \I_\theta (\meseq) -  \Big( \frac{\beta N}{2\d}\log N ) \Big) \indic_{\s=0}\right|\le C \beta \chi(\beta) N+ N C_{\zeta}  ,\ee
where $C$ depends only on $\s$,  $\d$ and $\|\meseq\|_{L^\infty}$, and $C_\zeta$ depends only on  the quantity in \eqref{A4} and on $\zeta$.
\end{coro}

\subsection{Concentration bounds and minimal energy bound}

We  can directly deduce  from the above a first concentration bound, in the form of a control in exponential moments of the next-order energy.
\subsubsection{Concentration around the thermal equilibrium measure}
\begin{coro}[First concentration bound, in blown-up scale]
\label{cor521}
 Let $\mu$ be a bounded nonnegative density with $\int_{\R^\d} \mu=N$ and $\Q_\beta$ as in  \eqref{defQbu}, then, under the same assumptions as Corollary \ref{cor:boundlogK}, we have
\be \label{firstconcbound}
\left|\log \Esp_{\Q_\beta(\mu)}  \( \exp \frac\beta2\F(\XN, \mu)\) \right|\le C \beta \chi(\beta)N\ee
where $C$ depends only on $\d, \s$ and  $\|\mu\|_{L^\infty}$. 
\end{coro}
\begin{proof} It suffices to rewrite 
\begin{multline*}  \Esp_{\Q_\beta(\mu)}  \( \exp \frac\beta2\F(\XN, \mu)\)= \frac{1}{\K_\beta(\mu)} \int_{(\R^\d)^N} 
\exp\( - \beta \F(\XN, \mu) + \frac\beta2\F(\XN, \mu)\) d\mu^{\otimes N}(\XN)\\
=  \frac{\K_{\beta/2}(\mu) }{\K_\beta(\mu)} \end{multline*}  which we bound thanks to \eqref{boundlogK}.\end{proof} 
This estimate means that $\F(\XN, \mu)$ is typically bounded by $CN$ (modulo the $\chi(\beta)$ at low beta in the case  $\s\le 0$) i.e the next order energy is of the order of the number of points / of the blown-up volume occupied by the particles. This is to be compared with the leading order energy which scales like $N^2$.

We can then translate this into an estimate at the original scale.
For this, we need to restate the assumptions at that scale, using \eqref{scalingmm}.
\be\label{assumplbs2}
\text{There exists a set $\Lambda$ with piecewise $C^1$ boundary, s.t.  $\mu \ge m >0$ in $\Lambda$}\ee with 
\be\label{assgmm2}\frac{N^{-\frac\s\d}}{\mu(\Lambda^c)} \iint_{(\Lambda^c)^2} \g(x-y) d\mu(x) d\mu(y) - \frac{\log N}{\d} \mu(\Lambda^c) \indic_{\s=0} \ge - C\chi(\beta)  N.\ee

\index{thermal equilibrium measure}
\begin{coro}[First concentration bound -- thermal equilibrium measure version] For all \eqref{riesz}, 
for any $N$ and $\beta$, and any probability measure $\mu$ with bounded density,
and if $\s\le 0$ assuming in addition \eqref{assumplbs2}, \eqref{assgmm2} and \eqref{assumpbeta},  we have 
\be \label{expmom0}
\left|\log\Esp_{ \Q_{N,\beta}(\mu)} \( \exp \frac{\beta N^{-\frac\s\d}}{2}\(\F_N(\XN, \mu) +  \(\frac{N}{2\d}\log  N \)\indic_{\s=0} \) \) \right|\le C \beta \chi(\beta)  N.\ee 
If these assumptions are satisfied for $\mu=\mub$ \footnote{which we have seen is true in all cases $\s>0$ and in the $\d=2$  Coulomb case under the assumption  \eqref{assumpbeta} and \eqref{A1}--\eqref{A5}}, we have 
\be\label{expmom}
\left| \log \Esp_{\PNbeta} \( \exp \frac{\beta N^{-\frac\s\d}}{2}  \( \F_N(\XN, \mub)+ 
\(\frac{N}{2\d}\log N\) \indic_{\s=0}\) \) \right|\le C \beta\chi(\beta)  N\ee 
where $C>0$ depends only on $\s,\d$ and $\|\mu\|_{L^\infty}$, resp. $\|\mub\|_{L^\infty}$, and the constants in the assumptions.
\end{coro}

In view of  the results of Lemma \ref{prop:fluctenergy}, the bound \eqref{expmom0} shows that $\emp$ concentrates near $\mub$, with explicit bounds:

\begin{coro}[Concentration around the equilibrium measure - Local laws at the macroscopic scale]\label{coroconc} \index{concentration}
\mbox{}
\index{local laws}
Let  $\varphi$ be a  regular function, then, under the same assumptions, 
\be\label{ridicboundfluct}
\left|\log \Esp_{\Q_{N,\beta}(\mu) } \( \exp \frac{\beta}{C\|\varphi\|^2} N^{2-\frac\s\d}   \( \int_{\R^\d} \varphi \, d (\emp -\mu ) \)^2 \) \right|
\\
\leq C \beta \chi(\beta)     N ,
\ee
where $C$ depends only on $\s, \d$ and $\|\mu\|_{L^\infty}$, and 
$$\|\varphi\|= \begin{cases} \|\nab \varphi\|_{L^2} + \|\nab \varphi\|_{L^\infty} & \text{if } \s=\d-2\\ 
\|\varphi\|_{\dot{H}^{\frac{\d-\s}{2}}} + \| (-\Delta )^{\frac{\d-\s}{2}} \varphi\|_{L^\infty} & \text{if } \s\neq \d-2.\end{cases}$$
\end{coro}
\begin{proof} It suffices to insert \eqref{coulombfluct} or \eqref{rieszfluct} combined with \eqref{bornehnr} into \eqref{expmom}.
\end{proof} 

More precise concentration results but  in $W_1$ distance, using ``Coulomb-transport inequalities",  are obtained  in \cite{ChafaiHardyMaida} for a fixed $\beta$ regimes. Analogous concentration near $\mub$ in the small $\beta$ regime  is established in \cite{DPG1}.

We then also reexpress these concentration estimates in terms of the charge discrepancies of Section \ref{sec:discrepancy}.
Combining the result of  Lemma \ref{coronpball} with \eqref{bornehnr} and inserting into \eqref{expmom0}, we obtain  \index{discrepancy}
\begin{coro}[First discrepancy bound]\label{corodiscr} Let $B_R$ be any ball of radius $R$ in $\R^\d$, and let $D(B_R)$ be as in \eqref{defiD}. Under the same assumptions, 
we have 
\begin{multline}\left|\log \Esp_{\mathbb{Q}_{N,\beta}(\mu)}\( \exp\( \beta C^{-1}  N^{-\frac\s\d}
\frac{D(B_R)^2}{R^{\s}}\left|\min \(1, \frac{D(B_R)}{   \|\mu\|_{L^\infty(B_R)}R^\d}\) \right| \) \) \right|\\ \le C \beta \chi(\beta) N,\end{multline} for some $C>0$ depending only on $\d, \s $ and $\|\mu\|_{L^\infty}$. \end{coro}
Again, this gives the result for the Gibbs measure by using \eqref{splitzk}. 

 We will see in Chapter \ref{chaploiloc} a localized version of the results of Corollary \ref{coroconc} and \ref{corodiscr}, which we call {\it local laws}.

\subsubsection{Concentration around the equilibrium measure}
We now give the analogue statements relative to the equilibrium measure, relying on Corollary~\ref{cor5214}.

\begin{coro}[First concentration bound -- equilibrium measure version]\label{cor5219} For all \eqref{riesz}, assume  \eqref{A1}--\eqref{A4} so that $\meseq $ exists and is compactly supported. Assume also that $\meseq$ has a bounded density. Then for all $\beta>0$, 
  we have 
\begin{multline}\label{expmomeqm}
\left| \log \Esp_{\PNbeta} \( \exp \frac{\beta N^{-\frac\s\d}}{2}  \( \F_N(\XN, \meseq)+ 
\(\frac{N}{2\d}\log N\) \indic_{\s=0}+ N \sum_{i=1}^N\zeta(x_i) \) \)\right|\\ \le C \beta\chi(\beta)  N+ C_\zeta N,\end{multline}
where $C>0$ depends only on $\s,\d$ and $\|\meseq\|_{L^\infty}$, and $C_\zeta$ depends on \eqref{A4} and $\zeta$.
\end{coro}
\begin{proof}
We start from \eqref{rewritegibbs33} and \eqref{deftildeK} to write 
\begin{align*}
\Esp_{\PNbeta} \( \exp \frac{\beta N^{-\frac\s\d}}{2}  \( \F_N(\XN, \meseq) + N \sum_{i=1}^N\zeta(x_i) \)\)
= \frac{\tilde \K_{N,\beta/2}(\meseq,\zeta)}{\tilde \K_{N,\beta}(\meseq,\zeta)}.
\end{align*}
We next bound the log of the ratio in the right-hand side using \eqref{logz1eqm} and \eqref{lbtKmeq} and deduce the result.
\end{proof}

Similarly as Corollary \ref{coroconc} we obtain a bound for linear statistics.

\index{fluctuations}
\begin{coro}[Concentration around the equilibrium measure - Local laws at the macroscopic scale]\label{coroconcmeq}
\mbox{}
Let  $\varphi$ be a  regular function, then, under the same assumptions, 
\be\label{ridicboundfluctmeq}
\left|\log \Esp_{\PNbeta } \( \exp \frac{\beta}{C\|\varphi\|^2} N^{2-\frac\s\d}   \( \int_{\R^\d} \varphi \, d (\emp -\meseq ) \)^2 \) \right|
\\
\leq C \beta \chi(\beta)     N + C_\zeta N,
\ee
where $C$ depends only on $\s, \d$ and $\|\meseq\|_{L^\infty}$, and 
$$\|\varphi\|= \begin{cases} \|\nab \varphi\|_{L^2} + \|\nab \varphi\|_{L^\infty} & \text{if } \s=\d-2\\ 
\|\varphi\|_{\dot{H}^{\frac{\d-\s}{2}}} + \| (-\Delta )^{\frac{\d-\s}{2}} \varphi\|_{L^\infty} & \text{if } \s\neq \d-2.\end{cases}$$
\end{coro}
Of course an analogue to Corollary \ref{corodiscr} can also be written down.

\subsubsection{Bound on the minimal energy}

We finally note that Lemma  \ref{prominok} and Proposition \ref{pro642} provide an easy (probabilistic) way to prove an upper bound on the minimal modulated energy. Of course, this can be combined with the splitting formula \eqref{split0} or \eqref{splitting} to provide a bound for the minimum of $\HN$.
\begin{coro}\label{corobenergy}(Original scale) Let $\mu $ be a bounded probability  density over $\R^\d$. In the case $\s\le 0$, assume also that $\mu$ satisfies \eqref{assumplbs2}--\eqref{assgmm2} for $\beta=1$. Then 
we have 
\be \label{maxmin} \min \( \F_N(\cdot, \mu) + \( \frac{N}{2\d}\log N\)\indic_{\s=0}\) \le C N^{1+\frac\s\d}\indic_{\s\le 0}
\ee for some $C$ depending only on $\d $, $\|\mu\|_{L^\infty}$ and the constants in the assumptions.

(Blown-up scale)
Let $\mu $ be a bounded density such that $\int_{\R^\d} \mu=N$. 
In the case $\s\le 0$, assume also that $\mu$ satisfies \eqref{assumplbs}--\eqref{assgmm} for $\beta=1$. Then 
\be \label{maxmin} \min  \F(\cdot, \mu)  \le C N \indic_{\s\le 0}
\ee  for some $C$ depending only on $\d ,\s$, $\|\mu\|_{L^\infty}$ and the constants in the assumptions.
\end{coro}
\begin{proof}
Applying Lemma \ref{prominok}  and Proposition \ref{pro642} with $\beta=1$ we have 
$$\log \( \int \exp\(- N^{-\frac\s\d}\( \F_N(\XN, \mu) +\( \frac{N}{2\d} \log N \)\indic_{\s=0} \) \)d\mu^{\otimes N}(\XN)\) \ge - C N \indic_{\s\le 0}  $$ with $C$ depending only on $\d, V$. 
A mean-value argument ensures that there exists   $\XN$ such that 
$\F_N(\XN, \mu) \le C N^{1+\frac\s\d}\indic_{\s\le 0} $ for the same $C$.
\end{proof}
The bound \eqref{maxmin} can also be proven   (as first done in \cite{PetSer} for instance) by constructing a test configuration with the help of the subadditivity of Chapter \ref{chap:screening}, see Chapter \ref{chaploiloc} for more detail.

\section{Localization, separation and discrepancy}\label{sec5.5}
We now turn to other methods, in particular involving maximum principle ideas, that allow to obtain more refined estimates on localization, point separation and discrepancy. 
\subsection{Localization}\label{subsec:localization}

The splitting formula \eqref{split0}  indicates that $\zeta$ acts as an effective confinement potential, which vanishes in the drople, i.e~the set $\omega$ equal to the zero set of $\zeta$, which contains $\Sigma$ (with possibly strict inclusion).
The concentration result Corollary \ref{cor5219}  provides an average localization result. Indeed, insert the lower bound \eqref{minoF2} into \eqref{expmomeqm} we obtain 
\begin{coro}[Average localization]
Assume  \eqref{A1}--\eqref{A4} so that $\meseq $ exists and is compactly supported. Assume also that $\meseq$ has a bounded density. Then for all $\beta>0$, 
  we have 
\begin{equation}\label{expmomeqmzeta}
\left| \log \Esp_{\PNbeta} \( \exp \frac{\beta N^{1-\frac\s\d}}{2} \sum_{i=1}^N\zeta(x_i) \) \right|  \le C \beta\chi(\beta)  N+ C_\zeta N,\end{equation}
where $C>0$ depends only on $\s,\d$ and $\|\meseq\|_{L^\infty}$, and $C_\zeta$ depends on \eqref{A4} and $\zeta$.
\end{coro}
By Markov's inequality, this allows to bound from above the probability that $\sum_i \zeta(x_i)$ becomes large, which allows to control the distance to $\Sigma$ when estimates such as \eqref{assumpV5} and \eqref{assumpzeta6}, coming from the analysis of the (fractional) obstacle problem, are available.

In the two-dimensional Coulomb gas, a stronger result is proven in \cite{ameurloc}, we will discuss it below, see also 
 \cite{ChafaiHardyMaida} for the Coulomb case in any $\d \ge 2$.

We now prove perfect localization in the droplet for minimizers, i.e. that all the points belong to $\Sigma$, the support of $\meseq$. The simple argument (found in \cite{rns} in the two-dimensional case) relies on the maximum principle.
\begin{theo}[Perfect localization for minimizers] Assume $\s=\d-2$.
Let $\XN $ minimize $\HN$. Then for all $i$, $x_i\in \Sigma$, the support of $\meseq$.
\end{theo}
\begin{proof}
The main point is that for minimizers, each point is at the minimum of the potential generated by the other points. Indeed,  for any $x$, we may write that
$$\HN(\XN)\le \HN(\widehat \XN)$$ where 
the points of $\widehat{\XN}$ are defined by 
 $$\hat x_j= \begin{cases} x_j & \text{if} \ j\neq i\\
 x & \text{if } \ j=i.\end{cases}$$
 Spelling out the definition of the energy, this implies by direct comparison that 
 \begin{equation}\label{toaverage}
 \sum_{j\neq i} \g(x_i-x_j)+ NV(x_i) \le \sum_{j\neq i}  \g(x-x_j) +   NV(x) \ee
 or in other words 
 \be\label{hih}
 h_i(x_i) \le h_i(x)\ee where $$h_i:= \sum_{j\neq i} \g(\cdot -x_j)+ NV= \sum_{j\neq i} \g(\cdot -x_j)+ N(\zeta- h^{\meseq}+c) $$ using \eqref{defzeta}.
We then notice that in $\Sigma^c$, $\sum_{ j\neq i} \g(\cdot -x_j)$ is superharmonic and $h^{\meseq}$ is harmonic. Thus 
$\sum_{j\neq i} \g(\cdot -x_j)- N h^{\meseq} = \g*( \sum_{j\neq i} \delta_{x_j}- N \meseq) $, which is asymptotic to  $-\g$ at $\infty$ since $\meseq $ is compactly supported,  can only achieve its strict minimum relative to $\Sigma^c$ on $\partial \Sigma$ (after distinguishing the cases $\s \le 0$ and $\s>0$). Meanwhile $\zeta+c$ is also minimized in $\Sigma$.  So  we conclude that  $h_i$ is minimized in $\R^\d \backslash \Sigma$ on $\partial \Sigma$, and thus  if $ x_i\in \R^\d \backslash \Sigma$, we have a contradiction with \eqref{hih}. 
 \end{proof}
 In the case of general $\s\in (\d-2,\d)$, the result is still true and  the argument is similar, except we use the maximum principle for the operator $\div (\yg \nab \cdot)$ in $\R^{\d+1}\backslash (\Sigma \times \{0\})$, see \cite{PetSer} for the full proof.

\subsection{Isotropic averaging and consequences}\label{sec:isotropic}
\index{isotropic averaging}
We now continue with maximum-principle based arguments, and  present a result showing that points of  energy minimizers  are well-separated.  Separation results  for Fekete points but also subsequently for Riesz $\s$-energy minimizers have been established in very general geometries \cite{dahlberg,BrauchartDragnevSaff,hardinRSV}.  
The  result we present below was shown in the two-dimensional Coulomb case in \cite{aoc}, 
and recently extended to the low temperature regime $\beta \ge C \log N$
in \cite{ameurromero}, still in the 2D Coulomb case. 
A variant of the short proof below  was given  in  \cite{rns} in the $\d$-dimensional   Coulomb case and in \cite{PetSer} in the Riesz case $\s\in [\d-2,\d)$,   using a maximum-principle unpublished argument of Lieb \cite{liebpersonal}.
 We now present a reformulation of the argument that works in  the Coulomb case and relies on  {\it isotropic averaging}.   The isotropic  averaging method was introduced in the context of the Gibbs measure (with temperature) in \cite{leblehyper}, and made into a systematic and powerful method, that we discuss below, by Thoma in \cite{thoma}.
 \index{separation}
\begin{theo}[Point separation for minimizers in the Coulomb case by isotropic averaging] \label{separation} \mbox{} 
Assume $\s=\d-2$.
Assume $V$ is such that $(\Delta V)_+\in L^\infty$. Let $\XN$ be a minimizer of $\HN$. There exists an explicit constant $C>0$ depending only on $\d$  such that  
\be \min_{i\neq j} |x_i-x_j|\ge \frac{C}{\(N\|(-\Delta V)_+\|_{L^\infty} \)^{1/\d}}.\ee
\end{theo}
\begin{proof}
The core of the isotropic averaging is very similar to the monotonicity argument of Section \ref{sec:monoto}.
Let us use the same notation, in particular $\delta_x^{(\eta)}$. 
If $\XN$ minimizes $\HN$ then its energy is less than that obtained by isotropically averaging over the position of one point, call it $x_i$. In other words, averaging \eqref{toaverage} over $x$, we may write that 
$$
\sum_{j\neq i}(\g(x_i-x_j)- \int \g(x-x_j)   d\delta_{x_i}^{(\eta)} (x) \le N\int (V(x)  - V(x_i) )d\delta_{x_i}^{(\eta)}(x).
$$
Using that $\g*\delta_0^{(\eta)}= \g_\eta$ by definition and the radial nature of $\delta_{x_i}^{(\eta)}$, we deduce that 
\be \sum_{j\neq i} \g(x_i-x_j)-  \g_\eta(x_i-x_j) \le N\eta^2\| (\Delta V)_+ \|_{L^\infty} .\ee
We now recall that $\g_\eta \le \g$ and $\g_\eta(x) =\g(\eta)$ for $|x|\le \eta$, thus  the sum contains only nonnegative terms, and   restricting it to the points such that $|x_i-x_j|\le \hal \eta$ (supposing that such points exist) we obtain 
$$\sum_{j\neq i, |x_i-x_j|\le \hal \eta} \g(\frac{\eta}{2}) - \g(\eta) \le N \eta^2 \|(\Delta V)_+\|_{L^\infty} .  $$ We may now choose 
$$\eta< \begin{cases}  (2N \|(\Delta V)_+\|_{L^\infty})^{-1} & \text{if} \ \d=1\\
\( N \|(\Delta V)_+\|_{L^\infty}/( \log 2) \)^{-\hal} & \text{if} \ \d=2\\
\( \frac{ 2^{\d-1}-1 }{  N \|(\Delta V)_+\|_{L^\infty} }\)^{\frac{1}{\d}} & \text{if} \ \d\ge 3. 
\end{cases}$$
to obtain a contradiction, thus implying that there is no other point in $B(x_i, \hal \eta)$ than $x_i$.
\end{proof}
The fact that $V$ can contain a possibly singular arbitrary superharmonic part, without changing the estimate (because it does not change $(-\Delta V)_+$) is useful when considering particles subjected to an external potential itself generated by point charges, and for ``incompressibility estimates" in the Laughlin phase as in \cite{lry}, this is taken advantage of in \cite{thoma}.

The article \cite{thoma} provides a temperature-dependent version of the argument in the proof above by estimating quantitatively the entropic effect of the isotropic averaging (in other words, the change of volume in phase space due to it).
Thanks to these quantitative estimates, Thoma is able to obtain, for the Coulomb case in arbitrary dimension
\begin{itemize}
\item
``overcrowding estimates" \index{overcrowding} phrased at the blown-up scale that are valid up to the boundary and confirm the predictions of \cite{jlm} (see Section \ref{sec:leble} for the statement of the prediction). 
\begin{theo}[\cite{thoma}]\label{theric}
 Assume $\s=\d-2$, $\d\ge 2$.  For any $\beta$,  any ball $B_R$ of radius $R$, 
 if $Q \ge CR^\d+ C\beta^{-1}R^{\d-2}$, then 
 $$\PNbeta  \( \# \( X_N' \cap B_R\)\ge Q\) \le \begin{cases} \exp\( - \frac{\beta}{4} Q^2 \log \frac{Q}{R^2} + C \beta Q^2 + C Q\) & \d=2\\
 \exp\(- C \beta R^{2-\d} Q(Q-1)\) &  \d\ge 3.\end{cases}$$
 \end{theo}

\item estimates for the size of the minimal gap $\eta:=\min_{i\neq j} |x_i-x_j|$ between particles, which in the case $\d=2$ takes the form of the sharp estimate
\be\label{separationthoma}\PNbeta(\eta N^{1/\d} \le \gamma N^{-\frac{1}{2+\beta}}) \le C \gamma^{2+\beta} \quad \forall \gamma >0\ee where $C$ is independent of $N$ and one can let $\beta \to \infty $ as $N \to \infty$ to retrieve the result of Theorem \ref{separation}
which is analogous to the freezing regime of \cite{ameurrepulsion,ameurromero}. 
Note that in the case of the two-dimensional Coulomb gas at $\beta=2$, precise asymptotics for large gap probabilities are known \cite{forrestergap,charliergap}.

\item uniform upper bounds on the $k$-point correlation function $\rho_k$ as defined in \eqref{defrhok} which in the case $k=1$ provide so-called {\it Wegner estimates}.
\item non-number-rigidity in dimension $\d \ge 3$ 
and the possibility of number rigidity in dimension $\d =2$  in \cite{thoma2}.
\end{itemize}

Less sharp separation estimates in the bulk (with temperature)
 are also given in Corollary~\ref{coromindist}.
Note that a sort of opposite question to that of obtaining charge overcrowding probabilities is to estimate the probability of a hole (without any point). This involves potential theory and balayage of measure and  is addressed in \cite{adhikari,charlierhole}, see also \cite{ghoshnishry} for a review of such questions.

\index{limit point process}
A corollary of Theorem \ref{theric} is to give the existence of a limiting point process to the shifted blown-up configuration $\{\XN'-x\}$, for {\it any} $ x\in \R^\d$ (and not only for $x$ in the bulk).
Besides this result, the existence of limiting point process was only known in the one-dimensional logarithmic case (with the sine-$\beta$ process \cite{killipnenciu,valkovirag}) and the two-dimensional Ginibre ensemble case for which $V$ is quadratic and $\beta=2$, converging to the Ginibre point process. Recently, \cite{boursier23a} proved the convergence for the general one-dimensional Riesz case for all $\beta$.  

There is a large statistical mechanics literature (see  \cite{glm1,glm2,martin} and references therein) from the 70's on sum rules and various relations for correlation functions of interacting particle systems, in particular Kirkwood-Salzbourg, BBGKY, KMS,  DLR equations. These can be shown to be equivalent relations in the case of regular interaction kernels but in the case of singular interactions like the Coulomb one, the existence of solutions to these hierarchies was not known. The existence of a limiting point process, though up to subsequences, is thus important toward putting these ideas on firmer ground.

\subsection{Discrepancy, separation and localization estimates by complex analysis}\label{sec53}
\index{discrepancy}
Obtaining bounds on the charge discrepancies (also called Beurling-Landau densities in the context of sampling and interpolation theory)  as defined in \eqref{defiD} has always been one of the important goals of the analysis of Coulomb systems. We described in Section \ref{sec:discrepancy} how the electric formulation rapidly provides discrepancy estimates, which are however suboptimal. 
In the series of works \cite{aoc,ameurromero,marcecaromero}, tools from complex analysis and interpolation theory  are used, and the role of the electric formulation is played by Lagrange interpolation polynomials and  reproducing kernel estimates, which are however restricted to the two-dimensional situation. We now present the Lagrange  interpolation polynomial approach originating in \cite{ameurloc}. 

For any $i$, introduce the Lagrange interpolant
\be\label{defLj} L_i(x)= \prod_{j\neq i}\frac{(x-x_j)}{(x_i-x_j)} \frac{e^{-N V(x)}}{e^{-NV(x_i)}},\ee
for which one observes that $L_i(x_i)=1$ and $L_i(x_j)=0$ for any $j\neq i$.
 Also,  since we are in the two-dimensional Coulomb case, by \eqref{HN} we may rewrite $L_i$ as 
 \be \label{formulaL} L_i(x)= \frac{e^{- \HN(\widehat{\XN} )} }{e^{-\HN(\XN)}}\ee
 where the points of $\widehat{\XN}$ are as above defined by 
 $$\hat x_j= \begin{cases} x_j & \text{if} \ j\neq i\\
 x & \text{if } \ j=i.\end{cases}$$
Thus, if $\XN$ are Fekete points (minimizing $\HN$) then  $L_i(x)\le 1$ for every $x$.  Controls of $L_i$ express rigidity of the configuration. 
For such controls, the crucial fact is  that, under the assumption that $V$ is analytic, the function $L_i$ is also analytic away from the points. Using complex analysis estimates based on Cauchy's formula, which replace maximum principle arguments, this allows to obtain regularity for $L_i$, for instance  control $\|L_i\|_{L^\infty}$ and $\|\nab L_i\|_{L^\infty}$  by weaker norms of $L_i$.
For instance once $\|L_i\|_{L^\infty} \le C $ and $\|\nab L_i \|_{L^\infty} \le C N^{1/2}$ is proven, one can easily deduce separation of Fekete points as in Theorem \ref{separation}: indeed, for $\XN$ Fekete points, by the above remarks, for $i\neq j$,  
$$1= |L_i(x_j)- L_i(x_i)| \le \|\nab L_i\|_{L^\infty} |x_i-x_j| \le C N^{1/2} |x_i-x_j|$$
which yields the separation lower bound 
$|x_i-x_j|\ge N^{-1/2}/C$.

Let us now give a rough idea of  how the Lagrange interpolation idea can be used in the situation with temperature, the goal still being  to show controls on norms of $L_i$,  encoding some  rigidity of the configurations.

It follows from \eqref{formulaL} and definition \eqref{gibbs} that for any test function $\varphi$,  
\begin{multline}
\Esp_{\PNbeta}\( \varphi(x, x_i) L_i(x)^{\beta}\)
= \frac{1}{\ZNbeta} \int_{(\R^2)^N} \varphi(x,x_i) \frac{e^{-\beta \HN(\widehat{\XN} )} }{e^{-\beta \HN(\XN)}} e^{-\beta \HN(\XN)} d\XN\\=  \frac{1}{\ZNbeta}\int_{(\R^2)^N} \varphi(x,x_i) e^{-\beta \HN(\widehat{\XN} )}   d\XN\end{multline}
and integrating in $x$  we obtain 
\begin{align*} \int \Esp_{\PNbeta}\( \varphi(x,x_i) L_i(x)^{\beta}\)
 dx& =   \frac{1}{\ZNbeta}\int_{(\R^2)^N\times \R^2 }  \varphi(x,x_i) e^{-\beta \HN(\widehat{\XN} )}   d\XN\, dx\\
  & =  \frac{1}{\ZNbeta} \int_{(\R^2)^N\times \R^2 }\varphi(x,x_i) e^{-\beta \HN(\widehat{\XN} )}   d\widehat{\XN}\, dx_i
  \\
  & =    \int \varphi(x ,y) d\rho_N^{(1)}(x) dy \end{align*}
where $\rho_N^{(1)}$ is the one-point correlation function as defined in \eqref{defrhok}.

 Applying for instance with $\varphi(x,y)= \indic_{A}(y)$ one obtains
\be\label{indica} \Esp_{\PNbeta} \( \indic_{A}(x_i) \|L_i\|_{L^{\beta}(\R^2)}^{\beta}\)=|A|,\ee 
where $|A|$ denotes the measure of the set.

We now give an example of a maximum-principe based lemma from \cite{ameurloc}.
\begin{lem}\label{lemameur}Let $q$ be a holomorphic polynomial with degree $\le N-1$, then, letting $f=q e^{-NV}$ we have
\be \label{premam}
f(x)\le \|f\|_{L^\infty} e^{-N\zeta(x)}\ee and, letting $s > \max_{\overline{U}}(\Delta V)$ where $U$ is some bounded open neighborhood of $ \Sigma$, we have for any $x \in U$, 
\be\label{deuxam}|f(x)|^\beta\le e^{C s\beta} \dashint_{B(x, \frac1{\sqrt{N}}) } |f|^{\beta}\ee
where $\zeta$ is the function of \eqref{defzeta}, and $C$ depends only on $U$.
\end{lem}
\begin{proof}
Let us start with \eqref{premam}.
We let $u= \frac{1}{N}\log (\frac{|q|}{\|f\|_{L^\infty}})$ and  remark that, since the logarithmic of the modulus of a holomorphic function is harmonic,  $u$ is harmonic, hence
$$\Delta (u+h^{\meseq}-c) =-\cd \meseq \le 0$$
where $h^{\meseq}= \g* \meseq$ and  $c$ is the constant in \eqref{defzeta}. 
Moreover,  by the assumption on the degree of $q$, $u(x)\ge \log |x| +O(1)$ as $|x|\to \infty$, hence, since $h^{\meseq}$ behaves like $-\log |x|$ at $\infty$, we have 
$u+h^{\meseq}-c \ge O(1)$.  
But a superharmonic nonnegative function in the plane is constant by the maximum principle. So we have 
$u+h^{\meseq}-c = cst$. 
Moreover by definition of $f$ we have 
$q e^{-NV}\le \|f\|_{L^\infty} $ hence after taking the log of the modulus of both sides, $u \le V$ and thus $u+h^{\meseq}-c \le \zeta$  by \eqref{defzeta}. In particular $u+h^{\meseq}-c\le 0$ in $\omega$   so the constant above must be $\le 0$. It follows that 
$u+h^{\meseq}-c\le 0$ hence $u\le V- \zeta$,  hence the result  \eqref{premam} after taking the exponential.

For the second result \eqref{deuxam}, let $F(x)= |f(x)|^{\beta} e^{\hal s\beta |x|^2}$.
By definition of $f$ and choice of $s$, we have 
$$\Delta \log F(x)\ge -  \beta N \Delta V(x)+s\beta N\ge 0.$$
Using that $\Delta \log F= \frac{1}{F}\Delta F-\frac{1}{F^2} |\nab F|^2$, we find that $F$ is subharmonic in $U$. By the mean-value inequality for subharmonic functions it follows that for $N$ large enough 
$$F(x)\le \dashint_{B(x, N^{-1/\d})}F\le   e^{C s\beta} \dashint_{B(x,N^{-1/\d})} |f|^\beta $$ which implies the result.

\end{proof}
Applying this lemma to $q(x)=\frac{\prod_{j\neq i} (x-x_j)}{\prod_{j\neq i} (x_i-x_j)e^{-NV(x_j)}}$ and returning to \eqref{defLj} one immediately deduces that 
\begin{coro}
\label{coroameur}  Under the same assumptions, we have 
\be\label{coroa1}|L_i(x)|\le \|L_i\|_{L^\infty} e^{-N\zeta(x)}\ee
and for any $x \in U$, 
\be\label{coroa2}|L_i(x)|^\beta \le e^{Cs\beta}N\int_{B(x, \frac1{\sqrt{N}} )} |L_i|^\beta.\ee
\end{coro}

The relation \eqref{indica} allows to control the $L^\beta(\R^2)$ norm of $L_i$ with large probability: it suffices to find a bounded set $A$ that contains all the points except with small probability (this is provided by a first, weak, confinement), then argue that \eqref{indica} yields
$$\PNbeta(\|L_i\|_{L^\beta} >\lambda ) \le \PNbeta(\|L_i\|_{L^\beta} >\lambda \indic_{A}(x_i)) + \PNbeta (\indic_{A^c}(x_i))$$
with both terms in the right-hand side being small if $\lambda$ is large.
Then, once $\|L_i\|_{L^\beta}$ is bounded, up to a small probability event, \eqref{coroa2} allows to upgrade this into an $L^\infty$ bound $\|L_i\|_{L^\infty} \le C N^{\frac{1}{\beta}}$. We can then reinsert this into \eqref{coroa1} to obtain 
$$|L_i(x)|\le C e^{-N\zeta(x)}  N^{\frac{1}{\beta}}.$$
But since $L_i(x_i)=1$, taking the logarithm we deduce that for every $i$
$$\zeta(x_i)\le \frac{1}{\beta}\frac{\log N}{N}+ \frac{C}{N}$$ up to a small probability event. Together with \eqref{assumpV5} it provides the following
 strong localization result. 
\begin{theo}[Strong localization, \cite{ameurloc}]\label{thameurloc} Assume $\d=2$ and $\s=0$.
The points of $\XN$ belong to 
$$\left\{x, \dist(x, \Sigma) \le C \sqrt{\frac{\log N}{\beta N}}\right\}$$
except on an event with probability $o_N(1)$.\end{theo}
We refer to \cite{ameurloc} for a more general and precise statement which contains correction terms and is valid as long as $\beta \gg \frac{\log N}{N}$. Such results were known earlier for the $\d=1$ logarithmic case, at least if $V$ is quadratic. 

The same starting point (controls of $L^\beta$ norms of $L_i$, except with small probability) is used in \cite{ameurromero} and \cite{marcecaromero} to obtain strong discrepancy estimates, still in the two-dimensional Coulomb case. One may say that in these works, Lagrange interpolants are combined with   reproducing kernel representations and techniques from sampling and interpolation and spectral analysis of Toeplitz operators, while in \cite{thoma} this is replaced by isotropic averaging combined with 
 the electric formulation to deduce  control on  discrepancies.

 In \cite{marcecaromero} these techniques are pushed  to obtain  a new "freezing" regime for $\beta \ge c\log N$. More precisely, it is shown,  that when $\beta\ge c \log N$,   the orders  of  the separation and discrepancy are the same as that of energy minimizers (the $\beta = \infty$ case), while for $\beta$ of order $1$, they are known to fluctuate more   \cite{ledouxrider} while the point separation is not bounded below independently of $N$ (see for instance \eqref{separationthoma}).  
 
 In \cite{ameurmarcecaromero} the same results are shown in the 
the one-dimensional logarithmic case $\s=0$ in the case of quadratic $V$. Using known discrepancy and fluctuation results, the authors can in addition show that the freezing  regime is exactly   $\beta\ge c \log N$.

\chapter[Commutator estimate and dynamics]{The commutator estimate and application to dynamics}
\label{chap:commutator}

We continue here our study of the modulated energy $\F_N$ of Chapter \ref{chap:nextorder}, and make a detour through the study of dynamics of Coulomb and Riesz systems, in particular to the question of deriving mean-field limits for dynamics of the type \eqref{noise1}, \eqref{noise2} (with possibly $\beta=\infty$) via the \textit{modulated energy method}.

At the core of the method  is a functional inequality on $\F_N$  that we call  a commutator estimate, and which will  also be  important in Chapter \ref{chapclt} for studying fluctuations of Coulomb gases. 

We start by presenting the functional inequality in Section \ref{sec6.1}, and then turn to the dynamics in Section \ref{sec:dynamics}. Again we consider throughout the interaction range \eqref{riesz}.

\section{The functional inequality}\label{sec6.1}
\index{functional inequality}

For the questions mentioned above (dynamics, fluctuations), we need to  consider how $\F_N$ varies when the points are transported by a perturbation of identity $\Phi_t:=\id + t v(x)$, while $\mu$ is pushed-forward by the same map $\Phi_t$.
We  easily observe that 
\begin{multline}\label{13}
\frac{d}{dt}\Big|_{t=0} \F_N\( (\id + tv)^{{\oplus} N} (\ux_N), (\id + tv)\# \mu\)\\
=
\frac12\int_{(\R^\d)^2\backslash \triangle} (v(x)-v(y)) \cdot \nabla \g(x-y) 
d\Big(\sum_{i=1}^N\delta_{x_i} - N \mu\Big)^{\otimes 2}(x,y),
\end{multline}
where $(\id + tv)^{\oplus N}(\XN)= (x_1+tv(x_1), \dots, x_N+tv(x_N))$. More generally, 
 for any $n\ge 1$, \begin{equation}\label{15comm}
 \frac{d^n}{dt^n}\Big|_{t=0} \F_N( (\id+ tv)^{\oplus N} (\ux_N), (\id + tv)\# \mu)= \Ani_n(\XN, \mu,v ),\end{equation}
 where we let 
 \begin{equation}\label{16}
 \Ani_n(\XN, \mu,v):= \frac12\int_{(\R^\d)^2\setminus \triangle} 
\nabla^{\otimes n} \g(x-y):  (v(x)-v(y))^{\otimes n}  
d\Big(\sum_{i=1}^N\delta_{x_i} - N \mu\Big)^{\otimes 2}(x,y),
\end{equation}
where $:$ denotes the inner product between the tensors. We will discuss the control of such next order quantities in Section \ref{sec:nextordercomm} below.

The functional inequality we were referring to shows that $\Ani_1$, the first variation of $\F_N$ along the transport by a Lipschitz vector field $v$ is controlled by $\F_N$ itself. It first appeared in \cite{Serfaty2020} after a partial result in \cite{duerinckx}. It was then recognized in \cite{Rosenzweig2021spv} that it could be seen as a commutator estimate and thanks to this point of view it was  generalized in \cite{NRS2021} to a broader class of interactions. 
We now present the sharp version of the estimate as obtained in \cite{RosenSer2023}.  The notation used is that of Section~\ref{secloc}, in particular $\F_N^{\Omega}$ corresponds to the modulated energy localized in a subset $\Omega$. The estimate is {\it localized} in the sense that if the vector field $v$ is localized in $\Omega$, say a small region of $\R^\d$, then the variation of energy is bounded in terms of the localized energy $\F_N^{\Omega}$ only. This is unlike typical commutator estimates found in the harmonic analysis or PDE literature, and it will be essential for the analysis of fluctuations of Coulomb/Riesz gases at mesoscales.

\index{commutator estimates}
\begin{theo}[Sharp commutator estimate]\label{thm:FI}
There exists a constant $C>0$ depending only $\d$ and $\s$ such that the following holds. Let $\mu \in  L^\infty(\R^\d)$ be a probability density satisfying
\eqref{condmupourFN}. Let    $v:\R^\d\rightarrow\R^\d$  be a  Lipschitz vector field and $\Omega$ be a closed set containing a $3\lambda$-neighborhood of $\supp\, \nab v$ where $\lambda= (N \|\mu\|_{L^\infty})^{-1/\d} <1$.\footnote{In \cite{RosenSer2023}, we present a slightly more precise version where the definition of $\lambda$ is instead $(N\|\mu\|_{L^\infty(\Omega)} )^{-1/\d}$.}  For any pairwise distinct configuration $\XN \in (\R^\d)^N$, it holds that
\begin{multline}\label{main1}
\left|\int_{(\R^\d)^2\setminus\triangle}(v(x)-v(y))\cdot\nabla\g(x-y)d\Big(\sum_{i=1}^N\delta_{x_i} - N \mu\Big)^{\otimes 2}(x,y)\right| \\
\leq C\|\nabla v\|_{L^\infty}\Bigg( \F_N^{\Omega} (\XN,\mu) -  {\# I_\Omega\Big(\frac{\log \lambda}{2}\Big) }\indic_{\s=0} + {C \|\mu\|_{L^\infty}^{\frac{\s}{\d}} \# I_\Omega N^{\frac{\s}{\d}} }\Bigg),
\end{multline}
where $\F^\Omega_N$ is as in \eqref{Glocal}.
\end{theo}
In a first pass, the reader may simply take $\Omega=\R^\d$ and $\F_N^\Omega=\F_N$, which will suffice for the dynamics.

As seen in \cite{NRS2021}, \eqref{main1} can be seen as a commutator estimate, because, letting $f= 
\sum_{i=1}^N\delta_{x_i} - N \mu$ and ignoring the excision of the diagonal, we may rewrite the integral in the left-hand side of \eqref{main1} as 
$$\int v \cdot \nab (\g* f)-  \g *(\div  (v f) )= \langle f, \left[ v, \frac{\nab}{(-\Delta)^{\frac{\d-\s}{2}}}\right] f\rangle_{L^2}.$$ Although commutators of this type have been studied in the harmonic analysis literature (see \cite{SeegerStreetSmart} and references therein), the estimates there only apply to divergence-free vector fields, are not localizable, and do not quite provide what we need. 
In \cite{NRS2021}, this general commutator point of view was exploited and a  proof that covers  more general interactions $\g$ that have singularities of Riesz type (but without being necessarily  exactly the Riesz kernel)  for any $ \s \in [0, \d)$, was provided.

Here, we give the simplest proof, which works when considering all the exact Riesz kernels \eqref{riesz}.

\subsection{Proof outline}
The proof relies crucially on the electric formulation in extended space (we use the same notation as in Section \ref{secrieszcase}) and the notion of {\it stress-energy tensor},  a standard notion in mechanics and calculus of variations. 
\cite{stress-energy tensor}
\begin{defi}[Stress-energy tensor]
Given two (regular enough) distributions $f $ and $w$, we define the cross stress-energy tensor  as 
\begin{equation}\label{stresstensor}
[\nab h^f,\nab h^w]_{ij} := \yg\(\p_ih^f\p_jh^w +\p_jh^f\p_ih^w-\nabla h^f\cdot\nabla h^w\delta_{ij}\),
\end{equation}where $\delta_{ij}$ is the Kronecker symbol and  $h^f = \g*f$ is naturally extended to $\R^{\d+\k}=\{(x,y), x\in \R^\d, y \in \R\}$ if $\k=1$. \end{defi}

Strictly speaking, the stress-energy tensor associated to the potential $h^f$ (naturally extended to a function on $\R^{\d+\k}$)  is the quantity $[\nab h^f,\nab h^f]$, but it is convenient, as in \cite{Serfaty2020}, to view it as a bilinear function.

Our  interest in the stress-energy tensor stems from the relation 
\be \label{divT} \div [\nab h^f,\nab h^f]= \nab h^f \div(\yg \nab h^f)= -\cds \nabla h^f f ,\ee
which is true by direct computation if $f$ is sufficiently regular.  Here the divergence of the tensor is a vector, whose components are equal to the divergence of the rows/columns of the tensor. 
This way, nonlinear terms of the form $ \nab h^f f$ can be rewritten in divergence form. Moreover by definition \eqref{stresstensor}, the term $[\nab h^f,\nab h^f]$  is pointwise controlled  by the energy density: 
\be\label{pointwisec} |[\nab h^f,\nab  h^f]|\le \yg |\nab h^f|^2.\ee

Arguing formally, if one neglects the issue of the diagonal terms in both sides of \eqref{main1},  after desymmetrizing the left-hand side of \eqref{main1} and using that $\g$ is even, one can rewrite it  as 
$$\iint_{ \triangle^c}v(x) \cdot \nab \g(x-y) d f(x) df(y) - \iint_{\triangle^c} v(y) \cdot \nab \g(x-y) df(x) df(y) =
2\int_{\R^\d} v\cdot \nab h^f  df,$$
with $f= \sum_{i=1}^N \delta_{x_i}-1$. Using \eqref{divT} and an integration by parts, this can be reinterpreted as 
$\frac{2}{\cds}\int \nab v: [\nab h^f, \nab h^f ] $, and thanks to the pointwise control \eqref{pointwisec}, we have bounded the left-hand side of \eqref{main1} as follows:
\be \label{resulformel} \iint_{\R^\d\times \R^\d} (v(x)-v(y) ) \cdot \nab \g(x-y) df(x) df(y) \le
C \|\nab v\|_{L^\infty} \int_{\R^{\d+\k} } \yg |\nab h^f|^2,\ee where,  in view of \eqref{rewritF}, 
we formally recognize in the right-hand side the electric rewriting of $\F_N(\XN, \mu)$. Of course, the main problem is that this neglects the issue of the diagonal, and that the estimate needs to be properly \textit{renormalized} on both sides. The truncation method presented in Chapter \ref{chap:nextorder}, as well as all the properties demonstrated there, provide the right tool to do it. More specifically, the key  is  to take the \textit{point-dependent} truncation parameters $\rr_i$ (minimal distances) as defined in \eqref{defri} and to apply \eqref{resulformel} to $f= \sum_{i=1}^N \delta_{x_i}^{(\rr_i)}-N\mu$ instead. Then, the right-hand side equals  $\int_{\R^{\d+\k}} \yg |\nab h_{N,\rr}|^2$, which, crucially,  we are able to control by $\F_N$ itself (without needing to substract off a renormalization term!) thanks to \eqref{bornehnr}.
 Then, what remains to be done, which is the most delicate part of the proof,  is to evaluate the renormalization error of the left-hand side of \eqref{resulformel} (i.e. the error made when replacing $ \sum_{i=1}^N \delta_{x_i}$ by $ \sum_{i=1}^N \delta_{x_i}^{(\rr_i)}$)
  in terms of the $\rr_i$ and $\sum_i\g(\rr_i)$, which we also control thanks to \eqref{bgr}.
  
To make this strategy rigorous, let us start by    reformulating a useful identity from \cite[Lemma 4.3]{Serfaty2020}.

\begin{lem}\label{lem:commfost}
Let $v:\R^{\d}\rightarrow\R^{\d}$ be a Lipschitz vector field. Let us extend it trivially into a vector-field on $\R^{\d+\k}$, still denoted $v$,  by letting it depend only on the first $\d$ coordinates, and have vanishing last component  (we note that the extension has the same Lipschitz norm as the original vector field).
 For any test functions $f$ and $w$ in the Schwartz class of $\R^\d$, it holds that  \be
 \label{eq:commfost}
\int_{(\R^{\d})^2}(v(x)-v(y))\cdot\nabla\g(x-y)df(x)dw(y) = \frac{1}{\cds}\int_{\R^{\d+\k}}\nabla v: [\nab h^f,\nab h^w].
\ee
\end{lem}

Applying \eqref{eq:commfost} to $w=f$ and desymmetrizing the left-hand side, we obtain 
\be  \int_{\R^{\d+\k}} \nab v : [\nab h^f,\nab h^f]= 2\cds \int_{(\R^{\d})^2} v(x) \cdot \nab \g (x-y) df(x) df(y) \ee
where the right-hand side also equals  $2\cds\int_{\R^{\d}} v  \cdot \nab h^f df$ by definition of $h^f$.
Hence, we have obtained  the rigorous version of the  claimed relation \eqref{divT}.
 
 \subsection{Proof of Theorem \ref{thm:FI}}
First, we may assume that the points of $\XN$ are pairwise distinct, otherwise the right-hand side is $+\infty$.
Let us consider  $\vec{\eta}$ such that for every $i$, $\eta_i \le \rr_i$. 
Let 
\begin{equation}\label{defhni}
h_N^i (x)\coloneqq h_N(x)-\g(x-x_i),
\end{equation}
be the potential in $\R^\d$ generated by the configuration with $x_i$ removed and $\mu$, naturally extended to $\R^{\d+\k}$,
and observe in view of \eqref{formu23} that if the balls $\{B(x_i, \eta_i)\}_{i=1}^N$ are pairwise disjoint, then
\begin{equation}\label{eq:carHneta}
\nabla h_{N,\vec{\eta}}= \nab h_N^i  +\nab \g_{\eta_i} =\begin{cases} \nabla h_N & {\text{outside} \ \cup_{i=1}^N B(x_i, \eta_i)}\\
\nabla h_N^i & {\text{in}  \ B(x_i, \eta_i)}.\end{cases}
\end{equation}
From this property, it follows that
\begin{equation}\label{eq:energballs}
\int_{\R^{\d+\k} }\yg |\nabla h_{N, \vec{\eta}}|^2 = \int_{\R^{\d+\k} \setminus \cup_i B(x_i, \eta_i)} \yg
|\nabla h_{N} |^2 +\sum_{i=1}^N \int_{B(x_i, \eta_i)} \yg |\nabla h_N^i|^2.
\end{equation}

We now turn to the proof of the estimate \eqref{main1}. 

{\bf Step 1. Rewriting the left-hand side.}
 Desymmetrize the left-hand side of \eqref{main1}, we may rewrite it as 
\begin{align}
 I\coloneqq & \int_{(\R^\d)^2\setminus\triangle} ( v(x)-v(y))\cdot \nabla\g(x-y)d\Big( \sum_{i=1}^N\delta_{x_i} - N\mu\Big)^{\otimes 2} (x,y) \notag\\
\notag & =2 \sum_{i=1}^N  \int_{\R^{\d+\k}}  v(x_i) \cdot \nabla \g(x_i-x)d\Big(  \sum_{ j\neq i} \delta_{x_j} - N \mu\Big) (x) \\ & - 2N \int_{\R^{\d+\k}} v(x) \cdot \nabla \g(x-y) d\mu(x)d\Big(\sum_{i=1}^N \delta_{x_i} - N\mu\Big)(y) \notag\\
 & = 2\sum_{i=1}^N   v(x_i) \cdot \nabla h_N^i (x_i) - 2 N\int_{\R^{\d+\k}} v\cdot \nabla h_N d\mu. \label{eq:FIo1pre}
\end{align}
Using \eqref{eq:carHneta} and 
$h_N= h_{N, \vec{\eta}} +\sum_{i=1}^N\f_{\eta_i} (\cdot-x_i)$,
we next decompose $I$ as $\Term_1 + \Term_2 +\Term_3$, where 
\begin{align}\label{eq:FIo1T1}
 \Term_1 \coloneqq 2 \int_{\R^{\d+\k}} v\cdot \nabla h_{N, \vec{\eta}} \, d \Big( \sum_{i=1}^N \delta_{x_i}^{(\eta_i)} - N\mu\Big),
\end{align}
\begin{multline}\label{eq:FIo1T2}
\Term_2 \coloneqq  2 \sum_{i=1}^N \int_{\R^{\d+\k}} (v(x_i) - v) \cdot \nabla h_N^i d\delta_{x_i}^{(\eta_i)} \\
+2 \sum_{i=1}^N \int_{\R^{\d+\k}} (v(x_i)-v) \cdot \nabla \g_{\eta_i} (\cdot-x_i) d\delta_{x_i}^{(\eta_i)}  + 2N\sum_{i=1}^N \int_{\R^{\d+\k}} (v(x_i)-v) \cdot \nabla \f_{\eta_i} (\cdot-x_i) d\mu,
\end{multline}
and
\begin{multline}\label{eq:FIo1T3}
\Term_3 \coloneqq 2 \sum_{i=1}^N \int_{\R^{\d+\k}} v(x_i) \cdot \nabla h_N^i d\Big(\delta_{x_i} - \delta_{x_i}^{(\eta_i)}\Big) 
 -2 \sum_{i=1}^N\int_{\R^{\d+\k}} v(x_i) \cdot \nabla \g_{\eta_i} (\cdot -x_i) d \delta_{x_i}^{(\eta_i)} \\
- 2N  \sum_{i=1}^N\int_{\R^{\d+\k}} v(x_i) \cdot \nabla \f_{\eta_i} (\cdot-x_i) d\mu.
\end{multline}

{\bf Step 2. Showing that $\Term_3=0$.}
By  definition \eqref{defhni} of $h_N^i$, we may write
\begin{multline}\label{eq:FIo1T3pre}
\Term_3= 2\sum_{i=1}^N \sum_{j\neq i} \int_{\R^{\d+\k}} v(x_i) \cdot \nabla \g(\cdot-x_j) d\Big(\delta_{x_i} - \delta_{x_i}^{(\eta_i)}\Big)
\\
- 2N  \sum_{i=1}^N \int_{\R^{\d+\k}} v(x_i) \cdot \nabla \g(\cdot-y) d\mu(y) d\Big(\delta_{x_i} - \delta_{x_i}^{(\eta_i)}\Big) \\
- 2 \sum_{i=1}^N\int_{\R^{\d+\k}} v(x_i) \cdot \nabla \g_{\eta_i} (\cdot -x_i) d \delta_{x_i}^{(\eta_i)} 
- 2N  \sum_{i=1}^N\int_{\R^{\d+\k}} v(x_i) \cdot \nabla \f_{\eta_i} (\cdot-x_i) d\mu.
\end{multline}
Thanks to \eqref{eq:lemme} and \eqref{def:truncation}, we have for $i\ne j$,
\begin{align}
\int_{\R^{\d+\k}} \nabla \g(\cdot-x_j) d\Big( \delta_{x_i} - \delta_{x_i}^{(\eta_i)}\Big) =  \nabla \g(x_i-x_j)-  \nabla \g_{\eta_i} (x_i-x_j)= \nab \f_{\eta_i}(x_i-x_j).
\end{align}
The right-hand side vanishes because $ |x_i-x_j |>\eta_i$ for $j \neq i$ (by assumption that $\eta_i \le \rr_i$) and $\f_{\eta_i}$ vanishes outside of $B(0, \eta_i)$.  Thus, the first line of \eqref{eq:FIo1T3pre} vanishes. By the same reasoning, the second line of  \eqref{eq:FIo1T3pre} equals
\begin{align}
-2N \sum_{i=1}^N \int_{\R^{\d+\k}} v(x_i) \cdot \nabla \f_{\eta_i} (x_i-y)  d\mu(y),
\end{align}
thus it cancels with the last term on the third line of  \eqref{eq:FIo1T3pre}. It remains to show that for every~$i$,
\begin{align}
\int_{\R^{\d+\k}} v(x_i) \cdot \nabla \g_{\eta_i} (\cdot-x_i) d \delta_{x_i}^{(\eta_i)} =0.
\end{align}
This is true because by definition \eqref{defdeltaeta} and \eqref{divT}, 
\begin{align}
\nabla \g_{\eta_i} (\cdot-x_i) \delta_{x_i}^{(\eta_i)} &=-\frac{1}{\cds}  \nabla \g_{\eta_i} (\cdot-x_i) \div \left(|y|^\gamma \nabla \g_{\eta_i} (\cdot-x_i) \right)\notag\\
&= -\frac{1}{2\cds} \div[\nab \g_{\eta_i} (\cdot-x_i),\nab  \g_{\eta_i}(\cdot-x_i)].
\end{align}


{\bf Step 3. Estimating $\Term_1$.} Using Lemma \ref{lem:commfost} and that $\supp \, \nab v\subset \Omega$, we rewrite $\Term_1$ as
\begin{equation}
\Term_1 = \int_{\Omega\times \R^{\k}} \nabla v : [\nab h_{N,\vec{\eta}},\nab h_{N,\vec{\eta}}],
\end{equation}
where we use again the definition  \eqref{stresstensor}. It follows now from the pointwise bound \eqref{pointwisec} that 
\begin{equation}\label{eq:T1fin}
|\Term_1| \leq  C\|\nabla v\|_{L^\infty} \int_{\Omega\times\R^{\k}} \yg |\nabla h_{N,\vec{\eta}}|^2.
\end{equation}

{\bf Step 4. Estimating $\Term_2$.}  Note that  if $i$ is such that  $\dist(x_i,\supp\, \nab v)>\lambda$, the corresponding terms in $\Term_2$ vanish. For the remaining $i$, using the mean-value theorem on $v-v(x_i)$ and the explicit form of the probability measure $\delta_{x_j}^{(\eta_j)}$ given after Definition \ref{defi41}, we see that
\begin{multline}\label{eq:T21}
2 \sum_{i=1}^N \Big|\int_{\R^{\d+\k}} (v(x_i) - v) \cdot \nabla h_N^i \, d\delta_{x_i}^{(\eta_i)} \Big| \\
\le C\|\nab v\|_{L^\infty}\sum_{i , \dist(x_i,\supp\, \nab v) \le \lambda}\eta_i^{-\s}\int_{\pa B(x_i,\eta_i)} \yg |\nab h_N^i| d\mathcal{H}^{\d+\k-1},
\end{multline}
where $\pa B(x_i,\eta_i)$ is the sphere in $\R^{\d+\k}$ and $\mathcal{H}^{\d+\k-1}$ is the Hausdorff measure. Similarly,
\begin{align}\label{eq:T22}
2 \sum_{i=1}^N \Big|\int_{\R^{\d+\k}} (v-v(x_i)) \cdot \nabla \g_{\eta_i} (\cdot-x_i) d\delta_{x_i}^{(\eta_i)}\Big| \le C\|\nab v\|_{L^\infty}\sum_{i , \dist(x_i,\supp\, \nab v) \le \lambda}\eta_i^{-\s}.
\end{align}
Finally, recalling   that $\f_\eta$ is supported in $B(0, \eta)$, we have
\begin{align}\label{eq:T23}
2 \sum_{i=1}^N \Big|\int_{\R^{\d+\k}}(v- v(x_i)) \cdot \nabla \f_{\eta_i} (\cdot-x_i) d\mu\Big| \le C\|\nab v\|_{L^\infty}\|\mu\|_{L^\infty}\sum_{i , \dist(x_i,\supp \nab v) \le \lambda} \eta_i^{\d-\s}.
\end{align}
Combining \eqref{eq:T21}, \eqref{eq:T22}, \eqref{eq:T23} yields
\begin{multline}\label{eq:T2fin}
|\Term_2|\\ \le C\|\nab v\|_{L^\infty} \sum_{i , \dist(x_i,\supp\, \nab v) \le \lambda}\( \eta_i^{-\s} + N \|\mu\|_{L^\infty}\eta_i^{\d-\s}+\eta_i^{-\s}\int_{\pa  B(x_i,\eta_i)} \yg |\nab h_N^i| d\mathcal{H}^{\d+\k-1}\).
\end{multline}

{\bf Step 5. Conclusion.}
Combining the estimates \eqref{eq:T1fin} and  \eqref{eq:T2fin}, we have found that there exists a constant $C>0$ depending only on $\d,\s$, such that for every choice of $\vec{\eta}$ satisfying $\eta_i\leq\rr_i$, we have 
\begin{multline}\label{eq:prefin}
|I|\le C   \|\nabla v\|_{L^\infty} \Bigg[ \int_{{\Omega \times \R^{\k}}} |y|^{\gamma} |\nabla h_{N,\vec{\eta}}|^2 \\
\quad +
\sum_{i , \dist(x_i,\supp\, \nab v) \le \lambda}\( \eta_i^{-\s} + N \|\mu\|_{L^\infty}\eta_i^{\d-\s}+\eta_i^{-\s}\int_{\pa  B(x_i,\eta_i)} \yg |\nab h_N^i| d\mathcal{H}^{\d+\k-1}\) \Bigg]
\end{multline}
For each $t\in[\frac{1}{2},1]$, we apply this relation with $\eta_i = t\rr_i$ and then average both sides of the resulting inequality over $t\in [\frac12,1]$. The average of the last term on the right-hand side becomes 
\begin{equation}
\leq C \sum_{{i,\dist(x_i,\supp\, \nab v)\leq \lambda }}  \rr_i^{-\s-1}  \int_{B(x_i, \rr_i)}|y|^\gamma |\nabla h_N^i|.
\end{equation}
Applying  Cauchy-Schwarz's  inequality, we can also estimate 
\begin{equation}
\int_{B(x_i, \rr_i)}\yg |\nabla h_N^i| \leq  \(\int_{B(x_i, \rr_i)} |y|^\gamma|\nabla h_N^i|^2\)^{\frac12}\(\int_{B(0, \rr_i)} |y|^\gamma \)^{\frac12}.
\end{equation}
Since $\gamma>-1$ the last integral is convergent and bounded by  $C\rr_i^{\d+1+\gamma}$ if $\k=1$ or $C\rr_i^\d$ otherwise, which is always $C \rr_i^{\s+2}$ by \eqref{defgamma}. Using \eqref{eq:energballs} and Cauchy-Schwarz inequality, the average of the last term on the right-hand side of \eqref{eq:prefin} is thus bounded by 
\begin{align}
&C   \sum_{{i,\dist(x_i,\supp\, \nab v)\leq \lambda}}  \rr_i^{-\frac{\s}2}\Big(\int_{ B(x_i, \rr_i)}  |y|^\gamma|\nabla h_N^i|^2\Big)^{\frac12} \notag\\
&\leq C\Bigg(\sum_{{ i,\dist(x_i,\supp\, \nab v)\leq \lambda}} \rr_i^{-\s}\Bigg)^{\frac12} \Bigg(\int_{{\Omega \times \R^\k }}  |y|^\gamma|\nabla h_{N, \rr} |^2\Bigg)^{\frac12}\notag\\
&\leq C \Bigg( \sum_{{ i,\dist(x_i,\supp\, \nab v)\leq \lambda}}  \rr_i^{-\s} +\int_{{ \Omega }\times \R^{\k}}  |y|^\gamma|\nabla h_{N, \rr} |^2\Bigg).
\end{align}
Inserting this estimate into \eqref{eq:prefin}, we obtain, after averaging over $t\in [\hal, 1]$, 
\begin{multline}\label{eq:finapp}
|I|\le C   \|\nabla v\|_{L^\infty}\Bigg( \dashint_{t\in [\hal, 1]}\int_{{\Omega \times \R^{\k}}} \yg |\nabla h_{N,t\rr}|^2 (x)dx+ \int_{  \Omega \times \R^{\k}}\yg |\nabla h_{N,\rr}|^2 
\\ +\sum_{{ i, \dist(x_i,\supp \, \nab v)\leq \lambda }}  \rr_i^{-\s} 
+ N \|\mu\|_{L^\infty}  \sum_{{ i, \dist(x_i,\supp\,  \nab v)\leq \lambda }} \rr_i^{\d-\s}\Bigg). 
\end{multline}
 Recall from the statement of the proposition that $\Omega$ contains a closed $3\lambda$-neighborhood of $\supp \, \nab v$, so that the condition $\dist(x_i,\supp\, \nab v)\leq \lambda$ implies $x_i\in\Omega$ and $\dist(x_i,\pa\Omega)\ge 2\lambda$, thus for such points we have $\rr_i=\rrc_i$ in the notation \eqref{rrc}.  Using the estimates \eqref{eq:11}, \eqref{eq:14} on the right-hand side of \eqref{eq:finapp} and recalling $\rr_i\leq \lambda\le (\|\mu\|_{L^\infty} N)^{-\frac1\d} $, we conclude that 
\begin{equation}
|I|\leq C  \|\nabla v\|_{L^\infty} \Bigg( \F_N^{ \Omega} (\XN, \mu) - \Big(  \frac{\# I_\Omega  \log \lambda  }{2}\Big) \indic_{\s=0}  + C      \|\mu\|_{L^\infty}^{\frac{\s}{\d}} \# I_\Omega N^{ \frac{\s}{\d}}\Bigg),
\end{equation}
which completes the proof.

\subsection{Next order commutator estimates}\label{sec:nextordercomm}
In the study of fluctuations in Chapter \ref{chap:clt2}, we will need a second order version of the commutator estimate, controlling \eqref{15comm} with $n=2$ instead of \eqref{13}, in a localizable way.  This was first done in \cite{ls2} in the two-dimensional Coulomb case,  also subsequently in a nonlocalized way in  \cite{Rosenzweig2021spv},  and then revisited in \cite{Serfaty2020} in the Coulomb case, but the estimate is only sharp in the case $\d=2, \s=0$. 
In \cite{NRS2021}, second order commutator estimates  were also proven for all the Riesz cases, with suboptimal error terms. The proof relied on commutator estimates. Finally, estimates for all order $n$, with sharp error terms and which are localizable are obtained in \cite{RosenSer2023}.


We now state the  result from \cite{RosenSer2023}  valid for all \eqref{riesz}, and refer the reader to that paper for the proof, which is considerably more involved than that of Theorem \ref{thm:FI}.

\index{commutator estimates}
\begin{theo}[Sharp next-order commutator estimates \cite{RosenSer2023}] \label{prop:comparaison2} \mbox{}Let $\mu$ be a probability measure with a bounded  density, satisfying \eqref{condmupourFN}. Assume that $v:\R^\d\to\R^\d$ is smooth, that  $\Omega'$ is a ball of radius $\ell$ containing a $2\lambda$-neighboorhood of the support of $\nab v$ and $\Omega$  contains a $5\ell$-neighborhood of $\Omega'$, where $\ell >2 \lambda$. For any $n \ge 2$, we have for $N$ small enough,
\begin{multline}
\label{p42} |\Ani_n(\XN, \mu, v) | \\ \le   C
\Bigg( \sum_{p=0}^n(\ell \|\nab^2 v \|_{L^\infty})^p 
 \sum_{\substack{1\leq c_1,\ldots,c_{n-p} \\ c_i \ \mathrm{integer}\\  c_1+\cdots+c_{n-p} \le 2 n}} \lambda^{-(n-p)+\sum_{k=1}^p c_{n-k}} \|\nabla^{\otimes c_1} v\|_{L^\infty}\cdots\|\nabla^{\otimes c_{n-p}} v\|_{L^\infty}  
 \Bigg) 
\\
\times \Bigg(\F_N^{\Omega}(\XN,\mu)+ \#I_\Omega \frac{\log( N\|\mu\|_{L^\infty})  }{2\d } \indic_{\s=0} + C   \#I_\Omega \|\mu\|_{L^\infty}^{\frac\s\d} N^{\frac\s\d}  \Bigg).
\end{multline}
where  $C>0$  depends only on  $\d $, $ \s$ and $n$.
\end{theo}
The factor in front of the energy has the natural scaling in $\ell$. For instance, in a typical situation of application, $\|\nab^m v\|_{L^\infty}\le M \ell^{-m}$,  where $\ell$ is the lengthscale of variation of $v$. 
In that situation, using that $\ell \ge \lambda$,  the theorem yields
$$|\Ani_n(\XN, \mu, v) | \\ \le   C \ell^{-n}  \Big(\F_N^{\Omega}(\XN,\mu)+ \#I_\Omega \frac{\log( N\|\mu\|_{L^\infty})  }{2\d } \indic_{\s=0} + C   \#I_\Omega \|\mu\|_{L^\infty}^{\frac\s\d} N^{\frac\s\d}  \Big),$$
where $C$ depends on $M$.

We may note that  for the one-dimensional Coulomb case, it is immediate from the definition \eqref{16} that $\Ani_2\equiv 0$, since $\g''(x-y)= \delta_0(x-y)$.

\section{Application to dynamics}\label{sec:dynamics} \index{dynamics} \index{mean-field limits}
\subsection{Mean-field limits}
As mentioned above, the  commutator estimate has found two important applications, one is to central limit theorems for fluctuations in Coulomb gases, which we will see in Chapter~\ref{chapclt} and \ref{chap:clt2}, one is to dynamics, via the modulated energy method.

The question of proving mean-field limits for interacting systems is a classical one in statistical physics, and has attracted much attention in mathematics. We refer the reader to the surveys \cite{CD2021,jabinreview,JW2017survey,Golse2022ln} and references therein.

The question is to understand the limit as $N\to \infty$ of the empirical measure
 \be\label{munt}
\mu_N^t:= \frac1N \sum_{i=1}^N \delta_{x_i^t}\ee  associated to a solution $X_N^t:=(x_1^t, \dots, x_N^t)$ of a system of ODEs of the form 
\be \label{mfgf}
\left\{\begin{array}{l}
\dot x_i=-\displaystyle\frac{1}{N} \mathbb{M} \nab_{x_i}\mathcal H_N(x_1, \dots, x_N), \quad i=1, \dots, N\\ [3mm]
x_i(0)=x_i^0\end{array}\right.\ee
Here $\mathbb{M} $ is a fixed matrix that satisfies $\langle \mathbb{M}\xi , \xi \rangle \ge 0$. The case $\mathbb{M}=\id$ corresponds to the gradient-flow dynamics. The case where $\mathbb{M} $ is antisymmetric corresponds to conservative dynamics, including for instance the important ``point-vortex system" in fluids. The energy $\mathcal H_N$ is of the form  \eqref{HN} but for possibly general interaction potential $\g$ (more general evolutions are also considered). 
Studying the same evolutions with added noise or diffusion
\be \label{mfgfnoise}
d x_i=-\displaystyle\frac{1}{N} \nab_{x_i}\mathcal H_N(x_1, \dots, x_N) dt + \sqrt{\frac2 \theta} dW_i, \ee or 
\be \label{consnoise}
d x_i=-\displaystyle\frac{1}{N} \mathbb{M} \nab_{x_i}\mathcal H_N(x_1, \dots, x_N)dt + \sqrt{\frac2 \theta} dW_i, \ee
with $W_i$ being $N$ independent Brownian motions and $\theta> 0$ an inverse temperature,
 is also very interesting, as discussed in the introduction.

If the points $x_i^0$, which themselves depend on $N$, are such that $\mu_N^0$   converges  to some regular measure $\mu^0$, then a formal derivation leads to expecting that for $t>0$, $\mu_N^t$ converges to the solution of the Cauchy problem with initial data $\mu^0$ for the limiting continuity equation 
\be \label{limeq}
\partial_t \mu= \div( (\mathbb{M}\nab \g)*\mu) \mu), \ee
or in the case with noise 
\be \label{limeqnoise}
\partial_t \mu= \div( (\mathbb{M}\nab \g)*\mu) \mu)+\frac1 \theta \Delta \mu. \ee

Proving the convergence in law of the empirical measure is more or less equivalent (see \cite{CD2021}) to proving \textit{propagation of chaos} (a notion introduced by  Kac in kinetic theory). One says that there is propagation of chaos if, when particles are initially distributed according to a probability density $f_N^0(x_1, \dots, x_N)$  in a tensorized form 
$$f_N^0(x_1, \dots, x_N)= \mu^0(x_1)\dots \mu^0(x_N),$$ (in other words, initial particles positions are iid)
then  at later times, $f_N^t$ is {\it approximately} in tensorized form $\mu^t(x_1)\dots \mu^t(x_N)$.
The same questions also apply to second-order evolutions of the form \eqref{noise3}, which lead to Vlasov-type kinetic equations.

To  rigorously establish  convergence, various methods have been put forward: classical trajectorial methods \cite{mckean,sznitman} for the case with noise,  relative entropy method which works well in the presence of noise (see \cite{jabinwang,breschjabinsoler,BDJ} and \cite{lacker}, and the related method in \cite{HCR}, for the best results to date), and methods that rely on finding a good metric, such as a Wasserstein distance, to measure the distance from $\mu_N^t $ to $\mu^t$ and its time evolution.  Such methods have allowed to treat some singular interactions, but not as singular as the Coulomb, let alone super-Coulomb Riesz,  interaction, which remained open until recently.

 The crucial point is that the modulated energy $\F_N(\XN, \mu)$ acts as a good distance to measure the distance of the empirical measure to the expected limit, or to quantify a \textit{weak-strong uniqueness} result at the level of the mean-field equation \eqref{limeq}.
 This modulated energy method was introduced in \cite{Serfaty2017} in the closely related context of dynamics of Ginzburg-Landau vortices in \eqref{glhf} -- \eqref{gls}. It was then first adapted to the context of Coulomb and Riesz gases in dimension $\d\le 2$ in \cite{duerinckx}. The full Coulomb and Riesz cases with \eqref{rieszgene} without noise was finally treated in \cite{Serfaty2020} thanks to the commutator estimate of Theorem~\ref{thm:FI}. It gives  quantitative convergence of the empirical measure to the solution of \eqref{limeq}  in modulated energy. Let us state a main result.
 \begin{theo}[Mean-field convergence for Coulomb and Riesz first order dynamics]\label{theoMF}
Assume $\d-2\le \s<\d$. Assume $\mu^0$ is a probability density satisfying 
\be \label{condmuo}\int_{\R^\d} |\g_-| d\mu^0 <\infty,\ee and
 is such that $\mu^t$, solution of \eqref{limeq}  with initial data $\mu^0$, satisfies on some interval $[0,T]$,
\be\label{conditionsevol}
\sup_{t\in [0,T]}\|\mu^t\|_{L^\infty}<\infty, \quad \sup_{t\in [0,T]}\|\nab^{\otimes 2} \g* {\mu^t}\|_{L^\infty}<\infty,
\ee
then, letting $\XN^t$ be the solution to \eqref{mfgf}, 
 if $ \F_N(\XN^0, \mu^0)+ N \( \frac{\log N}{2\d}\) \indic_{\s=0}=o(N^2)$, we have
$$\frac{1}{N}\sum_{i=1}^N \delta_{x_i^t} \rightharpoonup \mu^t\quad \text{for every  } t \in [0,T],$$
with convergence in modulated energy or by Corollary \ref{coro453} in $H^{-\sigma}(\R^\d)$. More precisely we have  that \eqref{controlgronwall} holds in $[0,T]$.
\end{theo}
The condition \eqref{conditionsevol} boils down to a question of regularity of the solution to \eqref{limeq} and, if such a regularity is known, can then be reduced to a condition on the initial distribution $\mu^0$. We refer to \cite{Serfaty2020} for a discussion of the regularity results proven in the literature.

Examining the proof, particularly the last application of Gronwall's lemma that yields \eqref{controlgronwall}, thanks to the fact that the commutator estimate \eqref{main1} is made explicit in its $\mu$-dependence,  we obtain a uniform in time rate of convergence provided $\int_0^\infty \|\nab \g *{\mu^s} \|_{L^\infty} ds<\infty$, i.e.~provided the solution $\mu^t$ has sufficient decay as $t \to \infty$, which is a purely PDE question. We  know how to prove this in the setting of the torus \cite{ChodronRosenSer2022}, and also   in the case of added diffusion in the sub-Coulomb Riesz case in the whole space \cite{RosenSer2021}.
In both instances this   strong decay of the solution allows to obtain uniform in time convergence.
\smallskip 

The modulated energy method for dynamics was later improved, allowing to treat more general interactions \cite{NRS2021}, or relaxing assumptions on the regularity of the solution  \cite{Rosenzweig2022, Rosenzweig2021spv}.  Other applications to dynamics, including quantum dynamics, all relying on the commutator estimate Theorem \ref{thm:FI},  have been given in \cite{HkI2021,Rosenzweig2021ne, Porat2022,GP2021,menard,RosenSerlake}.

\subsection{Weak-strong uniqueness proof}
Let us now present the short proof of the weak-strong uniqueness principle for \eqref{limeq} as it will be a model for the main proof. We focus on the dissipative case $\mathbb{M}= \id$  (the conservative one is an easy adaptation) and on the Coulomb case for simplicity, the Riesz case is the same, using the extension procedure and adding the $\yg$ weight. 

Let $\mu_1$ and $\mu_2$ be two solutions to \eqref{limeq} and $h_i=\g* \mu_i$ the associated potentials, which solve
 \eqref{eqpoisson}. Let us compute 
\begin{align} \notag
\partial_t \int_{\R^{\d}} |\nab (h_1-h_2)|^2 &= 
 2 \cd \int_{\R^{\d}} (h_1-h_2)\partial_t (\mu_1-\mu_2) \\
\notag & =  2 \cd\int_{\R^{\d}} (h_1-h_2) \div(\mu_1 \nab h_1 - \mu_2 \nab h_2)\\
\notag & = -2 \cd \int_{\R^{\d}} (\nab h_1- \nab h_2)\cdot (\mu_1 \nab h_1 -\mu_2 \nab h_2)\\ \label{calc}
& = - 2 \cd \int_{\R^{\d}}|\nab (h_1-h_2)|^2 \mu_1 - 2 \cd \int_{\R^{\d}} \nab h_2\cdot \nab (h_1-h_2) (\mu_1-\mu_2).\end{align}
In the right-hand side, we recognize from \eqref{divT} the divergence of the stress-energy tensor $[\nab(h_1-h_2),\nab( h_1-h_2)]$, hence 
\begin{equation*}
\partial_t \int_{\R^{\d}} |\nab (h_1-h_2)|^2\le 2 \int_{\R^{\d}} \nab h_2 \cdot \div [\nab (h_1-h_2),\nab( h_1-h_2)]\end{equation*}
so, if $\nab^2 h_2$ is bounded, we may integrate by parts the right-hand side and bound it pointwise, thanks to \eqref{pointwisec},  by
$$2\|\nab ^2 h_2\|_{L^\infty} \int_{\R^{\d}} \left| [ \nab (h_1-h_2),\nab( h_1-h_2)]\right| \le 2 \|\nab ^2 h_2\|_{L^\infty} \int_{\R^{\d}} |\nab (h_1-h_2)|^2.$$ 
By Gronwall's lemma, we conclude that 
\be\label{wstrongu}
\int_{\R^\d} |\nab (h^{\mu_1^t} -h^{\mu_2^t} )|^2\le \exp\( C \int_0^t \|\nabla^2 (\g* \mu_2^s)\|_{L^\infty} ds\) \int_{\R^\d} |\nab (h^{\mu_1^0} -h^{\mu_2^0} )|^2
.\ee
Let us recall that 
$\frac{1}{2\cd}\int_{\R^\d} |\nab (h^{\mu_1^t} -h^{\mu_2^t} )|^2$ can also be rewritten as 
$$\iint_{\R^\d\times \R^\d} \g(x-y) d\(  \mu_1^t -\mu_2^t\) (x) d\(  \mu_1^t -\mu_2^t\) (y) $$
and is the (square of the) $\dot{H}^{-1}(\R^\d)$ semi-norm of $\mu_1^t-\mu_2^t$.
\eqref{wstrongu} constitutes a weak-strong uniqueness principle since it states that any solution ($\mu_1$ here) that starts close to a strong solution ($\mu_2$ here, for which $\|\nabla^2 (\g* \mu_2^s)\|_{L^\infty} $ is to be controlled), remains close to it on fixed time intervals.

\subsection{Time-derivative of the modulated energy} \index{modulated energy}
In the case of the empirical measure, the computation above is to  be performed on the modulated energy, which takes care of the Dirac singularities. 

To be able to use the electric formulation for  $\F_N(\XN^t, \mu^t)$ and use Theorem \ref{thm:FI}, we need \eqref{condmupourFN} to be satisfied by $\mu^t$. This is the reason for the assumption \eqref{condmuo}. Indeed, if $\s> 0$, the assumption of boundedness of $\mu^t$ in $L^\infty$ in \eqref{conditionsevol} and the decay of $\g$ suffice to ensure that \eqref{condmupourFN} holds. 
In the case $\s\le 0$, we claim that the assumption \eqref{condmuo} is propagated in time by \eqref{limeq}, i.e. 
\be\label{condsmut}
\int_{\R^\d} (-\g_-)d\mu^t<\infty,\ee
where we recall that $\g_-=\min(\g, 0)$.
Then we conclude  the desired condition as in Step 2 of the proof of Lemma \ref{1entraine2}.
To justify \eqref{condsmut}, it suffices to use \eqref{limeq},  integration by parts and the bounds \eqref{conditionsevol}
to find 
$$
\partial_t \int_{\R^\d} (-\g_-) \mu^t= \int_{\R^\d} \nab\g_- \cdot \mathbb{M}\nab h^{\mu^t}\mu^t
\le \|\nab h^{\mu^t}\|_{L^\infty} \int_{\R^\d} |\nab \g_-|\mu^t 
\le C_t \int_{\R^\d} (1-\g_-) \mu^t$$
by observing that $|\nab \g_-|\le C(1-\g_-)$. Using Gronwall's lemma then allows to deduce from \eqref{condmuo} that \eqref{condsmut} holds for each $t\ge 0$.

We next turn to the computation of the time derivative of the modulated energy, taking advantage of the laws of evolution.
\begin{lem}\label{lem31}
If $\XN^t$ is a solution of \eqref{mfgf} and $\mu^t$ solves \eqref{limeq}, then letting $\mu_N^t$ be as in \eqref{munt}, we have
\begin{multline}\label{f1}
\partial_t \F_N(\XN^t, \mu^t)\\
\le - \frac{N^2}{2} \iint_{\R^{\d}\times \R^{\d} \backslash \triangle} \mathbb{M}\(  \nab h^{\mu^t}(x)- \nab h^{\mu^t}  (y)\)\cdot  \nab \g(x-y) d (\mu_N^t - \mu^t)(x)d (\mu_N^t -  \mu^t)(y).
\end{multline} \end{lem}
\begin{proof}
We note that if $\s\ge \d-1$, $\nab \g$ is not integrable near $0$, so $\nab \g* \mu$ should be understood in a distributional sense and $\mu\nab (\g* \mu)=\mu \g* \nab \mu$ as well, assuming that $\mu$ is regular enough.
We may also check that this distributional definition is equivalent to defining $\nab h^\mu$ in principle value:
$$\nab h^\mu(x)= P.V. \int_{\R^{\d}\backslash \{x\}} \nab \g(x-y) d\mu(y).$$
Returning to  the definition \eqref{def:FN} and using  \eqref{mfgf} and  \eqref{limeq} and the symmetry of the problem, we have
\begin{align*}
& \partial_t \F_N(\XN^t, \mu^t)\\ &=
  N^2 \partial_t \iint\hal  \g(x-y) d\mu^t(x)d\mu^t(y)+ \partial_t \sum_{i\neq j}\hal \g(x_i^t-x_j^t)
-  N \partial_t \sum_{i=1}^N \int_{\R^{\d}} \g(x_i^t- y) d\mu^t(y)  
\\  &=    - N^2 \int_{\R^{\d}}  \nab h^{\mu^t} \cdot \mathbb{M} \nab h^{\mu^t}  (x) d\mu^t(x)     -\frac1N\sum_{i=1}^N   \langle \sum_{j\neq i}  \nab \g(x_i^t-x_j^t), \mathbb{M}     \sum_{k\neq i} \nab \g(x_i^t-x_k^t)\rangle \\
&
      \quad +   \sum_{j\neq i} \nab h^{\mu^t} (x_i^t)\cdot \mathbb{M}\nab \g(x_i^t-x_j^t)
       + N\sum_{i=1}^N  P.V. \int_{\R^{\d}   \backslash\{x_i^t\} }  \mathbb{ M} \nab h^{\mu^t} (x)\cdot \nab \g(x-x_i^t) d\mu^t(x) .\end{align*}
We then rewrite this as 
\begin{align*}
\partial_t \F_N(\XN^t, \mu^t) = &    - N^2 \int_{\R^{\d}}  \nab h^{\mu^t} \cdot \mathbb{M} \nab h^{\mu^t}   (x) d\mu^t(x)   -\frac1N\sum_{i=1}^N    \sum_{j\neq i}  \nab \g(x_i^t-x_j^t)\cdot  \mathbb{M}     \sum_{k\neq i} \nab \g(x_i^t-x_k^t)
\\
& + N^2 \int_{\R^{\d}}  \nab h^{\mu^t}(x) \cdot \int_{\R^{\d} \backslash \{x\}}  \mathbb{M}\nab \g(x-y) d\mu_N^t(y) d\mu_N^t(x)
\\
& + N^2 \int_{\R^{\d}}  P.V. \int_{\R^{\d}\backslash\{y\}}  \mathbb{ M} \nab h^{\mu^t} (x)\cdot \nab \g(x-y) d\mu^t(x) d\mu_N^t(y)
.\end{align*}
We recognize that the right-hand side can be recombined and symmetrized into
\begin{multline*} -N^2 \int_{\R^{\d}} P.V. \int_{\R^{\d} \backslash \{x\}}  \nab \g(x-y) d(\mu_N^t-\mu^t) (y)  \cdot \mathbb{M} \,  P.V. \int_{\R^{\d} \backslash \{x\}}  \nab \g(x-y) d(\mu_N^t-\mu^t) (y) d\mu_N^t (x) \\
-\frac{N^2}{2} \iint_{\triangle^c}  \mathbb{M}\( \nab h^{\mu^t}(x)- \nab h^{\mu^t}(y)\)  \cdot \nab \g(x-y) d(\mu_N^t-  \mu^t)(x) d(\mu_N^t-  \mu^t)(y),\end{multline*}
and since the first term is nonpositive by property of $\mathbb{M}$, we obtain the result.
\end{proof}
\subsection{Conclusion}
Combining \eqref{f1} with the commutator estimate \eqref{main1} applied to $v= \mathbb{M}\nab h^{\mu^t}$, we obtain 
\begin{equation}
\partial_t \F_N(\XN^t, \mu^t)
\le  C \|\nab^{\otimes 2} h^{\mu^t}\|_{L^\infty} \( \F_N(\XN^t, \mu^t) +N  \frac{\log (N\|\mu^t\|_{L^\infty})}{2\d} \indic_{\s=0} + C \|\mu^t\|^{\frac{\s}{\d}}_{L^\infty} N^{1+\frac{\s}{\d}}\)
\end{equation}
where $C$ depends only on $\s $ and $\d$.

We may now apply Gronwall's inequality to the quantity 
$$\Xi(t):=\F_N(\XN^t, \mu^t) + \(\frac{N }{2\d}\log N + \frac{N}{2\d} \log \sup_{s\in[0,t]} \|\mu^s\|_{L^\infty} \) \indic_{\s=0}+C_0  \sup_{ s\in [0,t]} \|\mu^s\|_{L^\infty}^{\frac\s\d} N^{1+\frac\s\d},$$ where $C_0$ is large enough that this quantity is nonnegative in view of \eqref{bornehnr}. If that quantity is initially small, it will remain small on any time interval on which $\mu^t$ and $\nab h^{\mu^t}$ are bounded.
More precisely, we may write 
\begin{equation}\label{controlgronwall}
\Xi(t)
 \le C 
\exp\( \int_0^t \|\nab^{\otimes 2} h^{\mu^s} \|_{L^\infty} ds\) \Xi(0). \end{equation}
In view of the assumptions, dividing by $N^2$, using that $\frac{\s}{\d}<1$, \eqref{bornehnr}  and Corollary \ref{coro453},
we obtain Theorem \ref{theoMF}.

\subsection{The case with noise: the modulated free energy method}\label{sec:dynnoise} \index{modulated free energy}
If we now consider the dissipative or conservative case with noise \eqref{mfgfnoise} or \eqref{consnoise},
the modulated energy method, as generalized  in \cite{NRS2021} to sub-Coulomb interactions, works in the case $\s<\d-2$, see \cite{RosenSer2021}: instead of considering $\F_N(\XN^t, \mu^t)$, one simply considers $\Esp(\F_N(\XN^t, \mu^t))$ and differentiates it in time. One can also bound higher order moments.

However, in the case $\s\ge \d-2$, this does not work. Bresch-Jabin-Wang  introduced instead in \cite{BJW} the {\it modulated free energy} which combines the relative entropy method and the modulated energy method. The point of view is then to consider $f_N^t$, the joint law of the solution to the SDE system \eqref{mfgfnoise} at time $t$. The modulated free energy is then defined as 
\be \label{modfree}
\mathcal F^\theta_N(f_N, \mu) =  \frac1\theta H_N(f_N\vert \mu^{\otimes N} ) + \int  \F_N(\XN, \mu) df_N(X_N),\ee
where the normalized relative entropy $H_N$ is defined by 
\be  H_N(f_N\vert \mu^{\otimes N})= N \int_{(\R^\d)^N} f_N \log \frac{f_N}{\mu^{\otimes N}} d\XN.\ee
Note that in units of $N^2$,  $\F_N$ is ``almost positive" by \eqref{minoF2}, while the relative entropy is nonnegative, and controls the convergence of the $k$-point marginals of  $f_N$ to those of $\varphi_N$ in total variation $(TV)$ distance  via the Csisz\`ar-Kullback-Pinsker
inequality 
$$\|\mu-\nu\|_{TV}^2\le 2 \int \mu \log \frac\mu{\nu}$$
combined with the subaddivity of entropy which yields 
$$\int f_N \log  \frac{f_N}{\mu^{\otimes N}}\ge \lfloor \frac{N}{k}\rfloor \int f_{N,k} \log \frac{f_{N,k}}{\mu^{\otimes k}} $$
where $\lfloor \cdot \rfloor$ denotes the integer part, and $f_{N,k}$ is the $k$-point marginal of $f_N$.
 Thus $\frac1{N^2}\mathcal F^\theta_N$ controls both relative entropy and modulated energy and metrizes the convergence of marginals of $f_N$ to $\mu^{\otimes k}$.
 
It is very important that \eqref{modfree} has the structure of a free energy, i.e.~an energy plus temperature times entropy. 
As seen from \cite{BJW} and reformulated in  \cite{RosenSer4}, another way of rewriting  \eqref{modfree} is as a relative entropy with respect to the {\it modulated Gibbs measure}
$$\mathbb{Q}_{N,\beta}(\mu) = \frac{1}{\K_{N,\beta}(\mu)} \exp\(- \beta N^{-\frac\s\d}\F_N (\XN, \mu) \)  d\mu(x_1) \dots d\mu(x_N), $$ with $\theta= \beta N^{1-\frac\s\d}$, which is exactly as defined in \eqref{defQ}.
 With this definition, elementary computations allow to check that 
 \be\label{Frelent} \mathcal F^\theta_N(f_N, \mu) =  \frac1\theta \( H_N(f_N\vert  \mathbb{Q}_{N, \beta} (\mu)) - N\log \K_{N,\beta}(\mu)\) ,\ee
  while in view  of \eqref{boundlogK} and \eqref{scalingK}, we can bound $$\frac{N}{\theta}|\log \K_{N,\beta}(\mu)|\le C  N^{1+\frac\s\d}\chi(\theta N^{\frac\s\d-1}) + \(\frac{N}{2\d} \log N\) \indic_{\s=0}\ll N^2 ,$$
thus  $\frac{1}{N^2}\mathcal F^\theta_N$ is equivalent to the suitably normalized  relative entropy with respect to the modulated Gibbs measure, up to a small and constant perturbation. Convergence in modulated free energy will thus yield convergence in relative entropy. 

As realized in \cite{BJW} and rephrased in \cite{RosenSer4}, thanks to its particular free energy structure, when differentiating in time $\mathcal F^\theta_N(f_N^t, \mu^t)$ with $\mu^t$ the mean-field limit solving \eqref{limeqnoise}, singular terms coming both from $\frac{d}{dt}(\theta^{-1} H_N)$ and from $ \frac{d}{dt}  \int  \F_N(\XN, \mu) df_N(X_N)$ exactly cancel and reveal only a nonpositive term and a commutator type term, as follows:
\begin{multline}
\label{eq:MFEdiss}
\frac{d}{dt}\mathcal F^\theta_N(f_N^t, \mu^t)  \leq - \frac{N }{\theta^2 }\int_{(\R^\d)^N }\left|\nab \log \frac{f_N^t}{\mathbb{Q}_{N,\beta } (\mu^t) } \right|^2 df_N^t\\
 -\frac{1}{2} \int_{(\R^\d)^N}\int_{(\R^\d)^2\setminus\triangle} (u^t(x)-u^t(y))\cdot \nabla\g(x-y) d\left(\sum_{i=1}^N\delta_{x_i} - N \mu^t\right)^{\otimes 2}(x,y)d f_N^t 
\end{multline}
where $u^t=\frac1\theta\nab \log \mu^t +\nab h^{\mu^t}$.
Discarding the nonpositive term  and  using just the commutator estimate of Theorem \ref{thm:FI}, together with the fact  that $\int \F_N df_N \le \mathcal F^\theta_N$,  this yields 
$$\frac{d}{dt}\mathcal F^\theta_N(f_N^t, \mu^t)  \leq C \mathcal F^\theta_N(f_N^t, \mu^t)  + o(N^2),$$
hence by Gronwall's lemma and the above remarks, we directly obtain
 local-in-time mean field convergence (in relative entropy) to the solution of \eqref{limeqnoise}. Using the commutator estimate requires a Lipschitz bound on $u^t$ which is delicate when working in the whole space, due to the necessary decay of $\mu^t$, however a self-similar transformation allows to reduce to a confined situation and obtain the needed bounds, which then even allows to conclude that global-in-time convergence holds \cite{RosenSer5}.

The negative term in the right-hand side of \eqref{eq:MFEdiss}  is the opposite of the   relative Fisher information with respect to the modulated Gibbs measure 
and, as observed in \cite{RosenSer4},  in case of existence of a uniform-in-time ``modulated'' Logarithmic Sobolev Inequality for $\Q_{N,\beta} (\mu^t)$,  it may be used to obtain an exponential decay of the modulated free energy, and what has been called \textit{generation of chaos} (i.e.~the probability becomes tensorized in large time, even if it is not initially), a term first coined by J. Lukkarinen. What we mean by uniform-in-time modulated Logarithmic Sobolev Inequality is that there exists a constant $C_{LS}>0$ such that for all $N\ge 1$, all $t\ge 0$ and all $f\in C^1((\R^\d)^N)$ we have  
\be \int_{(\R^\d)^N} f^2 \log \frac{f^2}{\int f^2 d\mathbb{Q}_{N,\beta}(\mu^t)} d\mathbb{Q}_{N,\beta}(\mu^t)\le C_{LS} \int_{(\R^\d)^N} |\nab f|^2 d \mathbb{Q}_{N,\beta}(\mu^t),\ee which allows to compare the modulated Fisher information with  the relative entropy appearing in \eqref{Frelent}, hence with $\mathcal F_N^\theta(f_N^t, \mu^t)$. 
Unfortunately, it seems very hard to prove such an inequality, except in the setting of $\d=1$
with some convexity assumptions (we refer the reader to \cite{RosenSer4}).

Even discarding the relative Fisher information term in \eqref{eq:MFEdiss}, one may still obtain uniform-in-time convergence, or uniform-in-time propagation of chaos and even generation of chaos, by taking advantage of the decay rate of the vector field $u^t$ in \eqref{eq:MFEdiss}. This is easier to do in the setting of the torus, and exponential decay of $u^t$ was proven, and used to conclude the uniform-in-time convergence in that context in \cite{ChodronRosenSer2022}, where the commutator estimate is extended  to the setting of the torus.

\part{Mesoscopic behavior}

\chapter[Screening]{The two energy quantities and the screening procedure}
\label{chap:screening}
\index{screening}
In previous chapters  we have reduced the study of the energy $\HN$ to that of the next order energy $\F_N$, and then given an electric formulation for $\F_N$ which replaces 
  pair interactions by an energy which  is local in space in terms of the electric field $\nabla h_N$. The question we now turn to  is  to show that this energy  (and similarly the free energy) is {\it almost additive} in space so that disjoint regions of space can be considered independently.
We can explain  the idea of how to do so by  using two energy quantities (in the spirit of Dirichlet-Neumann bracketing or the work of Armstrong-Smart for homogenization \cite{armstrongsmart}), one subadditive and one superadditive, which converge to each other on large scales.  This allows to quantify the defect of addivity of the energy or free energy over cubes.
To quantify this we will need to control the difference between these two energy quantities, which differ only by the boundary condition which is imposed.  This is where we use a {\it screening procedure} that allows to show that the effect of a specific boundary condition decays rapidly enough away from the boundary  to be negligible.

In this whole Part III, we restrict to the Coulomb case $\s=\d-2$ and work in blown-up scale, see Section \ref{sec:blowup} for definitions.
The adaptation of Part III to the other Riesz cases  \eqref{rieszgene} can be found in \cite{peilen} for the one-dimensional logarithmic case, and in \cite{PeilenSer} for the general case. 
The material in this chapter and the next originates in \cite{as} but with new simplifications.

\section{Dirichlet and Neumann problems}

Let us now  introduce the two local sub and super additive energy approximations, from \cite{as}.
Let us consider $U$ a subset of $\R^\d$ with piecewise $C^1$ boundary.  Most often, $U$ will be $\R^\d$,   a hyperrectangle or the complement of a hyperrectangle.
Although $N$ originally denoted the number of points in $\R^\d$ and defined the blown-up scale at which we are working,  we will also use the notation $N$ to denote the total number of points a system has in a generic set $U$ which may not be the whole space.

\smallskip

\subsection{Informal description}
\index{Neumann energy}
Given a nonnegative, bounded and integrable density $\mu$ over $U$, 
the first energy quantity  is obtained by solving for
\begin{equation}\label{defv}
\left\{\begin{array}{ll}
 -\Delta u = \cd\Big( \sum_{i=1}^{N} \delta_{x_i}- \mu \Big) &\ \text{in} \ U \\
 \frac{\pa u}{\partial \nu}=0 &\ \text{on} \ \partial U\\ [2mm]
 \nab u \to 0 &\ \text{at } \infty. \end{array}\right.
\end{equation}
Note that  this equation is solvable (and the solution is unique up to addition of a constant) if and only if $\mu(U)=N$, which means if and only if the system in $U$ is {\it neutral}. 
If $U$ is bounded, there is not condition ``at $\infty$". 
If $U$ is unbounded, we can find a solution by considering first $\g* (\sum_{i=1}^{N} \delta_{x_i}- \mu \indic_U)$ over $\R^\d$, and then subtracting off a harmonic function $w$ with prescribed Neumann boundary condition on $\partial U$ and such that $\nab w\to 0$ at $\infty$.
We will also later call  {\it screened system} such a system  with equation \eqref{defv} solved.
 
Neglecting the issue of renormalization, the first energy quantity is defined informally  as 
$$\F(\XN, \mu, U)= \frac1{2\cd}\int_{U}|\nab u|^2.$$
Unless ambiguous, we omit the dependence in $\mu$ in the notation and simply write  $\F(\XN, U)$ instead of $\F(\XN, \mu, U)$.
We note that  $\F(\cdot, \R^\d)$ coincides with $\F$ defined in~\eqref{Fbup}.

The second quantity is obtained by minimizing the energy with respect to all possible functions $u$ compatible with the points 
in the sense of satisfying $-\Delta u = \cd(\sum_{i=1}^N \delta_{x_i}-\mu)$, it naturally leads to   a superadditive energy and to solving a  Dirichlet problem:
\begin{equation}\label{defu}
\left\{\begin{array}{ll}
 -\Delta v = \cd\Big( \sum_{i=1}^{N} \delta_{x_i}- \mu \Big) &\ \text{in} \ U \\
 v=0 &\ \text{on} \ \partial U\ \text{and at } \infty. \end{array}\right.
\end{equation}
Neglecting again the issue of renormalization, the  second energy quantity $\G$  is defined informally as 
$$\G(\XN, \mu, U) = \frac1{2\cd}\int_U  |\nab v|^2.$$
Again, we will often simply write $\G(\XN, U)$.

One may check that for any given distribution $f$, 
the solution of the variational problem
\be
\label{3.4}
\min \Bigg\{ \frac{1}{2\cd}\int_{U} |\nab w |^2 , \quad 
 -\Delta w =  f \ \text{in} \ U\Bigg\}\ee
is achieved by the solution  of 
\be
\left\{\begin{array}{ll}
 -\Delta v = f &\ \text{in} \ U \\
 v=0 &\ \text{on} \ \partial U \ \text{and at } \infty. \end{array}\right.
\end{equation}
Indeed, it suffices to consider competitors of the form $w+t h$ with $h$ harmonic, solving  $\partial h/\partial \nu = g$ on $\partial U$ for any given $g$ of integral $0$ on $\partial U$. Writing 
that $\int_U |\nab (w+th)|^2 \ge \int_U |\nab w|^2$, integrating by parts and letting $t\to 0$, we find that $\int_{\partial U} wg =0$, and this being true for any $g$ of integral $0$, $w$ must be constant on $\partial U$. Since the problem is unchanged by addition of a constant, we may assume that $w=v$, thus $v$ minimizes the energy among all solutions of $-\Delta v= f$.

Applying this to $f= \cd \(\sum_{i=1}^N \delta_{x_i}- \mu\)$, we immediately deduce that for any $\XN$ such that $\F$ is defined, we have
\be \label{FlessG} \F (\XN, \mu, U) \ge \G(\XN, \mu, U).\ee

We will now see that $\F$ is good for pasting while $\G$ is good for restricting, leading to the property that $\F$ is subadditive and $\G$ superadditive.

The property is easy for $\G$.
Assume $U$ is the union of two sets $ U_1$, $ U_2$ with disjoint interiors and   piecewise $C^1 $ boundaries. If $\XN$ is a configuration in $U_1 $, $v_1$ the corresponding solution of \eqref{defu}  and $Y_{N'}$ a configuration in $U_2$, $v_2$ the corresponding solution of \eqref{defu},  and $v$ the solution of \eqref{defu}  in $U$ for the configuration $X_N\cup Y_{N'}$,
then 
\begin{align}\notag
\G(X_N\cup Y_{N'},  U)
& = \frac{1}{2\cd}\int_{U} |\nab v|^2 =\frac{1}{2\cd}
\int_{U_1}|\nab v|^2 +\frac1{2\cd} \int_{U_2} |\nab v|^2 \\ \label{add00}
& \ge\frac1{2\cd} \int_{U_1}|\nab v_1|^2 + \frac1{2\cd}\int_{U_2} |\nab v_2|^2=
  \G(X_{N},  U_1) + \G(Y_{N'},  U_2) ,\end{align}
where the inequality follows from  the minimality property of $v_1$, resp. $v_2$, seen above. So $\G$ is superadditive as claimed.

 It is convenient to also work with    ``electric fields" $E$ which are gradients of electric potentials and thus satisfy relations of the form
\be\label{dive}
-\div E= \cd\(\sum_{i=1}^n \delta_{x_i}-\mu\) .\ee
Any vector field, not necessarily a gradient, which satisfies such a relation, will be said to be compatible with $(\XN,\mu)$.

 The subadditivity of $\F$ relies on the following lemma  which exploits that the Neumann electric field (that is, the electric field with zero normal component) is the $L^2$ projection of any compatible electric field onto gradients.
  \begin{lem}[Projection lemma]\label{projlem0}
  Assume that $U$ is a bounded open of $\R^\d$ with piecewise $C^1$ boundary.
  Assume $E$ is a vector-field  and $u$ a function satisfying 
  \begin{equation}\label{eqe0}
   \left\{\begin{array}{ll}  
  \div E=  \Delta u & \text{in}  \ U\\
  (E -\nab u) \cdot \nu=0 & \text{on} \ \partial U.\end{array}\right.\end{equation}
  Then
  $$\int_{U} |\nab u|^2 \le \int_U |E |^2.$$ 
  \end{lem}
\begin{proof} It suffices to write
\begin{multline} 
\int_{U} |E|^2= \int_U |\nab u + E-\nab u|^2=
\int_U |\nab u|^2 + |E-\nab u|^2 + 2\nab u \cdot (E-\nab u) \\
\ge \int_U |\nab u|^2 - 2\int_U u \, \div (E-\nab u) = \int_U |\nab u|^2 \end{multline}
where we used \eqref{eqe0} and Green's formula.
\end{proof}

This way, the energy $\F(\XN, U)$ can be estimated from above by that of any vector field with same divergence, i.e.~after {\it relaxing} the condition of being a gradient, an idea already used in \cite{ACO} and \cite{ssgl}.

Let us now see how the projection lemma implies subadditivity.
Assume $U$ is the union of two sets $ U_1$, $ U_2$ with disjoint interiors and   piecewise $C^1 $ boundaries. If $\XN$ is a configuration in $U_1 $ and $Y_{N'}$ a configuration in $U_2$ with $\mu(U_1)=N$, $\mu (U_2)=N'$, then 
\begin{equation}\label{subad10}
\F(X_N\cup Y_{N'},  U) \le \F(X_{N},  U_1) + \F(Y_{N'},  U_2) .\end{equation}
Indeed, let $u_1$ and $u_2$ be the solutions to the Neumann problems and set $E_1= \nab u_1$, $E_2= \nab u_2$. 
 We have 
 \begin{equation}\label{dve}
 - \div E_1= \cd \Big( \sum_{i=1}^{N} \delta_{x_i} - \mu\Big)  \ \text{in} \ U_1 \qquad -\div E_2= \cd\Big(\sum_{i=1}^{N'}\delta_{y_i}- \mu\Big)\ \text{in} \ U_2 .\end{equation}
  We may now define $E= E_1\indic_{U_1}+ E_2\indic_{U_2}$ and note that it satisfies 
  \begin{equation}\label{eqq2}
 \left\{\begin{array}{ll}
 -\div E= \cd\Big(\sum_{p\in \XN\cup Y_{N'}} \delta_p - \mu\Big)
 &\quad \text{in} \ U\\
 E\cdot  n = 0 & \quad\text{on} \ \p U .\end{array}\right.\end{equation}
 Indeed, no divergence is created across $\partial U_1\cap \partial U_2$ thanks to the vanishing normal components on both sides. Thus, $E$ is compatible with $(X_N\cup Y_{N'}, \mu)$.
It then follows from Lemma \ref{projlem0} that 
$$\F(\XN \cup Y_{N'},  U) \le \frac1{2\cd}\int_U |E|^2  = \frac{1}{2\cd}\int_{U_1} |\nab u_1|^2 + \frac{1}{2\cd}\int_{U_2} |\nab u_2|^2 = \F(\XN, U_1)+ \F(Y_{N'}, U_2),$$
hence the claimed subadditivity.

We define corresponding partition functions.
\begin{defi}[Neumann Gibbs measure and partition function]
The Neumann partition function relative to~$U$, is defined  if~$\mu(U) =N$ by 
\begin{equation}
\label{defK7}
\K_\beta( \mu,U):=  N^{-N}\int_{U^N} 
\exp\left(- \beta \F(X_N,\mu,U) \right)
\, d\mu^{\otimes N} (X_N).\end{equation}
The  Neumann Gibbs measure is defined by \begin{equation}\label{defQ7}
d\Q_\beta( \mu,U)
:= 
\frac{1}{N^N\K_\beta( \mu,U)} 
\exp\left( -\beta \F(\XN,\mu, U)\right)
\, d\mu^{\otimes N} (X_N).
\end{equation}
\end{defi}
The notation agrees with \eqref{defKN2} and \eqref{defQbu} : when taking $U=\R^\d$ the definitions coincide, adopting the convention $\K_\beta(\mu, \R^\d)= \K_\beta(\mu)$ and $\Q_\beta( \mu,\R^\d)= \Q_\beta(\mu)$. Again we omit the $N$ in the notation since it can be recovered from $ \mu(U)$.

We may also consider in the same way the ``Dirichlet partition function"
\begin{equation}\label{defL}
\L_{N,\beta}( \mu,U):=  N^{-N}\int_{U^N} 
\exp\left(- \beta \G(X_N,\mu,U) \right)
\, d\mu^{\otimes N} (X_N).\end{equation}

In view of \eqref{FlessG} we have a comparison for the free energies as well:
whenever $\mu(U)=N$, we have 
\be \label{KlessL}
\L_{N,\beta}(\mu,U) \le \K_\beta(\mu,U).\ee

The subadditivity property has the following counterpart for the partition functions.
\begin{lem}[``Superadditivity" of Neumann partition functions] Assume $U$  is  partitioned into $p$ disjoint sets $Q_i$,  $i\in [1,p]$ which are such that $\mu(Q_i)=N_i$ with $N_i$ integer.
We have
\begin{equation}\label{superad2}
\K_\beta(\mu, U) \ge    \frac{N! N^{-N}}{{N_1}! \dots {N}_p! N_1^{-N_1} \dots N_p^{-N_p}}    \prod_{ i=1  }^p \K_\beta(\mu, Q_i).\ee
\end{lem}
\begin{proof}
It suffices to partition the phase space into sets of the form $\{x_{i_1}, \dots, x_{i_{N_j}} \in Q_j\}$ for each $j=1,\dots, p$, then to use  the subadditivity 
~\eqref{subad10}, or rigorously \eqref{subad1},  noting  that 
the number of ways to distribute $N$ points in the $p$ sets with ${N}_i$ points in each set is 
$
 \frac{N!}{{N}_1! \dots {N}_p! } $.
\end{proof}

When we consider a uniform density $\mu$, say $\mu=1$ without loss of generality, then the sub and superadditivity property will imply that 
\be \label{qt1} - \frac{\log \K_\beta(1, \carr_R)} {\beta R^\d} \ee is decreasing in $R$ 
while 
\be \label{qt2}- \frac{\log \L_\beta(1, \carr_R)}{\beta R^\d}\ee is increasing in $R$, when $\carr_R$ are cubes of quantized volume. Thus each of them will  have a limit, that we denote $f_\d(\beta)$, as it is a function of $\beta $ and $\d$ only.
Since on the other hand $\L_\beta\le \K_\beta$, it means that if we can show that these limits are the same, then $\frac{|\log \K_\beta(1,\carr_R)- \log \L_\beta(1,\carr_R)|}{\beta R^\d}$ will  bound from above the distance of each of them to $f_\d(\beta)$ and allow to get a rate of convergence for the two limits.
The goal is thus to show that 
\be \label{ratel}
\frac{|\log \K_\beta(1,\carr_R)- \log \L_\beta (1,\carr_R)|}{\beta R^\d}\to 0 \quad \text{as}\ R \to \infty,\ee
with a quantitative rate, which will bound the rate of convergence to the limit $f_\d(\beta)$.

The reason why \eqref{ratel} is true is that the effect of the boundary condition decays sufficiently fast away from the boundary  that it  only brings in  a surface contribution, which is negligible compared to the volume $R^\d$. In other words, we expect that \eqref{ratel} happens with rate $1/R$  (except that $\beta$ comes into play when it is small).
Moreover, since  \eqref{qt1}  and \eqref{qt2} converge rapidly to the same limit, it means that as $R$ gets large they become almost constant, in particular it means that the free energies $\log \K_\beta(1, \carr_R) $, superadditive, and $\log \L_\beta(1, \carr_R)$, subadditive, must both be almost additive.  
This property of {\it almost additivity of the energy} will be crucial.

Proving \eqref{ratel} requires work, and will rely on the {\it screening procedure}: that procedure consists in modifying arbitrary configurations and their potential (say $v$), into neutral configurations with the Neumann problem solved as in \eqref{defv}, without having added too much energy or entropy. Such modified configurations are called screened configurations. 
The screening procedure will create energy and volume errors that need to be precisely quantified. 
 
\subsection{Rigorous definitions} \index{electric formulation}
In order to deal with the renormalization properly, we
need to introduce a new modified version of the minimal distance \eqref{defrip} that is relative to the domain $U$, and that  makes the energy subadditive. In this part, we do not keep track of  $\|\mu\|_{L^\infty}$ dependence so we replace $\lambda$ by $1$ in \eqref{defrip} and  
 we let
 \index{truncation radii}
\begin{equation}\label{defrrc}
\rrh_i := \frac{1}{4}\min \( \min_{x_j\in U, j\neq i} |x_i-x_j|, \dist(x_i, \pa U), 1\).
\end{equation} 
This shrinks the radius of the balls when they approach $\partial U$, ensuring that all $B(x_i, \rrh_i)$ remain included in $U$ if $x_i \in U$.
\begin{defi}[Neumann electric energy in a domain $U$]
If $\mu (U)=N$, for a configuration $\XN$ of points in $U$ and $u$ as in \eqref{defv} we define\footnote{this definition is simpler than that in \cite{as}}
\begin{equation}\label{minneum}
\F(X_N,\mu,U) :=\frac{1}{2\cd}\( \int_{U}|\nab u_{\rrh}|^2 - \cd \sum_{i=1}^N \g(\rrh_i)\)  - \sum_{i=1}^N \int_{U} \f_{\rrh_i}(x - x_i) d\mu(x),
\end{equation}
where $u_{\vec{\eta}}$ is defined as in \eqref{formu2} and  $\f_\eta$ as in \eqref{def:truncation0}.

Analogously to \eqref{Glocal} we define a localized version of this energy in a measurable subset $\Omega$
\be \label{Glocal2} 
\F^\Omega(\XN, \mu, U) := \frac1{2\cd} \(\int_{\Omega\cap U} |\nab u_{\rrh}|^2 - \cd \sum_{i\in I_\Omega} \g(\rrh_i) \) - \sum_{i\in I_\Omega} \int_U \f_{\rrh_i} (x-x_i) d\mu(x) 
\ee
where  we let $I_\Omega = \{i, x_i\in \Omega\}$, and where this time
\begin{equation}\label{defrrc3}
\rrh_i := \frac{1}{4}\begin{cases} 
\displaystyle
\min \( \min_{ j\neq i} |x_i-x_j|, \dist(x_i, \pa U), 1\) & \text{if} \ \dist(x_i, \partial \Omega) \ge 2  \\ \min(1, \dist(x_i, \partial U)) & \text{if } \dist(x_i, \pa \Omega ) \le 1\\
\displaystyle t \min \( \min_{j\neq i} |x_i-x_j|, \dist(x_i, \pa U ), 1\)&\\
+(1-t) 
\min( \dist(x_i, \partial U ),1)& \text{if } \dist(x_i, \pa  \Omega)= 1+t 
, t\in [0,1].\end{cases}
\end{equation}
\end{defi}
We note that the definition  \eqref{defrrc3}  coincides with \eqref{defrrc} when taking $\Omega=\R^\d$.  
Also, the definition has been made so that the radii are continuous with respect to the location of the points, which will be useful in Chapter \ref{chap:derivW}.  Note that the points in $\Omega^c$ influence the value of $\F^\Omega$ via their truncated charge in $u_{\rrh}$ but have no effect on the value of $\rrh_i$ for $x_i \in \Omega$. This will be important in the definition Definition \ref{defibestpot} below.
Here, the  balls are enlarged to their largest possible values for points that approach the boundary of $\Omega$ (except for the part included in $\partial U$). This way, balls can potentially overlap the boundary of $\Omega$ and not be disjoint. This is the right choice to have the restriction property  analogous to  \eqref{locali}: 
\be \label{locali2}
\F (\XN, \mu, U) \ge \F^{\Omega}(\XN, \mu, U)+ \F^{\Omega^c} (\XN, \mu, U),\ee
which we will justify below in Lemma \ref{lem:contrdist1}.

Finally, we will also use another definition, similar to \eqref{rrc} but in the blown-up scale, which ignores $\partial U$ (hence the balls may overlap $\pa U$):
\begin{equation}\label{defrrc4}
\rrc_i := \frac{1}{4}\begin{cases} 
\min \( \min_{ j\neq i} |x_i-x_j|, 1\) & \text{if} \ \dist(x_i, \partial \Omega) \ge 2  \\ 1  & \text{if } \dist(x_i, \pa \Omega) \le 1\\
t \min \( \min_{ j\neq i} |x_i-x_j|, 1\)
+(1-t) & \text{if } \dist(x_i, \pa  \Omega)= 1+t
, t\in [0,1].\end{cases}
\end{equation}

In the rigorous treatment, it is not useful to define the Dirichlet (or minimal energy) yet, it will need to be defined in conjunction with a screenability property below.

\subsection{Projection lemma and consequences}

A truncated version of  an electric field $E$ can be defined just as for electric potentials : for any $E$ satisfying a relation of the form~\eqref{dive}, for any $\vec{\eta}$,  we let 
\be \label{eer}E_{\vec{\eta}}= E- \sum_{i=1}^n \nab \f_{\eta_i}(x-x_i).\ee


Next we state a projection lemma with renormalized vector fields. The proof is identical to that of Lemma \ref{projlem0}.
  \begin{lem}[Projection lemma]\label{projlem}
  Assume that $U$ is an open  subset of $\R^\d$ with piecewise $C^1$ boundary.
  Assume $E$ is a vector-field satisfying a relation of the form
  \begin{equation}\label{eqe}
   \left\{\begin{array}{ll}  
  -\div E= \cd\( \sum_{i=1}^N \delta_{x_i} -\mu\) &\quad \text{in}  \ U\\
  E \cdot \nu=0 & \quad \text{on} \ \partial U,\end{array}\right.\end{equation}
and $u$ 
  solves 
$$  \left\{\begin{array}{ll}  
  -\Delta u= \cd\( \sum_{i=1}^N \delta_{x_i}- \mu\)& \quad \text{in}  \ U\\
  \frac{\pa u}{\partial \nu}=0 & \quad \text{on} \ \partial U,\end{array}\right.$$ 
  and $$ u\(\frac{\pa u}{\partial \nu}- E \cdot \nu\) \to 0 \quad \text{ as} \  |x|\to \infty,\  x\in U.$$
  Then
  $$\int_{U} |\nab u_{\rrh}|^2 \le \int_U |E_{\rrh} |^2.$$ 
  \end{lem}
\begin{coro}[Subadditivity of $\F$] \label{corosubad} Assume $U$ is the union of two sets $ U_1$, $ U_2$ with disjoint interiors and   piecewise $C^1 $ boundaries. If $\XN$ is a configuration in $U_1 $ and $Y_{N'}$ a configuration in $U_2$ with $\mu(U_1)=N$, $\mu (U_2)=N'$, then 
\begin{equation}\label{subad1}
\F(X_N\cup Y_{N'},\mu,   U) \le \F(X_{N}, \mu, U_1) + \F(Y_{N'},\mu,   U_2) .\end{equation}
\end{coro}
\begin{proof}
 For~\eqref{subad1}, let $u$ and $u'$ be the solutions to the Neumann problems \eqref{defv} associated with the definition of $\F$ in~\eqref{minneum} and set $E= \nab u$, $E'= \nab u'$. 
 We have 
 \begin{equation}\label{dve}
 - \div E= \cd \Big( \sum_{i=1}^{N} \delta_{x_i} - \mu\Big)  \ \text{in} \ U_1 \qquad -\div E'= \cd\Big(\sum_{i=1}^{N'}\delta_{y_i}- \mu\Big)\ \text{in} \ U_2 .\end{equation}
  We may now define $E^0= E\indic_{U_1}+ E'\indic_{U_2}$ and note that it satisfies 
  \begin{equation}\label{eqq2}
 \left\{\begin{array}{ll}
 -\div E^0= \cd\Big(\sum_{p\in \XN\cup Y_{N'}} \delta_p - \mu\Big)
 &\quad \text{in} \ U\\
 E^0\cdot \nu = 0 & \quad\text{on} \ \p U \\
 E^0 \cdot \nu  \to 0 &\quad  \text{as } |x|\to \infty \end{array}\right.\end{equation}
 Indeed, no divergence is created across $\partial U_1\cap \partial U_2$ thanks to the vanishing normal components on both sides.
The result then follows from Lemma \ref{projlem}, noting that the renormalization does not interfere because the balls $B(x_i, \rrh_i)$ remain included in $U_1$, resp. $U_2$.
\end{proof}
  This proof tells us that screened configurations or screened electric fields can effectively be pasted together without creating  any additional divergence, and their energies can be simply added to produce an upper bound on the true energy. We will use that property repeatedly.

  Thanks to the subadditivity of the Neumann energy, letting  $\K_\beta(U, \mu)$ and $\Q_\beta(U, \mu)$ be defined via \eqref{defK7} and \eqref{defQ7}, we have that   \eqref{superad2} holds.

  The same argument  as in Corollary \ref{corosubad} allows to obtain the following. 
  \begin{coro}[The Neumann electric energy is larger than the regular one] \label{compFUF}
  For any $\mu, U$ as above, and any configuration $\XN$ in $U$, extending $\mu$ by $0$ in $U^c$ we have 
  \be\label{compFUF1} \F(\XN, \mu \indic_U)=\F(\XN, \mu \indic_U , \R^\d) \le \F(\XN, \mu, U),\ee
in particular it follows  from \eqref{lbF} that 
 \be \label{lbFneum}\F(\XN, \mu, U) \ge - C N \indic_{\s\ge 0} ,\ee
 with $C$ depending only on $\d, \s$ and $ \|\mu\|_{L^\infty}$.
  \end{coro}
  \begin{proof}
  Let $E_{\rrh}$ be as in \eqref{eer} where $E=\nab u$, with $u$ solution to \eqref{defv} arising in the definition of $\F(\XN, \mu, U)$.
  Let us extend $E$ and $E_{\rrh}$ by $0$ in $U^c$. Since the balls $B(x_i, \rrh_i)$ do not intersect $\partial U$ by definition \eqref{defrrc3}, both $E$ and $E_{\rrh}$ are normal to $\partial U$ and thus no divergence is created at the boundary and  the extended fields satisfy 
  \be\label{divEER}-\div E= \cd \(\sum_{i=1}^N \delta_{x_i}-\mu\) \quad \text{in } \ \R^\d, \qquad -\div E_{\rrh}= \cd \(\sum_{i=1}^N \delta_{x_i}^{(\rrh_i)}-\mu\) \quad \text{in } \ \R^\d.\ee
  Letting $h_N$ be the electric potential associated to $\XN$ and $\mu$ in $\R^\d$, as defined in \eqref{HNp}. We have $\div E_{\rrh} = \Delta h_{N, \rrh} $, and $E_{\rrh}$ and $h_{N, \rrh}$ have the same decay at infinity. Hence, applying the projection lemma,  Lemma \ref{projlem}, over $U$ we find that
  $$\int_{\R^\d} |\nab h_{N, \rrh}|^2 \le \int_{\R^\d} |E_{\rrh}|^2= \int_U |\nab u_{\rrh}|^2$$ and the claim  \eqref{compFUF1} follows immediately in view of the definition \eqref{minneum}.
    \end{proof}

  \subsection{Local energy  controls}
  We will repeatedly need the following lemma which is a variant in this context of Propositions \ref{34} and \ref{procontrolelocal}  and Lemma \ref{prop:fluctenergy}.


  \begin{lem}[Local energy controls]
 \label{lem:contrdist1}There exist $C, C_0, C_1>0$ depending only on $\d$ and $\|\mu\|_{L^\infty}$  such that 
for any configuration $\XN$ in $U$  and $u$ corresponding via~\eqref{defv}, and for any $\Omega \subset U$,
\be\label{15}\sum_{i\in I_\Omega} \g(\rrc_i)\le C_1  \F^{\Omega}(X_N,\mu, U)  +   C \# I_\Omega, \ee
  \be \label{14}
 \int_{\Omega} |\nab u_{\rrc}|^2 \le 4\cd  \( \F^{\Omega} (\XN, \mu, U) + C_0\#I_\Omega \), \ee
with $C_1=2$ in the case $\s=0$, $\rrc$ as in \eqref{defrrc4}; 
and,  if $\alpha_i = \rrc_i$ for all $i$'s such that $\dist(x_i, \pa\Omega) \le\alpha_i $,  we have
\be \label{pre11}
\F_N^\Omega(\ux_N, \mu) -  \mathcal F^{\vec{\alpha}}\ge
\hal \sum_{\substack{ i, j\in I_\Omega, i\neq j \\ \dist(x_i, \pa\Omega) \ge \alpha_i }} 
\(\g(x_i-x_j) - \g(\alpha_i)\)_+ ,
\end{equation}
where $$\mathcal F^{\vec{\alpha}}
= \frac{1}{2\cd}\( \int_{\Omega} |\nab u_{\vec{\alpha}}|^2 -\cd \sum_{i\in I_\Omega} \g(\alpha_i)- 2\cd \sum_{i\in I_\Omega} \int_{\R^\d} \f_{\alpha_i} (x-x_i) d\mu(x)\).$$

Moreover, we have
$$
\F (\XN, \mu, U) \ge \F^{\Omega}(\XN, \mu, U)+ \F^{\Omega^c} (\XN, \mu, U).$$

Let  $\varphi$ be a  Lipschitz function in $U$ with bounded support.  Let $\Omega$ be an open set    containing a $1$-neighborhood of  the support of $\varphi$ in $U$.  For any configuration $\XN$ in $U$, letting $u$ be defined as in~\eqref{defv} (resp.~$v$ as in~\eqref{defu}), we have
\begin{equation} 
\label{fluctuationsbu}
\left|\int_{\R^\d} \varphi \, \left(\sum_{i=1}^N \delta_{x_i} -  d\mu \right) \right|
\leq \frac{1}{\cd} \| \nab  \varphi\|_{L^2(\Omega)} \|\nab u_{\rrc}\|_{L^2(\Omega\cap U)}+
C \|\nab\varphi\|_{L^\infty(\Omega)} \# I_\Omega ,
\end{equation}  (and resp.~the same with $v_{\rrc}$ in place of~$u_{\rrc}$ if $\Omega\subset U$),
where $C$ depends only on $\d$ and $\rrc$ is computed with respect to any set containing $\Omega$.
  \end{lem}
  \begin{proof}
  The proof is an adaptation of those of Chapter \ref{chap:nextorder}, the main idea being to work with the electric field $\nab u$ extended by $0$ which brings us back to a full space situation.
  
{\bf Step 1: monotonicity.}   First we claim that Lemma \ref{monoto} can be reproven with no change in terms of electric fields (and at the blown-up scale): let $E$ be a vector field solving 
 \be \label{divED}
 -\div (\yg E) = \cds\( \sum_{i=1}^N \delta_{x_i} -  \mu \delta_{\R^\d}\) \quad \text{in} \ \Omega \times \R^{\k}\ee and let $E_{\vec{\alpha}}$, $E_{\vec{\eta}}$ be as in \eqref{eer} for $\vec{\alpha}$, $\vec{\eta}$ such that $\alpha_i \le \eta_i $ for all $i$.  Then, letting $I_N= \{i, \alpha_i\neq \eta_i\}$, if for each $i \in I_N$, $B(x_i, \eta_i) \subset \Omega$, we have 
 \begin{multline}\label{premono2}
 \int_{\Omega\times \R^{\k}} \yg |E_{\vec{\eta}}|^2 -\cds \sum_{i\in I_N}\g(\eta_i) -2 \cds \sum_{i\in I_N} \int_{ \Omega\times \R^{\k}}  \f_{\eta_i}(x-x_i)d\mu\\- \( \int_{ \Omega \times \R^{\k}}\yg |E_{\vec{\alpha}}|^2 - \cds \sum_{i\in I_N} \g(\alpha_i) -2\cds\sum_{i\in I_N} \int_{\R^\d} \f_{\alpha_i}(x-x_i)d\mu\) \le 0,\end{multline}
with equality if the $B(x_i,\eta_i)$'s are disjoint from all the other $B(x_j, \eta_j)$'s for each $i \in I_N$.  We carry out the same proof as in Lemma \ref{monoto} with $\nab u_{\vec{\eta}}$ replaced by $E_{\vec{\eta}}$ and $\nab u_{\vec{\alpha}}$ by $E_{\vec{\alpha}}$.
 We note that the proof of that lemma only relies on the fact that $\nab u_{\vec{\eta}}-\nab u_{\vec{\alpha}}=\sum_{i\in I_N} \nab \f_{\alpha_i, \eta_i} (z-x_i) $ which is still true for $E_{\vec{\eta}}-E_{\vec{\alpha}}$ and on the fact that $-\div (\yg \nab u_{\vec{\alpha}})= \cds\( \sum_{i=1}^N \delta_{x_i}^{(\alpha)}- N \mu \drd\)$ (and the same with $\vec{\eta}$) which is replaced by $-\div (\yg E_{\vec{\alpha}}) = \cds\( \sum_{i=1}^N \delta_{x_i}^{(\alpha)}-  \mu \drd\)$ from \eqref{divED}, \eqref{eer} and \eqref{eqpourfeta}.  We thus arrive at \eqref{premono2} without other changes than removing the $N$ factors in front of $\mu$.
 
 {\bf Step 2: rewriting of $\F^{\Omega}(\XN, \mu, U)$.} As in the proof of Corollary \ref{compFUF}, let us denote $E=(\nab u)\indic_U $ where  
$u$ is the solution of \eqref{defv} used in the definition of  $\F(\XN, \mu, U)$ in \eqref{minneum}.  Again the crucial point is that thanks to the Neumann boundary condition $E$ solves \eqref{divEER} in the full space.
From the  definition \eqref{Glocal2}, since $\f_{\rrh_i}(x-x_i) $ is supported in $B(x_i, \rrh_i) \subset U$, we may rewrite  
$$\F^{\Omega}(\XN, \mu, U)= \frac{1}{2\cd} \(\int_{\Omega\cap U} |E_{\rrh}|^2 - \cd\sum_{i\in I_\Omega} \g(\rrh_i) \) - \sum_{i\in I_\Omega} \int_{\R^\d} \f_{\rrh_i} (x-x_i) d\mu(x).$$
By the result \eqref{premono2}, equality case, applied to $E$  in $\Omega$, 
if $\vec{\alpha}$ is such that $\alpha_i\le \rrc_i$ with equality if $\dist(x_i, \pa \Omega) \le \rrc_i$,  
\be \label{FNeum2}
\F^{\Omega}(\XN, \mu, U)= \frac{1}{2\cd} \(\int_{\Omega\cap U} |E_{\vec{\alpha}}|^2 - \cd\sum_{i\in I_\Omega} \g(\alpha_i) \) - \sum_{i\in I_\Omega} \int_{\R^\d} \f_{\alpha_i} (x-x_i) d\mu(x).\ee

{\bf Step 3: control of small scale interactions and of minimal distances.} We can then copy without change  the proof of Lemma  \ref{lem:monoto} and Proposition \ref{procontrolelocal}, denoting $\mathcal F^{\vec{\alpha}}$ the quantity in the right-hand side of \eqref{FNeum2}.
The main differences are that we replace $\nab h_N$ by $E$, remove the $N$ factors in front of $\mu$ and replace $\lambda $ by $1$. 
 We thus obtain  \eqref{pre11}, \eqref{15} and \eqref{14}.
 
 {\bf Step 4: superaddivity for restriction.}
In view of \eqref{premono2}, changing the radii from $\rrh_i$ of \eqref{defrrc} to $\rrh_i$ relative to $\Omega$ in \eqref{defrrc3} can only decrease the computed value of $\F$. Splitting  then
$\int_U |\nab u_{\rrh}|^2$ into $\int_{U\cap \Omega} |\nab u_{\rrh}|^2+ \int_{U\cap \Omega^c} |\nab u_{\rrh}|^2$, we deduce the result  in view of the definition \eqref{Glocal2}.

 {\bf Step 5: proof of \eqref{fluctuationsbu}.}  The proof of \eqref{14} is similar to that of \eqref{coulombfluct}. Integrating~\eqref{defv} against $\varphi$ and using Green's formula, we  have 
\be\label{rel3}
\left|\int_\Omega   \varphi\,  d  \Big ( \sum_{i=1}^N \delta_{x_i} -   \mu\Big)\right|= 
\frac{1}{\cd}\left| \int_{\Omega\cap U} \nabla u \cdot \nab  \varphi\right|
.\ee
We may also bound, using \eqref{eq:intf} and $\rrc_i\le 1$, 
$$\left|\int_\Omega \nab \varphi \cdot \nab (u-u_{\rrc})\right|\le \|\nab\varphi\|_{L^\infty(\Omega)} \sum_{i\in I_\Omega} \|\nab \f_{\rrc_i}\|_{L^1} \le C \|\nab\varphi\|_{L^\infty(\Omega)} \# I_\Omega .$$
Meanwhile, by Cauchy-Schwarz,
$$\frac{1}{\cd}\left| \int_{\Omega\cap U} \nabla u_{\rrc} \cdot \nab  \varphi\right|
\le \frac{1}{\cd} \| \nab  \varphi\|_{L^2(\Omega\cap U)} 
 \| \nab u_{\rrc} \|_{L^2(\Omega\cap U)}  .
$$
Combining the above we obtain the result. The proof of the relation in terms of $v$ is the same.
  \end{proof}
  
  %
   
   \subsection{Free energy bounds}
   In this subsection, we show analogous bounds on  the free energy for the Neumann energy analogous to those obtained in Section \ref{sec523}. The upper bounds follow easily 
 from \eqref{lbFneum} and the lower bounds will follow analogously to  those of Section \ref{sec523} from the subadditivity combined with the following lemma. This is a much simpler proof than that of \cite{as}, which allows to get rid of  technical deterioration of $\rho_\beta$  for $\d\ge 5$ in that paper.  
   \begin{lem}[Green's function representation of the Neumann electric energy]
   Let $U$ be an open subset of $\R^\d$  with bounded and piecewise $C^1$ boundary  and $\mu$ a bounded  nonnegative density such that $\mu(U)=N$ is an integer.
   Let $G_U$ solve 
   \begin{equation}\left\{\begin{array}{ll}
   -\Delta_x G_U (x,y)=\cd(\delta_y(x) - \frac{1}{\mu(U)} \mu (x)) & \text{in} \ U\\
   \frac{\partial G_U}{\partial \nu}= 0  & \text{on} \ \partial U\end{array}\right.\ee
   and let
   $$H_U(x,y)= G_U(x,y)-\g(x-y).$$
   Then for any configuration  $X_{N}$ of points in $U$, we have 
   \begin{multline}\label{737} \F(X_{ N},\mu, U) = \hal \iint_{\R^\d\backslash \triangle } \g(x-y) d\( \sum_{i=1}^{N} \delta_{x_i} - \mu\indic_U\)(x) d\( \sum_{i=1}^{ N} \delta_{x_i} - \mu\indic_U\)(y) 
   \\
   + \hal \iint_{U\times U} H_U(x,y) d\( \sum_{i=1}^{ N} \delta_{x_i} - \mu\)(x) d\( \sum_{i=1}^{ N} \delta_{x_i} - \mu\)(y) .\end{multline}
   \end{lem}
   \begin{proof}
   We check that $G_U(\cdot, y)$ and $H_U(\cdot, y)$ are well-defined up to additive constants, and that $H_U$ is continuous. Arguing as in \eqref{hmubb}, we have that   $\g*\mu$ is well-defined as an $L^1_{\mathrm{loc}}$ function, thus for a.e. $y$, we may consider $v=\g*(\delta_y- \frac{1}{\mu(U)}\mu)$ and solve for $w=G_U-v$ which satisfies 
    \begin{equation}\left\{\begin{array}{ll}
\Delta w= 0 & \text{in} \ U\\
\frac{\pa w}{\partial \nu}= -\frac{\pa v}{\partial \nu}& \text{on} \ \pa U.\end{array}\right.\end{equation}
This can be done variationally since $\pa U$ is bounded.
Since $G_U$ is defined up to an additive constant, we may in addition require that 
\be\label{intGU}
 \forall y \in U, \ \int_U G_U(x,y) d\mu(x)=0.\ee
We may prove in the standard way that $G_U$ is symmetric, hence so is $H_U$, by writing
\begin{multline}
G_U(y_0,x_0)- G_U(x_0, y_0)\\
= \int_U G_U(x,x_0 ) \( -\frac1\cd\Delta G_U(x,y_0) +\frac{\mu(x)}{\mu(U)}\) -G_U(x,y_0) \(-\frac1\cd\Delta G_U(x,x_0) +\frac{ \mu(x)}{\mu(U)} \)dx=0\end{multline}
after using Green's theorem, the Neumann boundary condition and \eqref{intGU}.

We may then observe that the function $u$ of \eqref{defv} satisfies 
\be  u(x)= \int_U G_U(x,y) d\( \sum_{i=1}^{ N} \delta_{x_i} - \mu\)(y).\ee
  Inserting this into \eqref{minneum},  using Green's theorem, the boundary condition in \eqref{defv} and the fact that the balls $B(x_i, \rrh_i)$ are included in $U$,  we  find
   \begin{multline}
   \F(X_N, \mu, U)= - \frac1{2\cd} \int_U u_{\rrh}
 \Delta u_{\rrh}   - \hal \sum_{i=1}^N \g(\rrh_i)  - \sum_{i=1}^N \int_{U} \f_{\rrh_i}(x - x_i) d\mu(x)
   \\  =  \frac{1}{2} \iint_{U\times U}   G_U(x,y)   d\Big(\sum_{i=1}^{ N} \delta_{x_i}^{(\rrh_i)}  - \mu\Big)(x)    d\Big(\sum_{i=1}^{ N} \delta_{x_i}^{(\rrh_i)}  - \mu\Big)(y)
    - \hal \sum_{i=1}^N \g(\rrh_i)  - \sum_{i=1}^N \int_{U} \f_{\rrh_i}(x - x_i) d\mu(x).\end{multline}
    Splitting $G_U$ as $\g(x-y)+H_U(x,y)$, we then obtain 
    \begin{multline}\label{7144}
   \F(X_N, \mu, U)=  \frac{1}{2} \iint_{U\times U}   \g(x-y)   d\Big(\sum_{i=1}^{ N} \delta_{x_i}^{(\rrh_i)}  - \mu\Big)(x)    d\Big(\sum_{i=1}^{ N} \delta_{x_i}^{(\rrh_i)}  - \mu\Big)(y)
    - \hal \sum_{i=1}^N \g(\rrh_i)  
  \\  +\frac{1}{2} \iint_{U\times U}  H_U(x,y)   d\Big(\sum_{i=1}^{ N} \delta_{x_i}^{(\rrh_i)}  - \mu\Big)(x)    d\Big(\sum_{i=1}^{ N} \delta_{x_i}^{(\rrh_i)}  - \mu\Big)(y)- \sum_{i=1}^N \int_{U} \f_{\rrh_i}(x - x_i) d\mu(x).
    \end{multline}
    For the first line we use the fact that the balls $B(x_i, \rrh_i)$ are disjoint and the harmonicity of $\g$ away from $0$, and  the fact that $\g*\delta_0^{(\eta)} =\g_\eta$ as seen in Section \ref{sec412},  to write 
    $$\int \g(x-y)d \delta_{x_i}^{(\rrh_i)}(x) d \delta_{x_i}^{(\rrh_i)}(y)= \g(\rrh_i), \qquad
     \int \g(x-y)d \delta_{x_i}^{(\rrh_i)}(x) d \delta_{x_j}^{(\rrh_j)}(y)= \g(x_i-x_j) \ \text{if} \  i\neq j.$$
     We thus  obtain that 
     \begin{multline}
      \frac{1}{2} \iint_{U\times U}   \g(x-y)   d\Big(\sum_{i=1}^{ N} \delta_{x_i}^{(\rrh_i)}  - \mu\Big)(x)    d\Big(\sum_{i=1}^{ N} \delta_{x_i}^{(\rrh_i)}  - \mu\Big)(y)
    - \hal \sum_{i=1}^N \g(\rrh_i) \\
     = \frac{1}{2} \iint_{U\times U\backslash \triangle}   \g(x-y)   d\Big(\sum_{i=1}^{ N} \delta_{x_i}  - \mu\Big)(x)    d\Big(\sum_{i=1}^{ N} \delta_{x_i} - \mu\Big)(y)
\\- \iint_{U\times U\backslash \triangle} \g(x-y) d\Big(\sum_{i=1}^{ N} \delta_{x_i}^{(\rrh_i)}- \delta_{x_i}\Big) (x) d \mu(y).
   \end{multline} 
   Recalling (see \eqref{fconv}) that  $\g*\delta_0^{(\eta)}-\g =\g_\eta-\g =-\f_\eta$, we may rewrite the second line in the right-hand side as $ - \sum_{i=1}^{ N} \int_U \f_{\rrh_i}(x-x_i) d\mu(x)$, and
   we have thus established that 
\begin{multline}\label{7145} \frac{1}{2} \iint_{U\times U}   \g(x-y)   d\Big(\sum_{i=1}^{ N} \delta_{x_i}^{(\rrh_i)}  - \mu\Big)(x)    d\Big(\sum_{i=1}^{ N} \delta_{x_i}^{(\rrh_i)}  - \mu\Big)(y)
    - \hal \sum_{i=1}^N \g(\rrh_i) + \sum_{i=1}^{ N} \int_U \f_{\rrh_i}(x-x_i) d\mu(x) \\ 
    = \frac{1}{2} \iint_{U\times U\backslash \triangle}   \g(x-y)   d\Big(\sum_{i=1}^{ N} \delta_{x_i}  - \mu\Big)(x)    d\Big(\sum_{i=1}^{ N} \delta_{x_i} - \mu\Big)(y).\end{multline}
We next turn to the second line of \eqref{7144} and rewrite it as 
\begin{multline}\label{7146} \frac{1}{2} \iint_{U\times U}  H_U(x,y)   d\Big(\sum_{i=1}^{ N} \delta_{x_i}^{(\rrh_i)}  - \mu\Big)(x)    d\Big(\sum_{i=1}^{ N} \delta_{x_i}^{(\rrh_i)}  - \mu\Big)(y)\\
=  \frac{1}{2} \iint_{U\times U}  H_U(x,y)   d\Big(\sum_{i=1}^{ N} \delta_{x_i} - \mu\Big)(x)    d\Big(\sum_{i=1}^{ N} \delta_{x_i}  - \mu\Big)(y)
\\+\hal \iint_{U\times U} H_U(x,y) d\Big(\sum_{i=1}^{ N} \delta_{x_i}^{(\rrh_i)}- \delta_{x_i}\Big) (x) d\Big(\sum_{i=1}^{ N} \delta_{x_i}^{(\rrh_i)}  + \delta_{x_i} -2 \mu\Big)(y).
   \end{multline} 
Recognizing $\delta_{x_i}^{(\rrh_i)}-\delta_{x_i}$ as the Laplacian of $\f_{\rrh_i}(x-x_i)$ from \eqref{deltaf}  and integrating by parts in $x$, using that $-\Delta H_U(x,y)= -\cd \frac{\mu(x) }{\mu(U)}$ we may rewrite the last line as 
$$- \iint_{U\times U}\frac{ \mu(x)}{\mu(U)} \( \sum_{i=1}^{ N} \f_{\rrh_i}(x-x_i) \) d\Big(\sum_{i=1}^{ N} \delta_{x_i}^{(\rrh_i)}  + \delta_{x_i} -2 \mu\Big)(y)=0,
$$ by separation of variables, since $\mu(U)= N$. Inserting into \eqref{7146} and combining with \eqref{7145} and \eqref{7144}, we have obtained the result.\end{proof}

We next turn to the lower bound of $\log \K_\beta(\mu,U)$, defined in \eqref{defK7}.

\begin{prop}[Neumann free energy bound]\label{pro718}
 Let $U$ be an open subset of $\R^\d$  with bounded and piecewise $C^1$ boundary  and $\mu$ a bounded  nonnegative density such that $\mu(U)=N$ is an integer. If $\s\le 0$ and $U$ is unbounded, assume in addition that  \eqref{assumplbs}, \eqref{assgmm} and \eqref{assumpbeta} hold. 
Then
\be 
\label{bornesfiU}\left| \log \K_\beta(\mu,U) \right|\le C \beta \chi(\beta) N\ee
where $\chi$ is as in \eqref{defchibeta}, and $C>0$ depends only on $\d, \s, \|\mu\|_{L^\infty}$ and the constants in the assumptions.
\end{prop}
\begin{proof} The upper bound is straightforward from the definition \eqref{defK7} and \eqref{lbFneum}. For the lower bound, 
we follow the steps of the proof of Lemmas  \ref{prominok} and Proposition \ref{pro642}.
Assume first that $\s>0$.
Starting from~\eqref{defK7}  and using  Jensen's inequality, we may then write
\begin{equation*}
\log  \K_\beta(\mu,U)\ge - \frac{\beta}{(\mu(U))^N}    \int_{U^N}
  \F(\XN, \mu,U)   
 d\mu^{\otimes N} (X_N).
\end{equation*} 
We next insert \eqref{737} and argue as in \eqref{alignfbup} 
  to obtain
\be
\int_{U^N} \F(\XN, \mu,U)   d \mu^{\otimes N}(\XN)
 = -\frac{ N^{N-1}}{2} \iint_{U^2}\g(x-y) d \mu(x) d\mu(y),\ee after observing that the terms corresponding to the integrals of $H_U(x,y)$ cancel out.
 When $\s>0 $,  this yields $ \log \K_\beta(U, \mu)\ge 0$ hence the desired result.
 
 In the situation where $\s\le 0$, we may argue exactly as in the proof of Proposition \ref{pro642}: first we partition $\Lambda$ into cells $Q_i$ of size $O(R)$ with $\int_{Q_i}\mu=n_i$, then we replace the use of \eqref{commesubad} by \eqref{superad2}. Finally when integrating $\F(X_N, \mu,Q_i)$ within each cell, we notice that the contribution of the terms containing $H_{Q_i}$ in \eqref{737} cancel after integration, so we are left with simply the terms in $\g(x-y)$, and 
 $$\log \K_\beta(\mu,U) \ge \frac\beta2\sum_{i=0}^p \frac1{n_i} \iint \g(x-y) d\mu_i(x)d\mu_i(y) + O(\log n_i)$$ and we finish exactly as in the proof of Proposition \ref{pro642}.
\end{proof}

Using \eqref{lbFneum}, we have a lower bound for $\min \F(\cdot,  \mu,U)$, and from \eqref{bornesfiU} applied to, say, $\beta=1$ we get that the exists a configuration such that a converse inequality holds, thus we have the following. 

\begin{coro}[A priori bounds for minimizers]
Let $U$ and $\mu$ be as in Proposition \ref{pro718}, under the same assumptions, defining
\be \label{defEi}
\mathsf{E}_\infty ( \mu,U):= \min_{\XN} \F( \XN, \mu,U),\ee
 we have 
\be \label{bornesfieU}
|\mathsf{E}_\infty(\mu,U) |\le C N,\ee
where $C>0$ depends only on $\d, \s, \|\mu\|_{L^\infty}$ and the constants in the assumptions.
\end{coro}
In view of \eqref{subad1} we also have the subbaditivity property: if $U$ is the disjoint union of $U_1$ and $U_2$, 
\be \label{subad2}
\mathsf{E}_\infty(\mu,U)\le \mathsf{E}_\infty(\mu, U_1)+\mathsf{E}_\infty(\mu,U_2).\ee
     \section{The screening procedure}
  \subsection{Motivation and heuristics}
  \index{screening}

Since  Neumann boundary conditions  allow to compute the energy subadditively over disjoint sets, this motivates the screening procedure, first introduced in \cite{ssgl} using ideas of \cite{ACO}. It  consists in taking an arbitrary configuration in a cube (or rectangle) or its complement, with reasonably well controlled energy, and producing from it a configuration with Neumann boundary condition solved.  The issue is to do so without changing  the configuration too much -- it will be possible to modify the configuration only in a boundary layer -- and without adding too much energy. During the procedure, points will be deleted, the total number of points will change so as to achieve neutrality, and this modifies the phase-space volume of the family of configurations. We will also show that this change of volume can be well-controlled, which allows to give a probabilistic screening (or screening at the level of Gibbs measures): given a family of configurations we will  produce a whole family of configurations, which will allow, roughly,  to bound $\L_{N,\beta}$ in terms of $\K_\beta$.

\begin{rem}
\label{twosit}
The two main situations we need to treat are:
\begin{enumerate} 
\item the case where $U=\R^\d$ and $\mu$ is a positive density with compact support or with convergent tails and  
$\int_{\R^\d} \mu=N$. We will denote $\Lambda $ a set where $\mu \ge m>0$ for some positive constant $m$ (either the support of $\mu$, or its ``essential support" if it has tails). 
\item the case where $U$ is a hyperrectangle (or disjoint union of hyperrectangles), we consider the Neumann partition function on $U$, and $\mu$ is a uniform density, say $1$. Then we take $\Lambda=\R^\d$.
\end{enumerate}
\end{rem}
To perform the screening, we need to be in a region where the background density $\mu$ is not too small, because it is the negative background charge $-\mu$ which is used to neutralize and screen possible charge excesses. Thus, in the first case above, screening will only be possible in ``the bulk" i.e.~a region where $\mu$ is bounded from below by some constant $m>0$ (and the error estimates will depend on $m$).  Truly, we will have to be a small distance away from the set $\{\mu <m\}$ for results to hold. 
In the second case $\mu=1$ up to the boundary, and the Neumann boundary condition holds there, the screening will preserve the Neumann boundary condition and  there will be no difficulty in applying the  screening procedure up to the boundary.

We have to perform two variants of the screening: an ``inner screening" when $\Omega=Q_R$ and an ``outer screening" when $\Omega=U \backslash Q_R$. Both are entirely  parallel, so we present below the inner screening. Here $Q_R$ is a hyperrectangle with sidelengths in $[R, 2R]$, and  for any $t \in \R$ we will denote
 by $Q_{R+t}$ the hyperrectangle of same center as $Q_R$ and sidelengths increased by $t$.

The set  $\Omega$ needs to be  ``quantized" in the sense that $\mu(\Omega)$ is an integer, equal to $ \mn$, which is also the optimal number of points. 
Assume we are given a configuration $X_n$ of $n$ points in $\Omega= Q_R\cap U$, and let $u$ be an associated potential satisfying a relation of the form 
$$-\Delta u= \cd\( \sum_{i=1}^n \delta_{x_i}- \mu\) \quad \text{in} \ \Omega$$
for instance $u$ is the function solving \eqref{defu}, and $E=\nab u$.

The goal of the (inner) screening is to produce a family of configurations $Y_\mn$ having $\mn$ points in $\Omega$, coinciding with $X_n$ except in a boundary layer near $\pa \Omega$, for which we can solve 
$$\left\{ \begin{array}{ll}
- \div E^0= \cd(\sum_{i=1}^{\mn} \delta_{y_i} - \mu) & \text{in} \ \Omega\\
E^0 \cdot \nu =0 & \text{on} \ \partial \Omega\end{array}\right.$$
The electric field $E^0$ will also coincide with $E$ except in a boundary layer near $\pa \Omega$. We call $\Old$ (like old) the interior set where the configurations  $X_n$ and $Y_\mn$ and their electric fields coincide, and by $\New $ (like new) the boundary layer where $X_n$ and $E$ are to be deleted and  replaced  by  configurations that have the correct number of points. A recent feature of the construction, which appeared in \cite{as}, is to sample these new points according to a Coulomb Gibbs measure in $\New$.

We say that the configuration $X_n $ has been screened because the resulting electric field $E^0$ can be extended to $0$ outside $\Omega$ without any jump in the normal component, as if the system of point charges + background charge $\mu$ was not generating any field outside $\Omega$.
 The screened electric field $E^0$ may not be a gradient, however thanks to Lemma \ref{projlem} its energy provides an upper bound for computing $\F(Y_\mn,\Omega)$. The goal of the construction is to show that we can build $E^0$ and $Y_\mn$ without adding too much energy to the  original one.
 
 Not all configurations are screenable: configurations that have for instance too many points very near the boundary, and no points inside, cannot be transformed into neutral configurations by changing the configuration in a boundary layer only. A reasonable energy bound rules out such pathological behavior and will be the ``screenability condition" \eqref{screenab}. 
 
 The boundary layer size depends on two lengthscales of choice $\ell $ and $\tilde \ell$, both $<R$.
The lengthscale $\tilde \ell$ represents the distance over which one needs to look for a good contour near $\pa \Omega$ by a mean-value argument. For outer screening, this implies that we need a bit of buffer space to perform the screening argument.
  We will denote by $\bar S(X_n)$ the energy of the best screenable electric field $\int |E_\rr|^2$ in a $\tilde \ell$-neighborhood of $\pa \Omega$.  By a mean-value argument we find some contour $\Gamma$ at distance $\le \tilde \ell$ from $\pa \Omega$ so that the trace of $E\cdot \nu$ on $\Gamma$, denoted $g$, satisfies $\int_{\Gamma} |g|^2 \le C \bar S(X_n)\tilde \ell^{-1}$.
For the given configuration $X_n$, we denote by $\Old$ the set enclosed by $\Gamma$ and by $\New $ its complement. We let $n_\Old$ be the number of points of $X_n$ that belong to $\Old$.  The part of the  configuration $X_n$ which belongs to $\New$ is discarded, to be replaced by a configuration $Z_{\mn - n_\Old}$, in such a way that $Y_\mn$, union of $X_n $ restricted to $\Old $ and $Z_{\mn-n_\Old}$ has exactly $\mn$ points, as desired.

The other lengthscale $\ell\le \tilde \ell$, represents the distance to $\Gamma$ needed to ``absorb" the possibly nonzero boundary condition $g$ and replace it by a vanishing one.  That size needs to be large enough depending on $\bar S(X_n)/\tilde \ell$ and the screening will only be possible if $\bar S(X_n)/\tilde \ell$ is small enough compared to $\ell $, this is the screenability condition. 

Finally the lengthscale $\eta<\ell$ corresponds to a buffer point-free zone at distance $\le \eta$ from $\Old$ in which no points are placed. In order to ensure small entropy errors, this lengthscale will need to be taken small when $\beta$ gets small.

The ``absorption" of $g$ and its replacement by $0$ Neumann boundary data (or zero boundary normal component for $E$) is done by splitting $\New$ 
 into cells  $\mathcal C_k$ of sidelength $ \ell$ where we  solve appropriate elliptic problems and estimate the energies by elliptic regularity estimates.  More precisely  the building block is to solve 
\begin{equation}
\left\{\begin{array}{ll}
-\Delta u_k = \cd\( \sum \delta_{z_i}-\mu\) & \text{in} \ \mathcal C_k
\\
\frac{\pa u_k}{\partial \nu} = \tilde g & \text{on}\,  \pa \mathcal C_k
\end{array}\right.\end{equation}
where $\tilde g$ is  equal to $g$ on the parts of $\mathcal C_k$ which intersect $\Gamma$ (if any) and $0$ otherwise, and $z_i$ are points to be sampled in $\New$, leaving a point-free layer of width $\eta$ near the boundary of $\Old$.

 Pasting together  the electric fields $\nab u_k$ obtained  in each $\mathcal C_k$ produces an  electric field which is compatible with $Z_{\mn-n_\Old}$, allowing to evaluate the energy $\F$ in $\New$ 
 via Lemma \ref{projlem}, i.e. bounding it by $\sum_k |\nab u_k|^2. $

The evaluation of the energy in each cell $\mathcal C_k$  follows the idea of \cite{ACO}  of doing it in two parts by decomposing $u$ into $h_1+h_2$ where 
\begin{equation}
\left\{\begin{array}{ll}
-\Delta h_1 = \cd\( \sum \delta_{x_i}- \mu - m_k\) & \text{in} \ \mathcal C_k
\\
\frac{\pa h_1}{\partial \nu} = 0 & \text{on} \,  \pa \mathcal C_k
\end{array}\right.\end{equation}
 where $m_k$ is a constant chosen so that this problem is solvable and won't be too large (in particular we can keep $\mu+ m_k >0$) as soon as the initial energy is not too large; and 
 \begin{equation}
\left\{\begin{array}{ll}
-\Delta h_2 = \cd m_k& \text{in} \ \mathcal C_k
\\
\frac{\pa h_2}{\partial \nu} = \tilde g & \text{on} \, \pa \mathcal C_k
\end{array}\right.\end{equation}
 which will absorb the boundary condition.  This will amount to replacing in the boundary layer $\New$ the reference measure $\mu$ by a modified measure $\tilde \mu$.
 By elliptic estimates, the energy cost is directly related to $\int_{\Gamma} |g|^2$, itself controlled by $\bar S(X_n)/\tilde \ell$.
 Note that some complications arise due to points that are very close to $\Gamma$ whose balls $B(x_i, \rr_i)$ cut through $\Gamma$.

 The energy error term (the sum of the old energy with $\F(Z_{\mn-n_\Old }, \New)$) can the  be bounded by  the number of such cells 
 plus $\bar S(X_n) /\tilde \ell$.
This produces an error on the free energy of order
 $$ C\beta R^{\d-1}\ell+ C \frac{ R^{\d-1}}{\ell}.$$
  Optimizing this error over $\ell<R$  leads to stopping at $\ell= \frac{1}{\sqrt{\beta}}$  and $R > C\max(1, \frac{1}{\sqrt\beta})$.
  This is how our {\it minimal lengthscale} $\rho_\beta$ of order $\max(1, \frac{1}{\sqrt{\beta}})$ appears.
  When $\beta $ is small this becomes large, and screening cannot be obtained at smaller lengthscale.
 In physics, $\rho_\beta$ corresponds to the {\it Debye screening length}.
 
 \begin{figure}[h!]
\begin{center}
\includegraphics[scale=0.9]{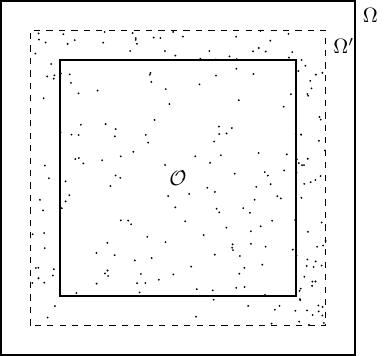}
\caption{Setup for the screening}
\end{center}
\label{figureecrantage}
\end{figure}

\subsection{Screening statement} \label{sec:descriscri}

The following result can be found in \cite{as}, prior versions are found in \cite{ssgl,rs,PetSer,lebles}.

Let us  first start with the geometry and screenability condition.
\begin{defi}[Quantized hyperrectangles of lengthscale $R$]\label{defQR}
We define $\mathcal Q_R$ as the  set of closed hyperrectangles $Q$ in $\R^\d$ whose sidelengths belong to $[R, 2R]$ and are such that $ \mu(Q) $ is an integer.
\end{defi}
We also denote $\carr_r(x)$ the cube of sidelength $r$ centered at $x$. 
The presence of the set $\Omega'\subset \Omega$ is meant to be able to take a configuration in a set $\Omega'$, screen it and extend it to a slightly larger set $\Omega$.  Since the configuration and its electric field will be completely discarded in a boundary layer near $\partial \Omega'$, it need only  satisfy \eqref{eqsp} in a possibly slightly smaller set than $\Omega'$, which will be useful later in conjunction with the best screenable energy $\G_U$. By slightly smaller, we mean that the   distance between the nested sets $\Omega'', \Omega' $ and $\Omega$ is at most $2\tilde \ell$.

    There are two variants of the construction: one, used to prove the local laws, starts from the control on the energy  in $Q_R$ only, and finds a good boundary which is the boundary of a cube; the other, used once local laws are known, starts from the local  control of the energy on cubes of size $\rb$ to  obtain improved error estimates, and uses a good boundary which is piecewise affine in order to have local controls on the boundary.  In the first variant, one uses the bounds on $S$  below, and in the second variant the bounds on $S'$.

\begin{defi}[Screenability] \label{defscreen} Assume $U$ is either $\R^\d$ or a  finite disjoint union of  hyperrectangles with parallel sides, belonging to $ \mathcal Q_R$
and all included in $\Lambda$;  or the complement of such a set. 
Assume $\mu$ is a bounded density satisfying $ \mu\ge m>0  $ in 
$\Omega= Q_{R}\cap U $ (inner case), resp. $\Omega= U\backslash Q_R$ (outer case) where $Q_R$ is a hyperrectangle of sidelengths in $[R,2R]$ with sides parallel to those of $U$, and such that 
 $\mu(\Omega)=\mn$, an integer.
Let $\ell$ and $\tilde \ell$ be such that  $R\ge \tilde \l \ge \l \ge 1$. In the outer case, we also assume that if $U = \R^\d$, 
 \be Q_{R+\tilde \ell}  \subset\Lambda
 \end{equation}
 and in the case where $U$ is a disjoint union of hyperrectangles with parallel sides, that 
 the faces of  $\partial  Q_{R}   $  are at distance $\ge 2 \tilde \ell$ from their respective  parallel faces of $\partial U$.
 
 Let $\Omega''\subset\Omega'\subset \Omega$ and  $Q_{R-\tilde \ell } \cap U \subset \Omega''$ in the inner case, respectively $ U\backslash Q_{R+\tilde \ell}\subset \Omega''$ in the outer case.  
Let $X_n$ be a configuration of points in  $ \Omega'$ and let  $w$ solve
 \be \label{eqsp}
 \begin{cases} 
 -\Delta w= \cd \( \sum_{i=1}^n \delta_{x_i}-\mu\)  &  \text{in} \  \Omega''\\
 \frac{\partial w}{\partial \nu}= 0 & \text{on} \ \partial U \cap  \Omega''.\end{cases}
  \ee
We denote  if $\Omega= Q_R \cap U$,
\be \label{definitionsannex0}
S(X_n,w) =  \int_{(Q_{R-\tilde \l}\backslash Q_{R-2\tilde \l})\cap U} |\nab w_{ \rrh}|^2  \qquad S'(X_n,w)=\sup_x \int_{(Q_{R-\tilde \l}\backslash Q_{R-2\tilde \l}) \cap \carr_{ \l}(x)\cap U} |\nab w_{ \rrh}|^2,\ee 
respectively if $\Omega= U \backslash Q_R$, 
\be \label{definitionsannex2}
S(X_n,w) =  \int_{(Q_{R+2\tilde \l}\backslash Q_{R+\tilde \l} )\cap U } |\nab w_{\rrh}|^2 
\qquad S'(X_n,w)=\sup_x \int_{(Q_{R+2\tilde \l}\backslash Q_{R+\tilde \l}) \cap \carr_{\l}(x)\cap U} |\nab w_{ \rrh}|^2,\ee where $\rrh$ is defined as in \eqref{defrrc3}.
We say that $w$ is inner screenable (resp. outer screenable) if 
\be \label{screenab}
 \l^{\d+1}\ge C\min \(\frac{S(X_n,w)}{\tilde \ell},S'(X_n,w)\),\ee for some $C>0$  depending only on $\d$ and $m$.
 \end{defi}
 We can now define the best screenable potential and  its energy.
 There are two complications in the definition compared to the formal definition of the Dirichlet energy in Section~\ref{sec1}. The first one is that we have to retain the Neumann boundary condition on $\pa U$  when we treat the case of Neumann partition functions (as in item 2) of Remark~\ref{twosit}). The second is that we have to add to the definition the condition of being screenable. During the bootstrap proof of the local laws, the bounds on the energy obtained at prior scales are sufficient to ensure inner  screenability in small boxes, however we cannot know for certain that outer screenability also holds on the sole basis of these energy controls (because the needed screenability is in terms of the  boundary size of $U\backslash Q_R$ which when $R$ is relatively small is much smaller than the volume of $U\backslash Q_R$).  
 
We may now define  the appropriate notion of the quantity $\G$.
 \begin{defi}[Best screenable potential and energy] \label{defibestpot}
 With the same notation as above and $\rrh$ as in \eqref{defrrc3}, given a configuration $X_n$ of points in $\Omega$,
 we let 
 \begin{multline}
 \label{innernrj}\G_U^{\mathrm{inn}}(X_n, \mu, \Omega) =\min  \Big\{  \frac{1}{2\cd} \( \int_{\Omega}  |\nab w_{\rrh} |^2 - \cd \sum_{i=1}^n \g(\rrh_i) \) - \sum_{i=1}^n \int_{\Omega }\f_{\rrh_i} (x-x_i) d\mu(x),\\
  w \, \mathrm{inner  \ screenable\ satisfying\  a\  relation\ of\  the\ form } \\
  \left\{\begin{array}{ll}
  -\Delta w= \cd\(\sum_{i=1}^n \delta_{x_i}- \mu + \sum_{j} \delta_{x_j}^{(\eta_j)}\)& \text{in} \ \Omega\cap U\\ \frac{\partial w}{\partial \nu} = 0 & \text{on} \ \partial U\cap \partial \Omega\end{array}\right.
  \\
 \mathrm{ where}\   x_j \notin \Omega, \ \eta_j \le \frac14 \min (1, \dist(x_j, \partial U)), \ w_{\rrh}= w-\sum_{i=1}^n \f_{\rrh_i}(\cdot -x_i), \ \rrh\ \mathrm{as\  in\ \eqref{defrrc3}} 
  \Big\}, \end{multline}
 respectively, 
 \begin{multline}
 \label{outernrj} \G_U^{\mathrm{out}}(X_n,\mu, \Omega) =\min\Big\{  \frac{1}{2\cd} \( \int_{\Omega}  |\nab w_{\rrh} |^2 - \cd \sum_{i=1}^n \g(\rrh_i) \) - \sum_{i=1}^n \int_{\Omega }\f_{\rrh_i} (x-x_i) d\mu(x),\\
  w \,\mathrm{outer \ screenable \ satisfying\ a\ relation\ of \ the \ same \ form \ in \ } \Omega\Big\}. \end{multline}
 By the direct method in the calculus of variations, one may check that the minima are achieved.
 We also define 
 \begin{align}\label{bestS}
&  \bar S(X_n)= \inf \{ S(X_n, w), w \ \text{achieving the min in $\G_U^{\mathrm{inn}}(X_n, \Omega) $, resp. $\G_U^{\mathrm{out}} (X_n, \Omega) $}\} 
 \\ \label{bestS2}  & \bar S'(X_n)= \inf \{ S'(X_n, w), w \ \text{achieving the min in $\G_U^{\mathrm{inn}}(X_n, \Omega)$, resp. $\G_U^{\mathrm{out}} (X_n, \Omega) $}\} 
 .\end{align}
\end{defi}
\begin{rem}\label{rem725}
The introduction of $\sum_j \delta_{x_j}^{(\eta_j)}$ in the equation solved by $w$ is meant to account for the possibility of points outside $\Omega$ whose smeared charges could overlap $\Omega$. That additional charge can only be supported in a layer of distance $\le \frac14$ from $\partial \Omega$, so that $w$ solves 
\eqref{eqsp} in 
$\Omega''= \Omega \backslash \{ x,\dist(x, U\backslash \Omega) \le \frac14\}$, which allows to apply the screening procedure to $w$ in  $\Omega$.
 Note also that the radii $\rrh_i$ depend only on the points in $\Omega$, i.e.~are determined by $X_n$ only, so $\G_U^{\mathrm{inn/out}}$ is a well-defined function of $X_n$.

\end{rem}
When $u$ is inner/outer screenable, it is a competitor  in the definition of $\G_U^{\mathrm{inn/out}}$, thus we have the following.
\begin{lem}\label{lemrestri}
Let $u $ be the solution of \eqref{defv} used in the definition of \eqref{Glocal2}. If $u$ is inner screenable, resp.~outer screenable, then 
\be \F^{\Omega}(\XN, \mu, U) \ge \G_U^{\mathrm{inn/out}}(\XN|_{\Omega}, \mu,\Omega) .\ee
\end{lem}

The following proposition is proved in the appendix. 
\begin{prop}[Screening]\label{proscreen} Let us use the same assumptions and notation as in Definition \ref{defscreen}. Given $\eta\ge 0$ such that 
\be \label{conditionsureta}
 \eta \le \ell \frac{m}{4\|\mu\|_{L^\infty}},\ee
there exists   $C>5$ depending only on $\d,m$ and $\|\mu\|_{L^\infty}$  such that the following holds. Let $X_n  $ be a configuration of points in  $ \Omega'$ and let  $w$ solve
 \eqref{eqsp} in $\Omega''\subset \Omega'$ and be screenable in the sense of \eqref{screenab}.
   There exists
   a set $\Old $ such that $Q_{R-2\tilde \ell}\cap U \subset \Old \subset Q_{R-\tilde \ell }\cap U$ in the case of inner screening (resp. $   U\backslash Q_{R+2\tilde \ell}\subset \Old \subset U\backslash Q_{R+\tilde \ell }$ for outer screening), a subset~$I_\pa\subseteq \{1,\ldots,n\}$
    and a positive  measure  $\tilde \mu$ supported  in 
   $\New_\eta:=\{x\in \New, \dist(x, \Old) \ge \eta\}$ with 
    $\New:= \Omega \backslash \Old$ (all depending on $X_n$) such that the following holds:
\begin{itemize}
\item $\N$ being  the number of points of $X_n$ such that $B(x_i, \rrh_i)$ intersects $\Old$, we have 
\be \label{bornimp}
\tilde \mu(\New)=\tilde \mu(\New_\eta)= \mn-\N,\qquad |\mu(\New)-\tilde \mu(\New)|\le C\(R^{\d-1}+ \frac{S(X_n,w)}{\tilde \ell}\)
\ee
\be
\label{mmut2}  \|\mu -\tilde \mu\|_{L^\infty(\New_\eta)} \le \frac{m}{2},\qquad 
\int_{\New_\eta} (\tilde \mu-\mu)^2 \le C \frac{S(X_n,w)}{\l\tilde \ell}+ C \frac{\eta^2}{\ell} R^{\d-1}
\ee
\item we have $\# I_\pa\le C \frac{S(X_n,w)}{\tilde \ell}$
\item   for any configuration $Z_{\mn-\N}$ of $\mn -\N$ points  in $\New_\eta$, the configuration $Y_{\mn}$ in $\Omega$ equal to the union of the points $x_i$ of $X_n$ such that $B(x_i, \rrh_i)$ intersects $\Old$  and the points $z_i$ of  $Z_{\mn - \N}$ (which thus has $\mn$ points) satisfies
  \begin{multline} \label{nrjy}
\F(Y_\mn, \mu, \Omega)\le \frac{1}{2\cd} \( \int_{\Omega'}  |\nab w_{\rrh} |^2 - \cd \sum_{i=1}^n \g(\rrh_i) \) - \sum_{i=1}^n \int_{\Omega'}\f_{\rrh_i} (x-x_i) d\mu(x)
\\+  C\( \frac{\l S(X_n,w)}{\tilde \l}  +   R^{\d-1} \tilde \ell+  \F(Z_{\mn-\N} ,\tilde\mu, \New_\eta)   + |\mn-n|
+
\sum_{(i,j) \in J} \g(x_i-z_j) \)
\end{multline} 
where the index set $J=J(X_n)$ in the sum is given by
\begin{equation*}
J:=\left\{ (i,j) \in I_\partial \times \{1,\ldots, \mn-n_\Old\}  \,:\, |x_i-z_j|\le \rrh_i \right\}. 
\end{equation*}
\end{itemize}
Moreover, if $\ell^{\d+1}> C\frac{S(X_n,w)}{\tilde \ell}$ in \eqref{screenab}, we can take $ \Old$ to be equal to $Q_t\cap U$ for some $t \in [R-2 \tilde \ell, R-\tilde \ell]$ (resp. $U\backslash Q_t $ for some $t \in [R+\tilde \ell, R+ 2\tilde \ell]$), while if not, the boundary of $\Old $ is in general piecewise affine and made of  facets parallel to the faces of $Q_R$ of sidelengths bounded above and below by constants times $\ell$, all included in some $Q_{t+\ell}\backslash Q_t$ for $t \in [R- 2\tilde \ell , R-\tilde \ell -\ell] $, resp.  $t\in [R+\tilde \ell, R+2\tilde \ell-\ell]$.
\end{prop}

Applying the proposition to $w$ achieving the min in \eqref{innernrj}, resp. \eqref{outernrj}, we directly obtain the  upper bound 
\begin{multline}\label{ubapscreening}
\F(Y_{\mn}, \mu, \Omega) - \G_U^{\mathrm{inn/out}}(X_n,\mu, \Omega')\\
\le  C\( \frac{\l S(X_n,w)}{\tilde \l}  +   R^{\d-1} \tilde \ell+  \F(Z_{\mn-\N} ,\tilde\mu, \New_\eta)   + |n-\mn|
+
\sum_{(i,j) \in J} \g(x_i-z_j) \)\end{multline}
with the right-hand side equal to the various screening errors (boundary energy, new added energy,  interactions of points near the boundary of $\Old$). 

Once this result is established one may tune the parameters $\l, \tilde \l$ to obtain the best results.
For instance, at the beginning we may only know that $\int_{Q_R} |\nab w_{\rrh}|^2 $ is bounded by $O(R^\d)$, we then bound $S(X_n,w)$ and $S'(X_n,w)$ by $O(R^\d)$, optimize the right-hand side of~\eqref{nrjy} and choose $\ell \le\tilde \ell$ satisfying the constraints and obtain 
$$\F(Y_{\mn},  \Omega) \le \G_U(X_{n} , \Omega) + C (R^{\d-\sigma}+  |n-\mn|),$$
for some $\sigma>0$, i.e.~we get an error which is smaller than the order of the energy. The error $|n-\mn|$ can be controlled via the energy on a slightly larger domain, and shown to be negligible as well.

At the end of the bootstrap argument in the next chapter, we will know that the energy and points are well distributed down to say, scale $C$.
This means that we then know that (for good configurations) $\bar S'(X_n)$ is controlled by $\tilde \l^\d$ and $\bar S(X_n)$ by $R^{\d-1}\tilde \l$. The condition~\eqref{screenab} is then automatically satisfied and  we can  thus take $\l=C$, $\tilde \l=C$, and  we may also control $|n-\mn|$ by $O(R^{\d-1})$ to obtain a bound 
$$\F(Y_\mn, \Omega)\le \G_U(X_{n} , \Omega)+ C R^{\d-1}$$
i.e.~with an error only proportional to the surface, the best one can hope to achieve by this approach.

\begin{rem}If one is not interested in obtaining $\beta$-independent estimates, one may simply choose $\eta= 1$ in the above, then since $\rrh_i \le \frac 14$, the point-free zone $\New \backslash \New_\eta$ guarantees that the set $J$ is empty.
\end{rem}

The screening is also possible in the Riesz cases $\d-2<\s<\d$ using the dimension extension procedure, it was first done in \cite{PetSer}. The extension involves additional  technical difficulties in the construction.

The screening procedure is instrumental to show the local laws of Chapter \ref{chaploiloc}, and also  free energy expansions and  CLT results for fluctuations of linear statistics in Chapters \ref{chapclt} and \eqref{chap:clt2}.

  \chapter[Local laws and almost additivity]{Local laws and almost additivity of the free energy} \label{chaploiloc}
  
  In this chapter, we continue to focus  on the Coulomb case. 
The goal is twofold: \\
(i) prove the almost additivity of the free energy as outlined in the previous chapter, thanks to the screening procedure.\\ 
(ii) provide analogues of the concentration bounds of \eqref{firstconcbound} but localized at mesoscopic and microscopic scales, i.e.~prove that 
\be \label{loclawaprouver}\left| \log \Esp_{\Q_\beta(U,\mu)} \( \exp \frac\beta2\F^{\carr_R} (\XN, \mu,U)\)\right|\le \mathcal C  \beta \chi(\beta) R^\d\ee
where $\Q_\beta$ is as in \eqref{defQ} and $\F^{\carr_R}$ as in \eqref{Glocal2}. This  means 
 that the share of the energy in a cube $\carr_R$ of size $R$ is controlled by the volume of that cube, as long as $ \rho_\beta \le R \le N^{1/\d}$ with 
 \index{minimal scale}
  \be \label{defrhobeta}
 \rho_\beta = C \max\(1, \sqrt{\frac{\chi(\beta) }{\beta}}\)\ee the minimal scale or Debye length scale already alluded to at the end of the previous chapter. Since we are discussing things in blown-up coordinates, this corresponds to a control down to the (temperature-dependent) microscale $\rho_\beta N^{-1/\d}$ in original coordinates.

  This local law will allow to control the number of points that can fall in a small cube and provide discrepancy estimates, and the local energy control will also be crucial in conjunction with the use of Theorem \ref{thm:FI} to obtain control of fluctuations in the next chapter. Such controls can be viewed as a manifestation of {\it rigidity} of the Coulomb gas down to the minimal scale.

The local law down to microscale allows to deduce that the energy in a boundary layer of microscopic size around $\pa \carr_R$ is proportional to the surface $\rho_\beta R^{\d-1}$ which gives an optimal screening error of that order, and in view of the heuristics given in the previous chapter, yields almost additivity of the free energy up to such boundary terms, see Proposition \ref{proaddi}, and in particular, as announced in the previous chapter,  that $\frac{\log \K_\beta( \carr_R,1)}{\beta R^\d}$ converges to a limit at speed $R^{\d-1}$, see Theorem \ref{th1}, with the correction due to $\rho_\beta$ blowing up when $\beta $ gets small. 

%


These local laws, first obtained in \cite{as}, extend to  Coulomb interactions in arbitrary dimension some results  obtained for the two-dimensional Coulomb case at fixed $\beta$ and down to mesoscales  in \cite{loiloc} (for electric energy and point discrepancy) and \cite{bbny}  (for point discrepancy), it also extends them {\it down to the microscale} and to possibly $N$-dependent $\beta$.
 We recall that  better local laws on the number of points (but not on the energy $\F$) valid up to the boundary in all Coulomb cases, were recently proven by Eric Thoma in \cite{thoma}, see Section \ref{sec:isotropic}. 
 
The extension of the local laws \eqref{loclawaprouver} to general Riesz cases \eqref{rieszgene} are to be found in \cite{PeilenSer}. The one-dimensional logarithmic case, or $\beta$-ensembles case, has on the other hand been much studied. Local laws for the electric energy (down to microscale) can be found in \cite{peilen}. 
Local laws concerning the number of points,  Stieltjes transform, as well as description of particle spacings, local statistics and their universality,  have been the object of many papers, we refer to \cite{joha,shch,borotguionnet,BorGui2,BMP22,BekFigGui,bey1,bey2,lambertledouxwebb,bl,bls} and references therein.

\section{Method and heuristics}\label{sec8.1}
The goals stated above are accomplished jointly by a procedure of bootstrap on scales, originally introduced in this context in the case $ \d=2$  in  \cite{loiloc}, also taking inspiration from \cite{armstrongsmart}, which 
 is akin to a  renormalization argument in statistical physics. The result of \cite{bbny} also relied on a  bootstrap on scales, but using  loop equations in place of the screening procedure.
 
 At the largest scale $R= N^{1/\d}$, the  law \eqref{loclawaprouver} is already known from \eqref{firstconcbound}. Assuming it is true down to some scale $2R$, we wish to show it is also true for scale $R$, without any deterioration of the constant $\mathcal C$ in \eqref{loclawaprouver}.  The control at scale $2R$ ensures  that configurations  have a well-controlled energy and that most of them are thus screenable, and it will allow to control the screening errors and show that they are $\ll R^\d$. This will allow to   show that the two free energy quantities $\log \K_\beta(\carr_R) $ and $\log \L_\beta(\carr_R)$  of the previous chapter are very close, which in turn will imply the desired estimate \eqref{loclawaprouver} at scale $R$.
 The closeness of the  free energy quantities will also naturally imply the almost additivity of the free energy, as hinted to in the previous chapter.  The bootstrap procedure has to stop at scale $\rho_\beta$,  the Debye length, that we call minimal scale (by analogy with \cite{armstrongsmart}),   below which the screening errors become as large as the volume, thus deteriorating $\mathcal C$. 
  
We note that at the regular scale $\rho_\beta$ becomes $\rb N^{-1/\d}$, which is  $\chi(\beta)^{1/2} \beta^{-1/2} N^{-1/\d}= \chi(\beta)^{1/2} \theta^{-1/2}$, with $\theta$  as in \eqref{deftheta}. Thus the same  lengthscale for rigidity as the macroscopic one in Section \ref{secthermal} appears (they are exactly equal if $\d\ge 3$ for which $\chi(\beta)=1$). 
We conjecture that this lengthscale is sharp and that \eqref{loclawaprouver} is optimal, in particular that local laws do not hold below $\rb$ because particle numbers behave in a Poissonian way below that scale. 
  
 
 Again we are in the situation of Remark \ref{twosit}. 
 In the first situation of that remark, the screening procedure requires a bit of ``buffer space" to be executed and  the local laws will only be valid in the bulk, at a distance (still much smaller than $N^{1/\d}$) from the boundary of the ``essential support of $\mu$" (a set where $\mu\ge m>0$). In the second situation where we consider a hyperrectangle with Neumann boundary condition, results are valid up to the boundary.

Let us now get  into more detail of  
 the heuristics.
We wish to obtain  the local law \eqref{loclawaprouver}, where $\carr_R$ denotes  a {\it closed} cube of radius $R$ (not necessarily centered at the origin).
 We know from \eqref{firstconcbound} that this holds at the largest scale $R=N^{\frac1\d}$. Assume we know it holds for all cubes of size $\ge 2R$ sufficiently far from the boundary, with the same constant $\mathcal C$, and let us try to show it holds for down to $R$ with the same constant $\mathcal C$.

 To do so let us try to estimate 
 \be
 \Esp_{\Q_\beta(U,\mu)} \( \exp\( \frac\beta2 \F^{\carr_R}(\XN, \mu)\) \)  = 
 \frac{\displaystyle\int_{(\R^\d)^N} \exp\(-\beta \F(\XN, \mu) + \frac\beta2\F^{\carr_R} (\XN, \mu) \) d\mu^{\otimes N} } { \displaystyle\int_{(\R^\d)^N} \exp\(-\beta \F(\XN, \mu)  \) d\mu^{\otimes N} }.\ee
 In the numerator we will bound below $\F^{\carr_R} (\XN, \mu) $ by $\G^{\mathrm{inn}}(\XN|_{\carr_R}, \carr_R)$ by simple comparison as in Lemma \ref{lemrestri} and the same for $\F^{\carr_R^c}(\XN, \mu)$ (we omit the $\mu$-dependence in the notation). We bound from below  the denominator by the integral over configurations obtained by gluing together configurations in $\carr_R$ and $\carr_R^c$ with Neumann boundary conditions. Let us denote by $\mn$ the optimal number of points in $\carr_R$ that is $\mn= \int_{\carr_R} \mu$ (assumed to be an integer), and for a configuration $X_N$ in $\R^\d$, let us denote by $n$ the actual number of points that fall in $\carr_R$.
Using for now the notation \eqref{defL},  we thus bound
 \begin{multline}
 \Esp_{\Q_\beta(U,\mu)} \( \exp\( \frac\beta2 \F^{\carr_R}(\XN, \mu)\) \) \\  \le 
 \frac{  \displaystyle \sum_{n=0}^N    \left( \begin{array} {l} N \\ n \end{array}\right) 
\displaystyle \int_{(\R^\d)^N} \exp\(-\hal \beta \G^{\mathrm{inn}} (\XN|_{\carr_R} ,  \carr_R) \) \exp\(-\beta \G^{\mathrm{out}} (\XN|_{ \carr_R^c},  \carr_R^c) \)  d\mu^{\otimes N}  (\XN) }   {   \left( \begin{array} {l} N \\ \mn \end{array}\right)   \displaystyle \int_{(\carr_R)^\mn} \exp\(-\beta \F(\cdot, \mu, \carr_R) \) d\mu^{\otimes\mn} \int_{\(\carr_R^c\)^{N-\mn}} \exp\(- \beta \F(\cdot, \mu, \carr_R^c) \) d\mu^{\otimes (N-\mn)}   }   \\
 \le \frac{  \displaystyle \sum_{n=0}^N    \left( \begin{array} {l} N \\ n \end{array}\right) \L_{n,\beta/2} (\mu, \carr_R) \L_{N-n,\beta}(\mu, \carr_R^c) }{ \left( \begin{array} {l} N \\ \mn \end{array}\right)
\K_\beta(\mu,\carr_R) \K_\beta(\mu,\carr_R^c)} .
 \end{multline}
Heuristically,  the sum in the numerator concentrates near $n=\mn$, otherwise the cube is very unbalanced and the energy is too large by discrepancy estimates such as \eqref{disc10bu}--\eqref{disc1bu} in view of the law \eqref{loclawaprouver} at scale $2R$.
 We are left with having to evaluate
 $$\frac{\L_{\mn, \beta/2}(\mu,\carr_R)}{\K_{\beta/2}(\mu,\carr_R)} 
 \frac{\L_{N-\mn,\beta}(\mu,\carr_R^c)}{\K_\beta(\mu,\carr_R^c)}  \frac{  \K_{\beta/2}(\mu,\carr_R)}{\K_\beta(\mu,\carr_R) }.$$
To evaluate the last ratio, we may make use of the Neumann free energy bounds \eqref{bornesfiU}.
On the other hand, the first two ratios can be shown to be close to 1 by screening since screening precisely allows to compare $\L_\beta$ and $\K_\beta$. Moreover, the local law \eqref{loclawaprouver} at scale $2R$ provides the bounds needed for the screening (by providing good energy bounds for most configurations).
 We will then obtain 
 $$\log \Esp_{\Q_\beta(U, \mu)} \( \exp\( \frac\beta2 \F^{\carr_R}(\XN, \mu)\) \) \le C\beta \chi(\beta) R^\d + o(R^{\d})$$ 
 which yields the desired estimate if $\mathcal C$ was chosen large enough to start with.  
Once this is obtained, we know that local laws hold down to scale $\rho_\beta$ and we can use this to find that the difference between $\log \L_\beta$ and $\log \K_\beta$ is only a surface error, hence the almost additivity of $\log \K_\beta$ up to this surface error.

 
 

\section{The case of minimizers}\label{sec8.2} \index{energy minimizers}
Let us now present rigorous results, starting with the case of energy minimizers (or $\beta=\infty$) as an easier illustration of the method.
We will treat in parallel the two situations described in Remark~\ref{twosit}.

The idea is a simple comparison: 
let $\nab h$ be the electric field for a minimizer $X_N^0$ of $\F(\cdot, \mu, \R^\d)$, i.e.~achieving $\mathsf{E}_\infty(\mu, \R^\d)$ as in \eqref{defEi}, and let $\carr_R$ be a cube of size $R$ included in the (blown-up) essential support of the measure $\mu$ (a bit far from the boundary) then if the energy in $\carr_{2R}$ can be well bounded in terms of $R$, we can perform inner screening relative to $\carr_R$ and replace it by the electric field associated to a minimizer of $\F(\cdot,\mu, \carr_R)$ inside $\carr_R$ while adding only $O(R^{\d-1})$ to the energy. Comparison then allows to deduce a bound on $\F^{\carr_R} (X_N^0, \mu, \R^\d)$. 

Thanks to the comparison between the sub and superadditive energies, we can get a convergence with a rate in the case of the uniform density $\mu=1$, second case in Remark~\ref{twosit}.



\begin{theo}[Uniform distribution of energy and points down to microscale for minimizers]
\label{th4}\mbox{}
Assume $\d\ge 1$ and $\s=\d-2$.
\begin{enumerate}
\item  (Neumann problems in cubes)
Let $\carr_R$ be a closed cube of size $R$ with $R^\d$ an integer. There exists $C>0$ depending only on $\d$ such that 
\be\label{mn1}
\left|\frac{\mathsf{E}_\infty(1,\carr_R)}{R^\d}-\mf(\infty)\right|\le \frac{C}{R}\ee
where $\mf(\infty)\in \R$ and  moreover, 
 if $X_{R^\d}$ is a minimizer for $\mathsf{E}_\infty(\carr_R,1)$, for any cube $\carr_L(x)\subset \carr_R$ such that $L \ge C$,  we have 
\be \label{mn2}
\left|\int_{\carr_L(x)} \sum_{i=1}^{R^\d} \delta_{x_i} - |\carr_L| \right|\le C L^{\d-1}\ee
and  the energy is uniformly distributed in the sense that 
\be \label{mn3}\F^{\carr_L(x)}(X_{R^\d}, 1,\carr_R)= L^\d \mf(\infty)+ O(L^{\d-1}).\ee

\item
 (Minimizers of the Coulomb  energy at the usual scale). 
Assume that $\mu $ is a bounded probability density  supported in some compact set $\Sigma$ and  satisfies $ \mu \ge m>0 $ in  $\Sigma$ (resp. let  $\mu= \meseq$). There exists $C>0$ depending only on $\d , m , \|\mu\|_{L^\infty}$ such that the following holds.

 Let $\XN^0$ minimize $\F_N(\XN, \mu)$ (resp. minimize $\HN$). 
Let  $R \ge C N^{-1/\d}$ and $Q_R(x)$ be a hyperrectangle  of sidelengths in $[R,2R]$ centered at $x$ satisfying  $N \int_{Q_R} \mu \in \mathbb{N}$, and 
\be \label{borninfdist}\dist(Q_R(x) , \partial \Sigma)  \ge     C N^{ \frac{-2}{\d(\d+2)} } \ee
we have
\begin{equation}\label{loclawpointsmin}
\left| \int_{Q_R(x)}  \sum_{i=1}^N \delta_{x_i}-  N \int_{Q_R(x)} d\mu \right|\le C\(N^{\frac{1}{\d}} R\)^{\d-1},\end{equation}
and 
\be\label{mn4} \left|\F^{Q_{RN^{1/\d}} (x)}((\XN^0)',  \mu')-  N \mf(\infty) |Q_R|\right|\le 
 C(N^{\frac1\d} R)^{\d-1}.\ee
\end{enumerate}
\end{theo}

\begin{rem} We will give in Corollary \ref{corofw}  a variational interpretation for $\mf(\infty)$ 
$$\mf(\infty)=\min\mathbb{W}(\cdot, 1) $$
where $\mathbb{W}$ is the ``jellium  renormalized energy".
\index{jellium energy}

The explicit rate in~\eqref{mn3} is an improvement compared to \cite{rns,PRN}. As in \cite{rns}, we can also prove with the same method the same results on minimizers and the minimum of the renormalized energy ${\mathbb{W}}(\cdot, 1)$ of \cite{ssgl,rs,PetSer}. For instance the limit as $R \to \infty$ that defines ${\mathbb{W}}(\cdot, 1)$ can be shown to be $\mf(\infty)$ with rate $1/R$: the upper bound is by periodization of a minimizer for $\mathsf{E}_\infty$  while the lower bound is obtained as in ~\eqref{minofk} to be combined with~\eqref{mn1}.\end{rem}
\begin{proof}
{\bf Step 1: bootstrap.} In order to accommodate both situations of the theorem, we work at the blown up scale, and consider 
that  $\mu $ satisfies $0<m\le \mu  $ in $\Sigma$, and  let $X_N^0$ be a minimizer of  $\F(\cdot, \mu, U)$ among configurations with $N= \mu(U)$ points, where $U$ is either a cube $\carr_R$ or $\R^\d$. Let $u$ be the electric potential satisfying \eqref{defv}.
We claim that if  $L\ge C$  then
\be \label{aprouve} \F^{\carr_L (x)} (\XN^0, \mu, U)+ C_0 \# I_{\carr_L(x)} \le C L^\d \ee  for some $C>0$ depending only on $\d$ and $\mu$, and $C_0$ the constant in \eqref{14}.

This is proven by a bootstrap. First  we may reduce to proving this for hypercubes 
with faces parallel to the boundary of $U$ in instance (1). 
We know that \eqref{aprouve} is true for the maximal scale $L=R$ in instance (1)  or $L=N^{1/\d}$ in instance (2) by \eqref{bornesfieU}.  Assume this is true for some $\tilde L$ i.e.~assume 
\be\label{hypboot}
 \F^{\carr_{\tilde L}(x)} (\XN^0, U)+ C_0 \# I_{\carr_{\tilde L}(x)}
  \le \mathcal{C} {\tilde L}^\d ,\ee 
 we need to show it is true for $ \frac34\tilde L \ge L  \ge \tilde L/2$. We have emphasized the constant $\mathcal{C}$ which must be independent of $L$, so we need to prove that  the same result holds for $L$ with the same constant $\mathcal{C}$.
Let us reduce  to $Q_{ L} $ such that $\mn := \int_{Q_L\cap U} \mu$ is an integer, and denote by 
$n=  \#I_{Q_L} $.

First by~\eqref{hypboot},  we have from
\eqref{14}, \eqref{disc10bu}--\eqref{disc1bu} that 
\be \label{mn}  |n-\mn| \le C L^{\d-1} + C\sqrt{\mathcal{C}} L^{\d-\hal} ,\ee
and moreover that 
$$ \int_{\carr_{\tilde L} } |\nab u_{\rrc}|^2 \le  4\cd\(\F^{\carr_{\tilde L}(x)} (\XN^0, U)+ C_0 
\# I_{\carr_{\tilde L}(x)}\)
 \le 4\cd \mathcal{C} {\tilde L}^\d .$$
We then apply the outer screening (with respect to $Q_L$) of   Proposition \ref{proscreen} with $S(\XN^0,u) \le \mathcal C {\tilde L}^\d$,  and we make the choices of $\ell \le \tilde \ell \le L $ that are the smallest possible to satisfy \eqref{screenab}
\footnote{Note that in the definition of $S(X_n,w)$,  $w_{\rrh}$ can be changed into $w_{\rrc}$ as in \eqref{defrrc4} since $\rrh$ and $\rrc$ coincide in the region of integration}  and obtain small  errors in \eqref{nrjy}.  For instance we choose 
$\ell=  L^{\frac{\d}{\d+2}} $,  $\tilde \ell= M    L^{\frac{\d}{\d+2}} $, with $M\ge 1$ large enough. The condition $\ell \le \tilde \ell \le L$ is satisfied as long as $L$ is larger than some constant depending on $M$.
 The separation from the boundary  conditions will be satisfied by assumption (see below).
 The screenability condition \eqref{screenab} is satisfied for the potential $u$ since 
$\tilde \ell \ell^{\d+1} = M \ell^{\d +2}\ge S(X_N^0,u)$ in view of \eqref{hypboot} and $\tilde L \le 2L$,  if $M$ is chosen sufficiently large in terms of $\mathcal C$. 
  This implies that the best inner and outer screenable potentials have an even smaller energy \eqref{innernrj} and \eqref{outernrj}, and are thus screenable as well.

 We use Proposition \ref{proscreen} with  $Z_{\mn-\N}$ minimizing $\F(\cdot, \tilde \mu, \mathcal N)$ where $\tilde \mu$ is given by the result of the proposition (recall that that minimum is bounded by the order of the volume, see~\eqref{bornesfieU}).
  The proposition or \eqref{ubapscreening} applied to minimizers of \eqref{innernrj} and \eqref{outernrj} thus yields  in view of~\eqref{mn} and~\eqref{hypboot}
\be \label{111}\mathsf{E}_\infty(\mu,Q_L^c) \le \G_U^{\mathrm{out}}(\XN^0|_{Q_L^c},Q_L^c)  +  C\( \frac{  \mathcal C}{M} L^\d 
+  M L^{\d-1} L^{\frac{\d}{\d+2}}  + \sqrt{\mathcal{C}} L^{\d-\hal}  \)
.\ee 
Choosing $M$ large enough, the error in the right-hand side can be made $\le \hal \mathcal C L^\d$ when $L$ is large enough, and thus combining \eqref{subad2},  and~\eqref{111},  it follows that 
$$\F(\XN^0, U)= \mathsf{E}_\infty(\mu,U)  \le \mathsf{E}_\infty(\mu, Q_L)+ \mathsf{E}_\infty(\mu,Q_L^c) 
\le \mathsf{E}_\infty(\mu, Q_L)+  \G_U^{\mathrm{out}}(\XN^0|_{Q_L^c}, Q_L^c)   +\hal \mathcal C L^\d.
$$
On the other hand, by \eqref{locali2} and Lemma \ref{lemrestri} (which applies since $u$ is screenable), we have
$$\F^{Q_L} (\XN^0, U)+ \G_U^{\mathrm{out}}(\XN^0|_{Q_L^c}, Q_L^c)\le \F(\XN^0, U).$$
Hence, combining the two,   we deduce that if $L $ is large enough (depending on $\mathcal{C}$)
$$ \F^{Q_L} (\XN^0, \mu,U)\le  \mathsf{E}_\infty(\mu,Q_L)+ \hal \mathcal{C} L^\d.$$
In view of~\eqref{mn} we have as well  $n\le \hal \mathcal{C} L^\d$ if $L$ is large enough.
With \eqref{bornesfieU} applied in $Q_L$, this concludes the proof of~\eqref{aprouve}.

Let us examine the  distance condition to the boundary which the screening proposition requires.  If $\Sigma$ is a proper subset of $U$, at each step we need a safety distance $\tilde \ell$. But we chose $\tilde \ell =  M L^{\frac{\d}{\d+2}}$. At the first iteration $L$ is of order $R$, respectively $N^{1/\d}$, so 
$\tilde \ell = M R^{\frac{\d}{\d+2}}$.
 At further iterations,  we need $Q_{L+\tilde \ell} $ to be included in $Q_{\tilde L}$ so we need a further distance  $ CL^{-\frac\d{\d+2}}$, and  since $L$ gets multiplied by a factor $\in [\hal,\frac34]$ at each step, summing the distances over the iteration still gives a condition of the form $ \dist \ge M R^{\frac{\d}{\d+2}}$ at the blow-up scale, hence condition \eqref{borninfdist} at the original scale. 

In the case where $U=\carr_R$  and we can take $\Sigma =\R^\d$, 
we need that $Q_L(x)$ has sides parallel to those of $\carr_R$, satisfying that the faces of $\partial( \carr_L(x) \cap \carr_R) $ not belonging to $\partial \carr_R$ are at a distance $\ge C R^{\frac{\d}{\d+2}}$ from the parallel faces of $\partial\carr_R$ for some $C>0$. This condition is however not really restrictive, as in the first step we can always include a given $Q_L$ into one satisfying the required condition and  $L \in (\frac78,\frac14)$. We can then repeat the argument during the bootstrap and obtain the result up the boundary. 
\smallskip

{\bf Step 2: local laws.}
Now that we know~\eqref{aprouve} down to scale $C$, we can use it to obtain
\be \label{erreursn}
|n-\mn|\le C L^{\d-1}.\ee
Following the relations after \eqref{111}, we find  
the improvement $$  \F^{Q_L} (\XN^0, \mu, U) \le \mathsf{E}_\infty(\mu, Q_L)+ CL^{\d-1}$$
and also that  \eqref{erreursn} holds.
By screening, i.e.~by Proposition \ref{proscreen} applied in $Q_L$ to $\XN^0|_{Q_L}$ and $u$,  we also have
\be \label{minofk}
\mathsf{E}_\infty(\mu,Q_L)\le \F^{Q_L} (\XN^0, \mu, U)+ CL^{\d-1},\ee  so  we conclude that
\be \label{815}
\F^{Q_L} (\XN^0,\mu,U)= \mathsf{E}_\infty(\mu,Q_L)+O(L^{\d-1}),\ee
with the $O$ depending only on $\d,m$ and $\|\mu\|_{L^\infty}$.
\smallskip 

{\bf Step 3: Conclusion.}
Let  $U= \carr_{2R}$ with $R^\d$ an integer, and $\mu=1$, and  $X_{\bar N}^0$ be a minimizer of $\F(\cdot , 1,\carr_{2R})$, splitting $U$ into $2^\d$ cubes of size $R$, denoted $\carr_R^i$,  and using \eqref{locali2}, we have 
$$ \mathsf{E}_\infty(1,\carr_{2R}) = \F(X_{\bar N}^0, 1, \carr_{2R}) \ge
\sum_i \F^{\carr_{R}^i} (X_{\bar N}^0,1, \carr_{2R}).$$
Inserting the result \eqref{815}, it follows that 
\be \mathsf{E}_\infty(1,\carr_{2R})\ge 2^\d  \mathsf{E}_\infty(1,\carr_R) +O(R^{\d-1})\ee
and combining with \eqref{subad2}, we arrive at 
\be \frac{\mathsf{E}_\infty(1,\carr_{2R})}{|\carr_{2R}| }=\frac{  \mathsf{E}_\infty(1,\carr_R) }{|\carr_R|} +O(R^{-1}).\ee
Applying this iteratively to $2^kR$ and summing the series, we deduce that 
$$\mf(\infty):= \lim_{R\to \infty} \frac{\mathsf{E}_\infty(1,\carr_R) }{|\carr_R|} $$ exists and that \eqref{mn1} holds. 
We also then obtain \eqref{mn3} from \eqref{815} and \eqref{mn2} from \eqref{erreursn}.

For the case of~(2),  we apply the results of the previous steps with $U=\R^\d$, $\mu=\mu'$,  and then a blow-down, using \eqref{scalingF}. The relation \eqref{mn4} is  a rewriting of~\eqref{815}   and~\eqref{loclawpointsmin} follows from~\eqref{erreursn}. \end{proof}

\section{Local laws with temperature}
\subsection{Statement} 
We now turn to the question of proving \eqref{loclawaprouver} with temperature $\beta>0$. This is more difficult since we have to deal with many configurations at once.
The following statement is meant again to either be applied with $U=\R^\d$, giving information on the Gibbs measure $\Q_\beta(\R^\d, \mu)$, or to $U$ a hyperrectangle and $\Lambda= \R^\d$ (in which case condition \eqref{conddist0} is empty) giving the local law for the Neumann Gibbs measure $\Q_\beta(U, \mu)$ in a hyperrectangle. Since we have made some simplifications compared to \cite{as}, we obtain a better result than in that paper, in particular some assumption is removed and $\rho_\beta$ (defined in \eqref{defrhobeta}) is improved, and the result is valid for $\d=1$.

In the theorem, we place the same assumptions as in Proposition \ref{pro718} to ensure the a priori bound \eqref{bornesfiU}. 
We recall that  by \eqref{rewritegibbs3} the original Coulomb gas Gibbs measure is equal to $\Q_\beta(\R^\d, \mub')$ for $\mub'$ the blown-up thermal equilibrium measure, and that as discussed after \eqref{assumplbs}--\eqref{assgmm}, 
these assumptions are satisfied for $\mub'$ in $\Lambda=\Sigma'$ provided \eqref{A5} holds and $\theta\ge \theta_0>0$. 
Thus the theorem below applies to \eqref{gibbs} as soon as $\theta>\theta_0$ and \eqref{A1}--\eqref{A5} hold.
\index{minimal scale}
\index{local laws}

\begin{theo}[Local laws]
\label{th3} 
Assume $\d\ge 1$ and $\s=\d-2$. Let $U$ be an open subset of $\R^\d$ with bounded and piecewise $C^1$ boundary and $\mu$ be a bounded nonnegative density such that $\mu(U)=N$ is an integer.  Assume  that  $\mu \ge m>0 $ in a set $\Lambda$.  
If $\s\le 0$ and $U$ is unbounded, assume in addition that \eqref{assumplbs}, \eqref{assgmm} and \eqref{assumpbeta} hold. 
There exists a constant $C>0$ depending only on $\d, m$, $\|\mu\|_{L^\infty} $ and the constants in the assumptions, such that the following holds.

Let   $\carr_R(x)$ be a closed a cube of size $R\ge \rb$ centered at $x$ (where $\rho_\beta$ as is in \eqref{defrhobeta}) with 
\be
\label{conddist0}
\dist(\carr_R(x) , \pa \Lambda\cap U) \ge d_0\ee
where 
\be \label{defd00}
d_0:= 
C \max\( \(\frac{N^{\frac1\d} }{ \max(1, \beta^{-\hal} \chi(\beta)^{\frac12})}\)^{-\frac23}   N^{\frac1\d} ,   N^{\frac{1}{ \d+2}}  \) , 
 \ee 
 we have, for $C_0$  the constant in \eqref{14},
  \begin{equation}\label{locallawint0}
  \left|\log \Esp_{\Q_{\beta}(\mu,U) } \( \exp\( \hal \beta\(  \F^{\carr_{R}(x)} (\cdot,  \mu,U)   + C_0 \# I_{\carr_R (x)}  \)  \)\)
  \right| \le 
C   \beta\chi(\beta ) R^\d .
  \end{equation}
  \end{theo}
  
\subsection{Proof of the local laws with temperature}\label{sec8.4}

When studying Gibbs measures, we  need to show that given a set of configurations with well-controlled energy, we may screen them and sample new points in $\New $  to obtain a set with large enough volume in which~\eqref{nrjy} holds.  This is possible  and  yields comparison   of partition functions (reduced to screenable configurations) as stated in the following proposition from \cite{as}.

Here the quantity $\ep_e$ corresponds to the energy error while $\ep_v$ corresponds to the volume error. We want the volume errors to be bounded by $O(\beta)$ times the volume, which is   more difficult to obtain when  $\beta$ is small.

\begin{prop}\label{pro42}
With the same assumptions and notation as in Proposition \ref{proscreen}, assume in addition that 
$\tilde \ell \ge \max(\beta^{-\frac{1}{\d-\s}} \indic_{\s\le 0},1)$.
Let us define the set $\mathcal D_{s,z}$ to be 
\begin{equation}
\label{defA}
\mathcal D_{s,z}= \left\{X_n \in \Omega^n,\bar S(X_n) \le s \ \text{and} \  \bar S'(X_n) \le z
 \right\}\end{equation}
where $\bar S, \bar S'$ are as in~\eqref{bestS}, resp.~\eqref{bestS2}, relative to $\Omega$. For any  number $s$  such that 
\be\label{constr2}
 \l^{\d+1}\ge  C \min ( \frac{s}{\tilde \l} , z) ,\ee  
 and \be\label{constr3}s< c \tilde \l^2 R^{\d-1}\ee for some $c>0$ small enough (depending only on $\d, m, \|\mu\|_{L^\infty}$), 
 there exists $\alpha, \alpha'$ satisfying $\alpha +n-\mn\ge 1$,
 \be \label{cara} \left|\frac{\alpha'}{\alpha}-1\right|\le C\( \frac{1}{\tilde \ell}+\frac{s}{\tilde \ell^2 R^{\d-1}}\), \qquad  \frac1C \tilde \ell R^{\d-1}\le \alpha \le C \tilde \ell R^{\d-1}\ee
such that letting  
 \be \ep_{e} := C\( \frac{s\l}{\tilde \ell} + R^{\d-1}\tilde \ell \chi(\beta) +|n-\mn|\) \ee 
 and 
 \begin{itemize}
 \item if $\ell^{\d+1} \ge C \frac{s}{\tilde \ell}$ in \eqref{constr2}
 \begin{multline}\label{epv}
 \ep_{v}:=  C\( \frac{s}{\ell \tilde \l}+\eta R^{\d-1} +\log \frac{\tilde \ell}{\eta}\)\\+\mn-n+\alpha-\alpha'  + (\mn-n-\alpha) \log \frac{\alpha}{\alpha'} - (\alpha+n-\mn+\hal) \log \(1+\frac{n-\mn}{\alpha}\)  + \hal \log \frac{n}{\mn}
 \end{multline}
 \item  otherwise 
 \begin{multline}\label{epv2}
 \ep_{v}:=  C\( \frac{s}{\ell \tilde \l}+\eta R^{\d-1} +\log \frac{\tilde \ell}{\ell}  + \frac{\ell}{\eta} \log \frac{R}{\ell} \)\\+\mn-n+\alpha-\alpha'  + (\mn-n-\alpha) \log \frac{\alpha}{\alpha'} - (\alpha+n-\mn+\hal) \log \(1+\frac{n-\mn}{\alpha}\)  + \hal \log \frac{n}{\mn}
 \end{multline}

 \end{itemize}
   we have that if $\eta$ satisfies \eqref{conditionsureta}
\begin{multline}\label{resscreen}
\frac{1}{n!}  \int_{ \mathcal D_{s,z}} \exp\(-\beta \G_U^{\mathrm{inn/out}}(X_n, \mu, \Omega) \)d\mu^{\otimes n} (X_n)
\\ \le  C \exp\(  \beta \ep_e+ \ep_v \) 
\frac{\mn^\mn}{\mn!} \K_\beta(\mu, \Omega)\int_{\Omega^\mn}  \exp\(-\beta \F(Y_\mn, \Omega) \)d \mu^{\otimes \mn} (Y_\mn)  ,\end{multline}
where $\G_U^{\mathrm{inn/out}}$ is as in Definition \ref{defibestpot} and $\K_\beta$ as in \eqref{defK7}.
Here $C>0$ depends only on $\d, m, \|\mu\|_{L^\infty}$.
\end{prop}

\begin{proof} Again, we consider the inner screening case, the outer case being parallel.
The set $\mathcal D_{s,z}$ consists of configurations with controlled energy, which satisfy the screenability condition \eqref{screenab} as soon as \eqref{constr2} holds.
Let us first consider the first situation in which \eqref{screenab} is satisfied with $S(X_n, w) /\tilde \ell$. In that case, the set $\Old (X_n)$ (we emphasize here for a moment the dependence on~$X_n$) is of the form $Q_t \cap U$. 
We may split $\mathcal D_{s,z}$ into a disjoint union 
 $\cup_k \mathcal E_k $ 
 where 
 $$\mathcal E_k= \{ X_n\in \mathcal D_{s, z}, \Old (X_n) = Q_t, t \in [R-2\tilde \ell + k \eta , R-2\tilde \ell + ( k+1) \eta)\}.$$
 There are  $O(\tilde \ell/\eta)$ such sets that are nonempty.
Thus, for each $X_n\in \mathcal D_{s,z}\cap \mathcal E_k$ with $s,z $ satisfying~\eqref{constr2}, the screening construction of Proposition~\ref{proscreen} can be applied to the best screenable potential achieving the minimum in \eqref{bestS}. This  provides a number~$\N(X_n)$ and  a set~$\Old (X_n)$ (we emphasize again their dependence on~$X_n$). When screening, we delete~$n-\N$ points in the configuration, those that fell outside of~$\Old$, there are~$\binom{n}{\N}$ ways of choosing the indices of the points that get deleted. In  terms of volume of configurations, this loses at most~$\mu(\New)^{n-\N}$ volume. In addition we glue each~$X_n|_{\Old}$  with~$\mn-\N$ points of~$Z_{\mn-\N}=(z_1, \dots , z_{\mn-\N})$ in $\New_\eta$, there are~$\binom{\mn}{\N}$ ways of choosing the indices for the gluing, resulting in configurations~$Y_\mn$ in~$\Omega$  satisfying~\eqref{nrjy} and coinciding with $X_n$ in $\Old$. 
Since there is an $\eta$-sized point-free layer in $\New$ near $\partial \Old$, this guarantees that
 two configurations in each $\mathcal E_k$ have their $\partial \Old$ at distance $\le \eta$ of each other, hence 
 for each $k$ the same configuration $Y_{\mn}$ cannot be produced twice from configurations in $\mathcal E_k$ (this is the  reason for this point-free zone).

We then integrate the choices of $(z_1, \dots , z_{\mn-\N})$ with respect to the measure $\mu$ restricted to $\New$, and  after summing over  the $O(\frac{\tilde \ell}{\eta})$  possible values of $k$, we deduce that 
\begin{align}
\label{depart}
& \int_{ \Omega^\mn}\exp\(-\beta \F(Y_\mn, \mu, \Omega) \) d\mu^{\otimes \mn}(Y_\mn)
\\ & \quad \notag
\geq \frac{\eta}{C\tilde \ell}
\int_{ \mathcal D_{s,z}}\int_{ \New_\eta(X_n)^{\mn-\N}} \exp\bigg[-\beta \G_U^{\mathrm{inn/out}}(X_n , \Omega)- C\beta
\bigg( \frac{s\l}{\tilde \ell} + R^{\d-1}\tilde \ell+ \F(Z_{\mn-\N}, \tilde \mu(X_n), \New_\eta(X_n))
\\ & \qquad\qquad\qquad\qquad\qquad\qquad\qquad\qquad\notag
  + |\mn-n| 
+\sum_{(i,j) \in J}   \g(x_i-z_j)  \bigg) \bigg] 
\\ & \qquad \qquad\qquad\qquad\qquad\qquad\qquad\qquad\qquad\notag
\times \frac{\binom{\mn}{\N}}{\binom{n}{\N}}  \frac{1}{\mu(\New)^{n-\N} } \, d\mu|_{\New_\eta}^{\otimes (\mn-\N)} (Z_{\mn-\N})\, d\mu^{\otimes n}(X_n) .
\end{align}
We will need the following that controls the free energy added in the screening layer in the same spirit as the free energy a priori bounds of Lemma \ref{prominok} and Proposition \ref{pro718}.
\begin{lem}\label{lemclaim44}
For each $X_n\in \mathcal D_{s,z}$, we have 
\begin{multline} \label{claim44}
\int_{\New_\eta^{\mn-\N}} 
\exp\bigg(-C\beta \bigg( \F(Z_{\mn-\N}, \tilde \mu,\New_\eta)  +\sum_{(i,j) \in J}   \g(x_i-z_j)\bigg) \bigg) d\mu|_{\New_\eta}^{\otimes (\mn-\N)} (Z_{\mn-\N}) \\ 
\geq 
(\mn- \N)^{\mn-\N}
 \exp\( \mu(\New_\eta)-\tilde \mu(\New_\eta)  - C\(  \beta \chi(\beta)  R^{\d-1} \tilde \ell + \frac{s}{\ell \tilde \ell}+ \frac{\eta^2}{\ell}   R^{\d-1}\)  \).
\end{multline}
\end{lem}
Before giving the proof of~\eqref{claim44}, let us use it to obtain the proposition. By Stirling's formula, we have
\begin{align}
\label{414}
& \log  \( \frac{ (n-\N)!    }{ (\mn-\N)!}  \frac{(\mn-\N)^{\mn-\N}}{\mu(\New)^{n-\N} }\) 
\\ & \qquad \notag
\geq \mn -n   + (n-\N) \log \frac{n-\N}{\mu(\New)}   + \hal \log \frac{\mn  (n-\N)}{n  (\mn-\N)} - C. 
\end{align}
Combining~\eqref{claim44}--\eqref{414} and inserting into~\eqref{depart}, we obtain, for a constant~$C$ depending only on~$\d, m$ and $\|\mu\|_{L^\infty}$,
\begin{align*}
\lefteqn{
\frac1{\mn!}
\int_{ \Omega^\mn}\exp\(-\beta \F(Y_\mn, \mu,\Omega) \) d\mu^{\otimes \mn}(Y_{\mn})
} \quad & 
\\ & 
\geq 
\exp\( -C \beta\( \frac{s\l}{\tilde \ell} +  R^{\d-1} \tilde \l \chi(\beta)+ |\mn-n|\) -C \frac{s}{\ell \tilde \l} -C\frac{\eta^2}{\ell}   R^{\d-1} -  \log \frac{\tilde \ell}{\eta} \)
\\ & \qquad \times 
\frac1{n!}\int_{ \mathcal D_{s,z}} 
\bigg[ \exp\(-\beta \G_U^{\mathrm{inn/out}}(X_n, \Omega)  + \mn -n  +  \mu(\New_\eta)- \tilde \mu(\New_\eta)  \)
\\ & \qquad\qquad\qquad \times 
\exp\( (n-\N) \log \frac{n-\N}{\mu(\New)} + \hal \log \frac{\mn  (n-\N)}{n  (\mn-\N)}-C \)\bigg] d\mu^{\otimes n}( X_n).
\end{align*}
We may next use a mean-value argument to obtain, for some  configuration $X_n^0\in \Omega^n$,
\begin{align*}
\lefteqn{\frac1{\mn!}
\int_{ \Omega^\mn}\exp\(-\beta \F(Y_\mn, \Omega) \) d\mu^{\otimes \mn}(Y_{\mn}) 
}  & 
\\ &
\geq 
\exp\bigg[ \mn  -n  +\mu(\New_\eta(X_n^0))-\tilde \mu(\New_\eta (X_n^0))+  (n-\N(X_n^0)) \log \frac{n-\N(X_n^0)}{\mu(\New(X_n^0))}   
\\ & \qquad 
+ \hal \log \frac{\mn  (n-\N(X_n^0))}{n  (\mn-\N(X_n^0))}-C 
  -C \beta\( \frac{s\l}{\tilde \ell} +  R^{\d-1} \tilde \l \chi(\beta)+ |\mn-n|\)   -  C\frac{s}{\ell \tilde \l}   -C\frac{\eta^2}{\ell}   R^{\d-1} - \log \frac{\tilde \ell}{\eta} \bigg]  
\\ & \qquad \times \notag
\frac{1}{n!}\int_{ \mathcal D_{s,z}} \exp\(-\beta \G_U^{\mathrm{inn/out}}(X_n, \Omega) \) d\mu^{\otimes n} (X_n).
\end{align*}
We then let $\alpha = \tilde \mu(\New (X_n^0))$ and $\alpha'= \mu(\New (X_n^0))$. We note that in view of~\eqref{bornimp}, we have $\alpha + n-\mn= n-n_{\Old(X_n^0)}\ge 0 $ and we may even reduce to the situation where this is $\ge 1$ otherwise the corresponding term in Stirling's formula should be $0$. 
We also note that  by construction of $\tilde \mu$, $\mu(\New_\eta(X_n^0))- \tilde \mu (\New_\eta(X_n^0))= \alpha'- \alpha-\int_{\New\backslash \New_\eta} \mu= \alpha'-\alpha +O( \eta R^{\d-1}).$
In view of~\eqref{bornimp}, we have that~\eqref{cara} holds and we may rewrite the first exponential terms as 
 \begin{multline*}\exp\(\mn -n  + \alpha'-\alpha  +O(\eta R^{\d-1})+  (n-\mn + \alpha) \log \frac{n- \mn+\alpha}{\alpha'}   + \hal \log \frac{\mn  (n-\mn+\alpha)}{n  \alpha}\)\\
 =  \exp\(\mn -n  + \alpha'-\alpha +O(\eta R^{\d-1}) +  (n-\mn + \alpha) \log \frac{\alpha}{\alpha'}   + \( n-\mn+\alpha+\hal\) \log \(1+ \frac{n-\mn}{\alpha}\)  \)   .\end{multline*}
Rearranging terms and recalling that $\eta \le \ell$, we obtain the proposition in that case.

Finally, we consider the second case where \eqref{screenab} is not satisfied with $\frac{S(X_n, w)}{\tilde \ell} $. In that case, there are $O(\frac{\tilde \ell}{\ell})$ choices of strips of the form $Q_{t+\ell}\backslash Q_t$ where $\partial \Old$ can lie. For each of them, there are $O(\frac{R^{\d-1}}{\ell^{\d-1}})$ facets forming $\partial \Old$, and $O(\ell/\eta)$ choices for each so that we know their location up to an error $\eta$. This allows to partition $\mathcal{D}_{s,z}$ into  $O\( \frac{\tilde \ell}{\ell}   \( \frac{R^{\d-1}}{\ell^{\d-1}}\)^{ \ell/\eta}\)$ sets $\mathcal E_k$, such that two configurations in each $\mathcal E_k$ have their $\partial \Old$ at distance $\le \eta$ of each other (this then ensures thanks to the point-free $\eta$-layer that two such configurations cannot be screened into the same configuration).
The rest of the proof is identical as in the first case, except for the factor 
$ \frac{\tilde \ell}{\ell}   \( \frac{R^{\d-1}}{\ell^{\d-1}}\)^{ \ell/\eta}$ replacing $\frac{\eta}{\tilde \ell}$. 
This adds an extra $\log \frac{\tilde \ell}{\ell} +C \frac{\ell}{\eta} \log \frac{R}{\ell}$ to the volume error, and the proof is complete.

\end{proof}
 
\begin{proof}[Proof of Lemma \ref{lemclaim44}]
Applying Jensen's inequality, we find
\begin{multline*}
\int_{\New_\eta^{\mn-\N}} 
\exp\bigg(-C \beta \bigg(\F(Z_{\mn-\N}, \tilde \mu,\New_\eta)   +\sum_{(i,j) \in J}   \g(x_i-z_j)\bigg) \bigg) \, d\mu|_{\New_\eta}^{\otimes (\mn-\N)} (Z_{\mn-\N}) 
\\
= \!\int_{\New_\eta^{\mn-\N}} \!\exp\bigg[\!\!-\!C\beta \bigg( \F(Z_{\mn-\N}, \tilde \mu,\New_\eta) \!+\!\sum_{(i,j) \in J}   \g(x_i-z_j) \bigg)+ \!\sum_{i=1}^{\mn-\N} \!\log \frac{\mu}{\tilde \mu} (z_i)  \bigg] d\tilde\mu^{\otimes (\mn-\N)} (Z_{\mn-\N}) 
\\
\geq 
\tilde \mu(\New_\eta)^{\mn-\N} 
\exp\Bigg[  \tilde \mu(\New_\eta)^{\N-\mn} 
\int_{\New_\eta^{\mn-\N} } \bigg( -C\beta\bigg( \F(Z_{\mn-\N}, \tilde \mu,\New_\eta)
 +\sum_{(i,j) \in J}   \g(x_i-z_j)\bigg)
\\+ \sum_{i=1}^{\mn-\N} \log \frac{\mu}{\tilde \mu} (z_i)  \bigg) \,
d\tilde \mu^{\otimes (\mn-\N)} (Z_{\mn-\N}) \Bigg]
\end{multline*}where we recall  that $\tilde \mu(\New_\eta)=\mn-\N$. We then use the same proof as in Lemma \ref{prominok}, Proposition \ref{pro642} and Proposition  \ref{pro718}.  The term $\sum_{(i,j) \in J}   \g(x_i-z_j)$ adds a contribution 
\begin{equation*}- C \beta  (\mn - \N)^{\mn- \N}  \sum_{i\in I_\pa}\int_{|z-x_i|\le \rrc_i}       \g(x_i-z)  d\tilde \mu(z)
    \ge - C \beta  (\mn - \N)^{\mn- \N}   \# I_{\pa}
\end{equation*}
and, by $\#I_\pa \le Cs/\tilde \ell$ and~\eqref{constr3}, we conclude that 
\begin{align*}
&
\int_{\New_\eta^{\mn-\N}} \exp\bigg(-C \beta \bigg(\F(Z_{\mn-\N}, \tilde \mu,\New_\eta)   +\sum_{(i,j) \in J}   \g(x_i-z_j)\bigg) \bigg) d\mu|_{\New}^{\otimes (\mn-\N)} (Z_{\mn-\N})
\\ & \qquad 
\ge   (\mn-\N)^{\mn-\N} \exp\bigg(    \int_{\New_\eta} \tilde \mu \log \frac{\mu}{\tilde \mu} - C \beta R^{\d-1}\tilde \l (1+|\g( R )|\indic_{\s\le 0}) \bigg).
\end{align*}
In the case $\s\le 0$ i.e.~$\d=1,2$, in view of the fact that~$\tilde \ell \ge \beta^{-\frac{1}{\d-\s}}$, we see from its construction  that~$\New_\eta $ can be partitioned into disjoint nondegenerate cells of size $\max(1,\beta^{-\frac{1}{\d-\s}})$ in which $\tilde\mu$ integrates to an integer. Using superadditivity as in the proof of Propositions \ref{pro642}  and~\ref{pro718}, 
we can improve this into 
\begin{multline}\label{claim45}\int_{\New_\eta^{\mn-\N}} \exp\bigg(-C \beta \bigg(\F(Z_{\mn-\N}, \tilde \mu,\New_\eta)   +\sum_{(i,j) \in J}   \g(x_i-z_j)\bigg) \bigg) d\mu|_{\New_\eta}^{\otimes (\mn-\N)} (Z_{\mn-\N})
\\ \ge
(\mn- \N)^{\mn-\N}
 \exp\(    \int_{\New_\eta} \tilde \mu \log \frac{\mu}{\tilde \mu}- C \beta \chi(\beta)  R^{\d-1} \tilde \ell  \).
\end{multline}
Using then~\eqref{mmut2},~\eqref{constr3} and~\eqref{bornimp} we have $|\frac{\mu}{\tilde \mu} -1|< \hal$  in $\New_\eta$  if $c$ is chosen small enough and thus by Taylor expansion
\be\label{413}\int_{\New_\eta} \tilde \mu \log \frac{\mu} {\tilde \mu} = \int_{\New_\eta} \mu - \tilde \mu + O\( \int_{\New_\eta} \frac{|  \mu- \tilde \mu|^2}{\tilde \mu} \)  = \mu(\New_\eta)- \tilde \mu(\New_\eta) + O \( \frac{s}{\ell \tilde \l}+\frac{\eta^2}{\ell}  R^{\d-1}\).   \ee
Together with \eqref{bornimp} and \eqref{claim45} this concludes the proof.
\end{proof}

\begin{rem} \label{remerr}
When summing the contributions over $\Omega$ where $n$ points fall and $U\backslash \Omega $ where $N-n$ points fall, the errors of~\eqref{epv} compensate and add up to a well bounded error. More precisely, 
if  $\alpha, \alpha'$, respectively $\gamma, \gamma'$ satisfy ~\eqref{cara} then for every $n$ we have
\begin{multline}\label{425}
\alpha-\alpha'+ (\mn-n-\alpha) \log \frac{\alpha}{\alpha'} - (\alpha+n-\mn+\hal) \log \(1+\frac{n-\mn}{\alpha}\)  + \hal \log \frac{n}{\mn} 
\\+ \gamma'-\gamma+ (n-\mn-\gamma) \log \frac{\gamma}{\gamma'} - (\gamma+\mn-n+\hal) \log \(1+\frac{\mn-n}{\gamma}\)  + \hal \log \frac{N-n}{N-\mn} \\ \le  C\( \frac{R^{\d-1}}{\tilde \ell} + \frac{s^2}{\tilde \ell^3 R^{\d-1}}\).\end{multline}
\end{rem}
\begin{proof}
First we notice that since the expressions arising here  originate in Stirling's formula,  they can be restricted to the case of $\alpha+ n-\mn \ge 1$, $\gamma+\mn-n\ge 1$, $n\ge 1$ and $N-n\ge 1$ (all the quantities involved are integers). 

We then study the expression in the left-hand side of~\eqref{425} as a function of the real variable $n$ (with the above constraints).  Differentiating in $n$, we find that it achieves its maximum when  
\begin{multline*} \log \frac{\gamma \alpha'}{\gamma'\alpha}- \log \( 1+\frac{n-\mn}{\alpha}\)  + \frac{1}{2(\alpha+ n-\mn)} + \log \( 1+ \frac{\mn-n}{\gamma}\)  -\frac{1}{2(\gamma+\mn-n)}\\+ \frac{1}{2n}-\frac{1}{2(N-n)} 
  =0. \end{multline*}
Using $\alpha + n-\mn\ge 1$, $\gamma+\mn-n\ge 1$, $n \ge 1$, $N-n\ge 1$ and~\eqref{cara} we deduce that 
$$\left| \log \( 1+ \frac{\mn-n}{\gamma}\)- \log \( 1+\frac{n-\mn}{\alpha}\)\right|\le C$$
and thus 
$$ \frac{  1+ \frac{\mn-n}{\gamma}}{1+\frac{n-\mn}{\alpha}}   \ \text{is bounded above and below}$$
and it follows easily in view of~\eqref{cara} that $|n-\mn|\le C  \tilde \ell R^{\d-1}$. 
To find the maximum of~\eqref{425} it thus suffices to maximize it for such $n$'s. But for such $n$'s we may check that 
$\hal \log \( 1+\frac{n-\mn}{\alpha}\) $, $\hal \log \(1+ \frac{\mn-n}{\gamma}\)$, $\log \frac{n}{\mn} $ and $\log \frac{N-n}{N-\mn}$
 are all bounded by a constant depending only on $\d, m, \|\mu\|_{L^\infty}$, hence it suffices to obtain a bound for 
 \begin{multline}\label{426}
\alpha-\alpha'+ (\mn-n-\alpha) \log \frac{\alpha}{\alpha'} - (\alpha+n-\mn) \log \(1+\frac{n-\mn}{\alpha}\)  
\\+ \gamma'-\gamma+ (n-\mn-\gamma) \log \frac{\gamma}{\gamma'} - (\gamma+\mn-n) \log \(1+\frac{\mn-n}{\gamma}\).  \end{multline}
 Differentiating in $n$, we find that this expression is maximal exactly for 
 $$  1+\frac{\mn-n}{\gamma}  =\frac{\gamma'\alpha}{\gamma\alpha'}  \(1+\frac{ n-\mn}{\alpha}\)  \Leftrightarrow n=\mn+ \frac{\frac{\gamma}{\gamma'}- \frac{\alpha}{\alpha'}} {\frac{1}{\gamma'}+\frac{1}{\alpha'} }$$
 Inserting this into~\eqref{426} we find  that the expression is then equal to 
 \begin{multline*}\alpha-\alpha' - \alpha \log \frac{\alpha}{\alpha'}-\alpha \log  \( 1+\frac{n-\mn}{\alpha}\)-\gamma \log    \( 1+  \frac{\mn -n }{\gamma}\) +(\mn-n) \log \frac{\gamma'}{\gamma}  \\= O\( \frac{R^{\d-1}}{\tilde \ell} + \frac{s^2}{\tilde \ell^3  R^{\d-1}}\)  \end{multline*}
 where we used a Taylor expansion and ~\eqref{cara}.
\end{proof}

The next corollary allows to compare the Neumann partition function to the ``Dirichlet" partition function reduced to good (and screenable) configurations.

\begin{coro}\label{coro43}With the same assumptions and notation as in the previous proposition and Definition \ref{defscreen},
there exists $C>0$ depending only on $\d, m, \|\mu\|_{L^\infty}$ such that the following holds. Let  $M \ge 1$ and 
$$\mathcal B_n=\left\{ X_n \in \Omega^n, \sup_x \int_{  \{ 
(\pa \Omega)_{-2\tilde\l}\cap \carr_L(x)}|\nab w_{\rrc}|^2\le M \chi(\beta)L^\d\right\}$$
where $w$ is a minimizer for \eqref{innernrj}, resp. \eqref{outernrj}, relative to $\Omega$, and  $(\pa \Omega)_{-2\tilde\l}$ denotes $Q_{R-\tilde \l}\backslash Q_{R-2\tilde \l}\cap U $ if $\Omega =Q_R\cap U$ and $Q_{R+2\tilde \l}\setminus Q_{R+\tilde \l}\cap U $ if $\Omega= U\setminus Q_R$.
If \be\label{condsurL} R>L > CM\max(  \chi(\beta) ,\beta^{-\frac{1}{\d-\s}}\indic_{\s\le 0})  ,\ee  and
$\dist(Q_R, \pa\Lambda \cap U)\ge L$, we have
\begin{multline}\label{eqsulk}
\frac1{n!} \int_{\mathcal B_n} \exp\(-\beta \G_U^{\mathrm{inn/out}}(X_n, \Omega) \)d\mu^{\otimes n} (X_n) 
\\ \le  C\frac{\mn^\mn}{\mn!} \K_\beta(\Omega,\mu) \exp\Bigg[  \beta \(  C R^{\d-1}L  \chi(\beta)M +|n-\mn|   \)   +  \frac{C M \chi(\beta) R^{\d-1}}{  L}  - \log \min(1, \beta)  \\
+ \mn-n +\alpha-\alpha'+ (\mn-n-\alpha) \log \frac{\alpha}{\alpha'} - (\alpha+n-\mn+\hal) \log \(1+\frac{n-\mn}{\alpha}\)  + \hal \log \frac{n}{\mn} 
\Bigg],\end{multline}
with $\alpha, \alpha'$ satisfying 
 $$  \left|\frac{\alpha'}{\alpha}-1\right|\le C \frac{   \chi(\beta) }{L } , \qquad  \frac1C L  R^{\d-1}\le \alpha \le C L R^{\d-1}.$$
 \end{coro} 
\begin{proof} 
If $X_n$ in $\mathcal B_n$ then 
 $$\bar S (X_n) \le \frac{R^{\d-1}}{L^{\d-1}} M  \chi(\beta) L^\d, \quad \bar S'(X_n) \le M  \chi(\beta) L^\d .$$
 using the definition~\eqref{bestS} or~\eqref{bestS2}.
  We check that 
   setting  $\ell=\tilde \ell = L$ and $s= M \frac{R^{\d-1}}{L^{\d-1}}  \chi(\beta)L^\d$ and $z= M \chi(\beta) L^\d$ we  have  that if~\eqref{condsurL} holds, then up to making the constant larger in~\eqref{condsurL}, in view of \eqref{defchibeta},~\eqref{constr2} and~\eqref{constr3} hold. Choosing then $\eta= \frac{\min(1, \beta) L m }{4\|\mu\|_{L^\infty}}$,  \eqref{conditionsureta} is satisfied and   the result follows by applying the result of Proposition~\ref{pro42}. 
\end{proof}
The next goal is to select $s, \l,\tilde \l$ to optimize the errors made in Proposition \ref{pro42}. This way
we  obtain the main result of this section, which implements the heuristic idea described in the first section of this chapter, and shows that one can run the bootstrap procedure and show that if one has good energy controls at some scale, one can deduce   control at  half scales.


In all the rest of the paper, we will denote the event that $\XN$ has $n$ points in $\Omega$ by
 \be 
\label{Bndef} \mathcal A_n:=\{ \XN \in U^N, \# I_\Omega = n\}.
\ee
The next proposition allows to leverage on the screening for a given number of points $n$ falling in $\Omega$. The work in the  main proof will then consist in conditioning on $n$.
\begin{prop}[Bootstrap] \label{proboot}
Assume $U$ is either $\R^\d$  or a  finite disjoint union of disjoint hyperrectangles all included in $\Lambda$ with parallel sides belonging  to $ \mathcal Q_\rho$ for some $\rho \ge \max(1,
\beta^{-\frac{1}{\d-\s}} \indic_{\d\le 2})$, or the complement of such a set. 
Let $\mu$ be a density such that $\mu \ge m>0 $  in the set $\Lambda$ and $\mu(U)=N$ is an integer.
Let $C_0$ be  the constant of 
~\eqref{14}. 
  
There exists a   constant  $C>0$  depending only on $\d, m$ and  $\|\mu\|_{L^\infty}$ such that the following holds.
Assume  that $Q_R$ is a hyperrectangle of sidelengths in $[R,2R]$ with sides parallel to those of $U$, that $\mu (Q_R \cap U)=\mn$ and $Q_R \cap U \subset \Lambda$.
  Assume that  there exists a cube 
$\carr_L$ of size $L$ such that 
  \be\label{locallaw}
  \left|\log \Esp_{\Q_\beta (\mu,U)} \( \exp \(\frac{\beta}{2} \( \F^{ \carr_L} (\cdot ,\mu, U)+ C_0 \# I_{\carr_L})  \)    \) \)\right| \\
  \le \mathcal C \beta \chi(\beta) L^\d\ee with $\mathcal C>1$, 
and assume  that 
    $\carr_L$ contains  $ Q_{R+2\tilde \l} \cap U$ 
 with $$   L\ge R \ge \hal L,$$  
\be \label{condRL}
    R>C '\max(1, \beta^{-\hal} \chi(\beta)^{\frac12}) 
\ee and
 \be\label{deftl} \tilde \l=   
  C''  \max  \( \(\frac{R }{ \max(1, \beta^{-\hal} \chi(\beta)^{\frac12})}\)^{-\frac23}   R,   R^{1-\frac{2}{\d+2}}\) \ee 
for  some $C', C''$ large enough,  both depending only on $\d$, $m,\|\mu\|_{L^\infty}$  and $\mathcal C$.
 Assume in addition that    \be\label{assumpdist}
 \dist (Q_R\cap U , \partial \Lambda \cap U) \ge \tilde \l. \ee
Then there exists a sequence $\gamma_n$ satisfying
\be\label{boundgamma}
\sum_{n=0}^N  
\gamma_n\le \exp\(- \mathcal C \beta \chi(\beta)R^\d\)
\ee such that we have
\begin{multline}
\label{resLK}
\Esp_{\Q_\beta(\mu,U)} \(  \exp \(  \frac{\beta}{2} \( \F^{Q_{R-2\tilde \ell}}(X_N,\mu, U) \) \indic_{\mathcal A_n}\)   \)
\\ \le
\gamma_n + \frac{\K_{\beta/2} (\mu, Q_R) }{\K_\beta (\mu,Q_R)} \exp\(
\beta\( \frac{ \mathcal C }{ 4} \chi(\beta) R^\d  
 +|n-\mn| + \frac{C_0}{2}n\)  \).
\end{multline}
\end{prop}
Once one has obtained local laws down to the minimal scale $\rb$,  Corollary \ref{coro43} will allow to improve the error term and bound it by $R^{\d-1}$.

\begin{proof}
{\bf Step 1: the case of excess energy}. We denote $\Omega= Q_R\cap U$ and $\omc= Q_{R-2\tilde \ell}\cap U$. 
Recalling the definition of~$\mathcal{A}_n$ in~\eqref{Bndef}, and letting  $M\ge 1$ be a constant to be determined below, we define 
\begin{equation*}
\mathcal B_n:= \Big\{ X_N \in \mathcal{A}_n, \  \int_{\carr_L \backslash \omc } |\nab u_{\rrc} |^2  \le  M \mathcal{C} \chi(\beta) L^\d\Big\},\end{equation*} where for each configuration, $u$ is defined as solving \eqref{defv} over $U$ and $\rrc$ is as in \eqref{defrrc3} relative to $\carr_L$. 
 If $X_N \in \mathcal B_n^c$ then 
 \be \label{pcas}
 \int_{\carr_L\backslash \omc  } |\nab u_{\rrc}|^2  >  M \mathcal{C} \chi(\beta) L^\d,
 \ee  
 and  view of the definition \eqref{Glocal2} and \eqref{14} we then have that 
 \begin{multline*}
 \F^{\carr_L} (\XN, \mu, U) + C_0 \#I_{ \carr_L} 
\ge \F^{\omc} (\XN,\mu, U) +  \frac{1}{4\cd} \int_{\carr_L\backslash \omc} |\nab u_{\rrc}|^2
\\ \ge \F^{\omc} (\XN,\mu, U) 
 + \frac{M \mathcal C \chi(\beta) L^\d}{4\cd}.
 \end{multline*}
  We deduce that 
\begin{align*}
 & 
\Esp_{\Q_\beta(U,\mu)} \( \exp \(\frac{\beta}{2} \( \F^{ \carr_L} (\cdot ,\mu, U)+ C_0 \# I_{\carr_L}  \) \)\indic_{\mathcal B_n^c}   \)
\\ & \qquad 
\ge \  \exp\(\frac{\beta}{2} \frac{M \mathcal{C}  \chi(\beta)L^\d} {4\cd}\) \Esp_{\Q_\beta(U,\mu)} \( \exp \(\frac{\beta}{2}   \F^{ \omc} (\cdot ,\mu, U)       \)\indic_{\mathcal B_n^c}\).
\end{align*} 
It follows  from \eqref{locallaw} that 
\be \label{controlbad}
\Esp_{\Q_\beta(U,\mu)} \( \exp \(\frac{\beta}{2} \F^{ \omc} (\cdot ,\mu, U)     \) \indic_{\mathcal B_n^c}\)\le  \gamma_n,
\ee
with 
$\sum_{n=0}^N
\gamma_n \le   \exp\(- \mathcal C \beta \chi(\beta)R^\d\)$,
 provided~$M$ is chosen large enough, depending only on $\d, m, \|\mu\|_{L^\infty}$.
We henceforth fix~$M$.
\smallskip

{\bf Step 2: the case of good energy bounds.}\\
We now wish to estimate the same quantity in the event $\mathcal B_n$.  First we note that   being in $\mathcal B_n$ implies a control of $S(X_n, u)$  defined as in \eqref{definitionsannex0} or \eqref{definitionsannex2} by $M \mathcal C \chi(\beta) L^\d$.

We next wish to choose   $\tilde \l=\tilde \epsilon  R$ with $\tilde \epsilon <\frac14 $ to be determined later, and  $\ell=\epsilon\tilde \ell$ with $0\le \epsilon\le 1$, satisfying
\be\label{defl}
  \l^{\d+1} \ge  \frac{CM\mathcal C \chi(\beta) L^\d }{ \tilde \l}    \ee with $C$ as in~\eqref{constr2}.
This way, choosing $s= M\mathcal C \chi(\beta) L^\d$, the screenability condition~\eqref{constr2} is verified for $u$.
To apply Proposition \ref{pro42}, we also need 
\be  \label{condtl}\max \(\beta^{-\frac{1}{\d-\s}} \indic_{\d\le 2},1\) \le \tilde \epsilon R
\ee
  and 
\be \label{435b} 
M \mathcal C \chi(\beta)  L^\d< c\tilde \ell^2 R^{\d-1}.\ee
Using~\eqref{locali2}, Lemma~\ref{lemrestri} (which applies since $u$ is screenable) and~\eqref{14},  we have 
\begin{align*}
  \frac{\beta}{2}  \F^{ \omc} (X_{N} , U)   -\beta \F(X_{N}, U) & \le 
 \frac{\beta}{2}\F^{\omc} (X_{N}, U)   -\beta \F^{\Omega}(X_{N}, U) 
 - \beta\F^{U\backslash \Omega} (X_{N}, U)\\
  & \le  \frac{\beta}{2} \F^{\omc} (X_{N} , U) -\frac{\beta}{2} \F^{\Omega}(X_{N}, U)  -\frac{\beta}{2} \F^{\Omega}(X_{N},U) - \beta\F^{U\backslash \Omega} (X_{N}, U)
  \\  & \le  -\frac{\beta}{2} \F^{\Omega \backslash  \omc} (X_{N}, U)    -\frac{\beta}{2} \F^{\Omega}(X_{N},U) - \beta\F^{U\backslash \Omega} (X_{N}, U)
  \\
& \le - \frac{\beta}{2} \F^{\Omega}(X_{N},U)  - \beta \F^{U\backslash \Omega} (X_{N}, U)  + \frac{\beta}{2} C_0 n
\\ & \le  - \frac{\beta}{2} \G_U^{\mathrm{inn}}(X_{N}|_{\Omega} , \Omega) - \beta \G_U^{\mathrm{out}} (X_{N}|_{U\backslash \Omega} 
, U\backslash \Omega)  + \frac{\beta}{2}C_0n .\end{align*}
Thus,
\begin{align*}
\lefteqn{
\Esp_{ \Q_\beta(\mu,U)}  \( \exp \(\frac{\beta}{2} \F^{ \omc} (\cdot , U) \) \indic_{\mathcal B_n}\)
} \quad & 
\\ & 
= \frac{1}{N^N \K_\beta(\mu,U)} \int_{\mathcal B_n} \exp\(  \frac{\beta}{2} \F^{\omc}(X_N,U) -\beta \F (X_N,U)  \) d\mu^{\otimes N} 
\\ & 
\le  \frac{1}{N^N  \K_\beta(\mu,U)}   \frac{N!}{n!(N-n)!}  \int_{\Omega^n \cap \mathcal B_n^-} \exp\( - \frac{\beta}{2} \G_U^{\mathrm{inn}}(\cdot , \Omega)    + \frac{\beta}{2} C_0 n\) d\mu^{\otimes n}
\\ & \qquad 
\times \int_{(U\backslash \Omega)^{N-n}\cap \mathcal B_n^+ } \exp\( - \beta \G_U^{\mathrm{out}}(\cdot , U\backslash \Omega)   \)d\mu^{\otimes (N-n)}.\end{align*}
Inserting~\eqref{resscreen} applied in $\Omega $ (with $\beta/2$  instead of $\beta$) and in $U\backslash \Omega$ and using Remark \ref{remerr}, we deduce that, for $\eta$ satisfying \eqref{conditionsureta},
\begin{align*}
\lefteqn{
\Esp_{\Q_\beta(\mu,U)}  \( \exp \(\frac{\beta}{2} \( \F^{ \omc} (\cdot , U)  \)  \) \indic_{\mathcal B_n}\)
} \quad & 
\\ & 
\leq
  \frac{1}{N^N  \K_\beta(\mu,U)}   \frac{N!}{\mn!(N-\mn)!}    C \mn^\mn (N-\mn)^{N-\mn} \K_{\beta/2}(\Omega,\mu)  \K_{\beta}(\mu, U\backslash \Omega) \exp\(  \beta \ep_e+ \ep_v + \frac{\beta}{2} C_0 n\)
  \end{align*} 
with  \be \ep_{e} := C\(\ell \frac{ M \mathcal C \chi(\beta) L^\d }{\tilde \ell}+ R^{\d-1} \tilde \ell \chi(\beta)+ |n-\mn|\) 
\ee and 
\be \ep_v := C\(\frac{ M \mathcal C \chi(\beta) L^\d  }{\ell \tilde \ell}+ \frac{R^{\d-1}}{\tilde \ell}+ \frac{ ( M \mathcal C \chi(\beta) L^\d )^2}{\tilde \ell^3 R^{\d-1}}   +\eta R^{\d-1} 
+\log \frac{\tilde \ell}{\eta}
 \),\ee where we used the choice $s:=M\mathcal C \chi(\beta) L^\d$.
   We may also bound from below  $\K_\beta(\mu,U)$  using ~\eqref{superad2} applied with the sets $\Omega $ and $U\backslash \Omega$, which yields
   $$ \frac{N! \mn^{\mn} (N-\mn)^{N-\mn} }{N^N  \K_\beta(\mu,U) \mn! (N-\mn)!}  \K_{\beta/2}(\mu,\Omega)\K_{\beta} (\mu, U \backslash\Omega) 
   \le  \frac{\K_{ \beta/2}(\mu,\Omega)}{\K_{\beta} (\mu,\Omega)}.$$ 
   Inserting into the above, we obtain that 
    \begin{equation}\label{438}
\Esp_{\Q_\beta(\mu,U)}  \( \exp \(\frac{\beta}{2}  \F^{ \omc} (\cdot , U)     \) \indic_{\mathcal B_n}\)
\\ 
\leq C   \frac{\K_{ \beta/2}(\mu,\Omega)}{\K_{\beta} (\mu,\Omega)} \exp\(  \beta \ep_e+ \ep_v + \frac{\beta}{2} C_0 n\).
\end{equation} 
   
We now search for the smallest~$\tilde \l$ and $\eta$ such that the terms of~$\beta \ep_e+\ep_v$ (except those involving~$n$ and $\mn$) are $\le  \beta \chi(\beta) \frac{\mathcal C}{8}R^\d$ 
that is 
 \begin{align*} 
\left\{   
\begin{aligned}
& C\frac{ M\mathcal C \chi(\beta) L^\d \ell}{\tilde \ell} \le \frac{\mathcal C }{8} \chi(\beta)  R^\d,
\\&
 CR^{\d-1}\tilde \ell \chi(\beta) \le \frac{\mathcal C }{8} \chi(\beta) R^\d,\\
 &  C \frac{M\mathcal C \chi(\beta) L^\d}{\ell \tilde \ell} \le \frac{\mathcal C }{8}\beta  \chi(\beta) R^\d,
 \\
& C \frac{R^{\d-1}}{\tilde \ell}\le \frac{\mathcal C }{8} \beta \chi(\beta) R^\d,
 \\ &C \frac{(M \mathcal C \chi(\beta) L^\d )^2}{\tilde \ell^3 R^{\d-1}} \le \frac{\mathcal C }{8}\beta \chi(\beta) R^\d\\
 & \eta R^{\d-1}  \le  \beta\frac{\mathcal C }{8} \chi(\beta) R^\d\\
& \log \frac{\tilde \ell}{\eta}\le
\frac{\mathcal C }{8} \beta\chi(\beta) R^\d,
\end{aligned}\right.
\end{align*}
and also~\eqref{defl},~\eqref{condtl},~\eqref{435b}, and \eqref{conditionsureta} are satisfied.
 With our choice  $R \le L \le 2R$ and $\ell =\epsilon \tilde \ell$, $\tilde \ell=\tilde \epsilon R$, after direct computations we find that, leaving aside the conditions on $\eta$,  these reduce to:
  \begin{align*} 
\left\{   
\begin{aligned}
&  
C \epsilon M \le \frac18,
\\ &
C \tilde \epsilon \le \frac{\mathcal C}{8} ,
\\ &
\frac{ C M }{\epsilon \tilde \epsilon^2} \le \frac{\beta}{8} R^2,
\\ &
\frac{C}{\tilde \epsilon} \le \frac{\mathcal C}{8} \beta \chi(\beta) R^2,
\\ & CM^2\mathcal C\chi(\beta) \le  \frac{\beta}{8} R^2 \tilde \epsilon^3,
\\ &
\epsilon^{\d+1}\tilde \epsilon^{\d+2} R^2 \ge C M \mathcal C \chi(\beta) ,
\\ &
\tilde \epsilon R \ge \max\(\beta^{-\frac1{\d-\s}}\indic_{\d\le 2}, 1\) ,
\\ &
CM \mathcal C \chi(\beta) <\tilde \epsilon^2  R,
\end{aligned} 
\right. 
\end{align*}
    for some constant $C>0$ large enough, and depending only on $\d,m$ and $\|\mu\|_{L^\infty}$.
    We can  take $\epsilon$ small enough that the first condition is realized.
    We can then make the other conditions realized by requiring 
    $$\left\{ \begin{aligned} &R^2\min( \beta \chi(\beta)^{-1},1) \ge C'\\
    & \tilde \epsilon= C''\max\((R^2 \beta \chi(\beta)^{-1})^{-1/3},  R^{-\frac{2}{\d+2}}\)
    \end{aligned}\right.$$
    where $C', C''$ are large enough constants 
     depending on the other parameters.
     Choosing then $\eta= \min(1, \beta) \frac{m}{4\|\mu\|_{L^\infty} } \ell\le  \min(1, \beta) \frac{\tilde \ell} {4}$,  and inspecting the definition \eqref{defchibeta}, we find that in all dimensions the conditions on $\eta$ are also satisfied, up to making the constants larger if necessary. 
     
    Combining ~\eqref{438} with~\eqref{controlbad}, we obtain the result. 
\end{proof}

We now can complete the proof of the local laws.
\begin{proof}[Proof of Theorem \ref{th3}]
We first note that it is enough to prove the local law result in  hyperrectangles $Q_R\in \mathcal{Q}_R$ (as in Definition \ref{defQR}), with sides parallel to those of $U$ and even more generally in $Q_{R-2\tilde \ell}$ if $\tilde \ell<\frac14R$, with $R\ge \rb$. Also, if $U$ is a hyperrectangle, thanks to a similar reasoning, we may ignore the condition in Definition \ref{defscreen} about the faces of $\partial (Q_R \cap U)$ being sufficiently far from the parallel faces of $U$ (see also the discussion in the proof of Theorem \ref{th4}) . 

 Indeed, thanks to the lower bound on $\mu$,  general cubes  of size $R\ge \rb$ can be covered by a finite number of such hyperrectangles. 
 The proof then proceeds by a bootstrap on the scales:
we wish to show that if 
 \be\label{hyprec}
 \log \Esp_{\Q_\beta(\mu,U)} \( \exp \(\frac{\beta}{2} \( \F^{\carr_L(x) }(\cdot, U)+ C_0 \#I_{\carr_L(x)}\) \)\) \le \mathcal C \beta\chi(\beta) L^\d ,\ee
 for any $\carr_L(x)$  sufficiently far from $\partial \Lambda$, then  if $ \frac34 L \ge R \ge \hal L$, and as long as $R$ is large enough, we have
 \be\label{hyprecfin}
 \log \Esp_{\Q_\beta(\mu,U)} \( \exp \(\frac{ \beta}{2} \( \F^{Q_{R-2\tilde \ell}}(\cdot, U)+C_0 \#I_{  Q_R} \) \)\) \le \mathcal C \beta\chi(\beta) R^\d.  \ee 
   By iteration, this will clearly imply the result: indeed rewriting 
   $ \Esp_{\Q_\beta(\mu,U)} \( \exp  \(\frac{ \beta}{2} \( \F(\cdot, U) \) \)\)$ as $ \frac{\K_{\beta/2}( \mu,U)} {\K_{\beta}(\mu,U)} $ as in the proof of Corollary \ref{cor521}, and using \eqref{bornesfiU}, 
     we have that~\eqref{hyprec} holds for  $L\ge \hal N^{\frac1\d}$ up to changing $\mathcal C$ if necessary.  
Without loss of generality, we may now  assume for the rest of the proof that  $L \le \hal N^{\frac1\d}$.

 To make sure that the constants are independent of~$\beta $ and~$R$,  we have used the notation~$\mathcal C$, and we wish to prove~\eqref{hyprecfin} with the same constant~$\mathcal C$ as in~\eqref{hyprec}. 
 In the sequel, unless specified, {\it all constants~$C>0$ will be independent of~$\mathcal C $}, i.e.~they may depend only on~$\d,m$ and~$\|\mu\|_{L^\infty}$.
 
   \smallskip

 Let us now consider~$Q_R\in \mathcal Q_R$, denote~$\mn=\mu(Q_R\cap U)$  and as previously, denote by~$\mathcal A_n$ the event that~$X_N$, a configuration of~$N$ points in~$U$,  has~$n$ points in~$Q_R\cap U$.
We wish to control 
\begin{align*}
&
\Esp_{\Q_\beta(\mu,U)} \( \exp \(\frac{\beta}{2} ( \F^{Q_{R-2\tilde \ell} }(\cdot, U)+C_0 n) \)\)
\\ & \qquad
=\sum_{n=0}^N  \exp\left( \frac{\beta}{2} C_0 n\right)
 \Esp_{\Q_\beta(\mu,U)} \( \exp \(\frac{\beta}{2} ( \F^{Q_{R-2\tilde \ell} }(\cdot, U) )\indic_{\mathcal A_n} \)\).
\end{align*}
The terms in the sum for which $n$ is close to $\mn$, more precisely $|n-\mn| \le K R^{\d-\hal}$ are easily treated using~\eqref{resLK}. The terms for which $|n-\mn|> KR^{\d-\hal}$  will be handled separately and controlled by energy-excess considerations. 

To apply Proposition \ref{proboot} we need $Q_{R+\tilde \ell}$ to be included in a cube $\carr_L$ in which the local laws hold and at distance $\ge \tilde \ell $ as in~\eqref{deftl} from $\partial \Lambda$.
At the first iteration,
$L$ is of order $N^{\frac1\d}$ and $ R \ge \hal L$ so we need 
$$
\dist(Q_R,\partial \Lambda) \geq 
 C''  \max  \( \(\frac{N^{\frac1\d} }{ \max(1, \beta^{-\hal} \chi(\beta)^{\frac12})}\)^{-\frac23}   N^{\frac1\d} ,   N^{\frac{1}{ \d+2}}\) .
 $$
At further iterations, to have $Q_{R+\tilde \ell} $ be included in $\carr_L$,  we need a further distance of  
$$ C''  \max  \( \(\frac{R }{ \max(1, \beta^{-\hal} \chi(\beta)^{\frac12})}\)^{-\frac23}   R,   R^{1-\frac{2}{\d+2}}\) .$$  Since $R $ is multiplied by a factor in $[\hal , \frac34]$ at each step,  summing  the series over the iterations gives a total distance 
$$C''' \max  \( \(\frac{N^{\frac1\d} }{ \max(1, \beta^{-\hal} \chi(\beta)^{\frac12})}\)^{-\frac23}   N^{\frac1\d} ,   N^{\frac{1}{ \d+2}}\) $$ hence  a condition of the form~\eqref{conddist0} suffices. 
 \smallskip

{\bf Step 1: the bad event}. We claim that  in the bad event  $|\mn -n| > KR^{\d-\hal} $, we have 
\be\label{claim80}
 \F^{Q_{R+3}}(\XN, U) - \F^{Q_R}(\XN, U) \ge C R^{1-\d} |\mn - n|^2 - C \#I_{Q_{R+3}\backslash Q_R}\ee
where $C>0$ depends only on $\|\mu\|_{L^\infty}$ and $\d$.
Assuming this, and changing $C_0$ to the larger constant in~\eqref{claim80} if necessary,  we then write 
\begin{multline}\label{510}\Esp_{\Q_\beta(\mu,U) }\( \exp \(\frac\beta2 (\F^{Q_R }(\cdot, U)+  C_0 n) \)   \indic_{\mathcal A_n} \)
\\
\le \Esp_{\Q_\beta(\mu,U) }\( \exp \( \frac \beta2 (\F^{Q_{R+3} }(\cdot, U)  +   C_0 \#I_{Q_{R+3}\backslash Q_R }  + C_0n) \) \indic_{\mathcal A_n}   \) \exp\(- \frac{\beta}{2} C R^{1-\d}|\mn -n|^2 \)\\
\le  \Esp_{\Q_\beta(\mu,U) }\( \exp \( \frac \beta2 (\F^{Q_{R+3} }(\cdot, U)  +   C_0 \#I_{Q_{R+3}}) \) \indic_{\mathcal A_n}   \) \exp\(- \frac{\beta}{2} C R^{1-\d}|\mn -n|^2 \)  .
\end{multline} Since $ L \le 2R$ and $|\mn-n| > K R^{\d-\hal}$, we now see that if we choose $K = C \sqrt{\mathcal{C} \chi(\beta)}$ where $C>0$ is large enough and depends only on $C, C_0$ and $\d$,
the exponent in the second term in the right-hand side is at most $-  \mathcal C \beta \chi(\beta) L^\d$.

On the other hand, using Lemma \ref{lemrestri},~\eqref{locali2} and~\eqref{14}, we may check that 
$$\F^{Q_{R+3}}(\cdot, U)+C_0 \#I_{Q_{R+3}}\le \F^{\carr_L} (\cdot,U)+C_0 \#I_{\carr_L},$$ hence in view of 
~\eqref{510} and the assumption that~\eqref{hyprec} satisfied in a cube $\carr_L$ containing $Q_{R+3}$, we may  then bound 
\begin{multline}\label{stp1}\sum_{n, |\mn -n|> KR^{\d-\hal} } \log \Esp_{\Q_\beta(\mu,U) }\( \exp \(\frac \beta2 (\F^{Q_R }(\cdot, U)+ C_0 n) \) \indic_{\mathcal A_n}    \)
\\ \le \exp\(-\mathcal C \beta \chi(\beta) L^\d\)  \sum_{n=0}^N \Esp_{\Q_\beta(\mu,U) }\( \exp \( \frac\beta2 (\F^{\carr_L}(\cdot, U)  + \beta  C_0 \#I_{\carr_L } )\) \indic_{\mathcal A_n}   \)\le 1.
\end{multline}

To prove the claim, in view of~\eqref{disc10bu} we may  write 
\be\label{bf45}C \int_{Q_{R+2}\backslash Q_{R+1}} |\nab u_{\rrh}|^2 \ge CR^{1-\d} \( |n-\mn| - C\|\mu\|_{L^\infty}R^{\d-1}\)^2\ge c R^{1-\d} |\mn -n|^2\ee if $K$ is chosen large enough (depending on $\d$ and $\|\mu\|_{L^\infty}$),
where $c>0$ is a constant depending only on $\d,m$ and $\|\mu\|_{L^\infty}$.
In view of~\eqref{locali2} we have 
\be\label{fqr3}
\F^{Q_{R+3}}(X_N, U) - \F^{Q_R}(X_N, U) \ge \F^{Q_{R+3}\backslash Q_R}(X_N, U).\ee
By~\eqref{14}, we may write that 
$$ C\F^{Q_{R+3}\backslash Q_R}(X_N, U) \ge \int_{Q_{R+3} \backslash Q_R} |\nab u_{\rrc}|^2 - C \mathcal \#I_{Q_{R+3}\backslash Q_R}$$ where $u_{\rrc}$ is computed with respect to $Q_{R+3}\backslash Q_R$. But by definition $ \int_{Q_{R+3} \backslash Q_R} |\nab u_{\rrc}|^2$ is larger than 
$ \int_{Q_{R+2} \backslash Q_{R+1}} |\nab u_{\rrh}|^2$ with this time $\rrh$ computed with respect to $U$, which is bounded below by~\eqref{bf45}. 
Inserting into~\eqref{fqr3} we thus conclude~\eqref{claim80}.

 \smallskip

\noindent 
{\bf Step 2: the good event.}
We next  consider the terms for which $|\mn -n |\le KR^{\d-\hal}$.
For those, we may apply 
Proposition \ref{proboot} (at least if $R >C$ with $C$ made large enough). We need to assume~\eqref{condRL}. In view of~\eqref{resLK} we  may thus write  
\begin{align*}
\lefteqn{
\sum_{|n-\mn|\le KR^{\d-\hal}}  
 \Esp_{\Q_\beta(\mu,U)} \( \exp \(\frac{\beta}{2} \( \F^{Q_{R-2\tilde \l} }(\cdot, U) +C_0n \) \)\indic_{\mathcal A_n} \)
 } \qquad \  & 
 \\ & 
 \le  \sum_{n=\mn-KR^{\d-\hal}}^{\mn+ KR^{\d-\hal}} \exp\left( \beta
 \( \frac{ \mathcal C }{ 4} \chi(\beta) R^\d   +|n-\mn|
 + C_0 n\)    \)   \frac{\K_{\beta/2}(\mu, Q_R)}{\K_\beta(\mu, Q_R)}  
+ \gamma_n\exp\left( \frac{\beta}{2} C_0 n \) .\end{align*} 
Recalling the choice of $K$ as $C\sqrt{ \mathcal C \chi(\beta)}$ and using that $\mn= \mu(Q_R) \le \|\mu\|_{L^\infty} R^\d$,  we have that if 
$|n-\mn|\le KR^{\d-\hal}$ and $R \ge \mathcal C \chi(\beta)$, we have $KR^{\d-\hal} \le C R^\d$ and  $n\le CR^\d$, with $C$ depending only on $\d, m, \|\mu\|_{L^\infty}$. 
But $R \ge \mathcal C \chi(\beta)$ is satisfied as soon as \eqref{condRL} holds (up to changing $C'$ if necessary).
\smallskip

Using~\eqref{boundgamma} and using \eqref{bornesfiU} to bound the ratio of partition functions, we deduce that, for every $R$ satisfying \eqref{condRL}, 
\begin{align*}
& \sum_{n=\mn-K R^{\d-1/2}}^{\mn+ KR^{\d-1/2}}   \Esp_{\Q_\beta(\mu,U)} \( \exp \(\frac{\beta}{2} \( \F^{Q_{R-2\tilde \l} }(\cdot, U) +C_0n \)\)\indic_{\mathcal A_n} \)
 \quad
\\ & \quad 
\leq C R^\d \exp\(\beta \( \frac{ 3 \mathcal C }{ 8} \chi(\beta) R^\d 
  + C_0 CR^\d \) \)   \exp\( C\beta \chi(\beta)R^\d \)  
+\exp\(\beta C_0 C R^\d -  \mathcal C \beta \chi(\beta) R^\d\).
\end{align*} 
Making $\mathcal C$ larger if necessary (compared to the constants $C_0$, $C$ appearing here) we deduce 
\begin{multline}
\label{bonterm}
 \sum_{n=\mn-K R^{\d-1}}^{\mn+KR^{\d-1}} \Esp_{\Q_\beta(\mu,U)} \( \exp \(\frac{\beta}{2} \( \F^{Q_{R-2\tilde \l} }(\cdot, U) +C_0n\) \)\indic_{\mathcal A_n} \) 
\\ 
\le  \exp\( \beta \frac{\mathcal C}{2} \chi(\beta) R^\d      + C \log R    \).
\end{multline} The logarithmic term can then  be absorbed  using that $R\ge \rb$.
\smallskip

Combining~\eqref{stp1} and~\eqref{bonterm}, we conclude that 
~\eqref{hyprecfin} holds. This  yields that for any $ \carr_R(x)$ satisfying  ~\eqref{conddist0},  the estimate~\eqref{locallawint0} holds.

\end{proof}
 
 \section{Consequences of the local laws and almost additivity}
 \subsection{Direct corollaries}
Since the electric  energy $\F$ controls discrepancies and linear statistics, we can immediately deduce controls on such quantities. These discrepancy controls are not expected to be sharp, but they already provide a form of rigidity of the particles numbers. We refer to Section~\ref{sec:isotropic} for the better discrepancy estimates due to Thoma and obtained by isotropic averaging, and to Section \ref{sec:leble} for the best-to-date two-dimensional hyperuniformity (or variance of the number of points) result of Lebl\'e \cite{leblehyper}.

\begin{coro}[Discrepancy controls]\label{coro61}
\index{discrepancy}
Assume the hypotheses of Theorem \ref{th3} for $\carr_R(x)$ with $R\ge \rb$ and let $B$ be a ball such that $2B \subseteq \carr_R(x)$. 
There exists $C>0$ depending only on $\d,m$ and $\|\mu\|_{L^\infty}$ such that letting $$D:= \int_B \( \sum_{i=1}^N \delta_{x_i}- d\mu\right) ,$$ we have
\be \label{loclawpoints0}
 \log \Esp_{\Q_\beta(\mu,U)} \Bigg(       \exp \Bigg( \frac{\beta}{C } \chi(\beta)^{1/3} R^{2(1-\d)} \rb^{\d-2/3}  D^2  \  \Bigg) \Bigg)  \le    C  \beta \chi(\beta )  \rb^{\d}  ,
 \ee
and 
\be\label{loclawpoints00}\log \Esp_{\Q_\beta(\mu,U)} \( \exp\(  \frac{\beta}{C} \frac{D^2 }{R^{\d-2}} \min \(1, \frac{|D|}{R^\d}\) \) \) \le C \beta \chi(\beta ) R^\d.  
\ee
\end{coro}
\begin{proof}
We may suppose $x=0$. 
First, we observe that by choice of $C_0$ and~\eqref{14} we  have for any  $R\ge \rb$,
\be \label{fini} \log \Esp_{\Q_\beta(\mu,U)} \( \exp \( \frac{1}{2C} \beta \int_{\carr_R}| \nab u_{\rrc}|^2 \) \) 
\le C  \beta \chi(\beta)R^\d  \ee
where $\rrc$ is computed with respect to $\partial \carr_R$.
We next may use either first~\eqref{disc10bu}--\eqref{disc1bu} or second~\eqref{disc30}--\eqref{disc3}  (after suitable blow-up) to deduce from this a control of the discrepancy, after noting that these inequalities apply as well to $u$ instead of $h_N$.

In the first way we cover  $\carr_{R+2}\backslash \carr_{R-2}$ by at most $O( (R/\rb)^{\d-1} )$ cubes  $Q_k$ of size $\rb$.  Applying~\eqref{fini} for the cubes $Q_k$ and using 
the generalized H\"older inequality 
\be\label{gholder}
\Esp(f_1 \dots f_k) \le \prod_{i=1}^k \Esp(f_i^k)^{\frac{1}{k}},\ee which can be proved by induction, 
 we find 
\be \label{tryu}
\log \Esp_{\Q_\beta(\mu,U)} \( \exp\( C^{-1} \beta (\frac{R}{\rb})^{1-\d}\int_{\carr_{R+\rb}\backslash \carr_{R-\rb}} |\nab u_{\rrc}|^2 \) \) \le C\beta \chi(\beta)\rb^\d,
\ee for some constants $C$ depending only on $\d ,m$ and $\|\mu\|_{L^\infty}$.
In view of~\eqref{disc10bu}--\eqref{disc1bu}, we  then have that for all $1\le \delta \le \rb$, 
$$\left|\int_{\carr_R} \sum_{i=1}^N \delta_{x_i}- d\mu\right|^2  \le C \|\mu\|^2_{L^\infty} R^{2(\d-1)}\delta^2 + C \frac{R^{\d-1}}{\delta}  \int_{\carr_{R+\rb}\backslash \carr_{R-\rb}} |\nab u_{\rrc}|^2  .$$
Choosing $\delta = (\chi(\beta) \rb)^{1/3}$ and inserting into~\eqref{tryu}, we find ~\eqref{loclawpoints0}.

In the second way, we simply bound $\int_{B_{2R}}|\nab u_{\rrc}|^2$ using~\eqref{fini}.
Inserting into~\eqref{disc30}--\eqref{disc3} directly yields~\eqref{loclawpoints00}.
\end{proof}

The second corollary is a control on linear statistics for Lipschitz functions. It follows from the combination of~\eqref{fini} and
\eqref{fluctuationsbu} applied in $\R^\d$.
In Chapter~\ref{chapclt}, we will see how to obtain a better control by the transport method, but under a stronger regularity assumption  on the test function.

\index{fluctuations}
  \begin{coro}[Linear statistics control]
Under the same assumptions as Theorem \ref{th3}, if $\varphi$ is a  $1$-Lipschitz function  supported in $\carr_R(x)$, then 
\begin{equation}\label{loclawphi}
\left|\log \Esp_{\Q_\beta(\mu,U)}\( \exp \frac{\beta }{C R^\d}\(  \int_{\R^\d} \varphi( \sum_{i=1}^N \delta_{x_i}-\mu) \)^2      \)\right|\le
  C \beta \chi(\beta) R^{\d},\ee
for some  $C>0$ depending only on $\d,m$ and $\|\mu\|_{L^\infty}$.
\end{coro}

The final corollary is a control of minimal distances, direct consequence of~\eqref{15} and ~\eqref{locallawint0} applied with $R=\rb$.
\index{separation}
\begin{coro}[Minimal distance control]\label{coromindist}Under the same assumptions as Theorem \ref{th3},
for any point $x_i$ of the (blown-up) configuration at distance $\ge d_0$ from $\partial \Lambda \cap U$,  denoting
$$\rr_i=\min\( \min_{j\neq i} |x_i-x_j|, \frac14\) $$ we have 
\be
\label{loclawdistmin0}
\left| \log \Esp_{\Q_\beta(\mu,U)} \( \exp \left( \frac{\beta }{2} \g(\rr_i  )  \right) \) \right|
\leq   C \beta\chi(\beta)\rb^\d.
\ee
\end{coro}

The fact that ~\eqref{loclawpoints0} gives a bound on all the moments of the number of points in a compact set centered at $x$ satisfying~\eqref{conddist0}    yields that if $x$ is far enough from $\pa \Lambda$, the law of the point configuration $\{x_1-x, \dots, x_N-x\}_N$ converges as $N \to \infty$, up to  extraction of a subsequence, to a limiting point process with simple points and finite correlation functions of all order. 
As mentioned in Section \ref{sec:isotropic}, 
the overcrowding estimates of Theorem \ref{theric} from \cite{thoma} give an easier proof of subsequential convergence to a limiting point process, 
which in additions works up to the boundary. \index{limit point process}


\subsection{Almost additivity of the free energy} \index{almost additivity}

As explained in Section \ref{sec8.1}, the reasoning of the bootstrap proof of the local laws also yields the almost additivity of the free energy up to surface errors. 
 
     \begin{prop}[Almost additivity of the free energy] \label{proaddi} Assume $\d\ge 1$ and $\s=\d-2$. Let $U$ be an open subset of $\R^\d$ with bounded and piecewise $C^1$ boundary and $\mu$ be a bounded nonnegative density such that $\mu(U)=N$ is an integer.  Assume  that  $\mu \ge m>0 $ in a set $\Lambda$.  
If $\s\le 0$ and $U$ is unbounded, assume in addition that \eqref{assumplbs}, \eqref{assgmm} and \eqref{assumpbeta} hold. 
      Assume $\hat U$ is a subset of $ \Lambda$ at distance $\ge d_0$ from $\pa \Lambda$ with $d_0$ as in~\eqref{defd00}, and is a disjoint union of $p$ hyperrectangles $Q_i$ belonging to $\mathcal{Q}_{R}$, with $R \ge \rb$  satisfying 
\be 
\label{rrb} 
R \ge \rb+ \( \frac{1}{\beta\chi(\beta)} \log \frac{R^{\d-1}}{\rb^{\d-1}}\)^{\frac1\d}.
\ee
Then there exists $C>0$, depending only on $\d, m$ and $\|\mu\|_{L^\infty}$, such that 
\begin{align}
\label{subad3}
& \left| \log \K_\beta (\mu, \R^\d) 
- \left( \log \K_\beta (\mu,\R^\d\backslash \hat U)+ \sum_{i=1}^p \log \K_\beta (\mu, Q_i) \right) \right|
\\ & \qquad \notag
\leq C p  \(       \beta R^{\d-1}     \rb   \chi(\beta) +\beta^{1-\frac1\d} \chi(\beta)^{1-\frac1\d} \(\log  \frac{R}{\rb}\)^{\frac1\d}   R^{\d-1}  \).
\end{align}
     If $U$ is a subset of $\Lambda$ equal to a disjoint union of $p$  hyperrectangles $Q_i$ belonging to $\mathcal{Q}_{R}$, with $R \ge \rb$ satisfying \eqref{rrb}, $N_i= \mu(Q_i)$, then we have,      \be
     \label{subad4} 
     \left| \log  \K_\beta(\mu,U) -
     \sum_{i=1}^p \log \K_\beta (\mu,Q_i) \right| \leq C p \Bigg(    \beta R^{\d-1}      \chi(\beta) \rb  +
            \beta^{1-\frac1\d} \chi(\beta)^{1-\frac1\d} \(\log  \frac{R}{\rb}\)^{\frac1\d} R^{\d-1}  \Bigg), 
     \ee with $C$ as above. 
      \end{prop}
   \begin{proof}
We only need to prove upper bounds for $\log \K_\beta(\R^\d)$ and $\log \K_\beta(U)$, since the matching lower bounds are direct consequences of~\eqref{superad2}, Stirling's formula and the control~\eqref{lrob} below.
  
\smallskip

We recall that by Theorem \ref{th3} the local laws hold down to scale $\rb$ in $U=\R^\d$. In particular, for any cube~$\carr$ in~$\hat U$ of size~$r \ge \rb$,  we have 
     \be \label{lolocd} \log \Esp_{\Q_\beta(\mu, \R^\d)} \( \exp \( \frac{1}{2C} \beta \int_{\carr}| \nab u_{\rrc}|^2 \) \) 
\le C r^\d \beta \chi(\beta).\ee 
      Let $Q_1$ be the first rectangle in the list, and let us denote by $n$ the number of points a configuration has in $Q_1$ and by $\mn = \mu(Q_1)$. Let us also define 
      $$\hat Q_1:=\{x\in Q_1, \dist (x, \partial Q_1) \le r \}$$ and 
     $$\mathcal B:= \left\{X_N\in (\R^\d)^N \,:\,   |n- \mn|\le \ep , \quad  \sup_x \int_{\hat Q_1\cap \carr_{r}(x)} |\nab u_{\rrc}|^2 \le M \chi(\beta) r^\d
        \right\}$$ where we let
     $$\ep:=M\( R^{\d-1} \sqrt{\chi(\beta) \rb} \)$$ 
     and $M>0$ is to be selected below.  The first condition~$|n- \mn|\le \ep$ in the definition of~$\mathcal{B}$ has large probability in view of~\eqref{loclawpoints0}.
    For the second condition, by a covering argument we have $O( \frac{R^{\d-1}}{r^{\d-1}})$ conditions to satisfy and each of them has probability
     of the complement bounded by $\exp\(- \frac{M }{C}\beta \chi(\beta)r^\d\)$ if $M$ is large enough, in view of
              ~\eqref{lolocd}.
               Using a union bound  we thus  have 
               $$\Q_\beta(\mu) [\mathcal B^c]\le  C \frac{R^{\d-1}}{r^{\d-1}}  \exp\(- \frac{M}{C} \beta \chi(\beta)r^\d\)$$
               and this is $\le \hal$ if 
               $$  C\frac{R^{\d-1}}{r^{\d-1}}  \exp\(- \frac{M}{C} \beta \chi(\beta)r^\d\)\le \hal$$
               so we choose 
               \be \label{defir}
               r= \rb+ \( \frac{1}{\beta\chi(\beta)} \log \frac{R^{\d-1}}{\rb^{\d-1}}\)^{\frac1\d}\ee
               which satisfies the condition if $M$ is large enough.
               In view of the definitions \eqref{defKN2} and \eqref{defQbu}, it follows that 
     $$N^{-N} \int_{\mathcal B^c} \exp\left( -\beta \F( \cdot)  \right)\, d\mu^{\otimes N} 
     =  \K_\beta(\mu) \Q_\beta(\mu)[ \mathcal B^c] \le 
      \hal \K_\beta (\mu). $$
          We thus have
\begin{multline*}
\frac{N^N }{2} \K_\beta(\mu)
\leq 
\int_{\mathcal B}\exp\(-\beta \F(\cdot,\mu)\)d\mu^{\otimes N}
\\ 
\leq 
\sum_{n=\mn-\ep  } ^{\mn+\ep} \frac{N!}{n!(N-n)!}     
\int_{\mathcal B} \exp\(-\beta \G_{\R^\d}^{\mathrm{inn}}(\cdot, Q_1) \)d\mu^{\otimes n}   \int_{\mathcal B} \exp\(-\beta \G_{\R^\d}^{\mathrm{out}}(\cdot, \R^\d\backslash Q_1) \)d\mu^{\otimes (N-n)},
\end{multline*}
where for the second line we subdivided the event over the possible values of $n$ and  applied Lemma~\ref{lemrestri}.

We now apply the results of Corollary \ref{coro43} with $L=r$ to $Q_1$ and $\R^\d \backslash Q_1$, combined with Remark \ref{remerr}. For that we check that~\eqref{condsurL} is satisfied since $ r \ge \rb$, and obtain
\begin{align*}
\K_\beta(\mu)  
& \leq  
2 \K_\beta (\mu, Q_1)\K_\beta(\mu,\R^\d \backslash Q_1) \sum_{n=\mn-\ep}^{\mn+\ep} \frac{N! N^{-N}}{\mn!(N-\mn)!} \mn^\mn  (N-\mn)^{N-\mn}  
\\ & \qquad 
\times\exp\( C \beta \(R^{\d-1} r  \chi(\beta)M+\ep  \)-\log \min(1 ,\beta)   +  \frac{M^2 \chi(\beta)^2 R^{\d-1}}{  r} \).
\end{align*}
Next, using Stirling's formula we have 
      $$ \frac{N! N^{-N} \mn^{\mn} (N-\mn)^{N-\mn}  }{\mn!(N-\mn)!   } \le C  \sqrt{ \frac{ N}{ 2\pi \mn  (N-\mn)}}\le C$$
and we deduce
\begin{multline*}   \K_\beta(\mu) 
\le \log   \K_\beta(\mu,Q_1)+\log \K_\beta(\mu,\R^\d\backslash Q_1)\\+C+ \log \ep
+  C\beta \(M R^{\d-1} r  \chi(\beta) +\ep\)- \log \min(1,\beta)
+  \frac{M^2 \chi(\beta)^2 R^{\d-1}}{  r} .
\end{multline*}
Since \be\label{eqrf}
r \ge \rb \ge \max(1, \chi(\beta)^{\hal}\beta^{-\hal} ) \ge 1\ee we have $\frac{\chi(\beta)}{r}\le \beta r$ so   we may absorb the last term into $\beta M R^{\d-1}r \chi(\beta)$. Also, 
 since $ r \ge \rb \ge 1$ and $\chi(\beta)\ge 1$,  by definition of $\ep $ we may absorb  $ \ep$ into $M R^{\d-1} r \chi(\beta)   $.
       Since $R \ge \rb \ge \sqrt{\chi(\beta)}$, we  have $\ep=M R^{\d-1} \sqrt{\chi(\beta) \rb}\le C R^\d$, so
inserting the definition of $r$, 
we have obtained
\begin{multline*}
\log  \K_\beta(\mu)  
\le  \log   \K_\beta(\mu,Q_1)+\log \K_\beta(\mu,U\backslash Q_1)  \\+ C\(  \log  R   -\log \min(1, \beta)    +  \beta R^{\d-1}     \rb \chi(\beta) + 
            \beta^{1-\frac1\d}\chi(\beta)^{1-\frac1\d} R^{\d-1} \(\log  \frac{R}{\rb}\)^{\frac1\d}  \) .
\end{multline*}
Finally, 
 since $R \ge \rb \ge C\chi(\beta)^{\hal} \beta^{-\hal} $ we have that, for every $R \ge \rb$,
\be \label{lrob}
\log R \le  C \beta \chi(\beta)\rb R^{\d-1},
\ee  and in view of \eqref{defchibeta} we also have $-\log \min (1, \beta) \le C \beta R^{\d-1}\rb \chi(\beta)$, which allows us to absorb the $\log R$ and $\log \beta$ terms into the others. 

\smallskip

We may now iterate this by bounding $\log   \K_\beta(\mu,\R^\d \backslash \cup_{i=1}^j Q_i)   $ in the same way thanks to the local laws up to the boundary 
 of Theorem   \ref{th1}.  This yields~\eqref{subad3}. 
 
 The proof of~\eqref{subad4} is analogous, using that the local laws hold up to the boundary for $\Q_\beta(\mu,U)$.
\end{proof}

 Fixing $\mu $ to be the constant $1$ and restricting to $R$ such that $R^\d$ is an integer, we can then assert that 
 $-\frac{\log\K_\beta(1,\carr_R)}{\beta R^\d}$ has a finite limit as $R \to \infty$, which we denote $\mf(\beta)$. Moreover, the convergence is at speed $R^{-1}$.  This is the analogue with temperature of Theorem \ref{th4}(1). 
 \index{free energy expansion}
 \begin{theo}[Free energy expansion, Neumann jellium case]\label{th1}
 Assume $\d \ge 1$ and $\s=\d-2$.
 There exists a function $\mf: (0, \infty) \to \R$  and a constant $C>0$ depending only on $\d$ such that 
  \be\label{bornesurf}
 -C\le \mf(\beta)\le C \chi(\beta)\ee 
 \be\label{bornesurfp} 
 \text{$\mf$ is locally Lipschitz in $(0,\infty)$ with} \  |\mf'(\beta)|\le \frac{C\chi(\beta)}{\beta}
 ,\ee 
 and such that 
 if $R^\d$ is an integer we have
  \be \label{1.26} \left| - \frac{\log \K_\beta(1,\carr_R)}{\beta R^\d}-\mf(\beta)\right|\le C  
 \( \chi(\beta)\frac{\rb }{R}   + \frac{\beta^{-\frac1\d} \chi(\beta)^{1-\frac1\d} }{R}\log^{\frac1\d} \frac{R}{\rb}  \)
 \ee
 where $\rb$ is as in \eqref{defrhobeta}.
  \end{theo}
 
 Note that in dimension $1$ an explicit formula for $\mf(\beta)$ in terms of the first eigenvalue of a Fr\"obenius operator was provided in \cite{kunz}. In particular, that formula implies that $\mf(\beta)$ is analytic in $\beta$ and thus the one-dimensional jellium has no phase transitions. 
 We will use this fact for the proof of the CLT in Chapter \ref{chap:clt2}.
 
 By scaling, if instead $\mu=m$ a constant, we obtain that letting $Q'=m^{\frac{1}{\d}}Q$ is a hyperrectangle,
$$\F(\XN, m,  Q)= \begin{cases}
m^{\frac\s\d}\F(m^{\frac1\d}\XN, 1, Q')  & \text{if} \ \s\neq 0, \\
\F(m^{\frac1\d}\XN, 1, Q') - \frac{m|Q|}{2\d} \log m & \text{if} \ \s=0.\end{cases}$$
Thus,  we have that 
$$\K_\beta( m,Q)= m^{-m|Q|} \K_{\beta m^{\s/\d} }( 1,Q') \exp\(\frac{\beta}{2\d}|Q|m\log m \indic_{\s=0}\),$$
and we deduce from \eqref{1.26} that 
  \be \label{formem}\lim_{R\to \infty} -\frac{\log \K_\beta(m,\carr_R)}{ \beta R^\d} = m^{1+\frac\s\d} \mf(\beta m^{\frac\s\d}) + \( \frac1\beta- \frac1{2\d}\indic_{\s=0}\) m \log m.\ee
 
 This then allows to deduce the free energy for a general $\mu$ by using the almost additivity after splitting the support of $\mu$ into hypercubes small enough that $\mu$ is almost constant in each.


\begin{proof}[Proof of Theorem \ref{th1}]
Let us first start by treating the case of a cube $\carr_R$ with  $R^\d$ integer.
In view of~\eqref{superad2} and Stirling's formula, we have 
$$\frac{1}{\beta} \log \K_\beta(1,\carr_{2R})\ge  O\(\frac{\log N}{\beta}\) + \frac{2^\d}{\beta} \log \K_\beta(1,\carr_R).$$
Thus, denoting $\phi(R)= \frac{\log\K_\beta(1,\carr_R)}{\beta R^\d}$, this means that 
$$\phi(2R)  \ge \phi(R)+ O\(\frac {\log R}{\beta R^\d}\) $$
and summing these relations we have
$$\phi(\infty) \ge \phi(R)+ O\( \sum_{k=1}^\infty\frac{\log R }{\beta 2^k R^\d}\), $$
that is, 
\be \label{ubb}\phi(R) \le \phi(\infty)+ O\(\frac{\log R}{\beta R^\d}\).\ee
On the other hand, 
in view of~\eqref{subad4}, 
we have
\be\label{subad23}
\K_\beta(1,\carr_{2R})  \le    2^\d   \log \K_\beta(1,\carr_R) + C\beta R^\d \(
 \frac{\chi(\beta) }{R} \( \rb + \beta^{-\frac1\d} \chi(\beta)^{-\frac1\d}\log^{\frac1\d} \frac{R}{\rb}\) \), \ee
that is,
 \begin{equation*}
  \phi(2R) \le  \phi(R) + C \(  \frac{\chi(\beta) }{R} \( \rb + \beta^{-\frac1\d} \chi(\beta)^{-\frac1\d}\log^{\frac1\d} \frac{R}{\rb} \)\)  .
\end{equation*} Summing these relations, we conclude just as above that 
  \be\label{lbb}\phi(\infty) \le \phi(R) + O\(  
 \frac{\chi(\beta) }{R} \( \rb + \beta^{-\frac1\d} \chi(\beta)^{-\frac1\d}\log^{\frac1\d} \frac{R}{\rb}\)   \).\ee 
Denoting by 
$-\mf(\beta)$ the value $\phi(\infty)$ and recalling~\eqref{lrob}, we have the desired bound   by 
 combining~\eqref{ubb} and \eqref{lbb}.

 \smallskip
 
 In view of~\eqref{majok} and~\eqref{minologk2d} applied with $\mu=1$ and $U=\carr_R$, we also have 
 $- C \chi(\beta) \le  \phi(R) \le C $ with $C$ independent of $\beta$, 
 which implies that $  -C \le \mf(\beta)\le C\chi(\beta)$.

\smallskip
 
To check that $\mf$ is locally Lipschitz, let us observe that 
\begin{align*}
\log \frac{\K_{\beta+\delta}(1,\carr_R)}{\K_\beta(1,\carr_R)}
&
= \log\Esp_{\Q_\beta(1,\carr_R)} \( \exp\left( -\delta \F(\cdot,1,\carr_R) \right) \)
\\ & 
\leq
\frac{2 |\delta|}{\beta} \log \Esp_{  \Q_\beta(1,\carr_R)} \( \exp\left( \hal \beta \F(\cdot,1, \carr_R) \right) \) 
\\ &
\le C |\delta |\chi(\beta) R^\d,
\end{align*}
using H\"older's inequality and~\eqref{locallawint0}. 
Dividing by $\beta R^\d$ and sending $R\to\infty$ yields~\eqref{bornesurfp}.  
\end{proof}

\chapter{The transport method and free energy expansions}
\label{chapclt}

\index{fluctuations}
\index{transport method}

In this chapter, we return to the normal scale, we continue to specialize in the Coulomb case, and wish to analyze fluctuations of linear statistics i.e.~quantities of the form 
\be \label{defFluct} \Fluct_\mu (\xi):= \sum_{i=1}^N \xi(x_i) - N \int_{\R^\d} \xi d\mu,\ee for $\xi$ a regular enough function, and 
where $\mu$ is the reference measure, typically the  equilibrium or thermal equilibrium measure, satisfying \eqref{condmupourFN}.
 We recall that a first, nonoptimal, bound on fluctuations was given in \eqref{ridicboundfluct}. We will now see how to obtain a finer bound.

The starting point is to reexpress the Laplace transform (or moment generating function) of $\Fluct_\mu(\xi)$.  This approach was pioneered by Johansson \cite{joha}.  Indeed, to show that a random variable is often bounded, it suffices to show that its Laplace transform is. To show that a random variable converges to a Gaussian, it suffices to show that its Laplace transform converges to that of a Gaussian.

By definition of the Gibbs measure \eqref{gibbs}, letting $V_t:= V+t\xi$, we have
\begin{multline} \label{ratioz}
\Esp_{\PNbeta} \( e^{-\beta t N^{1-\frac{\s}{\d}}   \sum_{i=1}^N \xi(x_i)  }    \)\\=
\frac{1}{\ZNbeta(V)} \int \exp\(- \beta N^{-\frac\s\d} \Big( \hal \sum_{i\neq j} \g(x_i-x_j)+ N \sum_{i=1}^N (V+t \xi)(x_i) \Big)\)
=    \frac{\ZNbeta(V_t)}{\ZNbeta(V)},\end{multline}
thus, understanding fluctuations via the Laplace transform leads to understanding ratios of partition functions.
To do so, we may first use the splitting with respect to the equilibrium measure or the thermal equilibrium measure, as in 
 \eqref{rewritegibbs33} or \eqref{splitzk}, and we are thus led to comparing reduced  partition  
 functions $\K_{N,\beta}$.   This brings us to the question of estimating the change in reduced partition function, or in free energy, under a perturbation of the external potential $V$, or equivalently under a perturbation of the reference measure.

 The evaluation of the ratio of partition functions will be done in two different ways:\\
 (i) the first way is to  view the perturbed measure as the push-forward of the reference measure  by a transport map. For that we introduce a transport method, which allows to evaluate the change in partition function when making a  perturbation of the equilibrium measure.
 \\
 (ii) the second way is to estimate the partition function for general equilibrium measures by using the almost additivity of Chapter \ref{chaploiloc}, partitioning the domain into small cubes in which the equilibrium measure is almost constant, and using (i) to estimate the error made in each cube.\\
 Comparing the two ways of estimating the partition function then allows to deduce an improved estimate.

 In this chapter, we start with introducing the transport method, which allows to evaluate the change in partition function and other quantities under the variation induced by a transport, which we call a ``transport calculus". The quantities from Chapter \ref{chap:commutator} will naturally appear in the form 
 $$\frac{d}{dt}\log \K_{N,\beta}(\mu_t)= -\beta N^{-\frac\s\d}\Esp_{\Q_{N,\beta}(\mu_t)}\(\Ani_1(\XN, \mu_t,\psi_t)\)$$
 where $\mu_t$ is the transport of $\mu_0$ along the flow of $\psi_t$.
The term $\Ani_1$, which is defined in \eqref{16}  involves a singular kernel, except in  very special cases in $\d=1$ such as  log gases that we will discuss below,  and 
 the commutator estimate of Theorem \ref{thm:FI} will play a major role in estimating it. This part is not restricted to the Coulomb case, but works in the Riesz case as well. 
 
 We will then show how to deduce free energy expansions for general (nonhomogeneous) equilibrium measures, taking advantage of the almost additivity of the energy. This is where we restrict to the Coulomb case.
 
In the next chapter, we will leverage on the two expansions of the free energy
to study fluctuations and obtain a central limit theorem (CLT) in one and two dimensions.

The transport method to obtain the CLT was pioneered in \cite{ls2}. It  has also been used in \cite{bls} for the one-dimensional log case, \cite{lebleimrn} for the sine-$\beta$ process, in \cite{leblezeitouni} to 
obtain a  local CLT for the two-dimensional Coulomb gas, and  in \cite{lambertleblez} to  analyze  the maximum of the log-gas potential.
Finding the appropriate transport map is equivalent to the inversion of the ``master operator" in work on one-dimensional log gases such as \cite{borotguionnet,BorGui2}. 
The transport method and the terms $\Ani_1$ that appear via it are also a substitute for  the Dyson-Schwinger equations in these contexts. 

We present here the method with an  improved version over the more recent presentation of \cite{s2}, including adding the treatment of the Coulomb $\d=1$ case.

 \section{Transport calculus}

We recall the definition of  general reduced Gibbs measures  \eqref{defQ}.


We wish to understand, with the help of the quantities of  \eqref{16} in Chapter~\ref{chap:commutator}, the variations of energy, free energy and other quantities along a transport, which we call a transport calculus.
For that we will exploit the Eulerian/Lagrangian correspondence.
\begin{defi}[Push-forward of a measure] If $X$ and $Y$ are measurable spaces, and $\mu$ a measure on $X$, $\Phi$  a measurable map from $X$ to $Y$, the push forward of $\mu$ by $\Phi$ is defined as the measure on $Y$ such that  $\Phi\#\mu(A)= \mu (\Phi^{-1}(A))$ for any set measurable set $A$ in $Y$.
In particular, if $f$ is a measurable function on $Y$, we have 
\be \int_Y f(y)\, d(\Phi\#\mu)(y) = \int_X  f(\Phi(x)) d\mu(x).\ee\end{defi}

Let $\psi_t: \R^\d \to \R^\d$ be a Lipschitz vector field depending continuously on a ``time" parameter $t\in [0,1]$. Let us define the flow $\Phi_t: \R^\d \to \R^\d$ to be the solution to 
\be \label{defflow}
\left\{ \begin{aligned}
&\frac{d \Phi_t}{dt} (x)= \psi_t (\Phi_t(x))\\
& \Phi_0(x)=x.\end{aligned}\right. \ee
This flow is well-defined for $t\in [0,1]$ by standard ODE theory. Moreover, it is standard and easy to check that if $\mu$ is a probability density then the push-forward 
\be \label{defmutflow}\mu_t := \Phi_t \# \mu\ee solves 
\be\label{continuityeq} \partial_t \mu_t + \div (\psi_t \mu_t)= 0 .\ee
Indeed, let $\varphi$ be a test-function, by definition of the push-forward and by \eqref{defflow}, we have 
$$\int \varphi \partial_t \mu_t = \partial_t \int \varphi d\mu_t = \partial_t \int \varphi (\Phi_t(x)) d\mu(x)= \int \nab \varphi (\Phi_t(x)) \cdot \psi_t(\Phi_t(x)) d\mu(x)= \int \nab \varphi  \cdot \psi_t d\mu_t.$$
This is the Eulerian formulation, as opposed to the Lagrangian formulation \eqref{defflow} which follows ``particles trajectories.''

The next proposition provides the main results of differentiation of energy and free energy along a transport in terms of the quantities of Chapter \ref{chap:commutator}. We recall the notation $\Ani_n$ from \eqref{15comm}, in particular that 
\begin{multline*} \frac{d^n}{dt^n}\Big|_{t=0} \F_N( (\id+ tv)^{\oplus N} (\ux_N), (\id + tv)_\# \mu)= \Ani_n(\XN, \mu,v )
\\
= \frac12\int_{(\R^d)^2\setminus \triangle} 
\nabla^{\otimes n} \g(x-y):  (v(x)-v(y))^{\otimes n}  
d\Big(\sum_{i=1}^N\delta_{x_i} - N \mu\Big)^{\otimes 2}(x,y),
\end{multline*}
 and $\K_{N,\beta}$ and $\Q_{N,\beta}$ from  \eqref{defKN}, \eqref{defQ}.
 We will also make use of the H\"older semi-norms as in \eqref{defholder} $$  |\varphi|_{C^\alpha} = \sup_{x,y} \frac{|\varphi(x)- \varphi(y)|}{|x-y|^\alpha} .$$

\begin{prop}[Transport calculus] \label{propcalctrans1} Let $\mu$ be a probability density on $\R^\d$ satisfying \eqref{condmupourFN}. 
Let  $\psi_t$, $t \in [0,1]$,  be a Lipschitz vector field, and $\Phi_t$ solve \eqref{defflow}, and let  $\mu_t$ be as in \eqref{defmutflow}. Let us abuse notation by also denoting $\Phi_t(\XN)= (\Phi_t(x_1), \dots, \Phi_t(x_N))$ for any $\XN\in (\R^\d)^N$.
For any $t\in [0,1]$, we have
\be \label{derivft}\frac{d}{dt} \F_N(\Phi_t(\XN),\Phi_t\#\mu)= \Ani_1(\Phi_t(\XN ), \mu_t, \psi_t),\ee   
\be \label{derivani} 
\frac{d}{dt} \Ani_1(\Phi_t(\XN), \mu_t, \psi_t)= \Ani_2(\Phi_t(\XN), \mu_t, \psi_t) ,\ee
or more generally, for any $n \ge 0$, 
\be \label{derivFn}\frac{d^n}{dt^n} \F_N(\Phi_t(\XN),\Phi_t\#\mu)= \Ani_n (\Phi_t(\XN), \mu_t, \psi_t)).\ee
Moreover,
\be \label{derivlogkt}\frac{d}{dt}\log \K_{N,\beta}(\mu_t)= -\beta N^{-\frac\s\d}\Esp_{\Q_{N,\beta}(\mu_t)}  \( \Ani_1(\XN , \mu_t, \psi_t)\).\ee
Letting $G(\XN, t)$ be a general function of $\XN $ and $t$, we have
\begin{align}\nonumber \frac{d}{dt} \Esp_{\Q_{N,\beta}(\mu_t) } \( G(\XN, t)\) =& -\beta N^{\frac\s\d} \Cov_{\Q_{N,\beta}(\mu_t) }  \left[ G(\XN,t), \Ani_1(\XN, \mu_t, \psi_t)\right]\\ \label{derivgt} & + \Esp_{\Q_{N,\beta}(\mu_t) }\( \nab_{X_N} G(\XN,t)\cdot (\psi_t)^{\otimes N} + \partial_{t} G(\XN,t) \),
\end{align}
where $\Cov$ denotes the covariance.
\end{prop}
As an example of application, combining \eqref{derivgt} and \eqref{derivlogkt}, we may find that 
\begin{multline}\label{deriv2kt}
\frac{d^2}{dt^2} \log \K_{N,\beta}(\mu_t)\\
= (\beta N^{-\frac\s\d} )^2 \Var_{\Q_{N,\beta}(\mu_t)}  
\( \Ani_1(\XN, \mu_t, \psi_t)\) -\beta N^{-\frac\s\d} \Esp_{\Q_{N,\beta}(\mu_t)} \( 
\Ani_2(\XN, \mu_t, \psi_t)\),\end{multline} where $\Var$ denotes the variance.

Here, the fact that we have reduced  (thanks to the use of the thermal equilibrium measure) to partition functions expressed relative to $\mu^{\otimes N}$ and not the Lebesgue measure, is very convenient for the proof, see in particular the crucial computation \eqref{formul} below. 

\begin{proof}
By semi-group property of the flow $\Phi_t$, we have 
\be \label{DPhit}\Phi_{t+h} (x)=  (\id+ h \psi_t) ( \Phi_t (x))+ o(h),\quad \text{as} \ h \to 0,\ee
and also 
\be \label{movmut}\mu_{t+h} = (\id + h \psi_t) \# \mu_t +o(h),\quad \text{as} \ h \to 0.\ee
Thus by definition  \eqref{15comm}, we find that \eqref{derivft} holds. The proof of \eqref{derivani} and \eqref{derivFn} is analogous.

Let $\phi_h$ be a diffeomorphism from $\R^\d$ to $\R^\d$ (extended into one from $(\R^\d)^N$ to itself). By \eqref{defKN} and the change of variables $y=\phi_h(x)$ (or by definition of the push-forward)  we have 
\begin{align}\notag
\frac{\K_{N,\beta}(\phi_h\#\mu)}{\K_{N,\beta}(\mu)} &=
\frac{1}{\K_{N,\beta}(\mu)} \int_{(\R^\d)^N} \exp\(-\beta N^{-\frac\s\d}\F_N(\XN, \phi_h\#\mu) \)d(\phi_h\#\mu)^{\otimes N}(X_N)
\\  \notag & = \frac{1}{\K_{N,\beta}(\mu)}\int_{(\R^\d)^N} \exp\(-\beta N^{-\frac\s\d}\F_N( \phi_h(\XN), \phi_h\#\mu)\) d\mu^{\otimes N}(X_N)\\
& \label{formul} = \Esp_{\Q_{N,\beta}(\mu)} \(  \exp\(-\beta N^{-\frac\s\d}(\F_N(\phi_h(\XN),\phi_h\#\mu)-\F_N(\XN, \mu)) \)\).\end{align}
Applying to $\mu=\mu_t$ and $\phi_h= (\id + h \psi_t)$, letting $h \to 0$, since \eqref{movmut} holds we obtain  in view of \eqref{derivft} the relation \eqref{derivlogkt}.

Let us now turn to \eqref{derivgt}. We have 
\begin{align}\notag
\frac{d}{dt} \Esp_{\Q_{N,\beta}(\mu_t) } \( G(\XN, t)\)
 = &- \frac{1}{\K_{N,\beta}(\mu_t)^2}\frac{d}{dt} \K_{N,\beta}(\mu_t) \int_{(\R^\d)^N} e^{-\beta N^{-\frac\s\d} \F_N(\XN, \mu_t)} G(\XN,t) d\mu_t^{\otimes N} (\XN)
\\ \label{aliQN}
& +  \frac{1}{\K_{N,\beta}(\mu_t)} \frac{d}{dt}\int_{(\R^\d)^N} e^{-\beta N^{-\frac\s\d} \F_N(\XN, \mu_t)} G(\XN,t) d\mu_t^{\otimes N}(\XN). \end{align}
The first term in the right-hand side is equal to 
$$\(- \frac{d}{dt} \log \K_{N,\beta}(\mu_t)\) \Esp_{\Q_{N,\beta}(\mu_t)} (G(\XN, t))$$
for which we insert \eqref{derivlogkt} to find it is equal to 
$$ \beta N^{-\frac\s\d}  \Esp_{\Q_{N,\beta}(\mu_t)} \( \Ani_1(\XN, \mu_t, \psi_t)\)  \Esp_{\Q_{N,\beta}(\mu_t)} (G(\XN, t)).$$
 By a change of variables as above, the second term in \eqref{aliQN} is equal to 
$$ \frac{1}{\K_{N,\beta}(\mu_t)}  \frac{d}{dh}_{|_{h=0} }\int_{(\R^\d)^N}e^{-\beta N^{-\frac\s\d} \F_N (\phi_{h}(\XN),  \phi_{h} \#\mu_t)    }  G( \phi_{h}(\XN), t+h)  d\mu_t^{\otimes N}(\XN)
$$and in view of \eqref{derivft}, this is 
$$\Esp_{\Q_{N,\beta}(\mu_t)} \( - \beta N^{-\frac\s\d} \Ani_1(\XN, \mu_t, \psi_t)  G(\XN, t) \) + \Esp_{\Q_{N,\beta}(\mu_t)}\( \nab_{X_N} G(\XN,t) \cdot  (\psi_{t})^{\otimes N} + \partial_t G \) .$$
Combining all these relations, we find \eqref{derivgt}.
\end{proof}

\begin{rem}[The case of $\beta$-ensembles]
In the particular case of one-dimensional log gases $\d=1,\s=0$, we have 
$$\Ani_1(\XN, \mu, \psi)= \iint_{\R\times \R\backslash \triangle} \frac{\psi(x)-\psi(y)}{x-y} d\( \sum_{i=1}^N \delta_{x_i}-N\mu\)(x) d\( \sum_{i=1}^N \delta_{x_i}-N\mu\)(y) 
$$
and  the kernel involved in the integration $ \frac{\psi(x)-\psi(y)}{x-y}$,  is regular if $\psi$ is regular enough, contrarily to all other cases with $\s\ge 0$ for which it is always singular. 
Moreover, one may rewrite  the above as 
\be\label{ani11d}
\Ani_1(\XN, \mu, \psi)= \iint_{\R\times \R} \frac{\psi(x)-\psi(y)}{x-y} d\( \sum_{i=1}^N \delta_{x_i}-N\mu\)(x) d\( \sum_{i=1}^N \delta_{x_i}-N\mu\)(y)-\sum_{i=1}^N \psi'(x_i)\ee and the double integral on the right-hand side can be seen as a bilinear statistics with respect to the regular test function $\frac{\psi(x)-\psi(y)}{x-y}$. 
As we will see  for instance in \eqref{laplace1}, fluctuations  can themselves be reexpressed via ratio of partition functions, and fluctuations for regular functions are (in this case) of order $1$ with respect to $N$. Thus, integrating \eqref{derivlogkt}, after finding the appropriate transport field $\psi_t$ \footnote{which is equivalent to inverting a ``master operator" as alluded to above.} we can deduce that 
\begin{multline}\label{logK11d}\log \K_{N,\beta}(\mu_1)-\log \K_{N,\beta}(\mu_0)= \beta \int_0^1 \Esp_{ \Q_{N,\beta}(\mu_t)} \( \sum_{i=1}^N \psi_t'(x_i)  \) dt+O(1)\\= 
\beta N\int_0^1\int_{\R} \psi_t'(x) d\mu_t(x) dt +O(1),\end{multline}
which already gives the correct leading order to the free energy difference. In addition, since free energy differences encode fluctuations (again, via \eqref{laplace1} or variants), the form of the  leading order term allows to deduce the leading order (of order 1) of the expectation and variance of fluctuations of regular linear statistics. By a Fourier trick which changes double fluctuations into single fluctuation, one can then plug back this estimate for fluctuations into \eqref{ani11d} to obtain the next order in the expansion of $\Ani_1$, i.e.~an expansion to $o(1)$, hence an expansion to $o(1)$ of \eqref{logK11d}, which in turns provides the order $1/N$ term in the expansion of fluctuations. This can be iterated to arbitrary order, providing expansions to all powers of $1/N$ of both free energy and fluctuations, provided $\psi$ (hence in fact $V$) has enough derivatives. This is described in more detail in \cite[Appendix A]{bls} and is a reinterpretation of what is accomplished in the ``topological recursion'' of \cite{borotguionnet}.
\end{rem}

With this result at hand, we can take advantage of the commutator estimate of \index{commutator estimates} Theorem~\ref{thm:FI} which, in view of \eqref{16}, states  that\footnote{We note that $\lambda$ in that theorem and its proof can simply be replaced by $N^{-1/\d}$} 
\be\label{resthFI} |\Ani_1(\XN, \mu, \psi) |\le C\|\nab\psi\|_{L^\infty} \(\F_N(\XN, \mu)+  \( \frac{N\log N }{2\d}\)\indic_{\s=0} +C N^{1+\frac\s\d}\)\ee
with $C$ depending only on $\d, \s$ and $\|\mu\|_{L^\infty}$.
In view of Proposition \ref{propcalctrans1} this allows to bound the variation of the energy (or free energy) along a transport by the energy itself. 
Moreover, since the commutator estimate can be localized, we can combine this with the local laws to obtain localized bounds.

Here is an example of use where we deduce a control on  the variation of energy along a transport via an ODE argument.
\begin{coro}\label{corocontrenergyt}
Assume $\mu$ is a bounded probability density satisfying \eqref{condmupourFN}.  
 Let  $\psi_t$, $t \in [0,1]$ be a Lipschitz vector field, and $\Phi_t$ solve \eqref{defflow}, and $\mu_t$ be as in \eqref{defmutflow}.
Let $C_0$ be chosen large enough in view of \eqref{eq:14} so that \be \label{defXi}  \Xi(t): = \F_N^{\Omega}  (\Phi_t(\XN), \mu_t) + \( \frac{\# I_N}{4} \log N \) \indic_{\d=2} + C_0 \#I_{\Omega} N^{\frac{\s}{\d}} \ge 0.\ee
Assume that  $\Omega$ contains a $2N^{-1/\d}$-neighborhood of the support of $\psi_t$   and that $\Phi_t$  maps $\Omega$ to itself  for every $t\in [0,\tau]$.
Then we have \be \label{controlenergyt}  \forall t \in [0,\tau],  \qquad \Xi(t) \le \exp\( C\int_0^t  |\psi_s|_{C^1} ds\) \Xi(0),\ee where $C$ depends only on $\s,\d$.
\end{coro}
\begin{proof} First we note that $\Phi_t$ coincides with the identity map outside $\Omega$ and that $\mu_t $ defined by \eqref{defmutflow} coincides with $\mu$ outside $\Omega$, for each $t \in [0,\tau]$. 
Examining the definition of $\F_N^{\Omega}$ in \eqref{Glocal} and comparing to \eqref{fnmeta2} (the equality case with $\eta_i=\rr_i$), and using the fact that $\Phi_t=\id$ in $\Omega^c$, we deduce that $$\Xi'(t)= \frac{d}{dt}  \F_N(\Phi_t(\XN), \mu_t). $$
In view of \eqref{derivft} we thus obtain 
$$\Xi'(t)= \Ani_1(\Phi_t(\XN), \mu_t, \psi_t)$$
and 
combining with \eqref{resthFI},  we deduce
 $$\Xi'(t) \le C |\psi_t|_{C^1} \Xi(t),$$ with $C>0$ depending only of $\d$ and $\s$, 
after making $C_0$ larger if necessary.
 Applying Gronwall's lemma to the function $\Xi$, we deduce the result. 
\end{proof}

We finish with stating the adaptation of the transport calculus when working instead with the standard equilibrium measure and the definitions \eqref{deftildeK}, \eqref{deftildeQ}.
 The main difference  is that a Jacobian term appears. 
\begin{lem}[Transport calculus, usual equilibrium measure] \label{propcalctrans2}
Let $\mu$ be a probability density satisfying \eqref{condmupourFN}. Let  $\psi_t$, $t \in [0,1]$ be a Lipschitz vector field, and $\Phi_t$ solve \eqref{defflow}, and $\mu_t$ be as in \eqref{defmutflow}. Let $\zeta_t : \R^\d\to \R$ be space-time Lipschitz. 
We have, for any $t\in [0,1]$,
\begin{multline} \label{derivlogkteqx}\frac{d}{dt}\log \tilde \K_{N,\beta}(\mu_t,\zeta_t)= \Esp_{\Q_{N,\beta}(\mu_t,\zeta_t)}  \Bigg(-\beta N^{-\frac\s\d} \(\Ani_1(\XN , \mu_t, \psi_t) +N \sum_{i=1}^N \partial_t \zeta_t(x_i)+ \nab \zeta_t \cdot \psi_t(x_i) \)\\+  \sum_{i=1}^N  \div \psi_t (x_i)   \Bigg).\end{multline}
\end{lem}
\begin{proof}Let $\phi_h$ be a diffeomorphism from $\R^\d $ to $\R^\d $ (extended into one from $(\R^\d)^N$ to itself) with nonnegative Jacobian. 
By the change of variables $y_i=\phi_h(x_i)$, we have 
\begin{align} \label{formulx} 
&\frac{\tilde\K_{N,\beta}(\phi_h \# \mu_{t},\zeta_{t+h})}{\tilde\K_{N,\beta}(\mu_t,\zeta_t)} \\
\notag &=
\frac{1}{\tilde \K_{N,\beta}(\mu_t,\zeta_t)} \int_{(\R^\d)^N} e^{-\beta N^{-\frac\s\d}\Big(\F_N(\phi_h(\XN), \phi_h\#\mu_t) + N \sum_{i=1}^N \zeta_{t+h}(\phi_h(x_i))  \Big) } \det D\phi_h (\XN) d\XN \\   \notag & = \Esp_{\Q_{N,\beta}(\mu_t,\zeta_t)} \(  e^{-\beta N^{-\frac\s\d}\( \F_N(\phi_h(\XN),\phi_h\#\mu_t)-\F_N(\XN, \mu_t) + N \sum_{i=1}^N \zeta_{t+h}(\phi_h(x_i)) - \zeta_t(x_i)\)+\sum_{i=1}^N \log \det D\phi_h(x_i) } \) .\end{align}
Applying  to $\phi_ h = I + h \psi_t$, letting $h \to 0$ and using 
 \eqref{derivft} we deduce  \eqref{derivlogkteqx}.
\end{proof}

\section{Application: Lipschitzness of the free energy}

Let start by defining a blown-down version of the Neumann partition function 
of \eqref{defK7}. Let $U$ be an open subset of $\R^\d$ with piecewise $C^1$ boundary.  Let $\mu$ be a nonnegative density with $ N\int_U \mu = \mn$, an integer. 
Following \eqref{scalingF} we let $$
\F_N(\XN, \mu, U):= N^{\frac\s\d}\F(\XN', \mu', U')- \(\frac{\mn}{2\d}\log N\)\indic_{\s=0} $$
with $\F$ as in \eqref{minneum} and $(\XN', \mu', U')$ the blown-up quantities $\XN'= N^{1/\d} \XN$, $\mu'(x)= \mu(xN^{-1/\d}) $ and $U'= N^{1/\d} U$.
We then let 
\be \label{defKbd}
\K_{N,\beta}(\mu, U) := \int_{U^N} \exp\( -\beta N^{-\frac\s\d} \F_N(\XN, \mu, U)\) d\mu^{\otimes N} (\XN),
\ee
and note that it coincides with \eqref{defKN} when $U=\R^\d$ so that $\K_{N, \beta}(\mu)= \K_{N, \beta}(\mu, \R^\d)$. We also denote 
\be \label{defd0p}
\bar d_0:=
C N^{-\frac1\d} \max\( \(\frac{N^{\frac1\d} }{ \max(1, \beta^{-\hal} \chi(\beta)^{\frac12})}\)^{-\frac23}   N^{\frac1\d} ,   N^{\frac{1}{ \d+2}}  \) \ee the blown-down of the distance $d_0$ of \eqref{defd00}.

In Theorem \ref{th1}, we proved the convergence of the Neumann partition function for  growing cubes in the case where the background measure $\mu$ is constant. In order to understand the free energy for nonhomogeneous background, we want to proceed by approximation, reducing to the case of a constant one after showing that 
 $\log \K_{N,\beta}(Q, \mu)$   varies little if $\mu$ varies little. 
 
 The method consists in using   an explicit transport $\psi_t$ that transports $\mu_0$ into $\mu_1$ as in \eqref{defmutflow} and using the transport calculus.
 We start by an elementary lemma that allows us to work with linear interpolants within this framework. 
 \begin{lem}\label{1221}Let $\mu_0$ and $\mu_1$ be two probability densities on $\R^\d$ and let  $\mu_s= (1-s) \mu_0 + s\mu_1$ be their linear interpolant. 
 Assume that for  $s \in [0,1]$,  $\psi_s$ is a Lipschitz vector field on $\R^\d$, depending continuously on $s$, and satisfying
 \be \label{divpsimu}
- \div (\psi_s\mu_s)= \mu_1-\mu_0.\ee
 Then defining $\Phi_s$ as in \eqref{defflow}, we have that $\mu_s = \Phi_s \# \mu_0$.
 \end{lem}    
\begin{proof}
Let $\bar \mu_s = \Phi_s \# \mu_0$ as defined in \eqref{defmutflow}. The argument consists in showing that by uniqueness, we must have $\mu_s= \bar \mu_s$.
As seen in \eqref{continuityeq}, $\bar \mu_s $ solves 
$$\partial_s \bar \mu_s+ \div (\psi_s  \bar \mu_s) =0.$$
On the other hand, $\mu_s$ solves 
$$\partial_s \mu_s = \mu_1-\mu_0= - \div (\psi_s \mu_s).$$
So $\mu_s$ and $\bar \mu_s$ solve the same linear continuity equation. Since $\psi_s$ is Lipschitz, this equation enjoys uniqueness (for instance take $u_s=\mu_s-\bar \mu_s$ and check that $\frac{d}{ds}\int u_s (x)\varphi(\Phi_s(x))dx=0$ for all test functions $\varphi$), hence $\bar \mu_s= \mu_s$, as desired.
\end{proof}

We use this   to evaluate the change in  $\log \K_{N,\beta}(\R^\d, \mu)$ when $\mu$ varies only in a hyperrectangle $Q_\ell$. The change is controlled by the energy via the commutator estimate, which is then  combined with the local laws to control the energy when  on mesoscales. For that reason, we restrict to the Coulomb case and place the same assumptions as for the local laws. 
 
\begin{prop}[Variation of free energy -- global case]\label{lemcompdesk}
Assume $\d\ge 1$ and $\s=\d-2$. 
Assume $C\ge \ell \ge \rb N^{-1/\d}$. Let $\mu_0,\mu_1$   be two nonnegative Lipschitz densities bounded above and below by positive constants in a set $\Lambda$. If $\s\le 0$, assume in addition that \eqref{assumplbs}, \eqref{assgmm} and \eqref{assumpbeta} hold for $\mu_0$ and $\mu_1$. 
Let $Q_\ell$ be  a hyperrectangle of sidelengths in $[\ell, 2\ell]$ included in $\Lambda$. If $\ell \ll 1$ as $N \to \infty$, assume also that  $\dist(Q_\ell , \pa \Lambda) \ge\bar d_0$ as in \eqref{defd0p}.

Assume that 
$$\mu_1- \mu_0= -\Delta \xi$$
for some $C^3$ function $\xi$, and that $Q_\ell$ contains a $2N^{-1/\d} $-neighborhood of the support of $\xi$.


Then 
\be\label{compdesk}
|\log \K_{N,\beta}(\mu_1,\R^\d)-\log \K_{N,\beta}(\mu_0,\R^\d)|
\le C  \beta \chi(\beta) N\ell^\d  \(  (|\mu_0|_{C^1}+|\xi|_{C^3})  |\xi|_{C^1}+ |\xi|_{C^2}\)
 . \ee
where $C$ depends only on $\d$ and the upper and lower bounds for $\mu_0$ and $\mu_1$ and the constants in the assumptions.
\end{prop}

\begin{proof}
Consider the linear interpolant $\mu_s= (1-s) \mu_0+s\mu_1,$ and 
$$\psi_s:= \frac{\nab \xi}{\mu_s}.$$
By definition, $\psi_s$ is Lipschitz, and $Q_\ell$ contains a $2N^{-1/\d}$-neighborhood of its support, and  $\psi_s$ solves 
\eqref{divpsimu}.
We can estimate
\be \label{419}|\psi_s|_{C^1}
\le C \(\left\| \frac{1}{\mu_s}\right\|^2_{L^\infty} |\mu_s|_{C^1}    |\xi|_{C^1}+   \left\|\frac{1}{\mu_s}\right\|_{L^\infty}  |\xi|_{C^2}\) \le C \( (|\mu_0|_{C^1}+|\xi|_{C^3})     |\xi|_{C^1}+ |\xi|_{C^2}\)
\ee where $C$ depends on $\d$  and the upper and lower bounds for $\mu_0$ and $\mu_1$.
By combining Lemma~\ref{1221} and \eqref{derivlogkt} in Proposition \ref{propcalctrans1},
we obtain
\be\label{dtl}
\frac{d}{ds} \log \K_{N,\beta}(\R^\d, \mu_s)= 
\Esp_{\Q_{N,\beta}( \mu_s)} \( - \beta  N^{-\frac\s\d}  \Ani_1(X_N,  \mu_s,\psi_s) \).\ee
Inserting  the commutator estimate of Theorem \ref{thm:FI}, we obtain 
\begin{multline*}
\left|\frac{d}{ds} \log \K_{N,\beta}(\R^\d, \mu_s)\right|\\
\le C |\psi_s|_{C^1} \Esp_{\Q_{N,\beta}( \mu_s)}\(  - \beta  N^{-\frac\s\d}\(
\F_N^{Q_\ell} (\XN, \mu_s) +  \( \frac{\#I_{ Q_\ell} \log N }{2\d}\)\indic_{\s=0} +C \#I_{ Q_\ell}N^{\frac\s\d}\)\).\end{multline*}
 Inserting 
\eqref{419}, the rescaling formula as in \eqref{scalingF}  and using the local laws \eqref{locallawint0} which hold in $Q_\ell$ (if $\ell $ is not small they hold automatically by \eqref{expmom0}), 
 we deduce that 
$$
 \left|\frac{d}{ds} \log \K_{N,\beta}(\R^\d, \mu_s)\right|\le C    \beta \chi(\beta) N\ell^\d \(  (|\mu_0|_{C^1}+|\xi|_{C^3})|\xi|_{C^1}+ |\xi|_{C^2}\). $$
  Integrating between $0$ and $1$ gives the result.
 \end{proof}

\begin{prop}[Variation of free energy -- Neumann cube case]\label{lemcompdeskcube}
Assume $ \rb N^{-1/\d}\le \ell \le C$.
Let $\mu_0,\mu_1$ be two Lipschitz  densities bounded  above and below by positive constants in $Q_\ell$, a hyperrectangle of sidelengths in $[\ell, 2\ell]$ with $N\mu_0(Q_\ell)=N\mu_1(Q_\ell)=\mn$ an integer.
Then 
\begin{multline}\label{compdeskcube}
|\log \K_{N,\beta}(Q_\ell, \mu_1)-\log \K_{N,\beta}(Q_\ell, \mu_0)|\\
\le C \beta \chi(\beta) N\ell^\d \(\ell^2  ( |\mu_0|_{C^1}+  |\mu_1-\mu_0|_{C^1} )  |\mu_1-\mu_0|_{C^1} 
+ \ell  |\mu_1-\mu_0|_{C^1}\),\end{multline}
where $C$ depends only on $\d$ and a lower bound for $\mu_0$ and $\mu_1$. \end{prop}
\begin{proof}
Since we are working with the Neumann energy in a cube, we need to find a transport that preserves the cube and solves \eqref{divpsimu}. For that  we let $\xi$ solve 
\be \label{solxi}
\left\{
\begin{array}{ll}
-\Delta \xi= \mu_1-\mu_0& \text{in} \ Q_\ell\\ [2mm]
\frac{\pa \xi}{\pa \nu}=0 & \text{on} \ \pa Q_\ell .\end{array}\right.
\ee
By elliptic regularity and scaling we have  
$$|\xi|_{C^1}\le C \ell^2|\mu_1-\mu_0|_{C^1}, \quad |\xi|_{C^2}\le C \ell|\mu_1-\mu_0|_{C^1}.$$
Consider, for $0\le s \le 1$, the linear interpolant
$ \mu_s= (1-s) \mu_0+ s\mu_1$.
Setting $$\psi_s:= \frac{\nab \xi}{\mu_s},$$ 
we  have $$-\div (\psi_s\mu_s)= \mu_1-\mu_0,$$ i.e.~\eqref{divpsimu} is satisfied, and 
\begin{multline} \label{418}|\psi_s|_{C^1}\le C \(\left\| \frac{1}{\mu_s}\right\|^2_{L^\infty} |\mu_s|_{C^1}    |\xi|_{C^1}+   \left\|\frac{1}{\mu_s}\right\|_{L^\infty}  |\xi|_{C^2}\) \\
\le C \( \ell^2  \(  |\mu_0|_{C^1} + |\mu_1-\mu_0|_{C^1}\)  |\mu_1-\mu_0|_{C^1} 
+ \ell  |\mu_1-\mu_0|_{C^1}\),
\end{multline}
where $C$ depends only on $\d$ and a lower bound for $\mu_0$ and $\mu_1$.
The rest of the proof is identical to the previous one.
 \end{proof}

 \section{Free energy expansions for inhomogeneous density}\label{secfree}
 \index{free energy expansion}
 
 The results of Proposition \ref{proaddi} allowed to evaluate $\log \K_{N,\beta}(\mu)$ by almost addivity, cutting the region into cells in which $\mu$ is almost constant.  
 Since the only available formula \eqref{formem} is for uniform densities (provided by Theorem \ref{th1} where the function $\mf(\beta)$ was introduced),   in each cell we  need to compare $\log \K_{N,\beta}(\mu, Q_\ell)$ with $\log \K_{N,\beta}(\dashint_{Q_\ell} \mu, Q_\ell)$ where $\dashint \mu$ is the average of $\mu$. This can be done  via Proposition  \ref{lemcompdeskcube}. In order to have a good precision, we need the cells of the partition to be small, while in order to have small additivity errors in Proposition \ref{proaddi}, we need them to be large. 
 Optimizing the effect of these two types of error (additivity error and approximation error) over the size of the partitioning cells leads to choosing cells of size $N^{\frac{1}{2\d}}$ in blown up scale, and we get after some direct but tedious computations  (for which we refer to \cite[Section 6]{s2}) the following result.
 
\begin{prop}[Free energy expansion for general density in a hyperrectangle]\label{coro24} Assume $\s=\d-2$.
 Let $R$ be such that $RN^{1/\d}$ satisfies \eqref{rrb}.  Let $Q $ be a hyperrectangle of sidelengths in $(R, 2R)$. Let $\mu$ be a  Lipschitz  density bounded above and below by positive constants in $Q$,  and assume $N\int_{Q} \mu=\mn$ is an integer.  Then,
 \begin{multline}\label{expzcasgb}
\log \K_{N,\beta}( \mu,Q)=  
  - \beta N \int_{Q} \mu^{1+\frac\s\d} \mf(\beta \mu^{\frac{\s}{\d}} ) - \frac{\beta}{2\d}N\(
\int_{Q}\mu \log \mu\) \indic_{\s=0} + \(\frac{\beta}{2\d}\mn \log N \) \indic_{\s=0}\\+ O\(\beta \chi(\beta) NR^\d \mathcal{R}(N, R, \mu)\)
\end{multline}
where we let 
\be\label{defRRR1}\mathcal{R}(N, R, \mu):=
\max\(  x(1+|\log x|), (y^{\frac12} +y) (1+|\log y|^{\frac1\d} ) \)\ee
after setting
$$ x:= \frac{\rb}{R N^{\frac1\d}} ,\quad y:= \frac{ \rb |\mu|_{C^1}  }{N^{\frac{1}{\d}} },$$
and the  $O$ depends  only on $\d$ and the upper and lower bounds for $\mu$.
\end{prop}
What is important here is that we get an explicit  error rate, valid for broad temperature regimes. The quantity $x$ is small by \eqref{rrb}, the estimate is interesting when  $y$ is  small too.  If $\beta$ is fixed then so is $\rb$ and the error rate obtained is $R^\d N^{1-\frac1{2\d}} \log^{\frac1\d} N$.

We deduce the following.
\begin{coro}[Relative expansion, local version] \label{th1b}
Let $\mu_0$ and $\mu_1$ be  two nonnegative Lipschitz densities  bounded above and below by positive constants in a set $\Lambda$, and  coinciding outside $Q_\ell$, a hyperrectangle included in $\Lambda$  of sidelengths in $(\ell, 2\ell)$ with $C\ge \ell\ge \rb N^{-1/\d}$ that satisfies $\dist(Q_\ell, \pa \Lambda) \ge \bar d_0$. Assume $N\int_{Q_\ell} \mu_0=N\int_{Q_\ell}  \mu_1=\mn$ is an integer. We have
\begin{multline}\label{relatexpansion}
\log \K_{N,\beta}( \mu_1, \R^\d)- \log \K_{N,\beta}(\mu_0,\R^\d)
= 
  - \beta N  \int_{Q_\ell} \mu^{1+\frac\s\d} \mf(\beta \mu_1^{\frac{\s}{\d}} ) -\frac{\beta}{2\d}N\(
\int_{Q_\ell}\mu_1 \log \mu_1 \) \indic_{\s=0}
  \\+ \beta N \int_{Q_\ell}  \mu_0^{1+\frac\s\d} \mf(\beta \mu_0^{\frac{\s}{\d}} ) +\frac{\beta}{2\d}N\(
\int_{Q_\ell} \mu_0 \log   \mu_0\)  \indic_{\s=0}
\\+O\( \beta \chi(\beta) N\ell^\d\(\mathcal R(N, \ell, \mu_0)+ \mathcal R (N, \ell,  \mu_1)\)\)
\end{multline}
where $\mathcal R$ is as in Proposition \ref{coro24}, and the $O$ depends only on $\d$ and the upper and lower bounds for $\mu_1$ and $ \mu_0$ in $Q_\ell$.
\end{coro}
\begin{proof}
We may apply \eqref{subad3} (in rescaled form) to both $\mu_0 $ and $ \mu_1$ and subtract the obtained relations  to get that 
\begin{multline*}
\log \K_{N,\beta}(\mu_1, \R^\d)- \log \K_{N,\beta}( \mu_0, \R^\d)
=\log \K_{N,\beta}(\mu_1, Q_\ell)- \log \K_{N,\beta}(\mu_0,Q_\ell)\\
+O  \(    N^{1-\frac1\d}   \beta \ell^{\d-1}     \rb   \chi(\beta) +N^{1-\frac1\d}\beta^{1-\frac1\d} \chi(\beta)^{1-\frac1\d} \(\log  \frac{\ell N^{\frac1\d}}{\rb}\)^{\frac1\d}   \ell^{\d-1}  \).
\end{multline*}
Inserting the result of \eqref{expzcasgb} applied to $\mu_0$ and $ \mu_1$, we deduce 
\begin{multline*}
\log \K_{N,\beta}(\mu_1)- \log \K_{N,\beta}( \mu_0)=  - \beta N \int_{Q} \mu_1^{1+\frac\s\d} \mf(\beta \mu_1^{\frac{\s}{\d}} ) -\frac{\beta}{2\d}\( N
\int_{Q}\mu_1 \log \mu_1 \) \indic_{\s=0}
  \\+ \beta N \int_{Q}  \mu_0^{1+\frac\s\d} \mf(\beta \mu_0^{\frac{\s}{\d}} ) +\frac{\beta}{2\d}N
  \(
\int_{Q} \mu_0 \log   \mu_0\) \indic_{\s=0}
\\+O\( \beta \chi(\beta) N\ell^\d\(\mathcal R(N, \ell, \mu_0)+ \mathcal R (N, \ell,  \mu_1)+     N^{-\frac1\d} \ell^{-1}     \rb +\beta^{-\frac1\d} \chi(\beta)^{-\frac1\d} \(\log  \frac{\ell N^{\frac1\d}}{\rb}\)^{\frac1\d}   \ell^{-1}N^{-\frac1\d} \) \).\end{multline*}
Checking  that $\beta^{-\frac1\d} \chi(\beta)^{-\frac1\d}\le \rb$ by \eqref{defrhobeta} and \eqref{defchibeta}, by definition of $x$ we see that we may absorb the last error terms into $\mathcal R$.\end{proof}

\begin{rem}\label{rem2dci}
A crucial point here, that will be used in the proof of the CLT in Chapter~\ref{chap:clt2} is that for $\s=0$, the terms $-\beta N \int_{Q}\mu_1^{1+\frac\s\d} \mf(\beta \mu_1^{\frac{\s}{\d}}) $ and 
$\beta N \int_{Q}\mu_0^{1+\frac\s\d} \mf(\beta \mu_0^{\frac{\s}{\d}}) $ simply cancel out. 
This is a manifestation of the conformal invariance of the Coulomb operator in two dimensions.
In contrast, for all other dimensions, a density-dependent  effective temperature  $\beta \mu^{\s/\d}$ appears.
\end{rem}

Returning to the splitting formula \eqref{splitting}, we may now obtain a general expansion for the free energy or minimal energy.
For this it suffices to partition the space into regions with quantized mass, use the almost additivity of the free energy of Proposition  \ref{proaddi}, the expansion just obtained,
 and control the contributions of the tails by the cruder bounds of \eqref{bornesfiU}.
 
 The result is given in the Coulomb case, but let us emphasize that the same formula is proven for all Riesz cases $\s \in [\d-2,\d)$, albeit with a less precise error term, see  \cite{lebles}. 
This provides the precise next order term (order $N$ term) in the (free) energy expansion, with an explicit power error rate. 
In the one-dimensional logarithmic case, free energy expansions formulae are known, even up to arbitrary order, for quadratic potential $V$ thanks to Selberg's formula, or for regular enough $V$'s, see in particular \cite{borotguionnet,BorGui2,shch}. Other than this one-dimensional setting, prior results are restricted to  the $\beta=2$  two-dimensional Coulomb case: the most recent results obtain  expansions to all order  in radial situations   including the possibility of a multi-connected droplet \cite{byunkangseo,ameurcharliercronvall2}. \index{determinantal point process}
 Wiegmann and Zabrodin \cite{zabrodinwiegmann} made a prediction for the order 1 term in the free energy expansion in the general $\beta$ two-dimensional case, another derivation via formal free field (or path integral) computations can be found in \cite{klevtsov}.  A more precise prediction is  also made in \cite{shakirov,cantellez} for $\beta $ equal to $2$ and nearing $2$, the next order term being of order $\sqrt N$, and predicted to be independent of $\meseq$.   Recently, this has been proved  in the $\beta=2$ case for some specific non radially symmetric potential \cite{byun}, see also \cite{klevtsovrandom,klevtsovmamarinescu} in the Riemannian setup.
Finally \cite{tellezforrester,jancomanificatpisani} provide finite effect corrections to the free energy. 
Assembling all the above predictions, the prediction for formula \eqref{energyexpconj}
 in the two-dimensional Coulomb case is 
 $$\log \ZNbeta = A_\beta N^2 +\frac14 N\log N + B_\beta N + C_\beta \sqrt{N} + D_\beta \log N + E_\beta+o(1)$$
 with 
\begin{align*}
 A_\beta&=- \beta N^2 \I(\meseq)\\
B_\beta&= \beta\mf(\beta)+(1-\frac\beta4\indic_{\s=0} ) \int_{\Sigma} \meseq\log \meseq \\
 C_\beta& =  k \frac{4}{3\sqrt\pi} \log \frac\beta2\\
 D_\beta& = \hal - \frac{\chi}{12}\\
 E_\beta&= -\hal \log \( \frac{\det_{\zeta}(-\Delta_{\R^2} \backslash \Sigma)}{\det_{\zeta}(-\Delta_{\R^2} )}\) 
+ c(\beta)+ d(\beta) \int_{\pa\Sigma}\frac{ \partial \log \Delta V}{\partial \nu} \\ &+ \frac{(\frac{\beta}4-1)^2}{\beta} 
\( \frac1{4\pi} \int_{\R^2}|\nab (\log \Delta V)^\Sigma|^2  \) 
\end{align*}
where $k$ is the number of connected components of $\Sigma$, $\chi$ is the Euler characteristic of $\Sigma$,  $B_\beta $ involves an entropy term called in this context \textit {Mabuchi functional}, $E_\beta$ from \cite{zabrodinwiegmann} is based on the zeta regularized determinant of the Laplacian and involves a \textit{Liouville functional} from conformal field theory, where $f^\Sigma$ denotes the harmonic extension of $f$ outside $\Sigma$, and $c$ and $d$ are unknown functions.

We are here confirming the next-to-leading order, or order $N$,  term $B_\beta$ (Note that we are expressing it via the thermal equilibrium measure but we could as well express it via the usual equilibrium measure, as was done in \cite{lebles}.  The reader interested in this can refer to that paper or use  the definition \eqref{defEtheta} and the results of Theorem \ref{th1as}.  
  
In the general Riesz case, other than \cite{lebles} there are few results as well, except results in \cite{BHS2} on Riesz-energy minimizers on the torus. 
In contrast to all the results in the literature, we emphasize that the errors in the formulas below are independent of $\beta$ as long as $\theta>\theta_0$ where $\theta$ is as in \eqref{deftheta}, which allows to reach very high temperature regimes.
The formula for the general Riesz case proven in \cite{lebles} takes the same form, but is  expressed in terms of $\meseq$ and a quantitative error rate is not provided.


\begin{theo}[Free energy / minimal energy expansion]\label{thglob2}
 Assume $\s= \d- 2$ with $\d \ge 2$.
Assume \eqref{A1}-\eqref{A5}, $\theta \ge \theta_0>0$,  and that the equilibrium measure $\meseq$ is Lipschitz  on its support (if $\beta=\infty$), resp. that the thermal equilibrium measure is Lipschitz.  
We have
\begin{multline}\label{expansionminimizers} \min \HN(\XN) = N^{2} \mathcal E(\mu_V) - \frac{1}{2\d} (N \log N) \indic_{\s=0} + \frac{N}{2\d} \( \int \mu_V \log \mu_V \) \indic_{\s=0}
\\+ N^{1+\frac\s\d} \mf(\infty)
 \int_{\R^\d} \mu_V^{1+\frac\s\d} + O (N^{1+\frac\s\d-\gamma})
 \end{multline}
 and 
\begin{multline}
\label{expvar}
 \log \ZNbeta=-\beta N^{2-\frac{\s}{\d}}\mathcal E_\theta(\mub) +\frac{\beta}{2\d} (N\log N) \indic_{\s=0} -N  \frac{\beta}{2\d} \( \int_{\R^\d} \mub \log \mub\)
  \indic_{\s=0} \\
   - N \beta \int_{\R^\d} \mub^{1+\frac\s\d} \mf(\beta \mub^{\frac\s\d} ) +O(\beta \chi(\beta) N^{1-\gamma})
   \end{multline}  
where $\gamma>0$ depends only on $\d$, 
and the $O$ depends  on $\d$,  upper and lower bounds for $\meseq $ in $\Sigma$ and $|\meseq|_{C^1(\Sigma)}$, respectively $\theta_0$, 
an upper bound for $\mub$, and a lower bound for $\mub $ and control for the Lipschitz norm of  in $\{x\in \Sigma, \dist(x, \pa \Sigma) \ge \max( \theta^{\ep-\hal}, \bar d_0) \}$ (provided for instance by \eqref{boundsmub}) and in dimension $2$ a bound like \eqref{intromut}.
  \end{theo}
  
We also recall that it will be proven in  Corollary \ref{corofw}, that 
$$\mf(\infty)= \min \mathbb{W}( \cdot, 1)$$
while a variational formula will be given for $\mf(\beta)$ in Corollary \ref{corointervar}
\be \beta \mf(\beta)=\min \mathcal I^1_\beta.\ee

\begin{proof} We give the proof for the case with temperature, the case of minimizers is similar, working with $\meseq$ instead of $\mub$.
First, as verified after \eqref{assgmm}, the thermal equilibrium measure $\mub$ satisfies \eqref{assumplbs} and \eqref{assgmm} for  $\Lambda=\Sigma$, the support of the equilibrium measure $\meseq$ and $\mub(\Sigma^c)\le C \beta^{-1/2} N^{-1/\d}$ by Theorem \ref{th1as}. 
Let us define 
\be\label{defocs}\hat \Sigma =\{x\in \Sigma, \dist(x, \pa \Sigma) \ge  \bar d_0 \}.\ee One may check that $\theta^{\ep-\hal}$ can be made smaller than $\bar d_0$ thus by \eqref{boundsmub} we have uniform bounds for $|\mub|_{C^1(\hat \Sigma)}$ and we also have that the local laws hold in $\hat \Sigma$.
Using Lemma \ref{tiling}, we  then   partition most of  $\hat \Sigma$ into hyperrectangles $Q_i$ of sidelengths  of order  $\rb N^{-1/\d}\le r\le \bar d_0 $, such that $N\int_{Q_i}\mut=\mn_i$ is an integer.  We keep only the hyperrectangles that are inside $\hat\Sigma$. This way the local laws are satisfied in $U:=\cup_i Q_i$ and \eqref{subad3} applies. Moreover 
 \be\label{masseext} \mut(\R^\d\backslash U) \le C (r+ \beta^{-1/2} N^{-1/\d}).\ee

We apply \eqref{subad3} to $\mut$ and combine it with the result of Proposition \ref{coro24} in each $Q_i$ to obtain
\begin{multline}\label{preres}
\log \K_{N,\beta}( \mut)= -\beta N \int_{U}  \mut^{1+\frac\s\d} \mf(\beta \mut^{\frac{\s}{\d}} ) - \frac{\beta}{2\d} N\(
\int_{U}\mut \log \mut\) \indic_{\s=0} \\   - \frac{\beta}{2\d} N \mut(U) (\log N) \indic_{\s=0}+\log \K_{N,\beta}(\mut, \R^\d\backslash U)
+ O\(\beta\chi(\beta) N |U|\mathcal{R}\(N, r, \mut\) \) 
\end{multline}
where  we can absorb the errors in \eqref{subad3} into the $\mathcal R$.
To bound $\log \K_{N,\beta}(\R^\d \backslash U, \mut)
$ we use  \eqref{bornesfiU} and obtain (after rescaling) 
\begin{multline*}\left|\log \K_{N, \beta}(\mut, \R^\d \backslash U)+ \frac{\beta}{4} N\mut(U^c)( \log N) \indic_{\d=2}
 \right|\\
  \le C \beta \chi(\beta) N \mut( \R^\d \backslash U)\le C\beta \chi(\beta) (r+ \beta^{-1/2} N^{-1/\d}).\end{multline*}

It remains to bound 
$$ -\beta N \int_{U^c}  \mut^{1+\frac\s\d} \mf(\beta \mut^{\frac{\s}{\d}} ) -\frac{\beta}{4}  N
\(\int_{U^c}\mut \log \mut \) \indic_{\d=2} ,$$
for this we may use the bound for  $\mf$ given in \eqref{bornesurf} and \eqref{masseext}. 

In dimension $\d=2$ we bound $\int_{U^c}\mut \log \mut$ by  a bound provided for instance by 
\eqref{intromut}.
We conclude that 
\begin{multline*}-\beta N \int_{\cup_i Q_i}  \mut^{1+\frac\s\d} \mf(\beta \mut^{\frac{\s}{\d}} ) - \frac{\beta}{4} N
\( \int_{\cup_i Q_i}\mut \log \mut\) \indic_{\d=2} \\= -\beta N \int_{\R^\d}  \mut^{1+\frac\s\d} \mf(\beta \mut^{\frac{\s}{\d}} ) -\frac{\beta}{4}N\(
\int_{\R^\d}\mut \log \mut\) \indic_{\d=2} + O\(  N\beta  \chi(\beta)(r+ \beta^{-1/2} N^{-1/\d})\).\end{multline*}
Inserting this  into \eqref{preres}, there remains to optimize over $r$ to obtain the result.
\end{proof}


\chapter{Analysis of fluctuations}\label{chap:clt2}
\index{fluctuations}
  In this chapter, we return to our initial purpose in the previous one which is to analyze the fluctuations via the Laplace transform as in \eqref{ratioz}. 
  In the general Coulomb case, this will lead us to improved controls of fluctuations, compared to Corollary \ref{coroconc}, which are sharp in two dimensions.  
   In the one and two-dimensional (Coulomb) case, by comparing two ways of estimating the ratio of partition functions in \eqref{ratioz}, 
   one via almost additivity and approximation, and one by transport, we are able to deduce a better error on the expansion, which allows to precisely evaluate the terms in \eqref{ratioz}.
   This leads to a full  central limit theorem for fluctuations in that case, where we show that the fluctuations of smooth test functions converge to a specific Gaussian, and the electric field 
converges to the Gaussian free field. 

Prior to this such results were known in the extensively-studied one dimensional logarithmic case  $\d=1$, $\s=0$, under various regularity assumptions on $V$, for test functions that live at the macroscale \cite{joha,borotguionnet,BorGui2,shch,bls,lambertledouxwebb} and at mesoscales \cite{bl,BMP22,peilen}, see also \cite{boursier2023clt} for the $\d=1$ Riesz case on the circle.
In dimension $\d=2$, it was first proven in \cite{ridervirag} in the determinantal case  \index{determinantal point process} where $\beta=2$, with $V$ quadratic, and in \cite{ahm,ahm2} still for $\beta=2$ but for  $V$ analytic.  
 This was extended to all $\beta$ and all mesoscales $\ell \ge N^{-\alpha}$, $\alpha<\hal$ 
 in \cite{ls2} (which includes the boundary case) and in \cite{bbny2}.

 We will describe the result from \cite{s2} valid for all $\beta$   provided $\theta\gg 1$ i.e. $\beta \gg N^{-1+\frac\s\d} $, and valid down to the  microscale, more precisely to the {\it minimal scale}, i.e. for  $\ell \gg\rb$. Such very large (and also very small) temperature regimes were not treated before. An exception is the one-dimensional $\beta$-ensemble case in the borderline regime $\beta= cN^{-1}$ for which the limit point process is Poissonian \cite{allezdumaz,nakanotrinh1,nakanotrinh2,hardylambert,lambertpoisson}.
  
 The result we will present for the one-dimensional Coulomb case $\d=1$, $\s=-1$ is  new.

  \section{Improved control of fluctuations}

To deduce the convergence of the Laplace transform of \eqref{defFluct} from  \eqref{ratioz}, the correct scaling is to choose a $t$ which depends on $N$ and tends to $0$ as $t \to 0$ (for instance $t=1/N$ if $\s=0$), thus it will be important to think in terms of linearization as $t \to 0$.
  
  To understand the ratio of partition functions $Z_{N,\beta}(V_t)/Z_{N,\beta}(V)$ in \eqref{ratioz}, we proceed by using the splitting, either with respect to the thermal equilibrium measure as in \eqref{splitzk} (this is what is done in \cite{s2}), or with respect to the usual equilibrium measure  as in \eqref{rewritegibbs3} (this is what was originally done in \cite{ls2}).
    The former has the advantage of working for a broad temperature regime, including for small $\beta$ and of giving the simple expressions of Proposition~\ref{propcalctrans1}, without volume change terms. The disadvantage compared to the latter is that the transformation of the thermal equilibrium measure under perturbation of $V$ is less simple than that of  the regular equilibrium measure, a problem which will be solved by making an approximation, and that it requires more regularity of $V$ as well as appealing to the estimates for $|\mub|_{C^\sigma}$ provided by Theorem \ref{th1as}. The approach with the regular equilibrium measure has the disadvantage of the volume change terms in Proposition \ref{propcalctrans2}, but the advantage that the equilibrium measure change is completely explicit in the bulk and that there are no approximation errors related to it.

Here the Coulomb and Riesz cases differ a lot: indeed in the Coulomb case the perturbed equilibrium measure $\meseq$ is easy to compute when $\xi$ is supported in $\Sigma$ (the support of $\meseq$) :   it is $\mueqt=\meseq+\frac{t}{\cd } \Delta \xi $  (see \eqref{densmu0}).
In the Riesz case, the perturbed equilibrium measure depends on the perturbation $\xi$ in a nonlocal way, and thus the construction of an appropriate transport is much more delicate,   this is done in  \cite{PeilenSer}.

The Coulomb  case where $\xi$ is not supported in $\Sigma$   i.e.~has a support intersecting $\partial \Sigma$ is also much more delicate: one needs to understand how $\partial \Sigma$ is displaced under the perturbation. This is described precisely (in all dimensions) in \cite{serser}. The PDE approach used there replaces  Sakai's theory used in \cite{ahm}, which is restricted to the two-dimensional analytic setting. Doing so allowed us to treat the case where the support of $\xi$ may overlap $\pa \Sigma$ (the first instance was in \cite{ahm} in the case of $V$ analytic and $\beta=2$), we refer to \cite{ls2}. To treat that case, the use of the regular equilibrium measure is much more convenient. 

We note the very interesting physics results of \cite{cardosostephan} that show that the edge behavior of the Coulomb gas is different from the bulk one: they predict ``freezing at the edge"  \index{crystallization} and observe much stronger oscillation of the density near the edge, as if the effective temperature was larger there than in the bulk. Related are predictions of slow decay of correlations at the boundary made in \cite{forresterjancovici} and justified in the determinantal case $\beta=2$ with Szeg\"o kernel formulae  in \cite{Ameur2022,ameurcharliercronvall,ameur2023random}.

We start here by presenting the proof with the thermal equilibrium measure, and  explain how to use the approach via the regular equilibrium measure  in Section~\ref{seceqmeasure}. 
\subsection{Assumptions}

We assume $\s=\d-2$, and continue to assume  \eqref{A1}--\eqref{A5}, $\theta >\theta_0$.
We assume in addition that $V\in C^5$ or even $C^8$ (here we are not trying to minimize the regularity assumption) and \eqref{caf} holds, so that the results of Theorem \ref{th1as} hold.
 
The result will be valid down to  the (temperature dependent) microscale,  \index{minimal scale} i.e.~for test functions supported in a cube of size  $\ell$ which may depend on $N$,  such that 
\be\label{ass1}
 \rb N^{-\frac1\d} \ll \ell \le C  \quad \text{as} \ N\to \infty\ee 
 where $\rb $ is as in \eqref{defrhobeta}. In view of \eqref{deftheta} and since $\rb\ge C \beta^{-1/2}$ by \eqref{defrhobeta}, this implies in particular that $C\ge \ell \gg \theta^{-1/2}$ hence $\theta \gg1$.
 
They will also be valid away from a boundary layer (we recall we treat here only the interior case and refer to \cite{ls2} otherwise), i.e.~where the local laws of Chapter~\ref{chaploiloc} hold:
\be\label{defocs}
\hat{\Sigma}:=\{x\in  \Sigma, \dist(x, \pa \Sigma) \ge \bar d_0\}\ee
with $\Sigma$ the support of $\meseq$ and $\bar d_0$ is as in \eqref{defd0p}.

We recall that from \eqref{boundsmub} we have uniform bounds on $\mub$ in $C^k$ in $\hat\Sigma$, provided $V\in C^{k+4}$, and $\mub$ is bounded below in $\hat \Sigma$ by a constant $m>0$ depending only on $\meseq$ and $\theta_0>0$.


\subsection{Preliminary results}

As explained above, we  assume that $\xi$ is supported in $\hat \Sigma$.

We do not have an exact formula for the perturbed thermal equilibrium measure $\mut^{V_t}$ with perturbed potential $V_t$, however \cite{ascomp} provides an expansion up to arbitrary inverse powers of $\theta $ (the large effective temperature of \eqref{deftheta}) already stated in \eqref{corrections}. It will suffice for our purposes to retain only the leading order term in $1/\theta$, however a more precise description, requiring more regularity of $\xi$ can be obtained by using next order terms, we refer to \cite{s2} for that treatment.

Our first step is thus to replace $\mub^{V_t}$  by the approximation 
\be \label{defnut}
\nut := \mub+ \frac{t}{\cd}\Delta \xi.\ee
Observe that if $\xi$ is supported in $\hat{\Sigma}$  where  $\mub\ge m>0$, $\nut$ is also a probability density, as long as  $|t||\xi|_{C^2}$ is smaller than a constant depending only on  $\d$ and $m$.

  Instead of splitting the energy with respect to $\mut^{V_t}$ we can provide a splitting with respect to $\nut$.
  The proof of the following lemma is presented below in Section \ref{lemaux}.
  \begin{lem}\label{lemepsi} If $t|\xi|_{C^2}$ is small enough (depending on $m>0$ above), 
  we have 
    \begin{multline} \label{laplace1}
\Esp_{\PNbeta} \( e^{-t\beta N^{1-\frac\s\d} \sum_{i=1}^N \xi(x_i)}\) 
\\= e^{ -\beta N^{2-\frac\s\d} \( \mathcal E_\theta(\nut)- \mathcal E_\theta(\mut)\) }\frac{\K_{N,\beta}(\nut)}{\K_{N,\beta}(\mut)} 
 \Esp_{\Q_{N,\beta}(\nut)} \(\exp\( -\theta \Fluct_{\nut} (\ep_t) \) \)
\end{multline}
where 
\be\label{defept}  \ep_t:= \g*\nu_\theta^t + V+ t\xi + \frac{1}{\theta}\log \nu_\theta^t-c_\theta \ee with $c_\theta$ as in \eqref{eqmb},
we have  that $\ep_t$ is supported in the support of $\xi$  and  if $V\in C^4$, 
\be \label{estept0}\|\ep_t\|_{L^\infty} \le   C\frac{t}{\theta}   |\xi|_{C^{2} } ,
\ee
and if in addition $V\in C^5$, 
\be\label{estept} 
|\ep_t|_{C^1}\le   C\frac{t}{\theta}   (|\xi|_{C^2 }+ |\xi|_{C^3 })  ,
\ee
where $C>0$ depends only on $\d, \s, m$ and the norms of $V$.
\end{lem}
Examining \eqref{laplace1}, three things thus need to be done:
\begin{itemize}
\item[1)]  evaluate the limit as $t\to 0$ of $\exp\( - \beta N^{2-\frac\s\d} (\mathcal{E}_\theta (\nut) - \mathcal{E}_\theta (\mut) )\)$,  this is done in Lemma~\ref{lemleadord}
\item[2)] control $\frac{\K_{N,\beta}(\nut)}{ \K_{N,\beta}(\mub)} $ by Proposition~\ref{lemcompdesk}
\item[3)] show that the last expectation, really the approximation error,  is close to $1$ by a priori rough estimates on fluctuations from Chapter \ref{chap:nextorder}.
\end{itemize}

For the first item we will prove in Section \ref{lemaux} the following,  by explicit computations.
\begin{lem}\label{lemleadord}
We have
\be\label{513} \mathcal{E}_\theta(\nut)- \mathcal{E}_\theta(\mut)- t\int_{\R^\d} \xi d\mut = - \frac{t^2}{2\cd}\int_{\R^\d}|\nab \xi|^2  
 + \frac{t^2}{2\theta\cd^2} \int_{\R^\d} \frac{|\Delta \xi|^2}{\mut} + O\( \frac{t^3}{\theta} \int_{\R^\d}\frac{|\Delta\xi|^3}{\mut^2} \)
\ee and \be\label{version2}
\left|\mathcal{E}_\theta(\nut)- \mathcal{E}_\theta(\mut)- t\int_{\R^\d} \xi d\mut\right|
 \le Ct^2 |\supp \nab \xi| \(  |\xi|_{C^{1}}^2   +\frac{ |\xi|_{C^{2}}^2}{\theta}\) ,
\ee
where $O$ and $C$ are universal.
\end{lem}
For the second item we observe that by \eqref{defnut} and Proposition \ref{lemcompdesk} we have
\be \label{difflk}
 \left|\log \frac{\K_{N,\beta}(\nut)}{ \K_{N,\beta}(\mub)} \right| 
 \le  C  \beta \chi(\beta) N\ell^\d \(  (|\mub|_{C^1(Q_\ell)}+| t| | \xi |_{C^3}) |t| | \xi|_{C^1} + |t| | \xi|_{C^2}\). \ee

For the third item we have the following, whose proof relies on the local laws and its consequence \eqref{loclawphi},  see proof in Section \ref{lemaux}.
\begin{lem}\label{lemthirditem} If $t|\xi|_{C^2}$ is small enough (depending on $m>0$ above), we have
\be\label{laplace10a}\left| \log  \Esp_{\Q_{N,\beta}(\nut)} \( \exp \(-\theta \Fluct_{\nut}(\ep_t) \)\)
\right|
\le
C \sqrt{\chi(\beta)} \beta N^{1+\frac1\d} \ell^\d|\ep_t|_{C^1} + C \theta N\ell^\d  |\ep_t|_{C^1}^2\ee
and 
\begin{equation}\label{laplace2a}\left| \log  \Esp_{\Q_{N,\beta}(\nut)} \( \exp \(-\theta \Fluct_{\nut}(\ep_t) \) \)\right|
\le
C \|\ep_t\|_{L^\infty} \theta  N \ell^\d+  C\|\ep_t\|_{L^\infty}^2 \theta N \ell^{\d-2}.
\end{equation}
\end{lem}

\subsection{Main result on fluctuations control}\label{sec1013}

Inserting the result of  \eqref{estept0}, \eqref{version2}, \eqref{difflk} and \eqref{laplace2a}  into \eqref{laplace1}, we have proven the following result from \cite{s2}. 
\begin{theo}[Control of fluctuations, general Coulomb case]\label{th10.1}Let  $\s=\d-2$. Assume  $V\in C^5$, \eqref{A1}--\eqref{A5}, \eqref{caf} hold, $\theta>\theta_0>0$,  and $\xi\in C^{3}$.
 Assume that a hyperrectangle $Q_\ell\subset \hat \Sigma$, with $\ell$ satisfying \eqref{ass1}, contains a $2N^{-1/\d}$-neighborhood of $ \supp\,\xi$ (if it does not hold that  $\ell\ll  1$, assume only $Q_\ell \subset \Sigma$).
Then, letting $\Fluct_{\mub}(\xi)= \sum_{i=1}^N \delta_{x_i}-N\mub$, for any $t\in \R$ such that $|t||\xi|_{C^2}$ is smaller than a positive constant depending only on $\d$ and $m$,  we have 
\begin{multline}
\left|\log \Esp_{\PNbeta} \( e^{ \beta t N^{1 -\frac{\s}{\d}} \Fluct_{\mub}( \xi) }\) 
\right|\\
\le  C|t|N\ell^\d\( |\xi|_{C^2}+\beta \chi(\beta)(| \xi|_{C^1}+|\xi|_{C^2})\)+
  Ct^2N\ell^\d \(  \beta N^{\frac2\d}    |\xi|_{C^1}^2  +\beta \chi(\beta)| \xi |_{C^3}  | \xi|_{C^1}
 \) ,
\end{multline}
where the constants depend only on the constants in the assumptions and upper and lower bounds for $\mub$ in $Q_\ell$, but not on  $N$, $\beta$, $\xi$ or $t$.
\end{theo}
\begin{proof} We insert all the announced elements   into \eqref{laplace1} to find
\begin{multline}
\left|\log \Esp_{\PNbeta}\( \exp(-\beta t N^{1-\frac\s\d} \Fluct_{\mub}(\xi)\)     \right|\\
  \le C \beta N^{2-\frac\s\d} \( t^2 \ell^\d \( |\xi|_{C^1}^2 +   \frac{|\xi|_{C^2}^2}{\theta } \)\)
 +  C  \beta \chi(\beta) N\ell^\d \(  (|\mub|_{C^1(Q_\ell)}+ |t| | \xi |_{C^3}) |t| | \xi|_{C^1} +| t| | \xi|_{C^2}\) \\
 +C   |t|   |\xi|_{C^{2} }   N \ell^\d+  C 
  t^2   |\xi|_{C^{2} }^2
 \frac{  N \ell^{\d-2}}{\theta}.
 \end{multline}
Using that  $|t||\xi|_{C^2}$ is smaller than a constant  and $\ell \le C$ to absorb some terms, inserting $\s=\d-2$, using \eqref{deftheta}, $\s=\d-2$ and \eqref{ass1} to find $\theta \ell^2 \ge \chi(\beta) \max(1,\beta) \ge 1$, we obtain \begin{multline}
\left|\log \Esp\( \exp(-\beta |t| N^{1-\frac\s\d} \Fluct_{\mub}(\xi)\)     \right|\\
  \le Ct^2  \beta N^{2-\frac\s\d}  \ell^\d  |\xi|_{C^1}^2  +C |t| N \ell^\d|\xi|_{C^2}
 + C  \beta \chi(\beta) N\ell^\d \(  (1+ |t| | \xi |_{C^3}) |t| | \xi|_{C^1} +| t| | \xi|_{C^2}\) 
  \end{multline}
hence the result.
\end{proof}

We may then optimize over $t$ and as an illustration, we give the following corollary in the macroscopic case $\ell=1$, obtained with  $t= -\min (N^{-1}, N^{-\hal -\frac1\d}) $. This allows to treat the case where $\xi(x)=\xi_0(x\ell^{-1})$ for some fixed $\xi_0$.
\begin{coro}[Macroscopic scale] Under the same assumptions, if $\d\ge 2$, we have
\be
\left|\log \Esp_{\PNbeta} \( e^{\beta  N^{ -\frac{\s}{\d}}| \Fluct_{\mub}( \xi)| }\) 
\right|\le  C \(  |\xi|_{C^2}+  |\xi|_{C^1} + N^{\frac2\d-1} |\xi|_{C^1}^2 + N^{-1} |\xi|_{C^1}|\xi|_{C^3}\)  ,
\ee
while if $\d=1$, we have 
\be
\left|\log \Esp_{\PNbeta} \( e^{ \beta  N^{ \frac{1}{2}} |\Fluct_{\mub}( \xi)| }\) 
\right|\le  C \(N^{-\hal}(  |\xi|_{C^2}+  |\xi|_{C^1}) +  |\xi|_{C^1}^2 + N^{-2} |\xi|_{C^1}|\xi|_{C^3}\)  ,
\ee
where the constant depends on $\d, \mub $ and $\beta$, but not on $N$ or $\xi$.
\end{coro}
In the meso/microscales, we apply instead to $t=-\ell^2 \min \((N\ell^\d)^{-1}, (N\ell^\d)^{-\hal -\frac1\d})\) $ and obtain 
\begin{coro}[Mesoscopic and microscopic scale]
Let  $\s=\d-2$. Assume  $V\in C^5$, \eqref{A1}--\eqref{A5}, \eqref{caf} hold, $\theta>\theta_0>0$.
Assume that $\xi \in C^3$ is supported in a ball of radius $\ell$ included in $\hat{\Sigma}$  and $|\xi|_{C^k} \le M \ell^{-k}$ for all $k \le 3$.  If $\d\ge 2$ we have 
\be
\left|\log \Esp_{\PNbeta} \( e^{ \beta ( N^{ \frac1\d}\ell)^{-\s}| \Fluct_{\mub}( \xi) |}\) 
\right|\le  C(M+M^2)  ,
\ee
while if $\d=1$,  we have
\be
\left|\log \Esp_{\PNbeta} \( e^{ \beta ( N\ell)^{\hal} |\Fluct_{\mub}( \xi) |}\) 
\right|\le  C(M+M^2)  ,
\ee
where $C$ depends only on $\d, \mub$ and $\beta$, but not on $N$ or $\xi$.
\end{coro}
These  results express a strong form of rigidity of the Coulomb gas: when $\beta $ is fixed the fluctuations are much smaller than those of a Poisson point process, since they do not need to be normalized by $\sqrt{N}$.  
Note that a control under the sole assumption that $\nab \xi \in L^2$, but valid only for $\beta\le 2$, is obtained in \cite{berman2}.

The results are particularly simple in the two-dimensional  Coulomb case for which $\s=0$, they say that fluctuations of functions living at any scale larger than the  microscale are bounded. This turns out to be sharp as we will see in the next sections. We already can see that the order of fluctuations in other dimension is more subtle, and we do not claim that the bounds obtained here are optimal.

In the physics literature, the papers \cite{jlm,lebo} (see also \cite{martin,martinyalcin}) contain  a well-known prediction of  an order $N^{1-1/\d}$ for the variance of the number of points in a domain (see Section \ref{sec:leble} below), however there is no prediction for the order of fluctuations of  smooth linear statistics. 
On the other hand  \cite{chatterjee,gangulysarkar} obtained bounds on the number of points and linear statistics for the  hierarchical Coulomb gas  model, a simplified model introduced by Dyson, and the full Jancovici-Lebowitz-Manificat law was established for that model in \cite{nishryyakir}.
In \cite{chatterjee} the  order of fluctuations of smooth linear statistics was speculated upon ($N^{1/3}$ vs. $N^{1/6}$) with supporting arguments from the example of orthogonal polynomial ensemble treated in \cite{bhardy} in favor of $N^{1/3}$, and finally it was shown in \cite{gangulysarkar} to be in $N^{1-2/\d}$, again still for the hierarchical model instead of the full model and for $\beta$ of order $N^{1/3}$.

\section[Central limit theorems]{Central limit theorem in the one and two-dimensional Coulomb cases}

To go beyond and really understand the limit of  Laplace transform of fluctuations, we need to return to the identity obtained in \eqref{laplace1}.
 The variance term already appears explicitly in the quadratic term of the right-hand side of \eqref{laplace1} as can be seen in \eqref{513}.
  That term indicates that in order to have a finite limiting variance, one should take 
 $t =\tau N^{-1+\frac{\s}{2\d}},$ with $\tau $ of order $1$, thus  a small $t$ as $ N \to \infty$.  
  In the mesoscales, we will consider that $\xi= \xi_0(\frac{\cdot}{\ell})$ for some fixed $\xi_0$. 
 Then one should take 
 $ t= \tau N^{-1+\frac{\s}{2\d}} \ell^{-\frac\s{2}} = \tau \ell^{\d-\s} (N^{\frac1\d}\ell)^{\frac\s2-\d}$. 
The question is then to understand more precisely  the term $\log \frac{\K_{N,\beta}(\nut)}{\K_{N,\beta}(\mub)} $ that appeared in \eqref{laplace1}, since the other terms appearing there have precise enough expansions or bounds.
A first interpretation  of this ratio is via \eqref{dtl}, which already allowed in the previous chapter to bound the term with the help of the commutator estimate, but a second one is provided by the expansion \eqref{expzcasgb} which was obtained  by leveraging the almost additivity.
Since neither of these estimates is sufficient on its own, the idea is to compare them to obtain a stronger estimate.

\subsection{Statement of results}
Let us now state the main results, starting with the two-dimensional case.
The two-dimensional case is special as the order $N$ term in the relative expansion of \eqref{relatexpansion}  does not involve the function $\mf$, as seen in Remark \ref{rem2dci}. This makes it so that the relatively unknown function $\mf$ does not appear at all in the proof or in the limiting quantities.

\index{central limit theorem}
\begin{theo}[CLT in dimension 2 for possibly small $\beta$]
\label{th2} 
Let $\d=2$, $\s=0$.   Assume  $V\in C^{8}$,  \eqref{A1}--\eqref{A5}, $ \theta>\theta_0>0$  and \eqref{caf} hold, assume that  $\xi\in C^{6}$,  and $Q_\ell\subset \hat \Sigma$ is a hypercube of sidelength $\ell$ which contains a $2N^{-1/\d}$ neighborhood of the support of $\xi$ (if it does not hold that $\ell \ll 1$ then only assume $Q_\ell \subset \Sigma$). Assume also that $\xi $ takes the form $\xi_0(\frac{x-\bar x}{\ell})$ for some fixed function $\xi_0\in C^6(\R^\d)$.
Assume $N^{\frac1\d}\ell  \rb^{-1}\to \infty$ as $N \to \infty$\footnote{which implies $\theta \gg 1$.}, and\footnote{the  assumption allows $\beta$ to tend to $+\infty$ as $N\to \infty$ but not at too fast of a rate}
 \be\label{condsup3d2}
\beta^{\hal} \ll (N^{\frac1\d}\ell)^{\frac54-\frac\d2}\log^{-\frac1{2\d}}(N^{\frac1\d}\ell).\ee
Then for any fixed $\tau$, we have \be \label{723}\left|
\log \Esp_{\PNbeta} \(\exp\(-\tau \beta^\hal   \Fluct_{\mub}(\xi) \)\)+
\tau \mathrm{mean}(\xi)   - \frac{ \tau^2}{2}  \mathrm{var}(\xi) \right|\to 0 \quad \text{as} \  \frac{N^{\frac1\d}\ell }{\rb} \to \infty\ee
with 
\be\label{variance}
\mathrm{var}(\xi)=
\frac{ 1}{\cd}  \int_{\R^\d} |   \nab \xi_0|^2 
\ee and 
$$\mathrm{mean}(\xi)=  -\frac{\beta^\hal }{4\cd}\displaystyle \int_{\R^\d} 
   \log \mu_\theta\,    \Delta \xi .$$
\end{theo}
Note that  the convergence rates are independent of $\beta$, which means that $\beta$ can be dependent on $N$, the only condition being $N^{\frac1\d}\ell\gg \rb$ and \eqref{condsup3d2}.
The covariance structure found in \eqref{variance}  characterizes the convergence to a Gaussian free field, see \cite{kangmakarov} for definitions and reference on this conformal field theory context.  More precisely we have shown the following. 

\begin{coro}Under the same assumptions, 
  $\beta^{1/2}\( \Fluct_{\mub}(\xi)+ \frac{1 }{4 \cd} \int_{\R^\d} ( \Delta \xi   ) \log \mut\) $  converges \footnote{\label{note2} The convergence is in the sense of convergence of the Laplace transforms, which implies convergence in law  but is in fact a bit stronger.}  as $N \to \infty $ to a Gaussian  of mean $0$ and variance  $ \frac1{\cd}\int_{\R^2} |\nab \xi_0|^2 $. Moreover, letting 
$h_N= \g* \( \sum_{i=1}^N \delta_{x_i} - N \mub\)$ as in \eqref{def:hnmu}, we have that  $\beta^{\hal} h_N$ converges to the Gaussian free field as $N \to \infty$.
\end{coro}

These results  were first shown for fixed $\beta$ (independent of $N$)  and  for mesoscales $\ell\ge N^{-\alpha}$, $\alpha<\hal$, in \cite{ls2,bbny2}. 
It is important in all such results that $\xi$ is supported in one connected component of the droplet $\Sigma$, for otherwise testing the fluctuations against $\xi$ would allow to  count the number of particles in each component, while that number may fluctuate by an $O(1)$. The correct result in such a case must then include  an extra discrete Gaussian in the limit, as seen in the one-dimensional ``multi-cut" log gas case in \cite{shch,BorGui2} as well as in the two dimensional Coulomb gas case \cite{ameurcharliercronvall}.

The case where $\Sigma$ has one connected component (to avoid the subtlety mentioned just above) but the support of $\xi$ may intersect $\pa \Sigma$ was only treated in \cite{ahm,ahm2} in the case $\beta=2$ and in \cite{ls2} for all $\beta$ (we refer the interested reader to that paper) and one gets  instead a limiting variance $\frac1{\cd}\int_{\R^2}|\nab \xi^\Sigma|^2 $ where $\xi^\Sigma$ denotes the harmonic extension of $\xi$ outside $\Sigma$.  This was also recently extended to fractal contours in the context of determinantal point processes in \cite{zlin}.

The extension to all $\ell \gg N^{-1/2} \rho_\beta$ and to possibly small $\beta$ was obtained in \cite{s2}. 
Further corrections to the variance \eqref{variance}, in power of $1/\theta$, are provided there. 
The assumptions of regularity on $V$ and especially on $\xi$ are certainly not optimal, it is an interesting open question to find what the minimal regularity on $\xi$ needed is for the result to hold. The application of the second order commutator estimates \eqref{p42} is the main bottleneck. In  \cite{s2}, we only required $V\in C^7$ and $\xi \in  C^4$ because we used a different version of \eqref{p42}, the new version allows for a slightly simpler proof and can better generalize to higher dimension.

In radial setups, one may hope to compute more things: for instance there are  physics predictions on next order cumulants  \cite{schehrandco}, and studies of  fluctuations in the case $\beta =2$ when $\Sigma$ has more than one connected component \cite{Ameur2022,ameurcharliercronvall, ameur2023random}.  This shows that the  soft edge situation ($V$ smooth) is  quite different from the  hard edge  ($V=\infty$ outside $\Sigma$) one, in particular fluctuations may  be non-Gaussian even though the droplet is connected.

When $\beta $ is so large that \eqref{condsup3d2} fails, it is likely that the CLT is not true as one expects crystallization \index{crystallization} and fluctuations near a lattice rather than GFF-like fluctuations, see Chapters~\ref{chap:renormalized} and \ref{chap:derivW}. Instead we can normalize $\Fluct(\xi) $ differently to obtain a convergence result. 
\begin{theo}[Low temperature and minimizers in dimension 2] \label{thlowt2}Let $\d=2$, $\s=0$.
  Assume  the same as in the previous theorem on $V$, $\xi$, $\ell$. Assume that $\beta \gg 1$  and $N^{\frac1\d}\ell \gg 1$ as $N \to \infty$. Then we have (in the sense of convergence of the Laplace transforms)
$$\lim_{N\to \infty} \( \Fluct_{\mub}(\xi)+  \frac{1}{4\cd}   \displaystyle \int_{\R^\d} 
(\log \mut)  \Delta \xi \)= 0.$$ 
\end{theo}

The case of minimizers of $\HN$ corresponds to $\beta=\infty$ and can be obtained by simply letting $\beta \to \infty$ in the case with temperature since the constants are independent of $\beta$. 
\begin{coro}Under the same assumption,  
assume that $\XN$ minimizes $\HN$   and $N^{\frac1\d}\ell \gg 1$ as $N \to \infty$. Then we have 
$$\lim_{N\to \infty} \Fluct_{\meseq}(\xi)=-  \frac{1}{4}   \displaystyle \int_{\R^\d} 
\frac{\Delta \xi }{\cd } \log \meseq.$$ 
\end{coro}

Note that this generalizes \cite{ls2} and also completements the results on minimizers in \cite{aoc,rns,PetSer,PRN}, see Theorem \ref{th4}.

In the one-dimensional Coulomb case, the function $\mf(\beta)$ appears in the relative free energy expansion \eqref{relatexpansion}. We are able to use the analyticity of $\mf$ proven in \cite{kunz} to still deduce a similar result for fixed $\beta$. The fact that the bounds on derivatives of $\mf$ may degenerate as $\beta\to 0$ is the reason why we do not treat the high temperature regime in this setting.

\index{central limit theorem}
\begin{theo}[CLT in dimension 1]
\label{thdim1} 
Let $\d=1$, $\s=-1$.   Assume  $V\in C^{5}$, \eqref{A1}--\eqref{A5}, $ \beta>\beta_0>0$  and \eqref{caf} hold, assume that  $\xi\in C^{3}$,  and $Q_\ell\subset \hat \Sigma$ is a hypercube of sidelength $\ell$ which contains a $2N^{-1/\d}$ neighborhood of the support of $\xi$ (if it does not hold that $\ell \ll 1$ then only assume $Q_\ell \subset \Sigma$). Assume also that $\xi $ takes the form $\xi_0(\frac{x-\bar x}{\ell})$ for some fixed function $\xi_0\in C^4(\R^\d)$.
Assume $N^{\frac1\d}\ell  \to \infty$ as $N \to \infty$ and 
 \be\label{condsup3d21d}
\beta^{\hal} \ll (N^{\frac1\d}\ell)^{\frac54-\frac\d2}\log^{-\frac1{2\d}}(N^{\frac1\d}\ell).\ee
Then for any fixed $\tau$, we have 
\be \label{7231d}
\lim_{ N^{1/\d } \ell\to +\infty}   \ \log \Esp_{\PNbeta} \(\exp\(-\tau \beta^\hal  (N\ell)^{\hal} \Fluct_{\mub}(\xi) \)\)=  \frac{\tau^2}{2\cd} \int_{\R^\d}|\nab \xi_0|^2 .\ee
In other words, $\beta^{\hal}(N\ell)^{\hal}\Fluct_{\mub}(\xi) $ converges to a Gaussian of mean zero and variance $ \frac{1}{\cd} \int_{\R^\d}|\nab \xi_0|^2$.
\end{theo}

\begin{theo}[Low temperature and minimizers in dimension 1] \label{thlowt1d}Let $\d=1$, $\s=-1$.   Assume  $V\in C^{5}$, \eqref{A1}--\eqref{A5} and \eqref{caf} hold. Assume
 that  $\xi\in C^{3}$,  and $Q_\ell\subset \hat \Sigma$ is a hypercube of sidelength $\ell$ which contains a $2N^{-1/\d}$ neighborhood of the support of $\xi$,  with $\ell$ satisfying $N^{-1/\d}\le \ell\le C$, and \eqref{bornesxi} holds. 
Assume $\beta \gg 1$  and $N^{\frac1\d}\ell \gg 1$ as $N \to \infty$. Then  we have (in the sense of convergence of the Laplace transforms)
$$\lim_{N\to \infty} (N\ell)^{\hal}\Fluct_{\mub}(\xi)  = 0.$$ 
\end{theo}

\subsection{Proof of the CLT results}

In view of \eqref{derivlogkt} going further to obtain the CLT requires to precisely estimate $\Esp_{\Q_{N, \beta}(\mut)}(\Ani_1)$, which is what was called in \cite{ls2} the {\it anisotropy term}. 
 We do it by comparing with a  second way of estimating $\log \K(\nut)-\log \K(\mub)$ which is by almost additivity estimates as in Chapter~\ref{chaploiloc}.
The core of the proof is an improvement of 
\eqref{compdesk}, which relies on comparing that relation with the result \eqref{relatexpansion} obtained by almost additivity.

For that we write that Corollary \ref{th1b} provides an expansion of the form 
\be \label{2eway0}
\log \K_{N,\beta}( \mu_t)- \log \K_{N,\beta}(\mu_0)= N\beta\( \mathcal{Z} (\beta, \mu_t)-\mathcal{Z}(\beta, \mu_0) \) + O(\beta \chi(\beta) N\ell^\d \mathcal R_t)\ee
where $\mathcal R_t$ is the rate that  corresponds to the error rate in \eqref{expzcasgb} for the measure $\mu_t$, to which we add that for the measure $\mu_0$.
The form of $\mathcal{Z}$ is 
\be\label{formemZ} \mathcal{Z}(\beta, \mu)= - \int_{\R^\d} \mu^{1+\frac\s\d} \mf(\beta \mu^{\frac\s\d}) - \frac{1}{2\d}\( \int_{\R^\d}\mu \log \mu\) \indic_{\s=0}.\ee  
We will also denote  by $\mathcal{B}(\beta, \mu_0, u)$ the derivative at $t=0$ of the function $\mathcal{Z} (\beta, \mu_t)$ when $\frac{d}{dt}|_{t=0} \mu_t= u$. For instance in view of \eqref{formemZ}, we compute that 
if $\d=2$,
\be\label{B2d}
 \mathcal{B}(\beta, \mu_0, u)= - \frac{1}{4}  \int_{\R^\d}  u \log \mu_0\ee
 while if $\d=1$,
  \be\label{B1d}    \mathcal{B}(\beta, \mu_0, u)= \int_{\R^\d} u \frac{\mf'(\beta \mu_0^{-1}) }{\mu_0^2}.\ee
We also need to assume a quantitative differentiation property of the form
 \be\label{diffmZ}\mathcal{Z}(\beta, \mu_t)- \mathcal{Z}(\beta, \mu_0)= \mathcal B(\beta, \mu_0, u)+ O(t^2 \ell^\d \|u\|_{L^\infty}^2) \ee when $t\ell^{-2} \le \frac{1}{2\cd}m$, with $O$ depending only on $\d$ and $\s$.
By straighforward computation this condition holds for $\d=2$. It holds in other dimensions if we know a bound for $\mf'' $. This is in turn only known in one dimension,  uniformly for $\beta>\beta_0$, thanks to the work of \cite{kunz} that proves that $\mf $ is in fact analytic. In higher dimensions, $\mf$ may not always be twice differentiable, values of $\beta$ for which is fails to be correspond to phase-transitions.

\begin{prop}[Improved relative expansion and exponential moments of $\Ani_1$]\label{propmomani} Assume $\s=\d-2$. 
Assume that $\mu_0$ is a $C^4$ probability density over $\R^\d$, $\xi \in C^6$ (or  $\mu_0 \in C^1$ and $\xi\in C^3$ if $\d=1$) , $Q_\ell$ is a ball of radius $\ell\ge N^{-1/\d}$   contaning a $2N^{-1/\d}$-neighborhood of the support of $\xi$ and $\mu_0\in C^2(Q_\ell)$ with $\mu_0 \ge m>0$ in $Q_\ell$. Assume also  that there exists a constant $M$ such that for $k\in[0, 6]$, we have
\be \label{bornesxi}|\xi|_{C^k}\le M \ell^{-k}.\ee
Let then 
\be \mu_t= \mu_0+ \frac{t}{\cd}\Delta \xi.\ee
Assume that  \eqref{diffmZ} holds with $u = \frac{1}{\cd}\Delta \xi$, and that the local laws \eqref{locallawint0} hold for $\Q_{N,\beta}(\mu_0)$ in $Q_\ell$.
Let 
$ \alpha =  \( \max_{|s| \le \ell^2}  \mathcal R_s\)^{\hal}   .$
For every $t $ such that $|t|\ell^{-2}\le \hal \alpha $ and $|t| \le \frac{\cd m}{\|\Delta \xi\|_{L^\infty}}$, we have
 \begin{multline} \label{meilleureexp}\log \frac{\K_{N,\beta}(\mu_t)}{\K_{N,\beta}(\mu_0)}= t N  \beta  \mathcal{B}(\beta, \mu_0, \frac{1}{\cd}\Delta \xi) 
 \\ + O \(t^2 \beta\chi(\beta) N\ell^{\d-4}  +  |t| \beta \chi(\beta) N\ell^{\d-2} 
  \( \max_{|s|\le \ell^2} \mathcal R_{s}   \)^{\hal} 
\)
\end{multline}
and 
\begin{multline}\label{expmomani1}
\Bigg|\log  \Esp_{\Q_{N,\beta}(\mu_0)}  \Bigg( \exp\Bigg(t  
 \beta N^{-\frac\s\d}\Ani_1 (\XN, \mu_0, \frac{1}{\cd} \frac{\nab \xi}{\mu_0} 
 )       - t \beta N      \mathcal{B} (\beta, \mu_0,\frac1{\cd}\Delta \xi )
     \Bigg) \Bigg)\Bigg| \\
  \le  C|t|\ell^{-2}  \beta \chi(\beta) N\ell^\d 
   \( \max_{|s|\le \ell^2} \mathcal R_{s}   \)^{\hal} .
\end{multline}
with constants that depend on $\d$, $M$,  the bounds on $\mu_0$, $\beta_0$ if $\d=1$,  but not $N$,  $\ell$,  $t $.
\end{prop}
The main point here is that compared to \eqref{2eway0} we have gained a factor of $t$ in the error terms, which will be very small when $t$ is very small.
\begin{proof}
\noindent
{\bf Step 1. Pushing the expansion to next order.}
\\ 
Let us revisit the proof of 
Lemma \ref{lemcompdesk}. For $|s|\le |t|$, we   define  $\psi_s= - \frac{1}{\cd}\frac{\nab \xi}{\mu_s}$  where $\mu_s= \mu_0 +\frac{s}{\cd}\Delta \xi$, and let $\Phi_s$ be given by \eqref{defflow}. The assumptions $\mu_0 \ge m$ and  $|t| \le \frac{\cd m}{\|\Delta\xi\|_{L^\infty}}$ ensure that $\mu_s$ remains a probability density. 

 By Lemma \ref{1221}, we have $ \mu_t= \Phi_t \# \mu_0$.
Applying the change of variables (or transport) \index{transport method}  \eqref{formul} with $\mu=\mu_0 $ and $\phi_h = \Phi_t$, we find 
\be \label{1264}\frac{\K_{N,\beta}( \mu_t)}{\K_{N,\beta}(\mu_0)} = \Esp_{\Q_{N,\beta}(\mu_0)}  \( \exp (-\beta N^{-\frac\s\d} ( \F_N(\Phi_t(\XN), \Phi_t \# \mu_0 ) - \F_N(\XN, \mu_0))\) .\ee
Previously we have estimated this by the integral of $\Ani_1$ (using \eqref{derivft}) and the commutator estimate, Theorem \ref{thm:FI}. In order to go beyond, we can get a more precise estimate by using Taylor's integral formula to second order, and writing  in view of \eqref{derivft} and \eqref{derivani}, that
\be  \F_N( \Phi_t(\XN),  \Phi_t \# \mu_0 ) - \F_N(\XN, \mu_0)
= t\Ani_1 (\XN, \mu_0, \psi_0) + \int_0^t(t-s) \Ani_2(\Phi_s(\XN), \mu_s, \psi_s) \, ds.
\ee
Using the explicit form of $\psi_s$ and the linearity of $\Ani_1$ in its last variable, we obtain
\be \label{rewriteFdiff}
\F_N( \Phi_t(\XN),  \Phi_t \# \mu_0 ) - \F_N(\XN, \mu_0)
 =- t \Ani_1 \(\XN, \mu_0, \frac{1}{\cd} \frac{\nab \xi}{\mu_0} \) + \int_0^t (t-s)\Ani_2 \(\Phi_s(\XN) , \mu_s,  \psi_s \)\, ds.\ee
 
 {\bf Step 2. Inserting the second order commutator estimate.} \index{commutator estimates}
If $\d\ge 2$, the second order commutator estimate \eqref{p42}  bounds $\Ani_2 (\Phi_s(\XN) , \mu_s,  \psi_s )$ by the energy, with factors depending on the norms of $\psi_s$ and $\mu_s$. If $\d=1$ and $\s=-1$, $\Ani_2$ is simply zero by definition \eqref{16}, as remarked after Proposition \ref{prop:comparaison2}, so no further regularity of $\xi$ is needed.
To estimate these factors for $\d \ge 2$ we observe that for $k\le 4$, we have
\be \label{obser} |\psi_s|_{C^k} \le C \ell^{-k-1}\ee
where $C$ depends on $M$, the lower bound $m$ for $\mub$ in $\hat\Sigma$, and the bounds on norms on $\mub $ in \eqref{boundsmub}.
 Indeed, we first observe that by definition of $\mu_s$, we have for $\sigma \le 4$,
 \be \label{bornemus} 
 |\mu_s|_{C^\sigma}\le C (1+ |s| |\xi|_{C^{\sigma+2}} ) \le C  (1+ |t| \ell^{-\sigma -2})\le C \ell^{-\sigma} \ee where we used that $|t| \ell^{-2}\le 1$.
 Arguing as for \eqref{419}, by definition of $\psi_s$ we then easily deduce \eqref{obser}.
 With this and  the assumption $\ell \ge  N^{-1/\d}$  we find that the factor in  the right-hand side of \eqref{p42}  applied to $v=\psi_s$ 
is bounded by  $ C  \ell^{-4} $  and  we  thus obtain
$$\left|\int_0^t (t-s)\Ani_2 \(\Phi_s(\XN) , \mu_s,  \psi_s \)\, ds\right|
\le C t  \ell^{-4}     \int_0^t  \Xi(s) ds,$$
where $\Xi$ is as in \eqref{defXi}. To bound $\Xi(s)$ we use  Corollary \ref{corocontrenergyt},  \eqref{obser} and $|t| \ell^{-2} \le C$, and conclude that 
$$\left|\int_0^t(t-s) \Ani_2 \(\Phi_s(\XN) , \mu_s,  \psi_s \)\, ds\right|
\le C t^2  \ell^{-4}    \Xi(0)   . $$
Inserting into \eqref{rewriteFdiff}, we have obtained 
$$
\F_N( \Phi_t(\XN),  \Phi_t \# \mu_0 ) - \F_N(\XN, \mu_0)\\
 = -t \Ani_1 \(\XN, \mu_0, \frac{1}{\cd} \frac{\nab \xi}{\mu_0} \)  + O(  t^2 \ell^{-4}  \Xi(0)) $$
 and inserting into \eqref{1264} yields
 $$ \frac{\K_{N,\beta}( \mu_t)}{\K_{N,\beta}(\mu_0)} = \Esp_{\Q_{N,\beta}(\mu_0)}  \( \exp \(\beta N^{-\frac\s\d}
 \( t \Ani_1 \(\XN, \mu_0, \frac{1}{\cd} \frac{\nab \xi}{\mu_0} 
 \)  + O(   t^2  \ell^{-4}     \Xi (0)) \) \)\) .
 $$
 Inserting next the local laws \eqref{locallawint0} that hold in $Q_\ell$, this can be rewritten 
 \be \label{1eversion}
 \frac{\K_{N,\beta}( \mu_t)}{\K_{N,\beta}(\mu_0)} = \Esp_{\Q_{N,\beta}(\mu_0)}  \( \exp \(\beta N^{-\frac\s\d}
  t \Ani_1 \(\XN, \mu_0, \frac{1}{\cd} \frac{\nab \xi}{\mu_0} \)
   + O(    t^2  \beta \chi(\beta)  N \ell^{\d-4}   )   \) \) .
 \ee

{\bf Step 3. Comparison of the two formulas by H\"older trick and control of exponential moments of $\Ani_1$.}

The measures   $\mu_t $ and $\mu_0$ differ  only in $Q_\ell$ and their difference integrates to $0$, moreover since their densities are bounded below, 
up to changing $\ell$ by at most $O(N^{-1/\d})$, we may without loss of generality assume that $N\int_{Q_\ell} \mu_0=N\int_{Q_\ell} \mu_t$ is an integer, so we may apply Corollary \ref{th1b} 
to find  \eqref{2eway0}. Inserting \eqref{diffmZ}, we obtain 
\be\label{2eway}
\log \frac{\K_{N,\beta}( \mu_t)}{ \K_{N,\beta}(\mu_0)}= t \beta N\mathcal B(\beta, \mu_0,\frac{1}{\cd} \Delta \xi)
 + O\(\beta N \ell^\d t^2 |\xi|_{C^2}^2 \) +
O(\beta \chi(\beta) N\ell^\d \mathcal R_t),\ee
where $O$ depends  on $m$ and also on $\beta_0$ such that $\beta>\beta_0$ if $\d=1$.
We can now compare the two relations \eqref{1eversion} and \eqref{2eway}. The right-hand side of \eqref{2eway} consisting of constants, using  \eqref{bornesxi} to absorb one error term into the other, the comparison implies that 
\begin{equation*} \Esp_{\Q_{N,\beta}(\mu_0)}  \(\exp\(t L +
O\(\beta \chi(\beta) N\ell^\d  ( \mathcal R_t+ t^2 \ell^{-4}    )\) \)\)=1 .\end{equation*}
where we let 
\be\label{defLL}
L:=  \beta N^{-\frac\s\d}\Ani_1 \(\XN, \mu_0, \frac{1}{\cd} \frac{\nab \xi}{\mu_0} 
 \)      -\beta N \mathcal B(\beta, \mu_0,\frac{1}{\cd} \Delta \xi)
  . \ee
  We wish to  identify the terms of order $t$, in the regime where $t\ell^{-2} $ is small.
To  do so, let us see this relation in the form 
$$\Esp_{\Q_{N,\beta}(\mu_0)}  \( e^{ t L + \Error} \)=1.$$
Using the Cauchy-Schwarz inequality, we may write
\begin{multline*}\Esp_{\Q_{N,\beta}(\mu_0)} \( e^{\hal tL }\) = \Esp_{\Q_{N,\beta}(\mu_0)}  \( e^{\hal ( t L + \Error)  -\hal  \Error} \)\\\le  \( \Esp_{\Q_{N,\beta}(\mu_0)}  \( e^{  t L + \Error} \)\)^{\hal}  \(\Esp_{\Q_{N,\beta}(\mu_0)} \( e^{  -\Error} \)\)^\hal =\(\Esp_{\Q_{N,\beta}(\mu_0)} \( e^{  -\Error} \)\)^\hal= e^{-\hal\Error} .\end{multline*} 
We thus deduce that for all $t$ such that $t\ell^{-2}$ is small enough, we have 
\be\label{borninfani} \log  \( \Esp_{\Q_{N,\beta}(\mu_0)}  \( e^{\hal tL}   \)\) 
  \le O\(\beta \chi(\beta) N\ell^\d  ( \mathcal R_t+ t^2 \ell^{-4}    )\).
\ee 
The idea is now to apply this relation to  $t$ as large as possible, then  deduce information for smaller $t$  by H\"older's inequality. 

First, we apply \eqref{borninfani} to  $t= \pm \alpha \ell^2$ with $\alpha$ chosen such that 
$$ \alpha =  \( \max_{|s| \le \ell^2}  \mathcal R_s\)^{\hal} .$$ Without loss of generality we may always assume $\alpha$ is small enough that $ \alpha \ell^2 \le \frac{\cd m}{\|\Delta \xi\|_{L^\infty}}$.
This way, we find
$$ \log   \Esp_{\Q_{N,\beta}(\mu_0)}  \( e^{\pm\hal \alpha \ell^2L} \)  
   \le O\(\beta \chi(\beta) N\ell^\d     \max_{|s| \le \ell^2} \mathcal R_{s}\) 
. $$
Then by H\"older's inequality, we deduce that  for $|t|\le \hal \alpha \ell^2$ we must have
\begin{multline} \label{concl1}\log   \Esp_{\Q_{N,\beta}(\mu_0)}  \( e^{ t   L  }\)  
  \le \frac{2|t|}{  \alpha \ell^2}  \log  \Esp_{\Q_{N,\beta}(\mu_0)}  \( e^{\pm\hal \alpha \ell^2 L }\) \\
  \le C  |t|\ell^{-2}  \beta \chi(\beta) N\ell^\d    \( \max_{|s|\le \ell^2} \mathcal R_{s}   \)^{\hal}
. \end{multline}
In order to obtain a reverse inequality we may observe that by H\"older's inequality 
we have 
$$1\le \Esp (e^{t L} ) \Esp(e^{-tL} ) $$ 
thus applying \eqref{concl1} to $-t$, we obtain the converse inequality and reinserting \eqref{defLL} we can assert that  for $|t|\le \hal \alpha \ell^2$,
we have
\begin{multline}
\Bigg|\log  \Esp_{\Q_{N,\beta}(\mu_0)}  \Bigg( \exp\Bigg(t  
 \beta N^{-\frac\s\d}\Ani_1 (\XN, \mu_0, \frac{1}{\cd} \frac{\nab \xi}{\mu_0} 
 )     - t \beta N     \mathcal B(\beta, \mu_0 , \frac{1}{\cd} \frac{\nab \xi}{\mu_0} )
     \Bigg) \Bigg)\Bigg| \\
  \le  C|t|\ell^{-2}  \beta \chi(\beta) N\ell^\d 
   \( \max_{|s|\le \ell^2} \mathcal R_{s}   \)^{\hal} .
  \end{multline} The main gain here is that we have been able to multiple the error rate $\mathcal R$ by the small factor $t$. Thus we know the exponential moments of $\Ani_1$ with good precision.

  Inserting this relation back into \eqref{1eversion} and using H\"older's inequality again to separate terms,   we obtain the  expansion \eqref{meilleureexp}.
\end{proof}

\begin{proof}[Proof of the main theorems]
We will run the proof for the one-dimensional and two-dimensional cases together as they only differ in localized places.

To obtain the main CLT, the proper choice of $t$ in \eqref{laplace1} is
\be \label{choixt1}
t= \tau \ell^2 \beta^{-\hal } (N^{\frac1\d}\ell)^{-1-\frac{\d}{2}}
\ee
for high temperatures (Theorems \ref{th2} and \ref{thdim1}) and 
\be \label{choixt2}t=\tau \ell^2 \beta^{-1} (N^{\frac1\d}\ell)^{-1-\frac\d2}
\ee
for low temperatures (Theorems \ref{thlowt2} and \ref{thlowt1d}). 
We always assume that $|t|\le \frac{\cd m}{\|\Delta \xi\|_{L^\infty} } $, so that   $\nut$ remains a probability density.

We note that by assumption on the form of $\xi$, \eqref{bornesxi} holds.
 This is the only fact about $\xi$ that we will use, except for rephrasing the variance at the very end.

  We may now return to \eqref{laplace1} and insert \eqref{meilleureexp} applied to $\mu_0=\mub$ and $\mu_t= \nut$, \eqref{laplace10a} and \eqref{513}, this yields 
  \begin{multline}\label{preconcl}
  \log \Esp_{\PNbeta}\( e^{- t \beta N^{1-\frac\s\d} \sum_{i=1}^N \xi(x_i)}\) 
  =  t \beta N^{2-\frac\s\d} \int_{\R^\d} \xi d \mub+ \frac{\beta N^{2-\frac\s\d} t^2 }{2\cd}\(  \int_{\R^\d} |\nab \xi|^2    - \frac1{\theta \cd}  \int_{\R^\d} \frac{|\Delta \xi|^2}{\mut} \)
 \\+\frac{t N  \beta}{ \cd} \( 
- \frac{1}{4} \(  \int_{\R^\d} \Delta \xi \log \mub\) \indic_{\d=2}   +     \(\int_{\R^\d} \Delta \xi \frac{\mf'(\beta \mub^{-1}) }{\mub^2}\)\indic_{\d=1}\)
 + \Error_1+\Error_2+\Error_3.\end{multline}
  Here $\Error_1$ is the error term in \eqref{513}, with our assumption \eqref{bornesxi} it is $O(t^3  \beta N^{2-\frac\s\d} \theta^{-1} \ell^{\d-6} ),$ $\Error_2$ is the error term in \eqref{laplace10a}, using \eqref{estept}  it is 
$$\Error_2= O\( t
  \sqrt{\chi(\beta)} \beta N^{1+\frac1\d}\ell^{\d-3} \theta^{-1}   
 + t^2 \theta^{-1} N\ell^{\d -6} \).$$  Finally $\Error_3$ is the error term in \eqref{meilleureexp}.
  We now insert
  $\theta =  \beta N^{1-\frac\s\d}$, \eqref{choixt1} and \eqref{defRRR1}. We check that the choice of $t$ of \eqref{choixt1} satisfies $|t|\ell^{-2}\le\hal \alpha$, for $N$ large enough, under the assumption $\ell N^{1/\d} \rho_\beta^{-1} \to \infty$ (which is stronger than $\ell N^{\frac1\d} \to \infty$).  Indeed,  in this regime, $x,y $ in \eqref{defRRR1} tend to $0$, and $\mathcal R \to 0$ at an algebraic rate, more precisely, we can bound 
 \be \label{maxRs}
 \max_{|s|\le \ell^2} \mathcal R_s \le C \(\frac{N^{\frac1\d}\ell}{\rho_\beta} \)^{-\hal} \( \log \frac{ N^{\frac1\d}\ell}{\rho_\beta}\)^{\frac1\d}.\ee

    We then obtain that, in that limit, 
  $$\Error_1= O\( \tau^3 \ell^\d N  \beta^{-\frac32}  (N^{\frac1\d}\ell)^{-3-\frac{3\d}{2}}  \)= O(\tau^3 (N^{\frac1\d}\ell)^{\d-3-\frac{3\d}{2} } \beta^{-\frac32} ).$$We note that in view of \eqref{defchibeta} for both $\d=1$ and $\d=2$ (and $\s=\d-2$) the assumption $\ell N^{1/\d} \rho_\beta^{-1} \to \infty$ implies $ N^{1/\d}\ell \beta^{3/4} \to \infty$ in dimension 1, and $N^{1/\d}\ell \beta^{1/2}$ in dimension 2, and 
  \be \Error_1\le \begin{cases} O\( ( \ell N^{1/\d} \rb^{-1})^{-\frac{7}{2}}\)& \text{if} \ \d=1\\
  O \(( \ell N^{1/\d} \rb^{-1})^{-4}\) & \text{if} \ \d=2.\end{cases}\ee

  For the second term, we use that $\chi(\beta) \beta^{-1} \le  \rho_\beta^2 $ and $\s=\d-2$ to obtain
  \begin{multline*}
  \Error_2= O\(  \sqrt{\chi(\beta)} N^{\frac\s\d+\frac1\d} \ell^{\d-3}
   ( \tau \ell^2 \beta^{-\hal} (N^{\frac1\d}\ell)^{-1-\frac\d2}  )
   + C  N^{\frac\s\d}  \ell^{\d-6} \beta^{-1}  ( \tau \ell^2 \beta^{-\hal} (N^{\frac1\d}\ell)^{-1-\frac\d2}  )^2
  \)\\=
  O\( \tau \rho_\beta   (N^{\frac1\d}\ell)^{\frac\d2-2} + C \tau^2    \Big(\frac{N^{\frac1\d}\ell}{\rho_\beta}\Big)^{-4}\) =o(1).
  \end{multline*}
 For the third error term,  using \eqref{maxRs}, we obtain
   \begin{multline*}
   \Error_3= O\( 
   \tau^2  \beta^{-1}(N^{\frac1\d}\ell)^{-2-\d}   
   \beta\chi(\beta) N\ell^\d   + \tau  \beta^{-\hal} (N^{\frac1\d}\ell)^{-1-\frac\d2}    \beta \chi(\beta) N\ell^{\d}
    \( \max_{|s|\le \ell^2} \mathcal R_{s}   \)^{\hal} 
 \)
  \\ = O\(\tau^2 \beta \rho_\beta^2   (N^{\frac1\d}\ell)^{-2}      +  \tau \beta^{\hal}  \chi(\beta) (\rb)^{\frac\d2-1} \(\frac{N^{\frac1\d}\ell}{\rho_\beta} \)^{\frac\d2-\frac54} \( \log \frac{ N^{\frac1\d}\ell}{\rho_\beta}\)^{\frac1{2\d}}  \) .
   \end{multline*}
  When $\d\le 2$,  this tends to $0$ algebraically as soon as $\beta \le 1$, and if $\beta \ge 1$ (then $\rho_\beta=1$) we use the assumption \eqref{condsup3d2}. This is where we change the definition of $t$ to get convergence in the high $\beta$ regime.
   
We thus see that all error terms tend to $0$ at an algebraic rate when $\tau$ is fixed and $ N^{\frac1\d}\ell \gg \rho_\beta,$ if $\d\le 2$. The limitation to convergence in higher dimension is only due to the poor precision of $\mathcal{R}_s$.

  Inserting the expression \eqref{choixt1} into \eqref{preconcl}, we have  thus found 
\begin{align*}
 &  \log \Esp_{\PNbeta}\( \exp\( -  \tau \ell^2 \beta^{-\hal} (N^{\frac1\d}\ell)^{-1-\frac\d2}  \beta N^{1-\frac\s\d} \Fluct_{\mub}(\xi)\)\) 
  \\ & 
  =   \frac{ N^{1+\frac2\d} \tau^2 \ell^4  (N^{\frac1\d}\ell)^{-2-\d}  }{2\cd}
  \(  \int_{\R^\d} |\nab \xi|^2    - \frac1{\theta \cd}  \int_{\R^\d} \frac{|\Delta \xi|^2}{\mut} \)\\ 
  & +
  \frac{\tau \ell^2  N (N^{\frac{1}{\d}} \ell)^{-1-\frac\d{2}}  \beta^{\hal} }{ \cd} \( 
- \frac{1}{4} \(  \int_{\R^\d} \Delta \xi \log \mub\) \indic_{\d=2}   +     \(\int_{\R^\d} \Delta \xi \frac{\mf'(\beta \mub^{-1}) }{\mub^2}\)\indic_{\d=1}\)+o(1).\end{align*}
Moreover, by assumption on $\xi$, we have 
\be \int_{\R^\d} |\nab \xi|^2    - \frac1{\theta \cd}  \int_{\R^\d} \frac{|\Delta \xi|^2}{\mut} = \ell^{\d-2}\int_{\R^\d}|\nab \xi_0|^2 +O\(|\xi_0|_{C^2} \frac{\ell^{\d-4}}{\theta}\)\ee
and since, as noted after \eqref{ass1} we have $\ell\gg \theta^{-1/2}$, the second term is $o(\ell^{\d-2})$.
When $\d=2$, we have thus obtained that 
\be
  \log \Esp_{\PNbeta}\( \exp\( -  \tau \beta^{\hal} \Fluct_{\mub}(\xi)\)\) 
  = \frac{  \tau^2  }{2\cd} \int_{\R^\d} |\nab \xi_0|^2  +   \tau  \frac{\beta^{\hal}  }{4 \cd}  \int_{\R^\d} \Delta \xi \log \mub  +o(1).
\ee
This implies the result of Theorem \ref{th2} since  the inverse Laplace transform of a Gaussian is a Gaussian.

In the case $\d=1$, 
by change of variables, we check that
  $$\frac{\tau \ell (N^{\frac{1}{\d}} \ell)^{-\frac1{2}}  \beta^{\hal} }{ \cd}     \int_{\R} \Delta \xi \frac{\mf'(\beta \mub^{-1}) }{\mub^2}= \frac{\tau (N^{\frac{1}{\d}} \ell)^{-\frac1{2}}  \beta^{\hal} }{ \cd}     \int_{\R} \Delta \xi_0(y) \frac{\mf'(\beta \mub(\bar x+\ell y)^{-1} )}{\mub(\bar x +\ell y)} dy,$$   
   and since $N^{1/\d}\ell \to +\infty$, this linear term tends to $0$.
   We have then obtained for $\d=1$ that  
\begin{align}\notag
  \log \Esp_{\PNbeta}\( \exp\( -  \tau  \beta^{\hal} (N^{\frac1\d}\ell)^{\hal}  \Fluct_{\mub}(\xi)\)\) 
  &  =   \frac{ \tau^2 \ell^{2-\d}  }{2\cd}
  \(  \ell^{\d-2}\int_{\R^\d}|\nab \xi_0|^2 +o(\ell^{\d-2}) \)  +o(1)\\
  &  =   \frac{ \tau^2  }{2\cd} \int_{\R^\d}  |\nab \xi_0|^2  +o(1),\end{align}
and we conclude that Theorem \ref{thdim1} holds as well.
   
   The proof for the low temperature case Theorems \ref{thlowt2} and \ref{thlowt1d} is similar: we choose instead \eqref{choixt2} which is equivalent to taking  $\tau= s \beta^{-\hal}$ and obtain instead
  \begin{multline*}
  \log \Esp_{\PNbeta}\( \exp\( -  s  (N^{\frac1\d}\ell)^{1-\frac\d2}  \Fluct_{\mub}(\xi)\)\) 
  =   \frac{ s^2 \beta^{-1}  }{2\cd} \int_{\R^\d}  |\nab \xi_0|^2\\  
  -\frac{s  }{ 4\cd}  
\(  \int_{\R^\d} \Delta \xi \log \mub\) \indic_{\d=2}   +  \frac{s (N^{\frac1\d}\ell)^{-\hal} }{\cd}  
\(  \int_{\R} \Delta \xi_0(y) \frac{\mf'(\beta \mub(\bar x+\ell y)^{-1} )}{\mub(\bar x +\ell y)} dy\)
\indic_{\d=1}  +o(1). \end{multline*}
\end{proof}

\begin{rem}
The strategy of proof could work as well in higher dimensions, but  encounters two obstacles. The first is that the proof
requires to know a $C^2$ bound on $\mf$. This can be hypothetized away, as a sort of no phase-transition assumption, since phase transitions correspond to loss of regularity of the free energy and $f_\d$ is the thermodynamic limit, i.e.~free energy per unit volume, also defined as the pressure.  
  The second is that the error rate $\mathcal R$ needs to be quantitative and tend to $0$ at a certain rate in order to compensate for the $(N^{\frac1\d}\ell)^{\d-2}$ factor that appears in the 
 third error. 
 Provided we know such a rate, and assume the regularity 
 of $f_\d$ we thus can obtain by the same proof a CLT in higher dimension. The rate obtained in Proposition~\ref{coro24}, which we do not believe to be optimal,  is however not sufficient. \end{rem}
   
   \begin{rem} We note that the local laws are needed only to obtain the CLT at mesoscopic scales $\ell \ll 1$.\end{rem}

\subsection{Approach via the usual equilibrum measure} \label{seceqmeasure}
Working with the usual equilibrium measure  avoids having to analyze precisely the thermal equilibrium measure and thus having to use  the results of Theorem \ref{th1as}, in particular the $C^k$ bounds for the thermal equilibrium measure  \eqref{boundsmub}. The downside is that the results are less precise when $\beta $ gets small, and do not allow to go down to the threshhold scaling $\theta=\theta_0$, i.e.~$\beta = N^{\frac\s\d-1}\theta_0$ for some fixed $\theta_0$.

When working with the equilibrium measure, we use instead of $\K_{N,\beta}$ and $\Q_{N,\beta}$ the quantities $\tilde \K_{N,\beta}(\mu, \zeta)$ and $\tilde \Q_{N,\beta}(\mu, \zeta)$ in \eqref{deftildeK}, \eqref{deftildeQ}. One may check that the analysis of Chapters~\ref{chap:screening} and  \ref{chaploiloc}, i.e.~the screening, almost additivity and local laws, holds as well for fixed $\beta$ when using $\Q_{N,\beta}(\mu,\zeta)$. It was first done this way in \cite{lebles}.

One can also argue that  in view of \eqref{splitzk} and \eqref{rewritegibbs33}, we have 
$\tilde \Q_{N,\beta}(\meseq, \zeta)=\Q_{N,\beta}(\mub)=\PNbeta$, hence in view of Theorem \ref{th3}, the local laws hold for $\tilde \Q_{N,\beta}(\meseq, \zeta)$, under the sole assumption that $\mub\ge m>0$, which is a much milder fact to check than \eqref{boundsmub} -- a local uniform convergence of $\mub $ to $\meseq$  in $\Sigma$ as $\theta \gg 1$ suffices.

Let us now continue explaining the approach under this assumption that local laws hold for $\PNbeta$ hence $\tilde \Q_{N,\beta}(\meseq, \zeta)$.
We assume for simplicity that the test function $\xi $ is supported in  $\Sigma$. When working at mesoscales, we assume that $Q_\ell\subset \hat\Sigma$ (as in \eqref{defocs}) contains a $2N^{-1/\d}$-neighborhood of $\supp\, \xi$.

In the case where one works with $\meseq$,  in view of \eqref{rewritegibbs} and with obvious notation we have, letting still $V_t=V+t\xi$,
\begin{multline} \label{laplace0x}
\Esp_{\PNbeta} \( e^{-\beta t N^{1-\frac\s\d}  \sum_{i=1}^N \xi(x_i)  }    \)
=    \frac{Z_{N,\beta}(V_t)}{\ZNbeta(V)}\\
 = \exp\( - \beta N^{2-\frac\s\d} \(\mathcal{E}^{V_t} (\mu_{V_t}) - \mathcal{E}^V (\mu_V) \)  \)\frac{\tilde \K_{N,\beta}(\mu_{V_t},\zeta_{V_t})}{\tilde \K_{N,\beta}(\mu_{V}, \zeta_{V} )}.
\end{multline}

Because $\xi$ is supported in $\Sigma$, by Theorem \ref{theoFrostman} and the unique characterization of the equilibrium measure, one may check that the perturbed equilibrium measure is simply equal to 
$$\mu_{V_t}= \meseq+ \frac{t}{\cd} \Delta \xi,$$and that $\zeta_{V_t}=\zeta_{V}$.
This is the reason why the interior case is easier to treat than the case where the support of $\xi$ overlaps $\Sigma^c$, then the support of the equilibrium measure changes (see \cite{serser}).  Again we refer to \cite{ls2} for the treatment of that more general situation.  

We are in a situation where $\zeta$ is independent of $t$.
 Moreover,  defining $$\mu_0= \meseq, \quad \mu_s= \mu_0+  \frac{s}{\cd}\Delta \xi, \quad \psi_s= - \frac{\nab \xi}{\cd \mu_s},$$ we have by Lemma \ref{1221} that
$$\mu_s=\Phi_s \# \mu_0,$$  with $\Phi_s$  as in \eqref{defflow}.

Moreover, since for $|s|\le |t|$, we have $\mu_s = \mu_{V_s}$, the measure $\tilde \Q_{N,\beta}(\mu_s, \zeta)$ is equal to $\PNbeta$ for the potential $V_s$, and by the assumption made above, it also satisfies the local laws.

To take limits in \eqref{laplace0x}, let us now examine the ratio of the reduced partition functions via the variant of the transport calculus in Lemma \ref{propcalctrans2}.

\begin{lem} Let $\mu_s, \psi_s$ and $\xi$ be as above. We have   
\begin{multline}\label{premdevlogk0x}
\log \frac{\tilde \K_{N,\beta}(\mu_t, \zeta)}{ \tilde \K_{N,\beta}(\mu_0,\zeta)} =-
    N \(\int_{\R^\d}\mu_t \log \mu_t - \int_{\R^\d} \mu_0\log \mu_0\) \\+
\int_0^t   \Esp_{\tilde\Q_{N,\beta} (\mu_s,\zeta) }\(  - \beta N^{-\frac\s\d} \Ani_1(\XN, \mu_s, \psi_s)  + \Fluct_{\mu_s}\(  \div  \psi_s \)  \)\, ds\end{multline}
    and 
    \begin{multline}
    \label{deuxdevlogk0x}
    \log \frac{\tilde\K_{N,\beta}(\mu_t, \zeta)}{ \tilde\K_{N,\beta}(\mu_0,\zeta)} =-
    N \(\int_{\R^\d}\mu_t \log \mu_t - \int_{\R^\d} \mu_0\log \mu_0\) \\+
    \Esp_{\tilde\Q_{N,\beta}(\mu_0,\zeta_0)} \Bigg(  \exp\Big(-\beta N^{-\frac\s\d}\( 
  \Ani_1 \(\XN, \mu_0, \psi_0 \) + \int_0^t(t-s) \Ani_2 (\Phi_s(\XN) , \mu_s,  \psi_s )\)
\\ +\int_0^t \Fluct_{\mu_0} ((\div \psi_s) \circ \Phi_s) \, ds  \Big) \Bigg) .
    \end{multline}
    \end{lem}
\begin{proof}
Since $ \mu_s=\Phi_s \# \mu_0$, integrating \eqref{derivlogkteqx}
we obtain
$$
\log \frac{\tilde\K_{N,\beta}(\mu_t, \zeta)}{\tilde \K_{N,\beta}(\mu_0,\zeta)}
 = \int_0^t \Esp_{\tilde\Q_{(\mu_s,\zeta)} } \( - \beta N^{-\frac\s\d} \Ani_1(\XN, \mu_s, \psi_s) + \sum_{i=1}^N \div  \psi_s(x_i)  \) \, ds. $$
We may rewrite 
$$ \sum_{i=1}^N \div  \psi_s(x_i) =N \int_{\R^\d} \div   \psi_s\,  d\mu_s + \Fluct_{\mu_s} ( \div \psi_s).$$
On the other hand, we observe that, using integration by parts, 
\begin{multline}\label{deriventropiex}
\frac{d}{ds}\int_{\R^\d} \mu_s \log \mu_s =
\int_{\R^\d}\log \mu_s \frac{d}{ds}\mu_s =-
\int_{\R^\d}\div (\psi_s  \mu_s)   \log \mu_s =  \int_{\R^\d} \psi_s \mu_s \cdot \nab  \log \mu_s \\
=\int_{\R^\d}  \psi_s \cdot \nab \mu_s  = - \int_{\R^\d}\div \psi_s \, d\mu_s .
\end{multline}
  Assembling these relations, we find that \eqref{premdevlogk0x} holds.

Let us now write  \eqref{formulx}  with $\mu_0$ in place of $\mu_t $ and $\Phi_t$ in place of $\phi_h$,  recalling that $\Phi_t\# \mu_0= \mu_t  $, using that $\Phi_t$ is supported where $\zeta_{t}$ and $\zeta_0$ vanish, and  inserting   \eqref{rewriteFdiff}, we find
\begin{align*}
&\frac{\tilde\K_{N,\beta}(\mu_{t},\zeta)}{\tilde\K_{N,\beta}(\mu_0,\zeta)} =
\Esp_{\tilde\Q_{N,\beta}(\mu_0,\zeta_0)} \Bigg(  \exp\Big(-\beta N^{-\frac\s\d}\( 
 t \Ani_1 \(\XN, \mu_0, \psi_0 \) + \int_0^t (t-s)\Ani_2 (\Phi_s(\XN) , \mu_s,  \psi_s )\, ds\)
\\ &\qquad \qquad+\sum_{i=1}^N \log \det D\Phi_t(x_i)  \Big) \Bigg) .
\end{align*}
Using \eqref{DPhit} we may also compute that $\frac{d}{ds} \log \det D\Phi_s= (\div \psi_s) \circ \Phi_s$, hence 
\begin{align*}
\sum_{i=1}^N \log \det D\Phi_t(x_i)& =\int_0^t \sum_{i=1}^N \div \psi_s(\Phi_s(x_i)) ds \\
& = \int_0^t  \int_{\R^\d} N \div \psi_s(\Phi_s) d\mu_0 + \Fluct_{\mu_0} (\div \psi_s\circ \Phi_s) ds \\&
= \int_0^t \int_{\R^\d} N \div \psi_s \, d\mu_s  + \Fluct_{\mu_0} (\div \psi_s\circ \Phi_s)  ds\\ &
= -N \(\int_{\R^\d} \mu_{t}\log \mu_{t}- \int_{\R^\d} \mu_0\log \mu_0\)  + \int_0^t \Fluct_{\mu_0} (\div \psi_s\circ \Phi_s)  ds
\end{align*} where we used \eqref{deriventropiex}.
 We conclude that \eqref{deuxdevlogk0x} holds.

\end{proof}

At this point, inserting this relation into \eqref{laplace0x} we are in a very similar situation as  that of \eqref{laplace1}. First, we need to  estimate 
  $\mathcal E^{V_t}(\mu_t)- \mathcal E^V (\mu_V)$ 
  and  to
bound $\Fluct_{\meseq}(\div \psi_0)$. 
The former is done via  the following explicit estimate, proved in Section \ref{lemaux}.
 \begin{lem}\label{lemtexi}
 Assume that $\xi$ is $C^2$ and  supported in $\Sigma = \{\zeta=0\}$ (see notation in \eqref{defzeta}), then 
 \be \label{termesexplicites}
 \mathcal E^{V_t} (\mu_{V_t}) - \mathcal E^V (\mu_V)= t \int_{\R^\d} \xi \, d\meseq- \frac{t^2}{2\cd} \int_{\R^\d} |\nab \xi|^2.\ee\end{lem}
  
  For the latter,  we have the following, obtained by the rough fluctuations bounds of 
 \eqref{coulombfluct} combined with \eqref{bornehnr} (see Section \ref{lemaux} for the proof).
 Using that
 \begin{align}
 \label{psic1}
 & |\psi_t|_{C^1} \le C\( (|\meseq|_{C^1} + |t| |\xi|_{C^3} )|\xi|_{C^1} +|\xi|_{C^2}\)\\
 \label{psic2}
  & |\psi_t|_{C^2} \le C \( (|\meseq|_{C^2} + |t| |\xi|_{C^4} )|\xi|_{C^1} + (|\meseq|_{C^1} +| t| |\xi|_{C^3} ) |\xi|_{C^2} +|\xi|_{C^3}\),\end{align} we obtain 
 \begin{lem}\label{lem1031}Under the same assumptions,
 \begin{multline} \label{laplacenew}
 \left|\int_0^t \Esp_{\Q_{(\mu_s,\zeta)}} (\Fluct_{\mu_s}(\div \psi_s)) \right|\\
  \le 
 C|t|
  \sqrt{\chi(\beta)}  N^{\frac12+\frac\s{2\d}} \ell^\d  
  \( (|\meseq|_{C^2} +| t| |\xi|_{C^4} )|\xi|_{C^1} + (|\meseq|_{C^1} + |t||\xi|_{C^3} ) |\xi|_{C^2} +|\xi|_{C^3}\)
  .   \end{multline} \end{lem}
  
Here we need that $\psi_s$ is $C^2$, which is implied by the conditions $\xi \in C^3$, $\meseq \in C^2$. In view of \eqref{densmu0} the regularity on $V$ needed is $C^4$.
\begin{rem}
We have not tried to optimize over the regularity of $\psi$ here, the estimates of \eqref{coulombfluct} are in fact valid with less regularity on $\div \psi$.
\end{rem}

To estimate the ratio in \eqref{laplace0x}, we may first use \eqref{premdevlogk0x}.  We directly estimate (as in \eqref{B2d}) that 
\be\int_{\R^\d} \mu_{V_t}\log \mu_{V_t}- \int_{\R^\d} \mu_V \log \meseq = 
\frac{t}{\cd}\int_{\R^\d}  \Delta \xi (\log \meseq ) +O( t^2 |\xi|_{C^2}\ell^\d).\ee
One may  then obtain a bound on $\int_0^t \Esp_{\Q_{(\mu_s,\zeta)} }\(  - \beta N^{-\frac\s\d} \Ani_1(\XN, \mu_s, \psi_s) \) $ as in \eqref{difflk}, but with $\mub$ replaced by $\mu_V$, by simply inserting the commutator estimate of Theorem \ref{thm:FI} combined with the local laws. The only regularity needed to apply the commutator estimate  is that $\psi_s $ be Lipschitz, which is implied by  $\xi \in C^2$ and $\meseq \in C^1$.

Using \eqref{psic1} and \eqref{psic2}, 
 we then obtain the following control on fluctuations.

\begin{theo}[Control of fluctuations, Coulomb case, equilibrium measure]Let $\d\ge 1$. Assume   \eqref{A1}--\eqref{A5}. 
Assume that a cube $Q_\ell$,  with $\ell$ satisfying \eqref{ass1},  in which local laws hold contains a $2N^{-1/\d}$-neighborhood of $ \supp\,\xi$, or take $\ell=1$. Assume $\xi \in C^3(Q_\ell)$ and $\meseq \in C^2(Q_\ell)$.
Then, for any $t$ such that $|t||\xi|_{C^2}$ is small enough, we have 
\begin{multline}
\left|\log \Esp_{\PNbeta} \( e^{ \beta t N^{1 -\frac{\s}{\d}} \Fluct_{\meseq}( \xi) }\) 
\right|\le
 \\ Ct
  \sqrt{\chi(\beta)}  N^{\frac12+\frac\s{2\d}} \ell^\d  
  \( (|\meseq|_{C^2} + t |\xi|_{C^4} )|\xi|_{C^1} + (|\meseq|_{C^1} + |t| |\xi|_{C^3} ) |\xi|_{C^2} +|\xi|_{C^3}\)
\\+ C|t| \beta \chi(\beta) N\ell^\d \( (|\meseq|_{C^1} + |t| |\xi|_{C^3} )|\xi|_{C^1} +|\xi|_{C^2}\) 
+C( t^2 |\xi|_{C^2}N\ell^\d  + t^2  \beta N^{2-\frac{\s}{\d}} \ell^\d |\xi|_{C^1}^2 ),
\end{multline}
where the constants depend on $\d, \s, m, \|\meseq\|_{L^\infty} $, but not on $\xi$, $\beta$, $N$  or $t$.
\end{theo}
These estimates are as good as those of Theorem \ref{th10.1} when $\beta$ is fixed, but not when $\beta \to 0$ as $N \to \infty.$ One can then derive similar corollaries.

To prove central limit theorems, we instead insert \eqref{deuxdevlogk0x} into \eqref{laplace0x} and still use \eqref{termesexplicites} and \eqref{laplacenew}. We use again Proposition \ref{prop:comparaison2} to bound the $\Ani_2$ terms. The evaluation of   $\Esp(\exp(-\beta N^{-\frac\s\d} \Ani_1(\XN, \mu_0, \psi_0))$ we use should be done as in Proposition \ref{propmomani} except working with $\tilde \Q_{N,\beta}(\mu_0, \zeta)$ instead of $\Q_{N,\beta}(\mu_0)$. This way, choosing $t=\tau \beta^{-1} N^{-1} $, one can arrive, as in \cite{ls2}, to a two-dimensional CLT in the form 
$$ \log \Esp_{\PNbeta}\( e^{-\tau  \Fluct_{\meseq} (\xi)}\) \to -\tau \mathrm{mean}(\xi) +\frac{\tau^2}{2} \mathrm{var}(\xi)$$
with 
$$\mathrm{mean}(\xi)= \begin{cases} (\frac1\beta-\frac14  ) \frac{1}{\cd} \displaystyle\int_{\R^\d}( \log \meseq )\Delta \xi_0,&\  \ell=1\\
0 & \ell \to 0\end{cases}
 \qquad \mathrm{var}(\xi) = \frac1{\beta   \cd}\int_{\R^\d} |\nab \xi_0|^2, $$
 valid for fixed $\beta$, any $\ell \gg N^{-1/\d}$,  and requiring only that $\meseq\in C^2$ (hence $V \in C^4$ suffices)  and $\xi\in C^4$.
 One can check that this result is consistent with Theorem \ref{th10.1} in view the difference between $\meseq$ and $\mub$ given by \eqref{corrections}.

\subsection{Proof of auxiliary lemmas}\label{lemaux}

\begin{proof}[Proof of Lemma \ref{lemepsi}]
With the quantities introduced, proceeding as in the proof of the splitting formula \eqref{split0} except splitting with respect to the probability $\nut$ instead of $\mut$ and using \eqref{defept}, we find 
\begin{multline}
\mathcal H_N^{V_t} (X_N)=  N^2 \mathcal E(\nut) + N \int_{\R^\d} (\g*\nut + V_t ) d\( \sum_{i=1}^N \delta_{x_i} - N \nut\) + \F_N(X_N, \nut)
\\
=  N^2 \mathcal E(\nut) + N \int_{\R^\d} (- \frac{1}{\theta} \log \nut + \ep_t) d  \( \sum_{i=1}^N \delta_{x_i} - N \nut\) + \F_N(X_N, \nut)\\
= N^2 \mathcal E_\theta(\nut) + \F_N(X_N, \nut) - \frac{N}{\theta} \sum_{i=1}^N \log \nut (x_i)  + N \int_{\R^\d} \ep_t \,d \( \sum_{i=1}^N \delta_{x_i} - N \nut\).
\end{multline}
Inserting into the definition of the Gibbs measure \eqref{gibbs}, we obtain \eqref{laplace1}.
We now prove that $\ep_t$, which measures how far $\nut$ is from solving the same equation as $\mub^{V_t} $ (see \eqref{eqhmub}), is a small function as $\theta \gg 1$.
Notice that $\ep_t$ is supported in $\supp  \, \xi$ and that 
\be \label{hnutmut}
\g* \( \nu_\theta^t- \mut\)  = - t \xi.\ee 
Since $\g*\mut+ V + \frac{1}{\theta} \log \mut=c_\theta$ by \eqref{eqmb} and by definition \eqref{defnut}, we deduce that 
\begin{equation}\label{hnut}
\ep_t= \g* \nu_\theta^t+ V+t\xi+ \frac{1}{\theta}\log \nu_\theta^t-c_\theta= \frac1\theta\log \frac{\nu_\theta^t}{\mut} =
\frac{1}{\theta} \log \( 1+\frac{t}{\cd} \frac{\Delta \xi}{\mut} \).
  \end{equation}
With direct computations and \eqref{boundsmub}, it follows that \eqref{estept0} and \eqref{estept} hold.
\end{proof}


\begin{proof}[Proof of Lemma \ref{lemleadord}] We have  
\begin{multline*}
\mathcal{E}_\theta(\nut)- \mathcal{E}_\theta(\mut)\\=
\(\hal \iint_{\R^\d\times \R^\d} \g(x-y)d\nut(x)d \nut(y)- \hal \iint_{\R^\d\times \R^\d}  \g(x-y) d\mut(x) d\mut(y)+\int_{\R^\d} V_td\nut- \int_{\R^\d} V d\mut  \)
\\+ \frac1\theta \(\int_{\R^\d}  \nut \log \nut -\int_{\R^\d} \mut \log \mut\)
\\
= \hal \iint_{\R^\d\times \R^\d}  \g(x-y) d(\nut-\mut)(x)d(\nut-\mut)(y) + \iint_{\R^\d\times \R^\d}  \g(x-y) d(\nut-\mut)(x) d\mut(y) \\+ \int_{\R^\d} V d(\nut-\mut) + t \int_{\R^\d} \xi d\mut+ t \int_{\R^\d} \xi d(\nut-\mut)  + \frac1\theta \(\int_{\R^\d}  \nut \log \nut -\int_{\R^\d} \mut \log \mut\) \\
= \hal \iint_{\R^\d\times \R^\d}  \g(x-y) d(\nut-\mut)(x)d(\nut-\mut)(y) +\int_{\R^\d} (\g*{\mut}+ V+ \frac{1}{\theta} \log \mut) d(\nut-\mut) \\+ t\int_{\R^\d} \xi d\mut+ t\int_{\R^\d} \xi d(\nut-\mut) + \frac{1}{\theta} \int_{\R^\d} \nut(\log \nut-\log \mut) .\end{multline*}
The second term of the right-hand side  vanishes by characterization of $\mut$ in \eqref{eqhmub}, and we are left with 
\begin{multline*}
\mathcal{E}_\theta(\nut)- \mathcal{E}_\theta(\mut)- t\int_{\R^\d} \xi d\mut\\=
\frac{1}{2\cd} \int_{\R^\d} |\nab (\g*({\nut}-{\mut}))|^2 + t\int_{\R^\d} \xi d(\nut-\mut) + \frac{1}{2\theta} \int_{\R^\d} \mut \( \frac{\nut}{\mut}-1\)^2 
+  O\(\frac{1}{\theta} \int_{\R^\d} \(  \frac{\nut}{\mut}-1\)^3 \mut\) 
\end{multline*}
where we Taylor expanded the logarithm. 
We then use \eqref{defnut} to see that 
$$  |\nab (\g*({\nut}-{\mut}))|^2= t^2|\nab \xi|^2 $$
and $$\frac{\nu_\theta^t }{\mut}=1+ t \frac{ \Delta \xi}{\cd\mut}.$$
We thus find \eqref{513}. Alternatively we can Taylor expand the log only to first order and get instead  a bound by $$
 C t^2  \( \int_{\R^\d} |   \nab \xi |^2+
 \frac{1}{\theta}\int_{\R^\d} \mut \left|\frac{ \Delta \xi}{\mut} \right|^2\)$$ from which we deduce \eqref{version2}. \end{proof}

\begin{proof}[Proof of Lemma \ref{lemthirditem}]
By Theorem \ref{th3}, local laws and concentration hold for $\Q_{N,\beta}(\nu_\theta^t)$ in $\hat\Sigma$ where $\nut$ is bounded below provided $t|\xi|_{C^2}$ is small enough.  A rescaling of \eqref{loclawphi} yields that  for any $\varphi$ such that $\|\nab \varphi\|_{L^\infty} \le N^{\frac1\d}$,
$$\left|\log \Esp_{\Q_{N,\beta} (\nu_\theta^t)} \( \exp \frac\beta{CN\ell^\d}\( \Fluct_{\nut} ( \varphi ) \)^2 \)\right|\le C 
\beta \chi(\beta) N^{-\frac\s\d} \ell^{\d} \|\nab \varphi\|_{L^\infty}^2$$
We may then apply this to $\varphi=\sqrt{C}\ell^{\frac\d2} N^{\frac{1}{\d}+\hal} \sqrt{\lambda} \ep_t$. Thus,  for any $\lambda$ such that $\sqrt{\lambda C} \ell^{\frac\d2}N^{\hal} |\ep_t|_{C^1} \le 1$ (which ensures that $\|\nabla \varphi\|_{L^\infty} \le N^{1/\d}$),  using  also Young's inequality to write  
$$\theta \int \ep_t \, d\( \sum_{i=1}^N \delta_{x_i} - N \nut\)\le \theta \lambda \( \Fluct_{\nut}  (  \ep_t)\)^2 + \frac{\theta}{4 \lambda},$$
we have
$$\log \Esp_{\Q_{N,\beta}(\nut)} \( \exp\(\theta  \Fluct_{\nut} ( \ep_t )\)\)
\le C \lambda \beta  \chi(\beta) N^{1+\frac{2-\s}{\d}}\ell^{2\d}| \ep_t|_{C^1}^2 + \frac{\theta}{ 4\lambda}$$
and optimizing over $\lambda\le |\ep_t|_{C^1}^{-2}(N\ell^\d)^{-1}$ we find \be\label{laplace10}\left| \log  \Esp_{\Q_{N,\beta}(\nut)} \( \exp \(-\theta \Fluct_{\nut}(\ep_t)
 \)\)
\right|
\le
C \sqrt{\chi(\beta)} \beta N^{1+\frac1\d} \ell^\d|\ep_t|_{C^1}+C \theta N\ell^\d  |\ep_t|_{C^1}^2\ee
hence the result \eqref{laplace10a}.

We next turn to proving \eqref{laplace2a}. This time we bound 
$$\left| \Fluct_{\nut} (\ep_t )\right|\le \|\ep_t\|_{L^\infty} ( \#I_{Q_\ell} + N \ell^\d)
$$where $\#I_{Q_\ell}$ denotes the number of points in each configuration that fall in the set $Q_\ell$ containing  the support of $\xi$.
We can in turn bound from above 
$$\#I_{Q_\ell} \le N\int_{Q_\ell} d \nut + D(x,C\ell)$$
where $B(x, C\ell)$ is a ball that contains $Q_\ell$ and $D(x, \ell) =\int_{B(x,C \ell)}  \sum_{i=1}^N \delta_{x_i} - N d\mu$.
Arguing as before, we write 
$$\theta \|\ep_t\|_{L^\infty} D(x,C \ell)\le  \|\ep_t\|_{L^\infty}  \(D^2(x, C\ell) \beta N^{\frac2\d-1} \ell^{2-\d}  \lambda +  \frac{\theta  N\ell^{\d-2} }{4\lambda}\)
$$
and thus using a rescaling of \eqref{loclawpoints00} and \eqref{deftheta}, we find, 
$$\log \Esp_{\Q_{N,\beta}(\nut)}\(\exp\(  \theta \|\ep_t\|_{L^\infty} D(x, C\ell) \) \) \le  C \|\ep_t\|_{L^\infty} \lambda \beta \chi(\beta) N\ell^\d+ \frac{\beta \|\ep_t\|_{L^\infty} N^{1+\frac2\d} \ell^{\d-2}}{4\lambda}.
$$Optimizing over $\lambda \le \|\ep_t\|_{L^\infty}^{-1}$ we find 
$$\log \Esp_{\Q_{N,\beta}(\nut)}\(\exp\( \theta \|\ep_t\|_{L^\infty} D(x, C\ell) \) \) \le  C \|\ep_t\|_{L^\infty} \sqrt{\chi(\beta)} \beta N^{1+\frac1\d} \ell^{\d-1} +  C\|\ep_t\|_{L^\infty}^2 \beta N^{1+\frac2\d} \ell^{\d-2}.$$
After observing that 
$\sqrt{\chi(\beta)} N^{-\frac1\d}\ell^{-1}\le 1$ by  \eqref{ass1} and \eqref{defrhobeta}. It follows that \be\label{laplace2}\left| \log  \Esp_{\Q_{N,\beta}(\nut)} \( \exp \(-\theta \Fluct_{\nut}( \ep_t)\) \)\right|
\le
C \|\ep_t\|_{L^\infty} \theta  N \ell^\d+  C\|\ep_t\|_{L^\infty}^2 \theta N \ell^{\d-2},
\ee
hence the result.
\end{proof}

\begin{proof}[Proof of Lemma \ref{lemtexi}]
 By definition  \eqref{definitionI},  we compute that
 \begin{multline*}\mathcal E^{V_t} (\mu_{V_t}) - \mathcal E^V (\mu_V)=
 \hal\iint_{\R^\d\times \R^\d} \g(x-y) d\mu_{V_t} (x) d\mu_{V_t} (y)\\ -\hal \iint \g(x-y) d\mu(x) d\mu(y) + \int_{\R^\d} Vd(\mu_t-\mu)+ t \int_{\R^\d}\xi d\mu_t.\end{multline*}
 Inserting that $\mu_{V_t}= \mu_V + \frac{t}{\cd} \Delta \xi $  and that $\g * \Delta \xi =  - \cd\xi$, we obtain 
  \begin{align*} \mathcal E^{V_t} (\mu_{V_t}) - \mathcal E^V (\mu_V)
&  =\frac{t}{\cd} 
 \iint_{\R^\d\times \R^\d} \g(x-y) d\mu_{V} (x) d\Delta \xi  (y)  +\frac{t}{\cd} \int_{\R^\d} V \Delta \xi+ t \int_{\R^\d} \xi d\mu_V \\
 & + \hal \frac{t^2 }{\cd^2}\iint_{\R^\d\times \R^\d} \g(x-y) \Delta \xi(x) \Delta \xi(y)+ \frac{t^2}{\cd}   \int_{\R^\d} \xi \Delta \xi\\ &
 = \frac{t}{\cd } \int_{\R^\d} (h^{\mu_V}+ V) \Delta \xi - \frac{t^2}{2\cd} \int_{\R^\d} \xi \Delta \xi + \frac{t^2}{\cd} \int_{\R^\d} \xi \Delta \xi + t \int_{\R^\d} \xi d\mu_V
 \\ &
 = \frac{t}{\cd} \int_{\R^\d} (\zeta +c)  \Delta \xi + \frac{t^2}{2\cd} \int_{\R^\d} |\nab \xi|^2 + t \int_{\R^\d} \xi d\mu_V.
 \end{align*}
 Using that $\xi$ is supported in the set $\{\zeta = 0\}$, we conclude that 
 \eqref{termesexplicites} holds.\end{proof}

  \begin{proof} [Proof of Lemma \ref{lem1031}]
Since  local laws and concentration hold for $\Q_{N,\beta}(\mu_s,\zeta)$ in $\hat\Sigma$ where $\mu_s$ is bounded below,  a rescaling of \eqref{loclawphi} yields, after application of Jensen's inequality, that  for any $\varphi$,
$$\left| \Esp_{\Q_{N,\beta} (\mu_s,\zeta)} \(  (\Fluct_{\mu_s} (\varphi)) ^2 \)\right|\le C 
\chi(\beta) N\ell^\d N^{\frac\s\d} \ell^{\d} \|\nab \varphi\|_{L^\infty}^2$$
We may then apply this to $\varphi=\div \psi_s$. Thus,  for any $\lambda$, since we have 
$$| \Fluct_{\mu_s}( \div \psi_s)  |\le  \lambda \( \Fluct_{\mu_s} (\div \psi_s) \)^2  + \frac{1}{4 \lambda},$$
we also have
$$\left| \Esp_{\Q_{N,\beta}(\mu_s,\zeta)} \(  \Fluct_{\mu_s} (  \div \psi_s )\)
\right|
\le C \lambda  \chi(\beta) N^{1+\frac{\s}{\d}}\ell^{2\d}| \div \psi_s|_{C^1}^2 + \frac{1}{ 4\lambda}$$
and optimizing over $\lambda$ we find \be\label{laplace10}\left|   \Esp_{\Q_{N,\beta}(\mu_s,\zeta)} \( \Fluct_{\mu_s}(\div \psi_s) \)
\right|
\le
C \sqrt{\chi(\beta)}  N^{\frac12+\frac\s{2\d}} \ell^\d|\div \psi_s|_{C^1}\ee
hence the result.
\end{proof}

\section{Nonsmooth test-functions}\label{sec:leble}
Understanding fluctuations for less smooth test-functions, in particular for indicator functions of sets, which correspond to fluctuations of number of points or charge fluctuations, is generally harder. Such fluctuations are expected to be larger than those for regular test functions. It is generally not known what the threshold of regularity to have, say, boundedness of fluctuations, is. 
However, in the one-dimensional logarithmic case this is proven: the variance of  of the number of points in an interval is logarithmic  in the size of the interval \cite{najnudelvirag}.

It was conjectured for a long time in the statistical mechanics literature \cite{martin,martinyalcin,lebo,LWL,jlm,torquato} that the two-component Coulomb gas is ``hyperuniform" (in the terminology of \cite{torquato} for all $\beta>0$, i.e.~the charge fluctuations in a ball has  much smaller variance than that of a Poisson point process, i.e.~than the area (after zooming).
This was proven in \cite{leblehyper} in the following.
\index{hyperuniformity}

\begin{theo}[Hyperuniformity of the 2D Coulomb gas \cite{leblehyper}]
\label{ththomas}
Let $\d=2, \s=0$, and $\beta>0$.
Let $\{x_N\}_N$ be a sequence of points in $\Sigma $ with $\dist(x_N, \partial \Sigma) \ge c>0$  and $R=R_N$ such that $R_N \gg N^{-1/2}$ as $N \to \infty$.
Then, under $\PNbeta$ the variance of the number of points in $B(x_N,R)$ is $o((N^{1/2}R)^2)$.
\end{theo}
In the determinantal case $\beta=2$ this was already known at the level of the Ginibre point process \cite{shirai,osadashirai,fenzllambert} and at the level of the Coulomb gas \cite{charliergap,ameurcharlier}, with the stronger bound $O(N^{1/2} R)$ corresponding to the perimeter, which is conjectured to be the variance order for all $\beta>0$. In fact the more precise 
$$\Var(\#\{\XN \cap \Omega \})= C N^{1/2}|\partial \Omega|$$ is conjectured, see \cite{martinyalcin}. 
The proof of Theorem \ref{ththomas} follows the overall road map introduced in  \cite{nsv}  in the context of zeroes of Gaussian Analytic Functions. Its implementation
relies on all the tools presented so far (electric formulation, transport, screening) but also new ideas such as isotropic averaging and the analysis of ``subsystems".

In dimension 2, 
precise conjectures, the so-called Jancovici-Lebowitz-Manificat laws, are also given for the deviations of charge discrepancies in balls in \cite{jlm}. They assert that, asymptotically, and in blown-up scale,
$$\log \PNbeta\(  D(x,R) \ge R^\alpha\) \simeq - R^{\varphi(\alpha)}+o(1) \quad \text{as}\  R \to \infty$$
where $D(x,R)$ is the discrepancy in $B(x,R)$ and
$$\varphi(\alpha)= \begin{cases} 2\alpha-1 & \text{if}\ \hal<\alpha\le 1,\\
3\alpha-2& \text{if} \ 1\le \alpha\le 2,\\
2\alpha & \text{if} \ \alpha\ge 2.\end{cases}$$
We saw in Section~\ref{sec:isotropic} that part of these conjectures were proved in \cite{thoma}. It was  also proven very recently  in \cite{nishryyakir} that they hold for the hierarchical Coulomb gas studied in \cite{chatterjee,gangulysarkar} and alluded to in Section \ref{sec1013}.

\part{Microscopic behavior and local limits}

\chapter{The jellium renormalized energy}
\label{chap:renormalized}
In this last part, we return to the general Riesz case \eqref{riesz}. 

We work in blow-up coordinates and see how we can derive an infinite volume version of the next order energy $\F$, called ``renormalized jellium energy.''\index{jellium energy}
This function was first introduced in \cite{ssgl} in the context of Ginzburg-Landau vortices, which corresponds to the two-dimensional Coulomb case. The definition and the name were inspired by the work of  \cite{bbh} on Ginzburg-Landau vortices. In \cite{rs},  a definition based on smearing the charges  was provided for the general Coulomb case, and in \cite{PetSer} it was generalized to the Riesz case $ \d-2\le \s<\d$  based on the truncation procedure. We now present a new and simpler definition, relying on the idea of letting the truncation depend on the point as in Chapter \ref{chap:nextorder}. This definition avoids the need to introduce an extra parameter $\eta$ and take a double limit as in \cite{rs,PetSer}.

\section{Motivation}
Let us return to the electric formulation for $\F$, via the electric potential. In the blown-up setting, we   have 
from \eqref{HNp} and \eqref{44}
\be \label{dhe}- \div (\yg \nab  {h_N}) = \cds\(\sum_{i=1}^N \delta_{x_i'}- \mu' \drd\)\quad \text{in} \ \R^{\d+\k}
\ee
where $\mu' $ is the blown-up  reference probability measure (typically either $\mu_V $ or $\mub$) around some origin $x_0$,  $\mu'(x')= \mu( x' N^{-\frac1\d}+x_0)$.
Let us assume  that $\F(\XN', \mu') \le C N$, which is true for typical configurations by \eqref{firstconcbound}. Then for most blow-up centers $x_0$, the number of points  becomes infinite  and they fill up the whole space, with  their local density remaining bounded. In such a situation $\sum_{i=1}^N\delta_{x_i}$  converges to a distribution 
$\C\in \config(\Rd)$, where 
for $A$ a Borel set of $\Rd$ we denote by $\config(A)$ the set of (possibly infinite) locally finite point configurations in $A$ or equivalently the set of non-negative, purely atomic Radon measures on $A$ giving an integer mass to singletons (see \cite{dvj2}). 
We mostly use $\C$ for denoting a point configuration in $\config(\R^\d)$ and we will write $\C$ for $\sum_{p \in \C} \delta_p$ and also $\mathcal C(U) $ for $\int_U \sum_{p\in \C} \delta_p$. Note that points could appear with multiplicity. We say that $\C$ is simple if no point of the configuration has multiplicity $>1$.
 
Taking $N\to \infty$ in \eqref{dhe} and assuming that the density $\mu$ is continuous,  we expect to obtain  a relation of the form
\be \label{dH} -\div (\yg\nab H)= \cds\(\sum_{p \in\C} \delta_p- m\drd\) \ \quad \text{in} \ \R^{\d+\k}
\ee with $m=\mu(x_0)$.
This corresponds to what is called in physics  a  {\it jellium}, that is an infinite configuration of discrete positive charges in a uniform neutralizing bath, with uniform density $m$. This is also called Uniform Electron Gas and  used as a toy model for  solid  matter  seen as a gas of electrons in a uniform background charge representing the atomic nuclei density.

The goal of this chapter is to define a Coulomb/Riesz jellium energy for such a system. 
It  is a priori not clear how to do this because of the infinite size of the system and  because of the lack of local charge neutrality of the system. Computing the sum of pair interactions between the charges in the jellium leads to a priori divergent sums. 
The definition we present instead replaces it with (renormalized variants of) the extensive quantity $\int \yg |\nab H|^2$  (see \eqref{formalcomputation} and the comments following it).

The  name \textit{renormalized} energy  reflects the fact that the integral of $ \yg |\nab H|^2 $ is infinite, and is computed in a renormalized way by first applying a truncation and then removing the appropriate divergent part $\cds \g(\eta)$, as in Chapter \ref{chap:nextorder}.

\section{Definitions and first properties}

As in Chapter \ref{chap:screening} we consider more generally the electric field  $E=\nab H$  and, forgetting that $E$ comes from  a gradient, we  wish to define the energy associated to an electric field $E$.

The reason for considering the electric field as the variable is that it will be the object on which we have compactness.
\begin{defi} \label{def1211}For any $m\ge 0$, we say that an electric field  $E$ is compatible with $(\mc{C}, m)$ if it satisfies  an equation of the form 
\be \label{eqclam}
-\div (\yg E)= \cds\(\C- m\drd\)\quad \text{in} \ \R^{\d+\k} \ee
where $\C\in \config(\R^\d)$.
We denote by $\Elec(\C, m)$ the set of such vector fields, by $\Elec_m$ the union over all configurations for fixed $m$, and $\Elec=\cup_{m>0} \Elec_m$.

The knowledge of $E$ and $m$ suffices to recover the configuration, via $$\sum_{p\in \C} \delta_p= \frac{1}{\cds}\( m\drd- \div (\yg E)\),$$
thus for any $E \in \Elec_m$, there exists a unique underlying configuration $\C$ such that $E$ is compatible with $(\C, m)$. We denote it $\conf_m(E)$.\end{defi}

Note that if $E\in \Elec(\C,m)$ then so does $E+ X$  whenever $\yg X$ is divergence-free. In other words, by formulating in terms of electric fields, we have not retained the information that $E$ was a gradient. Also, when $\gamma=0$ for instance, we can add to $E$ any   gradient of a harmonic function and still satisfy \eqref{eqclam}. In other words, when taking $N \to \infty$, we have lost the ``boundary conditions at infinity" or far field.

We now   follow the same steps as in Chapter \ref{chap:nextorder} and Chapter \ref{chap:screening} to define an energy in this infinite volume setting. The difficulty is that the energy will be computed over growing cubes, and care needs to be taken near the boundary of the cubes: in order to provide an energy that is bounded below, the definition needs to  avoid favoring accumulation of points near the boundary of the cubes.

Given a configuration $\C$ of points in $\R^\d$ and a density $m>0$, we  first define if $\s\ge 0$
\be\label{defrC}
\rr_p=\begin{cases}
\frac14\min_{q\in \C, q\neq p} \( |q-p|, m^{-1/\d}\)& \text{if $p$ is a simple point of}\ \C\ \text{or} \  \s<0\\
0 &\text{if $p$ appears with multiplicity and $\s\ge 0$}, \end{cases}\ee
 the (truncated) minimal distance to nearest neighbors in the infinite configuration.
 Given a closed cube $\carr_R$, analogously to \eqref{defrrc4} we define the minimal distance relative to the cube 
 by 
 \be \label{defrrt}
 \rrc_p=  \begin{cases}\rr_p & \text{if } p\in \carr_R, \dist(p, \pa \carr_R) \ge  2m^{-1/\d} \\
  \frac{m^{-1/\d}}4 &   \text{if } \dist(p, \pa \carr_R) \le  m^{-1/\d} \ \text{or} \ p\notin \carr_R\\
  t\rr_p+(1-t) \frac{m^{-1/\d}}{4}& \text{if } p \in \carr_R, \dist(p, \pa \carr_R)= (1+t) m^{-1/\d} , t\in [0,1].
  \end{cases}\ee 
The reason for this definition is that, as in Chapter \ref{chap:screening}, it makes the energy superadditive over cells. The definition also makes $\rrc$ a continuous function of the point configuration.

Let $\C$ be a point configuration, for $m\ge 0$  let $E$ be in $\Elec(\C,m)$. For any $\veta$, family of positive numbers  indexed by $\C$,  we define the truncation of $E$ with parameters $\veta$ as
\begin{equation} \label{defEeta1}
E_{\veta}(x) := E(x) - \sum_{p \in \C} \nabla \f_{\eta_p}(x-p),
\end{equation}
where $\f_{\eta}$ is as in \eqref{def:truncation}. In particular we define $E_{\rr}$ to be 
$ E(x) - \sum_{p \in \C} \nabla \f_{\rr_p}(x-p)$.
Note that here there are  contributions from points that lie outside $\carr_R$. 
 We
  observe that 
\be\label{divee}
-\div(\yg  E_{\rr}) = \cd\( \sum_{p\in \C} \delta_p^{(\rr_p)} - m\drd\)\quad \text{in} \ \R^{\d+\k}.\ee This procedure is exactly the same, at the level of the electric fields, as the truncation procedure described in Chapter \ref{chap:nextorder}.

\subsection{Monotonicity property and lower bound} \index{monotonicity}
For $E \in \Elec(\C, m)$ and $\carr_R$ the {\it closed} cube centered at $0$ and of sidelength $R$, we let 
\begin{multline} \label{defFcarrr}
\mathcal F^{\carr_R} (E,m):= \\
\begin{cases}\  \displaystyle \frac{1}{2\cds}\displaystyle \int_{\carr_R \times \R^\k } \yg |E_{\rrc}|^2 - \displaystyle \hal \sum_{p \in \C\cap \carr_R}  \g(\rrc_p) -   m \displaystyle\sum_{p \in  \C \cap \carr_R}  \displaystyle\int_{\R^\d} \f_{\rrc_p} ( x-p) dx    & \text{if all $\rr_p>0$ for $p\in \carr_R$}
\\  & \text{or if  $\s<0$}\\ 
+\infty & \text{otherwise}.\end{cases}
\end{multline}
Let us emphasize  that the contribution of points that lie on the boundary is counted in $\sum \g(\rrc_p)$. This is meant to make $\mathcal F^{\carr_R}$ lower semi-continuous. 

Let us next present a rewriting of Lemma \ref{lem:contrdist1}
in this context.
\begin{lem}[Monotonicity]\label{lemmonoto2}  Let $\carr_R$ be a closed cube of size $R$ and $\rrc$ be the minimal distance  of the configuration $\C$ relative to $\carr_R$ as in \eqref{defrrt}. 
For any $\veta$ such that $ \eta_p \ge \rrc_p$ for all $p$ with equality unless  $B(p, \eta_p) \subset \carr_R$,  we have 
\begin{multline}\label{reslemmono}
\hal \sum_{\substack{p\neq q\in\C \cap \carr_{R}\\ \dist(p, \pa \carr_R) \ge \eta_p  }} \( \g(p-q)- \g(\eta_p) \)_+ \\+  \frac{1}{2\cds} \int_{\carr_R\times \R^\k} \yg |E_{\veta}|^2 -\hal \sum_{p \in \C \cap \carr_R}  \g(\eta_p) - m \sum_{p \in \C\cap \carr_R}  \int_{\R^\d} \f_{\eta_p} ( x-p)   dx
 \le \mathcal F^{\carr_R} (E,m)    \end{multline}
 with equality if $\C$ is simple and all the $B(p, \eta_p)$'s are disjoint.\end{lem}

\begin{prop}For any simple configuration $\C$,  it holds that   
\be\label{bgrvi}
\sum_{p\in \C \cap \carr_R}\g(\rrc_p) \le  C_1 \( \mathcal F^{\carr_R}(E,m) + \(C_0 m^{\frac\s\d}+ (\frac{1}{\d} \log m) \indic_{\s=0} \) \C(\carr_R)\) \ee 
and
\be \label{bornehnrvi}
\int_{\carr_R\times \R^{\k}}\yg |E_{{\rrc}}|^2\le C\( \mathcal F^{\carr_R} (E,m) +
 \(C_0 m^{\frac\s\d}+ (\frac{1}{2\d} \log m) \indic_{\s=0} \)\C( \carr_R) \)  \ee
for some $C_1>0$ equal to $2$ in the case $\s=0$ and some $C_0>0$ depending only on $\d$ and $\s$. \end{prop}
We note in particular that 
\be\label{positivi} \mathcal F^{\carr_R} (E,m)  +\(( \frac{1}{2\d}\log m )\indic_{\s=0}+ C_0 m^{\frac{\s}{\d}} ) \C( \carr_R)\)\ge 0 .\ee
\begin{proof} It  follows the proof of Proposition \ref{procontrolelocal}.
Let us choose $\eta_p=\frac14 m^{-1/\d}$  for all $p$ in \eqref{reslemmono}, 
and observe that for each $p$ such that $\dist (p, \pa \carr_R) \ge \frac14 m^{-1/\d} $, by definition \eqref{defrrt} there exists $q\neq p$ such that 
$$(\g(p-q)- \g(\frac14 m^{-1/\d}))_+ \ge     (\g(4\rr_p)   - \g(\frac14 m^{-1/\d}) )_+ .$$ Using 
 that 
$\int |\f_{\rrc}|\le C_{\s,\d}\rrc_p^{\d-\s}$ by \eqref{fdmu} and $\rrc_p \le m^{-1/\d}$, we may thus write  that 
\begin{multline}\label{lb1}\hal \sum_{p\in\C \cap \carr_R, \dist (p, \pa \carr_R) \ge \frac14  m^{-1/\d}} ( \g(4 \rr_p)-\g(\frac14 m^{-1/\d}) )_+\le  \mathcal F^{\carr_R}( E, m)\\ - \frac{1}{2 \cds}\int_{\carr_R\times \R^{\k}}\yg |E_{\vec{\eta}}|^2 +\hal \sum_{p \in\C \cap \carr_R} \g(\frac14 m^{-1/\d} ) +  C m^{\frac{\s}{\d}} \C(\carr_R)  .\end{multline}
After rearranging terms, and adding back   $\g(\tilde \rr_p)$ for the $p$'s such that $ \dist(p, \pa \carr_R) \le \frac14 m^{-1/\d}$,  we obtain \eqref{bgrvi}. Inserting into the definition of $\mathcal F^{\carr_R}(E,m)$, we deduce \eqref{bornehnrvi}.
\end{proof}

\subsection{The jellium energy}\index{jellium energy}
 In the sequel, $\carr_R$ still denotes the {\it closed}  cube $[-R/2,R/2]^\d$.
\begin{defi}[Jellium renormalized energy for electric fields] Let $C_0$ be the constant of \eqref{bgrvi} and \eqref{bornehnrvi}.  Let $E\in \Elec_m$ i.e. satisfying \eqref{eqclam}. 
The Coulomb/Riesz renormalized energy of $E$ with background $m$ is defined by
\begin{equation}\label{defW}
\mc{W}(E,m) := \limsup_{R \ti} \frac{1}{R^{\d}}\(  \mathcal F^{\carr_R} (E,m)
+\(( \frac{1}{2\d}\log m )\indic_{\s=0}+ C_0 m^{\frac\s\d} \)
(\C( \carr_R ) -  mR^\d)\)
\ee
where $\mathcal F^{\carr_R}$ is as in \eqref{defFcarrr}.
\end{defi}

Note that  the terms added to $ \mathcal F^{\carr_R} (E,m)$ serve to control the number of points and use \eqref{positivi} but will eventually bring no contribution to the quantity as soon as we know that $\C(\carr_R) \sim mR^\d$, which we will show in Lemma \ref{neutralite0} below. 
\begin{rem}We note that the renormalized energy controls both the electric field and  the number of points. Indeed, in view of \eqref{bgrvi} and \eqref{bornehnrvi}, there exists a constant $C>0$ such that 
\be\label{contrWen}
\limsup_{R \to \infty} \frac{1}{R^\d}\(
\int_{\carr_R\times \R^\k} \yg |E_{\rrc}|^2 + \C( \carr_R)\)< C \( \mathcal{W}(E,m)+m\(C_0 m^{\frac\s\d}+(\frac1{2\d}\log m)\indic_{\s=0} \)  \).
\ee
We also  note that $\mathcal{W}$ is insensitive to a compact perturbation of $E$.\end{rem}
\begin{defi}[Jellium renormalized energy for infinite point configurations]\index{jellium energy}
Let $\C$ be a point configuration. We define the renormalized energy of $\C$ with background $m \ge 0$ as
\begin{equation}\label{de522}
\W(\mc{C},m) := \inf\{\mc{W}(E,m),  \ E \in \Elec(\C, m), \ E \ \text{is a gradient} \}
\end{equation}
with the convention $\inf (\varnothing) = +\infty$.
\end{defi}

In view of \eqref{positivi} and the definition, we have 
\begin{rem}\label{bb}
$\mathcal W(\cdot, m)$ and $\mathbb{W}(\cdot, m)$  are bounded below by a constant depending only on $\d, \s$ and $m$.
\end{rem}

We have the following
\begin{lem}\label{infatteint}
Let $\C$ be fixed. Two elements of $\Elec(\C, m)$  with finite energy which are gradients differ by a constant vector field. In particular the $\inf$ in \eqref{de522} is a $\min$.\end{lem}
\begin{proof} The proof is borrowed from  \cite[Lemma 2.3]{lebles} and relies on elliptic regularity theory. If $E_1$, $E_2$ are two gradient vector fields in $\Elec(\C, m)$ then $E_1-E_2=\nabla u$ with $u$ solving 
\be \label{diveqa}
 -\div (\yg \nab u)=0\qquad \text{in} \ \R^{\d+\k}.\ee
  Moreover, the finiteness of $\mathcal{W}(E_1, m)$ and $\mathcal{W}(E_2, m)$ directly imply that 
\be\label{82}\limsup_{R\to\infty} \frac1{R^\d}\int_{\carr_R\times \R^\k} \yg |\nab u|^2<\infty.\ee
The equation \eqref{diveqa} is divergence form equation with weight $\yg$ that belongs to the so-called  $A_2$-Muckenhoupt class, which makes it amenable to elliptic regularity theory. The result of \cite[Theorem 2.3.13]{FKS} 
gives that there exists $\lambda>0$ such that for any $X \in \R^{\d+\k}$
$$\mathrm{osc}(\nab_{\R^\d} u, B(X, r)) \le C \(\frac{1}{\int_{B(X, R)} \yg}\int_{B(X,R)} \yg |\nab_{\R^\d} u|^2\)^\hal  \(\frac{r}{R}\)^\lambda$$
where $\mathrm{osc} (u, B(X,r))= \max_{B(X,r)} u-\min_{B(X,r)}u$. In the Coulomb case for which $\k=0$ and $\gamma=0$, this is just standard regularity estimates for harmonic functions.
Inserting \eqref{82} and letting $R\to+\infty$, we deduce that $\mathrm{osc}(\nab_{\R^\d} u, B(X, r))=0$ which means that $\nab_{\R^\d} u$ is constant on every compact set of $\R^{\d+\k}$. In the case $\k=0$ this proves that $u$ is affine, and that $E_1$ and $E_2$ differ by a constant vector. In the case $\k=1$ this implies that $u$ is an affine function of the $\R^\d$ variables for each $y$. Writing $u(x,y)= a(y) \cdot x+b(y)$ and inserting back into \eqref{diveqa}, combining with the fact that $\int_{\R}\yg |\partial_y u|^2 dy$ is convergent, finally yields that $u$ is constant, hence the result.
\end{proof}

\subsection*{Scaling} Let us now record some natural scaling relations. 
If $E\in \Elec_m$, we define $\sigma_m E$ by 
\begin{displaymath}\label{defsigmam}
\sigma_m E := m^{-\frac{\s+1}{\d}} E\left(\frac{\cdot}{m^{1/\d}} \right). \end{displaymath}
We have $\sigma_m E\in \Elec_1$ and   have
\begin{equation} \label{scalingW} \begin{cases}
\mathcal{W} (E,m) = m ^{1+\frac\s\d} \mathcal{W} (\sigma_m E,1) & \text{if} \ \s\neq 0\\
\mc{W}(E,m)= m\mc{W}(\sigma_m E,1) - \frac{1}{2\d} m \log m  & \text{if} \ \s =0\end{cases}
\end{equation}  
and in the same way 
\begin{equation}\label{scalingWb}\begin{cases}
\mathbb{W}(\C, m)= m ^{1+\frac\s\d} \mathbb{W} (\sigma_m \C,1) & \text{if} \ \s\neq 0\\
\mathbb{W}(\C,m)= m\mathbb{W}(\sigma_m \C,1) -\frac{1}{2\d} m \log m  & \text{if} \ \s =0.\end{cases}
\end{equation}


\subsection{Discrepancy and neutrality}

The next property expresses that if the energy $\mc{W}(E,m)$ is finite then we control well the discrepancy in large boxes of the underlying system $(\C,m)$, implying that the system is on average neutral and the density of points on large boxes  converges to $m$. 
\begin{lem}[Asymptotic neutrality for configurations with finite energy] \label{neutralite0} Assume $\mc{W}(E,m) <\infty$ and let $\C$ be associated via \eqref{eqclam}.
 There exists $\ep>0$ depending only on $\d $ and $\s$, and $C>0$ depending only on $\d$, $\s$ and $m$, such that 
 \be \label{discrR}|\C( \carr_R)-m R^\d|^2\le  C R^{2\d - \ep} (1+  \mathcal{W}(E,m) ) .\ee
 In particular,
\be\label{reslemneutr}
\lim_{R\to  + \infty} \frac{\C(\carr_R)}{R^\d} = m.
\ee
\end{lem}

\begin{proof} By scaling, it suffices to prove it for $m=1$.  We 
argue exactly as in  Lemma \ref{coronp} (with $\mu$ instead of $N\mu$). Recalling that $\gamma= \s + 2-\k- \d$ from \eqref{defgamma}, we have $\gamma<1$, so we may find $\ep>0$ such that $\gamma+2 \ep<1$.  Taking  $\delta=R^{\gamma+\ep}$ in the proof of Lemma \ref{coronp}, we obtain that, letting 
$D(\carr_R)= \C(\carr_R)-  |\carr_R| $, 
\be\label{contnu} |D(\carr_R)|^2\le C R^{2(\d-1+\gamma+\ep)}+ \frac{ R^{\d+\gamma} }{R^{\gamma+\ep}}\int_{(\carr_{R+1}\backslash \carr_{R-1}) \times \R^\k} \yg |E_{\rrc}|^2 .\ee

In view of \eqref{contrWen} we have 
$$ \int_{\carr_{R+1} \times \R^{\d+\k}}\yg |E_{{\rrc}}|^2\le CR^\d ( \mathcal{W}(E,1) +1).$$
Inserting this and $\gamma+\ep<1-\ep$ into \eqref{contnu},  we obtain
$$ |D(\carr_R)|^2 \le  C R^{ 2\d-  \ep}+  C R^{2\d - \ep}   \mathcal{W}(E,1)  .$$
Dividing  by $R^\d$ and letting $R \to \infty$, we  get \eqref{reslemneutr} as well.\end{proof}
In view of the definition \eqref{defW}, we thus have
\begin{coro}
If $\mathcal W(E, m) <\infty$ then 
\be \label{limsupsansC}\mathcal W(E,m)= \limsup_{R \to \infty} \frac{1}{R^\d} \mathcal F^{\carr_R}(E,m).\ee
\end{coro}
The work \cite{gesandier} presents criteria for $\W$ to be finite in the case $\s=0$, $\d=2$. For instance, discrepancies in balls of radius $R$ that grow less than $R^{1-\ep}$ suffice.
Recently, \cite{huesmannleble} pushed the question further by examining, still in two dimensions, the link between hyperuniformity, finite $\W$ and Wasserstein distance to the Lebesgue measure,  for stationary point processes. In particular they show that finite $\W$ is equivalent to a certain quantitative hyperuniformity property. \index{hyperuniformity}

\section{The case of periodic configurations}

\begin{figure}[h!]
\begin{center}
\includegraphics[scale=0.9]{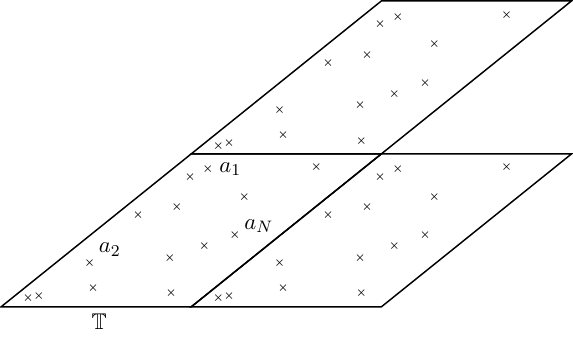}
\caption{Periodic configurations}\label{fig6}
\end{center}
\end{figure}

For periodic configurations, $\W $  can be computed and expressed  as a sum of pairwise periodized Coulomb or Riesz interactions between the points. By periodic configuration, we mean a configuration on the fundamental cell of a torus, repeated periodically, which can be viewed as a configuration of $N$ points on a torus (cf. Fig. \ref{fig6}).

\begin{prop}[Explicit form of the energy in the periodic case]\label{propperiodic} Let $a_1, \dots, a_N$ be $N$ points in a torus $\T$ of volume $n$ in $\R^\d$. Let $\C$ denote the infinite configuration on $\R^\d$ obtained by reproducing periodically $a_1,\dots, a_N$. 
If  $\s\ge 0$ and there is a multiple point, then for any $E \in \Elec(\C,1)$, we have  $ \mathcal{W}(E) = + \infty$. Otherwise, letting $H$ be  solution to 
\be\label{perH}
- \div (\yg \nab H ) = \cds\Big(\sum_{i=1}^N \delta_{a_i} -\drd \Big)\quad \text{in } \T\times \R^\k, \quad \int_{\T} H = 0,
\ee then seen as a periodic function in $\R^\d \times \R^\k$, 
any  periodic $E\in \Elec(\C,1)$   satisfies 
\begin{equation} \label{energieperiodique}
\mathcal{W}(E,1)\geq \mathcal{W}(\nabla H,1)
= \frac{\cds}{ 2 N} \sum_{i \neq j} G(a_i - a_j) + \frac{\cds}{2} \lim_{x \rightarrow 0} \Big(G - \frac{\g}{\cds}\Big)
\end{equation}
where $G$, the fractional Green function of the torus, solves 
\begin{equation}\label{527}
(-\Delta)^{\frac{\d-\s}{2}} G = \delta_0  - \frac{1}{|\T|}\ \mbox{ over } \T,\quad  \int_{\T} G  = 0.
\end{equation}

\end{prop}
Note the similarity with ``doubly periodic" Coulomb gases such as studied in \cite{forresterdoublyperiodic}.

\begin{rem}
1) The limit  appearing in this definition is called the Madelung constant of the torus.\\
2) In the case  $\d=1$ and $\s=0$,  an explicit formula for $G$ is available (see for instance \cite{borodinserfaty}):
$$G(x)= - \frac1{2\pi}\log \left|2\sin\frac{\pi x}{N}\right|.$$
For other values of $\s$ in $\d=1$, we have (see \cite{PetSer}) 
$$G(x)= 2\frac{N^{2\alpha-1}}{(2\pi)^{2\alpha}\Gamma(2\alpha) } \int_0^\infty \frac{ t^{2\alpha-1} (e^t \cos (\frac{2\pi}{N } x)-1) }{1-2e^t \cos (\frac{2\pi}{N} x) + e^{2t} }dt.$$
3) One can expand $G$ in Fourier series, then Eisenstein series appear, in particular there is a direct expression of $\W$ for a lattice configuration ($N=1$) in terms of the Eisenstein series of the lattice, see the proof of Theorem \ref{minimisationreseau} below. 
\end{rem}

One can also use these formulae valid in the periodic situation and extend them to nonperiodic configurations in order to provide alternative definitions of a Coulomb / Riesz jellium energy, and compare them with the definition in $\W$.  
This, and the comparison,  is done in the context of random point processes  \cite{leblepp}.  In \cite{borodinserfaty} this is used to compute $\W$ and its variants for some explicit point processes and use it as a measure of their rigidity.

\begin{proof} It suffices to consider the case of  simple points. The first inequality in \eqref{energieperiodique} follows from the projection lemma, Lemma \ref{projlem0}, adapted to the periodic setting: 
set $X=E-\nab H$ and observe that $X$ satisfies $\div (\yg X)=0$.
Then write 
$$\int_{\T\times \R^\k}\yg |E_\rr|^2 = \int_{\T\times \R^\k}\yg|\nab H_\rr+X|^2 = \int_{\T\times \R^\k} \yg|\nab H_\rr|^2 + \yg |X|^2 + 2\yg \nab H_\rr\cdot X$$
and the last term in the right-hand side vanishes after integration by parts since $\div (\yg X)=0$.

Let us now turn to the proof of the  equality in \eqref{energieperiodique}. Let $\eta<\min_{i=1}^n \rr_i$ and $\veta$ the vector in $\R^n$ of entries all equal to $\eta$.  Let $H$ be the solution of \eqref{perH}. As in Sections \ref{sec-extension} and \ref{secrieszcase}, the function $G$ solving \eqref{527} also satisfies 
\be -\div (\yg \nab G)= \cds \(\delta_0-\frac1{N} \delta_{\R^\d}\)\quad \text{in}\ \T\times \R^\k.\ee
 We then deduce  that $$H(x) = \cds \sum_{i=1}^N G(x-a_i)$$ with $G$ the Green function defined in the proposition, and thus $$ H_{\veta}(x)=\cds \sum_{i=1}^N G(x-a_i)- \sum_{i=1}^N \f_{\eta_i}(x-a_i).$$ Also $G=\frac{1}{\cds}\g + \phi$ with $\phi $ a continuous function. 
Since the $B(a_i,\rr_i)$ are disjoint, by definition \eqref{defW} and  the equality case in \eqref{reslemmono}, we have 
\begin{multline}
\mathcal W(\nab H,1)= \frac{1}{|\T|} \( \frac{1}{2\cds} \int_{\T\times \R^\k} \yg |\nab H_{\rr}|^2 -\hal \sum_{i=1}^N   \g(\rr_i) - \sum_{i =1}^N \int \f_{\rr_i} ( x-a_i)dx\)\\=
\frac{1}{|\T|}\(    
 \frac{1}{2\cds} \int_{\T\times \R^\k} \yg |\nab H_{\veta}|^2 -\hal \sum_{i=1}^N \g(\eta) - \sum_{i=1}^N  \int \f_{\eta} ( x-a_i)dx   \).\end{multline}

 Inserting  the above and using Green's formula, we have on the other hand
 \begin{multline}
\int_{\T\times \R^\k} \yg |\nab H_{\veta}|^2 = - \int_{\T\times \R^\k} H_{\veta} \, \div (\yg \nab H_{\veta})\\= \cds \int_{\T\times \R^\k}\(\cds \sum_{i=1}^N G(x-a_i)  - \sum_{i=1}^N\f_{\eta}(x-a_i)\) \Big(\sum_{j=1}^N \delta_{a_j}^{(\eta)} - \frac{1}{n}\delta_{\R^\d}\Big)(x)\\
=  \cds N (\g(\eta)+\cds \phi(0))  +\cds^2  \sum_{i\neq j} G(a_i-a_j) + \cds  \int_{B(0,\eta)} \f_{\eta}+o_\eta(1)
\end{multline}
where we used the disjointness of the balls, that $\f_\eta $ vanishes on $\pa B(0, \eta)$ where $\delta_0^{(\eta)}$ is supported, and that $\int_{\T} G=0$.
Letting $\eta \to 0$, in view of  \eqref{eq:intf}, we obtain \eqref{energieperiodique}.

\end{proof}

\section{Existence of minimizers}
In Corollary \ref{coroexmin} in the next chapter, we will prove that the minima of $\mathcal{W}(\cdot, 1)$ and $\W(\cdot, 1)$ are   achieved and are equal.
We now show the following, which relies on the screening procedure.
\begin{prop}\label{propperiodization}
For any $m>0$, any $\s, \d$ with $\s \in [\d-2,\d)$,  $\inf \mathcal{W}(\cdot, m)$ is finite. 
Moreover, $\F$ being as in \eqref{minneum},  we have 
\be\label{infFcr} \lim_{R\to \infty} \(\frac{1}{R^\d} \inf \F(\cdot,m, \carr_R)\) \le \min \mathcal{W}(\cdot, m)\ee
and 
\be\min \mathcal{W}(\cdot, m)= \lim_{R\to\infty} \min_{ E \text{ is } (R\mathbb{Z})^\d\text{-periodic}}  \mathcal{W}(E, m),\ee
with the limits taken along sequences of $R$'s such that $R^\d m$ are integers.
\end{prop}
\begin{proof} By scaling we reduce to the case $m=1$.
  We note that the periodic case of Proposition~\ref{propperiodic} implies that $ \inf \mathcal W(\cdot, 1)<+\infty$. 

 Let $(E_k)_{k\in \mathbb{N}} $ be a sequence such that 
 $$
 \mathcal{W}(E_k, 1) \le \inf \mathcal W(\cdot, 1) +\frac1k$$ and let $(\C_k)_k$ be the associated configurations.  In view of \eqref{defW} and Lemma \ref{neutralite0} we must thus have that there exists $C>0$ such that for all $R>1$
 \be \label{bpoints} | \C_k(\carr_R)|\le C R^\d\ee
 and 
 \be  \label{bnergy}\mathcal{F}^{\carr_R} (E_k, 1) \le \(\inf \mathcal W(\cdot, 1) +\frac1k +o_R(1) \) R^\d.\ee
 
 In the Coulomb case, we may now argue exactly as in the proof of Theorem \ref{th4} and  apply the screening result of Proposition \ref{proscreen} in any cube $\carr_R$ of quantized volume. It provides us with a screened configuration $X_R$ such that 
\be \label{fxr}\F(X_R,1, \carr_R) \le \mathcal{F}^{\carr_R} (E_k, 1)+o(R^\d)\le \(\inf \mathcal W(\cdot, 1) +\frac1k +o_R(1) \) R^\d\ee
 where $\F$ is defined in \eqref{minneum} in terms of the solution to the Neumann potential \eqref{defv} which we define by  $u_{k,R}$.  This proves \eqref{infFcr}. 
 
 The configuration $X_R$ can then be periodized into an infinite periodic configuration $\C_{k,R}$  on $\R^\d$, and $\nab u_{k,R}$ periodized into a periodic vector field $E_{k, R}\in \Elec(\C_{k, R}, 1)$ over $\R^\d$.
By periodicity and definition \eqref{defW} we have 
$$\mathcal{W}(E_{k, R}, 1) = \F(X_R, \carr_R).$$
In view of \eqref{fxr}, letting $k \to \infty $ and $R \to \infty$,  we deduce the inequality  
$$\lim_{R\to\infty} \min_{ E \text{ is}\ (R\mathbb{Z})^\d\text{-periodic}}  \mathcal{W}(E, 1)\le \inf \mathcal{W}(\cdot, 1) $$
while the converse inequality is trivial.

In the Riesz cases, we use instead the screening result of \cite{PetSer}.
\end{proof}

\begin{rem}
In the definition \eqref{defW} we have used cubes, where other shapes could be used (for instance balls, ellipsoids...). One may show, using screening and the method of the following chapter,   that while the value of $\mathcal{W} (E, m) $ may depend on the shape used, the minimum $\min \mathcal{W}$ does not, as long as the shapes remain nondegenerate.\end{rem}

\section{Minimization of $\W$ and the crystallization conjecture} \label{sec54}

\index{Abrikosov lattice}
\index{crystallization}

We have seen  that the minima of $\W$  can be achieved as limits of the minima over periodic configurations (with respect to larger and larger tori). On the other hand, Proposition \ref{propperiodic} provides  a more explicit expression for periodic configurations. 
In the one-dimensional case only, we know how to use this expression \eqref{energieperiodique}  to identify 
 the minimum over periodic configurations~:  a convexity argument, for which we refer to \cite[Prop. 2.3]{ss2} and \cite{lebleuniqueness}, shows that the minimum is achieved when the points are equally spaced, in other words for the lattice or crystalline distribution $\mathbb{Z}$. There is no uniqueness of minimizers since as we have seen, $\W$ is unchanged under a compact perturbation of the configuration, however a uniqueness result can be proven when viewing $\mathbb{W}$ as a function of stationary point processes, cf. \cite{lebleuniqueness,erbarleble}.

In higher dimension, determining the value of  $\min \W$  is a much more delicate question. The solution is known only in dimension 8 and 24, thanks to the resolution of the Cohn-Kumar conjecture by Cohn-Kumar-Miller-Radchenko-Viazovska \cite{ckrmv2}. 

To describe this conjecture, let us  define a point configuration $\C$ to be a nonempty, discrete, closed subset of Euclidean space $\R^\d$. For $p:\R_+\to \R$ any function, let the (lower) $p$-energy of $\C$ be 
\be \label{pener}  E_p(\C):= \liminf_{R \to \infty} \frac{1}{|\C \cap B_R|} \sum_{x, y \in \C \cap B_R, x\neq y}
p(|x-y|)\ee
where $B_R$ is the ball of center $0$ and radius $R$ in $\R^\d$. (Note this $E_p$ has nothing to do with the electric field $E$ encountered previously).
Finding the minimum of $E_p$ belongs to the wider class of (in general difficult) crystallization problems, see \cite{blanclewin} for a review.   Such questions are fundamental in order to understand
the  crystalline structure of matter.
 They also arise in  the arrangement of   Fekete points \cite{saffkuijlaars}.
One should immediately stress that there are very few positive results in that direction in the literature, in fact it is very rare to have a proof  that the solution to any minimization problem is periodic. Some exceptions include the two-dimensional sphere packing  problem, for which Radin \cite{radin} showed that the minimizer is the triangular lattice, and an extension of this by Theil \cite{theil} for a class of very  short range Lennard-Jones potentials.
The techniques used there do not apply to Coulomb  interactions, which are much longer range. Another positive result from \cite{bpt} shows the optimality of the triangular lattice 
for the question of minimizing the (normalized) $\|\sum_p \delta_{p} -1\|_{\mathrm{Lip}^*}$ over point configurations.

Let us now describe the Cohn-Kumar conjecture.
\begin{defi}
We say that  $p$ is  a completely monotone function of the  squared distance when $p(r)= g(r^2)$
with $g$ a smooth completely monotone function on $\R_+$ i.e. satisfying $(-1)^k g^{(k)}(r) \ge 0$  for all $r \ge 0$ for every integer $k\ge 0$.
\end{defi}
 This includes for instance Gaussians.

Let $\Lambda_0$ denote the triangular lattice $A_2$ in dimension $2$, the $\mathrm{E}_8$ lattice in dimension $8$ and the Leech lattice in dimension $24$, dilated so that their fundamental cell has volume 1.
 We do not give here the precise definitions of the $\mathrm{E}_8$ and Leech lattices, but suffice to say that these are Bravais lattices which means that they have the form $\sum_{i=1}^{\d} u_i \mathbb{Z}$ for some vectors $u_i \in \R^\d$, and that the triangular lattice in dimension 2 is the one spanned by two  vectors  of same norm forming an angle $\pi/3$, it is exactly what is called the Abrikosov lattice in the context of superconductivity, cf. Chapter \ref{chap:intro}.


\begin{conjecture}[{Cohn-Kumar~\cite{cohnkumar}}]
\label{conj.CK}
In dimension $\d=2,8$, resp.~$24$, the lattice $\Lambda_0$ is {\it universally minimizing} in the sense that it
  minimizes $E_p$ among all possible point configurations  of  density $1$  for all $p$'s that are  completely monotone functions of the squared distance.\end{conjecture}
  The conjecture is not true without the complete monotonicity assumption, as shown for instance in \cite{beterminpetrache}.
It  was proven in dimension 8 and 24 in \cite{ckrmv2} but remains open in dimension 2. 
In \cite{PetSercryst} it was shown that that conjecture implies the result on the minimum of $\mathbb{W}.$ 
\begin{theo}[\cite{PetSercryst}]
If the Cohn-Kumar conjecture \index{Cohn-Kumar conjecture} holds, then    $\Lambda_0$ achieves the minimum of $\mathbb{W}$, for $\d=2,8,24$.
\end{theo}
This implication relies on the reduction to the  periodic case for large tori of Proposition \ref{propperiodization} and the  following representation formula: for $\Lambda$ a given lattice of covolume $1$ and $n$ an integer
\be \label{Gint}
G_{n\Lambda} (x)=
\frac{1}{\Gamma(\frac{\d-s}{2})} \int_0^\infty \(  \sum_{ v \in n\Lambda}\Psi_t(x-v) -\frac{1}{N}\) t^{\frac{\d-s}{2}-1} \, dt\ee
 where $\Psi_t(x)$ is the standard heat kernel $(4\pi t)^{-\frac\d2} e^{-|x|^2/(4t)}$, and  $G_{n\Lambda}$ is the periodic fractional Green function on the torus of volume $n$. This way,  $G_{n\Lambda}$ is written as a superposition of heat kernels, which are completely monotonic functions of the square distance, to which the Cohn-Kumar conjecture applies.

In other dimensions, there does not always exist a universally minimizing lattice, i.e.~minimizers are expected to depend on the interaction. This is for instance the case for Riesz interactions in three dimensions, see \cite{blanclewin} and \cite{lewinsurvey}.
Identifying the minimizer remains an open question, even though it would suffice to be able to minimize in the class of periodic configurations with larger and larger period, using the formula \eqref{energieperiodique}. Common wisdom is that in sufficiently large dimension, minimizers should not be lattices. Indeed, counterexamples made with superpositions of two different lattices can be built, see  \cite{conwaysloane}.  

 In \cite{rns}, we  showed the equivalence between several ways of phrasing the  minimization of $\W$ in dimension $2$ over a finite size box~: minimization with prescribed boundary trace and minimization among periodic configurations. In all cases, we were  able to prove, in the spirit of \cite{ACO},  that the energy density and the points were uniformly distributed at any scale $\gg 1$, in good agreement with (but of course much weaker than!) the conjecture of periodicity of the minimizers. This was generalized to higher dimensional Coulomb cases in \cite{PRN}.  The Riesz case is harder and a conditional result is given  in the same paper.

\medskip

One question that is answered  is that of the minimization over the restricted class of pure lattice configurations, in dimension $\d=2$ only,
this means over   vector fields which are gradient of functions that are periodic with respect to a lattice $\mathbb{Z} \vec{u} + \mathbb{Z} \vec{v}$
 with $det(\vec{u}, \vec{v}) = 1$, corresponding to configurations of points that can be identified with $\mathbb{Z} \vec{u} + \mathbb{Z} \vec{v}$. 
\begin{theo}[The triangular lattice is the minimizer over lattices in 2D] \label{minimisationreseau} \mbox{}
The minimum of  $\mathcal{W}$ over this class of vector fields is achieved uniquely by the one corresponding to the  triangular (Abrikosov)   lattice $A_2$.
\end{theo}

When restricted to lattices, 
$\W$ corresponds to a ``height" of the associated flat torus in Arakelov geometry. With that point of view, 
the  result was already known since  \cite{osp}. The same  result was also obtained  in \cite{chen-oshita} for a similar energy.

We next give a sketch of the proof of Theorem \ref{minimisationreseau} from \cite{ssgl}, which is not very difficult thanks to the fact that  it reduces (as \cite{osp} does) to 
the same question for a certain modular function, which was solved by number theorists in the 50's and 60's.

\begin{proof}[Proof of Theorem \ref{minimisationreseau}] 
Proposition \ref{propperiodic}, more specifically \eqref{energieperiodique},    provides an explicit formula for the renormalized energy of such  periodic configurations. Denoting by $H_\Lambda$ the periodic  solution associated with \eqref{perH}, and expressing $G$ as a Fourier series, we find that 
\be\label{viaeis}
\mathcal{W}(\nab H_\Lambda,1) = \lim_{x \rightarrow 0} \left( \sum_{\vec{k} \in \Lambda^* \backslash \{0\}  } \frac{e^{2i \pi \vec{k} \cdot \vec{x}}}{4\pi^2 |\vec{k}|^2}  + 2\pi \log x\right).
\ee
By using either the first Kronecker limit formula (cf. \cite{lang}) or a direct computation, one shows that in fact
\begin{equation}\label{viaepsteinzeta}
\mathcal{W}(\nab H_\Lambda,1) =C_1 + C_2 \lim_{x \rightarrow 0, x > 0} \left( \sum_{\vec{k} \in \Lambda^* \backslash \{ 0\} } \frac{1}{|\vec{k}|^{2+x}} - \int_{\R^2}\frac{dy}{1 + |y|^{2+x} }\right),
\end{equation}
where $C_1$ and $C_2>0$ are constants.
The series $\sum_{\vec{k} \in \Lambda^* \backslash \{ 0\} } \frac{1}{|\vec{k}|^{2+x}} $  that appears is now the Epstein Zeta function of the dual lattice $\Lambda^*$. The first Kronecker  limit formula allows to pass from one modular function, the Eisenstein series, to another, the Epstein Zeta function. Note that both formulas \eqref{viaeis} and \eqref{viaepsteinzeta}, when $x \to 0$,  correspond to two different ways of regularizing the divergent series $\sum_{p \in \Lambda^*\backslash \{0\}} \frac{1}{|p|^2}$, and they are in fact explicitly related.

The question of minimizing $\mathcal{W}$ among lattices is then reduced to minimizing the Epstein Zeta function 
\begin{displaymath}
\Lambda \mapsto \zeta_{\Lambda} (x) : =  \sum_{\vec{k} \in \Lambda \backslash \{0\} 0} \frac{1}{|k|^{2+x}}
\end{displaymath}
as $x \rightarrow 0$. But results from \cite{cassels,rankin,ennola,ennola2,diananda,montgomery} assert that 
\begin{equation}
\zeta_{\Lambda} (x) \geq \zeta_{A_2} (x), \ \forall x > 0
\end{equation}
 and the equality holds if and only if $\Lambda = A_2$ (the triangular lattice).  Because that lattice is self-dual, it follows that it is the  unique minimizer.
 \end{proof}

As we have seen, the Cohn-Kumar conjecture implies that this triangular lattice does achieve the global minimum of $\W$.  
This was also conjectured in \cite{ssgl} on the basis that the triangular lattice   is observed in superconductors (it is then called  Abrikosov lattice in physics language) combined with the fact that $\W$ can be derived  as 
 the limiting minimization problem of the Ginzburg-Landau functional.
It was also proven in \cite{betermin} that this conjecture is equivalent to a conjecture derived by analytic continuation by  Brauchart-Hardin-Saff \cite{bhs} on the next order term in the asymptotic expansion of the minimal logarithmic energy on the sphere (an important problem in approximation theory, also related to Smale's 7th problem for the 21st century), which is obtained by  formal analytic continuation, hence by very different arguments. Moreover, the triangular lattice for the 2D Coulomb gas appears in \cite{alastueyjancovici}. All this reinforces the plausibility of this conjecture in dimension 2.

\begin{figure}[h!]
 \begin{center}
 \includegraphics[scale=0.5]{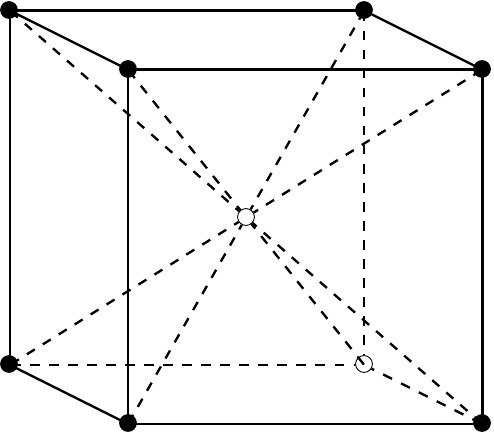}
 \includegraphics[scale=0.5]{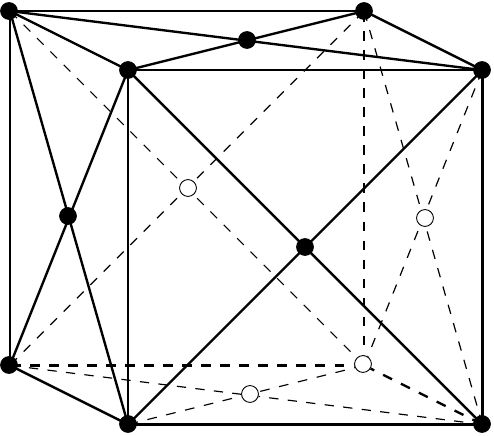}
\caption{BCC and FCC lattices}\label{fig8}
\end{center}
\end{figure}

 In dimension $\d\ge 3$ the  computation of the renormalized energy restricted to the class of lattices holds but the meaning  of \eqref{viaepsteinzeta} is not clear. The minimization of the Epstein Zeta function over lattices is then  an open number theoretic question (except in dimensions 8 and 24). 
 In dimension $3$, both the FCC (face centered cubic) and BCC (boundary centered cubic) lattices  (cf. Fig. \ref{fig8}) are expected to  play the role of the triangular lattice. It is only conjectured that FCC is a local minimizer in  \cite{sarns}, and so by duality BCC can be expected to  minimize $\W$ in the Coulomb case. As mentioned above, the Riesz case is different, as described in  \cite{blanclewin}  BCC is expected to be the global minimizer when $\s<3/2$, while it is FCC when $\s>3/2 $.

  \chapter[Minimizers via the jellium renormalized energy]{Microscopic characterization of minimizers via the jellium renormalized energy}
\label{chap:derivW}
\index{energy minimizers}
\index{jellium energy}
The main goal of this chapter is to obtain a next order asymptotic lower bound for the energy $\HN$. While a lower bound was already derived in  Corollary \ref{corlb}, we look here for a more precise lower bound by the jellium renormalized energy $\W$ of the previous chapter, as a way towards showing that $\W$  is  the infinite volume limit of $\F$. The general  lower bound valid for all configurations, in the spirit of $\Gamma$-convergence,  will be expressed in terms of 
local limits, which are point processes living at the microscale.  \index{limit point process}
This lower bound can then be complemented by upper bounds with the help of the screening procedure, thus
 providing a next order expansion and local limit description of energy minimizers. The main results is roughly that, after blow-up of a minimizer, the limit point configuration obtained minimizes $\W$. This recovers results first obtained in  \cite{ss1,ss2,rs,PetSer}.  \index{crystallization} This connects directly to the crystallization questions of the previous chapter. In particular, in dimensions 1, 8, 24 at least, minimizers should exhibit lattice patterns, and it is also expected to happen in dimension 2.  
 
In the case with temperature, the upper and lower bounds we obtain can be inserted into the Gibbs measure. This will be done in the next chapter,  and, when combined with an analysis of entropic effects, it allows  to derive a full Large Deviations Principle in terms of local point processes.  

\section{Tagged empirical field}\label{sec:tagged}
\index{empirical field}
Given $\XN = (x_1, \dots, x_N)$ in $(\R^{\d})^N$, we recall that we define $\XN'$ as the finite configuration rescaled by a factor $N^{1/{\d}}$, $\XN' = N^{1/\d} (x_1, \dots, x_N)$.

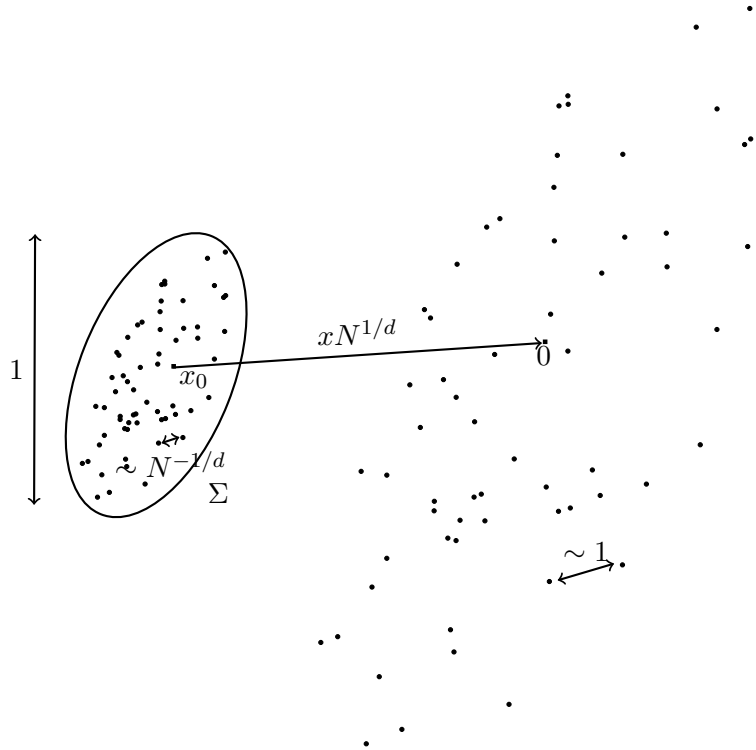
\begin{figure}[h!]

\begin{tikzpicture}[scale=0.65]

    \draw[thick, rotate around={-21.4:(2.93,0.02)}] (2.93,0.02) ellipse (1.56 and 3.05);
    
    \foreach \x/\y in {
        3.4873905/0.97596616, 12.416465/2.8353548, 1.881759/-0.64238524, 7.5995708/-2.019699,
        2.6447875/1.1015643, 9.888656/3.2121491, 3.0121381/0.95045507, 10.990707/2.7588215,
        2.2009015/-0.8849925, 8.556997/-2.7475212, 3.6333504/-0.70321816, 12.854344/-2.202198,
        1.7816882/-1.4033757, 7.2993574/-4.302671, 1.9834683/-2.3710744, 7.904698/-7.205767,
        3.3322291/0.73138, 11.950981/2.1015964, 2.519246/-0.77150905, 9.512031/-2.4070706,
        2.7063322/-2.200796, 10.07329/-6.6949306, 2.55571/1.0442115, 9.621424/3.040091,
        2.4703753/-0.7916594, 9.365419/-2.467522, 3.9967313/-0.43587658, 13.944487/-1.4001733,
        2.3349464/-1.8445178, 8.959132/-5.6260967, 1.7084861/-0.61662984, 7.079751/-1.9424331,
        2.2023432/-0.82007915, 8.561323/-2.552781, 3.3215678/-0.78059274, 11.918996/-2.434322,
        3.7675893/1.0029708, 13.257061/2.916369, 2.3495998/-0.112975724, 9.003093/-0.43147072,
        2.9786553/-1.3656317, 10.89026/-4.1894383, 4.109286/0.34808683, 14.282152/0.95171696,
        2.3114016/-1.693754, 8.888498/-5.173805, 3.4718027/-1.252508, 12.369701/-3.8500679,
        2.9619114/0.24532379, 2.3758307/-0.94904834, 9.081785/-2.9396884, 1.8307654/-2.012862,
        7.446589/-6.1311293, 2.474095/-0.27946866, 9.376578/-0.93094957, 4.316841/0.91198325,
        14.904816/2.6434062, 1.8817042/-1.2078063, 7.599406/-3.7159626, 2.9870129/0.45215577,
        10.915332/1.2639239, 3.1034417/1.9367944, 11.264618/5.7178397, 3.7726789/0.7739986,
        13.272329/2.2294521, 11.265007/0.5106919, 1.550455/-1.7407415, 6.605658/-5.3147683,
        4.331101/2.530727, 14.947595/7.4996367, 2.6094081/0.17824413, 9.7825165/0.44218883,
        2.0370505/-0.02844475, 8.065445/-0.1778778, 3.9694648/2.403753, 13.862687/7.118716,
        2.9573307/-0.72314864, 10.826285/-2.2619896, 2.3489385/-1.087221, 9.001108/-3.3542068,
        3.2688024/-0.6071938, 11.7607/-1.9141248, 3.0093002/1.3146701, 10.982194/3.8514667,
        4.3370485/1.6435814, 14.965439/4.8382006, 2.2640505/0.0071175094, 8.746445/-0.07119101,
        2.5437553/-0.9532312, 9.585559/-2.9522371, 3.0434859/1.8681333, 11.084751/5.511856,
        2.1765854/0.42617285, 8.48405/1.1859751, 1.4365168/-1.780025, 6.2638435/-5.432618,
        3.119937/-0.8657383, 11.314104/-2.6897585, 3.0401816/-0.88822204, 11.074838/-2.75721,
        3.4742808/1.539222, 12.377136/4.5251226, 2.741476/-0.5322949, 10.17872/-1.6894283,
        3.1056328/1.879119, 11.271191/5.5448136, 1.7431067/-2.469031, 7.183613/-7.4996367,
        2.1078687/-0.31805184, 8.2779/-1.046699, 3.0322888/1.5321805, 11.05116/4.5039983,
        4.296086/1.6055686, 14.84255/4.7241626, 2.3559263/0.79171646, 9.022072/2.282606,
        2.1356335/0.4831565, 8.361194/1.356926, 2.2935739/-1.0706358, 8.835015/-3.304451,
        4.1099553/1.8478192, 14.28416/5.4509144
    } {
        \filldraw (\x,\y) circle (1.2pt);
    }
    
    \draw[thick,->] (3.2595606,0.17542134) -- (10.739561,0.67542136) node[midway, above] {$x N^{1/d}$};
    \draw[thick,<->] (3.0395606,-1.3645786) -- (3.3995605,-1.2445786) node[midway, below] {$\sim N^{-1/d}$};
    \draw[thick,<->] (11.059561,-4.1645784) -- (12.199561,-3.8245788) node[midway, above] {$\sim 1$};
    \draw[thick,<->] (0.47956052,2.8954213) -- (0.45956054,-2.6245787) node[midway, left] {$1$};
    
    \node at (4.1924706,-2.3795788) {$\Sigma$};
    \filldraw[thick] (3.2595606,0.17542134) rectangle (3.2595606+0.05,0.17542134+0.05);
    \node at (3.6924708,-0.07) {$x_0$};
    \filldraw[thick] (10.779561,0.67542136) rectangle (10.779561+0.05,0.67542136+0.05);
    \node at (10.779561,0.42042133) {$0$};

\end{tikzpicture}
\begin{center}
\caption{An arbitrary blown-up configuration}\label{fig11}
\end{center}\end{figure}

One observable we wish to describe is  the \textit{tagged empirical field}  defined  as
\begin{equation}
\label{def:bEmp}
\bEmp_N[\XN] :=  \frac{1}{|\Sigma|} \int_{\Sigma} \delta_{\left(x,\,  \theta_{N^{1/{\d}} x} \cdot \XN' \right)} dx,
\end{equation}
where $\theta_x$ denotes the translation by $- x$. It is a probability measure on $\Sigma \times \config$, where $\Sigma$ is the support of the equilibrium measure and $\config$ is the space of  locally finite point configurations as in the previous chapter.
For any $x \in \Sigma$, the term $\theta_{N^{1/{\d}}x} \cdot \XN'$ is an element of $\config$ which represents the $N$-tuple of particles $\XN$ centered at $x$ and seen at microscopic scale (or, equivalently, seen at microscopic scale and then centered at $N^{1/{\d}} x$). In particular any information about this point configuration in a given ball (around the origin) translates to an information about $\XN'$ around $x$. We may thus think of $\theta_{N^{1/{\d}}x} \cdot \XN'$ as encoding the behavior of $\XN'$ around ``the tag" $x$. This terminology is inspired from classical works on random point processes \cite{georgii,georgiizessin}.
Limits of the tagged point processes $\bEmp_N[\XN]$ are naturally tagged point processes $ \bar{P}\in\mathcal {P}(\Sigma\times \config)$, whose first marginal is the normalized Lebesgue measure on $\Sigma$. As a consequence we may consider  the disintegration measures\footnote{We refer e.g. to \cite[Section 5.3]{AGS} for a definition.} $\{\bPst^x\}_{x \in \Sigma}$ of $\bPst$. For any $x \in \Sigma$, $\bPst^x$ is a probability measure on $\config$ and we have, for any $\Phi \in C^0\left(\Sigma \times \config\right)$ 
\begin{equation*}
\mathbb{E}_{\bPst}[\Phi] = \frac{1}{|\Sigma|} \int_{\Sigma} \mathbb{E}_{\bPst^x}[\Phi(x, \cdot)] dx.
\end{equation*}
Finally, the tagged point processes obtained at the limit will naturally be {\it stationary}, i.e. translation-invariant in their second entry.
We denote by $\probas_s(\config)$ the set of translation-invariant (or stationary) point processes. We also call stationary a tagged point process such that the disintegration measure $\bPst^x$ is stationary for (Lebesgue-)a.e. $x \in \Sigma$ and we denote by $\probas_s(\Sigma \times \config)$ the set of stationary tagged point processes.  

Finally, for any density $\mu$ on $\Sigma$,  we denote by $\probas_{s, \mu}(\Sigma\times \config)$ the set of stationary tagged point  processes such that 
for almost every $x\in \Sigma$, $\bP^x$  has intensity $\mu(x)$. 

We can also and will  consider local versions of the empirical field. 
Given $x_0$,   we define the ``local empirical field" averaged in a cube of microscopic scale size $R$ around $x_0$ by 
\be \label{defii}
P_N^{x_0,R} [X_N]:= \dashint_{\carr_R(N^{1/\d} x_0)} \delta_{\theta_{ x} \cdot  X_N'|_{\carr_R(N^{1/\d}x_0)} } dx\ee
where  $|_{\carr_R(N^{1/\d} x_0)}$ denotes the restriction of the configuration to $\carr_R(N^{1/\d} x_0)$, the cube of size $R$ centered at $N^{1/\d} x_0$. 
In other words we look at a spatial average at scale $R$ of the (deterministic) point process formed by the configuration.

To give a proper meaning to the convergences, we must specify the topology used. First we endow $\config$ with the topology induced by the topology of weak convergence of Radon measures, which makes it a metrizable  Polish space. This is a convergence against compactly supported continuous functions, which only allows to test local (i.e.~compactly supported) properties. 
 We endow $\probas(\config)$ and $\probas (\Sigma\times \config)$  with the usual topology of weak convergence of probability measures (for the Borel $\sigma$-algebra on $\config$).
This defines a notion of convergence corresponding to the weak convergence of probability distributions on $\config$. Another natural topology on $\probas(\config)$ is convergence of the finite distributions \cite[Section 11.1]{dvj2} --- sometimes also called convergence with respect to vague topology for the counting measure of the point process. These topologies coincide as stated in \cite[Theorem 11.1.VII]{dvj2}. 
We metrize it with an appropriate distance. For more detail see \cite{lebles}.

\begin{rem}
  Strictly speaking, elements of $\config$ are \textit{point processes} and elements of $\probas(\config)$ are \textit{laws of point processes}. However, in these notes, elements of $\config$ are called \textit{point configurations} (as above), a point process is defined as an element of $\probas(\config)$, and a tagged point process is a probability measure on $\Sigma \times \config$.
\end{rem}

\section{Energy lower bound in terms of the empirical field}
The first main result we wish to show is that the renormalized energy $\mathbb{W}$ of the previous chapter naturally arises as a $\Gamma$-liminf (in the terminology of Chapter \ref{chap:leadingorder}) of the next order energy $\F$ of \eqref{Fbup}, and can be seen as an infinite volume limit of $\F$. In order to see this concretely, it suffices to use 
 Fubini's theorem  to rewrite $\F(\XN', \mu')$ (where $\mu'$ is again the blown-up of $\mu$ defined by  $\mu'(x)= \mu(x N^{-1/\d})$) as 
 $$\F(\XN', \mu') \simeq \frac{1}{R^\d} \int_{\R^\d} \F^{\carr_R(z)}(\XN', \mu') dz$$
 where $R$ is any large scale (details will be given in the proof below) and $\F^{\carr_R}$ is as in \eqref{Glocal2}. 
 The idea is then to let $N \to \infty$ so that $\XN'$ converges to an infinite point configuration, while $\mu'$ converges to a uniform neutralizing background $m$, and then to let $R \to \infty$ so as to formally transform 
 $\lim_{R\to \infty}  \frac{1}{R^\d} \F^{\carr_R(z)}(\XN', \mu')
 $ into $\mathbb {W}(\C, m)$ by definition \eqref{defW} and \eqref{de522}. 
The way to properly take this double limit and obtain this way a lower bound is to use the empirical field process defined above, show its tightness, and then deduce the result by Fatou's lemma. 

In \cite[Chap.  5]{noteszurich} we describe in more detail how this can be seen as a special case of an abstract method for proving lower bounds on energies which depend on two scales for which $\Gamma$-convergence is known  at the small scale.


The   first main result we will prove is the following. 
For simplicity, we present the result in the setting of the whole space, but it would work without change when working with the Neumann energy in a set $U$.

Throughout we will use the notation
\be \label{bW} \overline \W(\bP,\mu):=|\Sigma|\int_{\Sigma} \(\int \W(\C, \mu(x)) d\bP^x(\C)\) dx\ee
for $\bP$ a tagged empirical point process in $\probas_{s}(\Sigma \times \config)$, where $\bP^x$ is the disintegration of $\bP$. 

\begin{prop}[Lower bound by the jellium renormalized energy]
 \label{prop:LowerBoundenergies} Let $\mu$ be a  bounded probability density on $\R^\d$ with compact support $\Sigma$ with $\pa \Sigma \in C^1$ and $\mu$ continuous in the interior of $\Sigma$.  
  Assume  $\{\mu_N\}_N$  is a uniformly bounded sequence of  probability measures on $\R^\d$    converging to $ \mu$ locally uniformly in $\Sigma$ as $N \to \infty$,  and let $\mu_N'$ be the blown-up densities $\mu_N'(x)= \mu_N(x N^{-\frac1\d})$.
    Let $\{\XN\}_N$ be a sequence of $N$-tuples of points in $\R^\d$, and let $X_N'= N^{1/\d}\XN$ be the blown-up points. 
    
   (Global version)  Assume that $N^{-1}  \F(\XN', \mu_N') $ is bounded independently of $N$, where $\F$ is as in \eqref{Fbup}. Then, up to extraction, the sequence $\{\bEmp_N[\XN]\}_N$  as in \eqref{def:bEmp} converges to some $\bP$ in $\probas_{s,\mu} (\Sigma \times \config)$, and we have
\begin{equation} \label{gliminf}
\liminf_{N \ti} \frac1N \F (\XN', \mu_N')  \geq \overline \W(\bP,\mu).
\end{equation} 

(Local version) Let $\{x_0^N \}_N$ be a sequence of points converging to $x_0$, a point in  the interior of $\Sigma$. There exists a constant $M>0$ depending only on $\d , \s$ and $\|\mu_N\|_{L^\infty}$ such that, if  $1\ll L \ll N^{1/\d}$ as $N \to \infty$ and if 
$$ \frac1{L^{\d}} \F^{\carr_L(N^{1/\d} x_0^N)} (\XN', \mu_N') + M \# \{ \XN' \cap \carr_L(N^{1/\d} x_0^N)\} $$ is bounded independently of $N$, then, up to extraction, the sequence $\{P_N^{x_0^N, L} [\XN]\}_N $ as in \eqref{defii} converges to some $P \in \probas_s (\config) $, and we have 
\begin{equation} \label{gliminfloc}
\liminf_{N \ti}\frac1{ L^{\d}} \F^{\carr_L(N^{1/\d} x_0^N)}  (\XN', \mu_N')  \geq \int \mathbb{W}(\C, \mu(x_0)) d P(\C).
\end{equation}
\end{prop}
Assuming that $\min \W(\cdot, m)$ is achieved, which we will prove below,  the quantity in the right-hand side of \eqref{gliminf} can be further bounded below by 
\be\label{bbeloww} 
\overline \W(\bP,\mu)
\ge \int_{\Sigma} 
\min \W (\cdot, \mu(x)) dx.\ee
But this expression can be further transformed by the scaling relation \eqref{scalingWb} 
which yields 
\be\label{minwreex} \int_{\Sigma} 
\min \W (\cdot, \mu(x)) dx= \int_{\Sigma} \mu(x)^{1+\frac\s\d}\min \W(\cdot, 1) -
\indic_{\s=0} \(\frac{1}{2\d} \int_{\Sigma} \mu(x) \log \mu(x)\, dx\).
\ee

\subsection{Tagged electric field process and compactness}
For the proof, we will use tagged electric field processes, which encode more information than just the configuration, since a configuration can be deduced from its electric field and not the converse.

We define an \textit{electric field process} as an element of $\probas(\Elec)$ (definition of $\Elec$ in \eqref{def1211}), usually denoted by $\Pelec$  (with $\mathrm{e}$ like electric) . We say that $\Pelec$ is \textit{stationary} when it is invariant under the (push-forward by) translations $\theta_x \cdot E := E(\cdot - x)$ for any $x \in \R^{\d} \subset \R^{\d} \times \{0\}^k$. We say that $\Pelec$ is compatible with $(P, m)$, where $P$ is a point process, provided $\Pelec$ is concentrated on $\Elec_m$ and the push-forward of $\Pelec$ by the map $\conf_m$ coincides with $P$ (see the notation in Chapter \ref{chap:renormalized}).

 Finally, we define a \textit{tagged electric field process} as an element of $\probas(\Sigma \times \Elec)$, usually denoted by  $\bPelec$, whose first marginal is the normalized Lebesgue measure on $\Sigma$. We say that $\bPelec$ is \textit{stationary} if for a.e. $x\in \Sigma$, the disintegration measure $\bar{P}^{\mathrm{e},x}$ is stationary (in the previous sense).  
 From the knowledge of a tagged electric field process $\bPelec$ we naturally deduce that of a tagged point process $\bar P$.
 
 For each configuration $\XN$ in $\R^\d$, we now precisely define  the tagged electric field process $P_N^{\mathrm{e}} [\XN]$ (then we will drop the $\XN$) by 
\be \label{taggedef}
P_N^{\mathrm{e}} [\XN](x,E):= \frac{1}{|\Sigma|}\int_{\Sigma} \delta_{(x,\theta_{N^{1/\d}x } \cdot E_N)}\, dx\ee
where $E_N= \nab u_N$  as in \eqref{defv} and $\theta_y$ is the action of translation by $y$.

We will need the following  tightness criterion for tagged electric field processes.

\begin{lem}[Compactness for sets with bounded electric energy and number of points] \label{lemcompact}
Let $\{E^{(n)}\}_{n}$ be a sequence of  vector fields in $\R^\d$, and  let $\{\mc{C}^{(n)}\}_n$ be a sequence of point configurations in $\R^\d$ such that 
\begin{equation}\label{eqstru}
-\div (\yg E^{(n)} )= \cds \left( \mc{C}_n - \mu^{(n)} \delta_{\Rd} \right) \text{ in } \R^{\d+\k},
\end{equation}
for some $\mu^{(n)} $ converging pointwise to a constant $m \ge 0$. 
Assume that 
there exists a constant $C_R$ such that 
\be\label{assborn}\sup_n \int_{\carr_R\times \R^\k} \yg |E^{(n)}_{\rrc}|^2+ |\C^{(n)} (\carr_{R}) | <C_R,\ee
where $\rrc$ is computed as in \eqref{defrrt} with respect to $ \carr_R$, the closed centered cube of sidelength $R$ in $\R^\d$.

Then there exists a vector field $E$, which is a gradient if the $E^{(n)}$'s are gradients, and  $\mc{C}$ a configuration in $\R^\d$,     satisfying
\begin{equation} \label{passerlimiteEnE}
-\div(\yg  E) = \cds \left( \mc{C} - m \delta_{\Rd} \right) \text{ in }  \R^{\d+\k},
\end{equation}
and such that, after extraction, for any compact set $K \subset \R^\d$,  $E^{(n)}$ converges weakly to $E $ in $L^p_{\loc}(K)$, for $p <\frac{\d+\k}{\s+1}$, 
$\C^{(n)}$ converges to $\C$ in $\config(K)$ and for every $R>0$, 
\begin{multline}\label{compiteE1}
\liminf_{n\to \infty}\(\frac1{2\cds}\int_{\carr_R \times \R^\k}\yg |E^{(n)}_{\rrc}|^2- \hal \sum_{p\in  \C^{(n)} \cap \carr_R} \g(\rrc_p) -\sum_{p\in \C^{(n)} \cap \carr_R} \int_{\R^\d} \f_{\rrc_p} (x-p)\mu^{(n)} (x) \, dx\)
\\
 \ge \mathcal{F}^{\carr_{R} } (E ,m) \end{multline} where $\mathcal{F}^{\carr_R}$ is defined in \eqref{defFcarrr}.
 
 Moreover, if $\{P^{(n)}\}_n$ is a sequence of probability measures on $L^p_{\mathrm{loc}}$, $p <\frac{\d+\k}{\s+1}$,  satisfying 
\be \forall R>1, \quad  \int \( \int_{\carr_R\times \R^\k} \yg |E_{\rrc}|^2+ \C( \carr_{R}) \) dP^{(n)}(E) <C_R\ee
 with $C_R$ independent of $n$, then  the sequence $\{P^{(n)}\}_n$ is tight in $L^p_{\mathrm{loc}}$ and any weak limit point $P$ satisfies that $P$-a.e.~$E \in \Elec_m$ and is a gradient. 
\end{lem}
\begin{proof} 
From \eqref{assborn}, after diagonal extraction we have that for every $k\in \mathbb{N}$, $\C^{(n)}$ converges in $\config(\carr_{k})$ to some $\C$, and $E_{\rrc}^{(n)}$ converges weakly in $L^2_{\yg}(\carr_k\times \R^\k)$ to some vector field $X$ such that, for every $R>0$, 
\be\label{liminf1} \liminf_{n\to \infty} \int_{\carr_R\times \R^\k }\yg |E^{(n)}_{\rrc}|^2\ge \int_{\carr_R}\yg |X|^2.\ee 
 On the  other hand, the bound on $\int\yg  |E^{(n)}_{\rrc}|^2$ 
and the argument of  Proposition \ref{procoer} provide  a bound on $E$ in $L^p(\carr_k )$ for  $1\le p<\frac{\d+\k}{\s+1}$, 
 hence we may find a weak limit point $E$ in $\cap_{k\in \mathbb{N}} L^{p}(\carr_k) $ which  satisfies \eqref{passerlimiteEnE}
 by taking the limit  as $n \to \infty$ in \eqref{eqstru} in the distributional sense. 
 In addition, if $E^{(n)}=\nab h^{(n)}$ for some $h^{(n)}$, by taking weak limits we also have that $E$ is a gradient.  Since  by definition \eqref{defEeta1} we have
 $$E_{\rrc}^{(n)} = E^{(n)} -\sum_{p\in \C^{(n)}}\nab \f_{\rrc_p} (\cdot -p),$$
 with $\rrc$ computed with respect to $\carr_R $, taking the weak $n\to \infty$ limit and using the continuity of $\rrc_p$ with respect to the configuration, we must have 
 $$X= E-\sum_{p\in \C}\nab \f_{\rrc_p}(\cdot-p) ,$$
and we recognize the right-hand side as equal to $E_{\rrc}$ (relative to  $\carr_R$). From \eqref{liminf1} we thus have for every $R>0$,
\be\label{liminf2} \liminf_{n\to \infty} \int_{\carr_R\times \R^k}\yg |E^{(n)}_{\rrc}|^2\ge \int_{\carr_R\times \R^\k}\yg |E_{\rrc}|^2.\ee 
The upper bound of the left-hand side implies that the right-hand side is finite, which we claim implies in the situation $\s \ge 0$  that $\rr_p$ cannot be $0$ for any point $p$ in $\C$. Indeed, if $\rr_p=0$ this means by definition \eqref{defrC} that the point $p$ comes with  multiplicity $q\ge 2$, and that $E_{\rr}= E$ in a neighborhood of $p$ (because then $\f_{\rr_p}=0$). We would thus have a vector field defined in a ball $B(p, r)$ for some $r>0$ satisfying 
$$-\div (\yg E)= \cds\( q \delta_p-m \drd \) \quad \text{in } B(p, r) \times \R^\k$$
 and such that $\int_{B(p, r)}\yg |E|^2 <\infty$.  
We may reason as in \eqref{1530}--\eqref{4180} that  if $r$ is small enough,  by Cauchy-Schwarz and Green's theorem we have 
\begin{multline}
\int_{B(p, r)} \yg |E|^2\ge\int_0^r \int_{\pa B(p, t)} \yg |E\cdot \nu|^2 \ge \int_0^r  \( \int_{\pa B(p, t)}  \yg E\cdot \nu\)^2 \frac{1}{\int_{\pa B(p, t)} \yg }  dt \\
 \ge c \int_0^r  \frac{1}{t^{\d-1+\k +\gamma}}  dt= c \int_0^r t^{-\s-1} dt  \end{multline} by \eqref{defgamma},
The right-hand side is always infinite if $\s \ge 0$, a contradiction.

We now know that if $\s \ge 0$,  $\rrc_p>0$ for the points of the limiting configuration in $ \carr_R$, thus  the definition of $\mathcal F^{\carr_R}(E, m)$ only involves the first case of \eqref{defFcarrr}. To prove \eqref{compiteE1}, in view of \eqref{liminf2}, it thus remains to check that in the case $\s\ge 0$ (otherwise these terms do not appear)  $-\hal \sum_{p \in  \carr_R} \g(\rrc_p)$ and $-m \sum_{p\in  \carr_R} \int \f_{\rrc_p} (x-p)$ are lower semi-continuous with respect to the convergence of the configuration, which they are by continuity of the definition of $\rrc$ and closedness of the cubes. 

We now turn to the statement about tightness.
For any $\ep>0$ and integer $k$, defining 
$$\mathcal K_{\ep, k}= \left\{ (x,E ) ,  \int_{\carr_k\times \R^\k}\yg |E_{\rrc} |^2 + \C(\carr_{k})  < \frac{ C 2^k C_k }{\ep}\right\}$$ for that same constant $C$, 
we have that $P^{(n)}( \mathcal K_{\ep, k}^c ) <\ep  2^{-k} $ hence $$P^{(n)} \(\( \cap_k \mathcal K_{\ep, k}\)^c \)= P^{(n)} \( \cup_k \mathcal K_{\ep, k}^c \) < \ep.$$ But $\cap_k \mathcal K_{\ep, k}$ is a compact set in $L^p_{\mathrm{loc}}$ by the above results. Thus we conclude that the sequence  is tight, and it satisfies the stated properties in view of the above.
 \end{proof}

\subsection{Proof of Proposition \ref{prop:LowerBoundenergies}}

{\bf Step 1. Fubini rewriting.}
For each configuration $\XN$, let $P_N^{\mathrm{e}} [\XN]$ be as in 
\eqref{taggedef}.
The idea of the Fubini rewriting can be explained by the following calculation, simplifying $\F$ into $\int_{\R^\d} |\nab u_N|^2$ where $E_N= \nab u_N$: by definition \eqref{taggedef} it holds that  for any $R>0$,  $\carr_R$ being the closed centered cube of sidelength $R$,
$$ \int  \( \dashint_{\carr_R} |E|^2 \) dP_N^{\mathrm{e}} [\XN](x, E)  = \frac{1}{|\Sigma|} \int_{\Sigma} \(\dashint_{\carr_R} |\nab u_N(N^{1/\d} x+ \cdot) |^2 \) dx .$$
Then, using the change of variables $z= N^{1/\d} x+y$ and Fubini's theorem, we may write 
\begin{align}\nonumber
 \int  \( \dashint_{\carr_R} |E|^2 \) dP_N^{\mathrm{e}} [\XN](x, E) &= \frac{1}{R^\d |\Sigma|} \int_{\Sigma} \(\int \indic_{ \carr_R} (y)  |\nab u_N(N^{1/\d} x+ y) |^2 dy  \) dx 
\\ \nonumber & \le  \frac{1}{R^\d |\Sigma|}  \int \( \int_{\R^\d}  |\nab u_N (N^{1/\d} x+ y) |^2 dx\)  \indic_{\carr_R}(y) dy\\ 
\label{fubinimodele}
&=  \frac{1}{N|\Sigma|}\int_{\R^\d} |\nab u_N|^2  
\end{align}
and the right-hand side is our simplified version of $\frac1{N|\Sigma|} \F(\XN, \mu_N')$.
Thus to bound from below $\frac1N \F(\XN, \mu_N' )$ it suffices  to take the limit of $P_N^{\mathrm{e}} [\XN]$ and use Fatou's lemma. 

\smallskip
 
 Let us now implement this precisely. 
The main point is that we want to work with positive and coercive quantities, in order to obtain tightness and use Fatou's lemma. To do so, we add to the energy a large constant times the number of points. 

For any $z\in \carr_R(0)$, using the superadditivity property  \eqref{locali} (equivalent to  \eqref{locali2} in the general Riesz case) we find that 
\be
\F(\XN',\mu_N') \ge \sum_{q \in (R\mathbb{Z})^\d }  \F^{\carr_R(z +q)} ( \XN', \mu_N') 
  \ee
  where by definition  \eqref{Glocal}
  \begin{multline}\label{Fmu'2} \F^{\carr_R(z )} ( \XN', \mu_N')\\
   =\frac1{2\cds}  \int_{\carr_R(z)\times \R^\k}\yg |\nab u_{N, \rrc}|^2 -\hal  \sum_{x_i' \in \carr_R(z)} \g(\rrc_i)  - \sum_{x_i' \in \carr_R(z) } \int_{\R^\d} \f_{\rrc_i} (x_i'-y) d\mu_N'(y) \end{multline} with $\rrc$ computed relative to $\carr_R(z)$ as in  \eqref{rrc} with $\lambda =1$ or as in  \eqref{defrrc4}.
Thus, after averaging over $z\in \carr_R(0)$, 
\begin{multline}\label{minoFmu''}
\F(\XN',\mu_N') \ge \frac{1}{R^\d}\int_{\carr_R(0)} \sum_{q \in (R\mathbb{Z})^\d}  \F^{\carr_R(z + q)} ( \XN', \mu_N')  \, dz
= \frac{1}{R^\d}\int_{\R^\d}   \F^{\carr_R(z)} ( \XN', \mu_N') 
\, dz ,
  \end{multline}
 and  arguing in the same way, for any $M> 0$, 
 \be\label{minoFmu'}
\F(\XN',\mu_N')+MN \ge  \frac{1}{R^\d}\int_{\R^\d}   \F^{\carr_R(z)} ( \XN', \mu_N') 
\, dz+ \frac{M}{R^\d} \int_{\R^\d}  \# \{\XN' \cap \carr_{R}(z) \}  
\, dz .
\ee
    By the local uniform convergence assumption on $\{\mu_N'\}_N$ and continuity assumption on $\mu$, we may write that for $x_i' \in \carr_R(z)$, for $R$ fixed and as $N \to \infty$, 
  $$ \int \f_{\rrc_i} (x_i'-y) d\mu_N'(y)=\( \mu(zN^{-1/\d}) +o_N(1) \) \int \f_{\rrc_i} (x_i'-y)dy .$$
 Using that $\int| \f_{\rrc_i}(x_i'-y)dy|\le C \rrc_i^{\d-\s} \le C $ by \eqref{eq:intf},  we may thus write 
  \begin{multline}\label{Fmu'3} \F^{\carr_R(z )} ( \XN', \mu_N') =\frac1{2\cds}  \int_{\carr_R(z)\times \R^\k}\yg |\nab u_{N, \rrc}|^2 -\hal  \sum_{x_i' \in \carr_R(z)} \g(\rrc_i)  \\- \mu(z N^{-1/\d})  \sum_{x_i' \in  \carr_R(z) } \int \f_{\rrc_i} (x_i'-y) dy 
  + o_N(1)  \C( \carr_R(z))   .\end{multline} 
Let us now define for $x \in \Sigma$, $E\in L^p_{\mathrm{loc}}$, $ p<\frac{\d+\k}{\s+1}$, 
\begin{multline}\label{deffNR} 
f_{N,R}(x,E):=\chi_N(x,E) \frac{1}{R^\d} \Bigg( \frac{1}{2\cds} \int_{\carr_R(0) \times \R^\k}\yg |E_{\rrc}|^2 -  \hal \sum_{p \in \C \cap \carr_R(0)} \g(\rrc_p) \\
-\mu(x) \sum_{p \in \C\cap \carr_R(0)}   \int \f_{\rrc_p} (p-y)dy
+M \C(\carr_{R})\Bigg)\end{multline}
and 
\begin{multline}\label{deffR}
f_{R}(x,E):=\chi(x,E)
 \frac{1}{R^\d} \Bigg( \frac{1}{2\cds} \int_{\carr_R(0) \times \R^\k}\yg |E_{\rrc}|^2 -  \hal \sum_{p \in \C \cap \carr_R(0)} \g(\rrc_p) \\
-\mu(x) \sum_{p \in \C\cap \carr_R(0)}   \int \f_{\rrc_p} (p-y)dy
+M \C(\carr_{R})\Bigg)
\end{multline}
 where $\chi_N$ and $\chi$ are defined as  
 $$ \chi_N(x, E):= \begin{cases} 1 & \text{ if $E= \nab u_N(N^{1/\d}x+\cdot)  $ }\\
 +\infty &\text{otherwise}\end{cases} $$
 and 
 $$ \chi(x, E):= \begin{cases} 1 & \text{ if $E$  is a gradient and s.t.~$-\div (\yg E)= \cds (\sum_{p\in \C}\delta_p-\mu(x) \drd)$ }\\ & \text{
 with only simple points in $ \carr_R$ if $\s\ge 0$}\\
 +\infty &\text{otherwise}\end{cases} $$
 and $\rrc$ is as in \eqref{defrrt}. 
 
Choosing $M>1+C_0+C\sup_N \|\mu_N'\|_{L^\infty}$ with $C$ depending only on $\d, \s$ and $C_0$ the constant of \eqref{bornehnrvi}, in view of that inequality,   there exists a constant $C>0$ such that
\be\label{bornifr}
 C f_{N,R}(x,E)  \ge \frac{1}{R^\d} \(  \int_{\carr_R\times \R^\k} \yg |E_{\rrc}|^2 + \C(\carr_{R}) \) \ge  0.\ee

In view of \eqref{taggedef}, after using Fubini's theorem as in \eqref{fubinimodele} we may  rewrite \eqref{minoFmu'} and \eqref{Fmu'3} as 
\be\label{fubi}
\F( \XN', \mu_N')+ M N \ge N |\Sigma| (1+o_N(1))\int  f_{N,R}(x,E)  \, dP_N^{\mathrm{e}} (x,E).\ee

  {\bf Step 2. Tightness and lower bound}.
 The upper bound assumed on $\F(\XN', \mu_N')$  and \eqref{fubi}  imply that 
 $\int f_{N,R}(x,E)\, dP_N^{\mathrm{e}} (x,E) $ is bounded independently of $N$. 
In view of  \eqref{bornifr}, it follows that 
$$\int \(  \int_{\carr_R\times \R^\k}\yg |E_{\rrc} |^2 + \C(\carr_{R}) \) d P_N^{\mathrm{e}} (x, E)< C  R^\d$$
where $C$ is a constant independent of $N$. By the second part of Lemma \ref{lemcompact}, we deduce that, up to extraction,  $P_N^{\mathrm{e}} \to \bPelec$ weakly for some tagged electric field process $\bPelec$. 
 
 In addition the results of Lemma \ref{lemcompact} and the definition of $\chi_N$ and $\chi$  ensure that if $x_N\to x$ and $E_N \to E$ (weakly in $L^p_{\mathrm{loc}}$), we have 
 $$\liminf_{ N\to \infty} f_{N, R}(x_N, E_N) \ge f_R(x, E).$$
Combining this with the weak convergence of $P_N^{\mathrm{e}}$ to $\bPelec$ we deduce that 
for each $R \ge 1$, we have 
$$\liminf_{ N\to \infty}\int f_{N, R}(x, E)dP_N^{\mathrm{e}} (x, E)  \ge\int f_R(x, E)d\bPelec(x, E),$$
in other words, returning to \eqref{fubi}, 
\be\label{fubi2}
\liminf_{N\to \infty}
\frac{1}{N} \F( \XN', \mu_N')+ M \ge  |\Sigma| \int f_{R}(x,E)\, d\bPelec(x,E).\ee 
Since $f_R(x, E)$ is defined to be $+\infty$ for $E\notin \Elec_{\mu(x)}$, we also deduce that 
$\bPelec$-a.e. $(x, E)$, we have  $E\in \Elec_{\mu(x)}$ and $E$ is a gradient.  In the same way, if $\s\ge 0$, for $\bPelec$-a.e. $(x, E)$ the points of $\conf_{\mu(x)}(E)$ (see Definition \ref{def1211})  are simple.
 \smallskip
 
 {\bf Step 3. Stationarity and conclusion.}
 To finish, let us justify that $\bPelec$ is stationary. 
 Observe that for every $x$, thanks to the assumed regularity of $\partial \Sigma$, we have 
 \be \lim_{\ep \to 0} \frac{|(\Sigma +\ep x)\triangle \Sigma|}{|\Sigma|} = 0 \ee  where $\triangle $ denotes the symmetric difference of sets. 
 Consider a test-function $\Phi \in C^0(\Sigma \times L^p_{\mathrm{loc}}) $ and $y \in \R^\d$. Since $\bPelec= \lim_{N\to \infty }  P_N^{\mathrm{e}}$, we may  thus  write
 \begin{multline*}\int \Phi(x, \theta_y E) \, d\bPelec = \lim_{N\to \infty}\frac1{|\Sigma|}\int_{\Sigma} \Phi(x, \theta_{N^{1/\d} x + y} E_N) \, dx
 \\=  \lim_{N\to \infty} \frac1{|\Sigma|} \int_{\Sigma} \Phi(x, \theta_{N^{1/\d} ( x + yN^{-1/\d} ) } E_N) \, dx
= \lim_{N\to \infty} \int \Phi(x, E) \, d P_N^{\mathrm{e}}= \lim_{N\to \infty} \int\Phi(x, E)\, d\bPelec,\end{multline*} hence $\bPelec$ is stationary.
 Pushing forward  by the map $E\mapsto \(- \frac{1}{\cds}\div (\yg E) + \mu_N'\drd\)$ naturally yields the convergence of $\bEmp_N[\XN] $, defined in \eqref{def:bEmp} to some $\bP\in \probas_s(\Sigma \times \config)$.


By stationarity of $\bPelec$, we now also have 
$$\int f_R (x,E) \, d\bPelec(x, E) = \int \( \lim_{R\to \infty}   f_R(x, E) \) \, d\bPelec(x, E).$$  Inserting into \eqref{fubi2}, we obtain 
\be\label{fubi3}
\liminf_{N\to \infty} \frac1N \F( \XN', \mu_N')+ M \ge  |\Sigma|  \int \( \lim_{R\to \infty}   f_R(x,E) \) \, d\bPelec(x, E)
.\ee 
 We may then rewrite $f_R$ as $\chi\( \frac{1}{R^\d} \mathcal F^{\carr_R}(E,\mu(x) ) + \frac{M}{R^\d}  \C(\carr_{R})\) $  in the notation of \eqref{defFcarrr}.
Thus
 by definition \eqref{defW}  and Lemma \ref{neutralite0},
 \be \label{limfRxE}
 \lim_{R\to \infty}   f_R(x,E) = \mathcal W(E,\mu(x)) +M \mu(x), \quad \bPelec(x,E)-a.e.,\ee hence  we have obtained
 \be\label{fubi4}
\liminf_{N\to \infty} \frac1N \F( \XN', \mu_N')+ M \ge  |\Sigma|  \int  \(\mathcal W(E,\mu(x)) +M\mu(x) \) d\bPelec(x, E)
.\ee 
 Since the first marginal of $\bPelec$ is the normalized Lebesgue measure on $\Sigma$ and since $\int \mu(x) dx=1$, it follows that 
 \be\label{fubi4}
\liminf_{N\to \infty} \frac1N \F( \XN', \mu_N') \ge  |\Sigma|  \int  \mathcal W(E,\mu(x))   d\bPelec(x, E)
.\ee Using the definition \eqref{de522}  we  get the  further lower bound
\be\label{fubi5}
\liminf_{N\to \infty} \frac1N \F( \XN', \mu') \ge  |\Sigma|  \int  \mathbb W(\C,\mu(x))   d\bPelec(x, E)\ee
and projecting this onto the tagged point processes $\overline P$, with \eqref{bW}   we obtain the result \eqref{gliminf}.
\smallskip

{\bf  Step 4. Localized version}.
We next turn to the localized version of the result. 
Given the cube $\carr_L(N^{1/\d} x_0^N)$ with $1\ll L \ll N^{1/\d}$, let now
$$P_N^{x_0^N,L,\mathrm{e}}(E):= \dashint_{\carr_L(N^{1/\d}x_0^N)} \delta_{\theta_x \cdot E_N} dx.$$
Instead of \eqref{minoFmu'} we  write that for any $R<L$, we have 
\begin{multline*}\F^{\carr_L(N^{1/\d}x_0^N)} (\XN', \mu_N') +M \# I_{\carr_L(N^{1/\d}x_0^N) } \\
\ge 
\frac{1}{R^\d} \int_{z \in \carr_{L-R}(N^{1/\d}x_0^N)}  \( \F^{\carr_R(z)} (\XN', \mu_N') + M  \#I_{ \carr_R(z) } \)dz.\end{multline*}  
With \eqref{deffR}, we may  rewrite this as 
$$\F^{\carr_L(N^{1/\d}x_0^N)} (\XN', \mu_N') +M \#I_{\carr_L(N^{1/\d}x_0^N) } \ge (L-R)^\d(1+o_N(1)) \int f_R(x_0^N, E) dP_N^{x_0^N,L-R}(E).$$
Since we assume that the left-hand side is bounded by $CL^\d$, if $L \gg R$ the same proof  as in Step 2 implies that $\{P_N(E)^{x_0^N,L-R,\mathrm{e}}\}_N$ is tight, hence has a subsequential limit $ P^{\mathrm{e}}$ which is proven as in Step 3 to be  stationary, and 
$$\liminf_{N\to \infty} \frac{1}{L^\d} \(\F^{\carr_L(N^{1/\d}x_0^N)} (\XN', \mu_N') +M
\#I_{\carr_L(N^{1/\d}x_0^N) }\) \ge 
\int f_R(x_0, E) d P^{\mathrm{e}}.$$
Letting $N \to \infty$, then  $R \to \infty$ with $R \ll L$,  and using \eqref{limfRxE}, we obtain 
$$
\liminf_{N\to \infty} \frac{1}{L^\d} \(\F^{\carr_L(N^{1/\d}x_0^N)} (\XN', \mu_N') +M\#I_{\carr_L(N^{1/\d}x_0^N) } \)\ge
\int  \mathcal W (E, \mu(x_0))  d P^{\mathrm{e}}(E) + M \mu(x_0)  .$$
Moreover, we note that when $R \ll L$, $P_N(E)^{x_0^N,L-R,\mathrm{e}}$ and $P_N(E)^{x_0^N,L,\mathrm{e}}$ have the same limit, $P^{\mathrm{e}}$. Projecting onto point processes we obtain that, $P$ being the (subsequential)  limit of $ P_N^{x_0^N, L} [\XN]$, 
\begin{multline*}
\liminf_{N\to \infty} \frac{1}{L^\d} \(\F^{\carr_L(N^{1/\d}x_0^N)} (\XN', \mu_N') +M\# I_{\carr_L(N^{1/\d}x_0^N) } \) \ge 
\int  \mathbb W (\C, \mu(x_0))  d P(\C) + M \mu(x_0) .\end{multline*}As above, in view of Lemma \ref{neutralite0}, we also have $L^{-\d} \# I_{\carr_L(N^{1/\d}x_0^N) } \to \mu(x_0)$ as $N \to \infty$ if the left-hand side is finite.   We can thus finally conclude that \eqref{gliminfloc} holds.

\section{Lower semi-continuity and existence of minimizers for $\W$}
With the same proof idea we can now finally give the proof of existence of minimizers of the renormalized energy.
\begin{coro}[Existence of minimizers for $\mc{W}$ and $\W$]\label{coroexmin}
The functions 
$\mc{W}(\cdot, 1)$
 and $\W(\cdot ,1)$  admit a minimizer and $\min_{\mathrm{gradients}}  \mathcal W(\cdot, 1) = \min \W(\cdot, 1)$. 
\end{coro}
\begin{proof}
The proof is analogous to the previous one: returning to the definitions \eqref{defW} and \eqref{de522}, let $E^n$ be a sequence of gradients in $L^p_{\mathrm{loc}}$, $p <\frac{\d+\k}{\s+1}$,
such that $\mathcal W(E^n, 1) \to \inf \W (\cdot, 1)$,
 let $\C^n$ be the associated configuration, 
 and let $R_n\to \infty$ be such that 
\begin{multline*}
\lim_{n\to \infty} \frac{1}{(R_n)^\d}\(\int_{\carr_{R_n}}\frac1{2\cd} |E^n_{\rrc}|^2 -\sum_{p\in  \carr_{R_n}} \g(\rrc_p)-  \sum_{p \in \carr_{R_n}  }  \int \f_{\rrc_p} (x-p)dx\)  + C_0\(\frac{ \C^n(\carr_{R_n})}{(R_n)^\d} - 1\)   \\
=\inf \W(\cdot, 1).
\end{multline*}
Defining $P_n$ to be  $$ \dashint_{\carr_{R_n}} \delta_{\theta_x  \cdot E^n}\, dx,$$
we obtain, exactly as in the proof above (noting that here it was important to make the definition  such that the quantity $\mathcal F^{\carr_R}$ is superadditive)
$$\inf_{\mathrm{gradients}} \mathcal{W} (\cdot, 1)= \inf \W(\cdot, 1)  = \lim_{n\to \infty} \mc{W}(E^n,1)  \ge\int \mc{W}(E,1)\, dP^{\mathrm{e}} (E) \ge \int \W(\C, 1) dP(\C),$$
where $P^{\mathrm{e}}$ is a stationary limit point (up to extraction) of $P_n$, and $P$ is its push-forward by $E\mapsto \cds ( -\div (\yg E)+\drd)$. Moreover, $P_n$ being supported on gradient vector-fields, $P$ also is. 
Thus minimizers must exist, and $P^{\mathrm{e}}$ must be concentrated on minimizers of $\mc{W}(\cdot, 1)$ over gradients  and $P$ on minimizers of $\W(\cdot, 1)$. We can prove in the same way the existence of minimizers of $\mathcal W(\cdot , 1)$ without restriction to gradients.
\end{proof}

Using again the same proof we obtain the following lemma which will be important for the proof of the Large Deviations Principle in the next chapter.
\begin{lem}
\label{corolscW}
The maps $$P\mapsto \int \W(\C,m)dP(\C), \qquad P\mapsto \overline \W(\bP,\mu), 
$$
are lower semi-continuous on the space $\probas_s(\config)$, respectively $\probas_s(\Sigma\times \config)$. In addition, their  sub-level sets are compact on these spaces. Thus, they are good rate functions in the sense of Definition \ref{ratefun}.
\end{lem}

Before proving the lemma, we need the following.
\begin{lem}[Lifting  stationary point processes to electric processes]\label{lemlifting}
Let $m\ge 0$ and $P\in \probas_s(\config) $ such that $\int \W(\C, m) d P(\C)<\infty$.  There exists $P^{\mathrm{e}}$ a stationary probability measure on $\Elec_m$, concentrated on gradients\footnote{we recall that a probability measure $P$ is concentrated on $S$ if for every $S'$, $P(S')=0$ if $S\cap S'=\varnothing$} such that the push-forward of $P^{\mathrm{e}}$ by $\conf_m$ equals $P$ and such that 
$$\int \mathcal W(E, m) dP^{\mathrm{e}}(E)= \int \W(\C, m) d P(\C).$$
\end{lem}
\begin{proof}
Let $P$ be as in the assumption. For $P$-a.e.~$\C$, the energy $\W(\C, m)$ is finite and according to Lemma \ref{infatteint} we may find a gradient electric field $E\in \Elec(\C,m)$
such that $\mathcal{W}(E,m)= \W(\C,m).$ Let $P^{\mathrm{e}}$ be the push-forward of $P$ by this map $\C\mapsto E$. 
It may happen that $P^{\mathrm{e}}$ is not stationary. In that case we consider a stationarizing sequence, namely a sequence of averages of translations of $P^{\mathrm{e}}$ over large hypercubes. Each element of this sequence is still compatible with $P$ (because $P$ is stationary) and has the correct energy. Any limit point of that sequence is stationary, has the correct energy, and is still compatible with $P$. \end{proof}

\begin{proof}[Proof of Lemma \ref{corolscW}]
We only prove the first result, the second one follows easily from the first using Fatou's lemma and the fact that $\W$ is bounded below (Remark \ref{bb}). 
Assume that $P_n \to P$ in $\probas_s(\config)$.  
We may assume without loss of generality that $\sup_n\int \W(\C,m)  dP_n(\C) <C.$
By Lemma \ref{lemlifting}, we may lift $P_n$ into  $P_n^{\mathrm{e}}$ which is stationary  and such that 
$$\int \W(\C, m) dP_n(\C)= \int \mathcal W(E,m) d P_n^{\mathrm{e}} (E)= \int 
\limsup_{R\to\infty} \frac{\mathcal{F}^{\carr_R}(E, m)}{R^\d}  dP_n^{\mathrm{e}} (E) $$ in view of the formula \eqref{limsupsansC}. 
Since $P_n^{\mathrm{e}}$ is stationary, we may also write that for all $R>0$, 
$$\int 
\limsup_{R\to \infty} \frac{\mathcal{F}^{\carr_R}(E, m)}{R^\d} 
 dP_n^{\mathrm{e}} (E) = \frac{1}{R^\d}\int 
 \mathcal{F}^{\carr_R}(E, m)  dP_n^{\mathrm{e}} (E) .
$$ 
Since we assumed $\int \W(\C, m) dP_n(\C) <C$, in view of \eqref{contrWen}, we have $$\frac{1}{R^\d} \int \(\int_{\carr_R\times \R^\k}  \yg |E_{\rrc}|^2 + \C( \carr_{R}) \) dP_n^{\mathrm{e}}(E) <C'.$$
By Lemma \ref{lemcompact}, we then deduce that $\{P_n^{\mathrm{e}}\}_n$ is tight, and we may assume, up to extraction that $\{P_n^{\mathrm{e}}\}_n$ converges to a limit $P^{\mathrm{e}}$, which is stationary and concentrated on gradients, hence we obtain the compactness of the sub-level sets. By compatibility, we must have that the push-forward of $P^{\mathrm{e}}$ by $\conf_m$ equals $P$. We may then conclude with the lower semi-continuity of $\mathcal{F}^{\carr_R}$ (application of \eqref{compiteE1} with $\mu^{(n)}=m$) and using the stationarity of $P$, that
 \begin{multline*}
 \liminf_{n\to \infty} \int \W(\C, m) dP_n(\C)= \liminf_{n\to \infty} \frac{1}{R^\d} \int 
 \mathcal{F}^{\carr_R}(E, m)  dP_n^{\mathrm{e}} (E) \\ 
 \ge \frac{1}{R^\d} 
\int  \mathcal{F}^{\carr_R}(E, m)  dP^{\mathrm{e}} (E)= \int \mathcal{W}(E , m) dP^{\mathrm{e}}(E)
 . \end{multline*}
Since $\int \mathcal{W}(E, m) dP^{\mathrm{e}}(E)
\ge \int\W(\C,m) dP(\C)$ by definition of $\W$, the proof of lower semi-continuity is complete.

\end{proof}

\section{Next order asymptotics for the minimal energy}

We may now complete Theorem \ref{th4} and conclude the next order asymptotics of $\HN$ at the level of minimizers. 
First we have 
\begin{coro}\label{corofw}
Assume \eqref{riesz}. With the notation of Theorem \ref{th4}, we have 
\be \label{egalfw}
\f_\d(\infty)= \min \W(\cdot, 1)= \lim_{R\to \infty} \frac{\mathsf{E}_\infty(1,\carr_R)}{R^\d}.\ee
\end{coro}
We present here the proof for $\s=\d-2$, which relies on the screening procedure presented in these notes, and remark that \eqref{mn1} provided a rate $1/R$ of convergence of the limit in \eqref{egalfw}.
In the case   $\s\in (\d-2,\d)$, it suffices to use the Riesz screening procedure of \cite{PetSer} instead since we do not need the quantitative convergence provided by \eqref{mn1}. 
\begin{proof}
One inequality can be obtained by comparing \eqref{mn4} to \eqref{gliminfloc} (for instance in the case $\mu_N=\mu=1$ in $\Sigma$).
The converse inequality is \eqref{infFcr} combined with Corollary \ref{coroexmin} and  \eqref{mn1}.
\end{proof}

We next turn to completing the analysis of  minimizers of $\HN$ or of $\F(\XN, \mu)$, which in view of the splitting formula \eqref{split0} are the same question. The result \eqref{gliminf} of Proposition~\ref{prop:LowerBoundenergies} combined with \eqref{bbeloww} and \eqref{minwreex} provides a lower bound. This lower bound is sharp because  a matching upper bound can be obtained thanks to the screening procedure: this consists in partitioning the support of $\mu$ into quantized hyperrectangles $Q_i$ of large microscopic size, and pasting in each of  them  screened  minimizers of $\W(\cdot, \mu_i)$ (as done for showing \eqref{infFcr}) where $\mu_i$ is the average value of $\mu$ in $Q_i$. One has to separately treat the boundary layer, which cannot be exactly tiled by hyperrectangles, via a rougher bound.
In the general Riesz case, it can be done in the same way, using the Riesz screening, see \cite{PetSer}. An additional difficulty arising in the Riesz case is that the screening procedure requires a lower bound on the density, here on $\mu_V$, but in Riesz cases, as seen in Remark \ref{remvanishingrate}    typical equilibrium measures vanish like $\dist(x, \partial \Sigma)^{1-\frac{\d-\s}{2}}$ as one approaches the boundary of their support. To deal with this, a boundary layer must be removed and treated separately with rougher estimates. We refer to \cite[Section 7]{PetSer} for details.

The fact that the upper and lower bounds match  implies that for minimizers there must be equality in \eqref{gliminf} and also in \eqref{gliminfloc}. This allows to identify the limits of the empirical fields $\bar P$ or $P$ as minimizers of the corresponding version of $\W$. We now recount all this in the following.

\index{energy minimizers}

\begin{theo}[The case of minimizers]
\label{theominifin}
Assume $\s\in [\d-2,\d)$. Assume $V$ satisfies \eqref{A1}--\eqref{A3} so that the equilibrium measure $\meseq$ exists and is compactly supported in $\Sigma$. Assume also that $\meseq$ is   H\"older continuous in $\Sigma$  and that  $\partial \Sigma$ is $ C^1$.
 In the Coulomb case $\s=\d-2$, also assume that $\mu_V (x) \ge m>0$ for all $x\in\Sigma$. 
We have 
\begin{multline}\label{devminh}
\min \HN = N^2 \mathcal{E} (\mu_V)- \frac{1}{2\d} (N \log N) \indic_{\s=0}\\+
N^{1+\frac\s\d}  \int_\Sigma \mu_V(x)^{1+\frac\s\d}\min \W(\cdot, 1) - \indic_{\s=0} \(\frac{N}{2\d}\int_{\Sigma} \mu_V(x)\log \mu_V(x)\)+o(N^{1+\frac\s\d}).\end{multline}
Moreover, if $\XN$ is a minimizer of $\HN$, then up to extraction, $\bEmp_N[\XN]$ as defined in \eqref{def:bEmp} converges to some $\bP\in \probas_{s,\meseq}(\Sigma \times \config)$
which is such that for almost every $x\in\Sigma$, the disintegration $ \bar P^x$ minimizes $P\mapsto \int \W(\C , \meseq(x)) dP(\C)$. 
\end{theo}
We note that in the Coulomb case $\s=\d-2$ we can obtain a more quantitative estimate of the $o(N^{1+\frac\s\d})$, as a power of $N$, see \cite{as}.
\begin{theo}[The case of minimizers, local result]\label{theominiloclaw}
Assume $\s=\d-2$ and the assumptions of the above theorem.
If $\XN $ is a minimizer of $\HN$, then for $R \gg 1$ as $N \to \infty$, for any $x_0^N \to x_0\in \Sigma $ such that $\dist (x_0^N, \pa \Sigma ) \ge C N^{-\frac{2}{\d(\d+2)}} $,  we have that, up to extraction, 
$P_N^{x_0^N, R}[\XN]\to P$ as $N \to \infty$ with $P$-a.e.~$\C$  minimizing $\mathbb{W} (\C, \meseq(x_0))$. Moveover,  
\be\label{egaliteloc}\lim_{N\to \infty} \frac{1}{R^\d} \F^{\carr_{R}(N^{1/\d}x_0^N)}(\XN',\mu_V')= \min  \mathbb{W}(\cdot, \meseq(x_0)).\ee
\end{theo}

\begin{proof}[Proof of Theorem \ref{theominifin} in the Coulomb case]
We note that \eqref{devminh} was already obtained as \eqref{expansionminimizers} in Theorem \ref{thglob2} assuming that $\meseq$ is Lipschitz on its support. The Lipschitz assumption there allowed to obtain a precise rate, but could be relaxed to just H\"older continuous. 
 
Next, combining 
\eqref{split0} and \eqref{scalingF}
we have 
\be \label{split32}\HN(\XN) = N^2 \I(\meseq) -\( \frac{N}{2\d}\log N\) \indic_{\s=0}+ N\sum_{i=1}^N \zeta(x_i)+  N^{\frac\s\d} \F(\XN',\meseq'),  \ee
and inserting the result of  Proposition \ref{prop:LowerBoundenergies} and using that $\zeta \ge 0$, we are led to
\begin{equation}\label{lowebd}
\HN(\XN) \ge N^2 \I(\meseq) -\( \frac{N}{2\d}\log N\) \indic_{\s=0} + N^{1+\frac{\s}{\d}}
\overline \W(\bP,\meseq)
+o(N^{1+\frac\s\d}) \end{equation}
where $\bP\in \mc{P}_s(\Sigma \times \config)$ is the limit (up to extraction) of $\bar P_N[\XN]$. Combining  this with \eqref{bbeloww} and \eqref{minwreex} provides the lower bound of \eqref{devminh}.
Comparing with the upper bound of \eqref{devminh} means that there must be equality in \eqref{lowebd}. It  follows that if $\XN$ minimizes $\HN$, the limit  points $\bP$ of the empirical field $\bar P_N[\XN]$ obtained above must minimize 
$\overline \W(\bP,\meseq)
$.  In particular this implies that for a.e. $x\in\Sigma$, $\bP^x$ minimizes $P\mapsto \int \W(\C, \meseq(x)) dP(\C).$

\end{proof}
\begin{proof}[Proof of Theorem \ref{theominiloclaw}] By \eqref{mn2} we already know that the number of points in $\carr_R(N^{1/\d}x_0^N)$ is controlled by $C R^\d$.  This yields the subsequential convergence of $P_N^{x_0^N,R}$ to some $P$,  with \eqref{gliminfloc}.  Moreover, \eqref{egaliteloc} is a consequence of \eqref{mn4} in Theorem \ref{th4}  combined with Corollary \ref{corofw}.
Thus, there must be    equality in  \eqref{gliminfloc}, hence   $P$-a.e. $\C$ minimizes $\mathbb{W}(\cdot, \meseq(x_0))$.
\end{proof}

We have obtained that the limit local processes, after some averaging at a large microscale, minimize $\W$. In view of the discussion of the previous chapter, and modulo a form of uniqueness of minimizers, we  expect these local limits to be lattices when in dimensions 1, 2, 8 and 24. 

\chapter{LDP for empirical fields}
\label{chap:ldp}
We now turn to the case with temperature and prove as a counterpart  of Theorems \ref{theominifin} and \ref{theominiloclaw} a global and local Large Deviations Principle for the push-forward of the Gibbs measure by the empirical field map, i.e.~at the level of   \eqref{def:bEmp} or \eqref{defii}, in terms of local point processes. This was first obtained in \cite{lebles}, but here the presentation is simplified by the new definition of $\W$ which avoids the use of two extra parameters in \cite{lebles}. We also prove the local version originally  found in  \cite{as}.
This will provide a variational interpretation for the free energy per unit volume (or pressure)  $\mf(\beta)$ introduced in Theorem \ref{th1} and a variational characterization of the sine-$\beta$ and Ginibre point processes.
Let us point out that in the hypersingular case $\s>\d$, an LDP in terms of local point processes was obtained in \cite{hlss}. Because of the divergent nature of the interaction, there is no equilibrium measure and no splitting formula in that case, instead the density effects, governing the behavior of the limiting empirical measure, play at the same order as the microscopic effects governing the local point processes.

While the energetic aspects are similar to the case of minimizers (lower bound via Proposition \ref{prop:LowerBoundenergies} and upper bound by screening), we now have to deal with the entropic effects. For that we need an analogue of the entropy to  use in  the way Sanov's theorem was used in Chapter \ref{chap:leadingorder}. The adapted notion of entropy is the specific relative entropy, and the first step will be to  show how it appears as a large deviations rate function for the reference measure.

\section{Specific relative entropy} \label{sec:defentropy}
\index{specific relative entropy}
\subsection{Definitions}
Let us start by defining the analogue of the entropy at the level of point processes, which is  the {\it specific relative entropy} with respect to the Poisson point process. It   can be found in  early papers on empirical fields, for instance \cite{FollmerOrey}, and we refer to the books \cite{seppalainen,velenik} for more detail.

 We recall that the Poisson point process with intensity $m$ is the point process characterized by the fact that for any bounded Borel set $B$ in $\R^\d$, we have
$$\mathbb{P}\( N(B)= n\)= \frac{(m|B|)^n}{n!}e^{-m|B|}$$
where $N(B)$ denotes the number of points in $B$. The expectation of the number of points in $B$ can then be computed to be $m|B|$, and one also observes that the number of points in two disjoint sets are independent, thus the points ``do not interact". 

For any $m \geq 0$, we denote by $\Poisson^m$ the (law of the) Poisson point process of intensity $m$ in $\Rd$, it is an element of $\probas_s(\config)$. Let $P$ be in $\probas_s(\config)$. We define the specific relative entropy of $P$ with respect to $\Poisson^1$ as
\begin{equation} \label{def:ERS}
\ERS[P|\Poisson^1] := \lim_{R \ti} \frac{1}{R^{\d}} \ent[P_{\carr_R}|\Poisson^1_{\carr_R}],
\end{equation}
where $P_{\carr_R}, \Poisson^1_{\carr_R}$ denote the restriction of the respective processes to the hypercube $\carr_R$. Here, $\ent[\cdot|\cdot]$ denotes the \textit{usual} relative entropy of two probability measures defined on the same probability space, namely
\begin{equation*}
\ent[\mu|\nu] := \int \log \frac{d\mu}{d\nu} \, d\mu
\end{equation*}
if $\mu$ is absolutely continuous with respect to $\nu$, and $+ \infty$ otherwise. 

\begin{lem} \label{lem:ERS} The following properties are known:
\begin{enumerate}
\item If $P$ is stationary, the limit in \eqref{def:ERS} exists.
\item The map $P \mapsto \ERS[P |\Poisson^1]$ is affine and lower semi-continuous on $\probas_s(\config)$.
\item The sub-level sets of $\ERS[\cdot| \Poisson^1]$ are compact in $\probas_s(\config)$ (it is a \textit{good} rate function).
\item We have $\ERS[P| \Poisson^1] \geq 0$ with equality if and  only if $P = \Poisson^1$.
\item
We have the following scaling results: $\sigma_m$ being as in \eqref{defsigmam},
\be
\label{scalingentropie}  \ERS[\Pst|\Poisson^1] = m\, \ERS[(\sigma_m \Pst)|\Poisson^1] + m\log m+1-m,\ee
and 
\be \label{scalingentropie2} 
\ERS[P|\Poisson^1]=  \ERS[P|\Poisson^m]+ m\log m+1-m.\ee
\end{enumerate}
\end{lem}
\begin{proof}
We refer to \cite[Chapter 6]{seppalainen} or \cite{velenik} for a proof. The first point follows from sub-additivity, the third and fourth ones from usual properties of the relative entropy. The fact that $\ERS[\cdot|\Poisson^1]$ is an \textit{affine} map, whereas the classical relative entropy is strictly convex, is due to the infinite-volume limit taken in \eqref{def:ERS}.   The scaling result is in \cite[Lemma 4.4]{lebles}.
\end{proof}

Next, if $\bar{P}$ is in $\probas_{s}(\Sigma \times \config)$ as defined in the previous chapter, given a density $\mu(x)$ over $\Sigma$, we define the tagged  specific relative entropy as
\begin{equation}
\label{def:bERS} \bERS[\bPst|\Poisson^\mu] := \int_{\Sigma} \ERS[\bPstx|\Poisson^{\mu(x)}] dx.
\end{equation}

\subsection{LDP for empirical fields without interaction} \label{sec:empwihtoutint}
\index{large deviations principle} \index{Sanov's theorem}
The main result we will use is a large deviation principle (recall the terminology from Section \ref{LDPsection})  for the tagged empirical field \eqref{def:bEmp}, when the points are distributed according to a reference measure on $(\Rd)^N$ where there is no interaction. This is a microscopic or  ``type III" (in the LDP jargon) analogue of Sanov's theorem for the empirical measures. 

\begin{prop}[Large Deviations for the reference measure] 
 \label{SanovbQN} Let $\{\mu_N\}_N$ be a sequence of probability densities on $\R^\d$ converging locally uniformly to $\mu$ in $\Sigma$, where  $\partial \Sigma\in C^1$, $\mu$ is continuous and bounded below by a positive constant in $\Sigma$. Then 
the push-forward of $\mu_N^{\otimes N}$ by the map $\XN\mapsto \bEmp_N[\XN]$ of \eqref{def:bEmp} satisfies a LDP at speed $N$ with good rate function $\bERS[\bP |\Poisson^{\mu}]$. In particular, for any $\bP\in \probas_{s,\mu}(\Sigma \times \config)$, we have
\be  \label{ldprefub}\limsup_{\ep \to 0} \limsup_{N\to \infty}\frac1N \log \mu_N^{\otimes N } \(\XN, \bEmp_N[\XN] \in B(\bP, \ep) \) \le - \int_\Sigma \ERS[\bP^x|\Poisson^{\mu(x)}]dx
\ee
and for any $\bP\in \probas_{s}(\Sigma \times \config)$ we have
\be \label{ldpreflb} \liminf_{\ep \to 0 } \liminf_{N\to \infty} \frac1N \log  \mu_N^{\otimes N }  \(\XN, \bEmp_N[\XN] \in B(\bP, \ep) \) \ge - \int_\Sigma \ERS[\bP^x|\Poisson^{\mu(x)}]dx
\ee
where the balls are taken for a distance metrizing the weak topology on $\probas(\Sigma\times \config)$.
\end{prop}

Early LDPs for empirical fields can be found in \cite{varadhansf,follmersf}, the  specific relative entropy is formalized in \cite{FollmerOrey} (for the non-interacting discrete case), \cite{georgii,Olla} (for the interacting discrete case) and \cite{georgiizessin} (for the interacting continuous case). In the light of these results, Proposition \ref{SanovbQN} is not surprising, but there are some technical differences. In our case, the reference measure $\mu_N^{\otimes N}$ is not the restriction of a Poisson point process to a hypercube and is not uniform. Moreover we want to study large deviations for tagged point processes (our \textit{tags} are not the same as the \textit{marks} in \cite{georgiizessin}) which requires an additional argument. These adaptations are largely drawing on ingredients from  \cite{lebles}. The main starting point is the following result from \cite{georgiizessin}.
\begin{prop}[LDP for Poisson's empirical field, \cite{georgiizessin}]\label{progz}
Let $\{\Lambda_N\}_N$ be a sequence of cubes increasing to $\R^\d$ and let $R_N$ be the push-forward of $\Poisson^1$ by the map 
$$\C\mapsto \frac{1}{|\Lambda_N|} \int_{\Lambda_N} \delta_{\theta_x \cdot \C} dx$$
where $\theta_x$ denotes the translation by $x$. Then $\{R_N\}_N$ satisfies a LDP at speed $|\Lambda_N|$ with rate function $\ERS[\cdot|\Poisson^1]$.\end{prop}
In \cite{lebles} it is adapted to the case of tagged point processes and the case of more general shapes than cubes. We recall that the $N$-point Bernoulli process in $\Lambda$ is the law of $N$ points chosen uniformly and independently in $\Lambda$. 
In particular the following is  proven in  \cite[Lemma 7.8]{lebles}.
\begin{lem}[LDP for Bernoulli point processes]\label{lemBernoulli}
Let $\Lambda$ be a compact set of $\R^\d$ with $C^1$ boundary and nonempty interior and let $S_N$ be the push-forward of the $N$-point Bernoulli process in $N^{\frac1\d} \Lambda$ by the map 
$$\C\mapsto \frac{1}{N|\Lambda|} \int_{N^{\frac1\d}\Lambda}  \delta_{(N^{-1/\d}x, \theta_x \cdot \C)} dx.$$
Then $\{S_N\}_N$ satisfies a LDP at speed $N$ with good rate function $\bar P \mapsto \int_{\Lambda} \ERS[\bP^x|\Poisson^{|\Lambda|^{-1}}] dx$.
\end{lem}
In \cite{lebles} the rate function is written as 
$\int_{\Lambda} \ERS[\bP^x|\Poisson^1] dx+ \log |\Lambda|-|\Lambda|+1$.
In view of \eqref{scalingentropie2} and viewing it as a process of intensity $m=|\Lambda|^{-1}$ we can also rewrite it exactly as
$$
 \int_{\Lambda} \ERS[\bP^x|\Poisson^1] dx+ \log |\Lambda|-|\Lambda|+1
 =
 \int_{\Lambda} \ERS[\bP^x|\Pi^m] dx .$$

We will use the following slightly more general version.
\begin{lem}\label{Bernoulligen}
Let $\Lambda$ be a compact set of $\R^\d$ with $C^1$ boundary and nonempty interior.
Also assume that  $\{n\}_N$ is a sequence of integers such that 
$$\lim_{N\to\infty} \frac{n}{N|\Lambda|}=m.$$ Let $S_N$ be the push-forward of the $n$-point Bernoulli process in $N^{\frac1\d} \Lambda$ by the map 
$$\C\mapsto \frac{1}{N|\Lambda|} \int_{N^{\frac1\d}\Lambda}  \delta_{(N^{-1/\d}x, \theta_x \cdot \C)} dx.$$
Then $\{S_N\}_N$ satisfies a LDP at speed $N$ with  good rate function $\bar P \mapsto \int_{\Lambda} \ERS[\bP^x|\Poisson^{m}] dx$.

Let $R_N$ be the push-forward of the $n$-point Bernoulli process in $N^{\frac1\d} \Lambda$ by the map 
$$\C\mapsto \frac{1}{N|\Lambda|} \int_{N^{\frac1\d}\Lambda}  \delta_{ \theta_x \cdot \C} dx.$$ Then $\{R_N\}_N$ satisfies a LDP at speed $N|\Lambda|$ with good rate function $\ERS[\cdot|\Poisson^{m}]$.
\end{lem}

Before we get to the proof of Proposition \ref{SanovbQN}, let us state a lemma that allows to reduce to the uniformly convergent situation.
\begin{lem}\label{lemsigmaeta}
Assume that $\Sigma_\eta\subset \Sigma$ is such that $|\Sigma\backslash \Sigma_\eta|\to 0$ as $\eta \to 0$. Then letting 
$$\bP_{N,\eta}[\XN]:=\frac{1}{|\Sigma_\eta|}\int_{\Sigma_\eta} \delta_{(x, \theta_{N^{1/\d}x} \cdot \XN') } dx,$$
we have that $$\dist_{\probas_s(\Sigma \times \config)} (\bP_{N,\eta}[\XN], \bEmp_N[\XN]) \to 0, \quad \text{as} \ \eta\to 0.$$
Let $\mathfrak{P}_{N,\eta}$ be the push-forward of some probability measure $\mathbb{P}_N$ on $(\R^\d)^N$ by $ \XN \mapsto 
\bP_{N,\eta}[\XN]$, respectively, $\mathfrak{P}_{N}$ be the push-forward of  $\mathbb{P}_N$ by $\XN \mapsto \bEmp_N[\XN]$. If $\mathfrak{P}_{N,\eta}$ satisfies a LDP at speed $a_N$ with good rate function $I_\eta$, $ I_\eta \to  I $ pointwise as $\eta \to 0$, and   $\mathfrak{P}_N$ is exponentially tight at speed $a_N$, then $\mathfrak{P}_{N}$ satisfies a LDP with good rate function $I.$\end{lem}
\begin{proof}
The first statement follows from the definition of $\bEmp_N[\XN]$ \eqref{def:bEmp} and the fact that the topology on $\probas_s(\Sigma \times \config)$ only allows to test against bounded functions. The second statement follows from the definition of LDPs after reducing to statements over balls by exponential tightness, see Corollary \ref{ldpboules}.
\end{proof}

\begin{proof}[Proof of Proposition \ref{SanovbQN}]
First, using Lemma \ref{lemsigmaeta}, we can work in a subset $\Sigma_\eta \subset \Sigma$ such that $\partial \Sigma_\eta$ is piecewise $C^1$, $|\Sigma\backslash \Sigma_\eta|\to 0$ as $\eta \to 0$ and 
$\mu_N\to \mu$ uniformly in $\Sigma_\eta$, as $N\to \infty$. We thus reduce to a situation where $\mu_N \to \mu$ uniformly, and will from now on drop the $\eta$ and  assume that we are in the uniform convergent situation in $\Sigma$.
\smallskip

{\bf Step 1. Reduction to the piecewise constant case.} 
Let $\eta >0$ (different from the $\eta$ just above). 
We may first  partition $\Sigma $ into cells $Q_i$ with piecewise $C^1$ boundary, $i=1, \dots, p$,  of diameter $\le \eta$ and aspect ratios bounded above and below,
requiring for instance that each $Q_i$ is included in a ball of radius $\eta$ and contains a ball of radius $\hal \eta$. 
We then let  $$ \mu_{N, \eta}= \sum_{i=1}^p  \indic_{Q_i}m_{N,i}, \qquad m_{N,i}:= \dashint_{Q_i}\mu_N, $$
and 
$$\mu_\eta= \sum_{i =1}^p \indic_{Q_i} m_i, \qquad m_i= \dashint_{Q_i} \mu$$
i.e.~$\mu_{N,\eta}$ is a piecewise constant approximation of $\mu_N$, with ``mesh size" $\eta$, and the same for $\mu_\eta$. By assumed uniform convergence of $\mu_N$ to $\mu$,  we have $\mu_{N,\eta} \to \mu_\eta$ uniformly. Since $\mu_\eta \to \mu$ as $\eta\to 0$, and since $\mu$ is bounded below in $\Sigma$, 
given $\delta>0$ we may choose $\eta>0$ small enough so that 
the Radon-Nikodym derivative of $\mu_{N,\eta}$ with respect to $\mu_N$ satisfies, for $N$ large enough, 
\be\label{mumueta}
1-\delta< \left|\frac{\mu_{N,\eta}}{\mu_N} \right|<1+\delta.\ee
 The  uniform convergence also implies  that 
\be \label{limmni}\lim_{N\to \infty} m_{N, i} = m_i.\ee

 We also let $\mn_i $ be an integer equal to $ N\int_{Q_i} \mu_N$ up to an error $\le 1$. We may choose them so that $\sum_{i=1}^p \mn_i=N$.
 
In view of \eqref{mumueta}, given an event $A$ we may write that
$$ \log \( (1-\delta)^N \mu_{N,\eta}^{\otimes N}(A)\)\le \log \mu_N^{\otimes N}(A) \le \log \( (1+\delta)^N \mu_{N,\eta}^{\otimes N}(A)\)$$ thus 
\be\label{mumueta1}(1-\delta)+ \frac{1}{N}  \log \mu_{N,\eta}^{\otimes N}(A)\le \frac1N 
\log \mu_N^{\otimes N}(A)\le (1+\delta) +  \frac{1}{N}  \log \mu_{N,\eta}^{\otimes N}(A),\ee
hence since in order to prove \eqref{ldprefub} and \eqref{ldpreflb} we need to   evaluate limits as $N\to \infty$ of $\frac1N\log \mu_N^{\otimes N}(A)$, it suffices to do so with $\mu_N$ replaced by  $\mu_{N,\eta}$, then let $\delta\to 0$. 
\smallskip

 {\bf Step 2. Lower bound.}   Let $\bP\in \probas_{s} (\Sigma\times \config)$.
Let $X_{\mn_i}$ be configurations of $\mn_i$ points in $Q_i$ and let  $\bEmp_{N,i}[X_{\mn_i}] $ be the associated tagged empirical process in $\probas_s(Q_i\times \config)$ as in \eqref{def:bEmp} with $\Sigma$ replaced by $Q_i$. Let $\bP_i$ be the restriction of $\bP$ to $Q_i$ i.e.~an element of $\probas_s( Q_i\times \config) $ obtained by restricting the first variable to $Q_i$. 
We note that $\frac{1}{(N \int_{Q_i} \mu_N)^{\mn_i}}(N  \mu_{N,\eta}|_{Q_i})^{\otimes \mn_i} $ can be identified with an $\mn_i$-point Bernoulli process in $Q_i$, and apply
 Lemma \ref{Bernoulligen} (after zooming by $N^{1/\d}$) with $n=\mn_i \to \infty$  to obtain, in view of  \eqref{limmni},  
that 
\be\label{1310}  \liminf_{\ep \to 0} \liminf_{N\to\infty} \frac{1}{N} \log \frac{(N \mu_{N,\eta}|_{Q_i})^{\otimes \mn_i} }{ (N \int_{Q_i} \mu_N)^{\mn_i}  } \( X_{\mn_i}, \bEmp_{N,i}[X_{\mn_i}]\in B(\bar P_i, \ep) \) \ge - \int_{Q_i} \ERS[(\bP_i)^x |\Poisson^{m_i}] dx .\ee
We can glue together  the $X_{\mn_i} $ to obtain a configuration $\XN$ in $\cup_{i=1}^p Q_i$.
Since $\bEmp_{N,i}[X_{\mn_i}]\in B(\bar P_i, \ep)$ for every $i$ and $p$ depends only on $\eta$, we find that 
$\bEmp_N[X_N] \in B(\bP, \varphi(\ep))$  where $\varphi(\ep)$ is a function that tends to $0$ as $\ep \to 0$,  which depends on $\eta $ but not on $N$.  We will use that notation throughout the proof, with a $\varphi$ that may change. 

Using that $\mu_{N,\eta}= \sum_{i=1}^p \mu_{N,\eta}|_{Q_i}$ and expanding 
$( \sum_{i} \mu_{N,\eta}|_{Q_i})^{\otimes N}$, retaining only the  terms that correspond to $\mn_i$ points for each $i\in [1,p]$, 
 we find
$$ \mu_{N,\eta}^{\otimes N} \ge \frac{N!}{\prod_{i=1}^p \mn_i!} (\mu_{N,\eta}|_{Q_i})^{\otimes \mn_i}.$$
We may thus write that
\begin{multline}  \frac{1}{N} \log (\mu_{N,\eta})^{\otimes N}\( X_N, \bEmp_N[X_N] 
\in B(\bP, \varphi(\ep)) \) \\ \ge 
\sum_{i=1}^p  \frac{1}{N} \log \frac{(N\mu_{N,\eta}|_{Q_i})^{\otimes \mn_i} }{ (N \int_{Q_i} \mu_N)^{\mn_i}}
 \( X_{\mn_i}, \bEmp_{N,i}[X_{\mn_i}]\in B(\bar P|_{Q_i}, \ep) \)+ \frac{1}{N}\log \( \frac{N!}{N^N} \prod_{i=1}^p \frac{  (N \int_{Q_i} \mu_N)^{\mn_i} }{ \mn_i!}  \).
 \end{multline}
  Using Stirling's formula, $| N \int_{Q_i} \mu_N-\mn_i|\le 1$ and $\sum_i \mn_i= N$ and \eqref{1310}, we deduce 
  that
  \begin{multline}   \liminf_{\ep \to 0} \liminf_{N\to\infty}\frac{1}{N} \log (\mu_{N,\eta})^{\otimes N}\( X_N, \bEmp_N[X_N] 
\in B(\bP,\varphi(\ep)) \)  \\
\ge - \sum_{i=1}^p \int_{Q_i} \ERS[(\bP_i)^x|\Poisson^{m_i}  ]dx.
\end{multline}
We next use \eqref{mumueta1}, the continuity of $
\int_{\Sigma} \ERS[\cdot|\Poisson^{m}]dx$ with respect to $m$, which is implied by 
\eqref{scalingentropie},  and let $\eta\to 0$ to obtain 
\begin{multline}   \liminf_{\ep \to 0} \liminf_{N\to\infty}\frac{1}{N} \log (\mu_{N})^{\otimes N}\( X_N, \bEmp_N[X_N] 
\in B(\bP, C\ep) \)  \\
\ge - \int_{\Sigma} \ERS[(\bP_i)^x|\Poisson^{\mu(x)}  ]dx, 
\end{multline}as desired.
\smallskip

{\bf Step 3. Upper bound}.
Let $\bP\in \probas_{s, \mu} (\Sigma\times \config)$, see  the definition in Section \ref{sec:tagged}. We know that for Lebesgue-a.e.~$x \in \Sigma$, $\bP^x$ has intensity $\mu(x)$, hence  we have   
\be \label{intens}
\int \#( \C\cap  \carr_1) d\bP^x(\C)= \mu(x).\ee

Let $\XN$ be a configuration of $N$ points with $\bEmp_N[\XN]\in B(\bP, \ep)$. We let $\bEmp_{N,i}$ be as above the  tagged empirical fields of the restriction of the configuration in $Q_i$. When $\XN$ is drawn from $(\mu_{N,\eta})^{\otimes N}$, these   are independent. Moreover, since $\bEmp_N[\XN]\in B(\bP, \ep)$,  then $\bEmp_{N,i}\in B(\bP_i, \varphi(\ep))$.
Indeed we remark that 
$$ \dashint_{\Lambda} \delta_{(x,\theta_{N^{1/\d} x} \cdot  X_N'|_{\Lambda} )} dx\  \ \mathrm{ and } \  \ \dashint_{\Lambda} 
\delta_{(x,\theta_{N^{1/\d}  x} \cdot  X_N' )} dx
$$ 
i.e.~the empirical fields with restricted or unrestricted configurations are close in the local topology when $\partial \Lambda$ is piecewise $C^1$.

We now wish to evaluate  the number of points $n_i(\XN)$ a configuration with similar tagged empirical field has in the set $Q_i$ by using \eqref{intens} and the closeness of $\bEmp_{N,i}$ to $\bP_i$.

The idea is to use the fact that by definition of $\bEmp_{N,i}$, 
\begin{multline}\label{Ccar1}
\int  \#( \C\cap  \carr_1) d\bEmp_{N,i}=\frac{1}{|Q_i|} \int_{Q_i} \#( \C\cap  \carr_1) \delta_{(x,\theta_{N^{1/\d}x}\cdot \XN')}(\C) dx\\
= \frac{1}{|Q_i|}\int_{Q_i}\#\{\theta_{N^{1/\d}x}\cdot \XN'\cap \carr_1\} dx= \frac{1}{|Q_i|} \int \# \{\XN \cap (x+\carr_{N^{-1/\d}})\} dx\end{multline}
and that by the fact that $\partial Q_i$ is piecewise $C^1$ and Fubini's theorem and $|\carr_{N^{-1/\d}}|=N^{-1}$, the right-hand side is  equal (up to a boundary error which is $o_N(1)$) to $|Q_i|^{-1}N^{-1}$ times the number of points of $\XN$ in $Q_i$.
We then wish to use the closeness of $\bEmp_{N,i} $ to $\bP_i$ to deduce, by comparing \eqref{Ccar1} and \eqref{intens} that 
\be\label{ninn}\frac{n_i(\XN)}{N|Q_i|}= \int_{Q_i} \mu(x) dx+o_N(1)+o_\ep(1).\ee
This is however not quite correct, because the $\ep$-closeness we know is in the local topology and we can only test against bounded and continuous local functions, and $\#\{\C\cap \carr_1\}$ is neither bounded nor continuous for the local topology. To remedy this, we can argue as in the proof of Lemma~7.8 in \cite{lebles} by first approximating this function by continuous ones (integrating against a smooth cutoff instead of the indicator of $\carr_1$) and then truncating the function at level $S$, and checking that the errors due to the truncation become negligible as $S \to \infty$.

We now deduce from \eqref{ninn}, the definition of $\mn_i$ and the convergence of $\mu_N$ to $\mu$   that 
\be \label{escardinal}
   |n_i(\XN)-\mn_i| \le N (o_N(1) +  o_\ep(1)):= N \rho_{N,\ep} .\ee

Again, $\frac{ (N \mu_{N,\eta}|_{Q_i})^{\otimes n_i} }{ (N \int_{Q_i} \mu_N)^{n_i}  }$ can be identified with an $n_i$-Bernoulli point process in $Q_i$, hence 
Lemma \ref{Bernoulligen}  gives this time 
\be \label{Pii} \frac{1}{N} \log \frac{ (N \mu_{N,\eta}|_{Q_i})^{\otimes n_i} }{ (N \int_{Q_i} \mu_N)^{n_i}  }\( X_{n_i}, \bEmp_{N,i}[X_{n_i}]\in B(\bar P|_{i}, \ep) \) \le - \int_{Q_i}  \ERS[(\bP_i)^x|\Poisson^{m_i} ] dx +o_N(1)+o_\ep(1).\ee
In view of \eqref{escardinal} we have 
\be\label{Pi2}\mu_{N,\eta}^{\otimes N} (\bEmp_N [\XN]\in   B(\bP, \ep))
 =  \sum_{\substack{n_i, \sum_i{ n_i}=N\\
 |n_i-\mn_i|\le N\rho_{N,\ep} }
 } \mu_{N,\eta}^{\otimes N} (B(\bP, \ep) \indic_{ \cap_{i=1}^p \{\# \{\XN \cap Q_i\}= n_i\} })\ee
but by independence and the above, we have
$$
   \mu_{N,\eta}^{\otimes N}(B(\bP, \ep) \indic_{ \cap_{i=1}^p \{\# \{\XN \cap Q_i\}= n_i\} })
  \le \frac{N!}{\prod_{i=1}^p n_i!} \prod_{i=1}^p ( \mu_{N,\eta}|_{Q_i} )^{\otimes n_i} (  \bEmp_{N,i}[X_{n_i}]\in B(\bP_i, \varphi(\ep))) .$$
 Inserting \eqref{Pii},  then reinserting into \eqref{Pi2}, we obtain 
 \begin{multline*}
\frac1N\log  \mu_{N,\eta}^{\otimes N} (\bEmp_N [\XN]\in   B(\bP, \ep))\\  \le  
\max_{\substack{n_i, \sum_i n_i=N\\ |n_i-\mn_i|\le N\rho_{N,\ep} }} \( -  \sum_{i=1}^p \int_{Q_i}  \ERS[(\bP_i)^x|\Poisson^{m_i} ] dx + \frac1N \log \frac{  N! \prod_{i=1}^p \mn_i^{n_i}   }{N^N\prod_{i\in  I} n_i!  }\) +o_N(1)+o_\ep(1)\end{multline*} 
 hence the desired result after using Stirling's formula,  letting $N \to \infty$, $\ep \to 0$ then $\eta\to 0$.

We note that the second statement (about non tagged processes) can be deduced from the first one (about tagged processes) by applying the forgetful map $\phi : \probas(\Sigma \times \config) \to \probas (\config) $ obtained by pushing forward by $(x,\C) \mapsto \C$, and using that the specific relative entropy is affine.

 \end{proof}

\section{LDP for empirical fields}
\index{empirical field}
\subsection{Statements and consequences on limit point processes}Let us introduce the rate function of \cite{lebles}, it is defined over the set of stationary point processes of intensity $m$
(equipped with the topology of weak convergence) by
\be \label{defFbeta}
\mathcal I_\beta^m (P):=\beta \int \W (\C,m) dP(\C)  + \mathsf{ent}[P|\Pi^m],\ee
where $\Pi^m$ is the (law of the) Poisson process of intensity $m$ over $\R^\d$, and $\mathsf{ent}$ is the specific relative entropy. In view of  Lemma \ref{corolscW} and \ref{lem:ERS}, it is a good rate function.

To minimize~\eqref{defFbeta} there is a competition (depending on $\beta$)
 between the energy term $\W$ which prefers ordered configurations (remember that $\W$-minimizing configurations are expected to be crystalline in low enough dimensions, as seen in Chapter \ref{chap:renormalized}) and  the relative entropy term which favors disorder hence  configurations that are more Poissonnian. The choice of temperature scaling that we made in~\eqref{gibbs} is  precisely the one for which these two competing effects are of comparable strength for fixed $\beta$.
 The limiting regime where $\theta$ of \eqref{deftheta} is constant is treated in \cite{dpg2} in a similar way.

\index{large deviations principle}
\begin{theo}[Global Large Deviations Principle]
\label{theoldpglob}
Assume $\s\in [\d-2,\d)$.
Let $\{\mu_N\}_N$ be a sequence of probability densities on $\R^\d$ satisfying \eqref{condmupourFN} and converging locally uniformly to $\mu$ in a compact set  $\Sigma$ such that $\partial \Sigma\in C^1$, $\mu$ is continuous and bounded below by a positive constant in $\Sigma$. Assume that $\beta$ is such that $\K_{N,\beta}(\mu_N)$ is a convergent integral for each $N$ large enough.
 
 Let $ \mathfrak{P}_{N,\beta} $ be the push-forward of  $\Q_{N,\beta}(\mu_N)$ by $\XN\mapsto \bEmp_N[\XN]$ as defined in \eqref{def:bEmp}.
 Then, assuming $\theta=\beta N^{1-\frac\s\d} \to +\infty$ as $N \to \infty$,  we have the following.
\begin{itemize}
\item
If $\beta $ is independent of $N$,  the sequence $\{ \mathfrak{P}_{N,\beta}\}_N$ satisfies a LDP at speed $N$ with good rate function 
$$\int_{\Sigma}\mathcal I_{\beta}^{\mu(x)} (\bP^x) \,  dx -\min \int_{\Sigma}\mathcal I_{\beta}^{\mu(x)} (\cdot) \, dx  .$$
Moreover, 
\be\label{expacas1} \lim_{N\to \infty} \frac{1}{N} \log \K_{N,\beta}(\mu_N)-\(\frac{\beta}{2\d} \log N \)\indic_{\s=0} 
= - \min_{\probas_{s,\mu} (\Sigma \times \config) }\int_{\Sigma}\mathcal I_\beta^{\mu(x)} (\bP^x) dx
 .\ee

\item
If $\beta \to 0$ as $N\to \infty$, then  $\{ \mathfrak{P}_{N,\beta}\}_N$ satisfies a LDP at speed $N$ with good rate function $$\int_{\Sigma}\mathsf{ent}[ \bP^x |\Poisson^{\mu(x)}]\, dx 
.$$

\item
If $\beta \to \infty$ as $N\to \infty$, then 
$\{ \mathfrak{P}_{N,\beta}\}_N$ satisfies a LDP  at speed $\beta N$ with good  rate function
 $$ \overline \W(\bP, \mu)  -\min_{\probas_{s,\mu}(\Sigma \times \config)} \overline \W(\bP, \mu) .$$ Moreover, 
 \be\label{expacas3}\lim_{N\to \infty} \frac{1}{\beta N} \log \K_{N,\beta}(\mu_N)- \( \frac{1}{2\d } \log N\) \indic_{\s=0}
= -\min_{\probas_{s,\mu}(\Sigma\times \config)}\overline \W(\bP, \mu) .\ee
 \end{itemize}
  \end{theo} 
We next state the analogous local result in the Coulomb case (valid for point processes averaged at any scale larger than microscopic). 
\begin{theo}[Local large deviations principle in the Coulomb case]
\label{theoldp} 
Let $\s=\d-2$. Assume $\mu$ is continuous and bounded below by a positive constant in  $\Sigma$. Let $R,\beta$ be such that $N^{\frac1\d}\gg R \gg \rb$ as $N \to \infty$ and $x_0^N \to x_0 \in \Sigma$ satisfies $x_0^N \in \Sigma$ and $\dist (x_0^N, \partial\Sigma) \ge \bar d_0$ as in \eqref{defd0p}. 
Assume that  $\{\mu_N\}_N$ be a sequence of probability densities on $\R^\d$ satisfying \eqref{condmupourFN},  bounded below by a positive constant in $\Sigma$,
 and  such that $\K_{N,\beta}(\mu_N)$ is a convergent integral for each $N$ large enough.
Assume that $\mu_N$ converges uniformly to $\mu$ in $\carr_R( N^{1/\d}x_0^N) $.

Let $ \mathfrak{P}_{N,\beta}^{x_0,R}$ be the push-forward of  $\Q_{N,\beta}(\mu_N)$ by $\XN \mapsto P_N^{x_0^N,R}[\XN]$ defined in \eqref{defii}.

Then we have the following.
\begin{itemize}
\item
If $\beta $ is independent of $N$,  the sequence $\{ \mathfrak{P}_{N,\beta}^{x_0,R}\}_N$ satisfies a LDP at speed $R^\d$ with good rate function 
$\mathcal I_{\beta}^{\mu(x_0)}-\min \mathcal I_{\beta}^{\mu(x_0)}  .$
Moreover,  if $\int_{\carr_R}\mu_N' $ is an integer, we have
 \be \label{gformul1}\lim_{N\to \infty} \frac{1}{R^\d} \log \K_\beta(\mu_N', \carr_R)= -  \min_{P\in \probas_s(\config) } \mathcal I_\beta^{\mu(x_0)} .\ee
\item
If $\beta \to 0$ as $N\to \infty$, then  $\{ \mathfrak{P}_{N,\beta}^{x_0,R}\}_N$ satisfies a LDP at speed $R^\d$ with good rate function $\mathsf{ent}[ P|\Pi^{\mu(x_0)}]
$. Moreover,  if $\int_{\carr_R}\mu_N' $ is an integer, we have
 \be \lim_{N\to \infty} \frac{1}{ R^\d} \log \K_\beta(\mu_N',\carr_R)=0 .\ee
\item
If $\beta \to \infty$ as $N\to \infty$, then 
$\{ \mathfrak{P}_{N,\beta}^{x_0, R}\}_N$ satisfies a LDP  at speed $\beta R^\d$ with good  rate function
 $\mathbb{W} (\cdot , {\mu(x_0)}) -\min {\mathbb{W}}(\cdot, {\mu(x_0)}) $.
 Moreover,  if $\int_{\carr_R}\mu_N' $ is an integer, we have
 \be\label{gformul3}
 \lim_{N \to \infty}\frac{1}{\beta R^\d}  \log \K_\beta(\mu_N',\carr_R)= -  \min_{P\in \probas_s(\config) }   \int \mathbb{W}(\C,\mu(x_0))   dP(\C)  .\ee
 \end{itemize}
  \end{theo} 
  
In view of \eqref{rewritePNbeta}, applying to $\mu_N= \mub$ and $\mu=\meseq$, with  $\Sigma= \supp \, \meseq$, and using  Theorem \ref{th1as} for the local uniform convergence of $\mu_N$ to $\mu$,
we obtain the following.
\begin{coro}
Assume the hypotheses of Theorem \ref{th1as}. 
Assume that the equilibrium measure $\meseq$  is compactly supported in $\Sigma$ and H\"older continuous in $\Sigma$, where $\partial \Sigma\in C^1$.
 Then  $\mathfrak{P}_{N,\beta} $,  the push-forward of  $\PNbeta$ of \eqref{gibbs} by $\XN\mapsto \bEmp_N[\XN]$, respectively $\XN \mapsto P_N^{x_0,R}[\XN]$, satisfies the results of the corresponding theorem above. \end{coro}

The theorem will be proven in the Coulomb case, but the statement is also correct in the Riesz case $\s \in (\d-2,\d)$. Note that a similar LDP for empirical fields is also proven in the hypersingular case $\s>\d$ in \cite{hlss}. That regime is quite different in the sense that there is no equilibrium measure and the macroscopic density optimizer  is determined implicitly via a local density approximation that involves an optimization over the microscopic distribution of points -- macroscopic and microscopic arrangement of points play at the same order in the energy. On the other hand, the short range nature of the interaction allows to easily get almost additivity of the energy without the need for the screening procedure.

The results above allow to assert that local point processes obtained for instance in \cite{thoma} must minimize $\mathcal{I}_\beta$. As a corollary, we also obtain that the known point processes, Ginibre for $\d=2,\s=0$ and $\beta=2$, sine-$\beta$ for $\d=1,\s=0$, and Riesz-$\beta$ obtained in \cite{boursier23a} in the case $\d=1,\s\in (0, 1)$  must minimize $\mathcal I_\beta$. 

\begin{coro}[Variational characterization of the classical point processes]
\index{Ginibre point process} \index{sine-$\beta$ point process}
The point processes sine-$\beta$ and Ginibre minimize the corresponding $\mathcal I_\beta^1$ among stationary point processes of intensity $1$. \end{coro}

  Applying to $\mu_N=\mu=1$ in $\Sigma$, comparing \eqref{expacas1} and \eqref{expacas3}  with \eqref{expvar} and using \eqref{scalingWb}, we find the analogue of \eqref{egalfw} in the case with temperature, relating $\mf(\beta)$ of  Theorem \ref{th1} with $\inf \mathcal{I}_\beta$.
  This way the formulae \eqref{gformul1}--\eqref{gformul3} are the same as \eqref{expzcasgb} except without the Lipschitz assumption on $\mu$.

  \begin{coro}[Variational interpretation of the pressure]\label{corointervar}
For any $\beta \in [0,+\infty)$   we have 
  \be \label{intervarf}
 \beta \mf(\beta)= \min_{P\in  \probas_s(\config) }\mathcal{I}_\beta^1(P),\ee and 
for any $m \ge 0$, 
\be\label{inFfb}
\min_{P \in  \probas_s(\config) }\mathcal{I}_\beta^m(P)= \beta m^{1+\frac\s\d} \mf(\beta m^{\frac{\s}{\d}}) +\(1-\frac{\beta}{2\d} \indic_{\s=0} \) m\log m.\ee
\end{coro}

If $\s\neq 0$ an effective temperature $\beta m^{\frac{\s}{\d}}$ depending on the density of points appears here (as well as every time the density dependence is kept explicit).

We recall that the understanding of the function $\mf(\beta)$ and its smoothness is one of the keys to understanding phase transitions. Note that \cite{KK} proposes explicit expressions for it.
We have provided here a variational interpretation for it, which offers another potential angle of analysis for it. 
  Unfortunately, we do not know if and when minimizers of $\mathcal I^1_{\beta}$ are unique (nonuniqueness would be another manifestation of a phase transition), as $\mathcal I_\beta^1$ is not in general convex. A notable exception is the one-dimensional logarithmic case, for which 
 uniqueness is proven in \cite{erbarleble}  by a displacement convexity argument combined with screening.
 
\subsection{Proof outline}

We will only prove Theorem \ref{theoldpglob} in the Coulomb case (and then we also assume $\mu_V\ge m>0$ on $\Sigma$ for simplicity), the result is also true in the Riesz case and requires using the Riesz screening procedure of \cite{PetSer}. We refer the interested reader to \cite{lebles}. 
Let us now give an idea of the steps of the proof, which are parallel to that of the minimizers case, Theorem~\ref{theominifin}. 

From \eqref{defQ} and \eqref{scalingF},  we may rewrite $\Q_{N,\beta}$ in blown-up scale as 
\be \label{rewrig2}\Q_{N, \beta}(\mu_N) = \frac{1}{N^N\K_\beta(\mu_N') } \exp\(- \beta \F(\XN', \mu_N') \) d(\mu_N')^{\otimes N}(\XN') .\ee

As usual, proving an LDP requires to prove exponential tightness (which in our case is easy thanks to the a priori bound on the number of points), prove a large deviations upper bound and a large deviations lower bound for balls, more precisely for 
$$\Q_{N,\beta}(\mu_N) \{\XN, \bEmp_N[\XN]\in B(\bP, \ep)\}.$$
The upper bound is quite straightforward with the results we already have at hand.
Indeed, inserting \eqref{gliminf} into \eqref{rewrig2} and combining it with the lower semi-continuity of Lemma~\ref{corolscW} we find that 
\begin{multline*}
\Q_{N,\beta}(\mu_N)\{\XN, \bEmp_N[\XN]\in B(\bP, \ep)\}\\
 \le \frac{1}{ \K_{\beta}(\mu_N')} 
\exp\(- N \( \beta \overline \W(\bP, \mu)   +o_{N,\ep}(1) \)  \)  \int_{\XN, \bEmp_N[\XN]\in B(\bP, \ep) } d(\mu_N)^{\otimes N} (\XN
),\end{multline*}
and we then obtain directly  in view of \eqref{ldprefub}  the following upper bound
\begin{multline}\label{mullb}
\log \Q_{N,\beta}(\mu_N) \{\XN, \bEmp_N[\XN]\in B(\bP, \ep)\} \\
\le -\log \K_{ \beta}(\mu_N') 
-N \beta  \overline \W(\bP, \mu)  
 - N \int_{\Sigma} \ERS[\bP^x|\Poisson^{\mu(x)} ]dx+ N o_{N,\ep}(1).\end{multline}
We note here that we are able to apply \eqref{ldprefub} thanks to the fact that we may restrict our attention to $\bP\in \probas_{s, \mu} (\Sigma \times \config)$  in view of Proposition \ref{prop:LowerBoundenergies}.
This is the needed upper bound -- the $\log \K_{\beta}(\mu_N')$ term is shown later to be the minimum  of the rate function. The competition between the energy  and entropy terms is straightforward.

The lower bound is much more delicate. It requires producing enough configurations whose empirical field  are in $B(\bP,\ep)$  and whose energy is bounded above by $ \overline \W(\bP, \mu)   $ up to $o(N)$. 
To do so, the starting point is to draw configurations at random from the law $\mu_N^{\otimes N}$. 
By  Proposition  \ref{SanovbQN}, the probability that the empirical field \eqref{def:bEmp} of such configurations drawn from $\mu_N^{\otimes N}$ resembles $\bP$ is like $\exp\(-N \int_\Sigma \ERS[\bP^x|\Poisson^{\mu(x)}]dx\)$, which is the desired volume estimate. But the control from above of  the energy of these random configurations in terms of $ \overline \W(\bP, \mu)  $ is delicate for two reasons: $\W$ is defined as a limit over increasing cubes, and  the background density varies which requires to localize the estimates. To do so, exactly as in the upper bound for Theorem \ref{theominifin}, we partition the domain into  large microscale hyperrectangles $Q_i$ in which $\mu_N'(Q_i)$ is integer, and draw configurations in each. We then screen the configurations in each $Q_i$, which  allows to bound from above the energy $\F$ by the sum of the Neumann energies over $Q_i$ thanks to \eqref{subad1}. 
The only data that we have is that $\bEmp_N[\XN]$ is close to $\bP$, and we need to use this to control the energy by $ \overline \W(\bP, \mu)  $, this requires an upper semi-continuity property.

The screening procedure modifies the original random configuration in a  boundary layer 
near the boundary of each $Q_i$. Since that boundary has small volume, the empirical field of the configuration does not get  modified much, so it remains close to $\bP$. On the other hand, modifying the configuration also modifies the volume estimate, but the relative error, which is handled by Proposition \ref{pro42}, is in the end shown to be small.
A delicate task is to handle the screenability condition since not all configurations are screenable.

 \subsection{Main proofs}
\begin{proof}[Proof of Theorem \ref{theoldpglob} in the Coulomb case]
In the proof we will abbreviate $\overline \W(\bP, \mu)  $ into $\overline \W(\bP)$.
\noindent
As outlined just above we may combine \eqref{rewrig2}, \eqref{gliminf},   Lemma \ref{corolscW}  and \eqref{ldprefub} to obtain \eqref{mullb}. This concludes the upper bound and we now turn to the lower bound.

{\bf  Lower bound. }

{\bf Step 1: Setup and partitioning.}
First, we note that in view of Lemma \ref{lemsigmaeta} we may reduce to the situation where $\mu_N\to \mu$ uniformly in $\Sigma$. Indeed, exponential tightness of $\Q_{N,\beta}(\mu_N)$ is an easy consequence of the fact that the total number of points in $\Sigma$ is bounded by $N$.

Let $\bP\in\probas_{s}(\Sigma \times \config)$ be such that $\overline \W(\bP) <\infty$.
We may lift $\bP$ into a stationary tagged gradient electric process $\bP^\mathrm{e}$ as in Lemma \ref{lemlifting}.
By stationarity of $\bP$, lifting and definition \eqref{defW}, we have for any $R>0$,
\be \label{Pex}\overline\W(\bP)= 
\int_{\Sigma } \int \mathcal{W}(E, \mu(x)) d\bP^{\mathrm{e},x}(E)dx= 
\int_{\Sigma}\int \frac1{R^\d} \mathcal F^{\carr_R} (E,\mu(x))    d\bP^{\mathrm{e},x}(E)dx .\ee
Indeed, Lemma \ref{neutralite0} and the stationarity imply that we can remove the 
$\frac{\C( \carr_R)}{R^\d}-\mu(x)$ term present in the definition of $\mathcal{W}$.

Given $R>1$ independent of $N$, let us apply Lemma \ref{tiling}  and
partition $\Sigma':= N^{1/\d}\Sigma$  so that    
$$\Sigma' = \cup_{i\in I} Q_i\cup \omega$$
where $Q_i$ are  hyperrectangles of size $\in [R, R+ C R^{1-\d}]$ included in $\Sigma'$ such that $\int_{Q_i} \mu_N': = \mn_i$, an integer,  and $\omega$ is a remaining boundary layer such that $|\omega|\le o(N)$.
  We let $$\Sigma'_{\mathrm{int}}= \cup_{i\in I} Q_i.$$ We would like to replace 
 the $Q_i$'s by translations of a fixed square. For that we may  find for each $i$ a square $\carr_i$ of center $x_i$ and sidelength exactly $R$ and included in $Q_i$.
 By Lemma~\ref{tiling} the difference between $R$ and the sidelength of $Q_i$ is bounded by $R^{1-\d}m^{-1}$. This way $|Q_i\backslash \carr_i|=o(R^\d)$. Also  we can pick $\carr_i$ in such a way that for any $\tilde \ell \ge C R^{1-\d}$ (with the same $C$ appearing in the sidelength of $Q_i$) we have 
 \be \label{taillegar}
 \{x\in Q_i, \dist (x, \partial Q_i)\ge \tilde \ell \}\subset \carr_i.\ee


For a configuration, we define its discrete average empirical field relative to the partition $Q_i$ by 
\be \label{defbqn}\bar D_N[\XN]:= \frac{1}{\#I }\sum_{i\in I} \delta_{(   N^{-1/\d} x_i, \theta_{x_i} \cdot X_N'|_{Q_i}  )}\ee
where we recall the $x_i$'s are the centers of the $\carr_i$.
\smallskip

{\bf Step 2. Comparing discrete and continuous averages.}
Since we will need to work with the discrete averages \eqref{defbqn} it is important to be able to show that they are close to the continuous averages. 

The idea is that if one knows that a discrete average of large hypercubes is very close to some point process $\Pst$, then the continuous average of much smaller hypercubes  is also close to $\Pst$ since it can be re-written using the discrete average up to a small error. More precisely for any fixed $\delta > 0$ establishing that a point process is in $B(\Pst, \delta)$ can be done by testing against local functions in $\Loc_k$ (the space of functions of $\config$ which are only functions of $\C \cap \carr_k$)  for some $k$ large enough. For $R,N$ large enough, an overwhelming majority of all translates of $\carr_k$ by a point in $\Sigma'$ is included in one of the hypercubes $Q_i$ (this follows from the definitions and the tiling).

For any such local function $f \in \Loc_k$ we have
\begin{equation} \label{discrtocont}
\frac{1}{|\Sigma'|} \int_{\Sigma'} f(\theta_{x} \cdot \mc{C}) \approx \frac{1}{\#I } \sum_{i\in I} \frac{1}{R^{\d}} \int_{Q_i} f(\theta_{x} \cdot \mc{C}) dx, 
\end{equation}
which allows us to pass from the assumption that the discrete average (in the right-hand side of \eqref{discrtocont}) of a configuration is close to $\Pst$ to the fact that the continuous average (in the left-hand side of \eqref{discrtocont}) is close to $\Pst$. These considerations are easily adapted to the situation of tagged point processes.
We conclude that if $\bar D_N[\XN]\in B(\bP, \ep)$ then $ \bEmp_N[\XN]\in B(\bP,  \varphi(\ep))$ for some function $\varphi(\ep) $ tending to $0$ as $\ep\to 0$,  and vice-versa.
\smallskip 

{\bf Step 3. Good controls for configurations near $\bP$.}

{\bf Substep 3.1.  Point and energy control.}
We will consider point configurations such that $\bar D_N[\XN]$ and $\bEmp_N[\XN]$ are close to $\bP$, more precisely are in some ball $B(\bP, \ep)$ for the weak local topology.

For that we may use Lemma \ref{lemsigmaeta} to replace $\bar P$ by its restriction to $\Sigma_{\mathrm{int}}= N^{-1/\d} \Sigma_{\mathrm{int}}'$, modification that we assume has been made from now on.

We first derive consequences of the fact that $\bar D_N[\XN]\in B(\bP, \ep)$.
We will use the following variant of \eqref{ldpreflb} for discrete average, which follows from \eqref{ldpreflb} in view of the conclusion of Step 2,
\be
 \label{ldprefdis}
\liminf_{\ep\to 0}\liminf_{N\to \infty} \frac1N \log (\mu_N)^{\otimes N} \(\XN, \bar D_N[\XN]\in B(\bP,\ep)\) 
 \ge - \int_{\Sigma} \ERS[\bP^x|\Poisson^{\mu(x)} ]dx.\ee

Secondly, the relation \eqref{discrR}  implies that 
\be \int |\C( \carr_R )-\mu(x) R^\d |^2 d\bP(\C) \le C R^{2\d-2\kappa}(1+ \overline{\W}(\bP)),\ee
for some $\kappa>0$ depending only on $\d$.
We may extend these relations to probabilities close to $\bP$.
If $\bar D_N[\XN] \in B(\bP, \ep)$, since we may test against bounded continuous local functions, approximating  the indicator function of $\carr_R$ by continuous functions, we 
 in particular deduce
\be\int   |\C(\carr_R)-\mu(x) R^\d|^2d \bar D_N[\XN](x,\C) \le CR^{2\d-2\kappa}(1+\overline{\W}(\bP)),\ee
which means by definition \eqref{defbqn} that 
\be\label{bnini0} \frac{1}{\#I} \sum_{i\in I}\left|\#\{\XN'\cap \carr_i\}- \mu'(x_i) R^\d\right|^2\le  CR^{2\d-2\kappa}(1+\overline \W(\bP)).\ee
By  uniform convergence of $\mu_N$ to $\mu$ and continuity of $\mu$, we also have 
$$|\mu'(x_i)R^\d - \mn_i|=\left|\mu'(x_i)R^\d - \int_{Q_i} \mu_N'\right|\le o_N(1) R^{\d},$$
and thus 
\be\label{bnini} \frac{1}{\#I} \sum_{i\in I}\left|\#\{\XN'\cap \carr_i\}- \mn_i \right|^2\le  CR^{2\d-2\kappa}(1+\overline \W(\bP)) + o_N(1) R^{2\d}.\ee

Thirdly, the relation
\eqref{Pex} ensures that  for any $R>0$ we have 
\be\label{FR1} \int \frac{1}{R^\d} \mathcal F^{\carr_R} (E, \mu(x))d \bP^{\mathrm{e}}(x, E)\le \overline \W(\bP),\ee
in fact we have equality.
We now need upper semi-continuity (for the local topology) and boundedness to be able  to replace $\bP^{\mathrm{e}}$ by $\bar D_N[\XN]$ in the inequality \eqref{FR1}. 

Let $M\ge \overline{\W}(\bP)+1$. 
First, we may truncate at level $M$ and write that 
\be\label{FR3} \int \(\frac{1}{R^\d} \mathcal F^{\carr_R} (E, \mu(x))\wedge M\) d \bP^{\mathrm{e}}(x, E)\le  \overline{\W}(\bP).\ee
Next, we note that by definition \eqref{defFcarrr}, and generalizing the definition in the obvious way to open cubes, 
either  $\mathcal F^{\carr_R}(E,m)=+\infty$ or
\be \label{Fcarcar}\mathcal F^{\carr_R}(E,m)= \mathcal F^{\overset{\circ} \carr_R}(E, m)
+r(\C,m)\ee
where $$r(\C,m) = - \hal \sum_{p\in \pa \carr_R}
\g(\rrc_p) -m\sum_{p\in \pa\carr_R} \int_{\R^\d} \f_{\rrc_p}(x-p),$$
in particular $\mathcal F^{\carr_R}(E,m)\le  \mathcal F^{\overset{\circ} \carr_R}(E, m)$.
Here we have isolated in $r$ the contribution of points that fall exactly on the boundary and which prevent the function $\mathcal F^{\carr_R}$ from being upper semi-continuous.  But  it is negligible: by the boundedness of $\rrc_p$ for $p \in \pa \carr_R$, $r$ is bounded by the number of points on the boundary, and 
in view of Lemma \ref{neutralite0} we must have that  $\int r(\C, \mu(x)) d\bP(\C)=O(R^{\d-\kappa}).$
Thus \eqref{FR3} can be changed into  \be\label{FR4} \int \( \frac{1}{R^\d} \mathcal F^{\overset{\circ}\carr_R} (E, \mu(x))   \wedge M\) d \bP^{\mathrm{e}}(x, E)\le  \overline{\W}(\bP)+O(R^{-\kappa}) .\ee

In addition, $ \mathcal F^{\overset{\circ}\carr_R}(E,m)$  coincides with
 the definitions of Chapter \ref{chap:screening} \eqref{minneum}  for gradient vector-fields,  while   $\bP^{\mathrm{e}}$ is concentrated on  gradient vector-fields (by definition and finiteness of $\overline\W(\bP)$). 

{\bf Substep 3.2. Preparing for screening.}
With the goal of applying  the screening procedure of Definition~\ref{defscreen} and Proposition~\ref{proscreen}, we introduce parameters $\max(\beta^{-\frac1{\d-\s}} \indic_{\s\le 0} ,1)<\ell\le \tilde \ell\le R$ such that 
\be \label{conditionselll}
\ell^{\d+1}\ge \frac{2\cd CMR^\d}{\tilde \ell}\ee for the constant $C$ of \eqref{screenab}.
More specifically we let $\tilde \ell = R^{1-\kappa}$ and $\ell = R^{1-2\kappa}$ with $\kappa $ small, where $R$ may depend on $\beta$. We may in particular choose $\kappa$ small enough (depending on $\d$) such that \eqref{conditionselll} holds.  In addition, making $\kappa$ smaller if necessary, we can assume it is the same $\kappa$ as in \eqref{bnini}. 

We will apply the screening with $\Omega=Q_i$  and $\Omega' =\carr_i$\footnote{The need to deal with $Q_i$ and $\carr_i$ being slightly different is the reason why the screening was including the possibility for the sets $\Omega$ and $\Omega'$ to be different.}, for that we need 
$\{x\in Q_i, \dist (x, \pa Q_i) \ge \tilde \ell\} \subset\carr_i  $ which in view of \eqref{taillegar} is guaranteed as long as 
$\tilde \ell\ge C R^{1-\d}$ which can be reduced to $\tilde \ell \ge C$ and with our choice, to $R$ being large enough.

Let us assume that $\mathcal F^{\overset{\circ} \carr_R}(E, \mu(x)) \le M R^\d$.
  Since we may assume that $E$ is a gradient,  then  $E=\nab w$ for some $w$ which is automatically inner screenable in $\carr_R$ according to Definition~\ref{defscreen}, in particular \eqref{screenab} is satisfied in view of the condition \eqref{conditionselll}. 
  We deduce from the definition \eqref{innernrj}  and \eqref{FR4}  that 
\be \label{FR12} \int \( \frac{1}{R^\d} \G_{\R^\d}^{\mathrm{inn}} (\C,\mu(x),\overset{\circ}\carr_R)  \wedge M\) d \bP(x, \C)\le  \overline \W(\bP)+O(R^{-\kappa}).
 \ee
  In the rest of the proof we will drop the $\mathrm{inn}$ superscript  and $\R^\d$ subscript and just write $\G$ instead of $\G_{\R^\d}^{\mathrm{inn}}$. 
 
 {\bf Substep 3.3. Upper semi-continuity of $\G$.}
  We now argue that  for any $m$, $\G(\cdot,m, \overset{\circ}\carr_R)  $ is upper semi-continuous within the class 
 of configurations such that $\G(\C,m, \overset{\circ}\carr_R) \le MR^\d$  with $M$ as above. Let $\C^k$ be a sequence of configurations in $\R^\d$  converging to $\C$ for the local topology. We may assume that ,  $\C$ has only simple points in $\overset{\circ}\carr_R$ otherwise  $\G(\C, m, \overset{\circ} \carr_R)=+\infty$ and the desired  result is true. 
  Let  $G$ be the Dirichlet Green's function of $\carr_R$, which is given by 
  \be \label{Kdiric} G(x,y)=\g(x-y)-\g(x-y^*)\ee
  with $y^*$ being a reflection of $y$ through the boundary of $\carr_R$. 
  
  Let
  $ w$ achieve the min in the definition \eqref{innernrj} relative to $\overset{\circ} \carr_R$,  and let us now change notation and denote by $p_1^k, \dots, p_n^k$  the points of $\C^k \cap \overset{\circ}\carr_R$ (for $k$ large enough, we can assume the number of points is constant independent of $k$). After extraction of a subsequence, we have $p_i^k \to p_i \in \carr_R$ as $k \to \infty$. 
  We then wish to build a competitor for the definition of 
 \eqref{innernrj} of $\G(\C^k, m,\overset{\circ} \carr_R)$.  Let
  $$w^k(x)= w_{\rrh}(x) +  \int_{\carr_R} G(x,y) d\( \sum_{i=1}^n \delta_{p_i^k}^{(\rrh_i^k)}   - \delta_{p_i}^{(\rrh_i)}\) (y) +\sum_{i=1}^n \f_{\rrh_i^k}(\cdot -p_i^k).$$
  We may check that $-\Delta (w^k-w) =\cd \( \sum_{i=1}^n \delta_{p_i^k}  - \delta_{p_i} \) $, hence
    $w^k$ is compatible with the points $p_i^k$ and admissible in the definition of $\G(\C^k, m, \overset{\circ}\carr_R)$ provided it is screenable. 
Moreover, by construction,  $w^k_{\rrh}=w_{\rrh}$ on $\pa \carr_R$, hence integrating  by parts, 
we find that 
\begin{multline*}
 \frac1{2\cd} \int_{\carr_R} |\nab w_{\rrh}^k|^2 - \frac1{2\cd} \int_{\carr_R} |\nab w_{\rrh}|^2
   = \frac1{2}\int_{\carr_R} (w_{\rrh}^k- w_{\rrh})\(  \sum_{i=1}^n \delta_{p_i^k}^{(\rrh_i^k)}+\delta_{p_i}^{(\rrh_i)}+2\sum_j \delta_{x_j}^{(\eta_j)} -2 m\)\\
 = \frac1{2} \int_{\carr_R\times \carr_R}  G(x,y)  d\( \sum_{i=1}^n \delta_{p_i^k}^{(\rrh_i^k)}  - \delta_{p_i}^{(\rrh_i)} \) (y) d\(\sum_{i=1}^n \delta_{p_i^k}^{(\rrh_i^k)}+\delta_{p_i}^{(\rrh_i)}+2\sum_j \delta_{x_j}^{(\eta_j)} -2 m\)(x).\end{multline*}
 for some $x_j\notin \overset{\circ}\carr_R$.
 Expanding and using the symmetry of $G$, we find that 
 \begin{multline*}
 \frac1{2\cd} \int_{\carr_R} |\nab w_{\rrh}^k|^2 - \frac1{2\cd} \int_{\carr_R} |\nab w_{\rrh}|^2
  \\ =\hal\sum_{i=1}^n \int G(x,y) d \delta_{p_i^k}^{(\rrh_i^k)}(x)d \delta_{p_i^k}^{(\rrh_i^k)}(y)- \int G(x,y) d \delta_{p_i}^{(\rrh_i)}(x)d \delta_{p_i}^{(\rrh_i)}(y)
   \\+
   \frac1{2}\sum_{i\neq l}  \int_{\carr_R\times \carr_R}  G(x,y)  d(  \delta_{p_i^k}^{(\rrh_i^k)}  - \delta_{p_i}^{(\rrh_i)} ) (y) d(\delta_{p_l^k}^{(\rrh_l^k)}+\delta_{p_l}^{(\rrh_l)}) (x)\\
   +
  \int_{\carr_R\times \carr_R}  G(x,y)  d\( \sum_{i=1}^n \delta_{p_i^k}^{(\rrh_i^k)}  - \delta_{p_i}^{(\rrh_i)} \) (y) d\Big(\sum_j \delta_{x_j}^{(\eta_j)} - m\Big)(x).\end{multline*}
   Since $p_i^k \to p_i$ and the points $p_i\in \overset{\circ}\carr_R$ are all distinct, $G$ being continuous away from the diagonal and the radii $\rrh$ being continuous functions of the point locations, these terms all converge to $0$  as long as $p_i, p_l\in \overset{\circ}\carr_R$. 
For   terms involving $p_i $ or $p_l$ that belong to $\pa \carr_R$ we use \eqref{Kdiric} to write  that
  $$\int G(x,y) d \delta_{p_i^k}^{(\rrh_i^k)}(y)= \int \g(x-y) d \delta_{p_i^k}^{(\rrh_i^k)}(y)- \int \g(x-y^*) d\delta_{p_i^k}^{(\rrh_i^k)}(y)= \g_{\rrh_i^k}(x-p_i^k)- \g_{\rrh_i^k} (x-(p_i^k)^*)$$  which then tends to $0$ uniformly as $k\to \infty$ if $p_i \in \pa \carr_R$. 
  We can thus conclude that 
  \be\label{limwk}
  \lim_{k\to \infty}   \int_{\carr_R} |\nab w^k_{\rrh}|^2 =\int_{\carr_R} |\nab w_{\rrh}|^2\ee
  and 
  \begin{multline*}  \limsup_{k\to \infty}\frac1{2\cd}   \int_{\carr_R} |\nab w^k_{\rrh}|^2-\hal \sum_{i=1}^n \g(\rrh_i^k) - m\sum_{i=1}^n \int \f_{\rrh_i^k}(x-p_i^k)  dx \\
  \le \frac1{2\cd}\int_{\carr_R} |\nab w_{\rrh}|^2-\hal \sum_{p_i \in \overset{\circ}\carr_R}^n \g(\rrh_i) - m\sum_{p_i \in \overset{\circ } \carr_R} \int \f_{\rrh_i}(x-p_i)  dx\le M R^\d, \end{multline*}
  since at worst when $p_i^k \to \pa \carr_R$ the points are not counted in the limit.
  It follows that  $w^k$ is screenable thanks to \eqref{conditionselll}, so $w^k$ is admissible in the definition of the minimum in \eqref{innernrj} for $\G(\C^k,m,\overset{\circ}\carr_R)$ and thus,  using that $w$ achieves the min in the definition of $\G(\C, m,\overset{\circ} \carr_R)$, we conclude that 
   \be\label{limwk2} \limsup_{k\to \infty} \G(\C^k, m, \overset{\circ} \carr_R) \le \G(\C ,m,\overset{\circ} \carr_R)\ee
which is the claimed upper semi-continuity.

    From this, we  deduce that 
     $\frac1{R^\d}\G(\cdot, \mu(x), \overset{\circ}\carr_R) \wedge M$ is upper semi-continuous.
      Finally, we conclude from \eqref{FR12} and this upper semi-continuity that if $\bar D_N[\XN]\in B(\bP, \ep)$, we have 
\be  \int \( \frac{1}{R^\d} \G (\C,\mu(x),\carr_R)  \wedge M\) d \bar D_N[\XN](x, \C)\le  \overline \W(\bP)+O(R^{-\kappa}) +o_\ep(1)   \ee
or in other words, by definition \eqref{defbqn},
\be\label{borneGmoy0} \frac{1}{\#I} \sum_{i\in I}  \frac{1}{R^\d}\G(\XN'|_{\carr_i} ,\mu'( x_i), \carr_i)\wedge M \le \overline \W(\bP)+O(R^{-\kappa}) +o_\ep(1)  .\ee

{\bf Step 4. Rectifying the background.}
We next wish to change $\G(\XN'|_{\carr_i} ,\mu'( x_i), \carr_i)$ into 
$\G(\XN'|_{\carr_i} ,\mu_N', \carr_i)$.
For that it suffices to add to $w$ achieving the min in the definition of $\G$ \eqref{innernrj}, the function $u$ solving 
$$\left\{\begin{array}{ll}
-\Delta u = \cds(\mu_N'-\mu'( x_i)) & \text{in}\  \carr_i\\
u=0& \text{on}\  \partial \carr_i.\end{array}\right.
$$
By elliptic estimates, the uniform convergence of $\mu_N$ to $\mu$ and the continuity of $\mu$,  we have that 
\be\label{ellest}
\|\nab u\|_{L^\infty(\carr_i)} \le C_R \|\mu_N'-\mu'( x_i)\|_{L^\infty (\carr_i)}\le C_R o_N(1).\ee
By definition of $\G$ we thus find that 
\be\label{porelim}
|\G(X_N'|_{\carr_i}, \mu_N', \carr_i)-\G(X_N'|_{\carr_i},  \mu'( x_i), \carr_i) |\le C \|\nab u\|_{L^2(\carr_i)}^2 +C
\|\nab u\|_{L^\infty(\carr_i)} \| \nab w_{\rrh}\|_{L^1(\carr_i)}.
\ee
Since we using $\G_U^{\mathrm{inn}}$ in the case $U=\R^\d$, the radii $\rrh$ coincide with those of \eqref{defrrc4}. The relation \eqref{14} can be checked to be valid for $\G$ as well as for $\F$, i.e.~ we have   the control 
$$\int_{\carr_i} |\nab w_{\rrh}|^2\le C  \(\G(\XN', \mu'(x_i), \carr_i) +C_0 \#\{X_N'\cap \carr_i\}\).$$
Combining with  \eqref{borneGmoy0}, \eqref{ellest}, \eqref{porelim}  and \eqref{bnini}, we thus deduce that 
\be \sum_{i\in I} |\G(\XN'|_{\carr_i} ,\mu_N', \carr_i)- \G(\XN'|_{\carr_i} ,\mu'( x_i), \carr_i)|\le o(N).\ee
In view of \eqref{borneGmoy0} and since, in view of the tiling procedure  $\#I $ is of order $\frac{N}{R^\d}$, 
 we have found that
\be\label{borneGmoy} \frac{1}{\#I} \sum_{i\in I}  \frac{1}{R^\d}\G(\XN'|_{\carr_i} ,\mu_N', \carr_i)\wedge M \le \overline \W(\bP)+O(R^{-\kappa}) +o_\ep(1) +o_N(1)  .\ee

{\bf Step 5. Screening.}
We are now in a position to apply the screening procedure to such configurations drawn near $\bP$.
For any $\XN $ such that $\bar D_N[\XN]\in B(\bP, \ep)$, we have all the properties described in the previous steps, and we proceed as follows. We let $I_1$ be the subset of $I$ such that $\G(X_N'|_{\carr_i}, \mu'(x_i), \carr_i)\le M R^\d$, in particular $X_N'|_{\carr_i}$ is screenable, and  let $I_2=I\backslash I_1$.
Moreover,  \eqref{borneGmoy} implies that $\#I_2/\#I \le \frac{\W(\bP)}{M}  $.

If $i\in I_1$, we apply Proposition \ref{proscreen} with  $\mu_N'$ as the reference measure, $\Omega =Q_i$ and $\Omega'=\Omega''=\carr_i,$  $\eta$ to be determined, and the $\ell , \tilde \ell$ chosen above.  This provides  a configuration $Y_{\mn_i}$ in $Q_i$, coinciding with  the restriction of $X_N'|_{\carr_i}$ in some set $\Old_i$ pasted with a configuration $Z_{\mn_i-n_{\Old_i}}$ in $\New_i= Q_i \backslash \{x, \dist(x,\Old_i)\le \eta\}$,  and a positive measure $\tilde \mu_i $ in $\New_i$,  such that 
\begin{multline*}
\F(Y_{\mn_i}, \mu_N', Q_i) \\ \le \G(X_N|_{\carr_i},\mu_N', \carr_i) 
+ C\(\frac{\ell MR^\d}{\tilde \ell} + R^{\d-1}\tilde \ell + \F(Z_{\mn_i-n_{\Old_i}} , \tilde \mu_i, \New_i) + |\mn_i-n_i|+ \sum_{k,j\in J_i}\g(x_k-z_j)\),\end{multline*} where $n_i$ is $\#\{\XN'\cap \carr_i\}$.
We emphasize here that the sets $\Old_i$ and $\New_i$ depend on $X_N'|_{\carr_i}$.

If $i\in I_2$, we let $\Old_i=\varnothing $, $\New_i = Q_i$ and $\tilde \mu_i=(\mu_N')|_{Q_i}$, i.e.~we delete the configuration in $Q_i$ and replace it with a generic configuration $Y_{\mn_i}^i=Z_{\mn_i}$.
Finally, we do the same for the configuration in $\R^\d \backslash \Sigma'_{\mathrm{int}}$ and replace it with a generic configuration of $\mn_0:= N-\sum_{i\in I} \mn_i$ points. We simplify  notation by writing $Q_0= \R^\d \backslash \Sigma'_{\mathrm{int}}$.

Pasting together the configurations obtained over all the $Q_i$'s gives a configuration $Y_N$ of $N$ points in $\R^\d$.  The set of all configurations obtained this way when $\XN $ varies in $A:=\{ \XN, \bar D_N[\XN]\in B(\bP, \ep)\}$ is denoted $A'$.

The important fact that we can check is that since $\#I_2=O(\frac1M \#I)$ and the configurations are unchanged in each $\Old_i$ subset of $Q_i$, whose area is $|Q_i|-o(R^\d)$, 
the total configuration is modified in only a vanishing fraction of the volume as $M \to \infty$ hence in view of this and of the result of Step 2,  the set $A'$ consists of configurations whose empirical field $\bEmp_N[Y_N]\in B(\bP, \varphi(\ep))$ for $N$ large enough and $M$ large enough, where $\varphi$ is some function tending to $0$ as $\ep \to 0$.
\smallskip

{\bf Step 6. Integrating and volume estimates.}
We next integrate the energy inequality
over the possible choices of $Z_{\mn_i-n_{\Old_i}}$ in each $Q_i$, with respect to the measure $(\mu_N') |_{Q_i}$, and then over the choices of the initial $\XN\in A$ as above.
This leads us to a calculation entirely similar to the proof of Proposition \ref{pro42}, in particular using the result of Lemma \ref{lemclaim44},
we may write that, taking into account the relabellings and multiplicity, 
\begin{multline}\label{lastlin}
\int_{A'}
\exp\(- \beta \F(Y_N,  \mu_N', \R^\d) \)d(\mu_N')^{\otimes N} (Y_N)
\\ \ge 
\int_{\XN\in A} \frac{\prod_{i\in I} n_i!}{N!}  \frac{N!}{\prod_{i\in I} \mn_i!} 
 \exp\( -\beta \sum_{i\in I_1}\Big( \G(\XN'|_{\carr_i}, \mu_N', \carr_i) +C
(\frac{MR^\d\ell}{\tilde \ell} +\chi(\beta) R^{\d-1}\tilde \ell+|\mn_i-n_i|) \Big)\)
\\
\times \prod_{i\in I_1} \frac{\mn_i!  (n-n_{\Old_i} )! }{n_i! (\mn_i- n_{\Old_i}!)}\frac{( \mn_i-n_{\Old_i})^{\mn_i-n_{\Old_i}}}{\mu_N'(Q_i\backslash \Old_i)^{n_i-n_{\Old_i}}} e^{\mu_N'(\New_i)-\tilde \mu_i(\New_i) -C( \frac{MR^\d }{\ell\tilde\ell}+\frac{\eta^2}{\ell} R^{\d-1} )-\log \frac{\tilde \ell}{\eta} } 
  d(\mu_N')^{\otimes n_i} (\XN'|_{Q_i})\\
 \times \prod_{i\in I_2}\frac{1}{\mn_i!}  \int_{Q_i^{\mn_i} } \exp\(-\beta \F(Y_{\mn_i}, \mu, Q_i)\) d(\mu_N')^{\otimes \mn_i} (Y_{\mn_i}) 
 \times \frac{1}{\mn_0!} \int_{Q_0^{\mn_0} } \exp\(-\beta \F(Y_{\mn_0}, \mu_N', Q_0)\) d(\mu_N')^{\otimes \mn_0} (Y_{\mn_0}) .
  \end{multline} 
In view of the a priori bounds \eqref{bornesfiU}, we can bound the last integrals, corresponding to $i\in I_2$, by $$\prod_{i\in I_2} \mn_i^{\mn_i }\frac{1}{\mn_i!} \K_\beta(Q_i,\mu_N') \le \exp\( \sum_{i\in I_2} ( \mn_i + C \beta \chi(\beta) R^\d) \) \le \exp\( C (1+\beta \chi(\beta)) \frac{N}{M}\), $$
where we used that in view of the tiling procedure, we have  $\#I\le C\frac{N}{R^\d}$. 
To control the last integral on the last line of \eqref{lastlin},  we can use again \eqref{bornesfiU} to obtain
$$\frac{1}{\mn_0!} \int_{Q_0^{\mn_0} } \exp\(-\beta \F(Y_{\mn_0}, \mu_N', Q_0)\) d(\mu_N')^{\otimes \mn_0} (Y_{\mn_0}) \le \exp \( C \beta \chi(\beta)o(N) \).$$
Choosing $\eta= \min(1,\beta) \frac{m}{4\|\mu_N\|_{L^\infty}} \tilde \ell$ where $m$ is a lower bound for $\mu_N$,  using   \eqref{bnini} to control $\sum_i |\mn_i-n_i|$, \eqref{bornimp} to control $\mu_N'(\New_i)-\tilde \mu_i(\New_i)$  and
 the choices $\ell =R^{1-2\kappa}$ and $\tilde \ell=R^{1-\kappa}$,   we arrive at 
\begin{multline}\label{arr4}
\int_{A' }
\exp\(- \beta \F(Y_N,  \mu_N', \R^\d) \)d(\mu_N')^{\otimes N} (Y_N)
\\ \ge  \exp\(- C\(\beta \chi(\beta) (NR^{-\kappa}+o(N)) + M NR^{-\kappa} + (1+\beta \chi(\beta)) \frac{N}{M}  +\frac{N}{R^\d}( \log R - \log \min(1, \beta)) \) \) \\
\times
\int_{\XN\in A}  \prod_{i\in I_2, i=0}  \frac{n_i!}{\mn_i!}    \prod_{i\in I_1} \frac{ (n_i-n_{\Old_i} )! }{ (\mn_i- n_{\Old_i}!)}
\frac{( \mn_i-n_{\Old_i})^{\mn_i-n_{\Old_i}}}{\mu_N'(Q_i\backslash \Old_i)^{n_i-n_{\Old_i}}} 
 \\ \exp\( - \beta \sum_{i\in I_1} \G(\XN'|_{\carr_i}, \mu'_N(x_i), \carr_i)  \) d(\mu_N')^{\otimes N}(\XN').
\end{multline}
First, for the contribution of the indices in $I_2$ and $i=0$, we may write that 
$\log \frac{n_i!}{\mn_i!} \le C |\mn_i-n_i|\log R$ and use \eqref{bnini} to control this by $N o_R(1)$.

Second, for the contribution of the indices in $I_1$,  by Stirling's formula, denoting as in the proof of Proposition \ref{pro42}, 
$\alpha_i= \tilde \mu_i(\New_i) $, $\alpha'_i= \mu_N'(Q_i\backslash \Old_i)$ and using that   by \eqref{bornimp}
\be \label{aai}\left|\frac{\alpha_i'}{\alpha_i}-1\right| \le C( \frac{1}{\tilde \ell}+\frac{ M R}{\tilde \ell^2} )\le C MR^{-\kappa}, \qquad \frac1C R^{\d-\kappa} \le \frac1C \tilde \ell R^{\d-1}\le \alpha_i\le C \tilde \ell R^{\d-1}\le C R^{\d-\kappa} \ee 
 and $\alpha_i+n_i-\mn_i = n_i-n_{\Old_i} \ge 0$,  
 we have
\begin{align*}
&\log   \prod_{i\in I_1} \frac{ (n_i-n_{\Old_i} )! }
{ (\mn_i- n_{\Old_i})!}
\frac{( \mn_i-n_{\Old_i})^{\mn_i-n_{\Old_i}}}{\mu_N'(Q_i\backslash \Old_i)^{n_i-n_{\Old_i}}} 
\\ & = \sum_{i\in I_1}
(\mn_i-n_i) + (n_i-n_{\Old_i}) \log (n-n_{\Old_i}) - (n_i-n_{\Old_i}) \log \alpha_i' \\ &+ \hal \log (n_i-n_{\Old_i}) -\hal \log (\mn_i-n_{\Old_i}) +O(1)\\ &
= \sum_{i\in I_1}
(\mn_i-n_i) + (\alpha_i+ n_i-\mn_i +\hal ) \log (1+\frac{ n_i-\mn_i }{\alpha_i}  )  +(\alpha_i+n_i-\mn_i)  \log \frac{\alpha_i }{\alpha_i'} +O(1).
\end{align*}
We next use \eqref{aai} and the convexity of the function $x \log x$, then \eqref{bnini} and $\#I\le C \frac{N}{R^\d}$, to obtain 
\begin{multline*}
\log   \prod_{i\in I_1} \frac{ (n_i-n_{\Old_i} )! }
{ (\mn_i- n_{\Old_i}!)}
\frac{( \mn_i-n_{\Old_i})^{\mn_i-n_{\Old_i}}}{\mu_N'(Q_i\backslash \Old_i)^{n_i-n_{\Old_i}}} \\
\ge -  C\Big(\sum_{i\in I_1}
|\mn_i-n_i| + R^{\d-\kappa} \sum_{i\in I_1} (1+ \frac{n_i-\mn_i}{\alpha_i} ) \log ( \frac{1}{\#I_1} \sum_{i\in I_1} (1+ \frac{n_i-\mn_i}{\alpha_i} ))\\
 +\sum_{i\in I_1}(  R^{\d-\kappa}+  |n_i-\mn_i| )  MR^{-\kappa}\Big)\\
\ge   - C N(1+M)R^{-\kappa} -o(N) - \frac{\sum_{i\in I_1} |\mn_i-n_i|^2   }{R^{\d-\kappa}  }  \ge - CN MR^{-\kappa}-o(N)\end{multline*}
for some $C$ that depends on $\bP$.
Inserting into \eqref{arr4}, and recalling that for configurations in $A'$, the continuous empirical field is close to $\bP$, 
we obtain that for $N$ large enough, after absorbing some terms, 
\begin{multline}\label{arr4b}
\int_{Y_N, \bEmp_N[Y_N] \in B(\bP, \varphi(\ep))}
\exp\(- \beta \F(Y_N,  \mu_N', \R^\d) \)d(\mu_N')^{\otimes N} (Y_N)
\\ \ge 
 \exp\(- C\(\beta \chi(\beta) NR^{-\kappa} + M NR^{-\kappa} + (1+\beta \chi(\beta)) \frac{N}{M}  -\frac{N}{R^\d} \log \min(1, \beta) \) \)
\\
\times \int_{\XN \in A}   
  \exp\( - \beta \sum_{i\in I_1} \G(\XN'|_{\carr_i}, \mu_N'(x_i), \carr_i)  \) d(\mu_N')^{\otimes N}(\XN').
\end{multline}
Combining with \eqref{borneGmoy} and using that $\#I R^\d =N(1+o(1))$ because the $Q_i$'s are almost hypercubes, we are led to 
\begin{multline}\label{arr5}
\int_{Y_N, \bEmp_N[Y_N] \in B(\bP, \varphi(\ep))}
\exp\(- \beta \F(Y_N,  \mu_N', \R^\d) \)d(\mu_N')^{\otimes N} (Y_N)
\\ \ge  \exp\(- \beta N (\overline \W(\bar P) +o_\ep(1))
- C\( (\beta \chi(\beta) +M)  NR^{-\kappa}  + (\beta \chi(\beta)+1) \frac{N}{M}  -\frac{N}{R^\d} \log \min(1, \beta)\) \)\\
\times \int_{A} d(\mu_N')^{\otimes N}(\XN') .
\end{multline}
{\bf Step 7. Conclusion by large deviations estimate.} Recalling that $A=\{ \XN, \bar D_N[\XN]\in B(\bP, \ep)\}$ and
combining   \eqref{ldprefdis}  with \eqref{arr5}, we obtain 
\begin{multline}\label{arr6}
\frac{1}{N} \log \( \frac{1}{N^N}
\int_{Y_N, \bEmp_N[Y_N] \in B(\bP, \varphi(\ep))}
\exp\(- \beta \F(Y_N,  \mu_N', \R^\d) \)d(\mu_N')^{\otimes N} (Y_N) \) 
\\ \ge  - \beta   \overline \W(\bar P) - \int_{\Sigma} \ERS[\bP^x, \Poisson^{\mu(x)}]\,dx +o_\ep(1)
\\- C\( (\beta \chi(\beta) +M)  R^{-\kappa}  + (\beta \chi(\beta)+1) \frac{1}{M}  -R^{-\d} \log \min(1, \beta)\),
\end{multline}
thus in view of \eqref{rewrig2}, we have obtained
\begin{multline}\label{arr6}
\frac{1}{N} \log \Q_{N,\beta}(\mu_N)  \left\{ Y_N, \bEmp_N[Y_N] \in B(\bP, \varphi(\ep))\right\}
\\
 \ge -\frac1N \log \K_{\beta}(\mu_N')   - \beta   \overline \W(\bar P) - \int_{\Sigma} \ERS[\bP^x| \Poisson^{\mu(x)}]\,dx\\
  +o_\ep(1)
  - C\( (\beta \chi(\beta) +M)  R^{-\kappa}  + (\beta \chi(\beta)+1) \frac{1}{M}  -R^{-\d} \log \min(1, \beta)\)
  .\end{multline}

{\bf Conclusion}.
Combining this lower bound with the upper bound \eqref{mullb}, 
letting $N \to \infty$, then $\ep \to 0$,  then $R\to \infty$,  and $M \to \infty$,  we then 
obtain, if $\beta$ is independent of $N$, 
 \begin{multline} - \beta   \overline \W(\bar P) - \int_{\Sigma} \ERS[\bP^x| \Poisson^{\mu_V(x)}]\,dx\\
 \le 
  \lim_{\ep \to 0} \liminf_{N\to \infty} \(
 \frac{1}{N} \log \Q_{N,\beta}(\mu_N)  \left\{ X_N, \bEmp_N[X_N] \in B(\bP, \ep)\right\}
+\frac1N\log \K_{\beta}(\mu_N') \) \\
\le \lim_{\ep \to 0} \limsup_{N\to \infty} \(
 \frac{1}{N} \log \Q_{N,\beta}(\mu_N)  \left\{ X_N, \bEmp_N[X_N] \in B(\bP, \ep)\right\}
+\frac1N\log \K_{\beta}(\mu_N') \) \\
\le  - \beta   \overline \W(\bar P) - \int_{\Sigma} \ERS[\bP^x| \Poisson^{\mu(x)}]\,dx.\end{multline}
If $\beta \to 0$ as $N\to \infty$ we need to take $R$ to be $\beta$-dependent in such a way that $R^{1-\kappa} \ge \beta^{-\frac1{\d-\s}}\indic_{\s\le 0}$. Since $\beta\chi(\beta)$ tends to $0$ faster than $\beta^{1/2}|\log \beta|$  by \eqref{defchibeta}, choosing $ R = \beta^{-1}$ for instance ensures all error terms tend to $0$ as soon as $\kappa<\hal$.  
We then obtain, after letting $N\to \infty$ then $\ep \to 0$ and $M \to \infty$,
 \begin{multline}  - \int_{\Sigma} \ERS[\bP^x, \Poisson^{\mu(x)}]\,dx\\ \le 
  \lim_{\ep \to 0} \liminf_{N\to \infty} \(
 \frac{1}{N} \log \Q_{N,\beta}(\mu_N)  \left\{ X_N, \bEmp_N[X_N] \in B(\bP, \ep)\right\}
+\frac1N\log \K_{\beta}(\mu_N') \) \\
\le \lim_{\ep \to 0} \limsup_{N\to \infty} \(
 \frac{1}{N} \log \Q_{N,\beta}(\mu_N)  \left\{ X_N, \bEmp_N[X_N] \in B(\bP, \ep)\right\}
+\frac1N\log \K_{\beta}(\mu_N') \)\\  \le  - \int_{\Sigma} \ERS[\bP^x| \Poisson^{\mu(x)}]\,dx.\end{multline}
If $\beta \to \infty$,  we obtain instead
 \begin{multline} -    \overline \W(\bar P) \le 
  \lim_{\ep \to 0} \liminf_{N\to \infty} \(
 \frac{1}{N\beta} \log \Q_{N,\beta}(\mu_N) \left\{ X_N, \bEmp_N[X_N] \in B(\bP, \ep)\right\}
+\frac{1}{N\beta}\log \K_{\beta}(\mu_N') \) \\
\le \lim_{\ep \to 0} \limsup_{N\to \infty} \(
 \frac{1}{N\beta} \log \Q_{N,\beta}(\mu_N) \left\{ X_N, \bEmp_N[X_N] \in B(\bP, \ep)\right\}
+\frac{1}{N\beta} \log \K_{\beta}(\mu_N') \) \le  -   \overline \W(\bar P) .\end{multline}

Exponential tightness of $\Q_{N,\beta}(\mu_N)$ is an easy consequence of the fact that the total number of points in $\Sigma$ is bounded by $N$.  As explained in Corollary \ref{ldpboules} and using the result of Step~2, it allows to 
 upgrade these results into results about arbitrary sets $A\in \mathcal{P}_s(\Sigma\times \config) $:
 \begin{multline} - \inf_{\overset{\circ}A} \( \beta   \overline \W(\bar P) +\int_{\Sigma} \ERS[\bP^x, \Poisson^{\mu(x)}]\,dx\)\\  \le 
 \liminf_{N\to \infty} \(
 \frac{1}{N} \log \Q_{N,\beta}(\mu_N) \left\{ X_N, \bEmp_N[X_N] \in A \right\}
+\frac1N\log \K_{\beta}(\mu_N') \) \\
\le \limsup_{N\to \infty} \(
 \frac{1}{N} \log \Q_{N,\beta}(\mu_N)\left\{ X_N, \bEmp_N[X_N] \in A\right\}
+\frac1N\log \K_{\beta}(\mu_N') \) \\ \le  - \inf_{\overline{A}\cap \probas_{s,\mu}(\Sigma \times \config)}\(  \beta   \overline \W(\bar P) +\int_{\Sigma} \ERS[\bP^x| \Poisson^{\mu(x)}]\,dx\) \end{multline}
and respectively the same for the other regimes. 
Applying to  $A$ equals to the whole space, and using \eqref{scalingK}  we obtain \eqref{expacas1} in the regime of fixed $\beta$, or 
$$\lim_{N\to \infty} \frac{1}{N} \log \K_{N, \beta} (\mu_N) + \( \frac{1}{2\d} \log N\) \indic_{\s=0} = 0$$  in the regime $\beta \to 0$ (which we already knew), 
 resp. \eqref{expacas3} if $\beta \to \infty$.

Inserting into the above relations,  we have obtained the full LDP results.  Note that the "goodness" of the rate functions is a consequence of   Corollary \ref{corolscW}  and Lemma \ref{lem:ERS}.
\end{proof}


    \begin{proof}[Proof of Theorem \ref{theoldp}] 
Let us consider $P$ a stationary probability measure on infinite point configurations with intensity $\mu(x_0)$, and $B(P, \ep)$ a ball for some distance that metrizes the weak topology.
We focus on  proving upper and lower bounds on $\log \mathfrak{P}_{N,\beta}^{x_0,R}(B(P,\ep)) $.
For simplicity, let us denote $\carr_R$ for $\carr_R(N^{1/\d}x_0^N)$.
\smallskip

{\bf Step 1: reducing to good number of points and good energy.}
Since $R$ is large enough, we may include $\carr_R$ in a hyperrectangle $Q_R$ such that $\mn=\mu(Q_R)$ is an integer and 
$|Q_R|-|\carr_R|=O(R^{\d-1})=o(R^\d)$.

Let us denote by $n$ the number of points a configuration has in $Q_R$. 
Since we assume $R \gg \rb \ge C\max\( \beta^{-\hal} \chi(\beta)^{\hal},1\)$, for $\kappa$ small enough we have from \eqref{defchibeta} that $R^{2-3\kappa}\ge \chi(\beta)$ in all dimensions, hence in view of the local laws in the form~\eqref{loclawpoints00} and~\eqref{fini}    we may write  that  for some $\kappa>0$
\be 
\Q_{N,\beta}(\mu_N)  \left\{ |n-\mn|\ge R^{\d-\kappa}\right\} \le \exp\left( -C\beta  R^{\d+\kappa} \right)
\ee
and 
\be  \Q_{N,\beta}(\mu_N)  \left\{ \sup_x  \int_{\carr_{R^{1+\kappa/\d }} } |\nab u_{\rrc}|^2 \ge C \chi(\beta) R^{\d+\kappa} \right\} \le \exp\left( -\chi(\beta) \beta R^{\d+\kappa} \right)
\ee
for some $C$  large enough independent of $R$ and $\beta$.
Hence we may restrict the study to the event 
$$\mathcal B=\left\{ |n-\mn|\le R^{\d-\kappa},\  \sup_x  \int_{\carr_{R^{1+\kappa/\d}}} |\nab u_{\rrc}|^2 \le  \chi(\beta)R^{\d+\kappa}\right\},$$ since the complement has a probability which is negligible in the speed we are interested in. In particular, such configurations are screenable. \smallskip

{\bf Step 2: upper bound.} We recall that $P_N^{x_0^N,R}$ is defined in~\eqref{defii}. Using~\eqref{locali2}  and Lemma~\ref{lemrestri} we have
\begin{align*}
\lefteqn{ 
 \mathfrak{P}_{N,\beta}^{x_0,R}(B(P, \ep)\cap P_N^{x_0^N,R}[\mathcal B] )
 } \quad & 
\\ &
= \frac{1}{ N^N\K_{\beta}(\mu_N')} \int_{\{P_N^{x_0^N,R}  [ \XN]\in B(P, \ep) \}\cap \mathcal B} \exp\left( -\beta \F(\XN',\mu_N',\R^\d) \right)\, d(\mu_N')^{\otimes N}(\XN)
\\ & 
\leq  
\frac{1}{ N^N \K_{\beta}(\mu_N')}  \\ & \  \int_{\{ P_N^{x_0^N, R}  [\XN]\in B(P, \ep)\}  \cap \mathcal B} \exp\left(-\beta \G_{\R^\d}^{\mathrm{inn} } (X_N'|_{Q_R},\mu_N', Q_R)
-\beta \G_{\R^\d}^{\mathrm{out}}(\XN'|_{Q_R^c}, \mu_N',Q_R^c ) \right)\,  d(\mu_N')^{\otimes N}(\XN).
\end{align*}
Splitting up the events as in the proof of Theorem \ref{th3} with $n$ being the number of points of the configuration which belong to $Q_R$, and using that   $P_N^{x_0^N,R}[\XN]$ depends only on the configuration in $\carr_R$ hence in $Q_R$, we may then write 
\begin{align}
\label{fac1}
\lefteqn{
\mathfrak{P}_{N,\beta}^{x_0, R}(B(P,\ep) \cap P_N^{x_0^N,R}[\mathcal B])
} \quad & 
\\ & \notag
\le \frac{1}{N^N \K_\beta(\mu_N')} \sum_{n=\mathrm{n}-R^{\d-\kappa} }^{\mathrm{n}+ R^{\d-\kappa}} \frac{N!}{n! (N-n)!} 
 \int_{\mathcal B_{n} \cap  (Q_R^c)^{N-n} } \exp\left( -\beta \G_{\R^\d}^{\mathrm{out}} (\cdot,\mu_N', {Q_R^c} ) \right)\, d(\mu_N')^{\otimes (N-n)}   
\\ & \qquad \notag
\times 
\int_{(Q_R)^n\cap \{ P_N^{x_0^N, R}  [N^{-1/\d} X_n]\in B(P, \ep)\} } \exp\left( -\beta \G_{\R^\d}^{\mathrm{inn} }(\cdot ,\mu_N', Q_R) \right)\,  d(\mu_N')^{\otimes n},
\end{align}
where $\mathcal B_{n}$ is $\mathcal B$ intersected with the event that $X_N'$ has $n$ points in  $Q_R$.
Then,~\eqref{eqsulk} applied with $L$ such that $ R \gg L \gg \rho_\beta$ and combined with Remark \ref{remerr} yields
\begin{align*}
\lefteqn{
 \int_{\mathcal B_{n} \cap  (Q_R^c)^{N-n} }\exp\(-\beta \G_{\R^\d}^{\mathrm{out}}(\cdot,\mu_N', {Q_R^c} ) \) d(\mu_N')^{\otimes (N-n)}   
} \qquad & 
\\ & 
\leq  \frac{(N-n)! (N-\mn)^{N-\mn}}{(N-\mn)!} \K_\beta(\mu_N',Q_R^c) \exp\( C (\beta \chi(\beta) +1) o(R^{\d})  \),
\end{align*}
with $C$ independent of $\beta$.

We next  apply  Lemma \ref{Bernoulligen} in $Q_R$ with $m=\mu(x_0)$ to obtain that
\begin{multline}
\label{appliclem}
 \frac{1}{|Q_R|} \log \(\frac{1}{ \mn^n} (\mu_N')^{\otimes n}
 \{ X_n \in (Q_R)^n , P_N^{x_0^N, R}[N^{-1/\d} X_n] \in B(P, \ep)\}\)
\\= -   \ERS[ P|\Poisson^{\mu(x_0)} ]+o_\ep(1)+ o_N(1) .\end{multline}
 Technically, we used that the  restriction of $(\mu_N'/\int_{Q_R} \mu_N')^{\otimes n}$ to $Q_R$ is asymptotic to an $n$-point Bernoulli process, the fact that $\mn= \mu(Q_R)$ and $|n-\mn|=o(R^\d)$.  
We may rewrite this as 
\begin{multline}
\label{appliclem2}
  \log  (\mu_N')^{\otimes n}
 \{ X_n \in (Q_R)^n , P_N^{x_0^N, R}[N^{-1/\d} X_n] \in B(P, \ep)\}
\\= - R^\d  \ERS[ P|\Poisson^{\mu(x_0)} ] + n \log \mn +R^\d(o_\ep(1)+ o_N(1) ).\end{multline}

Moreover, the same proof as that of Proposition \ref{prop:LowerBoundenergies} 
(applied to $\G_{\R^\d}^{\mathrm{inn}}$ instead of $\F$) yields that 
$$\liminf_{N\to \infty}\frac{1}{|Q_R|} \G_{\R^\d}^{\mathrm{inn}}(X_n, \mu_N',Q_R) \ge \int \W(\C, \mu(x_0)) dP'(\C)$$
where $P'$ is the limit, up to extraction, of $P_N^{x_0^N, R}[X_n]$. 
In addition, from the
lower semi-continuity of Lemma \ref{corolscW}, we deduce with the fact that $|Q_R|=R^\d+o(R^\d)$,  that if $ P_N^{x_0^N, R}[X_n] \in B(P, \ep)$ then 
$$\liminf_{N\to \infty}\frac{1}{R^\d} \G_{\R^\d}^{\mathrm{inn}}(X_n,\mu_N', Q_R) \ge \int  \mathbb{W}(\C,\mu(x_0)) dP(\C) -o_\ep(1).$$ 
Denoting $\W(P,m)$ for $\int \W(\C,m) dP(\C)$, 
combining this with \eqref{appliclem2} and inserting them into~\eqref{fac1} leads to 
\begin{multline}\label{lad2} 
 \mathfrak{P}_{N,\beta}^{x_0,R}(B(P,\ep)\cap P_N^{x_0, R}(\mathcal B))\\
 \le \exp\(  -  R^\d  \(\beta \W (P,\mu(x_0)) +   \mathsf{ent}[P|\Pi^{\mu(x_0)} ]+(1+\beta) o_{\ep,N}(1)+ C \beta \chi(\beta) R^{-\kappa}    \) \)   \\ 
 \times\sum_{n=\mn-R^{\d-\kappa}}^{\mn +R^{\d-\kappa}}  \frac{1}{N^N \K_\beta(\mu_N',\R^\d)}   \frac{N!}{ (N-\mn)!} 
     \frac{(N-\mn)^{N-\mn} \mn^n}{n!}\K_\beta( \mu_N', Q_R^c) 
.\end{multline}
On the other hand using~\eqref{superad2}, we have 
$$\K_\beta(\mu_N') \ge \frac{N! N^{-N}}{\mathrm{n}!(N-\mathrm{n})!  \mn^{-\mn}(N-\mn)^{-(N-\mn)}} \K_\beta(Q_R, \mu_N')\K_\beta ( \mu_N', Q_R^c),$$
 and inserting this into~\eqref{lad2},
we find 
\begin{align*}
\lefteqn{
\mathfrak{P}_{N,\beta}^{x_0,R}(B(P,\ep) \cap P_N^{x_0^N,R}(\mathcal B)) 
} \quad & 
\\ & 
\leq
\sum_{n=\mn-R^{\d-\kappa}}^{\mn +R^{\d-\kappa}} 
  \exp\(  - R^\d  \(\beta \mathbb{W}(  P,\mu(x_0)) +   \mathsf{ent}[P|\Pi^{\mu(x_0)} ]+(1+\beta)o_{\ep,N}(1)\)\)  \frac{\mn!}{n!} \mn^{n-\mn}\frac{1}{   \K_\beta( \mu_N', Q_R)} .
\end{align*}
We note that $\log \mn^{n-\mn}= O(R^{\d-\kappa}  \log R)=o(R^\d)$ and similarly  $\log \frac{\mn!}{n!} =o(R^\d)$  for $|n-\mn|\le R^{\d-\kappa}$, hence bounding the sum by the number of terms which is $O(R^{\d-\kappa})$ and reabsorbing this factor into the errors, we find the upper bound
\begin{multline} \label{bornesupldp}
\log 
\mathfrak{P}_{N,\beta}^{x_0,R}(B(P,\ep) \cap P_N^{x_0^N,R}[\mathcal B]) 
 \\
\leq
  - R^\d  \(\beta\mathbb{W}  (P,m) +   \mathsf{ent}[P|\Pi^m]+(1+\beta) o_{\ep, N} (1) \)- \log \K_\beta(\mu_N',Q_R)
\end{multline}
where we used that $ R \gg \rho_\beta$.
\smallskip

{\bf Step 3: lower bound.} We claim that given any $P $ such that 
$  \mathbb{W} (P,\mu(x_0)) +   \mathsf{ent}[P|\Pi^{\mu(x_0)}]$ is finite, we can construct a family $A$ of  configurations $X_{\mn}$ of $\mn$ points in $Q_R$ such that $P_N^{x_0^N,R}[X_{\mn}] \in B(P, \ep)$,  
\be \label{df1}\F(X_{\mn},\mu_N',Q_R) \le R^\d  \mathbb{W}(P,\mu(x_0))+o(R^\d)\ee
uniformly in $A$, and 
\be\label{df2}
 \log ( \mu_N')^{\otimes \mn}
 (A) 
\\= - R^\d  \ERS[ P|\Poisson^{\mu(x_0)} ] + \mn \log \mn +R^\d(o_\ep(1)+ o_N(1) )
.\ee
This follows the same steps as the lower bound in the proof of Theorem \ref{theoldpglob}, i.e.~it is done by sampling configurations whose local empirical field $P_N^{x_0^N, R}$ is close to $P$ and screening them in $Q_R$ which still keeps $P_N^{x_0^N, R}$  close to $P$.
It  is however an easier setting since we do not have to partition into rectangles.

We may thus write with the help of~\eqref{subad1}
\begin{align*}
\lefteqn{
\mathfrak{P}_{N,\beta}^{x_0^N,R}(B(P,\ep))
} \quad & 
\\ & 
= \frac{1}{N^N \K_\beta(\mu_N')} \int_{P_N^{x_0^N,R}(\XN) \in B(P, \ep) } \exp\left( -\beta \F(\XN,\mu_N',\R^\d) \right)\, d(\mu_N')^{\otimes N}(X_N)
\\ & 
\ge \frac{1}{N^N \K_\beta(\mu_N') } \frac{N!}{\mn! (N-\mn)! }\int_{(Q_R^c)^{N-\mn}}  \exp\left( -\beta \F(\cdot,\mu_N', Q_R^c) \right) d(\mu_N')^{\otimes (N-\mn)}(X_N)
\\ & \qquad \times \int_{A } \exp\left( -\beta \F(X_\mn, \mu_N', Q_R) \right)\, d(\mu_N')^{\otimes {\mn}}(X_\mn)
 \\ & 
 = \frac{\K_\beta(\mu_N',Q_R^c) }{N^N \K_\beta(\mu_N') } \frac{N! }{\mn! (N-\mn)!(N-\mn)^{-(N-\mn)} }\int_{A } \exp\left( -\beta \F(X_\mn, \mu_N', Q_R) \right)\, d(\mu_N')^{\otimes {\mn}}(X_\mn).
\end{align*}
But in view of  ~\eqref{subad3} we have 
$$\log \K_\beta(\mu_N')= \log \K_\beta ( \mu_N', Q_R)+ \log \K_\beta(\mu_N', Q_R^c)+o((1+\beta \chi(\beta) ) R^{\d})$$
so in view of \eqref{df1} and \eqref{df2} and  using Stirling's formula, we conclude that 
\begin{multline}\label{borninfldp}
\log \mathfrak{P}_{N,\beta}^{x_0,R}(B(P,\ep))
\\
\ge  - \log  \K_\beta (\mu_N', Q_R) - R^\d\(\beta \mathbb{W}(P,\mu(x_0)) 
+  \mathsf{ent}[P|\Pi^{\mu(x_0)} ]  +(1+\beta)o_{\ep,N}(1)\).\end{multline}

{\bf Step 4: conclusion}.
Exponential tightness at speed $R^\d$ follows from the fact that the number of points is essentially bounded by $CR^\d$ by Theorem \ref{th3}. Then by
 Corollary \ref{ldpboules}, we may upgrade the conclusions of the previous steps to a strong LDP result: for any Borel set $E$, it holds that, as~$N\to \infty$, 
\begin{align}
\label{concldp1}
\lefteqn{
\log \mathfrak{P}_{N,\beta}^{x_0,R}(E)
} \ \ & 
\\ & \notag
\leq 
-  R^\d \inf_{P\in \bar E} \(   \beta \mathbb{W}(P,\mu(x_0))  + \mathsf{ent}[P|\Pi^{\mu(x_0)} ] \) 
- \log \K_\beta( \mu_N', Q_R)   +(1+\beta) o(R^\d)
\end{align}
and
\begin{align}
\label{concldp2}
\lefteqn{
\log \mathfrak{P}_{N,\beta}^{x_0,R}(E)
} \ \ & 
\\ & \notag
\geq 
- R^\d\inf_{P\in\overset{\circ}{E}}  \(   \beta \mathbb{W}(P,\mu(x_0))  + \mathsf{ent}[P|\Pi^{\mu(x_0)} ] \)   - \log \K_\beta( \mu_N', Q_R)  +(1+\beta) o(R^\d).
\end{align}
Applying this relation to $E$ equal the whole space, we find 
$$  \log \K_\beta( \mu_N', Q_R)= -R^\d  \inf_{P\in \probas_s(\config) } \(  \beta \mathbb{W}(P,\mu(x_0))  + \mathsf{ent}[P|\Pi^{\mu(x_0)} ] \) +(1+\beta) o(R^\d) .$$
Reinserting into~\eqref{concldp1} and~\eqref{concldp2}, the stated LDP result follows if $\beta$ is fixed. The generalization to $\beta \to 0$ or $\beta\to \infty$ is straightforward from~\eqref{bornesupldp} and~\eqref{borninfldp}.
 This concludes the proof of Theorem \ref{theoldp}.
\end{proof}

  \appendix
\renewcommand{\theequation}{A.\arabic{section}.\arabic{equation}}


\chapter{Proof of the screening result}\label{appa}
\index{screening} 
The goal of this appendix is to prove the screening result of  Proposition \ref{proscreen}, we follow here closely \cite[Appendix C]{as}.
This follows from  adapting  and  optimizing  the procedure from \cite{ss1,rs,PetSer}, in particular \cite{PetSer} simplified to the Coulomb case.

Let us  first describe things informally, for the inner screening. 
Let $w$ solve \eqref{eqsp} and let $E=\nab w$ be the associated electric field,  which satisfies a relation of the form 
\be\label{dive}
\left\{\begin{array}{ll}
-\div E= \cd\(\sum_{i=1}^n \delta_{x_i}-\mu\) \quad &\text{in} \ \Omega''\\
E\cdot \nu= 0 & \text{on} \ \partial U\cap \Omega'' .\end{array}\right.\ee
Its truncated version of $E$ is defined  as in~\eqref{eer} by 
\be \label{eer2}E_{\rrh}= E- \sum_{i=1}^n \nab \f_{\rrh_i}(x-x_i)\ee
where $\rrh_i$ is as in~\eqref{defrrc3}. The precise choice of truncation is unimportant, we may use $\rr, \rrc$ or $\rrh$, since the configuration will be deleted in the boundary layer where  the definitions differ.

Given a configuration $X_n$  in $\Omega'$, together with its electric field $E$, and assume roughly that we control well its energy near the boundary of a hyperrectangle $Q_t$ with $t$ close to $R$. The goal of the screening is to modify the configuration $X_n$ and the electric field $E$ only outside of $Q_{t-1}$ and to extend them to a screened configuration $X^0$ and a screened electric field $\Escr$ in $\Omega= Q_R \cap U$  in such a way that 
$$\left\{ \begin{array}{ll}
- \div \Escr= \cd(\sum_{p\in X^0} \delta_p - \mu) & \text{in} \ \Omega\\
\Escr \cdot \nu =0 & \text{on} \ \partial U \cap \Omega\end{array}\right.$$
This implies in particular that the screened system is neutral, i.e the number of points of $X^0$ must be equal to $\mu(\Omega)$, an integer. We note that in the Neumann case where $\Omega$ can intersect $\partial U$, the desired boundary condition is already satisfied for the original field on $\partial U$, so there is no need to modify it near $\partial U$.

 The screened electric field $\Escr$ may not be a gradient, however thanks to Lemma \ref{projlem} its energy provides an upper bound for computing $\F(X^0, \Omega)$. The goal of the construction is to show that we can build $\Escr$ and $X^0$ without adding too much energy to that of the original configuration, which will allow to bound $\F(X^0, \Omega ) $ in terms of  $\G_U^{\mathrm{inn}}(X, \Omega)$.
In order to accomplish this, we will split the region to be filled into cells where we  solve appropriate elliptic problems and estimate the energies by elliptic regularity estimates. In order to ``absorb" and screen the effect of the possibly rough data on $\partial Q_t$, we need a certain distance $\l$, which has to be large enough in terms of the energy of $E$, this leads to the ``screenability condition" bound on $\l$, as previously mentioned.

\subsection{Finding a good boundary}

We focus on the outer screening proof,  the proof of the inner case is analogous (for details  of what to do near the corners, one may refer to \cite{rns}).

 Assume then that $\Omega=Q_R\cap U$. Since $U$ is assumed to be a disjoint union of parallel hyperrectangles, $\Omega$ is itself a hyperrectangle. 
 
We are given  $\tilde \l \ge \l \ge C$, a configuration $X_n$ in $\Omega'$, $E=\nab w$ with the notation $E_{\rrh}$ defined in~\eqref{eer2}.  
We recall  there are two variants of the construction depending on which term ensures the screenability condition is met.

     In the first case, 
 by a mean value argument we can find $\Gamma= \partial Q_t$ for some $t \in 
[ R-2\tilde \ell +2,R-\tilde \ell-2] $ such that 
\begin{equation}\label{bonbord}
\int_{\Gamma \cap U } |E_{\rrh}|^2 \le  \frac{S(X_n)}{\tilde \ell}
\end{equation}
and 
\begin{equation}\label{defMt}
  \int_{ (Q_{t+2}\backslash Q_{t-2})  \cap U}  |E_{\rrh}|^2 \le C   \frac{S(X_n)}{\tilde \ell} . \ee



    In the second case, using a mean-value argument we can find 
  $  t \in [R- 2\tilde \ell , R- \tilde \ell-\ell]$ such that 
  \be \label{bonbord1} \int_{ (Q_{t+\ell}\backslash Q_{t})  \cap U}  |E_{\rrh}|^2 \le C \frac{S(X_n)\ell}{\tilde \ell}\end{equation} and then, by a 
    covering argument and a mean-value argument in the strip $Q_{t+\ell}\backslash Q_t$, 
we can find a piecewise affine  boundary $\Gamma$, included in $Q_{t+\ell}\backslash Q_{t}$ for some $t \in [R- 2\tilde \ell +\ell+1, R- \tilde \ell-1]$, with faces parallel to those of $Q_R$, of sidelengths bounded above and below by constants times $\ell$, 
 such that 
\be\label{bonbord2}\int_{\Gamma\cap  U} |E_{\rrh}|^2 \le C \frac{S(X_n)}{\tilde \ell},
\qquad \sup_x \int_{\Gamma\cap U \cap \carr_\ell(x)} |E_{\rrh}|^2 \le C S'(X_n)   \end{equation}
   and 
   \be \label{bonbord3}\int_{\Gamma_1} |E_{\rrh}|^2 \le  C\frac{S(X_n)}{\tilde \ell},
 \ee
where $\Gamma_1$ denotes the $1$-neighborhood of $\Gamma$.

In both cases, we let $M= C \frac{S(X_n)}{\tilde \ell}$, and in the second case we let  $M_\ell= C S'(X_n)$, for the  largest $C$ appearing in the right-hand side.

We note that as soon as $\tilde \ell$ is large enough,   we only consider regions at distance $ \ge 1$ from $\pa \Omega$, so there is no difference between $\rrh$ and $\rrc$ there.

   
We denote by~$\mathcal O$ (like ``old")  the part of $\Omega $ delimited by $\Gamma$ and $\partial U$,  and by $\New$ (like ``new") the set $\Omega \backslash \Old$.  By construction, we have $\Old\subset\Omega''$. We keep $X_n$ and $E$ unchanged in $\Old$ and discard the points of $X_n$ in $\Old^c$ to replace them by new ones.
The good boundary $\Gamma$ may intersect some $B(x_i,\rrh_i)$ balls centered at points of $X_n$. These balls will need to be ``completed", i.e.,~the contributions of $\delta_{x_i}^{(\rrh_i)} \indic_{\New}$  retained.

\subsection{Preliminary lemmas}
We start with a series of preliminary results which will be the building blocks  for the construction of $\Escr$. 

\begin{lem}[Correcting fluxes on rectangles]\label{lem57}
Let~$H$ be a hyperrectangle of~$\R^\d$ with sidelengths in~$[\ell, C\ell]$ with~$C$ depending only on~$\d$.
 Let~$g\in L^2(\pa  H) $.
Then there exists a constant~$C$ depending only $\d$ such that the mean zero  solution of
\begin{equation}\label{eqnu}
\left\{\begin{array}{ll}
-\Delta h  = \int_{\partial H} g    & \text{in} \  H\\
\frac{\partial h }{\pa \nu}=g & \text{on} \ \pa H \end{array}\right.
\end{equation}
satisfies the estimate
\begin{equation}\label{estlcs2}
\int_{ H}  |\nab h|^2 \le  C \ell  \int_{\pa   H} |g|^2.
\end{equation}
\end{lem}
\begin{proof}
This is   \cite[Lemma 5.8]{rs}. 
\end{proof}
 
The next lemma serves to complete the smeared charges  which were cut into two pieces by the choice of the good boundary. 

\begin{lem}[Completing charges near the boundary]\label{chargesnearbdry}Let $\mathcal{R}$ be a hyperrectangle in $\R^\d$ of center $0$ and sidelengths in $[a,Ca]$ with $C$ depending only on $\d$. Let $F$ be a face of $\mathcal R$. Let $\{x_i\}_{i\in I}$ be points  contained in an $1/4$-neighborhood of $F$. Let $c$ be a constant such that
 \begin{equation}\label{CR}
c |F|=\cd\int_{{\mathcal R}}\sum_{i\in I } \delta_{x_i}^{(\rrh_i)}\,.
 \end{equation}
The mean-zero solution to 
\be \label{eqhuv}
 \left\{\begin{array}{ll}
  -\Delta h=\cd \sum_{i\in I}\delta_{x_i}^{(\rrh_i)}& \text{ in }{\mathcal R}\, ,\\[2mm]
  \frac{\pa h}{\pa \nu}=0& \text{ on }\pa {\mathcal R}\setminus  F\, ,\\[2mm]\frac{\pa  h}{\pa \nu}=c& \text{ on } F
\end{array}
\right.
\end{equation}
satisfies 
\begin{equation}\label{estboundary}
 \int_{{\mathcal R}}|\nab h|^2\le C \( (\# I)^{2} a^{2-\d} + \sum_{i\neq j } \g(x_i-x_j) +\sum_{i\in I} \g(\rrh_i)\)
\end{equation} where $C$ depends only on $\d$ and $a$.
\end{lem}
\begin{proof}Integrating  \eqref{eqhuv} over $\mathcal R$, we find that $c|F|\le \cd \#I$ hence $c\le C \#I a^{1-\d}$.

We then split $h=u+v$ where 
$$ \left\{\begin{array}{ll}
  -\Delta u=\cd \sum_{i}\delta_{x_i}^{(\rrh_i)} - c \frac{|F|}{|\mathcal R|}& \text{ in }{\mathcal R}\ \\[2mm]
  \frac{\pa u}{\pa \nu} =0& \text{ on }\pa {\mathcal R},\end{array}
\right.$$
and
 $$ \left\{\begin{array}{ll}
  -\Delta v=c \frac{|F|}{|\mathcal R|}& \text{ in }{\mathcal R}\ \\[2mm]
  \frac{\pa v}{\pa \nu}=0& \text{ on }\pa {\mathcal R}\setminus  F\ \\[2mm]
\frac{\pa v}{\pa \nu}=c& \text{ on } F.
\end{array}
\right.$$
The $v$ part is explicitly computable and has energy bounded by $C c^2 a^\d\le C (\# I)^2 a^{2-\d} $.
For the $u$ part, we observe that 
$$u=\cd\sum_{i\in I}\int G_{\mathcal R} (x,y) \delta_{x_i}^{(\rrh_i)}(y)$$ where $G_{\mathcal R}(x,y)$ is the Neumann Green function of the 
hyperrectangle with background $1$. From  \cite[Proposition A.1]{as}, we have 
$$G_{\mathcal R} (x,y) \le C \g(x-y)$$ hence we deduce the result.
\end{proof}

\subsection{Main proof}


\smallskip 
\noindent

We let $I_\pa $ be  the indices corresponding to the points of $X_n$ whose smeared charges touch $\Gamma$, i.e.
\begin{equation}\label{defl0}
 I_{\pa}=\left\{ i \in[1,n] :\ B(x_i,\rrh_i)\cap \Gamma \neq\varnothing\right\} 
\end{equation}
and define $$\N= \# I_{\pa} + \#\( \{i, x_i \in \Old\}\backslash I_{\pa}\).$$
The goal of the construction is to place an additional  
$\mn- \N$ points in $(Q_R\cap U)\backslash \Old$, where $\mn=\mu(Q_R\cap U)$, while leaving a point-free zone of thickness $\eta$.

By construction of $\Gamma$, we may  partition $(Q_R\cap U)\backslash \Old$ into hyperrectangles $H_k$ with sidelengths $\in [\ell/C,C\ell]$ for some positive constant $C>0$, and given $\eta\ge 0$, we let $H_k^\eta$ denote $\{x\in H_k, \dist(x, \Gamma) \ge \eta\}$. We build the $H_k$'s 
 in such a way that, 
 letting $m_k$ be the constant such that 
 \begin{equation}\label{defmi}
  m_k |H_k^\eta| = \frac1{\cd} \(  \int_{\Gamma \cap \pa  H_k}   E_{\rrh} \cdot{ \nu}-n_k\) +\int_{H_k\backslash H_k^\eta}\mu
 ,\end{equation}  with  $\nu$ denoting the outer unit normal to $\Old$ and $$n_k:= \cd\int_{H_k} \sum_{i \in I_\partial  } \delta_{x_i}^{(\rrh_i)},$$
 we have $\int_{H_k^\eta} (\mu+m_k ) \in \mathbb{N}$.
 This is possible if $|m_k|<\hal m$ (recall $\mu \ge m$) and  can be done by constructing successive strips as in Lemma \ref{tiling}, as soon as $\l>C>2\eta $ for some $C>0$ depending only on $\d$ and $m$.

  We will give below a condition for $|m_k|<\hal m  $.  Now define 
 \be \label{deftmu}
  \tilde \mu=\( \mu  \indic_{\{\dist(x, \Gamma) \ge \eta\}}+ \sum_k \indic_{H_k^\eta} m_k\).\ee It is a nonnegative density supported in $\New_\eta$.
  Since $$\N=-\frac{1}{\cd}\int_{\Gamma} E_{\rrh} \cdot \nu+ \frac1\cd \sum_k n_k +\int_{\Old } d\mu$$ and $\mn=\mu(\Omega)$, 
  in view of 
 ~\eqref{defmi}  we may check that 
  \be \label{intmut} \int_{\New}\tilde \mu= \int_{\New_\eta}\tilde\mu= \mn-\N.\ee
      \noindent
 {\bf Step 1: Defining $ \Escr$}.\\
 We define $\Escr$ by adding to $E$ a sum $ E_1 + E_2 +E_3$, some of these terms being zero except for $H_k$ that has some boundary in common with $\Gamma$, then denoted $F_k$. 
 
The first vector field contains the contribution of the completion of the smeared charges belonging to $I_{\pa}$.
We let
 $$E_{1}:=\sum_k \indic_{H_k} \nab h_{1,k}$$ where $h_{1,k}$ is the solution of
\begin{equation}\label{defh1}
\left\{\begin{array}{ll}
 -\Delta h_{1,k}=\cd \sum_{i\in I_\pa}\delta_{x_i}^{(\rrh_i)}& \text{ in } H_k
,\\[3mm]
 \frac{\pa h_{1,k}}{\pa \nu} =0& \text{ on }\pa H_k \setminus \Gamma \ ,\\[3mm]
 \frac{\pa h_{1,k}}{\pa \nu}=\frac{- n_k}{| F_k|  }    &\text{ on } F_k,
\end{array}
\right.
\end{equation}
We note that the definition of $ n_k$ makes this equation solvable.

The second vector field is defined to be $E_{2}= \sum_{k} \indic_{H_k} \nab h_{2,k}$ with  
\begin{equation*}
\left\{\begin{array}{ll}
   -\Delta  h_{2,k}= \cd m_k &  \text{ in } H_k\ ,\\[3mm]
  \frac{ \pa h_{2,k}}{\pa \nu}= g_k
  &\text{ on }\pa  H_k ,
\end{array}
\right.
\end{equation*} where we let $g_k=0$ if $ H_k$ has no face in common with $\Gamma$ and otherwise 
\begin{equation}\label{gi} g_k= - E_{\rrh}\cdot {\nu}    + \frac{n_k}{|F_k|} 
 \end{equation}
with $E_{\rrh} \cdot \vec{\nu}$ taken with respect to the outer normal to $\Old$.  We note that this is solvable in view of~\eqref{defmi}.

The third vector field consists in the potential generated by a sampled configuration $Z_{\mn - \N}$ in $\New_\eta$:  we let $E_3= (\nab h_3)\indic_{\New_\eta}$ where $h_3$ solves 
 \begin{equation}\label{defh4}
 \left\{\begin{array}{ll}
   -\Delta  h_3=\cd \left(\sum_{j=1}^{\mn-\N} \delta_{z_j}- \tilde \mu\right)& \text{ in } \New_\eta 
   \\[3mm]
 \frac{  \pa h_3}{\pa \nu}=0& \text{ on }\pa \New_\eta .
\end{array}
\right. 
\end{equation}

We note that this equation is solvable since~\eqref{intmut} holds.
We then define
$$
 \Escr=( E_1+E_2+E_3)\indic_{\New}+ E_{\rrh}\indic_{\Old}
+ \sum_{i,  B(x_i, \rrh_i) \cap \Old \neq \varnothing} \nab \f_{\rrh_i}(x-x_i)  
 $$
and  $Y_\mn= \{X_n, B(x_i, \rrh_i) \cap \Old \neq \varnothing\} \cup \{Z_{\mn-\N}\}$. 

We then let  $\bar\rr_i$ are the minimal distances as in~\eqref{defrrc} of $Y_\mn$. Note that  for the points near $\Gamma$, these may not correspond to the previous minimal distances for the configuration $X_n$ or $Z_{\mn-\N}$, which is why we use a different notation.

We note that  the normal components are always constructed to be continuous across interfaces,  so that no divergence is created there, and so, since $\Old\subset \Omega''$ where $w$ satisfies \eqref{eqsp},  
   $\Escr$ thus defined satisfies 
\be\label{divescr} \left\{\begin{array}{ll} - \div \Escr= \cd(\sum_{i\in Y_\mn} \delta_{y_i}-\mu) \quad & \text{in} \ \Omega\\
\Escr\cdot \nu= 0 & \text{on} \ \partial \Omega.\end{array}\right.
\ee

\smallskip 

\noindent
{\bf Step 2: Controlling $m_k$.}
First we control the $ n_k$.  Note that $$n_k\le n_k' :=\cd  \# \{i, B(x_i,\rrh_i) \cap H_k \neq \varnothing\},$$ and $n_k' \le (n_k')^2$ since $ n_k'$ is an integer. 
The results of Lemma \ref{coronp}  and \eqref{defMt} or \eqref{bonbord3}  allow to show that 
\begin{equation}\label{nalpha}n_k^2\le  (n_k')^2  \le C \int_{H_k}|E_{\rrh}|^2
  \le  C M_\l,\qquad   \sum_k n_k\le \sum_k (n_k')^2  \le C M. \end{equation}
We note that it follows that \be \# I_\pa \le \sum_k n_k' \le CM \le C \frac{S(X_n)}{\tilde \ell}.\ee
To control $m_k$ we write that in view of~\eqref{defmi},
\begin{equation}\label{a21}
|m_k| \le C \ell^{-\d}    \int_{\Gamma \cap \pa  H_k}  |E_{\rrh}| +|n_k|\l^{-\d}+\eta\ell^{-1} \|\mu\|_{L^\infty}  . \end{equation}
Using the Cauchy-Schwarz inequality  and~\eqref{bonbord} or \eqref{bonbord2},   we  bound 
\be\label{boundEecr}  \int_{\Gamma\cap \pa H_k} |E_{\rrh}|\le \ell^{\frac{\d-1}{2}}M_\l^{\hal}, \qquad  \sum_k\int_{\Gamma\cap \pa H_k} |E_{\rrh}|\le R^{\frac{\d-1}{2}}M^{\hal}.\ee
Combining with  \eqref{nalpha}, we conclude that 
\be \label{bmi}|m_k| \le C \ell^{-\frac{\d}{2}-\frac{1}{2}}M_\l^{\hal}+C \ell^{-\d} M_\ell^\hal + C \eta\ell^{-1}.\ee
The condition $|m_k|<\hal m$ is thus  implied by 
$$
C M_\l^{\hal} \l^{\frac{-  \d-1}{2}} <  \frac14 m \quad \text{and} \  \eta<\frac{1}{4\|\mu\|_{L^\infty}} \ell m.$$
The first condition is the second case of the screenability condition~\eqref{screenab}. The second is the condition on $\eta$.

As an alternate,  starting from \eqref{deftmu} and using \eqref{boundEecr},  \eqref{nalpha}, we can also bound 
\begin{equation}\left|\int_{\New} \mu - \tilde \mu\right| =
\frac1{\cd}\left|\sum_k  \(  \int_{\Gamma \cap \pa  H_k}   E_{\rrh} \cdot{ \nu}-n_k\) \right|
\le  CR^{\frac{\d-1}{2}} M^{\hal} + C  M 
\le C R^{\d-1}+ C \frac{S(X_n) }{ \tilde \ell}, \end{equation} using Young's inequality, \eqref{bonbord} or \eqref{bonbord2}, thus completing the proof of~\eqref{bornimp}.
In the same way,
we have 
$$\int_{\New_\eta} (\mu - \tilde \mu)^2=  \sum_k m_k^2 |H_k^\eta|,$$
 while, using Cauchy-Schwarz, we may also write  that 
$$ m_k^2 \le C \ell^{-2\d} \int_{\Gamma\cap \pa H_k} |E_{\rrh}|^2 \ell^{\d-1}+   Cn_k^2 \ell^{-2\d} +C \eta^2 \ell^{-2}$$
and thus using again \eqref{nalpha}, \eqref{bonbord} or \eqref{bonbord2}, we obtain 
$$\int_{\New} (\mu - \tilde \mu)^2 \le  C \ell^{-1} \int_{\Gamma} |E_{\rrh}|^2 + M \ell^{-\d} + C \eta^2\ell^{\d-2} \frac{R^{\d-1}}{\ell^{\d-1}}  \le  C \frac{S(X_n)}{\tilde \ell \ell}+ C \eta^2\ell^{-1} R^{\d-1}$$
 thus proving~\eqref{mmut2}.

 \smallskip

\noindent
{\bf Step 3: Estimating the energy of $\Escr$}.
To estimate the energy of $\Escr$ we need to evaluate $\int_{\Omega} |\Escr_{ \rrh}|^2$.
First, for $E_1$ we use  Lemma \ref{chargesnearbdry} and combine it with~\eqref{pre11} applied with $\alpha_i=\frac14 $ to bound $\sum_{p\neq q} \g(p-q)$ by the energy in a slightly larger set, thus we are led to  
\begin{equation*}
\int_{\New}  |(E_1)_{\rrh}|^2 \le C\(\sum_k  (n_k')^2  + CM\) \le C M,
\end{equation*}
where we have used~\eqref{defMt} or \eqref{bonbord3}, ~\eqref{nalpha}, and the geometric properties of 
$H_k$.

\smallskip

For $E_{2}$ we use 
 Lemma \ref{lem57} to get 
\begin{equation*}
 \int_{H_k} |E_{2} |^2 \le 
 C\l
\(\int_{\partial H_k\cap \Gamma} |E_{\rrh}|^2+Cn_k^2 \). 
\end{equation*}
Summing over $k$ and using~\eqref{bonbord} and \eqref{nalpha}, we obtain 
\begin{equation*}
 \sum_k \int_{H_k} |E_{2} |^2 \le 
  C\l M    .  \end{equation*}
  For $E_3$ we  use that,  by definition of $\F$  and using \eqref{eq:intf},
  \be \label{bornh3} \int_{\New_\eta } |\nab h_{3,\rrh}|^2 \le 2\cd  \F(Z_{\mn-\N}, \tilde \mu, \New_\eta)  + \cd \sum_{j=1}^{\mn-\N} \g(\rrh_j)    + C (\mn-\N) \ee 
  with the $\rrh_j$ defined relative to $\New_\eta$. In view of \eqref{premono2}, 
  the inequality still holds when $\rrh_j$ are replaced by larger balls.
     
We deduce that 
 \be
\int_{\Omega}|\Escr_{\rrh}|^2  
 \leq \int_{\mathcal{O} } |\nab w_{\rrh}|^2 + C\l  M   
 +  2\cd\F(Z_{\mn-\N}, \tilde \mu, \New)   + \cd \sum_{j=1}^{\mn-\N} \g(\rrh_j)
 + C (\mn-\N) .
 \ee
To estimate $\F(Y_{\mn}, \mu, \Omega)$ we use Lemma \ref{projlem}, the definition of $\F$ and \eqref{premono2}, which tells us that to go from the $\rr_i$ and $\rrh_j$  which lead to possibly intersecting balls,  to $\bar \rr$ the minimal distances of $Y_\mn$, we just  need to add  the new interactions $\sum_{(i,j) \in J} \g(x_i-z_j) $.
This yields 
\begin{align*}
\F(Y_\mn,\Omega)
&\leq
 \frac{1}{2\cd} \int_{\mathcal{O} } |\nab w_{\rrh}|^2    - \hal \sum_{i=1}^\mn \g( \rrh_i) -\sum_{i=1}^\mn \int_\Omega \f_{ \bar \rr_i}(y-y_i) d\mu(y)
+C
 \sum_{(i,j) \in J} \g(x_i-z_j)  \\ & \quad
   + C\l  M   
 + \F(Z_{\mn-\N}, \tilde \mu, \New) 
 +\frac{1}{2}\sum_{j=1}^{\mn-\N} \g(\bar \rr_j) + C(\mn-\N).
\end{align*}

It follows that 
\begin{align}
\label{concs}
\lefteqn{
\F(Y_\mn,\mu,\Omega)-\(  \frac{1}{2\cd} \int_{\Omega'}|\nab w_{\rrh}|^2- \hal\sum_{i=1}^{n} \g(\rrh_i) - \sum_{i=1}^n \int_{\Omega'} \f_{\rrh_i}(x-x_i) d\mu(x)\)} \quad & 
\\ & \notag
\leq - \frac{1}{2\cd}\int_{\Omega'\backslash \Old} |\nab w_{\rrh}|^2  +\hal \sum_{\{ i\in \{1,\ldots,n\}\,:\, x_i \notin \Old\}} \g(\rrh_i)  +
 C\sum_{(i,j) \in J} \g(x_i-z_j)+C \l M
\\ & \notag \quad  
+  \cd\F(Z_{\mn-\N}, \tilde \mu, \New)+ C (n-\N)+ C(\mn-\N).
 \end{align} 
On the other hand, since $\mathcal O $ contains $Q_{t-2}\cap \Omega $, we have in the first screening situation
\begin{align}\label{alig3}
\lefteqn{
\frac{1}{2\cd}\Bigg( -\int_{\Omega'\backslash \mathcal O} |\nab w_{\rrh}|^2+ \cd \sum_{\{ i\in \{1,\ldots,n\}\,:\, x_i \notin \Old\}}  \g(\rrh_i)\Bigg)  
}   \qquad  &
\\  \notag & 
\leq \frac{1}{2\cd} \int_{(Q_{t+2}\backslash Q_{t-2})\cap U } |\nab w_{\rrc}|^2 + \frac{1}{2\cd}\Bigg( \cd \sum_{\{ i\in \{1,\ldots,n\}\,:\, x_i \notin \Old\}}  \g(\rrh_i) -\int_{\Omega'\backslash Q_{t-2}} |\nab w_{\rrh}|^2\Bigg)
\\  \notag & 
\leq   \frac{M}{2\cd}+ C (n-\N), 
\end{align} 
where we bounded the second term in the right-hand side by  using \eqref{premono2}  to change $\rrh$ into $\frac14$ and then bounded $\sum \g(\frac14)$ for $x_i \notin \Old$  by the number of points not in $\mathcal O$.  In the second situation, we replace $Q_{t+2}\backslash Q_{t-2}$ by $Q_{t+\ell}\backslash Q_t$ and  use \eqref{bonbord1} instead.

Inserting \eqref{alig3}  into \eqref{concs} and using \eqref{defMt}, we find in all cases that 
\begin{multline*}
\F(Y_\mn,\mu,\Omega)-\(  \frac{1}{2\cd} \int_{\Omega'}|\nab w_{\rrh}|^2- \hal\sum_{i=1}^{n} \g(\rrh_i) - \sum_{i=1}^n \int_{\Omega'} \f_{\rrh_i}(x-x_i) d\mu(x)\)
\\ \le  C\l \frac{S(X_n)}{\tilde \ell} + C \F(Z_{\mn-\N}, \tilde \mu, \New) +
 C\sum_{(i,j) \in J} \g(x_i-z_j)+C( |n-\mn|+ |\mn-\N|).\end{multline*}
Using~\eqref{bornimp} and $\mu (\New) \le C \tilde \ell R^{\d-1}$  allows to  bound the last term on the right side,  and then we get~\eqref{nrjy}.





\newpage
 \begingroup
 
 \printindex

\bibliographystyle{alpha}
\bibliography{notescours2020}
\endgroup

\end{document}